\documentclass{aastex61}
\usepackage{lineno}
\usepackage{xspace}
\usepackage{soul}
\usepackage{color}
\usepackage{longtable}
\usepackage{appendix}
\usepackage{quoting} %
\usepackage{tocbasic}
\DeclareTOCStyleEntry[
  beforeskip=.9em plus 1pt,
  pagenumberformat=\textbf
]{tocline}{section}

\quotingsetup{font={bf}, leftmargin=3em, rightmargin=0in, vskip=1ex}

\shorttitle{3rd TDAMM Workshop White Paper}

\journalinfo{}

\begin{document}
\title{Multidisciplinary Science in the Multimessenger Era}

\author[0000-0002-2942-3379]{Eric Burns}
\affiliation{Department of Physics \& Astronomy, Louisiana State University, Baton Rouge, LA 70803, USA}

\author[0000-0003-2624-0056]{Christopher L. Fryer}
\affiliation{Center for Nonlinear Studies, Los Alamos National Laboratory, Los Alamos, NM 87545 USA}

\author[0000-0001-7863-1126]{Ivan Agullo}
\affiliation{Department of Physics \& Astronomy, Louisiana State University, Baton Rouge, LA 70803, USA}

\author[0000-0003-0123-0062]{Jennifer Andrews}
\affiliation{Gemini Observatory/NSF's NOIRLab, 670 N. A'ohoku Place, Hilo, HI 96720, USA}

\author[0000-0001-8525-3442]{Elias Aydi}
\affiliation{Department of Physics \& Astronomy, Texas Tech University, Box 41051, Lubbock, TX, 79409-1051, USA}

\author[0000-0003-4433-1365]{Matthew G. Baring}
\affiliation{Department of Physics and Astronomy - MS 108, Rice University, 6100 Main Street, Houston, Texas 77251-1892, USA}

\author[0000-0001-5393-1608]{Eddie Baron}
\affiliation{Planetary Science Institute, 1700 East Fort Lowell Road, Suite 106, Tucson, AZ 85719-2395 USA}

\author[0000-0001-9379-4716]{Peter G. Boorman}
\affiliation{Cahill Center for Astrophysics, California Institute of Technology, 1216 East California Boulevard, Pasadena, CA 91125, USA}

\author[0009-0005-0912-8304]{Mohammad Ali Boroumand}
\affiliation{Department of Physics \& Astronomy, Louisiana State University, Baton Rouge, LA 70803, USA}

\author[0009-0002-6989-1019]{Eric Borowski}
\affiliation{Department of Physics \& Astronomy, Louisiana State University, Baton Rouge, LA 70803, USA}

\author[0000-0002-4421-4962]{Floor S. Broekgaarden }
\affiliation{Department of Astronomy and Astrophysics, University of California, San Diego, La Jolla, CA 92093, USA}

\author[0000-0001-5525-089X]{Alan Calder}
\affiliation{Department of Physics \& Astronomy, Stony Brook University, Stony Brook, NY 11794, USA}

\author[0000-0002-0844-6563]{Poonam Chandra}
\affiliation{National Radio Astronomy Observatory, 520 Edgemont Rd, Charlottesville VA 22903}

\author[0000-0002-8179-1654]{Emmanouil Chatzopoulos}
\affiliation{Department of Physics \& Astronomy, Louisiana State University, Baton Rouge, LA 70803, USA}

\author[0000-0001-5403-3762]{Hsin-Yu Chen}
\affiliation{Department of Physics, University of Texas at Austin, Austin, Texas 78712, USA}

\author[0000-0003-3050-1298]{Kelly A. Chipps}
\affiliation{Physics Division, Oak Ridge National Laboratory, P.O. Box 2008, Oak Ridge TN 37831}

\author{Francesca Civano}
\affiliation{NASA Goddard Space Flight Center, 8800 Greenbelt Rd, Greenbelt, MD 20771}

\author[0000-0001-8822-8031]{Luca Comisso}
\affiliation{Department of Astronomy and Columbia Astrophysics Laboratory, Columbia University, New York, NY 10027, USA}

\author[0000-0001-9528-1826]{Alejandro Cárdenas-Avendaño}
\affiliation{Center for Nonlinear Studies, Los Alamos National Laboratory, Los Alamos, NM 87545 USA}

\author[0009-0002-7066-3988]{Phong Dang}
\affiliation{Department of Physics \& Astronomy, Louisiana State University, Baton Rouge, LA 70803, USA}

\author[0009-0004-0656-1850]{Catherine M. Deibel}
\affiliation{Department of Physics \& Astronomy, Louisiana State University, Baton Rouge, LA 70803, USA}

\author[0000-0003-0307-9984]{Tarraneh Eftekhari}
\affiliation{Center for Interdisciplinary Exploration and Research in Astronomy (CIERA), Northwestern University, 1800 Sherman Avenue, Evanston, IL 60201, USA}

\author[0009-0000-1860-9509]{Courey Elliott}
\affiliation{Department of Physics \& Astronomy, Louisiana State University, Baton Rouge, LA 70803, USA}

\author[0000-0002-2445-5275]{Ryan J. Foley}
\affiliation{Department of Astronomy and Astrophysics, University of California, Santa Cruz, CA 95064, USA}

\author[0000-0003-1087-2964]{Christopher J. Fontes}
\affiliation{Computational Physics Division, Los Alamos National Laboratory, Los Alamos, NM 87545 USA}

\author[0000-0002-8260-2229]{Amy Gall}
\affiliation{Smithsonian Astrophysical Observatory, Cambridge, MA 02138, USA}

\author[0009-0005-8353-7819]{Gwendolyn R. Galleher }
\affiliation{Computational Physics Division, Los Alamos National Laboratory, Los Alamos, NM 87545 USA}

\author{Gabriela Gonzalez}
\affiliation{Department of Physics \& Astronomy, Louisiana State University, Baton Rouge, LA 70803, USA}

\author[0000-0003-4315-3755]{Fan Guo}
\affiliation{Theoretical Division, Los Alamos National Laboratory, Los Alamos, NM 87545 USA}

\author[0000-0003-1878-2445]{Maria C. Babiuc Hamilton}
\affiliation{Department of Mathematics and Physics, Marshall University, Huntington, WV, 25755}

\author[0000-0001-9844-2648]{J. Patrick Harding}
\affiliation{Physics Division, Los Alamos National Laboratory, Los Alamos, NM 87545 USA}

\author{Joseph Henning}
\affiliation{Department of Physics \& Astronomy, Louisiana State University, Baton Rouge, LA 70803, USA}

\author[0000-0001-8087-9278]{Falk Herwig}
\affiliation{Astronomy Research Centre and Department of Physics \& Astronomy, University of Victoria, BC, Canada}

\author[0000-0002-9481-9126]{William Raphael Hix}
\affiliation{Physics Division, Oak Ridge National Laboratory, P.O. Box 2008, Oak Ridge TN 37831}
\affiliation{Department of Physics \& Astronomy, University of Tennessee, Knoxville, TN 37996}

\author[0000-0002-9017-3567]{Anna Y. Q. Ho}
\affiliation{Department of Astronomy, Cornell University, Ithaca, NY 14853, USA}

\author[0000-0003-2227-1322]{Kelly Holley-Bockelmann}
\affiliation{Department of Physics and Astronomy, Vanderbilt University, Nashville, TN, 37235}

\author[0000-0002-0476-4206]{Rebekah Hounsell}
\affiliation{University of Maryland, Baltimore County, 1000 Hilltop Cir, Baltimore, MD 21250}
\affiliation{NASA Goddard Space Flight Center, 8800 Greenbelt Rd, Greenbelt, MD 20771}

\author[0000-0002-0468-6025]{C. Michelle Hui}
\affiliation{NASA Marshall Space Flight Center, Huntsville AL 35812}

\author[0000-0002-1432-7771]{Thomas Brian Humensky}
\affiliation{NASA Goddard Space Flight Center, 8800 Greenbelt Rd, Greenbelt, MD 20771}

\author[0000-0001-6893-0608]{Aimee Hungerford}
\affiliation{Computer, Computational and Statistical Sciences Division, Los Alamos National Laboratory, Los Alamos, NM 87545}

\author[0000-0003-3318-0223]{Robert I. Hynes}
\affiliation{Department of Physics \& Astronomy, Louisiana State University, Baton Rouge, LA 70803, USA}

\author[0000-0002-1089-1754]{Weidong Jin}
\affiliation{Department of Physics and Astronomy, University of California, Los Angeles, CA 90095, USA}

\author[0000-0001-7252-3343]{Heather Johns}
\affiliation{P-4: Thermonuclear Plasma Physics, Los Alamos National Laboratory, Los Alamos, NM, 87544, USA}

\author[0000-0002-2383-1275]{Maria Gatu Johnson}
\affiliation{Plasma Science and Fusion Center, Massachusetts Institute of Technology, Cambridge, MA 02139, USA}

\author[0000-0002-3326-4454]{Bernard J. Kelly}
\affiliation{Center for Space Sciences and Technology, University of Maryland, Baltimore County, 1000 Hilltop Cir, Baltimore, MD 21250}
\affiliation{Gravitational Astrophysics Laboratory, NASA Goddard Space Flight Center, 8800 Greenbelt Rd, Greenbelt, MD 20771}

\author[0000-0002-6745-4790]{Jamie A. Kennea}
\affiliation{Department of Astronomy and Astrophysics, The Pennsylvania State University, University Park, PA 16802, USA}

\author[0000-0002-0873-1487]{Carolyn Kuranz}
\affiliation{Department of Nuclear Engineering and Radiological Sciences, University of Michigan, Ann Arbor, MI}

\author[0000-0001-5169-4143]{Gavin P. Lamb}
\affiliation{Astrophysics Research Institute, Liverpool John Moores University, IC2 Liverpool Science Park, 146 Brownlow Hill, Liverpool, L3 5RF, United Kingdom}

\author[0000-0002-8323-7104]{Kristina D. Launey}
\affiliation{Department of Physics \& Astronomy, Louisiana State University, Baton Rouge, LA 70803, USA}

\author[0000-0002-9854-1432]{Tiffany R. Lewis}
\affiliation{Department of Physics, Michigan Technological University, Houghton, MI 49931, USA}

\author[0000-0001-9200-4006]{Ioannis Liodakis}
\affiliation{Institute of Astrophysics, FORTH, N. Plastira 100, GR-70013 Vassilika Vouton, Greece}

\author[0000-0003-2367-1547]{Daniel Livescu}
\affiliation{Computer, Computational and Statistical Sciences Division, Los Alamos National Laboratory, Los Alamos, NM 87545}

\author{Stuart Loch}
\affiliation{Department of Physics, Auburn University, Auburn, AL 36849}

\author[0000-0002-6684-8691]{Nicholas R. MacDonald}
\affiliation{Department of Physics and Astronomy, The University of Mississippi, Oxford, MS, 38677}

\author{Thomas Maccarone}
\affiliation{Department of Physics \& Astronomy, Texas Tech University, Box 41051, Lubbock, TX, 79409-1051, USA}

\author[0000-0002-8472-3649]{Lea Marcotulli}
\affiliation{Yale Center for Astronomy \& Astrophysics, 52 Hillhouse Avenue, New Haven, CT 06511, USA}

\author{Athina Meli}
\affiliation{Department of Physics, North Carolina A\&T State University, Greensboro, USA}

\author[0000-0002-5358-5415]{Bronson Messer}
\affiliation{National Center for Computational Sciences, Oak Ridge National Laboratory, Oak Ridge TN 37831 USA}
\affiliation{Department of Physics \& Astronomy, University of Tennessee, Knoxville, TN 37996}

\author[0000-0002-2666-728X]{M. Coleman Miller}
\affiliation{Department of Astronomy, University of Maryland, College Park, MD, 20742}
\affiliation{Joint Space-Science Institute, University of Maryland, College Park, MD, 20742}

\author[0009-0009-3516-3567]{Valarie Milton}
\affiliation{Department of Physics \& Astronomy, Louisiana State University, Baton Rouge, LA 70803, USA}

\author[0000-0002-0491-1210]{Elias R. Most}
\affiliation{TAPIR, Mailcode 350-17, California Institute of Technology, Pasadena, CA 91125 USA}

\author{Darin C. Mumma}
\affiliation{Department of Physics \& Astronomy, Louisiana State University, Baton Rouge, LA 70803, USA}

\author[0000-0002-9950-9688]{Matthew R. Mumpower}
\affiliation{Theoretical Division, Los Alamos National Laboratory, Los Alamos, NM 87545 USA}

\author[0000-0002-6548-5622]{Michela Negro}
\affiliation{Department of Physics \& Astronomy, Louisiana State University, Baton Rouge, LA 70803, USA}

\author[0009-0005-0762-4507]{Eliza Neights}
\affiliation{Department of Physics, The George Washington University, 725 21st St NW, Washington, DC 20052}
\affiliation{NASA Goddard Space Flight Center, 8800 Greenbelt Rd, Greenbelt, MD 20771}

\author[0000-0002-3389-0586]{Peter Nugent}
\affiliation{Lawrence Berkeley National Laboratory, M.S. 50B-4206, 1 Cyclotron Road, Berkeley, CA, 94720-8139}

\author[0000-0003-1386-7861]{Dheeraj R Pasham}
\affiliation{MIT Kavli Institute for Astrophysics and Space Research, 70 Vassar Street, Cambridge, MA 02139, USA}

\author[0000-0001-6982-1008]{David Radice}
\affiliation{Department of Physics, The Pennsylvania State University, University Park, PA 16802, USA}

\author[0000-0001-5711-084X]{Bindu Rani}
\affiliation{NASA Goddard Space Flight Center, 8800 Greenbelt Rd, Greenbelt, MD 20771}

\author[0000-0002-3923-1055]{Jocelyn S. Read}
\affiliation{Nicholas and Lee Begovich Center for Gravitational-wave Physics and Astronomy, California State University Fullerton, Fullerton, CA 92831, USA}

\author[0000-0002-3855-5816]{Rene Reifarth}
\affiliation{Physics Division, Los Alamos National Laboratory, Los Alamos, NM 87545 USA}

\author{Emily Reily}
\affiliation{Department of Physics \& Astronomy, Louisiana State University, Baton Rouge, LA 70803, USA}

\author[0000-0003-2705-4941]{Lauren Rhodes}
\affiliation{Trottier Space Institute at McGill, 3550 Rue University, Montreal, Quebec H3A 2A7, Canada}

\author[0000-0001-8308-688X]{Andrea Richard}
\affiliation{Department of Physics and Astronomy, Ohio University, Athens, OH 45701}

\author[0000-0002-5294-0630]{Paul M. Ricker}
\affiliation{Department of Astronomy, University of Illinois, 1002 W. Green St., Urbana, IL 61801, USA}

\author[0000-0002-8866-7891]{Christopher J. Roberts}
\affiliation{NASA Goddard Space Flight Center, 8800 Greenbelt Rd, Greenbelt, MD 20771}

\author[0000-0003-1674-4859]{Hendrik Schatz}
\affiliation{Department of Physics Astronomy, Facility for Rare Isotope Beams, and Center for Nuclear Astrophysics across Messengers, Michigan State University, East Lansing, MI, 48824}

\author[0000-0002-8249-8070]{Peter Shawhan}
\affiliation{Department of Physics, University of Maryland, College Park, MD, 20742}
\affiliation{Joint Space-Science Institute, University of Maryland, College Park, MD, 20742}

\author[0000-0002-2427-5362]{Endre Takacs}
\affiliation{Department of Physics and Astronomy, Clemson University, Clemson, SC 29631, USA}

\author[0000-0001-5506-9855]{John A. Tomsick}
\affiliation{Space Sciences Laboratory, University of California, Berkeley, 7 Gauss Way, Berkeley, CA 94720-7450}

\author[0009-0006-8598-728X]{Aaron C. Trigg}
\affiliation{Department of Physics \& Astronomy, Louisiana State University, Baton Rouge, LA 70803, USA}

\author[0000-0002-7257-608X]{Todd Urbatsch}
\affiliation{X Theory Division, Los Alamos National Laboratory, P.O. Box 1663, Los Alamos, NM 87545}

\author[0000-0002-3305-4326]{Nicole Vassh}
\affiliation{TRIUMF, 4004 Wesbrook Mall, Vancouver, BC V6T 2A3, Canada}

\author[0000-0002-5814-4061]{V. Ashley Villar}
\affiliation{Center for Astrophysics Harvard \& Smithsonian, 60 Garden Street, Cambridge, MA 02138-1516, USA}
\affiliation{The NSF AI Institute for Artificial Intelligence and Fundamental Interactions}

\author[0000-0002-9249-0515]{Zorawar Wadiasingh}
\affiliation{Department of Astronomy, University of Maryland College Park, College Park, MD, 20742}
\affiliation{NASA Goddard Space Flight Center, 8800 Greenbelt Rd, Greenbelt, MD 20771}

\author[0000-0003-3630-9440]{Gaurav Waratkar}
\affiliation{Department of Physics, Indian Institute of Technology Bombay, Powai 400076, India}

\author[0000-0001-8401-030X]{Michael Zingale}
\affiliation{Department of Physics and Astronomy, Stony Brook University, Stony Brook, NY 11794-3800, US}

\newpage
\section*{Preface}
\label{sec:preface}
Time-domain and multimessenger (TDAMM) astrophysics was identified by Pathways to Discovery in Astronomy and Astrophysics for the 2020s \citep{national2021pathways}, the astrophysics Decadal, as a key scientific priority, promising to revolutionize our understanding of how the universe works. Time-domain astronomy is the study of the universe as it evolves over time. This includes transient events that explode and fade and temporal variation of persistent sources. Multimessenger astronomy refers to the use of observations of gravitational waves, neutrinos, cosmic rays, and dust with our existing capabilities in observing the electromagnetic spectrum to investigate questions whose answers have eluded us. Major new facilities have recently come online, and more are expected in the near future, which will upend our approach to the study of a large swath of astrophysics. 

\textbf{The 1st TDAMM Workshop:} In response to this Decadal priority, the NASA Physics of the Cosmos Program organized \href{https://pcos.gsfc.nasa.gov/TDAMM/}{\textit{Prioritizing the Science}}, in Annapolis, Maryland in August 2022. The workshop focused on a broad set of talks covering the multitude of sources of interest to TDAMM astrophysics. Following the meeting, the Scientific Organizing Committee wrote a \href{https://pcos.gsfc.nasa.gov/TDAMM/docs/TDAMM_Report.pdf}{summary white paper}. The scientists then produced a set of papers published in a \href{https://www.frontiersin.org/research-topics/56672/articles}{special journal issue} that was \href{https://assets.science.nasa.gov/content/dam/science/cds/researchers/nac/apac/2024/Gezari_tdamm_summary_apac_gezari_v2.pdf}{recently presented} to the Astrophysics Advisory Committee (APAC), a Federal Advisory Committee Act (FACA) committee responsible for advising NASA in the area of astrophysics. 

\textbf{The 2nd TDAMM Workshop:} At the first workshop there was a lively session on TDAMM infrastructure, referring to the hardware and software necessary to coordinate observations and access and analyze the wealth of data from all relevant facilities. NSF's National Optical-Infrared Astronomy Research Laboratory, with NASA involvement, organized \href{https://noirlab.edu/science/events/websites/MMA2023}{\textit{Windows on the Universe: Establishing the Infrastructure for a Collaborative Multi-messenger Ecosystem}} that was held in Tucson, Arizona in October 2023. The meeting contained a series of invited and contributed talks and a number of active discussions sessions to understand what specific investments and general guidelines were needed in this focus area. Following the meeting, the organizers and members of the interested community wrote a subsequent white paper \citep{ahumada2024windows}. This was reported to the Astronomy and Astrophysics Advisory Committee, a FACA committee advising NSF, DOE, and NASA, and informed a bespoke NSF call to address this focus area.

\textbf{The 3rd TDAMM Workshop:} The first conclusion of the first workshop noted the importance of TDAMM for other fields of physics. Although astronomy typically uses approximate models because they are sufficient to model observational data, this has begun to change in TDAMM. In the parlance of physics experiments, the use of time-domain and multimessenger observations to probe an underlying question or source is analogous to multidiagnostic experimental approaches, where the separate signals give complementary information for a greater overall understanding. Here, multiwavelength astronomy and spectral and polarimetric observations are additional diagnostics. This holistic approach to TDAMM requires the breadth of facilities and infrastructure investment considered at previous workshops, as well as theory and simulation integration of information from multiple related fields. 

These considerations led to the workshop \textit{Multidisciplinary Science in the Multimessenger Era} organized at Louisiana State University (LSU) in September 2024. The workshop was supported by LSU, by NASA's Physics of the Cosmos Program, by NSF Physics and NSF Astronomy as a Conference proposal through the Windows on the Universe, and by the Department of Energy's (DOE's) Los Alamos National Laboratory and the Center for Nuclear Astrophysics Across Messengers. In contrast to the first two meetings, which were driven predominantly by the Astrophysics Decadal, the scope was broadened to include the planning documents and priorities of additional fields of interest. 

The workshop opened with invited plenary talks to guide the extended small-group discussion sections. The key guiding questions presented ahead of the workshop for discussion were the following:
\begin{itemize} 
    \item Beginning with the questions outlined in the first time-domain and multimessenger white paper, what are the most important multidisciplinary questions of interest for time-domain and multimessenger astrophysics?
    \item What are the key measurements? How can we leverage current and forthcoming facilities? Do we need new ones? For astrophysical observations, are additional coordination recommendations needed beyond those in the second time-domain and multimessenger white paper?
    \item What advances are relevant for other fields of physics and national strategic priorities?
    \item How can multidisciplinary research be fostered? 
\end{itemize}

\newpage
\section*{Executive Summary} 
\label{sec:executiveSummary} 
\textit{Connecting Quarks with the Cosmos: Eleven Science Questions for the New Century} is a report delivered by the National Research Council shortly after the turn of the millennium \citep{national2003connecting}. To quote, ``No one agency currently has unique ownership of the science at the intersection of astronomy and physics; nor can one agency working alone mount the effort needed to realize the great opportunities. DOE, NASA, and NSF are all deeply interested in the science at this intersection, and each brings unique expertise to the enterprise. Only by working together can they take full advantage of the [scientific opportunities].''

Since this report, a revolution in astrophysics has occurred. We now regularly detect gravitational waves with LIGO and high-energy neutrinos by IceCube, facilities both supported by the NSF. The same events are seen through high energy photons by NASA's Fermi and Swift telescopes. The forthcoming Vera C. Rubin Observatory, jointly supported by the NSF and DOE, will produce videos of the universe at optical wavelengths. The characterization of astrophysical events with these facilities, and across the electromagnetic spectrum by telescope supported by all three agencies, allow for a more holistic method of understanding the physics of the cosmos. 

This time-domain and multimessenger approach is a priority science area, per the Astro 2020 Decadal \textit{Pathways to Discovery in Astronomy and Astrophysics for the 2020s} \citep{national2021pathways}. The promise of this focus area is also marked in the strategic planning document of several domains of science. Time-domain and multimessenger astronomy could answer a majority of the National Research Council's eleven questions before the half-century mark. These goals, along with others identified since this report, include mapping the complete origin of the elements, conducting precision cosmology across the universe, understanding extreme matter, observing mass extraction from black holes, testing unique predictions from quantum electrodynamics, and probing physics outside of equilibrium. 

However, this progress requires a change in approach. Historically, astrophysics has relied upon approximate models for interpretation, as these were sufficient to explain astrophysical data and consistent with results from terrestrial laboratories. The multi-diagnostic approach now used in astrophysics provides data beyond the fidelity of these models. Fortunately, transformational advancements in terrestrial experiments are beginning to recreate the extreme physics seen in the cosmos. The DOE's Facility for Rare Isotope Beams is revolutionizing the study of nuclear physics of relevance for time-domain and multimessenger astronomy. The DOE and NSF also operate high-energy and high intensity laser facilities which can now recreate plasma conditions found in astrophysical environments.

Because of these major (and dozens more) facilities, the creation of higher fidelity models for scientific advancement is predominantly a task of integration. When the knowledge is beyond the capability of any individual or institution, how can this be accomplished? The National Nuclear Security Administration, a branch of the DOE, applies an end-to-end approach to fulfill its national security mandate. This involves outlining all steps which occur in a problem and the creation of corresponding simulations. These simulations must be chained together, with the output from one informing the next, or through iteration between them. Each component is studied to understand its true error, allowing for full uncertainty quantification. The greatest sources of uncertainty are addressed predominantly through the use of existing components such as drawing on advances from open science or running new experiments at existing facilities. Sometimes this requires novel work, such as the development of new numerical approaches. This method has fostered new interdisciplinary fields, with some approaches now decades ahead of those used in astrophysics.

Maximizing the scientific return in time-domain and multimessenger astrophysics requires a similar end-to-end approach. This includes adoption of the techniques used by the National Nuclear Security Administration, a renewed focus on computational astrophysics, and engagement with other disciplines of physics. The funding agencies, professional societies, and respective communities must work to overcome structural barriers to foster multidisciplinary work. The motivation includes maximizing the scientific return of major US facilities and in developing generalist scientists for use in industry and the national security workforce. The majority of the needed investment can be done through alignment of existing programs, without requiring additional funds.

Within this context, priorities within the field of time-domain and multimessenger astronomy are evident. The sources most prepared for full end-to-end approaches are explosive transients, including supernovae, novae, and neutron star mergers. This is because of sustained investment in the area of nuclear astrophysics by the NSF and DOE which have fostered connections in the necessary communities and supported the multidisciplinary work to outline and address areas where the greatest progress can be made. However, a broader approach is needed to understand these events. Lastly, the success in this area will inspire emulation of this approach in other interdisciplinary areas of astrophysics, driving the application of this end-to-end approach on other sources in the future.

\newpage
\tableofcontents

\newpage
\section{Synthesis}
\label{sec:intro}
The preeminent starting document for this workshop is the Astrophysics 2020 Decadal, \textit{Pathways to Discovery in Astronomy and Astrophysics for the 2020s} \citep{national2021pathways}, as it recommended time-domain and multimessenger (TDAMM) astrophysics as a science priority. Much of the excitement in TDAMM science has been driven by two specific transients with multi-messenger observations:  a first transient with both neutrino and photon observations and a second transient with concurrent gravitational wave and photon observations.

In the first event, neutrino detectors (Kamiokande and IMB) observed the thermal ($\sim$MeV) neutrinos from the core-collapse of supernova (SN) 1987A.  This observation provided decisive proof that at least some SNe are produced in the collapse of the cores of massive stars.  The unexpectedly early emergence of nuclear gamma-rays drove the development of the convective engine, which is still the standard model for these sources. The Department of Energy's (DOE's) next generation neutrino facility, Deep Underground Neutrino Experiment (DUNE), will begin operation this decade. DUNE, and similar megaton scale detectors, will increase the detection rate of SNe through neutrinos from roughly once per century to roughly once per decade, and are also expected to detect the stochastic neutrino background from all core-collapse events in the universe.

The latest concurrent mulit-messenger detection occurred in 2017 with the joint gravitational wave (GW) and electromagnetic (EM) discovery of a binary neutron star merger.  This event has excited scientists across several disciplines.  In the United States, the ground-based interferometer is Laser Interferometer Gravitational-Wave Observatory (LIGO). LIGO and its international partners together comprise the International Gravitational-Wave Network (IGWN). With the forthcoming interferometer upgrades we expect regular detections of neutron star mergers through GWs this decade. Planned next generation detectors would detect tens of thousands of events each year. Similar to the use of neutrinos, GW observations provide diagnostic information on what is occurring in the hearts of these explosive transients, which is impossible to study with electromagnetic information. In the United States these facilities are supported by the National Science Foundation (NSF). 

Just as the expansion of observations across the EM spectrum gave new understanding of the universe, neutrino and GW detections have driven significant advances to our understanding of these transients and their underlying physics. A third key advancement in TDAMM science was the identification of an astrophysical contribution to high-energy neutrinos by NSF's IceCube Neutrino Observatory (IceCube). A fourth was the recent identification of the stochastic background of low-frequency gravitational waves by the international Pulsar Timing Array (PTA), whose US efforts are the North American Nanohertz Observatory for Gravitational Waves (NANOGrav) supported by the NSF and the Fermi Space Telescope supported by NASA. Future foundational advances in both areas will occur with the identification of unambiguous counterparts to these non-electromagnetic sources. For IceCube neutrinos there are several promising cases. For the PTA it is a matter of time. The European Space Agency (ESA),  will launch the Laser Interferometer Space Antenna (LISA) in the 2030s, which will observe GWs in between the PTAs and IGWN. National Aeronautics and Space Administration's (NASA's) contribution to LISA has been fully endorsed in the past two astro Decadals.

As there is yet to be a combined GW and neutrino detection, and other messengers (cosmic rays, dust, meteorites) are not yet associated to individual events, TDAMM science via concurrent observations is thus far only possible because of major electromagnetic facilities. NASA's gamma-ray missions provided the key early discovery of nucleosynthesis following the neutrino discovery in SN~1987A. The gamma-rays seen two seconds after the GWs in 2017 were observed by NASA's Fermi Space Telescope. Follow-up observations using ground-based optical telescopes supported by the NSF and DOE identified the precise position of this event, allowing characterization with NSF, DOE, and NASA facilities across the full electromagnetic spectrum. This also regularly occurs for the most promising neutrino counterpart candidates and other rare transients of interest. Optical and infrared observations are supported by all three agencies. Radio is predominantly the domain of the NSF. Ultraviolet, X-rays, and (low-energy) gamma-rays are the responsibility of NASA, with Fermi joined by the Neil Gehrels Swift Observatory (Swift) as the main workhorse machines. Very high energy gamma-rays are observed from the ground by DOE and NSF facilities. Cosmic rays detection is supported by all three agencies. Lastly, TDAMM science also involves the study of dust, meteorites, and ocean sediments from the cosmos, work which is done by chemists and geologists.  In many cases, these observations all contribute to understanding the same phenomena even though not of the same event.

The Astro Decadal mentions the transformative astrophysics enabled by the combination of these facilities as well as the forthcoming Legacy Survey of Space and Time (LSST) by the NSF and DOE's Vera C. Rubin Observatory (Rubin) as laying the foundation for this new era of TDAMM science. While TDAMM is a broad technique used across astrophysics, 
the Astro Decadal specifically highlights the priority area this decade as the application of these techniques to ``the study of neutron stars, white dwarfs, collisions of black holes, and stellar explosions'', emphasizing that this ``will reshape the understanding of topics as diverse as the origin of the carbon in bones and the metal in phones, the history of the expansion of the universe since the Big Bang, the life and death of stars, and the physics of black hole event horizons.''

The Decadal also states ``New, coordinated advances in several areas are required to unlock the workings of the dynamic universe.'' Specific recommendations include NSF investment in the Extremely Large Telescopes (ELTs), the next generation Very Large Array (ngVLA), next generation Cosmic Microwave Background (CMB) telescopes, upgrades to the existing GW interferometers and high energy neutrino telescopes, and technology towards the next generation of GW interferometers and NASA investment in Habitable Worlds Observatory. Investments have begun for all of these recommendations. There are two recommendations for telescopes which were not specific:
\begin{itemize}
    \item A suite of small and medium-scale ground and space-based observational facilities across the electromagnetic spectrum to discover and characterize the brightness and spectra of transient sources as they appear and fade away.
    \item Strong software and theoretical foundations to numerically interpret the gravitational wave signals from merging compact objects to extract new physics in the extremes of density and gravity, and ensure easy user access to the wealth of data on the dynamic universe and to model and interpret astronomical sources whose physical conditions cannot be replicated in laboratories on Earth.
\end{itemize}
These recommendations led to the creation of the TDAMM series of workshops, to work through the details of these Decadal recommendations. The first workshop focused on exploring the broad astrophysics questions in the umbrella of TDAMM. The second workshop focused on coordinated observations and corresponding software development. The findings of these workshops are reported in the preface and the status of the response is discussed in Section~\ref{sec:multidisciplinary_pastworkshops}. The focus of the fourth TDAMM workshop is explored in Section~\ref{sec:multidisciplinary_observingPlans}. This white paper, from the third workshop, is predominantly focused on the theory and analysis recommendation in the second bullet point above, and the necessary engagement with other fields of physics to advance TDAMM science. This requires an explanation of the relevant fields and their corresponding funding agencies. We begin with the former.

NASA is broken down into six mission directorates. Under the Science Mission Directorate are five divisions. The Astrophysics Division (APD) contains three key questions. ``How does the universe work?'' seeks to understand the physics of the cosmos, whose scope is a superset of TDAMM. The Biological and Physical Sciences Divison (BPS) is responsible for part of fundamental physics research in space. NSF has eight directorates. The Mathematical and Physical Sciences (MPS) directorate contains six divisions, including NSF Physics (PHY) and Astronomical Sciences (AST). PHY and AST both have foundational roles in TDAMM. TDAMM utilizes work from other areas of the NSF including MPS Chemistry (CHE), the directorate for Geosciences, and the Office of Advanced Cyberinfrastructure within the Directorate for Computer and Information Science and Engineering. DOE has three major branches, each with an undersecretary. The National Nuclear Security Administration (NNSA) is a semiautonomous agency responsible for military application of nuclear science. The strategic end-to-end approaches to solve complex, multiphysics problems applied in the NNSA are a guideline for how to approach complex TDAMM problems. The Office of Science and Innovation contains the Office of Science, which supports a broad range of programs. Among these are Nuclear Physics (NP), High Energy Physics (HEP), Fusion Energy Sciences (FES), and Advanced Scientific Computing Research (ASCR). While HEP contains the Cosmic Frontier, the direct astrophysics research area of the DOE, all of these programs are relevant for TDAMM science. 

Astroparticle physics is an interdisciplinary field of direct relevance for TDAMM, including of cosmic rays, cosmology, gamma-rays, and neutrinos. In addition to the Astro Decadal, another key planning document for astroparticle physics is the Particle Physics Project Prioritization Panel (P5) whose most recent 2023 report \textit{Exploring the Quantum Universe: Pathways to Innovation and Discovery in Particle Physics} \citep{2024arXiv240719176A}. This report prioritized the completion and enhancement of DUNE for its anticipated transformational results in neutrino physics and supports the upgrade of IceCube.

The field of gravity is closely aligned with astrophysics but is distinct, with its own community, culture, funding lines, hysicsrofessional society divisions, and organizations. LISA is space-based and thus governed by the astrophysics Decadal. IGWN and NANOGrav fall under the purview of NSF Physics and are not formally governed by the Astrophysics Decadal. Instead the future of ground-based gravitational wave detection is guided by a series of ad hoc committees, as well as the white papers produced by the relevant consortia. Despite the direct need and strong ties in astrophysics, astroparticle physics, and gravity for TDAMM, this multidisciplinary area falls into a strategic planning gap, where the key needs of TDAMM across areas are not necessarily considered holistically.

Explosive transients forge and eject new elements, altering their balance over cosmic time. Many electromagnetic signatures arise from the decay of radioactive isotopes produced in these events, either as reprocessed thermal emission or as direct nuclear lines. However, the temperatures, densities, and composition involved are only partially understood, and the isotopes being created are not fully known.  Tying yields to directly requires both and understanding of the physics driving these observables (including transport, atomic, dust formation, etc.) and a partnership with nuclear scientists. This field is governed by the 2023 Long Range Plan (LRP) for Nuclear Science \citep{aidala2023new} which has nuclear astrophysics woven throughout. It recommends multidisciplinary centers in the area of multimessenger science. The DOE currently supports the Center for Nuclear Astrophysics across Messengers (CeNAM) as a community organization entity, which arose as a successor to the NSF Physics Frontiers Center series of the Joint Institute for Nuclear Astrophysics (JINA) and the JINA Center for the Evolution of the Elements (JINA-CEE). Additional support includes the Network for Neutrinos, Nuclear Astrophysics, and Symmetries (N3AS; an NSF PHysics Frontiers Center), the Nuclear Physics from Multi-Messenger Mergers (NP3M; an NSF Focus Research Hub), as well as the Institute for Nuclear Theory (INT) supported by the DOE and the University of Washington. This multi-decade investment in organizing the community has built up a network of multidisciplinary scientists, helped advance the field in a strategic manner, created curated data for use by scientists in related fields, raised the visibility of nuclear science in other disciplines, and enabled entirely new lines of research at the interface of nuclear and astrophysics. As one example, this group created REACLib \citep[e.g.][]{cyburt2010jina} which makes nuclear information of interest to astrophysics easily accessible. This group was additionally an early supporter of this conference. DOE NP also pairs with ASCR to support Scientific Discovery through Advanced Computing (SciDAC) calls which can support large-scale efforts in nuclear astrophysics. The use of PFCs, SciDACs, and strategic community organization, as well as the broad support for nuclear astrophysics in the field of nuclear science at all levels, combine to form the best investment made in the support of multidisciplinary research in this area: the nuclear community is waiting on similar engagement from observational astrophysicists.

Nearly all of the EM and particle signals detected from these events are emitted from matter in a state of plasma, whose properties are again far beyond what occurs on Earth. The strategic document for plasma science in the US comes from a Decadal process, with the most recent being \textit{Plasma Science: Enabling Technology, Sustainability, Security, and Exploration (2021)} \citep{national2020plasma}. Astrophysics is the final frontier for plasma science, where extreme environments are beyond the current state of knowledge, whose understanding would strengthen the foundations of plasma science used in lasers and fusion. This Decadal notes the addition of NASA to the long-term DOE and NSF plasma partnership would advance fundamental science in astrophysical plasmas while addressing the needs of space missions, including of Heliophysics. There are also other programmatic efforts in the multidisciplinary area of plasma astrophysics including the \textit{Research Opportunities in Plasma Astrophysics: Report of the Workshop on Opportunities in Plasma Astrophysics} report \citep{bale2010research} and the American Physical Society (APS) Topical Group in Plasma Astrophysics (GPAP), now 25 years old.

The observation of lines in the spectra of the Sun was a key advancement in the development of atomic science. This field of research is now generally grouped into Atomic, Molecular, and Optical Physics which is governed by a Decadal process in the US, with the most recent report \textit{Manipulating Quantum Systems: An Assessment of Atomic, Molecular, and Optical (AMO) Physics in the United States} \citep{national2020manipulating}. The technology developed in AMO include powerful lasers uses in several fields of science, including inertial confinement fusion and the ground-based GW interferometers, as well as atomic lattice clocks used in GPS and fundamental physics research. These foundational technologies underlie research across several fields of physics and enabled new TDAMM frontiers including the GW interferometers. A key need in astrophysics from AMO are libraries of atomic and molecular lines and atomic collision data (excitation, ionization, recombination), allowing the mapping of observed lines in spectral observations to the yield of individual atoms and molecules. In TDAMM events, at least at early times, the temperatures are sufficiently high that typically atomic, not molecular, lines are of relevance. Knowledge of atomic lines at astrophysically relevant temperatures are only complete to Iron, which were mapped to fully explain the spectra seen in our Sun. However, a current priority in astrophysics is understanding the origin of heavier elements. The lack of complete atomic knowledge is limiting the scientific return of multiple facilities, including NASA's flagship James Webb Space Telescope (JWST). Such work is not a key priority of the full field of AMO. Thus, a concerted effort to build an atomic analog of REACLib is needed, and must be strategically supported by the astrophysics funding agencies in order to ensure long-term survival of this unique but niche area of critical expertise.

Fluid dynamics describes the flow of fluids. Of key interest to astrophysics is magnetohydrodynamics (MHD) which studies the flow of plasmas in the presence of electromagnetic fields and thus being relevant for relativistic outflows nearly all TDAMM events. Many codes now handle MHD in regimes where General Relativity is required (GRMHD). Higher fidelity methods include kinetic plasma simulations or calculations. Effects include instabilities, waves, and dynamo effect which can alter flow and affect magnetic fields, which are key to non-thermal signatures studied in astrophysical phenomena. Interactions can lead to nonlinear effects including turbulence, dissipation, heating, particle acceleration, and coherent structures, which are often neglected or poorly modeled in astrophysics. Additionally, our field has a possibly unique need of requiring polarimetric predictions. The study of fluids is wide-ranging but it does not have its own strategic planning process in the US. It is most appropriately discussed in the relevant context in the Plasma Decadals. 

In astrophysical sources it is impossible to isolate individual effects. Observations are recovering signatures affected by the combined effect of all of these physics at the same time. This has led to approaches and fields which sit between disciplines that have evolved into their own fields of study. The oldest of these in astrophysics is computation, with the first uses decades before the invention of the transistor. Astrophysics has often driven computation, including use of the first computers, development of modern supercomputing approaches, and now in handling and processing of massive data volumes. Key areas of modern computational approaches include multiphysics handling, scale bridging, and improving numerical methods. Nearly all understanding of astrophysical sources now requires computational modeling and simulation. Similar to fluid dynamics, the field is wide-ranging but does not have a strategic planning process.

Radiation transport is a multidisciplinary field which studies the propagation of radiation through a medium, including scattering, absorption, and emission. Much like fluid dynamics and computation, radiation transport is a technique that is broadly applicable, and does not have a strategic planning process. When a computational simulation of an event is finished, the output can be postprocessed by radiation transport codes to model the specific expected observational signatures. In astrophysics, it is the output of rad transport calculations which observations are compared with, allowing inference on the underlying system. Methods used within rad transport are often decades ahead of those used within astrophysics.

The newest multidisciplinary field of relevance is High Energy Density Physics\footnote{Note that this is distinct from the field of dense matter} (HEDP) which merges physics from several fields to explore a previously poorly understood regime. HEDP developed out of the National Nuclear Security Administration (NNSA). HEDP has now grown in the open sector due to the broad applicability of its approaches. For example, near the defined entry threshold for HEDP is the energy density involved in the formation of planets which motivated the Center for Matter at Atomic Pressures, an NSF Physics Frontiers Center. HEDP does not have a formal Decadal process, but has now completed two National Academy of Sciences (NAS) studies. The second, \textit{Fundamental Research in High Energy Density Science} \citep{national2023fundamental}, was charged with a dual mandate: ``(1) advancing the underlying science and (2) assuring 
that the workforce continues to maintain its high level of expertise in the future.''. A key objective in support of open HEDP science is workforce development of the unique capabilities needed by the NNSA. All objects of interest to TDAMM are well within the HEDP regime, but the relevant HEDP techniques have not been broadly adopted in astrophysics. This is despite the comparatively new capability of HEDP facilities emulating the conditions on the surface of white dwarfs, bringing ground-based facilities into the realm of compact objects for the first time. This is problematic for two reasons. First, astrophysical advances are unnecessarily lagging. Second, astrophysics is necessarily multidisciplinary, but our junior scientists are no longer developed to meet the generalist needs of astrophysics or the NNSA. The need for fostering and developing this field by DOE, NSF, and NASA was noted more than twenty years ago \citep{national2003connecting}. As their facilities can now emulate the conditions on the surface of white dwarfs, it is past time to engage.

Lastly, laboratory astrophysics is a noted interdisciplinary field highlighted explicitly or implicitly in the long-term planning documents of many fields of interest. This is the tailoring of ground-based experiments and data curation to provide insight on astrophysical questions, which underlies all of TDAMM science. For example, the generation of new knowledge of atomic, molecular, and nuclear lines, and the conditions where they arise, for interpreting astrophysical spectra arise from lab astro measurements. Lab astro, being an interdisciplinary field, also does not have a dedicated Decadal process; its strategic planning is typically done via ad hoc reports, including a recent one commissioned by the NSF and NASA in response to the Astro 2020 Decadal support for lab astro,  \textit{Enabling Cosmic Discoveries: The Vital Role of Laboratory Astrophysics} \citep{LATF_2024}, by the Laboratory Astrophysics Task Force (LATF).

With these $\sim$10 disciplines, all of which have their own questions, successes, and priorities, why should we bother with multidisciplinary studies? While there is no standing board nor long term planning process responsible for this question, the National Research Council convened a Committee on the Physics of the Universe 20 years ago, which delivered \textit{Connecting Quarks with the Cosmos: Eleven Science Questions for the New Century} \citep{national2003connecting}, which states:  ``No one agency currently has unique ownership of the science at the
 intersection of astronomy and physics; nor can one agency working alone
 mount the effort needed to realize the great opportunities. DOE, NASA, and
 NSF are all deeply interested in the science at this intersection, and each
 brings unique expertise to the enterprise. Only by working together can they
 take full advantage of the opportunities at this special time.
 Coordination and joint planning are essential. In some instances, two of
 the agencies, or even all three, will need to work together. In others, one
 agency may be able to close the gap between the disciplines of physics and
 astronomy.''

We note that this report is broadly focused on physics of the cosmos, of which TDAMM is only a subset; however, the needs are identical and the overlap substantial. We here list the eleven questions in the report and if they tie to the key questions of TDAMM, described below. 

\begin{enumerate}
    \item \textbf{What is Dark Matter?} Some proposed dark matter candidates can be probed with observations of TDAMM events. TDAMM advances on the concordance model of cosmology may provide new constraints on the properties of dark matter. However, this is not necessarily a key TDAMM question.
    \item \textbf{What is the Nature of Dark Energy?} The discovery of dark energy arose from TDAMM studies. Measuring its equation of state will require advanced TDAMM investigations. This is explored in Section~\ref{sec:questions_cosmology}. Major advancements are expected before this decade is out, but we will always seek greater precision in measurements to answer the questions which come next.
    \item \textbf{How Did the Universe Begin?} DOE, NASA, and NSF play a key role in these studies, but they are not a focus of TDAMM. However, TDAMM studies could probe whether gravity violates parity which may explain the excess of matter over antimatter at the beginning of time, which matches the original task of the NRC report.
    \item \textbf{Did Einstein Have the Last Word on Gravity?} TDAMM provides unique tests of gravity, especially in the strong regime. LIGO has already brought huge advancements in precision tests, but this remains an on-going effort which will expand with the launch of LISA. We highlight a unique test of gravity in Section~\ref{sec:questions_BZ}. 
    \item \textbf{What Are the Masses of the Neutrinos and How Have They Shaped the Evolution of the Universe?} Advancing our understanding of neutrino physics, including the mass ordering, is a key goal of DUNE. A fortuitous supernova could allow direct determination of the neutrino masses. All neutrino advances will feed back into concordance cosmology models by reducing assumptions or providing certainty, but this is not a TDAMM focus. 
    \item \textbf{How Do Cosmic Accelerators Work and What Are They Accelerating?} This is a direct and foundational TDAMM question. We explore this in Section~\ref{sec:questions_BZ} and in later sections. Resolution requires long-term and intentional effort across multiple fronts.
    \item \textbf{Are Protons Unstable?} DUNE and other forthcoming facilities will probe this question. TDAMM advances are unlikely to help in this area.
    \item \textbf{What Are the New States of Matter at Exceedingly High Density and Temperature?} This question is best probed through TDAMM studies and complementary ground-based experiments. Answering this question requires advances in both regimes. This is a key TDAMM focus and explored in Section~\ref{sec:questions_extremeMatter}. Major advances are expected in the coming decade.
    \item \textbf{Are There Additional Space-Time Dimensions?} This is a question which can be explored through TDAMM observations. However, these focus on additional large dimensions \citep{burns2020neutron}, rather than the additional small dimensions typically explored in the context of grand unified theories.
    \item \textbf{How Were the Elements from Iron to Uranium Made?} Every known or suspected source of heavy elements are key TDAMM sources. This is a focus of Section~\ref{sec:questions_elements}. This answer should be fully answered by the end of the next decade.
    \item \textbf{Is a New Theory of Matter and Light Needed at the Highest Energies?} This question is also best probed through TDAMM studies and the most extreme ground-based experiments possible. Answering this question requires advances in both regimes. This is a key TDAMM focus and explored in Section~\ref{sec:questions_photonSplitting}. Resolution requires long-term and intentional effort across multiple fronts.
\end{enumerate}

We note that the majority of the eleven questions at the intersection of physics and astronomy are best answered with specific TDAMM advancements. We also note the questions with the greatest progress are TDAMM-focused questions.


\subsection{Why we should Prioritize Multidisciplinary Science}
\label{sec:questions}
Despite the outstanding science questions at the intersection of physics and astronomy and success in particular areas such as cosmology, a broad and sustained coordinated effort has not occurred. TDAMM has both broad interest across fields, several new and forthcoming facilities, and is a comparatively confined scope. Thus it serves both as a unique area where interagency approaches are required, and as an opportunity to study whether such an approach proves greater than the sum of its parts. We below outline some of our own grand questions and motivation where progress require advances in TDAMM science. 

The eleven questions of the century at the intersection of physics and astronomy is a different scope than TDAMM science itself. As a result, we identify our own specific goals. These are sometimes broader than the related NRC questions, and sometimes narrower, but they are all well aligned.


\begin{quoting}
    \noindent Finding: DOE, NSF, and NASA all have a vested interest and major facilities of direct relevance to the major goals of TDAMM science. Strategic investment in community organization and direct integration efforts to foster multidisciplinary studies is required to answer these questions. This will ensure optimal use of major facilities, maximizes scientific return, and does so largely by aligning existing effort and priorities.
\end{quoting}

\subsubsection{The Origin of the Elements}
\label{sec:questions_elements}
The origin of the elements entered the realm of science from the realm of philosophy more than a century ago, with the modern breakdown roughly outlined in 1957 \citep{burbidge1957synthesis,cameron1957nuclear,cameron1957origin}. Modern understanding is accessible in textbooks or reviews \citep[e.g.][]{kobayashi2020origin,johnson2020origin,arcones2023origin}. Hydrogen, Helium, and some Lithium were forged in the Big Bang. The next few elements are generated from spallation of heavier elements. Carbon, Nitrogen, and about half of the heavy elements beyond iron are released in the winds of low and intermediate mass stars on their way to the formation of a white dwarf. 

All other elements arise from explosive transients and are key sources for TDAMM. New elements are forged at extreme temperatures where fusion can occur such as the interior of massive stars, which are released into the universe when the star dies. Fusion and nuclear capture reactions also produce new elements during the explosive death of stars, on the surface of the stellar remnants, or in the disruption of these remnants. Many are detailed below (Sections \ref{sec:sources_ccsn}, \ref{sec:sources_thermonuclear}, \ref{sec:sources_mergers}, \ref{sec:sources_novae}, \ref{sec:sources_xrayBinaries}, and \ref{sec:sources_magnetars}). 

A current goal in astrophysics is understanding the origin of about half of the heavy elements above iron that cannot be produced by long term neutron exposure in stellar interiors and have traditionally been attributed to the rapid neutron capture process (r-process), where capture occurs much more rapidly than decay. The ingredients were listed half a century ago \citep{burbidge1957synthesis,cameron1957origin}, being i) extremely dense, free matter, ii) with far more neutrons than protons, and iii) low neutrino luminosity (to preserve the neutron excess). This pointed to the formation or disruption of neutron stars. 

Neutron star mergers (Section~\ref{sec:sources_mergers}) are now generally believed to be one important site for nucleosynthesis of r-process elements, as predicted in \citet{eichler1989}. Here the merging of a neutron star with another neutron star or a black hole results in the disruption of at least one neutron star. The freed material is dense and neutron rich, and the event can proceed with sufficiently low neutrino luminosities, that elements including the actinides are thought to be produced. These new, heavy elements will radioactively decay towards the line of stability. For the first $\sim$days the plasma is opaque, trapping the emitting radiation and reprocessing it into a thermal signature which emits in the ultraviolet, optical, and infrared wavelengths over the first few hours, days, and weeks and is observable as a kilonova. At late times, when the ejecta has slowed, individual atomic lines may be isolated with observations of sufficiently sensitive infrared telescopes. Indeed, several groups have claimed the identification of lines from r-process elements, supporting the claim of neutron star mergers as the origin of at least some heavy elements.

The theoretical prediction of a kilonova signature following neutron star mergers, and the detailed confirmation of these predictions through the multi-messenger and multi-wavelength observations of the kilonova following the multimessenger GW and EM detection of a binary neutron star merger in 2017 is one of the great successes in multimessenger astrophysics and is a direct demonstration of the need for these various fields and the funding agencies to work together. These detections are now important for identifying more events and provide unique characterization information. The gathering, collation, and curation of nuclear data and application of nuclear reaction networks showed that these events could produce lanthanides and actinides. Accounting for atomic opacities for these elements, some measured in labs and many inferred from theoretical calculations, led to the prediction of a longer-lasting and red transient as the direct signature of heavy element nucleosynthesis. This work culminated in the first late-time observation of a kilonova with the James Webb Space Telescope, identifying a line complex which has been interpreted as arising from Tellurium, which would be direct proof of r-process nucleosynthesis of a specific heavy element. 

However, a much more complex picture of the origin of these elements has emerged in the last decade: there is some observational evidence for at least two different r-process sites, one operating on massive star timescales, and one on longer timescales. Indeed following the workshop, a paper claiming magnetar giant flares as this early r-process site were released \citep{patel2025direct}. Mapping out the rates of magnetar giant flares and neutron star mergers, both local rates and their evolution over cosmic time, as well as their respective elemental yield distribution is the last step in completely answering the first of the eleven NRC questions of the century.

\begin{quoting}
    \noindent Comment: The origin of the majority of the elements lies in explosive transients, a key focus in TDAMM science. Future work includes measurement of unambiguous elements from r-process sites, understanding r-process nucleosynthesis occurred in the universe, and identification of where additional nuclear processes are occurring in the cosmos. 
\end{quoting}

Moving beyond focusing on the origin of the elements, the next step is to understand the origin of the various isotopes. There is evidence for a process contributing exclusively to the lighter r-process elements, which may or may not be a neutron capture process, as well as for a continuum of neutron capture processes between the slow and the rapid, the so called intermediate neutron capture process that may take place in rapidly accreting white dwarfs \citep{Denissenkov2017}. As identified in the LRP \citep{NSAC-LRP-2023}, a key and timely goal is determining the origin of p-process elements. We speculate that COSI observations may help answer this question, but this requires strategic theory and simulation investment. Disentangling the relative contributions of these processes over the history of our Galaxy, identifying their astrophysical sites, and identifying the underlying nuclear reaction sequences, will be a major goal of astrophysics and nuclear physics in the coming decade. 

Splitting by funding agency, the NSF gravitational wave, optical, and infrared facilities are key to discovering and characterizing kilonova transients. These events have both thermal and non-thermal signature in ultraviolet, optical, and infrared wavelengths; thus, the NSF radio and NASA X-ray facilities have been required to allow the isolation and study of kilonova signatures. The NASA gamma-ray burst monitors have discovered the majority of events where kilonovae have been identified. The DOE support for nuclear astrophysics, both experimental and in simulation, allowed for nucleosynthesis predictions, which guided observations and the generation and improvement of models to allow for inference from the astrophysical observations. Currently, kilonovae models are not sufficient for robust inference. Much of the relevant nuclear physics of rare, unstable isotopes remains unknown and is only now coming into reach with new accelerator facilities. The DOE's flagship Facility for Rare Isotope Beams (FRIB) was built for this purpose, and the first results are now being published. Additionally, the lack of complete knowledge of the behavior of heavy elements from atomic physics, especially in non-local dynamic equilibrium (non-LTE), means the identification of the origin of the lines seen in kilonova is not unambiguous. Thus, lack of handling of non-LTE effects and investment in atomic physics is currently limiting the return of NASA's newest flagship. Sustained investment and the application of new techniques is absolutely required to advance this science.

However, this discussion and the priorities in astrophysics belies the breadth of science in this area. Similar results and approaches to the above have proven successful in the study of other transient classes. In addition to the r-processes, there are many nucleosynthetic pathways expected to occur in astrophysics including the i-process, the s-process, the p-process, the n-process, and the rp-process \citep{aidala2023new}. Sites for some of these processes are known, but not all. Identifying them is key to understanding the origin of the elements, and also is key information for understanding the physics which underlies a given source class. In every case the combined efforts from scientists and facilities supported across agencies are required for transformational advancement.

As one key opportunity coming this decade we highlight the forthcoming NASA Compton Spectrometer and Imager (COSI) mission. COSI will launch in 2027 as the first nuclear spectrometer launched in 25 years. Nuclear lines are among the cleanest diagnostics in astrophysics, beat only by non-electromagnetic signals, as they allow determination of the yield of individual isotopes and act as cosmic chronometers given their known half lives. COSI will detect a small sample of thermonuclear supernovae and novae. For thermonuclear supernovae the gamma-ray observations of nuclear lines and positron emission may be able to probe whether these events are the origin of p process elements. For novae the detection of multiple lines, utilized in concert with the removal of important nuclear physics uncertainties due to decades of investment in ground-based facilities \citep{aidala2023new}, will allow for precise measurements of interest for understanding the HEDP and related phenomena occurring in these events. However, making proper use of these observations and being informed for what other multiwavelength data is key for characterization requires immediate intentional and strategic investment in integration work.


\subsubsection{Cosmology}
\label{sec:questions_cosmology}
Cosmology is the study of the origin, evolution, and fate of our universe. Cosmology is perhaps the area where NASA, the NSF, and DOE best partner, utilizing both ground- and space-based instruments for ever more precise measurements of the Cosmic Microwave Background (CMB) and the expansion rate of the universe over time (the Hubble Parameter) through various means. The concordance model is $\Lambda$CDM, with $\Lambda$ representing dark energy as a cosmological constant and CDM the abbreviation for Cold Dark Matter. It reasonably explains all cosmological data in a self consistent manner, handling the precise CMB observables from early universe, the generation of large scale structure, and evolution of the expansion rate of the universe. Current goals include understanding what caused the epoch of inflation, solving the missing lithium problem, and understanding what dark matter and dark energy are composed of. 

Dark energy is modeled as a cosmological constant because it was the minimal modification required to explain the acceleration of the expansion rate of the universe and is allowed within General Relativity. Determining whether dark energy actually is a cosmological constant or if it evolves with cosmic time is a top priority in cosmology. The first results of the DOE Dark Energy Spectroscopic Instrument (DESI) show some evidence ($\sim3\sigma$) that dark energy is evolving \citep{adame2024desi}. The future results of DESI and from the ESA-led Euclid mission, the forthcoming NASA Nancy Grace Roman Space Telescope, and the NSF and DOE Vera C. Rubin Observatory will measure the dark energy equation of state and determine whether and how it evolves with time. These facilities are designed to precisely measure distances and expansion rates in several redshift bins which span the transition era, being the time between the original matter-dominated era of the universe and the current dark energy-dominated era. When exactly this occurs and how it occurs is a strong test of the equation of state of dark energy. If the initial DESI evidence is confirmed, this would falsify both $\Lambda$CDM and General Relativity, simultaneously breaking two standard models and providing crucial constraints on what could produce dark energy.

Distance measures in the universe are performed with standard objects or structures whose known or modeled intrinsic value can be compared with the measurement at Earth to infer distance. For example, sources with standard brightness can use the brightness measure at Earth to infer distance through the inverse square law. Standard sources were used to prove that other galaxies exist, provided additional evidence and precise measures of the (dark) matter content of the universe, and that the acceleration of the universe is accelerating They have and always will be foundational instruments in cosmology. Many standard sources are key TDAMM sources.

The most famous are thermonuclear supernovae (Section~\ref{sec:sources_thermonuclear}), observationally classified as type Ia supernovae, which were utilized for the discovery of dark energy. They can be detected from the reasonably nearby universe to when the universe was only $\sim$20\% of its current age. This allows utilization of a single source past both sides of the transition era. Studying these events, or improving their precision, is a key result for nearly all flagship ultraviolet, optical, or infrared facilities, and is a key source class for the forthcoming cosmology missions mentioned above.

There are two sources of uncertainty in the calibration of thermonuclear supernova. The first is the absolute calibration of the observing telescopes. For example, to ensure substantial progress, the absolute calibration requirement for Roman is 0.3\%. The second is the uncertainty in the calibration of the intrinsic brightness of type Ia supernovae. This is empirically derived from the cosmological distance ladder, where each ``rung'' uses one standard source to calibrate the next. For example, utilizing galaxies that have hosted at least one type Ia supernova as well as a population of well-characterized Cepheid variable stars. 

The belief in the reliability of calibrating thermonuclear supernovae can be seen in modern measures on Hubble Constant, which is the local expansion rate of the universe. Adam Riess was one of the scientists who won a Nobel Prize for the discovery of dark energy through observations of thermonuclear supernovae, which shifted the concordance cosmological model from CDM to $\Lambda$CDM. His group is claiming at $>5\sigma$ discovery significance a disagreement in the value of the Hubble Constant  \citep{riess2022comprehensive} when measured from their observations of type Ia in the nearby universe using type Ia observations as compared to measurements from the old universe using the Cosmic Microwave Background and additional information such as Baryon Acoustic Oscillations (BAO). If true this is falsification of $\Lambda$CDM and requires new physics. 

However, there is significant skepticism in the field on this result. There are numerous options considered to explain away the discrepancy including use of improperly calibrated cosmic ladder rungs, incorrect handling of dust, different progenitors having slightly different thermal emission, etc. Many of these explanations are only problems because we empirically calibrate thermonuclear supernova, i.e., we do not understand how they explode. If we did, we would avoid or minimize reliance on the cosmological ladder, we may be able to better model dust, and we could possibly account for differences in UVOIR signatures from different progenitors. The lack of understanding thermonuclear supernova will limit the scientific return of nearly all major UVOIR facilities, both past, present, and future. It also limits our greater conclusions, including the prevention of the use of type Ia with CMB+BAO studies to probe new physics. It also prevents advances in known physics, such as constraining the sum of the neutrino masses and a resolution to the mass hierarchy problem.

In addition to the new UVOIR facilities, we expect additional advancements in the study of cosmology. COSI will observe direct nuclear gamma-rays from a small number of thermonuclear supernovae, giving knowledge on progenitors for individual events and new information for understanding explosion mechanisms which may enable more precise use of type Ia for cosmology. DUNE will determine whether the neutrino mass hierarchy is normal or inverted \citep{abud2022snowmass}, reducing options in global cosmology fitting. LISA will give rates understanding of the merging of two white dwarfs, giving a direct handle on how many type Ia arise from the detonation of two white dwarfs vs one.

On a longer term horizon we can turn to standard sirens, where a quirk of General Relativity allows direct distance determination from gravitational wave observations of binaries. In these situations the calibration of the sources is irrelevant as it relies only on GR. This leaves only the absolute calibration of the GW detectors \citep{sun2020characterization} as a possible limitation; this sets a key requirement for future gravitational-wave interferometers. There are numerous approaches to utilize GW observations to measure the Hubble parameter; however, ``dark'' approaches, where only GW information is utilized, are unlikely to be sufficiently convincing alone to falsify a cosmological model. It is only multimessenger cases where the GWs measure the distance and the electromagnetic (EM) observations measure redshift where an unambiguous Hubble diagram can be constructed. 

The most common and well understood sources which are viable here are neutron star mergers. The number of events required for sub-percent precision cosmology require the third generation GW facilities, which would additionally allow for recovery deep into cosmic time. The distances required to recover the EM counterpart necessitate precise localizations of prompt gamma-ray burst emission and subsequent follow-up observations. Work has already begun on investigating possible sources of systematic error in these events. Sustained investment in multimessenger modeling is required to ensure these events achieve their cosmological promise, as their better standardization and detection throughout the universe will make them foundational standard sources for cosmology. The other possible source are merging massive or supermassive black holes whose electromagnetic counterparts are less well understood, but successful observation may enable larger statistical samples and provide additional distant events of key interest for some cosmological tests. These will be seen by LISA and possibly the PTAs. LISA and IGWN can be cross-calibrated through the joint observation (separated by years) of binary black hole mergers with $\sim$tens of solar masses.

\begin{quoting}
    \noindent Comment: TDAMM events will always be foundational to measuring distances in cosmology, and are of direct interest to all three funding agencies.
\end{quoting}

Regardless of the resolution to the Hubble Constant debate, and the outcome of the first precise tests of the equation of state of dark energy, the study of cosmology will continue, and it will continue to be a priority for all involved funding agencies. Future questions may include the composition of dark energy or why the universe is flat. For the latter, we currently must rely on fine tuning at all epochs of cosmic time to have our current shape of the universe. This seems unlikely to be the explanation. TDAMM events will always be foundational events for this field of study, and are of direct interest to all three funding agencies. A strategic theory and integration approach is key to maximizing the scientific return of the major facilities built for cosmology, which include flagship facilities of DOE, NASA, and NSF. This work requires large theory and computation investment, and would join active research in modeling radiation flows and require improvements in modeling matter in numerical relativity.

\subsubsection{Extreme Matter}
\label{sec:questions_extremeMatter}

Many astrophysical transients are produced in the formation or evolution of NSs.  NSs push our understanding of matter at extreme densities, above the nuclear equilibrium density of $2.7 \times 10^{14} {\, \rm g \, cm^{-3}}$, which thus also pushes our understanding of quantum chromodynamics~\citep{2001ApJ...550..426L}. This study becomes ever more important, as laboratory studies may be on the verge of yielding evidence about the composition and stiffness of matter beyond the nuclear equilibrium density.  \cite{1974PhRvL..32..324R} demonstrated the assumption of causality beyond a fiducial density sets an upper limit to the maximum mass of a NS.  However, theoretical studies of dense matter have considerable uncertainty in the high-density behavior of the equation of state largely because of the poorly constrained many-body interactions. These uncertainties have prevented a firm prediction of the maximum mass of a stable neutron star.

The composition of a neutron star chiefly depends on the nature of strong interactions, which are not well understood in dense matter. Most models that have been investigated can be conveniently grouped into three broad categories:  non-relativistic potential models, relativistic field theoretical models, and relativistic Dirac-Brueckner-Hartree-Fock models. In each of these approaches, the presence of additional softening components such as hyperons, Bose condensates, or quark matter can be incorporated. By measuring the radii and mass of neutron stars, the mass-radius relation can be used to constrain these models and, as these models improve, ultimately constrain the underlying physics of matter at extreme densities.  

But to tie measurements of the NS radius to its interior, we must also understand the NS crust.  Although the physics of the NS crust is often considered better understood than that of the NS interior~\citep{2012PhRvC..85c2801G}, understanding this nuclear pasta phase leveraging advances in materials-physics modeling is critical to our interpretation of observations~\citep[e.g.][]{2024arXiv240914482M}.  In addition to these models, astrophysicists must develop models that connect the observed emission (e.g. X-ray) to a neutron star radius.  To constrain quantum chromodynamics, astronomers and physicists must develop a deeper understanding of all of these fields, developing:  detailed equation of state models, materials models of neutron star crusts and emission models.

Fortunately, both laboratory experiments and astrophysical observations can be combined to probe this physics.  
Ground-based experiments have historically led the study of dense matter, with accelerators mapping the properties of the proton distribution in atoms as well as the equation of state of dense symmetric nuclear matter. Their modern study on asymmetric nuclear matter arises because heavy atoms have an excess of neutrons, necessary to overcome Coulomb repulsion in the nucleus. Here the neutron distribution is larger than that of the proton distribution, referred to as the neutron skin. The larger the neutron skin here, the larger the radius of a neutron star. The thickness of this skin was first accurately measured in a model-independent manner only recently \citep{adhikari2021accurate} and seems to show some tension with other methods.

NASA's Neutron Star Interior Composition Explorer (NICER) mission provided precise measures of the mass and radius of a small number of neutron stars in the Milky Way through incredible timing and spectroscopy capabilities which mapped the surface of these objects, accounting for GR effects. The  multimessenger constraints on the equation of state from the GW and EM observations of the binary neutron star merger gave yet another observational probe. Here the equation state can be probed through GW observations of the acceleration of the compact object inspiral due to matter effects, multimessenger classification of the merging objects and immediate remnant object, and more \citep{burns2020neutron}. Future GW interferometers may even be able to directly measure a phase transition from hadronic to quark-gluon matter during a binary neutron star merger \citep{marranghello2002phase}. 

Another probe of this physics can be constrained by observations of the cooling of neutron stars~\citep{2004ARA&A..42..169Y}.  Temperature measurements of newly formed neutron stars probe both conduction properties in the crust and alternative physics in the interior~\citep[e.g.][]{2025JHEAp..45..116P,2025JHEAp..45..371S}.  The formation process itself (e.g. the supernova explosion) also provides insight into this extreme physics.  Indeed, the highest temperature and density conditions in astrophysical phenomena occur in the bounce phase of stellar core-collapse.  These physics studies include probes of dark matter candidates (axions, dark bosons, etc.), neutrino oscillations and sterile neutrinos as well as the studies of dense matter equations of state~\citep[e.g.][]{2009PhLB..670..281F,2016PhRvC..94d5805R,2021PhRvL.126g1102C}. This physics alters the explosion energy, pulsar kicks and nucleosynthetic yields.  But we must disentangle these effects form the other uncertainties in the supernova/gamma-ray burst explosion mechanisms to probe this physics.

\begin{quoting}
    \noindent Comment: The proposed cosmic density ladder outlines an integrated approach to understanding the behavior of dense matter. This is likely to be one of the first NRC questions of the century to be answered:
\end{quoting}

Because dense matter is expected to follow the same underlying physics and the various approaches probe different but overlapping regions of parameter space, a ``density ladder'' has been proposed, analogous to the cosmic distance ladder \citep{piekarewicz2019neutron,reed2021implications}. This approach was endorsed by the LRP \citep{aidala2023new} because the advances of modern chiral effective field theory, current and forthcoming neutron skin measurements, additional heavy ion collision studies and new facilities like FRIB to study, and the aforementioned astrophysical probes will unlock precise understanding of QCD around suprasaturation densities. We note that most of these are effective probes of cold dense matter, whereas the birth or end of neutron stars can involve hot dense matter, providing yet another region of parameter space to probe.

However, greater integration of fields would be beneficial. There are now frameworks capable of inference on data from both astrophysics and nuclear physics \citep[e.g.][]{pang2021nuclear}; however, this relies on accurate posteriors from both fields. This is not an easy for subject matter experts in a given field, let alone across fields. In astrophysics some measures are more reliable than others. For example, inference on the equation of state of neutron stars based on the ejecta mass from a binary neutron star merger rely on the accuracy of kilonova models, which disagree on the ejecta mass value by an order of magnitude. Additionally, GW tests on the equation of state typically rely on phenomenological models (e.g. polytropes) rather than physics-informed models will become a significant limitation as observational fidelity and measurement precision improves. Intentional and strategic engagement across fields is necessary to properly and holistically handle data to understand dense matter. This includes advancing techniques within disciplines, such as improved numerical relativity handling of matter effects.

\begin{quoting}
    \noindent Comment: Neutron stars and magnetars will remain unique laboratories for probing other extremes of physics, such as color superconductivity and magnetized superfluids. While the integration of the relevant fields is nascent, the uniquely enabled science warrants investment.
\end{quoting}

The prior discussion in this section has focused on dense matter because the questions to answer are known, advances are expected, and the field is ripe for multidisciplinary studies. However, the title of the section is Extreme Matter because neutron stars also allow us to probe extremes of matter other than densities, namely a novel form of superconductivity and superfluidity \citep{bohr1958possible,baym1969superfluidity,lattimer2004physics,haskell2018superfluidity}. Here superconduction arises not from electrons forming Cooper pairs but from nucleon interactions. These contribute to neutron star glitches, magnetic fields, oscillations, and additional transient activity seen in magnetars, and are possibly associated with fast radio bursts \citep{2022ApJ...931...56L,2024Natur.626..500H}. Greater understanding of the astrophysical activity they power may advance understanding of these extreme physics behaviors. The discoveries to be made here will provide a deeper understanding of the physics of the cosmos, though possibly on a longer timescale than others.


\subsubsection{Energy Extraction from Black Holes}
\label{sec:questions_BZ}

A Penrose Process is a way to extract energy from a rotating black hole, named after Roger Penrose, who first unveiled its underlying principles in 1969 \citep{penrose1969gravitational}. Because of mass-energy equivalence, extraction of energy from a black hole results in a lowering of its mass. That is, a Penrose Process converts the mass of the black hole into energy, which is returned to the universe. Most Penrose Processes are utilized as thought experiments in gravity theory research; the Blandford-Znajek mechanism \citep{blandford1977electromagnetic} is an unusual Penrose Process as it is astrophysically motivated. More recently, magnetic reconnection has also been identified as another mechanism for energy extraction from rotating black holes \citep{CA21}. Proof that the Blandford-Znajek process is occurring in nature is one way to prove a Penrose Process occurs in reality. It is thus a path to testing whether mass can be rapidly extracted from a black hole. This also corresponds to probe as deep into a black hole as is thought to be possible.

\begin{quoting}
    \noindent Comment: Proof that the Blandford-Znajek process is occurring in nature is one way to prove a Penrose Process occurs in reality. It is thus a path to testing whether mass can be rapidly extracted from a black hole. This also corresponds to probe as deep into a black hole as is thought to be possible.
\end{quoting}

Non-rotating black holes have an event horizon radius determined by the Schwarzschild radius, $r_s = 2 G M / c^2$, where $G$ is the Gravitational constant, $M$ the mass of the black hole, and $c$ the speed of light. If the black hole is rotating the event horizon shrinks, down to half $r_s$ at maximal rotation. For rotating black holes an ergoregion develops that extends beyond the event horizon. Particles (or waves) that enter the ergoregion must co-rotate with the black hole, relative to distant inertial observers. Thus, they become entrained in the rotation and can gain energy from the black hole angular momentum. The Blandford-Znajek mechanism utilizes magnetic fields that thread the ergoregion and are thus spun up about the black hole poles. When accretion is occurring matter is channeled to the poles, resulting in collimated outflows, referred to as jets. Thus, the Blandford-Znajek mechanism converts the rotational energy of the black hole into kinetic energy driven along the axis of rotation, which is a type of $\Omega$-dynamo. 

The Blandford-Znajek mechanism was originally invoked to explain how to power Active Galactic Nuclei (AGN), but has been invoked or considered for nearly all jets which arise from black hole central engines, and are expected to be occurring in gamma-ray bursts. However, jets are powered by sources other than black holes, and thus there are other viable mechanisms. Additionally, there are jet-powering mechanisms for black hole engines which are all some form of an $\Omega$-dynamo which convert angular momentum in a plane into fast outflows along the polar region; for example, extracting the angular momentum of the disk into the jet power source. There are certainly situations which appear to require an $\Omega$-dynamo effect, but it is not yet clear if the rotational energy of the accretion disk is sufficient to power all known jets or whether the black hole rotational energy must also be tapped.


This is perhaps the most complex multiphysics problem in which a path to advancement can be seen. Numerous observational diagnostics can be observed in a wealth of sources to attempt to isolate individual questions. These include studying what jets can be formed by non-black hole engines. These can be probed by observations of accreting neutron stars and other Galactic objects (Section~\ref{sec:sources_xrayBinaries}), and multimessenger observations of neutron star mergers (Section~\ref{sec:sources_mergers}). Comparing the properties and limits (e.g. maximal energy) of these jets to those arising from black holes observed from the same sources as well as tidal disruption events (Section~\ref{sec:sources_TDEs}) and active galactic nuclei (Section~\ref{sec:sources_agn}) may elucidate key differences. Jet launch mechanisms can include neutrino annihilation driving a relativistic fireball, $\Omega$ dynamos from magnetars, accretion disks, or the Blandford-Znajek process. Each can alter the expected energy content of the jet, being Poynting flux dominating or kinetic energy dominated. Probes of jet energy content include modeling of multiwavelength signals, searches for associated neutrino emission, and polarization. The latter is key as it allows understanding the order and scale of magnetic fields, which tie to the jet launch mechanism, and provides unique information for modeling of these sources. Observations with NSF radio and optical polarimeters together with data from X-ray and gamma-ray energies by NASA's polarimeters, Imaging X-ray Polarimetry Explorer (IXPE) and COSI may be required for advancement. This then drives the development of computational approaches in fluid dynamics which make predictions on polarization signatures in different situations. One crucial measurement is the black hole spin, which can be probed through various means such as observational determination via X-ray reverberation mapping or theoretical modeling of allowable ranges for distinct source types will be key to separating contributions to power from Blandford-Znajek verse other mechanisms. 

Answering this question is one of the most difficult goals outlined in this paper. The physics extremes involved are still orders of magnitude beyond what could be probed in terrestrial laboratories in the coming decades. However, the scientific gain is enormous and incremental advancements can include major discoveries. The ergoregion is the deepest in a black hole that can be probed; thus, testing whether the Blandford-Znajek mechanism is operating in the universe is also a unique test of gravity. Answering this question also probes how the most powerful cosmic accelerators work and relates to what is being accelerated. This includes seeking the source of ultra-high energy cosmic rays, one of the eleven questions from the century NRC report \citep{national2003connecting}. These questions are of interest to DOE, NASA, and NSF. Advances require investment in theory and simulation, use of flagship ground-based facilities in HEDP and plasma physics, and coordinated observations with astronomical facilities.

\subsubsection{Photon Splitting}
\label{sec:questions_photonSplitting}
Quantum Electrodynamics (QED) is the quantum field theory for electrodynamics, describing how light and matter interact. QED is one of the most precisely tested and validated physics theories, but its domain of applicability in Nature has not been fully studied, particularly in its non-linear domain where photons, virtual particles and fields interact with themselves (See \cite{Gonoskov_RevMod_2022,fedotov.pr.2023} and reference cited therein). Indeed one of the Eleven Questions is: Is a New Theory of Matter and Light Needed at the Highest Energies? This report highlights two complementary areas to probe this question. 

The first is the advancements in ground-based plasma and HEDP experiments, which become more and more accessible with the fast development of high power laser technology, enabling the design, building, and operation of multi-PW laser facilities with peak intensities reaching $10^{24}$ W/cm$^2$ \citep{mp3report.2022.arxiv}. These facilities will allow the study of the interaction of high energy charged particles and photons with strong electromagnetic fields, giving rise to prolific electron-positron pair production, different types of electromagnetic cascades, and the transformation of electromagnetic field energy into the energy of secondary particles. 

\begin{figure}[ht]
\begin{center}
    \includegraphics[width=\textwidth]{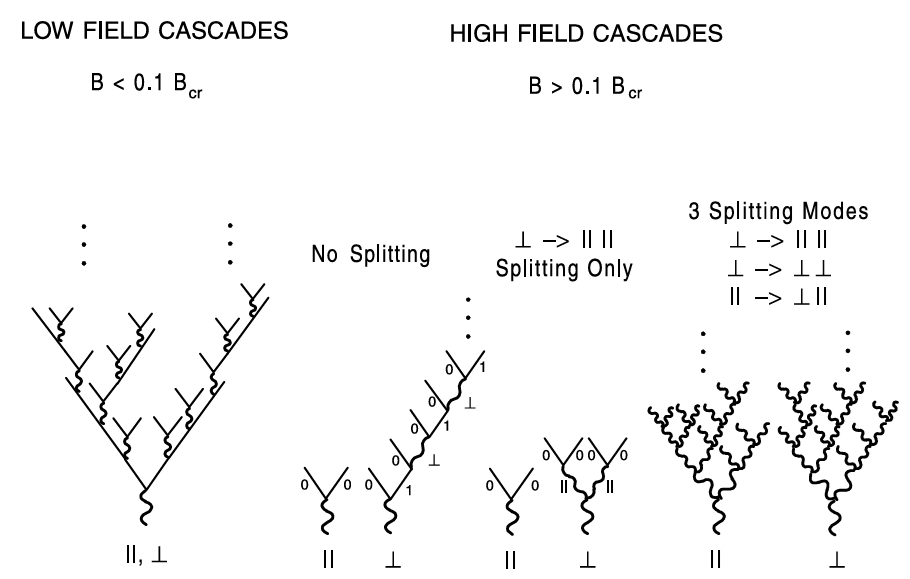}
    \caption{Pair cascades over a range of magnetic field strengths and orientations (parallel or perpendicular). Straight lines are electrons and positrons, with photons as wavy lines. Borrowed from \citet{2001ApJ...547..929B}}\label{fig:photonsplitting}
\end{center}
\end{figure}

The second is astrophysical soft gamma-ray observations of magnetars, which are neutron stars with magnetic fields up to around $10^{15}$~G, being well in excess of the quantum critical field of $4.4 \times 10^{13}$~G where exotic QED effects are expected to be both important and inevitable \citep{1992herm.book.....M,2006RPPh...69.2631H}. Photon splitting $\gamma\to\gamma\gamma$ is one such exotic phenomenon, wherein the presence of a strong field a photon may split into two \citep{1971AnPhy..67..599A} well below the magnetic pair creation ($\gamma \to e^+e^-$) threshold of 1 MeV \citep[e.g.,][]{Erber-1966-RvMP}. These two processes are permitted only in the presence of magnetic fields, and both possess rates that are extremely sensitive to the strength of the magnetic field {\bf B}, the angle $\theta_{\rm kB}$ that photons propagate relative to the field direction, and the energy of a photon. Importantly, Lorentz invariants and plasma parameters of magnetars sample a different regime than laser experiments, providing a complementary test of QED. 

\begin{quoting}
    \noindent Comment: Time-resolved spectropolarimetric studies of magnetars are one of the few methods to probe an untested prediction of quantum electrodynamics. 
\end{quoting}

The principal signature of the action of photon splitting in magnetars is the possible attenuation of hard X-ray photons in the 50~keV - 1~MeV band. This may best present itself in the apparent truncation around 200~keV of persistent hard X-ray tail emission, which is seen above 10~keV in around 10 magnetars \citep[e.g.,][]{Kuiper-2004-ApJ,Goetz-2006-AandA,denHartog-2008-AandA,Enoto-2010-ApJ,Younes-2017-ApJ}. Magnetospheric opacity due to photon splitting has been addressed in numerous papers \citep[e.g.,][]{HBG-1997-ApJ,BH-2001-ApJ,Hu-2019-MNRAS,Hu-2022-ApJ-opac}, with the general finding that it is active only above around 50~keV, and that it is very dependent on the magnetospheric locale and the observer's viewing direction. This translates into the splitting opacity being strongly dependent on the pulse phase of the emission \citep{Hu-2019-MNRAS}, so that with sensitive spectroscopy in the 50~keV-1~MeV band, one could in principle discern a cutoff energy that varies significantly with phase, spanning a range of a decade or more for select viewing perspectives.

In addition, if the emission is due to resonant inverse Compton scattering \citep[RICS;][]{BH-2007-ApandSS,Wadiasingh-2018-ApJ}, the highest energy photons are naturally preferentially emitted in the $\perp$ polarization state (X-mode).  In the limit of a weakly dispersive, magnetized QED vacuum, \cite{1971AnPhy..67..599A} posited polarization selection rules for $\gamma\to\gamma\gamma$, with the only permitted transition being $\perp\to\parallel\parallel$.  Thus, splitting opacity converts X-mode photons to O-mode ones (the $\parallel$ state). Accordingly, it is expected that unattenuated RICS emission will exhibit a dominance of X-mode photons below a splitting spectral turnover, while at energies in the turnover, splitting will yield a dominance of O-mode photons. This spectro-polarimetric transition is believed to be a hallmark signature of the action of photon splitting in magnetars, and accordingly underpins motivations for a future sensitive polarimeter in the hard X-ray or soft gamma-ray band \citep{2019BAAS...51c.292W}.

Given the importance of such a discovery, and the inherent difficulties, both approaches are necessary. The temporally-resolved spectropolarimetric studies in soft gamma-rays are the domain of NASA, and the ground-based experiments the responsibility of DOE. Fostering connections between these areas and all the science in between should be a priority.

\subsubsection{Out of Equilibrium Physics}
\label{sec:out_of_equilibrium}
Local thermodynamic equilibrium (LTE) is a term used to describe matter where it is assumed that all of the particles are in a steady state and fully coupled:  electron and ion energy distributions are described by a Maxwellian and photons are described by a Planckian (all with the same temperature - particles are fully coupled with each other).  In complete equilibrium, all reactions (e.g. nuclear or chemical reactions, atomic level excitation and de-excitation) are also in equilibrium (forward and reverse rates are equal).

There are very few applications where local thermodynamic equilibrium is valid and most astrophysics codes assume some aspect of the physics is out of equilibrium.  For example, stellar and supernova calculations typically include a time-dependent network to calculate nuclear burning.  But in these astrophysics applications, the reactants are assumed to be in equilibrium (and the burning rate is dictated by integrating over the Maxwellian distribution of Maxwellian energies).  Supernova light-curve calculations often capture the full energy distribution of the photons (using either continuous or multi-group solutions).  Although many supernova calculations assume that the atomic level states are in equilibrium, some apply their non-Planckian radiation spectrum and, in some cases, non-Maxwellian electron energy distributions, to study the atomic level states.  Because out-of-equilibrium solutions are computationally intensive, many studies assume equilibrium even when it is not valid.  For example, in calculations of both astrophysics and inertial confinement fusion typically use Eulerian hydrodynamics assuming Maxwellian distributions of the ions and electrons in conditions where an equilibrium pressure is not accurate.  

In the nuclear physics component of TDAMM, whenever radioactive beam facilities (FRIB) play a role in nuclear astrophysics there is non-equilibrium nucleosynthesis going on. Time-dependent solutions are key to understand observed abundance features, and the interplay between nuclear physics and astrophysics 3D macro-physics time scales is essential and a huge simulation and theory challenge. From a theory and simulation approach the 
Damks\"{o}hler number regime close to unity is not yet fully understood and only now serious explorations are starting. 

But a systematic approach to characterize the errors introduced by equilibrium assumptions has not been developed.  Indeed, many scientists are unaware of the equilibrium assumptions made in their studies.  For a broad range of fields, an analysis of the out-of-equilibrium effects is crucial.  When out-of-equilibrium effects are deemed important, solutions must be developed to adequately capture these effects.  The difficulty in modeling these conditions is that a fully, self-consistent study of all the possibly breaks in equilibrium for a given application is beyond the capability of most computational methods.  Identifying the most critical out-of-equilibrium features of any applied problem and constructing a solution that captures its effects to the needed accuracy is a problem that plagues scientists in all fields from fusion energy science to astrophysics.

\begin{quoting}
    \noindent Comment: The study of physics outside of equilibrium has become far more important in recent years. Adaptation of the techniques from these studies for use in interpreting TDAMM data is necessary in specific regimes. TDAMM can also provide unique data to test out-of-equilibrium approaches. 
\end{quoting}

One of the prime examples of out-of-equilibrium effects occurs in plasma physics.  In collisionless shocks, the ion and electron energy distributions will not be coupled (can not be described by the same temperature) and, indeed, may not be Maxwellian at all.  As we have discussed in Section~\ref{sec:disciplines_plasma}, a broad range of techniques have been developed to resolve the plasma physics on the small scale and to bridge these small-scale calculations to the full scale applications.  Identifying which techniques are best-suited for different applications and different conditions.

TDAMM Astrophysics applications are excellent laboratories of multi-scale, multi-physics integration of different operators, more often than not related through non-equilibrium (or non-LTE) physics coupling.  Scientists in a broad range of fields can use these astrophysics phenomena to test the robustness of their codes, and use TDAMM observations for validation.  Moving astrophysics into tackling how to handle non-LTE physics would bring us on par with other advancing fields in physics. Our capability of observing non-LTE, neutral plasmas provides a venue for these fields to test their approaches in a different regime, allowing for new mechanisms of validation.

\subsection{What must be done to Succeed}
\label{sec:what}
Major advancements in understanding or fully answering any of these complex problems can only be made with advances in multiple areas. This involves both capturing the necessary fidelity data in astrophysics and ground-based experiments and creating the necessary models for comparison. A great deal of effort is spent in gathering the proper data, both in the construction of new facilities and in the global astronomical coordination effort, where improvements are on-going. The ground-based experimentalists have continued to advance their own facilities and utilized them to answer their own priority questions. In both situations, this work is rarely guided by the needs of big picture which requires and understanding of multiple fields. Computational simulations must be able to match or utilize results from ground-based experiments and accurately model incredibly complex multiphysics problems. Generally, multiple stages of computational simulations are required, with the output of one forming the input of the other, and creating dependent chains. Lastly, the observed astrophysical data must be compared against the range of viable models.

\begin{quoting}
    \noindent Finding: A true end-to-end approach is needed for transformational understanding in the physics of the cosmos through TDAMM science. This is beyond the scope and effort possible by any individual discipline, facility, or agency.
\end{quoting}

This is ideally done in a coherent manner, where the observations and experiments performed are guided by the known or possible limitations of knowledge or the fidelity of applicable models. The models are iteratively improved as additional data is gathered. All of this is done in pursuit of answering the particular question of interest. That is, a true end-to-end approach is needed for transformational understanding in the physics of the cosmos through TDAMM science. However, this is beyond the scope and effort possible by any individual discipline, facility, or agency.

The NNSA is an entity that regularly and successfully executes such an approach. It is a semi-autonomous agency within the DOE responsible for military applications of nuclear science and predominantly operates through its three national laboratories, Los Alamos National Laboratory, Lawrence Livermore National Laboratory, and Sandia National Laboratory. The laboratories house some of the largest collections of scientists and engineers on the planet. The NNSA also operates a number of experimental facilities at various scales that are capable of performing experiments to probe new regimes of interest. When a key question is identified a team of relevant subject matter experts, possibly across laboratories, is assembled to determine the answer. Typically, the necessary work to be done is aligning existing knowledge, adapting experimental setups, or integrating codes. This is approximately 95\% of the required effort. The remaining 5\% requires new and novel work. 

The NNSA has a number of advantages, born out of necessity, in implementing this approach. First, it is a single entity built for this purpose, which houses the majority of the relevant expertise and facilities. Second, the budget of $\sim$\$24B ensures that the necessary work can be carried out. This is comparable to the total open science budget of DOE, NASA, and NSF combined. However, the NNSA has the disadvantage that it requires a unique workforce and must recruit from open science. This problem has drawn the attention of Congress, which has directed the NNSA to understand whether open science can help solve their workforce problem. This presents an opportunity.

Because of the fidelity of observational data now being gathered by major investments of DOE, NASA, NSF, and numerous other agencies, major progress in TDAMM requires the emulation of this approach. Numerous bifurcations that act against a unified approach prevent this from happening. There is no strategic plan to identify the full end-to-end work that must be performed to answer key questions. For example, the Astrophysics Decadal does not formally make recommendations for NSF Physics or the particle portion of DOE which are responsible for all gravitational wave and neutrino detectors. Further, the Astro Decadal process does not broadly engage with physicists whose work and expertise are directly relevant for TDAMM advances, such as the nuclear side of nuclear astrophysics. This is despite the fact that the annual expenditure and cost of the relevant assets for TDAMM are comparable to those of individual fields with their own formal strategic planning documents. In principle, the community could overcome the lack of strategic planning, but is divided by disciplines, funding agencies, professional societies, annual meetings, and advisory committees, as summarized in Table~\ref{tab:divisions}. Sufficient investment and community support to build bridges over these divides, along with a strategic approach, could maintain the US leadership in several disciplines of science. It could also create a pipeline of capable scientists and engineers to protect national security, revitalizing a national strategic use for astrophysics. The specific needs include generalist scientists and code developers, while astrophysics is increasingly producing only specialists and code users.

\begin{table}[!htbp]
    \centering
    \begin{tabular}{|r|c|c|c|c|c|}
    \hline
        Discipline & Key Society & Key Division(s) & Major Meeting & Planning Process & Key Agencies \\ \hline
        Astrophysics & AAS & HEAD & AAS Winter & Decadal & NASA, NSF, DOE \\ 
        ~ & APS & DAP & APS April & ~ & ~ \\ 
        Laboratory Astrophysics & AAS & LAD & AAS Summer & NAS (infrequent) & DOE, NASA, NSF \\ 
        Nuclear Science & APS & DNP & DNP Meeting & Long Range Plan & DOE, NSF \\ 
        Atomic Science & APS & DAMOP & APS March & Decadal & DOE, NASA \\ 
        Fluids and Turbulence & APS & DFD & APS March & - & DOE, NSF \\ 
        Radiation Transport & - & - & - & - & DOE \\ 
        Gravity & APS & DGRAV & APS April & - & NSF \\ 
        High Energy Density Physics & - & - & - & NAS (infrequent) & DOE \\ 
        Computation & APS & DCOMP & APS March & - & DOE, NSF \\ 
        Plasma Physics & APS & DPP, GPAP & APS March & Decadal & DOE, NSF \\ 
        High Energy Physics & - & - & APS April & P5 & DOE, NSF \\ \hline
    \end{tabular}
    \caption{A summary of the various divisions across disciplines. Astrophysics has two entries given the split between astronomy and astrophysics. The disciplines are additional divided by Federal Advisory Committee Act (FACA) committees and National Academies of Sciences (NAS) committees. The abbreviations are all defined in Section~\ref{sec:acroynmsAndDefinitions}.}\label{tab:divisions}
\end{table}

As a concrete example, we summarize what must be worked out for complete modeling of core-collapse supernova. There are seven phases which must be worked through and mapped into a self-consistent picture: the progenitor star, core collapse itself, shock breakout and passage, the photospheric phase, the nebular phase, the remnant phase, and presolar grains. The first six are the domain of telescopes and the central five the domain of TDAMM. Each of these five have new capabilities coming online recently or in the coming years: core-collapse can be studied directly by new neutrino and GW detectors, shock breakout by Einstein Probe, ULTRASAT and UVEX, photospheric phase by Rubin, the nebular phase by JWST, and the remnant phase by COSI. In order to consider all of these unique observables, we require simulations of every stage, creating a chain of simulations that must be linked together. The final output compared to the observations is limited by the weakest link in the chain. For example, if fluids are improperly handled in the progenitor star then the shock breakout model will be inaccurate. Each simulation stage requires the handling of multiple domains of physics, which change with each phase, thus requiring the tie-in to the other disciplines of science. Probably only such coordinated chains can ultimately develop the integrated multi-physics predictions that TDAMM observation interpretations require. Nearly all questions or astrophysical sources have a similar need. For example, while AGN do not significantly evolve on human timescales, multiple scales must be handled and integrated to understand their jets. Tidal disruption events are analogous to AGN and develop on $\sim$weeks timescale, giving separate handles on similar problems. Until this work is done, the best facilities to build or the best observing plans to follow to answer questions or provide new diagnostic information cannot be known with certainty.



\subsection{How to Foster Success}
\label{sec:findings}
How this can be done given the numerous divides that exist, especially between funding agencies, is not a solved problem. It certainly requires intentional, coordinated, and strategic effort from the funding agencies, from community members, and from additional institutions, including the advisory committees and the professional societies. Success means sustained investment over many years. As this general problem of coordination has not already been solved, despite formal recommendations \citep[e.g.][]{national2003connecting}, it seems unlikely that the ideal scenario will occur. However, many individuals have pursued this work over the long term with some success. Below we summarize our own findings in how this endeavor could be supported; however, any solution and approach to resolve the identified issues are certainly welcome. We also consider this workshop and white paper only one part of a chain of events in the pursuit of greater integration of aligned sciences, and look forward to other individuals building on this work and pursuing these grand coordinated and scientific efforts in the coming years. In astrophysics specifically the half-Decadal review is eminent. We do not comment on the specific TDAMM Decadal responses except to state that they are all of direct relevance as powerful diagnostics in TDAMM, and we support their continuation. The strategic response to the non-specific TDAMM recommendations have been explored in this TDAMM series of workshops; the findings from the first two and the status of the response is explored in Section~\ref{sec:multidisciplinary_pastworkshops}.

During the workshop, input was solicited on what discussion topics would be most fruitful. They included the following questions, which led to the set of topics explored below.
\begin{itemize}
    \item How can we foster multidisciplinary work? What can be done by the agencies? Professional societies? The community? 
    \item How can we develop early career scientists with multidisciplinary expertise?
    \item Do funding gaps exist (e.g., by scope, breadth, funding amount)? Do funding calls provide incentives to motivate multidisciplinary research.
    \item What programmatic changes could encourage greater collaboration? 
    \item What specific focused meetings should follow this? Where could/should they be done?
    \item What specific papers or archival representation would be most useful for multidisciplinary work?
\end{itemize}

Many of the findings listed below are rooted in changing the existing system in order to foster growth of multidisciplinary work in the community. In the vernacular of game theory, this is altering the rules to escape Nash Equilibrium. This is most commonly known in the Prisoner's dilemma. In this scenario two individuals, A and B, are being questioned by police for a crime they committed together. If both testify against each other, both serve two years. If only one testifies against the other, the testifier goes free and the other serves three years. If both remain silent then they serve only one year. While the optimal solution for them is coordination, that risks a greater punishment than testifying. We have inadvertently created a similar scenario in how astrophysics observing proposals are handled. If the community came together and set the best observing plans with the limited resources available for TDAMM observations and that data was made available to all groups interested then this would result in the maximal scientific return with these facilities. Instead, groups compete against each other for telescope time, sometimes proprietary, all to answer the same underlying question. This results in competing observations, observations which are not optimized with each other, groups analyzing limited data sets, and rushing to publication to be first. Breaking this dilemma is the focus of the fourth workshop, briefly summarized in Section~\ref{sec:multidisciplinary_observingPlans}. A similar programmatic alteration to foster larger collaborative efforts through authorship options is discussed in Section~\ref{sec:multidisciplinary_authorship}. While the need for coordination across agencies has been identified, and work is on-going, there is still little incentivization for the community to do so, as discussed in Section~\ref{sec:multidisciplinary_incentives}.


Beyond the sociological issues, technical problems still remain. As flagged in the Astro Decadal and previous white papers, investment in the appropriate analysis methods for TDAMM analysis can enable a great return for comparatively small investment, as explored in Section~\ref{sec:multidisciplinary_analysis}. A similar problem is the lack of true uncertainty which limits scientific return and prevents targeted improvements at the most poorly understood reason, Section~\ref{sec:multidisciplinary_multiphysics_UQ}. 

Fostering multidisciplinary science to include the broader community is necessary for sustained success. The curation of data from each discipline for use by scientists in the others is a place ripe for high reward from low investment, Section~\ref{sec:multidisciplinary_data}, which we note would also foster ease of access within each field. Workshops, meetings, and schools are where networks are built and students learn. These are fantastic mechanisms with broad opportunities which can be brought to bear on the TDAMM multidiscipline problem, Section~\ref{sec:multidisciplinary_meetings}. 

Following the example of the NNSA, the overwhelming majority of the work to be done requires aligning existing component pieces. We briefly mention major areas of support which can be used for multidisciplinary TDAMM studies and possible gaps in Section~\ref{sec:multidisciplinary_support}. The reason we that astrophysics is ready to move beyond approximate models in this area is because of the wealth of new facilities arising in the US and worldwide. We emphasize this and explore, should funding allow, what additional facilities would complete the global picture, Section~\ref{sec:multidisciplinary_facilities}. 

\subsubsection{Findings from the past TDAMM Workshops}
\label{sec:multidisciplinary_pastworkshops}

As this is the third workshop in a series, it is appropriate to consider whether the findings of the community in the previous workshops have resulted in positive developments in the field and to consider the findings in the multidisciplinary approach of this meeting. Here we repeat the summary findings of the first two workshops. For the first workshop, we summarize the bullets from the recent presentation to the APAC\footnote{\url{https://assets.science.nasa.gov/content/dam/science/cds/researchers/nac/apac/2024/APAC_Nov_2024_Agenda_Finalv2.pdf}}:
\begin{enumerate}
    \item Real-Time Cyberinfrastructure - Real-time transient detections; Software to do joint data analysis; Archive coordination 
    \item Theory Funding - Specific urgent topics; Interdisciplinary aspects w/ physics, lab Astro, cosmology; Precursor/preparatory science; High performance computing simulations
    \item TDAMM General Observer Facility - To streamline transient follow-up with NASA facilities; Reduce coordination burden from observers; Provide scheduling options; Assist with proposals preparation and submission; Manage funding
    \item NASA-NSF-International Coordination - Optimize observing schedules; Archives and alerts standardization; Joint proposals opportunities
    \item Continuity of Capabilities Across EM Spectrum - Wide-field, high-cadence imaging capabilities, especially in the UV, X-rays and gamma-rays; Rapid follow-up wide-field imaging and spectroscopic capabilities; In particular, wide-field, fast response and arcminute-scale localization X-ray and gamma-ray monitors are needed to replace the aging Swift and Fermi telescopes
    \item Crediting hidden figures - Data scientists, software/hardware developers, managers; Ensure appropriate rewards and recognition 
\end{enumerate}
  
For the second workshop, we edit (for brevity) the summary findings in the white paper \citep{ahumada2024windows}:
\begin{enumerate}
    \setcounter{enumi}{7}
    \item An endorsement of the Decadal recommended upgrade to the gravitational wave and neutrino facilities to NSF Physics and to the next-generation Very Large Array (ngVLA) for the future of radio to NSF Astronomy
    \item Emphasizing the future needs of optical and infrared as including both wide-field high-cadence surveys which cover the full night sky and community accessible spectrographs. Both are needed for the discovery and full characterization of rare transients, which are often the most useful objects for advancing understanding
    \item Infrastructure success requires the development of a viable software career path, both for recruitment of industry developers and support for industry programmers. Key steps include an enhancement of funding for software development in TDAMM, multi-year funding for sustainable jobs, greater emphasis on software by the funding agencies and the community, including fellowships and awards. 
    \item Similarly, the success of the broader software ecosystem is dependent on interoperability and sustained investment. The former is the only way distributed development can be integrated into a larger whole, especially between agencies or countries. The latter is necessary to allow development around core assets, rather than continually switching between the most recently developed platform. As an example, the broad adoption of GCN in multimessenger astronomy has grown because of the long-term investment by NASA. 
    \item Among the top recommendations was greater coordination between NASA and the NSF and international agencies. And additional support by the agencies for multi-facility development and observer facilities such as the NASA ACROSS and NSF AEON initiatives.
    \item Lastly, the final recommendation is the development of community-driven observing plans with exceptional events and minimal proprietary periods in other cases.
\end{enumerate}

NSF and NASA have made investments in the area of infrastructure. As this was the focus of the second workshop, we only briefly comment on the current status. This workshop informed a bespoke multi-year NSF call\footnote{\url{https://new.nsf.gov/funding/opportunities/mmc-wou-multi-messenger-coordination-windows-universe}} whose selections were announced in the fall of 2024. It is unclear whether the problem of sustainable investment has been addressed, though this is understandably limited given the time since the workshops. The NASA investment in GCN and its archives as well as the newer NASA ACROSS and NSF AEON initiatives should result in sustained investment in infrastructure that the community can build around. These steps have at least begun to address items 1, 3, and 12. The fourth workshop will help to address items 4 and 13. Given the limited time, it is not yet clear whether sustainable software development in the community is supported, but past steps are promising. We find that continued coordination between NSF and NASA in this area to be necessary to fulfill all findings of the past reports, including items 10 and 11. These investments also help develop a diverse workforce, item 6. 

However, one possible limitation of these past workshops is their somewhat singular focus on providing findings to only funding agencies. Although these institutions certainly wield a large influence in shaping the field, they are not alone in this capability. This workshop, and hopefully future workshops, will address the professional societies, journals, and other relevant institutions, where appropriate. We hope future workshops with adopt this approach, if it proves fruitful. These institutions have a role to play in item 6 and 7 above.

\begin{quoting}
    \noindent Finding: NASA and NSF have engaged with the community through the TDAMM series of workshops. They have been responsive to the community findings. This approach has been beneficial and should continue.
\end{quoting}

For facilities, the NSF and NASA have initiated work in all direct recommendations from the Decadal, including the ELTs, the ngVLA, the upgrades to the GW and neutrino telescopes, and HWO. NASA has met one of the key major investment needs identified in the first workshop: wide-field and spectroscopic capabilities in ultraviolet. The NSF has continued investment in wide-field optical and infrared monitors beyond only Rubin, including the extension of existing facilities. NASA Planetary will launch a wide-field infrared observatory this decade, with NASA Astrophysics covering TDAMM operations costs. We here support the finding for the need of spectrometers and polarimeters to characterize key transients of interest, as these are crucial diagnostics for answer the complex questions of TDAMM science.

In short, it is evident that commonalities have arisen in all three reports, which are further explored in the following sections. This include the few remaining key TDAMM facilities (Section~\ref{sec:multidisciplinary_facilities}), additional or focused theory funding (Section~\ref{sec:multidisciplinary_support}), and the development of joint analysis software (Section~\ref{sec:multidisciplinary_analysis}). It is also evident that NASA and the NSF are generally pursuing Decadal priorities and responding to the detailed community input gathered in these workshops and through other means.

\subsubsection{Community-Driven Observing Plans}
\label{sec:multidisciplinary_observingPlans}


The operational methodology of astronomical facilities varies greatly. Wide-field survey instruments like LIGO and Fermi observe huge swaths of the sky and their signal recovery is done via algorithms. Pointed telescopes operate either on predetermined observing plans or interrupt for Target of Opportunity observations of specific events of interest, determined either by the instrument team which operates them or through competitive proposal selection. These decisions are overwhelmingly made in isolation, often leading to concurrent observations with similar facilities and private data for key diagnostics including radio. Because these decisions are requested by observers who are competing with other scientists for limited resources they often ask for the minimal observations for a specific question, which can be insufficient for key questions of interest. Multi-epoch characterization observations are absolutely necessary for new understanding from the complex physics involved in these sources. For example, the NuSTAR observation of SN2014J, the nearest thermonuclear supernova in more than a decade, would have enabled more precise tests if it had been longer. Alternatively, multi-epoch spectroscopy of supernova with flagship facilities like Webb can perform tomography of elemental distributions by the temporal rise of atomic lines. Devoting significant observational resources to a single group may be unlikely, but is more justified if it were community planned and the data made available to all groups. 

The events which transform our understanding of a given source class are those which are detected by multiple diagnostics, which are often unusually nearby events with observations across and beyond the electromagnetic spectrum. For multidiagnostic studies this favors studying rare events in the nearby universe, favoring large fields of view/regard over depth. When such events occur one must hope that all relevant facilities will make observations at the appropriate time. One can check if pre-approved programs exist and whether the PI triggers their observation, or write dedicated director's discretionary time proposals for out-of-cycle review, which must be done under time pressure and while information is irrevocably lost. For GRB~221009A, a one-in-ten-thousand-year event, the JWST observations provide key insight but did not occur rapidly enough to isolate evolution, limiting the scientific return. 

\begin{quoting}
\noindent Comment: The Fourth TDAMM Workshop will develop community-driven observing plans for rare time-domain and multimessenger events. 
\end{quoting}

A better approach would be for community-defined observing plans which should enable greater devotion of resources from contributing instruments as they do not need to split allocations across proposing teams (for example, as occurred for Chandra observations of the multimessenger neutron star merger detection in 2017), more optimal use of those observations, and can consider the context of the capabilities of the fleet as a whole rather than instruments in isolation. Such an approach has recently been adopted for the ToO program of Rubin \citep{andreoni2024rubin}. This would seek consensus of theorists with guidance by observers and meeting limitations imposed by instrument teams. Such an approach will result in greater characterization of critical and rare transients, where missed or suboptimal observations may not be recovered in a lifetime. Data obtained through this program should be public immediately, fostering and rewarding groups which perform the best analysis and interpretation. The focus would be on the most time-critical set of observations for a given transient; additional observations can then be proposed for (either through ordinary TAC/GO proposal cycles or as DDT requests) with the knowledge that the early, time-critical observations will be taken on a best-effort basis, reducing the odds of a proposal being declined because of dependence on multiple observatories. If this approach proved beneficial for rare transients, it could be explored for more routine events. The process to determine these plans would involve a public review stage and community members would remain free to propose for observations not contained in the base plan.

We anticipate this approach greatly benefiting the scientific return of current facilities, providing a net gain for the agencies and instrument teams. However, we emphasize that this approach must include equivalent, and preferably enhanced, funding in order to properly support and incentivize the community.

We note that NASA's TDAMM/ACROSS initiative, in cooperation with other stakeholders, is organizing the 4th TDAMM workshop for the fall of 2025 organized around the theme of community-defined observing concepts for rare and important explosive transient events, and plans to organize preparatory discussions through the TDAMM Science Interest Group (SIG) beginning in the spring of 2025.

\subsubsection{Enhanced Referencing and Authorship}
\label{sec:multidisciplinary_authorship}
A core difficulty in TDAMM is the confluence of the large-scale collaborative approaches in several disciplines against the traditional small-group efforts in astronomy. The former include particle physics and nuclear science, as well as the astroparticle high-energy telescopes, neutrino telescopes, and the gravitational wave consortia. Such groups judge scientists based on production within large collaborations, which often results in few, if any, first author papers. The latter include most optical astronomers, as well as the broadband follow-up groups. These folks often judge scientists based on their number of first author papers. These ranking statistics do not mesh, and can cause difficulties in hiring, awards, and support.

A related issue is who is listed as first author in papers where two or more individuals truly do contribute equally. This becomes more common as larger numbers of authors contribute to a paper, which is necessary in multidisciplinary areas. This can also arise in true multidisciplinary work where crucial contributions can come from one or more experts in separate fields. 

Some efforts have been made to resolve these issues. These include marking when two authors jointly contribute to a given work, or through marking contact authors in large collaboration papers. However, these implementations are generally not tracked, and this can disadvantage individuals in their career. As a result, this can disincentive collaboration, which is a necessary component in multidisciplinary studies. 

This could be fairly easily implemented from a technical standpoint, if there were buy-in from specific parties. One such option could be initiated as a partnership between the NASA Astrophysics Data System, which tracks papers, individuals, and citation metrics, and the respective field journals operated by the US professional societies. These include the Astrophysical and Astronomical journals run by the American Astronomical Society and the Physical Review series run by the American Physical Society. Engagement with other citation tracking entities and additional astrophysics and related journals could begin once an initial setup was operational. 

\begin{quoting}
    \noindent Comment: Enhancements in the options and tracking of authorship would facilitate collaboration within astrophysics and between fields. Also, broader adoption of the NASA ADS proposal tracking scheme by additional facilities would provide a mechanism to reward non-proprietary data in astrophysics, and to generally support scientists who write proposals for data useful for multiple purposes.
\end{quoting}

\begin{quoting}
    \noindent Finding: The AAS and APS journals should work with the NASA ADS system to implement the technical tracking of these various authorship options.
\end{quoting}

One of the key points raised during the meeting was the importance of recognizing the efforts of successful General Observer (GO) and General Investigator (GI) proposals that enable data generation and dissemination for the broader scientific community. Specifically, we discussed the need for citable proposals linked to the publicly available observations when triggered through GO/GI programs. The centers managing GO/GI should enhance their systems to make accepted or funded proposals more easily citable. For example, when users download data, a citation that includes references to the proposal(s) that enabled the data collection should be automatically generated. 

Encouraging a culture where acknowledging the origin of data (and its associated proposals) becomes a standard practice (This could be endorsed and supported by Journals under the direction of the funding agencies). Journals, conferences, and institutions could play a role in promoting these norms.

\subsubsection{Incentives}
\label{sec:multidisciplinary_incentives}
The root of Nash Equilibrium is imbalanced incentives. A major revamp of the incentive structure in the field is needed in order to grow multidisciplinary research. Below we detail a few specific examples to make the problem concrete. A Pareto improvement is one which enhances the return of all individuals in a group, without harming any. Pareto optimality is when all such improvements have been made. However, in a zero-sum game, such as the measurement metric of community funding from allocation of fixed scientific budgets, all solutions are Pareto optimal. Instead the metric should be the scientific return given the allocation of the fixed budget. The amount of funding to the community is the same, but it can be tasked to greater purpose. Thus, we find that the incentive structure to reward scientists should be aligned with this goal, in order to make the best use of taxpayer funding. 

\begin{quoting}
    \noindent Finding: A major hurdle to fostering multidisciplinary studies is misaligned incentive structures.\\ 
\end{quoting}

\begin{quoting}
    \noindent All three agencies, and other field institutions, should emphasize reward mechanisms which support collaborative efforts, multidisciplinary proposals, and community organization. This includes scientists supported directly or through grant proposals by the funding agencies.
\end{quoting}

We have explored one such option at the individual / small group level for observers in Section~\ref{sec:multidisciplinary_observingPlans}. The problem of incentives arises elsewhere. Individual scientists or groups propose to existing calls and review criterion. Observers are often rewarded for proposing that a given observing will discriminate between two specific scenarios and can publish in high-rated journals with affiliated press if the observations are awarded. This incentivizes the use of approximate models where singular observations can discriminate, rather than the ideal scenario of incentivizing observations which would provide key constraints for higher fidelity models. Similarly, theorists may propose to link two chains of the larger puzzle because that is all that can be accomplished in a given cost or time cap. These same individuals will drift towards agencies who give the largest awards for their work, which come from calls in DOE and NSF. This leads to a disengagement in tying theory and models to NASA facilities, and results in a more limited scientific return as a result. 

At the facility level, it would be beneficial for TDAMM if agencies reviewed their support for the larger picture and not only for their own direct need. As one finite example we highlight the alerts released by the IGWN. Neutron star mergers can be detected in GWs before the objects actually merge. This detection and approximate localization information can be disseminated to the follow-up community and may allow EM telescopes to alter observing patterns or repoint to capture the earliest photons from these events, which are otherwise missed. For example, radio arrays can switch to wide-field observation modes, wide-field optical telescopes can preserve higher temporal resolution data, and NASA's Swift satellite could point its wide-field gamma-ray instrument towards the source. For the latter scenario this requires extremely rapid alerts from the IGWN, rapidly distributed to the Swift listeners, and fully automated commanding of the spacecraft all within $\lesssim$30~s \citep{tohuvavohu2024swiftly}. This end-to-end system is an example where without any component the system is not usable. While the alerts mechanism, Swift pipelines, and rapid commanding infrastructure (informed by a recent community report, \citealt{kennea2024time}) are ready, the IGWN alert timescale remains too slow for the needs of Swift. This delay limits investment in rapid automation of additional telescopes.

As the IGWN has evolved it has released additional information in its rapid alerts. Currently it releases the probability that a neutron star-black hole merger will have EM emission. This is based on approximate fitting to a single suite of simulated models, and fails to identify the cases when follow-up should actually be prioritized: when the predictions from different models disagree. This is how to test the equation of state of dense matter. Currently LIGO prioritizes non-TDAMM science because it is more likely to show success to motivate future upgrades and funds. Certainly this decision is understandable.

In both cases, reviews of LIGO in the US and IGWN more broadly would greatly benefit TDAMM if the larger picture were included. We use LIGO only as an example. They have continued to release additional information and have sought a fair balance with ensuring rewards and support for their own scientists. Generally, we find that reviews of major facilities for their support of the integrated TDAMM endeavor are likely to be beneficial for greater scientific return of US investments.

Because funding is the greatest incentive, a holistic review of funding sources, opportunities, and review criteria in the context of TDAMM may be beneficial. This could include funding options within and between agencies. Further, the agencies should consider whether reviews of employees at NASA centers, National Labs, or the NSF federally funded research and development centers are incentivizing the bigger picture.

\subsubsection{Analysis Methods}
\label{sec:multidisciplinary_analysis}

A related point to data curation is the ability to easily utilize the data for its intended purpose. This includes both access and analysis methods. Access can be either programmatic or through a user interface. Most archives provide some method for both. Many of these archives or related entities also develop and maintain analysis software for the data they house. Astrophysics is among the leading entities for ease of access of data, with NASA having the most mature archives. The current investment in this area will help ensure data is preserved and used for decades to come.

TDAMM analysis has unique requirements. We are often particularly interested in when a given observation is taken. It is often difficult to easily access the data of interest in a programmatic manner. While the astrophysics archives, coordinated through the International Virtual Observatory Alliance, have built code to programmatically access observations at a given position, the complementary option of doing the same with temporal information has lagged behind. This need has been identified by the archives and many are actively developing the code to provide this capability.

TDAMM analysis often also requires accessing data from multiple archives and analyzing them in a coherent manner. With the maturation of application programmatic interfaces the access should become a solved problem in the near future. However, analyzing data from multiple facilities is an understated problem. What is typically done is processing of data on individual facilities and performing multiwavelength or multimessenger analysis on the combined output. For example, converting from counts data to flux in several facilities across the electromagnetic spectrum and then performing a joint fit on the flux values from radio to gamma-rays. This is incorrect for multiple reasons. For example, in X-rays the energy-dependent flux values depend on the assumed model passed to the fitting engine because these telescopes have non-invertible response matrices. Performing spectral fits on imaging gamma-ray telescopes depends on models for background and signal. In both cases the inferred flux may change if a global fit were performed before processing the data.

\begin{quoting}
    \noindent Finding: A general-purpose analysis framework which properly handles data from all major TDAMM facilities is needed. This requires input from a broad range of expertise and must be done in a sustainable manner.
\end{quoting}

One approach to properly handling this data is the Multi-Mission Maximum Likelihood (3ML) \citep{vianello2015multi}. Here the fitting is done in a global maximum likelihood approach, where the summation of likelihood from each facility can allow a proper global fit. Data can be utilized if it exists in the typical wavelength-dependent manner. However, the general problem of lack of sustainable community-developed software remains. Some entity should be supported and tasked with building analysis software that is widely applicable and statistically accurate in TDAMM, where groups can contribute open source development or ensure their data is easily available in this method.

Additionally, high fidelity models to make use of the full set of TDAMM data do not necessarily exist. This may be a circular problem where they are not developed because i) no easily accessible dataset on a given event may exist, ii) no widely adopted fitting method is known, and iii) the models do not exist, so no individual problem is solved. The fitting methods can easily be developed and the data collated, which may motivate the development of more appropriate models.

\subsubsection{Multi-physics Uncertainty Quantification}
\label{sec:multidisciplinary_multiphysics_UQ}



Uncertainty quantification is difficult in most applications, but in TDAMM, where multiple physics effects all contribute to the evolution of a phenomenon, uncertainty quantification becomes extremely challenging.  This physics is studied via a broad range of messengers and combining these different observations also requires advancing existing analysis techniques.  Oftentimes, characterizing a given phenomena requires integrating multiple high-performance computing calculations along with detailed semi-analytic models.  All calculations must include approximations in the physics implementation and the coupling of different physics.  Testing must not only reduce the numerical errors but also identify the validity of the solutions.  Uncertainty quantification for these models relies on a broad range of methods that are both difficult and time consuming.  As such, most studies in astrophysics only do a subset of the tests described below.

Physics and engineering typically rely on a set of verification and validation techniques.  Traditionally, verification refers to testing a code to ensure that it is solving the implemented equations correctly.  These tests include:
\begin{itemize}
    \item Developing a set of standard analytic or semi-analytic solutions (typically for single physics components) and comparing codes to these solutions.  These test both the physics implementation and the numerical artifacts in a code (e.g. modeling the orbit of a binary system is a simple way to test the angular momentum conservation in a code).
    \item Conducting detailed code comparison studies.  Code comparison can occur naturally (multiple analyses of the same event) or through focused code comparison papers.  These latter focused studies often develop simplified, but application specific, problems that can be used in testing newly developed codes. 
    \item Focused laboratory experiments that can test the implementation of the physics in the codes.  These experiments test physics implementations and are, by nature, multiphysics.  More and more experiments are designed to test methods used to couple the physics.  They are in contrast to validation experiments that are designed to mimic the exact conditions of a phenomena.
    \item For computationally-intensive problems, the numerical hardware and framework can also lead to different results and these uncertainties must also be understood.
\end{itemize}
The astrophysical community has already developed a suite of tests for their codes based on analytic solutions, often leveraging work in the engineering and applied physics community.  For example, many codes test their results on Sedov blastwave and shocktube solutions.  the growth time in a simple Rayleigh-Taylor instability has also been used to test some astrophysics codes~\citep{2002ApJS..143..201C}.  A number of more application-specific analytic solutions have been developed and used to test codes~\citep[e.g.][]{1989ApJ...346..847C,1996ApJ...460..801F}.  As fewer and fewer astronomers focus on detailed analytic solutions, such application-specific comparisons are becoming more rare.

\begin{quoting}
    \noindent Finding: Complete uncertainty quantification is a necessary step for a successful end-to-end approach to TDAMM questions. The process of formally mapping stages to observations, determining the uncertainties at each step, and iteratively improving the specific areas of greatest need is the best path forward. Mapping of known unknowns is necessary to prioritize experiments and observations, and to determine where the remaining error may arise.  
\end{quoting}

A number of code comparison efforts have also been done and we can not describe all of these here.  Instead, we will discuss the specific example of transient light-curve solutions.  The astrophysical transient community developed a set of standardized tests and conducted a detailed comparisons of some of the existing codes~\citep{2022A&A...668A.163B}.  Comparisons provide both a way to test a code and determine the current uncertainty from different techniques.  In the case of supernova light-curves, the \cite{2022A&A...668A.163B} comparison showed that current codes continue to produce very different results, even on simplified toy supernova explosions before we can use these observations to constrain properties of the transient ejecta.  But detailed code comparison studies are extremely time consuming and are insufficiently supported in academia.  Multiple studies of a single event can lead to code comparisons in a less systematic manner.  For example, observations of the kilonova from the nearby neutron star merger GW170817 led to many analysis projects using different codes and different approaches.  They confirmed that light-curve calculations can differ by orders of magnitude in the inferred ejecta properties~\citep[see summary by][]{2018ApJ...855...99C}.

Validation studies are focused on testing whether a method or code is solving the correct physics.  These experiments are more difficult to develop because the physics must be identical to the phenomena being observed.  Few laboratory experiments will validate a code to solve astrophysics, but detailed, multi-messenger observations (typically from nearby events) also may be used to validate methods.  All experimental testing face their own limitations.
\begin{enumerate}
    \item Initial condition uncertainties:  Even laboratory experiments suffer from our lack of understanding of the detailed initial conditions from the experimental target to the drive used to power the experiment.  
    \item {Instrument Limitations:}  Detectors (either in laboratory or in space) have finite resolution in space, time and particle counts / spectrum energy.
    \item Oftentimes, experiments test only one or few aspects of the full application and the incorporation of multiple physics experiments compounds the total uncertainty. All sources of information should be propagated and weighted for uncertainty. 
    \item Sensitivity studies to different aspects of the physics provide insight into the validity of the chosen equations.  These sensitivity studies may also help identify errors in the code.
\end{enumerate}
Experimental studies typically require an iterative approach where both modeling and experiment drive improvements to both the experiments and theory.  These studies require multiple years (decade-long timescales) to make progress.  NNSA laboratories have developed strong experimental/code co-design programs, but such work requires dedicated support that isn't common in NSF and NASA funding opportunities.  Astrophysicists have leveraged existing experiments to test their codes~\citep[e.g.][]{2022A&A...668A.163B} but there was insufficient support to evolve the experiment to further test the astrophysics codes.

Astrophysics observations of a well-studied phenomena  also have the potential to validate the codes.  Especially if that phenomena has a wide range of diagnostics.  But one of the difficulties facing both astronomers and physicists with multi-diagnostic or multi-messenger observations is identifying how best to utilize these constraints.  One observation may be more dependent on one physics aspect.  Developing a systematic approach to combine information from all the verification and validation studies to address the uncertainties in a given application requires a detailed understanding of the fundamental physics, analytical understanding, data analysis, experimental and numerical uncertainties. No field has enough generalist scientists who are capable of bringing together a team of experts to under all facets of a problem along with its uncertainties.  Developing these generalists that can build these teams and providing funding sources that allow the formation of the broad teams required to truly analyze a problem is as critical for TDAMM science as it is for the national security problems at NNSA.  NNSA laboratories can provide insight on how to develop the framework for such systematic studies.

\subsubsection{Community Organization}
\label{sec:multidisciplinary_communityOrganization}
One mechanism to overcome structural barriers to multidisciplinary science is fostering greater community organization. This has been most successful in the nuclear side of nuclear astrophysics. Within the United States this began with the NSF Physics Frontiers Center (PFC) Joint Institute for Nuclear Astrophysics (JINA) in 2003. In 2014 there was a transition to the PFC Joint Institute for Nuclear Astrophysics - Center for the Evolution of the Elements (JINA-CEE). Michigan State University has played a key role in both of these PFCs, in part because it hosted NSF's National Superconducting Cyclotron Laboratory. In 2008 this facility was selected to be upgraded into the flagship DOE Facility for Rare Isotope Beams (FRIB), which began operations in 2022. Following the transition from NSF to DOE, the nuclear astro PFC transitioned to the Center for Nuclear Astrophysics across Messengers (CeNAM), supported by DOE's Nuclear Physics (NP) program. For more information see \url{https://www.jinaweb.org/}.

With now two decades of investment, the payoff is obvious.  This program has been instrumental in building connections across research groups and disciplines through regular meetings, schools, and network building. It has fostered the strategic development of a workforce useful for fundamental science as well as nuclear needs in industry. Because the composition of contributing scientists crosses disciplines, it has been key to data curation to sufficient fidelity to be accessible across disciplines. For example, these entities created, maintain, and update the Reaclib Database\footnote{\url{https://reaclib.jinaweb.org/}} which provides nuclear reaction rate data relevant for astrophysical applications. This database is utilized by computational scientists as inputs to models, astrophysical data analysts for interpretation of observations, and hardware specialists in the prioritization of experiments. Enabling science at the confluence of nuclear physics, astrophysics simulations, and astronomy observations has transformed the field and opened up entirely new lines of research. For example, after it had been considered settled knowledge that the trans-iron elements are produced by just the slow and rapid neutron-capture process multidisciplinary research enabled by JINA and JINA-CEE showed that especially in the early universe an additional intermediate neutron-capture regime in convective-reactive stellar environments was at work. This has since prompted more than two dozen nuclear physics experimental campaigns at rare-isotope facilities to measure $(n,\gamma)$ cross sections of unstable heavy elements, thereby adding new understanding to how the elements from iron to uranium were made. Through enabling and orchestrating multi-disciplinary approaches the PFCs and succession to CeNAM has directly led to numerous scientific discoveries and is well suited for the new era of nuclear astrophysics.

Similar research networks exist in other countries. Internationally these networks are brought together through the International Research Network for Nuclear Astrophysics (IRENA). IReNA is an NSF AccelNet Network of Networks, connecting 9 research networks across 4 continents for a greater whole. For more details see \url{https://www.irenaweb.org/}. It focuses on research in nuclear science, computation, and connections to astrophysics. 

One area where connections are lacking is with the astrophysics observers. Many optical observers are largely unaware of the massive sustained multidisciplinary effort which produces the models they use to interpret data. This contributes to the lack of knowledge of limitations of such models. However, these networks provide a known access point for interest astrophysicists. For example, the instrument team for NASA's forthcoming nuclear spectrometer mission, COSI, could engage with IRENA. 

This brief example demonstrations multiple advantages of community organization. These include identifying areas where connections are weak or data inaccessible to interested scientists, and fixing these problems. These include workforce development, network building, and direct scientific advancement. There are not similarly broad and mature efforts in other disciplines relevant for TDAMM. Not only is this limiting scientific return in astrophysics, it is preventing the full use of advances in other disciplines. 
 
\begin{quoting}
    \noindent Finding: The sustained investment in fostering community organization in nuclear astrophysics by the NSF and the DOE has been a great success. 
\end{quoting}
\begin{quoting}
    \noindent Comment: This approach should be adopted by other related fields to similarly multidisciplinary work, and broadened to enable the full end-to-end approach envisioned in this white paper.
\end{quoting}

We find that these efforts should be fostered across the disciplines of interest in the hope that they may be as productive as CeNAM and its predecessors. We additionally emphasize the continued need to invest in CeNAM, especially in the new era of multimessenger astronomy. These are perhaps the greatest investments the funding agencies could make to foster multidisciplinary science in general, but especially in the area of TDAMM.


\subsubsection{Curated Data}
\label{sec:multidisciplinary_data}

Data curation is a priority area in all fields of science. It is even more important for multidisciplinary studies as it must be made accessible to scientists who are not subject matter experts in the area. This is a particularly difficult problem. 

From the sustained community organization effort in nuclear physics, REACLib is a standardized and singular access point for reactions of interest to astrophysics \citep{cyburt2010jina}. This is an exemplar of data curation for multidisciplinary approaches. One could envision a similar product being built from sustained investment in atomic spectroscopy, though this database would need to handle additional information for non-LTE effects. For error quantification, in both areas, a code comparison of nuclear reaction networks and atomic opacity results from different groups would give a more direct characterization of the uncertainty from these lab astro inputs.

\begin{quoting}
    \noindent Finding: Data curation is an area where astrophysics is leading most scientific disciplines. For the specific needs of TDAMM, data curation to the degree it can be easily accessed by interested scientists from other disciplines is needed. While astrophysics should continue to invest in this area, its approaches should be adopted in other disciplines. As identified in the Astro 2020 Decadal, coordination of archives and intentional investment by the NSF are well motivated.
\end{quoting}

The state of data curation within astrophysics is varied, but we must similarly foster ease of access to data in other fields. The NASA archives fulfill the necessary step of data preservation and have helped to set data standards which have been widely adopted. Similar archives exist for some DOE and NSF assets. Many groups produce facility-specific catalogs such as the IGWN gravitational wave catalogs. However, given the interconnected nature of multimessenger, multiwavelength, and time-domain astronomy, this is insufficient: what is needed is an object-based community-driven transient catalog. There are object-specific catalogs such as SIMBAD\footnote{\url{https://simbad.u-strasbg.fr/simbad/sim-basicIdent=m33&submit=SIMBAD+search}} and the NASA Extragalactic Database\footnote{\url{https://ned.ipac.caltech.edu/}} which serve similar needs to other communities. The Transient Name Server was designed to fulfill this need for supernovae and does well in this endeavor, but it cannot scale to the needs of the new era and is incapable of handling data outside of optical and adjacent bands.

A new event-based transient catalog should be created. It should be capable of handling data from all wavelengths and messengers, being accessible both programmatically and through a user interface, be updateable by input from community members, and be open source. The ingestion of all data for a given transient for use in standard formats for theory and simulation work would save untold hours in busywork and remove a barrier to the development and use of complex multiphysics codes. The NASA Exoplanet Archive and the Open Supernova Catalog could serve as inspiration. This archive could serve as a single access point for TDAMM astrophysicists and include metadata of interest to scientists from other fields, such as the rates of transient events of interest, the expected operational timelines of major facilities, and provide summary information in an accessible manner for things like schools, conferences, and proposal calls across agencies. An advisory board should include scientists from other disciplines.

\subsubsection{Future Meetings and Workshops}
\label{sec:multidisciplinary_meetings}



Among the major hurdles to multidisciplinary science is the lack of regular large meetings. In astronomy this is typically the Winter meeting of the American Astronomical Society. The large annual meeting for physics disciplines depends on the field. The Division of Nuclear Physics of the APS has its own annual meeting, as do some others. However, the March and April APS meetings are the annual meetings for many divisions. Astrophysics belongs to the April meeting and is joined by Nuclear, Gravity, and Particle Physics, and has become a de facto TDAMM meeting. However, this misses several crucial fields with go to the April Meeting. 2025 and 2026 are an experiment, with the March and April APS Meetings being merged into a single venue. This provides an opportunity to utilize the broad APS Meetings as key venues for TDAMM science. Some joint sessions will be held at the 2025 meeting; however, these are organized by only two divisions, and the March and April meetings are being organized separately (though at the same venue). We find that TDAMM science would be best fostered by continuing to have merged March and April meetings of the APS, and making these the intentional place for TDAMM science and other multidisciplinary endeavors.

The large annual meetings provide the additional benefit of allowing interactions between the funding agencies and their broad communities. For example, both NASA and NSF have large contingents at the Winter AAS meetings. However, communities which do not attend these meetings are disadvantaged and not visible to headquarters. Within the AAS this includes Laboratory Astrophysics, which meets at the Summer AAS meeting. We find it would be beneficial to ensure the requisite visibility of lab astro either by shifting their AAS meeting to Winter or encouraging larger engagement of NASA and NSF at the Summer meeting. More broadly, non-astrophysicists generally do not attend AAS meetings. The lack of program officer attendance at their meetings has left them blind to the breadth of disciplines with avested interest in TDAMM. A necessary step to fostering multidisciplinary science is ensuring engagement of the relevant program officers at the joint APS meetings. 

While the broad meetings are key for communication, engagement, and outreach, it is small focused meetings and workshops which are most directly productive for TDAMM science. These include the TDAMM Workshop series, event-specific meetings, schools, retreats, and workshops. Unsurprisingly, we find the TDAMM Workshop series to be particularly beneficial for focused engagement of subject matter experts and the relevant program officers. While the fourth TDAMM workshop is now being planned, we find the series should continue for the foreseeable future. Future meetings could include broadening in scope to include additional agencies and institutions from the defense apparatus, greater ties with the space-based plasma communities including heliophysics, and focused meetings as the needs are identified.

\begin{quoting}
    \noindent Comment: Extended workshops are among the most powerful tools to foster multidisciplinary studies. These include building networks of interested scientists, tackling specific problems, identifying key areas of uncertainty and future projects, and often uncover unforeseen benefits. Numerous options for these meetings occur. 
\end{quoting}

\begin{quoting}
    \noindent Finding: Additionally, the use of large annual meetings is an opportunity to maintain connections and build momentum across fields. These opportunities should be utilized for fostering multidisciplinary science in TDAMM. The most appropriate venue for this are the large annual, multi-division APS meetings. TDAMM would be best served by a single annual meeting (i.e., the merging of the March and April meetings), and intentional growth of cross-division efforts.
\end{quoting}

Focus workshops are the likeliest area to have breakthroughs in specific needs. One example where promise currently lies includes multiphysics code development such as integration of transport and magnetohydrodynamics code or integration of models of different components in individual source classes. Utilizing these meetings to set interfaces between fields, quantify error in a specific area, or solving specific tasks between chains would be a fantastic use of these resources. Alternatively, experts from different fields can be brought together. One promising area could be fostering plasma astrophysics through engagement with those who have succeeded in plasma heliophysics and also with software engineers. Another area could be exploration of the agreement of nuclear reaction networks, which could be probed through calculations of specific elements. A similar approach for atomic opacities would be fruitful.

While many of these workshops occur, one typical problem is often the lack of sufficient experimental or observational expertise. As one example, an exploration of the uncertainties in nucleosynthesis could bring together nuclear experimentalists and theorists, transient engine studies, nuclear network calculations, and observational observers. This will identify the framework for a full system investigation to identify the future studies necessary as well as roadblocks and uncertainties in those studies. A similar end-to-end approach could be undertaken in the preparation for a forthcoming major mission or facility.

Numerous host options exist. In the United States these include the Aspen Center for Physics, the Kavli Institute for Theoretical Physics, the Institute for Nuclear Theory, and CeNAM which can host programs from a few days to a few months. International options include workshops held by the ECT, the Lorentz Center, the ISSI, The Niels Bohr Institute, the IPAM, the IPMU, and the Royal Society. We find that this community need is well served, both nationally and internationally, and emphasize the need for continued support. However, we note that some areas of need which could be addressed at a workshop are not necessarily the most attractive as they advance foundational understanding, rather than being the last step in science for discovery. Examples include code comparison or tests of agreement from nuclear and atomic calculation predictions. These topics are crucial to TDAMM success and beyond, their need is paramount. 

Schools are a fantastic mechanism to train interdisciplinary scientists. While many school opportunities exist, it is hard to find a summary list. As one example of how schools could be adapted for the needs of TDAMM, having nuclear graduate students attend astrophysics data workshops, or the reverse, will help each side better understand the other, and foster multidisciplinary scientists. A more in depth training opportunity could include full summer visits with scientists in complementary disciplines.

\subsubsection{Workforce Development}
\label{sec:multidisciplinary_workforce}


Astrophysics is facing a crisis.  The use of publicly available codes has made it easier for scientists to conduct studies and publish papers.  But fewer and fewer astrophysicists understand the physics, and more importantly, the physics implementation in these codes.  This has led to an increasing number of science results mistaking numerical artifacts for science or using codes in regimes where they are not viable, leading to a stagnation in astrophysical science.  Training in the fundamental physics and computational science is critical to the advancement of TDAMM science and this requires a robust understanding of physics, both in analytic and numeric regimes.  

Astrophysics has long claimed that it provides a gateway for young people to get interested in physics and engineering, ultimately providing the entryway for generalists that are so critical to technical industry and national security applications.  As with astrophysics, innovation and advancement requires a fundamental understanding of a broad range of physics and its implementation into numerical models.  No field in astronomy is better suited to this training than TDAMM science, with its complex problems that mimic many of those in industry and national security.  But TDAMM astronomy is only useful in recruiting scientists if it provides the basic training to do the work in these fields.  

\begin{quoting}
    \noindent Comment: The broad endeavor envisioned in this document would foster a new generation of generalist scientist, ready to solve outstanding and future problems for astrophysics, related disciplines, national defense, and industry.
\end{quoting}

DOE, in particular, NNSA laboratories, are continuously recruiting scientists with computational multi-physics and strong fundamental physics backgrounds.  Indeed, many of the leading scientists working in national security have been initially trained in astrophysics.  This skill set should be distinguished from code users or phenomenological scientists that make up most of the astrophysics graduates (who typically publish more papers and hence tend to obtain more astronomy grants and tenured positions).  Astronomy has been an important training ground for DOE laboratories but this is less true today as more and more astronomers are only trained in code use and not code understanding.  This trend must be reversed to make astronomy a useful training ground training scientists for DOE positions. In turn, this will provide a renewed physics capability in astrophysics. 

As universities tend to focus on scientists bringing in considerable grant money, the funding agencies are in an ideal position to affect change.  The current funding paradigm in astronomy (both NSF and NASA) must be modified to encourage the development of both broad scientists (with strong fundamental physics backgrounds) and computational experts.  Developing programs that require synergies between fields (astronomy, physics, numerics, mathematics) can provide strong encouragement to scientists and their universities.  Supporting workshops and schools to identify key issues that must be addressed by synergistic studies could help guide funding directions.  Some of this work can, or should, only be performed by senior scientists. This motivates an alteration to the approaches of PFCs or SciDACs which often only fund earlier career individuals, to ensure that true collaboration between fields occurs and is sustained. However, these mechanisms are also ideal for training of a strategic national workforce with unique capabilities, exposing those early career scientists to multiple career paths, and building networks to act as recruitment pipelines for the NNSA labs. 


\subsubsection{Support and Support Gaps}
\label{sec:multidisciplinary_support}

We foresee a few major funding needs for TDAMM. One is to build up community organization between fields, such as the role CeNAM plays for nuclear astrophysics. Similar existing and new institutions can then act as focal points for greater coordination between fields, to meet the full need of TDAMM science. At all levels we must have sufficient support for theory and simulation, as well as ensuring that the work performed is cognizant of all relevant physics.

The NSF is perhaps best suited to this overall picture. NSF Physics has PFCs which are large and multi-year, and were used to create the nuclear astrophysics coordination over two decades, and could be adapted to similar developments in other fields. The NSF also has the Accelerating Research through International Network-to-Network Collaborations (AccelNet) call which supports the international nuclear astrophysics group IRENA. NSF AST Astronomy and Astrophysics Grants (AAG) are non-prescriptive, allowing for proposals of any need identified by nearly any member of the community. The NSF CAREER award can similarly be used for multi-year purposes of a single group. However, solving an end-to-end TDAMM problem may require larger scope than a PFC or AccelNet can allow, as it requires real time investments from senior and junior scientists over many years.

The DOE SciDAC proposal calls are similar broad, multi-year efforts in multidisciplinary studies with a computational focus. They have been utilized with NP and ASCR to produce simulations which are broadly used in TDAMM, and guided major advancements in astrophysics. The DOE Office of Science Early Career Program is similar to the NSF CAREER program, with TDAMM-related awards in a number of focus areas. The Predictive Science Academic Alliance Program (PSAAP) is the primary mechanism by which the NNSA’s ASC program engages the U.S. academic community in advancing science-based modeling and simulation. 

NASA Astrophysics operates both the non-prescriptive Astrophysics Theory Program (ATP) and larger coordination efforts through the Theoretical and Computational Astrophysics Networks (TCAN) program. However, a single SciDAC or PFC exceeds the total budget of TCAN, and the ATP awards are generally smaller than corresponding grants from the NSF and DOE. This has limited engagement from scientists who handle large-scale multiphysics problems from focusing on NASA missions. The workshop identified the UVEX observations of shock breakout and COSI observations of nuclear gamma-rays from novae and thermonuclear supernovae as transformational advancements in astrophysics. However, in neither case were these events prime science goals for these missions. This is the result of the complex tradeoffs made in the preparation of proposals. However, in both cases, more mature theory predictions would have given more robust detection rate numbers and allow for comment on what the anticipated observations would return scientifically. Intentional investment in these areas, and recruitment of scientists to advance simulations to plan the next generation of NASA telescopes are needed. This requires bolstering of the NASA Astrophysics theory budgets. Additionally, a NASA CAREER award would be a fantastic way to allow new tenure-track, or similar, faculty to tackle general problems.

\begin{quoting}
    \noindent Finding: Each agency fulfills unique needs for the community. However, a full end-to-end approach may need strategic alignment of existing opportunities across agencies.
\end{quoting}

We reemphasize the need for coordination across agencies. As noted by the NRC in 2003, no agency can alone marshal the necessary expertise and resources to answer the most interesting questions between physics and astronomy, and this is still true in TDAMM. One possible mechanism may be alignment of a PFC with a SCiDAC and enhanced TCAN in order to fully answer one question, or fully understand one source.

\subsubsection{Facilities and Facility Gaps}
\label{sec:multidisciplinary_facilities}
For facilities, the multidisciplinary approach outlined in this report emphasizes the need for multidiagnostic approaches. In the language of astronomy, this is reemphasizes the pursuit of multimessenger astronomy, as well as complete coverage of the electromagnetic spectrum. For facilities, the NSF and DOE Rubin Observatory is nearing first light, with NASA's Roman Space Telescope shortly behind. The NSF has invested in planning and preparation work for the next generation neutrino and gravitational wave facilities as well as the ngVLA and the Extremely Large Telescopes (ELTs). The NSF has extended the Zwicky Transient Facility, ensuring wide-field optical discovery continues in the near-term future, which is complemented by numerous such telescopes coming online worldwide. NASA will launch the COSI, StarBurst, and ULTRASat missions around the next planned observing run of the gravitational wave network, begun work on the Habitable Worlds Observatory (HWO), and selected the Ultraviolet Explorer (UVEX) for both discovery and spectroscopy in the ultraviolet regime. NASA Planetary will launch the Near Earth Object Surveyor, which will complement Rubin in the infrared wavelengths, with a time domain component supported by NASA Astrophysics. In the longer term, NASA will launch the Laser Interferometer Space Antenna (LISA) Mission to study Galactic compact binaries as well as distant massive black hole binaries.

Internationally the advancement of the Cherenkov Telescope Array (CTA) at high energies promises a revolution at very high energies. China and ESA have recently launched Einstein Probe, which will revolutionize discovery of short-duration X-ray transients. These investments across and beyond the electromagnetic spectrum set the stage for a new age of astrophysics, and provide the diagnostics needed to learn the physics of the cosmos.

However, these are not the only facilities which must be considered. As previously mentioned, the study of neutrinos will be revolutionized with DUNE and the other megaton neutrino detectors. Similarly, the first results from FRIB and other new nuclear facilities are underway. The continued adoption of the high powered laser facilities and other user high energy density physics facilities for astrophysics will give an anchor to physics extremes which are only now possible to create on Earth. The sustained investment in atomic spectroscopy may allow us to utilize the full power of the James Webb Space Telescope for TDAMM. Adaptation of existing approaches in fluid dynamics and turbulence, computational science, radiation transport, and high energy density physics will advance our simulations and modeling to be able to handle the data from these diverse facilities. Further, the vast majority of telescopes can be used for TDAMM purposes. For example, the unique capabilities of NASA's flagships make them singularly capable of some TDAMM discoveries, such as direct proof of r-process nucleosynthesis. The combined investment in these facilities combined is measured in tens of billions of dollars.

\begin{quoting}
    \noindent Finding: The breadth of new facilities and current set of astronomical facilities represent a nearly complete program for TDAMM astrophysics. Additionally, some crucial TDAMM science is only possible with contemporaneous observations. Thus, we find that a strategic approach is required to maximize the scientific return of facilities across astrophysics and beyond. 
\end{quoting}

Some of the highlights mentioned above match formal recommendations and informal findings from past reports. For example, a key need identified in the first TDAMM workshop included a more capable ultraviolet discovery machine and spectrometer. This will be met by NASA's UVEX mission. The second workshop included the identified need that wide-field optical telescopes other than Rubin are still needed, particularly in the north hemisphere. The extension of their support for ZTF helps fulfill this need. 

In the short-term, there are two specific needs which have regularly appeared in multiple strategic and community TDAMM reports since 2020, which includes the NASA GW-EM Task Force Report\footnote{\url{https://pcos.gsfc.nasa.gov/gw-em-taskforce/GW-EM_Report_Final.pdf}}, the Decadal survey \citep{national2021pathways}, the 1st TDAMM Workshop Report\footnote{\url{https://pcos.gsfc.nasa.gov/TDAMM/docs/TDAMM_Report.pdf}}, and the 2nd TDAMM Workshop Report \citep{ahumada2024windows}. Identified in every report is the need for a new US-led high-energy transient monitor. Identified in a few report is the need for a rapid-response X-ray telescope. Currently Swift fulfills critical community needs for both of these capabilities. However, Swift is approaching reentry which will occur sometime in the next $\sim$7~years (this is probabilistic, given uncertainties in modeling and properties of the Solar maximum). Thus, it is likely that we now have a gap in these capabilities. 

Given the numerous high-energy monitors and X-ray telescopes, we very briefly detail the specific needs of TDAMM as identified here and in the reports which informed this workshop. For the high-energy monitor, the transient community outlined the needs in the Gamma-ray Transient Network Science Analysis Group report \citep{burns2023gamma}. These include a true all-sky gamma-ray monitor and a, extremely wide-field X-ray monitor with localization capability and a particularly soft low energy threshold. This meets key needs for neutron star mergers (multiple signatures), novae (early X-ray signature), supernovae (though earliest shock breakout), AGN (long-term monitoring), magnetars (complete flare coverage, fast radio burst counterparts), and provides total coverage for all externally identified transients. Despite the numerous high-energy monitors launched and planned, none will fulfill these crucial capabilities. NASA is currently investing in key X-ray detector technology development of interest for such a mission.

The other near-term need is for a rapid X-ray response telescope. This is needed for characterization of X-ray transients. This is true even when one does not care about the X-ray signal. For example, most neutron star mergers with precise localizations, with or without gravitational wave signals, are expected to be face-on. In order to study the quasithermal kilonova signal, one must subtract the afterglow, which requires multi-epoch X-ray observations (see Section~\ref{sec:sources_mergers}). There are numerous X-ray telescopes. The vast majority have response delays which are too great for the needs of TDAMM or do not have enough flexibility for multi-epoch observations. NICER can fulfill this role, without imaging capability, but is limited by the lifetime of the ISS. SVOM and Einstein Probe were designed for rapid response and are undergoing commissioning. While SVOM may not be sensitive enough for many cases, Einstein Probe is. Though this would enable only scientists of that team to make major discoveries, including those which rely on kilonovae. In the US, the proposed X-ray probe Advanced X-ray Imaging Satellite (AXIS) \citep{reynolds2023overview} could provide this capability, but that would require a faster response time than is currently required.

In the longer term, it is key to ensure that the technology is developed to continue to observe new diagnostics, particularly those which are most powerful. This most obviously arises in the completion of coverage of the neutrino and gravitational wave spectra. One particularly powerful capability for TDAMM would be a GW detector capable of observing the range around $\sim$1~Hz. This is not observable on Earth and above the range that LISA is sensitive to. This would allow detections of populations of white dwarf-white dwarf mergers and their associated thermonuclear supernovae (and separation of single degenerate events), precise localizations of neutron star mergers years in advance of merger, discovery of white dwarfs merging with neutron stars or black holes, and mapping properties of tidal disruption events. Fantastic.

In the context of powerful diagnostics and priorities identified in this report, we highlight a future capability which has not otherwise been discussed in US planning documents: a narrow-field gamma-ray spectropolarimeter. COSI will recover a small number of thermonuclear supernovae and nova. A more powerful telescope could uncover samples of these events. Nuclear lines are particularly powerful diagnostics, giving yields of individual isotopes, acting as cosmic chronometers, and often become detectable at comparatively early times in opaque plasmas. Their power relies on the decades of investment in nuclear astrophysics and corresponding simulations. Further, some focusing setups preserve polarization information. This could allow detailed characterization of numerous high-energy sources and jetted transients which are otherwise impossible. It would also enable the unique QED test of photon splitting through phase-resolved spectropolarimetric observations. NASA's technology gap list includes focusing gamma-rays, a key need for such a mission. 

We have discussed possible future facilities which would provide transformational information for TDAMM studies, we repeat that the vast majority of scientific gain in TDAMM can be done through aligning existing capabilities and platforms. Though if hundreds of millions of dollars does come available, we'd love to be the ones to spend it.

\subsection{Summary Findings for NASA, NSF, and DOE}
The findings in this report are generally found throughout this document in the relevant sections for contextual information. However, we here provide a summary of findings for the relevant funding agencies.

\textbf{NSF} PHY supports the ground-based GW interferometers and high-energy neutrino detectors. NANOGrav is leading the discovery of low-frequency GWs and should play a major role in the identification of the first (unambiguous) individual supermassive black hole binaries within the decade. IceCube sees the signature of the production of ultra-high energy cosmic rays, but its sources remain elusive. Without these facilities, there is no TDAMM era. AST supports optical, infrared, and radio telescopes at all scales. LIGO has revolutionized the study of several fields of interest, and its discoveries have been emphasized in the long term planning documents of half a dozen fields.  Rubin will come online this year and take nightly videos of the optical sky. The broad discovery and characterization EM facilities provide give context to the non-EM observations. 

Because of these major assets the NSF has played a leading role in integration efforts. This is likely because NSF has both PHY and AST, and PHY houses the relevant disciplines, and a recognition that combined efforts are greater than individual ones. The dedicated software infrastructure call was a key response to community needs identified through the TDAMM workshops and Decadal. Similarly, the MUSES cyberinfrastructure project was targeted to a specific integration need between a few disciplines in the study of dense matter. The non-prescriptive nature of the large AAG program has enabled community members to propose whatever need their individuall identify, and support these works, if warranted. Over longer timescales, the multidisciplinary PFCs focused in the area of nuclear astrophysics have led to great strides in TDAMM science, are continuing to develop several necessary integrated theory and simulation advances, and have outlined an approach we hope will be adopted to integrate astrophysics with other related disciplines. 

Within the NSF, the PFCs are the most appropriate venue for end-to-end approaches to TDAMM. They brought nuclear astrophysics into the the most mature field ready for multidisciplinary studies in TDAMM. Outside the NSF, the scientific return of NSF facilities is enhanced by results using NASA and DOE facilities. Advancements at the intersection of physics and astronomy is dependent on combining information from both PHY and AST and externally-run facilities into a greater whole. One possible limitation of the PFCs to the area of TDAMM is the requirement for the focus to fall directly under the purview of PHY. While certainly appropriate, the lack of a similar mechanism in AST may impede progress on integrating information from the PHY disciplines for discoveries within astrophysics. Additionally, successful emulation of the NNSA approach would need dedication of a higher proportion of time from more senior scientists, otherwise these individuals are unlikely to prioritize these projects and operate in a true multidisciplinary fashion.

\begin{quoting}
    \noindent Finding: The NSF supports facilities which enable the multimessenger era, as well as major electromagnetic facilities and multidisciplinary research efforts which make TDAMM science possible. Their successful approaches in nuclear astrophysics should be broadened for full end-to-end approaches, and adapted to other disciplines. One mechanism to enable this could be joint PHY and AST multidisciplinary efforts, or, ideally, those also supported by DOE and NASA. Such work may help develop generalist scientists needed to secure national defense.
\end{quoting}

Astrophysics is among the most visible sciences with enormous public interest, and is a recruitment pipeline for STEM education in the nation. Historically, astrophysics also produced uniquely capable multidisciplinary scientists. There generalists are of particular interest for both the NNSA, other areas of defense, and industry. At least within the NNSA, this pipeline has dwindled as astrophysicists have not been trained to their needs. End-to-end TDAMM approaches would help meet the need of astrophysics while also revitalizing the production of these generalists for advancement of the nation and securing the national defense.

\textbf{DOE} has multiple opportunities to strengthen tie to TDAMM astrophysics.  These ties will both advance TDAMM science and increase the impact of TDAMM observations by demonstrating the connection to fundamental physics.  In turn, TDAMM observations provide a new method to test both the physics and numerics studied by DOE.


Of the branches of DOE, the existing funding synergies for TDAMM science are within the office of the undersecretaries for Science and Innovation and the National Nuclear Security Administration (NNSA).  Within the office of the undersecretary for Science and Innovation, the Office of Science more closely resembles NSF programs and has the most connections to astrophysics, including High Energy Physics (HEP), Nuclear Physics (NP), and Advanced Scientific Computing Research (ASCR). Investments here are necessary to respond to the recommendations in the LRP and P5.  These programs have all funded (and continue to fund) specific astrophysics research including cosmic ray detectors such as the High Altitude Water Cerenkov detector (supported by HEP and NSF Physics) and the Fermi Gamma-ray Space Telescope (supported by HEP and NASA) as well as research in the supernovae and neutron star merger engines (supported by ASCR and NP).  Many dual-hatted astrophysicists also work in Basic Energy Science (BES) and Fusion Energy Science (FES) studying plasma physics.  Although the ties are extensive, they typically rely on a handful of scientists that work in both fields.  As with specific physics fields in NSF, astrophysics provides a public relations platform to highlight the excitement of these focused physics studies, building public awareness and attracting scientists to the field.

But TDAMM science has been intimately connected to the NNSA since its first director who, among other astrophysics projects, studied black holes (Oppenheimer-Volkoff equations).  NNSA scientists pioneered modern computational astrophysics and many of the computational advances in TDAMM science were associated with the NNSA laboratories.  NNSA mission science has strong parallels with astrophysical phenomena in that it requires the solution of complex problems.  NNSA laboratories have developed methods, using detailed multi-physics, and often scale-bridging, simulations to tie together focused physics experiments and full system observations with limited diagnostics to understand these complex problems.  The methodology of the approach to these national security problems can be directly applied to astrophysical applications, providing a means to further test the tools developed by NNSA.  As TDAMM science adopts this methodology, it demonstrates the approach to the general public, both confirming the importance of these efforts and training scientists with the broad skillsets required for national security.

NNSA laboratories tend to be isolated and, especially as new national security problems arise, tying to astrophysics will provide NNSA scientists with a broader understanding of the numerical tools available for a diverse set of problems.  As an example, as DOE focuses on better understanding of inertial confinement fusion, out-of-equilibrium physics effects become more important.  Ties to the heliophysics and space weather communities are already strong and these fields share insight into numerical advances in out-of-equilibrium plasma physics.  A similar gain can be obtained by having the atomic physics community and the implementation of atomic properties (e.g. Einstein coefficients) into a set of rate equations.  TDAMM scientists have implemented a number of simplifications and techniques to study the level-state distributions in astrophysical applications.  These methods could be adapted to inertial confinement fusion and other national security applications.

Simply put, what TDAMM astronomy gains from strong collaborations with DOE is a firmer base in specific fields of fundamental physics and training in a research approach designed to incorporate diverse physics needed to solve complex problems.  DOE benefits from the tests astronomy problems pose for DOE physics, codes and methodologies and a broader understanding of numerical methods for a diverse set of applications (improving the agility of DOE scientists, particularly those at NNSA laboratories).  Finally, TDAMM astronomy provides a way to highlight the exciting capabilities at DOE laboratories and these public relations can help ensure a long lasting pipeline of scientists to these national laboratories.

\begin{quoting}
    \noindent Finding: DOE has flagship scientific facilities which tie to TDAMM astrophysics. The multidisciplinary approaches, both within the NNSA and Office of Science, could be adapted to the needs of TDAMM. An expansion of training programs over a broad range of education levels, including additional student fellows, summer schools, workshops, and research programs, would foster the development of scientists which meet the needs of the DOE and industry. For direct scientific return, joint efforts with NASA and the NSF on multidisciplinary efforts would broaden the applicability of existing DOE investments in multidisciplinary studies in astrophysics.
\end{quoting}

The primary action item identified for DOE is to expand its training programs (increasing student fellows, summer schools, workshops, research programs) in holistic solutions to applied problems.  For DOE and industry, the nature of the applied problem is less important than ensuring the development of the next generation of generalist scientists.  TDAMM science provides ideal applications to train these scientists and better coordination of the training programs with TDAMM science needs will benefit both astronomy and DOE.  By working with academia, DOE can better advertise to academia the career opportunities for generalist scientists and build stronger collaborations with university professors.

\textbf{NASA:} In the US, NASA is largely responsible for astrophysical observations from space. X-ray, ultraviolet, and (soft) gamma-ray photons are only observable from above Earth's atmosphere. Space also affords a number of unique capabilities, such as continuous viewing. NASA's flagships have unmatched capabilities. As such, NASA has a foundational role to play in TDAMM astrophysics. NASA has provided responsible stewardship for these capabilities for decades; their missions have enabled major discoveries in every multimessenger transient to date.

NASA's missions unlock new capabilities which drive new multiphysics approaches. IXPE has driven the need for fluids and rad transport methods which can predict polarimetric signatures. ULTRASAT and UVEX are driving the need for rad hydro improvements for proper modeling of shock breakout. COSI will provide nuclear gamma-rays as transient diagnostics for novae and thermonuclear supernovae, being one of the most direct diagnostics possible. Habitable Worlds Observatory enables unique mapping of the origin of elements early in the history of the universe, which requires modernizing the ground segment of NASA flagship operations. As these elements may be necessary for, or increase the likelihood of, the development of complex life, this is well aligned with the mission focus.

\begin{quoting}
    \noindent Finding: NASA uniquely enables many major discoveries in TDAMM. However, these discoveries rely on use of multiple missions, NSF facilities, and modeling expertise typically supported by DOE. NASA's approach to TDAMM must be strategic, and coordinated with the other agencies. 
\end{quoting}

Because of these unique capabilities, and corresponding unique needs, we identify a few areas where NASA's approach could be better aligned to meet the global needs of TDAMM. The first: the fleet is more powerful than the individual missions. This is also true when done in the context of NSF and DOE facilities. A strategic approach to developing science, technology, and mission selection in the larger TDAMM context would enable greater scientific return.

One unique separation of responsibility within NASA is ownership of space-based studies of fundamental physics. Formally this responsibility appears to lie within BPS. However, many of their missions are focused on tests of fundamental physics which are performed directly on a spacecraft. NASA, typically in the area of TDAMM,  is the only agency which can enable some unique tests of fundamental physics. Famously, NASA's Fermi and NSF's LIGO combined to precisely measure the speed of gravity, killing a whole subfield of gravity. The proposed test of QED photon splitting, outlined above, can only be done by a NASA mission. LISA will further enable unique tests of fundamental physics. We request that APD and BPS ensure such science is considered.

We find that an enhancement in theory investment from NASA is necessary to meet the needs of TDAMM, and NASA's recommendations from the Astro Decadal. General investment in the area of TDAMM will allow informed prioritization of technology for the best future missions. Currently missions are selected with comparatively underdeveloped theory, at least for TDAMM science, and are not judged again in the process. This problem contributed to the lack of emphasis for primary science on shock breakout and super/nova in the UVEX and COSI proposals. Aligning theory and analysis support for forthcoming missions will help coevolve instrument choices, software development, and observational plans to ensure these facilities are best used, matching the codesign approach utilized in other fields. Having reviews ensure that the necessary theory and analysis investments are occurring is a mechanism to ensure science is truly driving mission design. Allowing for enhanced support for TDAMM theory and analysis for secondary science may be a viable mechanism to ensure the TDAMM uses of all missions are appropriate. For large-scale end-to-end approaches, TCANs must be larger. Without this investment, the relevant scientists will prioritize DOE and NSF calls, and NASA's Physics of the Cosmos program will not live up to its potential.

One unique area of need identified in this workshop is strategic support for atomic spectroscopy studies. This is required because this is not a priority in the field of AMO; initial steps are outlined in Section~\ref{sec:disciplines_atomic}. We highlight this need here because the missions whose observations are limited by understanding of atomic physics in certain regimes are predominantly NASA missions, including XRISM, JWST, and UVEX. The integration of atomic scientists in the development phases of XRISM demonstrates the success of this investment. However, for JWST and UVEX, these observations are not a critical driver, and this work is not supported by the missions. 


\subsection{Science Prioritization}
Time-domain and multimessenger astrophysics is a broad technique used to study a breadth of topics. While all of these events are worth study, we must operate in a finite funding environment. This requires prioritization. To quote the Astro 2020 Decadal: ``Within this discovery landscape, driven by improvements in gravitational wave and neutrino
detection, and upcoming facilities such as the Rubin Observatory, one priority area stands out: the
application of these new tools to the formation, evolution, and nature of compact stellar remnants such as
white dwarfs, neutron stars, and black holes, as probed by the gravitational wave signatures of their
mergers, together with rare explosive events that can be explored by the unique cadence and multi-color
sensitivity of the Rubin Observatory. Sensitive observations of high-energy neutrinos and charged
particles add new elements of discovery space, which will probe the universe’s most extreme particle
accelerators—New Windows on the Dynamic Universe.''. This description led the selection of source classes utilized at the workshop and in this document.

Within the context of multidisciplinary studies, the sources most ready for end-to-end approaches are explosive transients, including of supernova, nova, and neutron star mergers. The dominant reason for this is the sustained investment in nuclear astrophysics by the NSF and DOE, and the enormous work from the relevant communities in multidisciplinary models of these objects. This is further supported as each has forthcoming facilities which will enable regular detection of new diagnostics and these objects and their physics are of interest to all three funding agencies.

\begin{quoting}
\noindent Finding: Explosive transients are the source classes most prepared for full end-to-end approaches. These efforts should begin in earnest. Jetted transients have similar scientific promise, but must first have sustained investment to build the necessary community connections before an end-to-end approach can be undertaken. 
\end{quoting}

While this is detailed in the source sections below, we briefly comment on these statements. Core-collapse supernova have mature simulations. Their shock breakout signatures will be regularly observed for the first time with UVEX and ULTRASAT. Thermonuclear supernovae are foundational tools for cosmology. Their physics are closely related to those of novae, which serve as a local lab for many cosmological phenomena. The most difficult problem in modeling both of these sources is the same: reactive flows. A small number of each will be detected by COSI, allowing the unique use of nuclear gamma-ray information. Neutron star mergers are of interest to nearly every major facility, and the canonical multimessenger transients expected with the upgrades to LIGO. All of these objects tie to understanding the origin of the elements, and detailed understanding will resolve the possible contribution of neutron star mergers and supernova to the generation of the heaviest elements. These two source classes also give insight into the equation of state of neutron stars, playing a key role in the cosmic density ladder. These two NRC questions could have transformational understanding within a decade, given sufficient investment. Lastly, renewed understanding of these objects will enable greater precision in cosmology, both type Ia supernovae in the near term and neutron star mergers in the long term.

X-ray binaries, tidal disruption events, active galactic nuclei, and gamma-ray bursts are all events with accretion and jets. Future multiphysics of these sources would allow for understanding whether mass can be extracted from black holes, the origin of ultra-high energy cosmic rays, and the fastest material in the cosmos. These questions are equally as important as those probed by explosive transients; however, we are less well prepared to pursue them. This is for many, complex reasons. First, the physics involved are just more complicated. Terrestrial laboratories can barely emulate the best behaved portion of these objects. Second, the multidisciplinary communities are not as well established. This is driven in part by a lack of strategic and sustained effort in plasma astrophysics by the funding agencies; however, this is likely affected by the trend of plasma scientists to work in heliophysics, which have better data in a more well understood regime. As a result, the mapping of problems, dominant sources of uncertainties, and areas to target investment are not as mature. This is demonstrated by the lack of an unambiguous counterpart to IceCube neutrino sources. While several promising candidates have been detected, none of these scenarios have high fidelity modeling to prove self-consistent picture with the multiwavelength observations. Thus, the priority in these studies is in building up connections within and between the respective source class communities as well as interdisciplinary communities with plasma, HEDP, and fluids. Adoption of computational approaches other fields may also be fruitful.  

Magnetars ruin our neat picture. These objects are most closely aligned with the research of explosive transients. However, in addition to the uncertainty in dense matter, modeling must also contend with extreme magnetic fields and enhanced effects from superconduction and superfluidity in these stars. The scientific discoveries possible with magnetars may exceed the other source classes, including the aforementioned test of QED, but also in exploring extreme states of matter inaccessible in any other domain. However, the connections necessary to cohesively consider all of these effects do not exist. Magnetar studies will have knock-on effects in the approaches developed for explosive transients, can adopt computational approaches brought into astrophysics for plasma studies in jetted transients, and will of course continue with their own studies. Efforts which approach the key questions mentioned and foster multidisciplinary studies in this area should be supported.

\newpage
\section{Disciplines}
\label{sec:disciplines}
This section provides a concise overview of the various disciplines pertinent to TDAMM. Most of the disciplines mentioned have their own dedicated sections. However, laboratory astrophysics is an umbrella term for the application of ground-based experiments to the astrophysical endeavor. We combine these focused aspects into their larger respective disciplines. Astroparticle physics is subsumed under astrophysics. The sections were authored by subject matter experts in their respective fields, with the intention of being informative for scientists from other disciplines and fostering the pursuit of multidisciplinary studies.

\subsection{Astrophysics}
{\centering 
\textit{Contributors: Jennifer Andrews, Matthew G. Baring, Eddie Baron, Peter G. Boorman, Eric Borowski, Floor S. Broekgaarden, Eric Burns, Poonam Chandra, Emmanouil Chatzopoulos, Francesca Civano, Luca Comisso, Tarraneh Eftekhari, Ryan J. Foley, Gwendolyn R. Galleher, Fan Guo, J. Patrick Harding, William Raphael Hix, Kelly Holley-Bockelmann, Rebekah Hounsell, C. Michelle Hui, Robert I. Hynes, Weidong Jin, Heather Johns, Jamie A. Kennea, Gavin P. Lamb, Tiffany R. Lewis, Ioannis Liodakis, Nicholas R. MacDonald, Thomas Maccarone, Lea Marcotulli, Athina Meli, Bronson Messer, M. Coleman Miller, Matthew R. Mumpower, Michela Negro, Eliza Neights, Peter Nugent, Emily Reily, Lauren Rhodes, Paul M. Ricker, Christopher J. Roberts, John A. Tomsick, Aaron C. Trigg, V. Ashley Villar, Zorawar Wadiasingh}}\\

\label{sec:disciplines_astrophysics}
Astrophysics is among the oldest sciences. Its major discoveries have often contributed to the formation of new fields. At various times its major advances have been multidisciplinary discoveries, especially in the study of the physics of the universe. As other fields matured, the fidelity of their knowledge often exceeded the order of magnitude accuracy possible in astrophysics. However, the modern TDAMM approach enabled through several transformational facilities has prompted a renewed need for multidisciplinary approaches. 

The Astro Decadal prioritized the TDAMM study of compact objects \citep{national2021pathways}. Events that form or are powered by compact objects involve extreme energetics, densities, temperatures, timescales, and distances, and represent the final frontier of understanding in several domains of science. However, these events are typically affected by multiple extreme physics processes, preventing the development of isolated physics tests that is the focus of laboratory experiments.  These complications have often led to the acceptance of order-of-magnitude model accuracy.  But laboratory experiments also often suffer from uncertainties in physics beyond the target physics of the experiment.  Laboratory scientists overcome these constraints by combining multipled diagnostics and multiple experimental shots to study a given physical phenomena.

Astronomy has developed tools to probe the Universe in an attempt to understand the processes that produce the signals we detect. Observing the evolution of the behavior of the system is referred to as time-domain astronomy. Measuring small variation in brightness over a range of energies is taking a spectrum, where identification of elemental lines can give direct insight into the composition of the emitting region or measurement of the redshift to the object (the distance to the source). Multiwavelength astronomy provides insight through measurement of the light across the EM spectrum, including thermal and non-thermal emission processes. Some detectors can measure polarization, providing information on the geometry of the system or large-scale ordered magnetic fields. Radio interferometry can now achieve ultraprecise spatial resolution allowing for images of black holes and measurement of proper motion or extension of outflows. Multi-messenger astronomy involves combining information from light with complementary insight from GWs, neutrinos, dust, and meteorites. Modern studies combine the methods outlined above to gain insight into astrophysical phenomena and increase our knowledge of the universe. This approach is similar to some ground-based experiments, which add detectors to provide additional and complementary observations of singular experiments such as high-energy laser runs - a process referred to as adding diagnostics. 

TDAMM astronomy is the most promising avenue for fully understanding the physics of the cosmos. By combining information from forthcoming facilities with that from other disciplines in physics and related fields, we can move away from over simplified models and fully enhance our understanding of the Universe. Many TDAMM approaches rely on the science gained by contemporaneous observations of individual events, such as the study of neutron star mergers through GWs and light. In these cases, multiple diagnostics provide complementary information, enabling a more detailed understanding of what is happening. Obtaining the datasets of interest places extreme demands on observational facilities. For neutron star mergers identified via GWs, their initially poor localization on the sky requires large-scale, dedicated follow-up campaigns with wide-field surveys or clever targeting approaches to determine the origin, which requires filtering through large numbers of unrelated transients. Once the source is identified, telescopes across the electromagnetic spectrum must observe as rapidly as possible before key signatures fade away. Great efforts are devoted to success in this endeavor. Proper prioritization and use of these facilities require improved and accessible models. These are also required for optimal analysis of the data.

Just as with laboratory experiments where multiple shots and studies are conducted to study the same physical process, TDAMM science can also be done without concurrent observations. For example, the GW measurement of the stellar mass black hole spectrum is giving insights into how massive stars explode (Section~\ref{sec:sources_ccsn}). In searching for the origin of heavy elements, the variation in enrichment in old stars inferred from spectral observations as well as studies of isotope distribution in deep-sea cores both pointed to rare, high-yield events as the source, pointing to neutron star mergers. Holistic understanding of events through this approach have led to major discoveries, but require the connection of disparate information through complex and interlinked simulations. 

The new TDAMM era is possible because of a breadth of astrophysics and other facilities giving new insights across a range of physics of relevance for the physics of the cosmos. Making maximal use of these facilities requires a corresponding investment in theory and simulation to integrate the data into coherent understanding. Our finding for astrophysics is a minor alteration of a recommendation from the 2003 NRC Report \citep{national2003connecting}:
\begin{quoting}
\noindent Finding: The greatest scientific advances in physics and astronomy possible through time-domain and multimessenger astrophysics require an interagency initiative with the participation of DOE, NASA, and NSF. 
\end{quoting}

\subsection{Gravity}
{\centering 
\textit{Contributors: Ivan Agullo, Eric Burns, Hsin-Yu Chen, Alejandro Cárdenas-Avendaño, Kelly Holley-Bockelmann, M. Coleman Miller, Elias R. Most, David Radice, Jocelyn S. Read, Peter Shawhan, Gaurav Waratkar}}\\

\label{sec:disciplines_gravity}
Gravity is one of the fundamental forces and the only without a quantum field theory. Observationally, it plays a crucial role in our understanding of the universe. Gravitational waves (GWs) provide a novel means to measure distances, masses, tides, and merger rates of compact objects. This capability offers profound insights into astrophysical phenomena that would otherwise remain beyond our reach. Moreover, research in this field addresses fundamental questions in astrophysics and physics. It helps us explore the nature of dense matter, understand the structure of neutron stars, investigate the origins of r-process elements and gamma-ray bursts, unravel the nature of black holes (testing general-relativity in the strong field regime), and contribute to precision cosmology and the study of hierarchical structure formation in the universe.

Several major facilities are currently advancing our understanding of gravitational waves. The US ground-based interferometer, LIGO, is joined in IGWN by Virgo in Europe, KAGRA in Japan, and will deploy a detector in India. The next generation of telescopes include Cosmic Explorer in the US and the Einstein Telescope in Europe. Focused detectors provide characterization capability, such as the proposed NEMO in Australia targeted at $\sim$kHz frequencies to directly observing the merging of binary neutron stars. NANOGrav is joined by the European Pulsar Timing Array, the Indian Pulsar Timing Array Project (InPTA), and the Parkes Pulsar Timing Array for the international Pulsar Timing Array (iPTA). This consortium utilizes radio telescopes for a greater integrated whole. The frequency range between these two regimes will be broadly covered by LISA. In addition, emerging detector technologies, such as atomic interferometry and quantum crystals for low-frequency detection (in the millihertz to hertz range), and microwave cavities and levitated sensor detectors for high-frequency detection (in the megahertz range), promise to open new windows onto the gravitational-wave spectrum.

Gravitational-wave astronomy is relevant for multiple key questions identified both in the Astronomy 2020 Decadal Priority Science Area: New Windows on the Dynamic Universe and the 2023 Long Range Plan for Nuclear Science: Nuclear Astrophysics. Both key field documents highlight the following questions:
\begin{itemize}
\item {\bf What physics governs the lives and deaths of stars?} Remnant black holes and neutron stars are the endpoints of massive star lives. Ground-based facilities like LIGO, Virgo, and KAGRA provide a comprehensive survey of merging binaries of stellar-mass black holes and neutron stars, measuring the masses and spins of both the components. Improved detector sensitivities will increase chances of observing galactic transients, like those from accreting neutron stars, magnetar flares, or core-collapse supernovae. Space-based facilities like LISA make a census of the galactic compact binary population of neutron stars and white dwarfs. 
\item {\bf What is the nature of matter under extreme conditions?} The cold neutron-star equation of state, up to 5-6 times nuclear density, imprints on the gravitational-wave signal of neutron-star mergers through tidal interactions during orbital evolution as well as on the dynamic transition to merger. Post-merger gravitational-waves are sourced by any long-lived remnant after the stars collide, and are sensitive to the equation of state of hot dense matter along with more complex microphysics.
\item {\bf How are heavy elements produced in our universe?} Gravitational-wave observations are used to infer neutron-star merger rates, masses and spins of merging stars, and the equation of state of dense matter; redshift-dependent rates from upgraded observatories will also inform the merger delay time distribution. Joint kilonova detection will connect progenitor properties to outflow conditions. Together, these observations will inform the contribution of neutron-star mergers as a channel for r-process element production.
\item {\bf What fundamental physics we learn?} Gravitational-wave observations provide strong-field tests of general relativity. Binary mergers act as standard sirens for cosmology, offering well-calibrated luminosity distances. Electromagnetic redshift measurements from the same source event, or the redshifting of mass features in gravitational-wave populations, will enable high-precision cosmological measurements. The massive and supermassive black holes binaries observed by LISA and PTAs are the cornerstone of hierarchical structure formation. Future facilities like Cosmic Explorer could record early mergers of primordial black holes.
\end{itemize}
\begin{figure}
    \centering
    \includegraphics[width=1\linewidth]{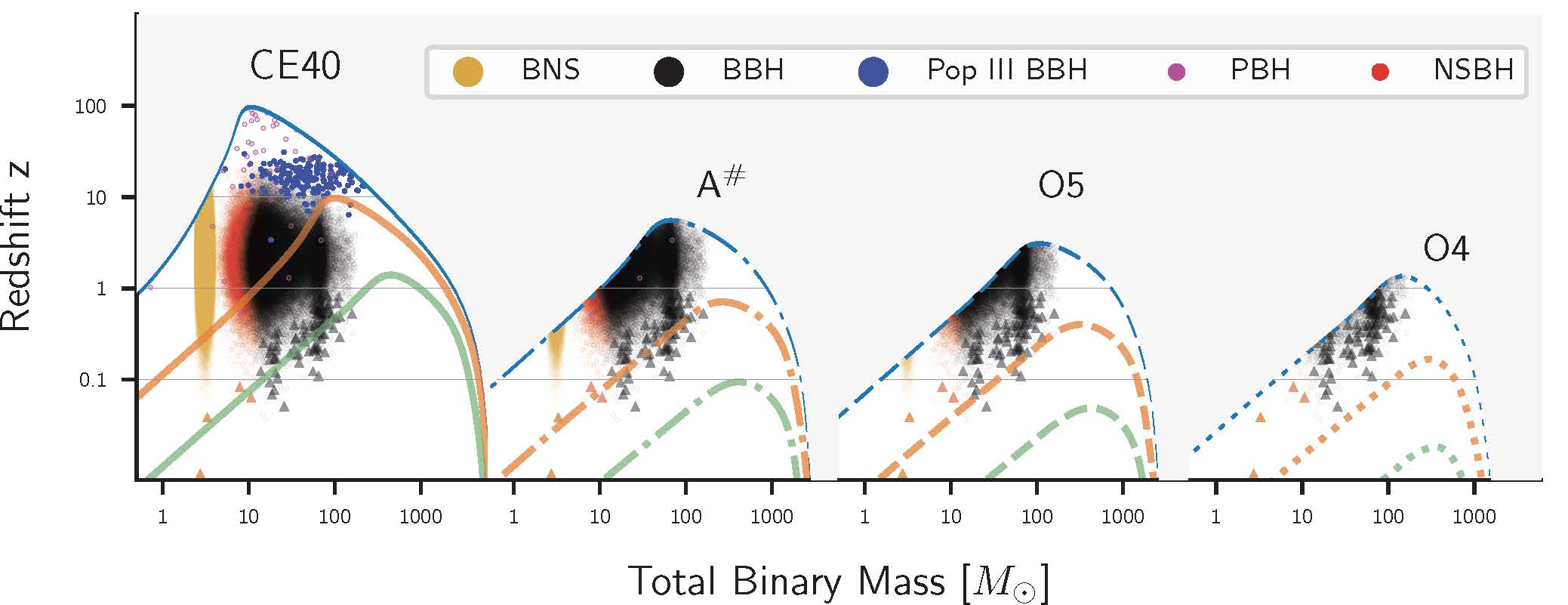}
    \caption{Reach of ground-based gravitational-wave observations of merger transients. Triangles mark events recorded before O4, which started in 2023. O5, A\#, and CE40 represent planned and nominal detector upgrades from the next few years until $\sim$2040. Figure from \citep{evans2023cosmic}.}
    \label{fig:gw-reach}
\end{figure}

\begin{quoting}
\noindent Finding: Gravitational wave astrophysics lies at the heart of major recent TDAMM advances and has fostered much of the current multidisciplinary ecosystem. The upgrades and next generation gravitational wave detectors are among the most important in driving future new discoveries.
\end{quoting}

Binary mergers are engines of kilonovae, gamma-ray bursts, and potentially some fast radio bursts. Observatories with sensitivity to merger events can identify (or rule out) progenitors of other transients out to the reach shown in Fig.~\ref{fig:gw-reach}. Gravitational-wave transient observations, especially with multiple facilities, provide distances (for mergers), sky locations, and progenitor properties for electromagnetic followup. 


It is important to emphasize that strong gravity plays a central role in many transients, in addition to mergers. Examples include CCSNe, accretion onto compact objects, magnetar flares, black-hole jets, etc. Even though no direct detection of gravitational-waves from non-binary sources has been announced yet, gravitational-wave astronomy is expected to play an increasingly important role in constraining the nature of some of these transients in the future. Moreover, tests of general-relativity with compact binaries, as well as Solar system tests of the theory, provide the foundation for the theoretical models employed to understand gravitating objects in general. Similarly, the data analysis techniques developed for gravitational-wave astronomy, such as matched filtering, are highly sophisticated and could find use in other fields, for example in the analysis of spectra and time-domain data. This synergy should be further explored.

To advance our understanding of complex astrophysical phenomena, we need comprehensive end-to-end simulations. These should span the entire process, from nuclear physics inputs to the modeling of kilonovae and gamma-ray bursts (GRBs), as well as from cosmological simulations to supermassive black hole mergers and their electromagnetic counterparts. While initiatives like Exascale Nuclear Astrophysics for Facility for Rare Isotope Beams (ENAF) by the DOE, and Network for Neutrinos, Nuclear Astrophysics, and Symmetries (N3AS) and Nuclear Physics from Multi-Messenger Mergers (NP3M) by the NSF are making strides in this area, there's a clear need for further collaboration. We require more projects that develop simulation infrastructures bridging these different phases, similar to NSF's Modular Unified Solver of the Equation of State (MUSES), which is developing a framework for the equation of state, but focusing on the simulation parts. End-to-end simulations must be followed by the construction of emulators and/or the calibration of semi-analytical models in order to perform simulation-informed interpretation of the observations. As part of this process, the validation of the simulations and uncertainty quantification are critical. The former should include code-to-code comparison for verification and, potentially, code validation using high energy density physics experiments.

Open data, simulation data and data from laboratory measurements and astronomical observations, is essential to interface different simulation codes in the context of a multi-step modeling effort and to interpret multi-messenger observations. However, the utility of this data is maximized only if there are open source tools and workflows enabling the community to process it.

While this report and much of the field is focused on the use of gravity as an observational tool, TDAMM science includes advancing our understanding of gravity. Many of the most precise tests of General Relativity have been performed by LIGO, some of which will be advanced by LISA. However, many of the most precise tests require multimessenger detections \citep[see, e.g., the summary in][]{burns2020neutron}.

Career development is another key consideration, particularly for scientists working at the intersection of different fields, such as nuclear physics and gravity. Opportunities such as fellowships for postdocs collaborating across multiple groups, like N3AS and NP3M, can foster this interdisciplinary work. However, we need a cultural shift in hiring and evaluation practices. This shift will be encouraged if funding agencies emphasize the importance of interdisciplinary research and commit to supporting it in the long term.

\begin{quoting}
\noindent Finding: New discoveries in gravity and TDAMM require and end-to-end solution. As this requires detailed work which is not typically the most visible, a visible shift to support this work from funding agencies through to universities is necessary. 
\end{quoting}

\subsection{Nuclear Science}
{\centering 
\textit{Contributors: Kelly A. Chipps, Phong Dang, Catherine M. Deibel, Joseph Henning, William Raphael Hix, Maria Gatu Johnson, Kristina D. Launey, M. Coleman Miller, Valarie Milton, Darin C. Mumma, Matthew R. Mumpower, Rene Reifarth, Andrea Richard, Hendrik Schatz, Nicole Vassh}}\\

\label{sec:disciplines_nuclear}

The TDAMM era of astronomy opens up a new window into the dynamic cosmos. Nuclear physics is at the heart of many of the time-variable sites of interest, including neutron star mergers, super- and hypernovae of all types, classical novae, other types of accreting white dwarfs, X-ray bursts, and neutron star phenomena.  Nuclear physics plays a key role in the underlying explosion engines, it shapes light curves, and it translates thermodynamic conditions into unique nucleosynthetic signatures observable through spectral features, Solar abundances, stardust, and possibly cosmic rays. There is a tremendous scientific opportunity emerging at the intersection of the advances in TDAMM astronomy, outlined in the Astro Decadal survey, and advances in nuclear physics, charted in the 2023 NSAC Nuclear Science Long Range Plan. With the new experimental and theoretical nuclear physics results expected in the coming decades, TDAMM astrophysics in concert with nuclear physics, atomic physics, and plasma physics has an opportunity to address some longstanding and fundamental questions in science. Through nuclear physics, TDAMM observations directly probe some of the most extreme environments in the cosmos and thus provide novel insights into not only the physical mechanisms behind various stellar explosions, but also the properties of dense cold and hot matter in regimes not accessible in the laboratory.  

In addition, TDAMM sites are connected to some of the long-standing open questions concerning the origin of the elements. In 2003 the NRC Committee for the Physics of the Universe released their study “Connecting Quarks with the Cosmos” identifying the origin of the heavy elements as one of the 11 science questions for the new century, and understanding TDAMM sites is central to this endeavor. Already multi-messenger observations of neutron star mergers have shed new light on this question, but its still unclear what elements neutron star mergers create. Additionally, mounting evidence, such as isotope studies of old stars, points to significant contributions from other yet unknown (likely TDAMM) sites. Addressing these open questions will have broad scientific impact. Much of the evolution of the cosmos is driven or shaped by the chemical and isotopic composition of the universe. The existence of life on earth hinges on the existence of the right amounts of carbon and oxygen, and radioactive heavy elements are key for heating Earth’s core and thus for creating the earth’s magnetic field shielding us from cosmic radiation. The evolution of the composition of the universe is defined by the exact magnitudes of nuclear reaction rates, nuclear masses, and the properties of the nuclear force – to quote Stan Woosley: nuclear physics is the DNA of the cosmos. Similarly, new insights in the properties of dense nuclear matter have far reaching consequences across astrophysics affecting black hole formation, neutron star masses, core collapse supernovae, and gravitational waves. 

\begin{figure}
    \centering
    \includegraphics[width=1\linewidth]{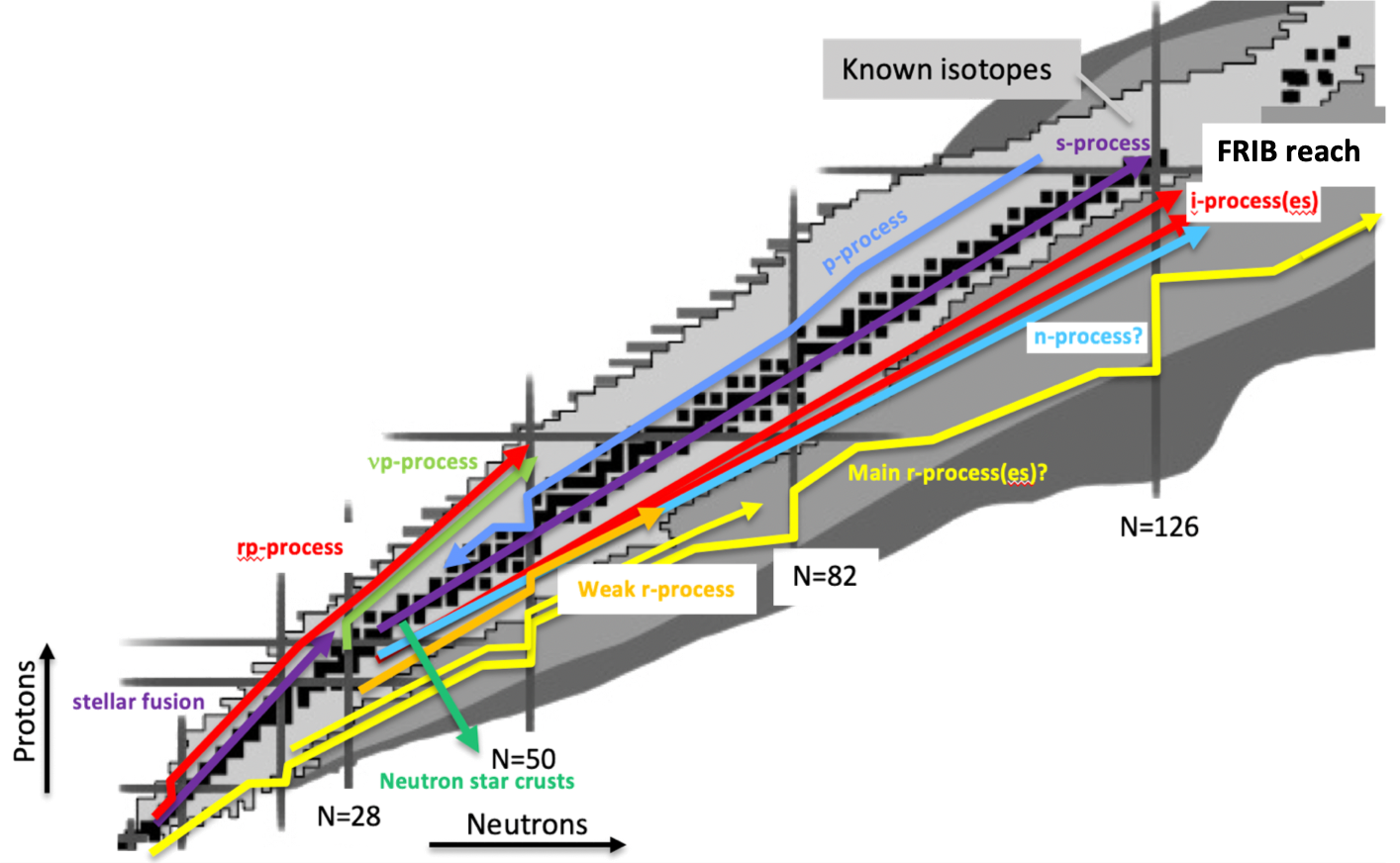}
    \caption{Schematic delineation on the chart of nuclides of the nuclear reaction sequences that power cosmic events and/or are responsible for the synthesis of the elements. Most of the reaction sequences involve unstable nuclei (stable nuclei are black) and most are related to TDAMM sites such as neutron star mergers (r-process), supernovae (r-process, weak r-process, n-process, p-process, $\nu$p-process), novae (rp-process), and neutron stars (rp-process, neutron star crust processes). Most of these processes are not well known in astrophysics, including TDAMM, but they are key to understand in nuclear astrophysics \citep[e.g.][]{NSAC-LRP-2023}}
    \label{fig:nuclear_chart}
\end{figure}

There is a broad range of specific TDAMM sites where nuclear physics shapes critical observables (Fig.~\ref{fig:nuclear_chart}). These and other examples have been highlighted in recent white papers \citep{JINAHorizons} and the 2023 NSAC Long Range Plan \citep{NSAC-LRP-2023}. Some are summarized here. The major new facilities of relevance for each source class are discussed in their respective sections in Section~\ref{sec:sources}.
\begin{itemize}
\item Core collapse supernovae: nuclear electron capture affects the dynamics of the core collapse, the nuclear equation of state drives the initial bounce that launches the shock and shapes neutrino and gravitational wave signals, and nuclear reactions create new nuclei via explosive nucleosynthesis, that define the observable composition of the ejecta. Nuclear reactions also create radioactive nuclei that define the light curve, and emit gamma-radiation that can be directly observed, for example with the upcoming COSI mission. Nuclear reactions in neutrino driven winds or in rapidly ejected material in jets may create heavy elements, in rare special cases such as strong magnetic fields, possibly up to uranium \citep{boccioli2024}.  Neutrinos drive the outflow, but also set the composition (electron fraction) and the yields can be used to probe out understanding of neutrino physics including oscillations and sterile neutrinos~\citep[e.g.][]{2006PhRvD..73i3007B}. 
\item Thermonuclear supernovae: Carbon fusion triggers the explosion, and a broad range of nuclear reactions create new stable and radioactive nuclei that shape composition observables and light curves. The nucleosynthesis is sensitive to the explosion mechanism \citep{seitenzahl2017}. 
\item Novae: Nuclear reactions during explosive hydrogen burning power the explosion and create freshly synthesized nuclei up to around Ca observable directly or as isotopic signatures in stardust. Novae may also create some radioactive isotopes that may be observable with the future COSI mission \citep{denissenkov2014}.
\item Neutron star mergers: The nuclear equation of state shapes the GW signal. Nuclear reactions in the rapid neutron capture process create radioactive heavy nuclei that power the electromagnetic kilonova counterpart, shape the light curve, and create element signatures in the spectra, observed for example by JWST \citep{thielemann2017}. Additionally telescopes like COSI will provide the possibility to observe specific MeV gamma emission lines from nuclear decays for potential nearby events. 
\item X-ray bursts: Explosive hydrogen and helium burning directly power frequently observed X-ray bursts and create new nuclei up to around Cd that may be observable as spectral signatures with NICER and future X-ray missions such as STROBE-X \citep{schatz1998}. 
\item Neutron star cooling in quasi-persistent transients: nuclear reactions in the crust of accreting neutron stars heat and cool the crust, which can then be observed to cool during quiescence. Anisotropies created by nuclear reactions may deform the neutron star leading to continuous gravitational wave radiation that may be detectable with future gravitational wave observatories. 
\item Magnetars and fast radio bursts: nuclear physics may play an important role in magnetic field evolution and various star quakes and outbursts, including observed oscillation spectra. Giant flares from magnetars may be an early source of r-process elements.
\item Other TDAMM sources: given the dynamic nature of TDAMM sources, it is likely that nuclear physics will play a role in many other sources, including the ones to be discovered. Possibilities include neutron star inspirals, accretion induced collapse events, or various types of accreting white dwarf systems. 
\end{itemize}

\noindent \textbf{Nuclear Physics Challenges for the TDAMM Era:} The nuclear physics needed for TDAMM sources poses particular challenges:
\begin{itemize}
    \item {\bf Rare isotope frontier:} The extreme conditions and short timescales of the events drive the nuclear reaction sequences away from the valley of stability (Figure~\ref{fig:nuclear_chart}). The radioactive (or rare) isotopes involved in the reaction sequences are challenging to produce in the laboratory and as such experimental information is very limited – in extreme cases such as neutron star mergers, reaction sequences involve nuclei that have never been observed experimentally \citep{thielemann2017}. Theoretical predictions are also challenging and not sufficiently accurate. This is due to both the fundamental challenges of the nuclear many body problem as well as the fact that theories have been traditionally developed and benchmarked with stable or near stable isotopes and cannot easily be extrapolated to more unstable nuclei. For reactions among two radioactive nuclei (such as the neutron captures on radioactive nuclei that take place in supernovae or neutron star mergers during nucleosynthesis), effective techniques and facilities to perform measurements remain to be developed \citep{reifarth2014}. 
    \item {\bf Low energy frontier:} A traditional challenge in nuclear astrophysics is the charged particle reactions in stars being too slow to be measurable in the laboratory at astrophysical energies \citep{wiescher2020}. Despite of the typically higher temperatures, this challenge also applies to some TDAMM sites. For example, the carbon fusion reaction triggering thermonuclear supernovae or certain types of X-ray bursts needs to be known at energies where cross sections are too low for direct measurements, and where unknown theoretical effects such as the formation of molecular clusters within the merging nuclei, prevent theoretical predictions. Another example is the rate for $\alpha$-capture on carbon, that has been shown to impact the black hole mass distribution inferred from gravitational wave observations. 
    \item {\bf Plasma frontier:} Also outstanding in the context of TDAMM nuclear physics are questions about plasma effects on nuclear reactions \citep{10.3389/fphy.2023.1180821}. This includes plasma electron screening \citep{10.3389/fphy.2022.942726} and reactants in excited states (thermally, through neutron interactions \citep{10.3389/fphy.2022.917229}, electron transfer, or electron capture) and other near-threshold effects \citep{10.3389/fphy.2022.1009489}. 
\end{itemize}

As such much of the nuclear physics needed for the TDAMM era remains unknown. However, in the coming decade with the developments outlined in the 2023 Nuclear Science Long Range Plan major advances that address these challenges and directly impact TDAMM science can be expected: 
\begin{itemize}
    \item {\bf In experimental nuclear science:} the startup of the FRIB rare isotope facility in the US in 2022, 31 years in the making, marks a major step in capability to produce many of the radioactive nuclei in TDAMM sites.  Upgrades at Argonne National Laboratory such as nu-CARIBU and the N=126 Factory will provide enhanced production capabilities for neutron rich radioactive nuclei in specific regions of key importance for kilonovae (and other r-process sites). Upgrades at US stable beam facilities as well as new underground accelerators in the US, Italy, and China will enable to push cross section measurements to lower energies \citep{best2016}. A new technique to measure neutron capture reactions on unstable nuclei using a heavy ion storage ring and a neutron spallation source is being developed at LANL \citep{RGH17}. And novel techniques using high powered lasers at NIF and University of Rochester open up opportunities to measure nuclear reactions in the extreme plasma and neutron flux environments encountered in TDAMM sites. For all these developments nuclear astrophysics has been a major motivation. 
    \item {\bf In theoretical nuclear science:} major advances for TDAMM science are on the horizon as well. Density functional theory has become a powerful tool to predict masses and nuclear structure across the entire chart of nuclides, and has also been applied to improve predictions of nuclear fission, which is of particular importance for kilonova observables where fission signatures may imprint themselves on light curves (i.e., affecting the fading rate) and the abundance patterns of the heavy elements produced. An important development for nuclear astrophysics has been the use of machine learning techniques in nuclear theory. One area where this has led to significant progress is uncertainty quantification, which is essential for using nuclear theory predictions for quantitative interpretation of observations. There have also been significant advances in reaction theory, both in theoretical techniques to describe reactions in the laboratory that can then be used as “surrogates” for reactions occurring in astrophysical environments, and in terms of understanding quantum effects that impact the low energy extrapolations of experimental data needed for astrophysical applications. Additionally recent work in nuclear astrophysics theory has made key progress in identifying links between potential observational signatures and specific nuclei, such as the influence of $^{254}$Cf on light curves. 
\end{itemize}

\textbf{What is Needed for the TDAMM Era of Nuclear Astrophysics}
\begin{itemize}
    \item {\bf Nuclear physics efforts:} The most critical need is to fully exploit the new experimental and theoretical opportunities now becoming available and outlined above. This includes more direct measurements of astrophysical reactions (including in some cases direct measurements in a plasma environment), taking advantage of \textit{ab initio} theoretical approaches possible for light nuclei to inform theoretical models for heavier nuclei, and the determination of reactions on excited states instead of the ground state (especially at the high temperatures in TDAMM sites many reactions occur among nuclei in excited states). In addition, a number of facility upgrades and new instrumentation outlined in the Nuclear Science Long Range Plan will be important for TDAMM science. These include at FRIB the new FRIB decay station (the 400 MeV accelerator upgrade that will greatly enhance the reach for heavy unstable nuclei), the High Rigidity Spectrometer, and the ISLA spectrometers, as well as at ANL the completion of the various upgrade projects. Strong support for smaller university laboratories, including the ones deep underground, will also be essential. In addition, the continued development of techniques for measurements of neutron capture rates in storage rings and for measurements of nuclear reactions in laser induced plasmas at universities and national laboratories will be needed. 
    \item {\bf Uncertainty quantification:} This is critical for enabling the quantitative interpretation of TDAMM observations. Uncertainty quantification is well established in the experimental community, but much remains to be done in quantifying uncertainties in theoretical predictions, and in extrapolations or corrections applied to experimental data. The latter is essential for propagating experimental uncertainties into astrophysical reaction rate data that are used in astrophysical models. Of importance are also correlations of uncertainties in theoretical predictions. For example, cases of models that predict properties for a large number of nuclei and where uncertainties from nucleus to nucleus can be strongly correlated, or predictions of different quantities using the same theory. Correlations of uncertainties with temperature and density are also important given the large thermodynamic variations encountered in TDAMM sites. The implementation of this physics into nuclear reaction networks is not straightforward and thus much more work must be done to understand the impact of these numerical uncertainties on network calculations.  Tying this to astrophysical observations requires understanding in detail the uncertainties in the explosions and the trajectory evolution (temperature/density versus time) for these explosions.  There are significant opportunities in bringing in machine learning and data science techniques.  
    \item {\bf Interdisciplinary workflows:} The new capabilities in experimental and theoretical nuclear physics cannot be effectively exploited to advance TDAMM science without effective interdisciplinary workflows. Experiments are expensive and beam time is highly competitive – a clear astrophysical motivation that directly links the nuclear physics to an observable and an actual observatory or mission is critical. Experiments also take a long time – the time from developing the idea to publishing a paper often ranges from 3-5 years, in cases where technical developments are needed it can be significantly longer. Experimental campaigns that provide the large scale data sets needed for many TDAMM applications can therefore take decades and need to be planned and optimized carefully to maximize scientific impact.
    
    At its most basic, there needs to be a work flow connecting nuclear physics, astrophysical models, and observations. A center piece of this work flow is the theoretical analysis of the connections between nuclear physics and observables, including sensitivity studies (which compare nuclear physics variations in the model to observables). In essence these studies determine how specific nuclear physics is linked to specific observables and as such they are key to understanding the role of nuclear physics in the cosmos. Such studies need to take into account the broad range of nuclear physics capabilities and uncertainties, and sensitivity to each observable across all messengers needs to be determined. They also need to be continuously updated as new insights into nuclear physics emerge, or new observations drive new questions. The lack of sensitivity studies is a significant impediment to rapid progress in nuclear astrophysics. 

    A particular challenge in this workflow are the interfaces. Databases for nuclear data for astrophysics play a key role connecting nuclear physics and models, however, owing to limited resources they are often not updated and do not contain complete information. Such databases are not simply compilations of experimental and theoretical nuclear data. Rather, astrophysical nuclear reaction rates are complex data products that combine multiple pieces of information from different experiments and theory and need to be carefully evaluated and produced on a case by case basis. Another challenge are computational limitations. Many TDAMM sites are not spherically symmetric and require 3D simulations. Advances in computational astrophysics have enabled the development of such 3D models with impressive fidelity, however, computing power and algorithms limit the implementation of the full nuclear physics. In most cases approximations such as trajectory postprocessing are employed that may not be appropriate.

    The interface between models and observations is also critical. Atomic physics can play an important role in making this connection. Limitations in atomic physics data, for example in kilonova models, or the analysis of the spectra of metal poor stars, directly limit the determination of nuclear signatures. In some areas, such as X-ray bursts, atomic physics has not been implemented sufficiently to determine even basic observational signatures.

    \item {\bf Strong connections to astrophysics and other fields:} The workflow discussion above makes it clear that very close connections between nuclear physics, astrophysics, astronomy, gravitational wave physics, cosmo-chemistry (stardust), atomic physics, and plasma physics are absolutely essential for nuclear astrophysics in the TDAMM era. 

    There is a history of strong ties among some of these fields from dedicated conference series such as Nuclei in the Cosmos, Nuclear Physics in Astrophysics, or the JINA/CeNAM Frontiers in Nuclear Astrophysics meetings. Centers such as the Joint Institute for Nuclear Astrophysics (JINA), its successor the Center for Nuclear Astrophysics across Messengers (CeNAM), the Institute for Nuclear Theory, N3AS, and similar centers abroad have been instrumental in creating close ties among the nuclear physics and astrophysics communities. This meeting is also a testament to these efforts. The International Research Network for Nuclear Astrophysics (IReNA) has connected many such international centers with JINA/CeNAM, joining unique areas of expertise from across the globe. 

    However, the new directions in the TDAMM era require the development of many new connections. In particular, closer connections between the nuclear science community and relevant ground based telescope or NASA missions would be highly beneficial, for example with COSI, JWST, LIGO, and others. The long timescales of experimental nuclear programs make it important to plan well ahead to ensure the important nuclear physics, or at least the instruments and techniques to obtain it, are in place when a new mission launches. Future missions provide a strong justification for nuclear physics experiments and theoretical developments, and nuclear physics can make significant contributions to the science case of an observational mission. It will also be important to foster closer ties to communities that were less well connected with nuclear physics in the past, such as the gravitational wave community as well as atomic and plasma physics. Centers will continue to play a critical role in creating the necessary connections. This has also been recognized in the 2023 Nuclear Science Long Range Plan that emphasizes explicitly the importance of interdisciplinarity and states “Multi-disciplinary collaborative centers built around nuclear experiment and theory will expedite discoveries and allow the field of nuclear science to lead the quest to understand the cosmos through novel observations”.
\end{itemize}

\begin{quoting}
    \noindent Finding: The 2023 Nuclear Science Long Range Plan and the Astro Decadal survey taken together provide an outstanding opportunity for major advances in TDAMM nuclear astrophysics over the next decade. Of particular importance will be support for the interdisciplinary workflows outlined above that will be essential for work in one field to have its full scientific impact in the other, and will greatly accelerate scientific discoveries. Especially for science at the interfaces between funding areas it can be difficult to obtain support as such work is not at the core of any field, but rather appears to be on the periphery.
\end{quoting}

\subsection{Plasma Physics}
{\centering 
\textit{Contributors: Luca Comisso, Fan Guo, Maria Gatu Johnson, Athina Meli, Elias R. Most, Zorawar Wadiasingh}}\\

\label{sec:disciplines_plasma}
Advances in plasma physics in astrophysics are tightly coupled with progress in laboratory (e.g. in fusion research and laser plasma research) and in the solar and space physics \citep{bale2010research,national2020plasma,baalrud2020communityplanfusionenergy}.  Typically, plasma physicists span all of these fields, using theories and numerical simulations tested in all applications.  TDAMM science can greatly benefit from strong ties with these applications.

Plasma processes in extreme astrophysical environments (relativistic, strong gravity, and/or strongly radiative) are ubiquitous in TDAMM. 
Plasma physics is at the center of nearly all TDAMM phenomena, providing essential tools for modeling the underlying plasma dynamics, nonthermal emissions, and multi-messenger signatures.  We provide several remarkable examples:
\begin{itemize}
    \item Magneto-hydrodynamics (MHD) and magnetic field generation (e.g. $\alpha-\Omega$ dynamos) are critical in transient engines.  For example, the Black Hole Accretion Disk engine (believed to be the standard engine behind gamma-ray burst afterglows), magnetically driven winds from neutron-star merger remnants (e.g., \citealt{Combi:2023yav,Most:2023sft,Kiuchi:2023obe}),  and rotation-driven supernovae.
    \item Radiation-dominated relativistic outflows (optically thin and thick) pair-loaded winds in high magnetic fields, such as those realized in magnetar giant flares \citep{2016MNRAS.461..877V}.
    \item Collisionless shocks and related instabilities and magnetic field amplifications are critical in particle acceleration behind the synchrotron emission sources for many astrophysical transients (gamma-ray burst afterglows, supernovae) as well as the origin of cosmic rays and high-energy neutrinos \citep[e.g.,][]{Marcowith2016}.  Indeed, observations of cosmic rays and high-energy neutrinos are essentially a probe of plasma physics properties in the extreme plasma conditions. 
    \item Magnetic reconnection explosively releases magnetic energy into bulk kinetic flows, heating, nonthermal particles that eventually lead to high-energy radiation. This process is important for explaining the emission of radiation from magnetars, relativistic jets, and flares from the vicinity of black holes \citep[e.g.,][]{Uzdensky2016,Guo2024}.
    \item Magnetized turbulence has been proposed to explain high energy activities in black hole accretion disks, coronae, and jets \citep[e.g.,][]{Comisso2019,Zhang2023}. 
    \item The acceleration of energetic charged particles in the above mentioned plasma processes has been proposed as an explanation for high-energy cosmic rays and radiation originating from black hole accretion disks, coronae, and jets \citep[e.g.,][]{Blandford1987,CFM2024}. 
    \item Charged particle transport within the Galaxy and the intergalactic medium.  The development of turbulent magnetic fields and the importance of global magnetic fields drives increasingly advanced transport methods to better interpret observations of cosmic rays \citep{RP2023}.
    \item Strong field QED \citep{2006RPPh...69.2631H,Gonoskov_RevMod_2022,fedotov.pr.2023} effects can be probed under the extreme conditions of intense magnetic fields found in neutron stars (magnetars) and certain black hole systems. Furthermore, a thorough understanding of plasma processes, and QED pair production, may be essential for explaining coherent emissions from neutron stars \citep{1971ApJ...164..529S} and FRBs \citep{2023RvMP...95c5005Z}. 
    \item Pulsars are prolific emitters of gamma-rays \citep{2023ApJ...958..191S} and are thus accelerators of cosmic rays, particularly relativistic positrons produced in pair cascades and possibly ions. Recent discoveries include GeV/TeV halos around middle-aged pulsars by HAWC, Fermi-LAT and LHAASO \citep{2017Sci...358..911A,2018A&A...612A...2H,2023ApJ...944L..29A,2021PhRvL.126x1103A}, and  the striking pulsed TeV emission of the Vela pulsar by HESS \citep{2023NatAs...7.1341H} arising from inverse Compton emission \citep{2018ApJ...869L..18H}. The plasma physics involved here are (1) plasma, particle acceleration and QED positron creation in pulsar magnetospheres and (2) how positrons transport from pulsars in the interstellar turbulence, reach Earth as energetic cosmic rays, radiate or annihilate as 511 keV emission.
    \item The interpretation of polarization measurements depends on a detailed understanding of plasma physics in the underlying environment \citep{Zhang2018,Zhang2020}. 
\end{itemize}
For many of these applications, much of the astrophysical work is done assuming simplified models. Diffusive shock acceleration (DSA), also known as Fermi acceleration, is often regarded as the dominant mechanism for particle acceleration in a wide range of astrophysical shocks, including relativistic shocks associated with afterglows of gamma-ray bursts (GRBs), active galactic nuclei (AGN) jets, and supernova remnants (SNRs). This process assumes that particles scatter elastically across a shock front, gaining energy with each crossing, which leads to the development of a power-law distribution in particle energies. Such models have historically provided a plausible explanation for observed non-thermal emissions across the electromagnetic spectrum. Consequently, simple power-law energy distributions for electrons are frequently adopted as a starting point in theoretical studies of relativistic shocks.

However, growing evidence suggests that these assumptions may be inadequate, especially when attempting to match the precision of modern observational datasets. For example, polarization studies, which serve as critical tools for probing the geometry and microphysics of magnetic fields, often rely on toy models that assume idealized and static field configurations. These models frequently fail to capture the complexity and time-dependence of the turbulent environments inherent in astrophysical shocks. While such approximations may yield qualitative insights, they often oversimplify the true underlying dynamics, leading to significant discrepancies when compared to high-resolution data from observatories like ALMA, the Event Horizon Telescope, and the Fermi Gamma-ray Space Telescope.

Recent advances in computational astrophysics, particularly through Particle-in-Cell (PIC) simulations, have exposed the limitations of fluid approaches. PIC simulations provide a fully kinetic treatment of plasma dynamics, enabling the exploration of both shock physics and magnetic field evolution with a kinetic physics description that is inaccessible to fluid-based (MHD) models. These simulations have revealed that mechanisms beyond DSA play a critical role in particle acceleration. For instance, magnetic reconnection, a process in which oppositely directed magnetic fields annihilate and release energy, emerges as a key contributor to particle acceleration in environments with strong turbulence and magnetic shear. This is particularly significant in relativistic shocks, where reconnection layers and plasmoids (small magnetic islands) are formed within the shock's downstream region, leading to rapid and localized particle energization \citep{Sironi2014,Guo2014,Nishikawa2020,Meli2023}.

In addition to reconnection, PIC studies have highlighted the importance of shock-driven instabilities, such as the Weibel instability, which generates small-scale magnetic fields in unmagnetized or weakly magnetized shocks \citep[e.g.][]{nishikawa2016evolution}. These fields contribute to particle scattering and acceleration, altering the electron energy spectrum from the simple power-law forms predicted by traditional DSA models \citep{Medvedev1999ApJ...526..697M,Kato2007ApJ...668..974K}. Furthermore, PIC simulations show that anisotropic particle distributions and non-thermal features, such as high-energy tails, emerge naturally in these environments, driven by the interplay between reconnection, turbulence, and shock-compression processes.

Polarization studies also benefit from these advances. The dynamic magnetic structures revealed by PIC models, such as plasmoids and turbulence-generated fields, lead to time-variable polarization signatures that differ markedly from those predicted by static models. These findings have been instrumental in interpreting the variable polarization seen in GRB afterglows and blazar jets \citep{Zhang2018}. For example, simulations suggest that the degree and angle of polarization can evolve rapidly as magnetic fields transition from highly ordered structures to more chaotic configurations through reconnection and turbulence.

In summary, while traditional models such as DSA and toy polarization frameworks have provided valuable foundations, they fall short of capturing the richness of the processes occurring in astrophysical plasmas. Advanced computational approaches, particularly PIC simulations, underscore the need for more sophisticated models that account for mechanisms like magnetic reconnection, turbulence, and kinetic-scale instabilities. These insights are critical for accurately interpreting the wealth of observational data now available and for advancing our understanding of high-energy astrophysical systems.


To interpret the increasingly sophisticated TDAMM observations, we must overcome the limitations of tools currently used in TDAMM science.  Most of the plasma-physics solutions rely on either of two primary techniques.  For dense plasmas, magneto-hydrodynamics solution are common and TDAMM scientists have led advances in these methods, moving from ideal to resistive MHD, or even two-fluid techniques.  Laboratory astrophysics is also pushing the development of detailed resistive MHD methods, and scientists in both fields have begun to work more closely to improve these techniques. Especially for multi-physics calculations, code comparison is critical.

In the context of TDAMM science, particle acceleration modeling relies heavily on the Particle-in-Cell (PIC) method, which has become the dominant computational tool for investigating kinetic-scale processes, which MHD approaches cannot address. PIC simulations are particularly effective at capturing the intricate dynamics of shock-driven particle acceleration, turbulence, and magnetic reconnection, making them indispensable for exploring the physics of high-energy plasmas. However, despite their strengths, PIC simulations have significant challenges. Numerical instabilities, such as grid heating and noise, often complicate the results. Additionally, practical constraints on the electron-to-ion mass ratio and the spatial and temporal scales that can be resolved impose limits on the fidelity and applicability of the method. These issues stem largely from the high computational cost required to simulate both the smallest kinetic scales and the largest system-wide structures simultaneously.

To overcome these limitations, researchers have used alternative kinetic modeling approaches, such as Vlasov solvers and hybrid fluid-kinetic models. These methods offer a trade-off between computational efficiency and the level of kinetic detail retained, providing opportunities to tackle problems that would otherwise be computationally prohibitive for PIC. Comparative studies of these models, alongside benchmarking efforts within the plasma physics community, are critical for improving uncertainty quantification (UQ) and ensuring that the results are robust and reproducible. Such efforts are key to building models that can reliably connect to observational data.

One of the biggest challenges in this field remains the vast disparity in scales. The kinetic processes captured by PIC simulations occur on microscopic scales, yet the phenomena of interest in astrophysical systems- such as supernova remnants, gamma-ray bursts, and AGN jets-span macroscopic distances, often measured in light years. For instance, while PIC simulations are typically limited to regions on the order of a few thousand Debye lengths, the shock fronts and turbulence in astrophysical environments encompass vastly larger scales. Bridging this gap is essential, and it demands the development of advanced scale-bridging techniques or subgrid models. These approaches aim to parameterize the effects of small-scale kinetic physics in a way that can be integrated into macroscopic fluid models, such as magnetohydrodynamics (MHD).

Recent progress in this area has been encouraging. Researchers are exploring methods that couple PIC simulations with MHD frameworks, allowing the two to work in tandem to capture both kinetic and macroscopic dynamics. Others are focusing on reduced kinetic equations that distill the essential physics of small-scale processes into a computationally tractable form. These innovations hold significant promise for extending the utility of PIC and related techniques to large-scale astrophysical phenomena. If successful, they could greatly enhance the ability of simulations to interpret and reproduce key observational features, such as particle energy spectra, polarization variabilities, or emission anisotropies.

While the PIC framework is extremely powerful, this method suffers from a number of numerical instabilities and requires limitations in both particle mass ratios and the size-scale for these calculations. For PIC methods, achieving high Lorentz factors realized for extreme astrophysical particle accelerators is currently out of reach. This is due to the requirement that large gyro-radii are contained within a simulation box, while simultaneously resolving the smallest scales related to the plasma skin depth. Therefore, extrapolation is currently necessary to map to real astrophysical systems. PIC methods are widely used in the field of plasma physics, especially in laser and beam driven particle acceleration communities. These fields, as well as the development of PIC methods, are mainly supported by NSF and DOE (WARPX, VPIC, and OSIRIS PIC codes). Thus, a closer collaboration regarding these methods is necessary. Other kinetic approaches exist and code-comparison will improve uncertainty quantification for these modeling methods.  To overcome the fact that the scale of these calculations is far smaller than the actual applications, advances in scale-bridging methods or subgrid models are critical.  New methods in scale-bridging \citep{Vaidya2018,Guo2020,Arnold2021,Seo2024} could lead to dramatic improvements in the power of these models, improving their ability to tie to astrophysical observations.  Advancements for several of the methods require breakthroughs in computation (e.g. porting to GPUs) This is especially true for connecting kinetic and MHD scales will strongly benefit from novel interpolation and matching techniques. Leveraging advances in AI to develop subgrid models based on large libraries of kinetic simulation will be a crucial aspect toward advancing this field.

Strong field quantum electrodynamics is a not so well explored corner of the Standard Model \citep{Gonoskov_RevMod_2022,mp3report.2022.arxiv,fedotov.pr.2023}. It is a theory of charged particle and photon interactions in the presence of strong electromagnetic fields, which can be encountered near compact astrophysical objects, like magnetars or black holes, or in terrestrial laboratories, in foci of high power lasers or in the interaction point of particle accelerators. Until recently, the parameters of the interaction that could have been achieved in terrestrial experiments were quite far from the astrophysical ones. However, with new laser facilities coming into operation or being designed (see \cite{Gonoskov_RevMod_2022} for a list of facilities) with peak intensities, which are expected to reach $10^{24-25}$ W/cm$^2$, and particle beams approaching 100 GeV \citep{mp3report.2022.arxiv}, not just the basic processes of strong field quantum electrodynamics can be explored, but complex plasma setups embedded in strong field environments. This can provide an important opportunity for TDAMM collaboration with DOE and NSF supported initiatives aimed at laser plasma and accelerator physics studies (see, e.g., \cite{mp3report.2022.arxiv}). Of particular interest are the already operating NSF funded ZEUS facility at the University of Michigan, and future NSF OPAL facility at LLE. However, these studies are complementary to astrophysical observations since the electromagnetic field Lorentz invariants and plasma parameters are generally very different for neutron stars and lasers interactions. In particular, QED Landau state transitions (virtual or real) are an important phenomenon in strongly magnetized neutron star plasmas, and other physical processes such as photon splitting come into play \citep{2006RPPh...69.2631H}.  

Methods development will span all plasma physics fields and it is critical that the work in TDAMM has strong ties to heliophysics, space weather, laboratory physics and fusion energy (with its ties to industry).  Strong ties with these different fields can help address a number of TDAMM-relevant questions:
\begin{itemize}
    \item What current models/tools exist that can be applied to our sources and studies? 
    \item What current models/tools exist in other fields that can be applied to our sources and studies?
    \item Based on what is used in other fields, what new studies are needed for TDAMM applications?
    \item What approximations/limitations are currently present in these approaches?  What advances are needed to overcome our lack of understanding?
    \item How can experiments improve our understanding?
    \item How can astrophysical observations improve our understanding of this physics?
\end{itemize}

Developing more collaborative approaches, e.g. NSF Frontier Centers, NSF Hubs, NASA TCAN, or privately funded initiatives, such as Simons Collaborations (SCEECS) would be desirable.  A metric for the success of these programs will be in how well techniques/methods/tests propagate between different disciplines. A critical element is that cooperation and integration be encouraged across currently disparate competitive and siloed plasma physics and astrophysics groups.

\begin{quoting}
    \noindent Finding: Plasma astrophysics must be utilized in all TDAMM sources. A strategic community organization effort, analogous to that done in nuclear astrophysics but including relevant fields in plasma science, would facilitate major discoveries in TDAMM science.
\end{quoting}





\subsection{Atomic Science}
{\centering 
\textit{Contributors: Christopher J. Fontes, Christopher L. Fryer, Amy Gall, Stuart Loch, Endre Takacs}}\\

\label{sec:disciplines_atomic}


Spectroscopy remains one of the most powerful tools in an astronomer’s toolkit. Since Isaac Newton first employed a simple prism to observe the Sun, spectroscopy has played a crucial role in advancing our understanding of the universe. Spectral analysis across different wavelength regions is essential for diagnosing the elemental and ionic composition, densities, velocities, and magnetic fields of the ejecta in astrophysical transients. For the purposes of this report, we refer to this field of study as traditional atomic physics.

In the context of TDAMM, this analysis provides critical insights into the temporal evolution of these events, enabling the study of dynamic processes such as shock propagation, material mixing, and energy dissipation. Doppler shifts and spectral broadening of these features probe the velocity distribution and kinematic properties of the ejecta, helping to uncover the underlying mechanisms driving these phenomena. Combining spectroscopic data with signals from gravitational waves, neutrinos, and other messengers, we can obtain a more complete picture of the physical conditions governing these transient events.
For example, follow-up EM spectral observations of neutron star mergers, combined with atomic analysis, revealed the presence of heavy r-process elements, linking such events to the production of the universe's heaviest elements. For supernovae, spectroscopy tracks the evolution of elements over time. Additionally, spectroscopy provides insights into time-dependent processes, leveraging atomic data to model ionization states, temperatures, and dynamics in non-equilibrium environments.

This science is grounded in a solid understanding of fundamental physics, including basic atomic, radiation, and plasma physics, along with the complex interplay between these fields. Key atomic physics requirements include energy levels, transition data (e.g., wavelengths, oscillator strengths) for identifying spectral features, collision data for modeling plasma conditions, and heavy-element opacities  that are crucial for modeling radiation transport in extreme environments like kilonovae. By examining how atomic transitions, energy level populations, collisional cross sections, and radiation processes influence plasma conditions in astrophysical transients, we can gain a deeper understanding of the spectroscopic signatures these events produce. While traditional atomic physics is not the primary focus of the Atomic, Molecular, and Optical (AMO) Physics Decadal Survey, \textit{Manipulating Quantum Systems: An Assessment of Atomic, Molecular, and Optical Physics in the United States} \citep{national2020manipulating}, it is thoroughly addressed by the Report of the Laboratory Astrophysics Task Force (LATF)\citep{LATF_2024}.   

The AMO laboratory astrophysics community, comprising theorists, experimentalists, and database curators, plays a critical role in supporting astrophysics. As astrophysics increasingly depends on AMO data for missions with enhanced sensitivity and resolution, LATF and related white papers continue to underscore the most urgent data needs and propose actionable solutions (e.g. \citep{Smith_2019,Kallman_2019,Nave_2019,Brickhouse_2018,Savin_2011}). While efforts, such as the occasional NASA Laboratory Astrophysics Workshops have been organized to align data production with astrophysical needs, funding constraints and institutional barriers continue to limit progress in critical atomic studies that underpin astrophysics and TDAMM science.

\begin{figure}
    \centering
    \includegraphics[width=0.49\textwidth]{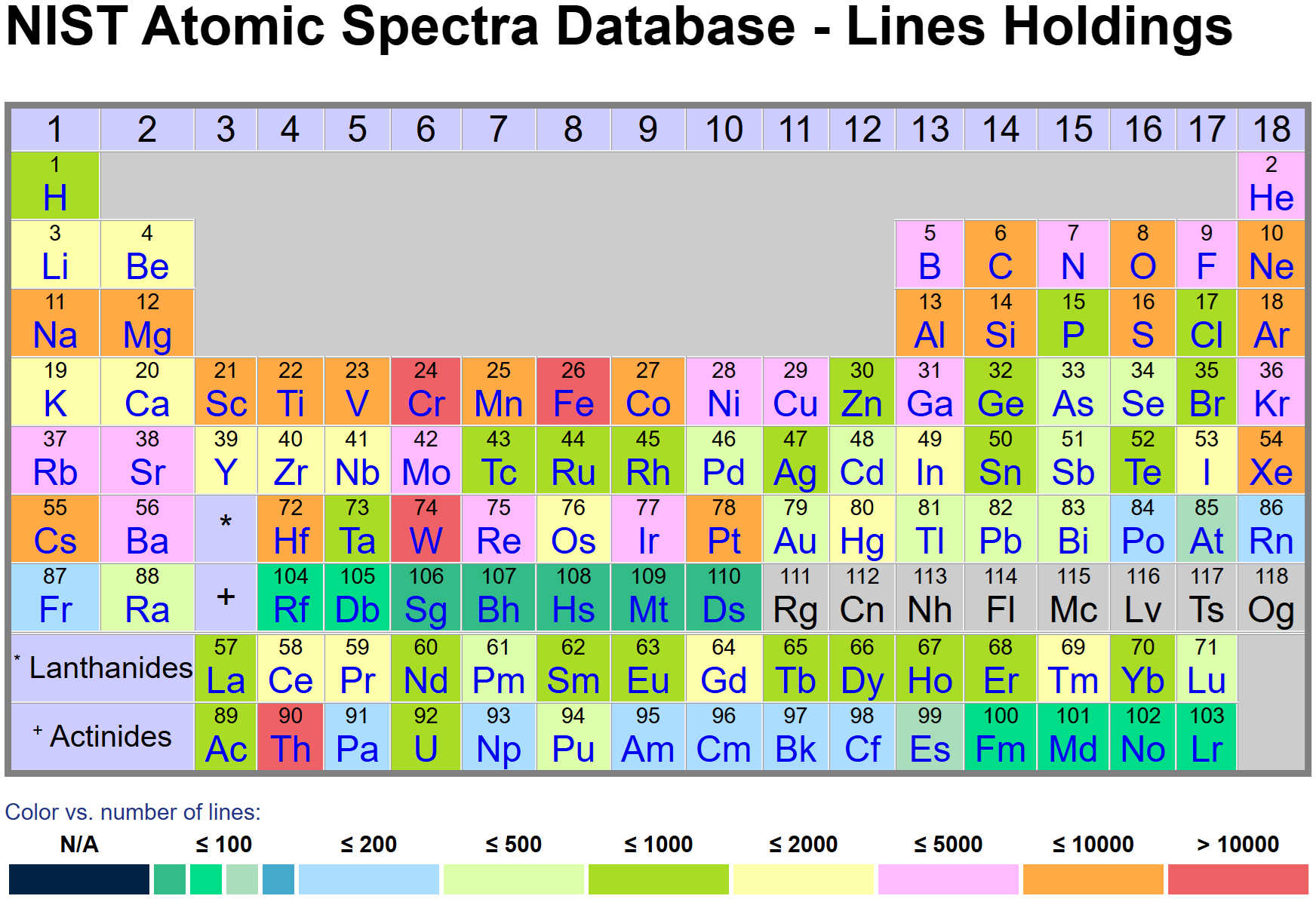} 
    \includegraphics[width=0.49\textwidth]{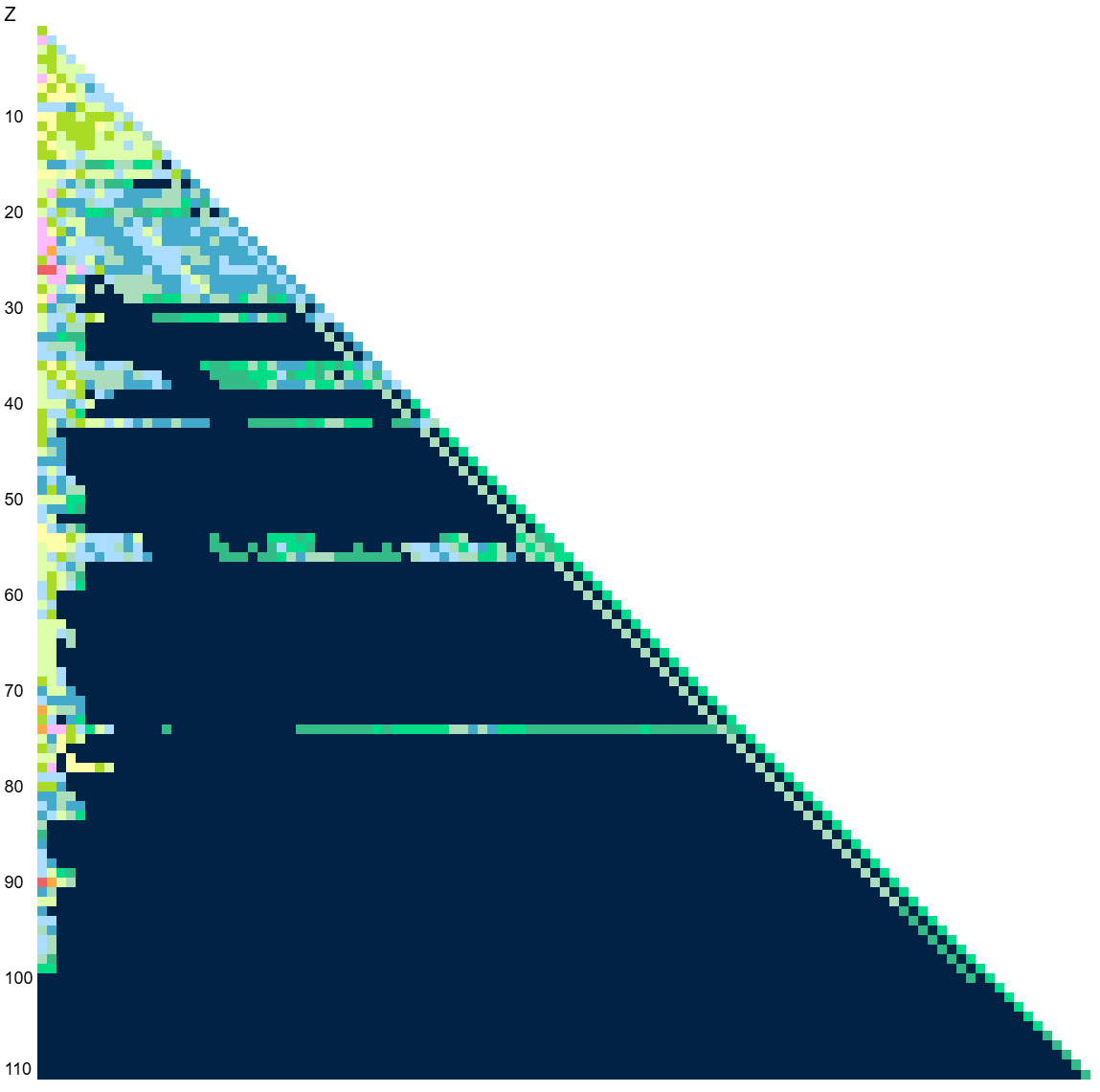} 
    \caption{\label{amo_nist_fig}
    Table and chart showing the number of spectral transitions in the NIST ASD database \citep[ASD][]{Kramida2024}. Of particular note are the large number of charge states relevant for TDAMM that have no or only a small number of transitions in the database.}
\end{figure}

\textbf{Challenge:} The challenge of AMO for TDAMM applications is in the wealth of high quality theoretical and experimental data that need to be generated and the relatively small workforce available. As mentioned above, this data includes accurate energies and wavelengths for element identification, opacities for radiation dominated events such as the early stages of kilonovae, and atomic collision cross sections with electrons and heavy particles for collisionally dominated plasmas as in the late stages of kilonova remnant plasmas. It should also be noted that the atomic workforce is highly motivated to support TDAMM and has much specialized knowledge in this field. As an example of the need for data, Figure~\ref{amo_nist_fig} shows the number of radiative transitions for the elements of the periodic table from the NIST Atomic Spectra Database \citep[ASD][]{Kramida2024}. It can be seen that many of the heavy elements generated in kilonova events have either no or very few measured emission rates.

An overview of the AMO community and available resources is given in the LATF report~\citep{LATF_2024}. In short, it notes that there is a significant network and infrastructure of experiments and code/theory expertise that has been built up over many decades. However, due to the field of AMO moving towards quantum computing and ultra-cold applications, there is a lack of people-power in almost all areas of this field. There is a current danger that the field is losing the critical mass of expertise required to keep it alive and growing, with the potential risk for a lack of relevant data to carry out TDAMM science.  In addition, it is not completely clear what specific atomic data are required for TDAMM science (see first bullet below). The availability of critically evaluated and curated AMO databases is a crucial bridge between the atomic and TDAMM communities. There is an urgent need for financial support of such databases, which often rely on grants from piecemeal funding sources. The disappearance of databases can severely hamper day-to-day spectroscopy research, as demonstrated by the unavailability of the NIST ASD during government shutdowns or the vanishing of the Oak Ridge National Laboratory atomic database due to lack of funding.

\begin{figure}[ht]
\begin{center}
    \includegraphics[width=\textwidth]{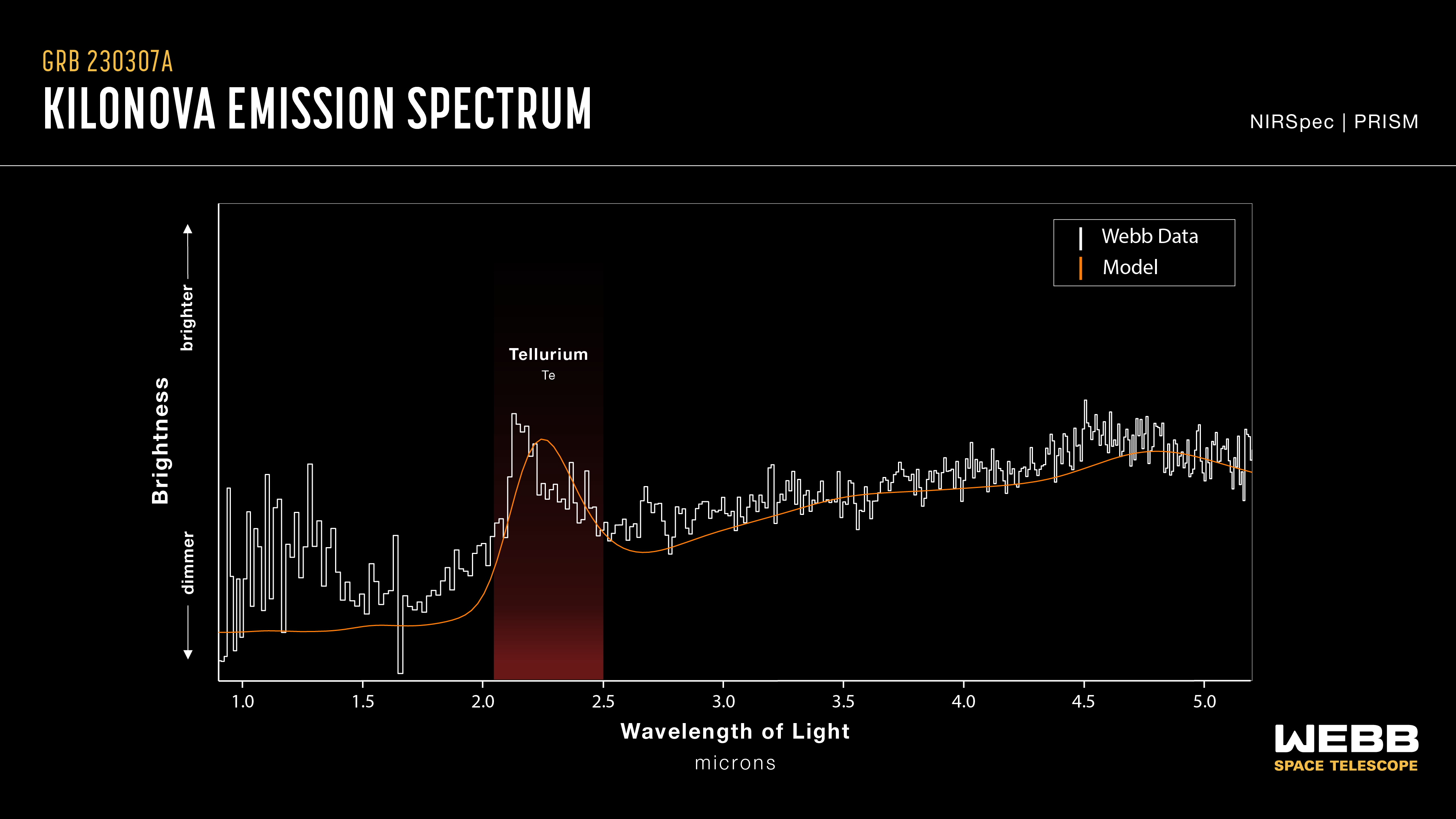} 
    \caption{
    The first JWST late-time infrared spectrum of a kilonova \citep{levan2024heavy}; figure from NASA, ESA, CSA, Joseph Olmsted (STScI). A line complex is evident, which was also seen in the kilonova following the multimessenger detection of a binary neutron star merger in 2017. The feature is likely a signature of Tellerium, which would be direct evidence of r-process nucleosynthesis. However, this identification is not totally self consistent with the broader observations and current expectations from atomic physics. Strategic investment in atomic physics is required to deliver on the promise of TDAMM astronomy and make use of the power of the JWST.}\label{te_fig}
\end{center}
\end{figure}

We offer the following findings, many of which overlap with the feedback contained in the LATF report:
\begin{itemize}
\item Community workshops where the atomic data users can meet with the atomic data experts, to allow the available resources to be best directed at the most urgent needs. These workshops could be part of the regular AAS and TDAMM meetings. A mechanism for the user community to request AMO data would be very useful.  For example, the elements being measured and calculated should be those that are predicted to be the most abundant in the TDAMM events. For many TDAMM applications, astronomers have developed numerical tools to calculate out-of-equilibrium physics, but these tools require different atomic physics data (e.g. Einstein coefficients) than what the atomic physics tends to provide.  
\item However, determining the specific interfaces and needs of the broad astrophysics community from atomic physics is no simple task. We recommend an extended workshop at the beginning of this effort, in order to agree on the interfaces between fields which will be built around.
\item There is an urgent need for long-term, stable financial support of critically evaluated and curated AMO databases. Support of AMO databases that allow experts to curate and maintain the databases, as well as methods to communicate the database information to the community, is strongly recommended. 
\item  Early career support should be expanded for traditional atomic physics researchers. See the following text about Workforce Development for more details. 
\item Joint funding of TDAMM research projects that connect astrophysics (NSF-AAG, NASA-APRA) to fundamental sciences, such as physics and chemistry (e.g., NSF-PHYS or NSF-CHEM), is encouraged. Such a joint approach would strengthen the connection between the atomic data producers and the TDAMM applications.
\item Atomic researchers should be embedded within space missions so that the atomic data needed for particular telescopes and detectors is in place in time for the mission observations. For example, the recent JWST spectral observations of tellurium features would have benefited from an initial investment in atomic support (See Fig.~\ref{te_fig} for details.). A successful example of this type of synergy is provided by the XRISM mission, which included funding for laboratory astrophysics from the start. A strategic investment to ensure the atomic data and analysis are sufficient to ensure we can analyze the data from the JWST, and soon UVEX, should be a NASA priority.

\end{itemize}

\textbf{Workforce Development:} The atomic and molecular physics community has a wealth of knowledge, advanced computational tools, databases, and specialized experimental equipment essential for producing the data needed in astrophysical spectroscopy. However, with senior researchers nearing retirement, there is a growing risk of losing a critical mass of expertise and personnel. The continued success of future astrophysical missions—especially those involving multi-messenger astronomy—depends on supporting the next generation of atomic researchers. Without targeted investment in early career scientists, the U.S. risks falling behind in this crucial field, reducing the scientific returns from upcoming missions. While there are opportunities for early career scientists, such as NSF CAREER, much of the funding opportunities are soft-money. However, the options for soft-money career scientists are limited and, in addition, this particular type of traditional atomic research is often not prioritized, making joint funding initiatives for early career researchers essential to maintaining U.S. leadership in this area.

\begin{quoting}
\noindent Finding: Traditional atomic physics has played a critical role in TDAMM science for many years. To enable further TDAMM breakthrough discoveries, we recommend concerted efforts to support workforce development, establish curated databases guided by community workshops, and integration of atomic researchers in mission development and operations. 
\end{quoting}

\subsection{Condensed Matter and Materials Science}
{\centering 
\textit{Christopher L. Fryer}}\\

\label{sec:disciplines_condensedMatter}
For the most part, most astrophysical phenomena are plasmas where the effects of condensed matter physics are unimportant.  But in both white dwarfs and neutron star crusts, condensed matter effects (e.g. crystalline structures) can potentially play a large role in the physical properties:  thermal coefficient, shear moduli, elasticity, stress, strain and strength tensors~\citep[see, for example][]{2008LRR....11...10C,Engstrom_2016}.  Dust formation and destruction also ultimately depends on the multi-body interactions requiring detailed condensed matter studies.  Condensed matter experts study the microscopic physical properties of matter that arise from electromagnetic forces between atoms and electrons.  These scientists also have developed a broad range of methods to scale these microscopic studies to macroscopic applications.  A number of computational tools have been developed to model this physics, most notably molecular dynamics~\citep{Kuksin01122005} and density functional theory techniques~\citep{2005MSMSE..13R...1M}.  Although these sophisticated methods have existed for decades in the fields of condensed matter, chemistry, and biology, most astrophysics estimates of condensed matter properties have used simplified approaches.  

For neutron star transients, materials properties can affect a broad range of observations.  The thermal conductivity of the neutron star crust can alter the time evolution of outbursts ranging from X-ray bursts to magnetar flares.  A series of molecular dynamics calculations have sought to better understand the materials properties affecting this thermal conductivity~\citep[e.g.][]{2009PhRvE..79b6103H}.  But this is only one material property studied in condensed matter.  Anisotropies, soft phonon modes, and elastic instabilities  could have significant effects on elasticity-related astrophysical observables such as magnetar flares~\citep{2011ApJ...727L..51P}, quasi-periodic oscillations~\citep{2005ApJ...628L..53I}, and possibly some pulsar glitches~\citep{2008LRR....11...10C}.  Strength properties of the neutron star crust will both dictate the nature and properties of magnetic field structures in magnetars and any interpretation of this emission from highly magnetized neutron stars will depend on our understanding of this physics.  For white dwarfs, the energy released in crystallization can dramatically alter the cooling timescale of the white dwarf with wide-ranging repercussions~\citep{2008ARA&A..46..157W}.  Numerical approaches in condensed matter continue to advance and, by leveraging these advances, astronomers can improve the analysis of phenomena occurring on neutron stars. 

\begin{quoting}
    \noindent Comment: Initiation of connections between astrophysicists and condensed matter and materials science are needed in order to develop multidisciplinary approaches at the intersection of these fields.
\end{quoting}

Molecular and dust production and destruction in astrophysical transients is important in both interpreting their infra-red emission and, ultimately, identifying the nature of the transient~\citep{1977ARA&A..15..267S}.  For example, in GRB~230307A, the 2 micron line could as easily be explained by CO as it could by TeIII~\citep{levan2024heavy}.  Dust can explain the required high opacity to explain the JWST observations beyond 3 microns.  Especially in the infra-red, understanding dust and molecular formation and optical features can be equally important to atomic studies.  And to do this accurately, astronomers must leverage tools used by condensed matter.  Although density functional theory has been used in star forming regions, only a few papers using modern condensed matter approaches have been used in TDAMM science~\citep[e.g.][]{2016P&SS..133...31M,2018MolAs..12....1M} and TDAMM dust astronomy could definitely benefit from closer ties to condensed matter experts.

\subsection{Computational}
{\centering 
\textit{Contributors: Emmanouil Chatzopoulos, Courey Elliott, Christopher L. Fryer, Gwendolyn R. Galleher , Fan Guo, William Raphael Hix, Kelly Holley-Bockelmann, Aimee Hungerford, Kristina D. Launey, Nicholas R. MacDonald, Athina Meli, Bronson Messer, Matthew R. Mumpower, Peter Nugent, Paul M. Ricker, Christopher J. Roberts, Todd Urbatsch, Michael Zingale}}\\

\label{sec:disciplines_computational}

Since the rise in computing in the 1940s, computational physics has become an important tool in understanding complex physical phenomena.  The complex problems in astrophysics and, in particular, astrophysical transients, are ideally suited to computational physics techniques.  NNSA laboratories spearheaded this research.  In the 1940s, methods to model transport, e.g. Monte Carlo~\citep{ulam47} , and hydrodynamic shocks~\citep{richtmyer48} were developed and applied to early computing resources at Los Alamos National Lab (LANL).  Stirling Colgate modeled the first supernova explosions at Lawrence Livermore National Laboratory~\citep{1961AJ.....66S.280C}.  In these early years, astrophysicists had strong ties to the national labs and these collaborative efforts led to dramatic progress in computational physics techniques for both national security and astrophysics applications~\citep{1997ASPC..123....3N}.

\begin{figure}
    \centering
    \includegraphics[width=1\linewidth]{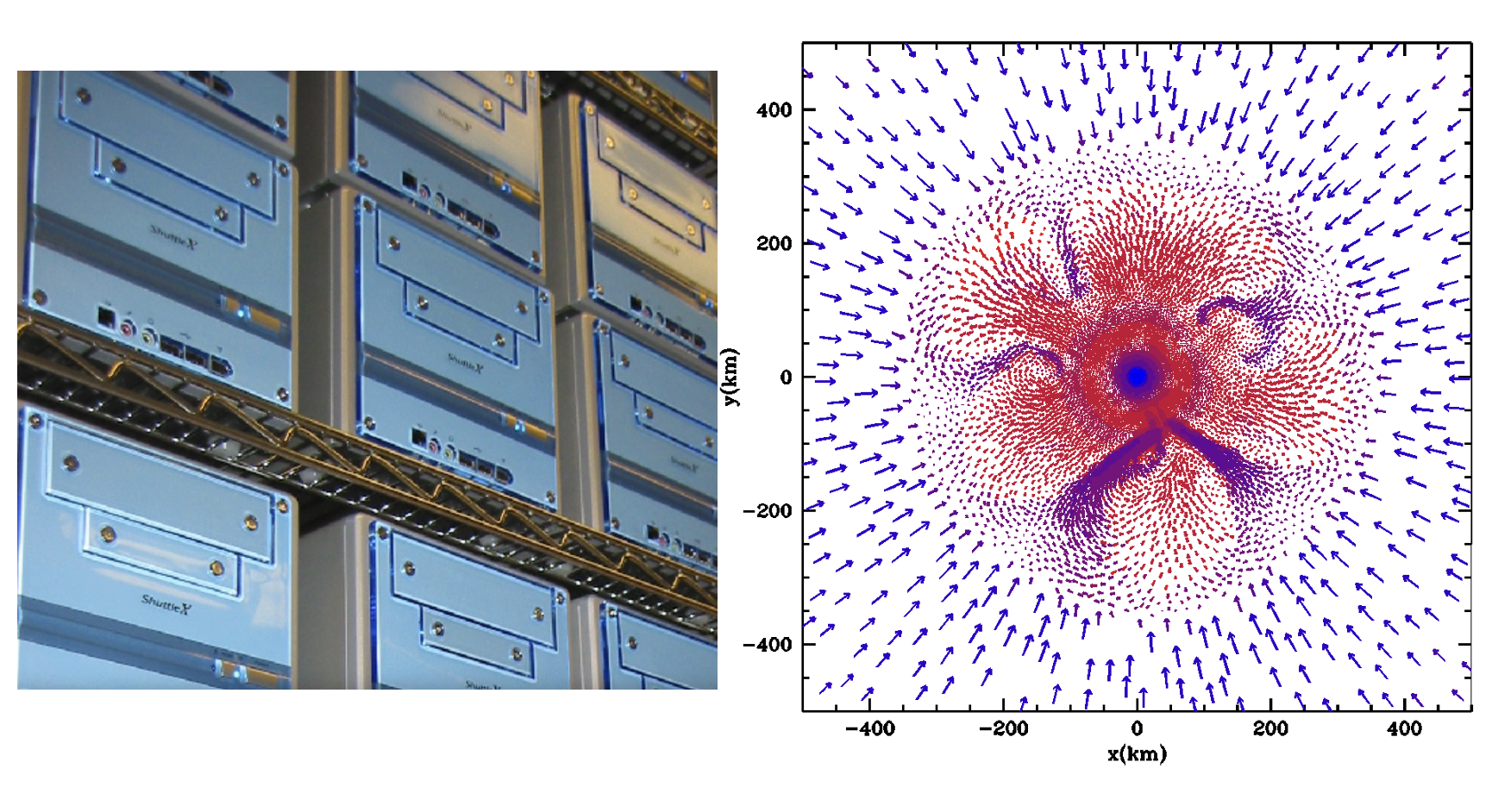}
    \caption{Image of the Space Simulator, first commodity machine in the top 100 of the top 500 list~\citep{Warren2003TheSS} and a slice of the 3-dimensional core-collapse supernova simulation it was designed to model~\citep{2002ApJ...574L..65F}.}
    \label{fig:comput}
\end{figure}

This coordination was not limited to numerical algorithms.  In an effort to build supercomputers more cheaply, scientists at NASA and NNSA laboratories worked together to develop what is now the standard for high performance computing clusters used for computational physics.  Initially termed Beowulf clusters, the idea of networking commodity-grade computers to build supercomputers is the framework behind all of the top 500 fastest computers in the world.  NASA scientists Thomas Sterling and Don Becker built the first such cluster in 1994.  In 1996, Mike Warren (LANL) and John Salmon (Caltech) built the first Beowulf cluster to appear on the top 500 list.  Figure~\ref{fig:comput} shows an image of the next generation of such a cluster alongside the supernova calculation the computer was designed to model.  

Computational astrophysicists have continued strong ties to DOE (often leading many of the computational efforts at these labs).  But many of these computational astrophysicists have lost their ties to observational astronomy.  In addition, the widespread use of computational modeling in a broad range of fields has allowed computational science to fracture where progress in computational algorithms in one field often do not disseminate to other fields.  For example, many astrophysicists are unaware of advances in radiation-hydrodynamics and transport techniques developed in nuclear engineering and at NNSA laboratories.

{\bf Tying to other Fields:}. Although the study of dynamic systems, including astrophysical transients, dominated the initial use of computational physics techniques, science is now entering an era where most fields are leveraging detailed modeling to advance science.  Many of these fields are directly tied to astrophysics and will be discussed in other sections of this paper.  For example, numerical methods for transport are studied by a variety of communities in nuclear power, fusion energy, space weather, z-pinch- or laser- driven experiments, and the animated film industry to name a few.  But a number of additional fields developing advanced numerical methods that could be leveraged in computational astrophysics exist.  These include material physics and biophysics modeling which have pioneered a broad set of methods to bridge the microscopic to macroscopic scales, capturing an increasing amount of the microscopic physics.  Such approaches could prove useful in astrophysical events with unresolved features.

The multi-physics nature of astrophysics mean that computational astronomers trained with an understanding of all these fields ultimately are ideal hires for a broad range of industry and national security applications.  Astronomy has claimed that it can attract the next generation of scientists needed to ensure the US remains a world leader (both in defense and in industry).  But this is only the case if we train scientists with a strong understanding of the numerics (and their relation to other fields) and physics.

{\bf TDAMM Challenges:}  Simulations allow us to peer into the unseen parts of these sources, interpreting the observations and revealing the physics that drives them.  Simulations connect the different measurements along the timeline of observations. The primary challenge in computational astrophysics is that, to be successful, we must couple multiple calculations to fully study a phenomena. For example, astronomers use a broad range of diagnostics are used to shape our understanding of core-collapse supernovae:  neutrinos from Supernova 1987A, early and late-time emission from the transient, compact (black hole or neutron star) remnants, ejecta remnants, nucleosynthetic yields from galactic chemical evolution, GW measurements of the black hole mass spectrum, etc.  But these observations span a huge range in space and time relative to the explosion.  Connecting all of this data can not be done by a single simulation.

Without these chains of simulations, our ability to employ observations from different phases of the event to achieve a more powerful constraint on the physics of the event is severely limited.  The connection between different messengers, at different times, lies at the heart of multi-messenger astronomy, making chains of simulations central to multi-messenger investigations.   However, precisely because the individual links in the chain employ different physics, and produce observables that span the EM spectrum and can include GW, neutrino and cosmic ray observables, the chains frequently extend beyond the boundaries of individual funding programs.  This makes coordination of the chains of simulations challenging.  Worse yet, coupling the different links requires utilizing advanced theory and analysis techniques.  For TDAMM science, computational science will have to bring together analytic, numerical, and analysis methods. There are many barriers to success:
\begin{itemize}
    \item Among this chain of required simulations are a set of focused physics calculations, e.g. calculations of atomic and nuclear data are crucial to provide the essential microscopic physics on which astronomical simulations rely.  While measurements of this atomic and nuclear data are desperately needed to ensure the realism of the simulations, computations often fill the many gaps in available data to allow the astronomical simulations to continue.
    \item Computational astrophysicists must be able to span a broad set of scientific fields and disciplines. Unfortunately, many links in the chain, both the microphysics and many of the numerical models, especially if they do not directly tie to observations, are often not supported by astronomy departments and funding agencies.    
    \item Open source codes are both a boon and weakness for astronomers.  By making a code open-source, science teams can more easily compare results.  The variety of codes in our community is a strength, since agreement across a variety of algorithms builds confidence in simulation results. Code comparison allows for determination of where the weak links are in the simulation chain, and where progress should be prioritized. However, as more and more astronomers rely on open-source codes rather than developing their own codes, we have reduced the number of computational-physics capable astronomers.  Because many  funding sources focus on publication count and not the ability to develop novel code methods, computational scientists struggle to get funding.  Since universities follow the funding, more and more universities are hiring code users, not code developers.  Astrophysics, a leader in computational methods in the past, is now being superseded by many other fields of science and engineering.  Unfortunately, this occurs as we enter an era where a detailed understanding of algorithms and simulation methods is essential.  As the increase in computing power slows, brute force methods to code development are no longer capable of solving the questions facing astronomy.  Astronomers need to rethink how they answer astrophysical questions, coupling high performance computing and analytic approaches.  The science goals of TDAMM cannot be met with a single simulation code and it is likely new approaches and new algorithms will need to be developed to target specific phases of evolution.  
    \item Many changes are also occurring from the perspective of the hardware and software side as well.  In the past decade, supercomputers have largely transitioned from being CPU-based to GPU-based (with the notable exception of the ARM A64fx CPUs). The community needs to be aware that computing architectures are continuing to evolve.  In 5--10 years, double precision support may be removed from hardware (driven by ML/AI trends).  We need to explore whether we can used mixed precision, emulating double precision in software where needed (for instance, rate equilibrium in networks). The needs of astrophysics must be considered in the construction of future supercomputers.
    \item Programming models are also changing.  Many astrophysics codes have shifted from Fortran to C++ to take advantage of C++ lambda-capturing methods of offloading (e.g., as in Kokkos).  Memory-safe languages (e.g. Rust) are also being promoted more, but have not yet taken root in High Performance Computing.  Porting codes to new languages is time-consuming and requires training students on new techniques.
    \item Most funding for computational astrophysics is not through astrophysics programs themselves (NNSA research, NSF Cyberinfrastructure, DOE PSAAP/SciDAC) plus a few non-profit mechanisms for small amounts of funding (like Linux Foundation's High Performance Software Foundation and Numfocus) that can support meetings and costs associated with Cyberinfrastructure platforms, etc.  NASA's open source funding is an exception.  But, for the most part, computational astrophysics is funded by agencies with alternative goals including advances in national security, energy applications, and industry.  Under these funding sources, solving astronomy applications is a secondary goal. 
    \item Multidisciplinary studies require continued dialog between fields in order to train scientists to have a complete understanding of the problem. Programmers are essential in these endeavors, including software engineers for high fidelity simulations and full stack developers for automated infrastructure work. There are barriers to the integration of these subject matter experts into the field, including restrictions on membership in the professional societies. These barriers should be removed.
\end{itemize}

{\bf Training Future Scientists:}  One of the justifications for astrophysical research is that the skills taught to astronomers strengthens the nation by training next generation STEM-field leaders in industry and national security.  Computational astrophysics has been at the heart of this claim, training the broad computational physics skills needed to lead research in a broad range of fields.  However, the reliance on open-source codes and lack of intentional code-development support has not only led to a stagnation in our advancement in understanding astrophysical phenomena, but also an end to the ability of astronomers to contribute greatly in industrial applications, weakening the claim that astronomy draws young minds into STEM fields for industry and national security.  The capability to do detailed, complex problems is one of the prime skill-sets needed in both industry and national security. Without these skills, astronomers are less attractive hires for research jobs in industry and national security.  

\begin{quoting}
\noindent Finding: Computational astrophysics is the connection point for all the fields of physics of relevance for TDAMM, and where models get compared to data. Rewarding code users but not code developers in astrophysics has led to stagnation, and we risk losing this required skillset. 
\end{quoting}
\begin{quoting}
\noindent Finding: The approaches needed for computational astrophysics are performed by some in our community, and widely within the NNSA. The NNSA has summer schools and visiting scientist programs to train these skills and build connections. Broad engagement and support from astrophysics in these programs is a path to renewing the leadership in computation in astrophysics and contributing to the NNSA workforce by making an additional career path evident to early career astrophysicists.
\end{quoting}

\subsection{Fluid Dynamics and Turbulence}
{\centering 
\textit{Contributors: Mohammad Ali Boroumand, Christopher L. Fryer, Falk Herwig, Daniel Livescu}}\\

\label{sec:disciplines_fluids}



\begin{figure}
    \centering
    \includegraphics[width=1\linewidth]{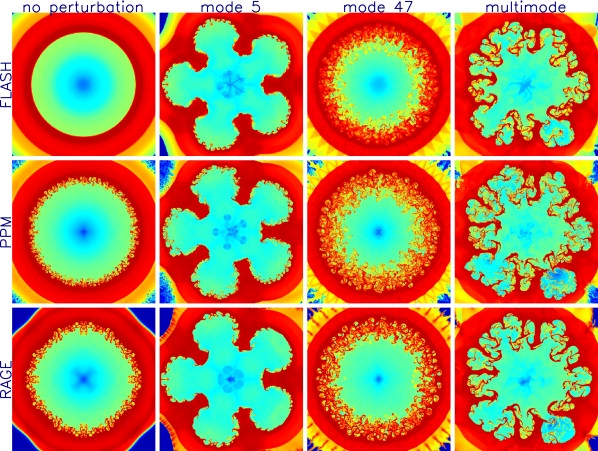}
    \caption{Comparison of 3 codes studying the implosion of a shell.  Although designed to test instabilities in the implosion of a National Ignition Facility capsule for Inertial Confinement Fusion, the physics is relevant to understand the growth of convective instabilities in a core-collapse supernovae~\citep{2014JCoPh.275..154J}.}
    \label{fig:turbulence}
\end{figure}

Fluid dynamics and turbulence play an important role in a broad range of STEM-field applications including electronics and cryogenic systems, aerospace engineering, the oil and gas industry, studying geophysical processes (e.g. groundwater flow), weather prediction, and astrophysics.  As such, work in fluids and turbulence is funded by industry as well as a wide range of NSF, DOE, and DOD funding agencies.  Fluid dynamics and turbulence play important roles in nearly every aspect of TDAMM science:  transient progenitors, instabilities in stellar winds (and the subsequent radiative reaction to these instabilities), reactive flows in thermonuclear explosions (novae, X-ray bursts, thermonuclear supernovae), convection in core-collapse supernovae and disk evolution. These features are typically hard to model for several reasons. For example, at the current grid level, most kinetic energy is subgrid and cannot be handled explicitly (Figure~\ref{fig:turbulence}).

These applications require models in different convective regimes.  
For stellar models, the convective turnover timescale is typically many orders of magnitude shorter than the evolutionary thermal or nuclear timescale. Yet, through global 3D stellar hydrodynamics simulations \citep[e.g.\ ][]{Herwig2023} it now becomes possible to investigate in a meaningful and accurate way the quantitative properties of 3D macro physics, such as mixing through rotation, convection, internal gravity waves and possible moderated by magnetic fields \citep{Leidi2023}.  While it remains difficult to model the burning phases in full hydrodynamics calculations there are now a growing set of multi-dimensional calculations modeling snapshots in time of this evolution. These snapshots can be connected through global 1D simulations supplemented with 3D-informed and calibrated models \citep{Anders2022,Mao2024}  and it will be possible to create through such large-scale fluid-dynamics simulations a realistic account of the 3D non-radial pre-supernova structure \cite[e.g.\ ][]{Andrassy2020}, a key starting point for any subsequent supernova TDAMM simulation pipeline \citep{Fields2020,Vartanyan2022}. Here, scientists can leverage techniques averaging the role of convection, e.g. Reynolds Averaged Navier Stokes (RANS) solutions (a more general approach than the mixing length theory commonly used).  At the other extreme, in the core-collapse supernova engine, the explosion timescale is only 5-10 times the convective turnover time.  Steady state solutions for this convection are not valid.  In addition, the growth timescale of the convection is critical.  This poses a different challenge to models because numerical viscosity (most astrophysics codes rely on Implicit Large Eddy Simulation (ILES) approaches where the viscosity depends sensitively on the spatial resolution) delays this convective  growth time, delaying the growth of convection which, in turn, significantly alters the fate of the explosion.

Many of the astrophysical applications require additional physics coupling to the hydrodynamic motion.  For example, in transients powered by nuclear fusion, nuclear burning is highly dependent on the geometry of the burning front which is determined by the nature of the turbulent convection.   For transient light-curves, the interaction of the radiation with clumpy winds, shell ejecta, companion stars can drive further turbulence and shocks in the outflow.  These shocks can dramatically alter the observed light-curves.  Magnetohydrodynamics in disks is critical in determining the disk viscosity and the generation of jets.  As with our core-collapse convection example, these simulations are typically under-resolved, limiting what we can quantitatively predict from these calculations.  

{\bf History:}  It was quickly realized that astrophysical plasmas, in particular in stars, should be turbulent and, as early as 1928, astronomy leveraged advances in meteorology and oceanography to understand and predict the behavior in stellar atmospheres \citep{1928MNRAS..89...49R}.  During this period, astrophysicists were well-connected with advances in these other fields~\citep[for a review, see][]{1949ApJ...110..329C}.  But as the different fields evolved and specialized, the connection between astrophysics and other communities studying turbulence weakened.  As such, when mixing length theory for stellar convection was developed, it did not leverage the advances at the time in RANS models used in weather prediction and engineering~\citep{2019ApJ...882...18A}.

{\bf Importance of Tying to Other Fields:}  The broad importance of turbulence and convection means that many fields have driven advances in the modeling of this physics.  Leveraging the techniques developed in these fields could lead to major advances in astrophysical modeling.  For example, ``[mixing length theory] is local, requires calibration, and has little connection to modern methods used by the turbulence community.''~\citep{2019ApJ...882...18A}. The RANS approach is much more general than the mixing length theory and can represent important turbulence flow effects, such as the cascade process that leads to time delay between kinetic energy and dissipation response and turbulence production by shear or pressure gradients, that cannot be captured by the latter. RANS has been the engineering workhorse in applications ranging from aerospace and auto industries to atmospheric and oceanic flows for many years.

Similarly, most astrophysics codes rely on implicit filters for large eddy simulations, but the broader turbulence community has developed a broad range of subgrid treatments that could potentially improve these models. A number of RANS extensions also exist that attempt to bridge the statistical and large edddy simulation descriptions that may provide solutions that are better tied to the underlying physics \citep{Towery2024}. Combustion and explosives modeling fields have also developed a number of new techniques to better capture reactive flows and these techniques may have applicability to thermonuclear transients \citep{NGL19}. 

In addition, various machine learning techniques have provided an additional boost in recent years to advance the accuracy, robustness, and speed of turbulence models. Integrating modern ML with physical modeling is the major challenge of what we call today Physics-Informed Machine Learning (PIML)~\citep{Karniadakis2021Physic-Informed,tian2023_lles}. In particular, in fluid dynamics, there has been significant PIML activity in recent years. 
Examples include embedding physical constraints into the closure model \citep{wang2017physics,tian2021physics} and 
PIML models infusing physical constraints into the neural networks \citep{mohan2020div}. 
Other efforts on turbulence modeling, such as those summarized in \citep{duraisamy2019turbulence} can span a great deal of aspects, e.g., from estimating the properties of the flow \citep{ali2024extracting}, to significant acceleration of the flow calculations \citep{DPNFB23} or the reaction network integration \citep{nouri24ske}.

Historically, turbulence models have been developed using fully resolved simulations of simple flows at  parameter ranges where such simulations are possible. 
This technique, called Direct Numerical Simulations, offers a wealth of information on par with experimental data, but with full control of initial and boundary conditions, as well as turbulence diagnostics. These simulations are very expensive and affordable on only the largest supercomputers. Due to their size (typically of hundreds of terabytes), the task of data management has proven daunting and, in
some cases, impossible. However, public databases, such as that hosted by Johns Hopkins University, offer access to multi-terabyte turbulence
datasets of canonical turbulent flows, including analysis tools, for users worldwide. Nevertheless, even relatively simple practical flows quickly become impossible to fully resolve on even the largest supercomputers. In engineering applications, it is common to invoke a mixing transition to justify
the relevance of finite scale range simulations to flows with much larger range of scales. Beyond this mixing transition, the fundamental turbulence features required by the modeling assumption no longer change. It is not known if such a mixing transition arises in complex astrophysical flows and whether relevant fully resolved simulations might become feasible in the near future. However, as we shall discuss in our section on high energy density physics, a number of experiments have been developed to test advances in astrophysics codes to validate the models and test various hypotheses, for example the mixing transition. The field of turbulence and associated convective modeling has advanced greatly and astrophysics is in an ideal position to leverage these new techniques.

Similar to other disciplines, TDAMM events occur in an extreme turbulence regime. Thus, astrophysics may be able to provide novel data for fluids scientists to utilize. This requires identifying signatures and diagnostics in astrophysics which can be utilized as validation or rejection of different modeling approaches, such as signatures of subgrid turbulence in hydro, combustion, or plasma waves. Additionally, TDAMM sources have some unique needs. Plasmas in TDAMM events often have strong magnetic fields. Modeling subgrid processes in flows with strong magnetic fields is not very mature. The modern area of multiwavelength polarization, in concert with other diagnostics, is a specific area where the needs of astrophysics can provide a new handle to understand the behavior of fluids and turbulence.

{\bf Challenges:}  Fluid dynamics and turbulence innovation spans a wide set of fields.  One of the primary challenges for astrophysicists in this field is to digest all of these advances and determine whether these techniques can be leveraged to help better solve astrophysical applications.  Many of the new methods are problem specific and may not be applicable to astrophysical problems.  Blindly applying methods from the aerospace or weather communities into astrophysics is unlikely to work.  Indeed, many attempts have been made to introduce a RANS prescription from the fluid dynamics community into stellar models but, given the stiffness of the stellar evolution calculations, these methods often lead to unphysical solutions \cite{nouri2019modeling}.    Similarly, developing subgrid models for astrophysical flow problems will also require a detailed understanding of both the stellar evolution codes and turbulence modeling strategies. Handling magnetic fields and generation of predictions for polarimetric signatures is a novel problem.

\begin{quoting}
\noindent Finding: Collaboration at the interface of astrophysics and fluids and turbulence presents significant opportunities for advancing our understanding in these fields. To fully capitalize on these opportunities, it is essential to cultivate a workforce of scientists who are proficient in both astrophysics and fluid dynamics. Current funding mechanisms, institutional support, and reward structures must be restructured to effectively promote the development of multidisciplinary scientists and to incentivize collaborative research efforts across these domains.
\end{quoting}

\subsection{High Energy Density Physics}
{\centering 
\textit{Contributors: Christopher L. Fryer, Aimee Hungerford, Heather Johns, Maria Gatu Johnson, Carolyn Kuranz, Todd Urbatsch}}\\

\label{sec:disciplines_hedp}




High energy density physics (HEDP) refers to physics at densities and temperatures beyond normal matter on Earth, encompassing a wide range of physics disciplines including plasma physics, turbulence and fluid dynamics, radiation transport, atomic physics, and nuclear physics.  HEDP lead studies in fusion energy science, environmental problems, national security science, planetary physics, and astrophysics.  HEDP encompasses many of the critical physics disciplines needed for TDAMM science.  HEDP scientists typically focus on laboratory experiments and HEDP science differs from the component physics fields in that most of these experiments require coupling multiple physics fields together to understand the experimental data.  Just as in astronomy, even focused laboratory experiments require the coupling of multiple physics fields to interpret the results.  Here we focus on the multi-physics nature of HEDP, deferring the discussion of individual physics disciplines to their appropriate section.

Astrophysicists conducting complex simulations have often worked closely with HEDP experimentalists and many astrophysical codes, e.g. FLASH~\citep{2002ApJS..143..201C}, validated their techniques against HEDP experiments.  With the formation of the laboratory astrophysics division in the American Astronomy Society, HEDP physics has begun to tie more tightly with the astrophysics community.  Areas of overlap with astrophysics include:
\begin{itemize}
   \item {\bf Radiation Physics:}  Theory for atomic physics, radiation transport, hydrodynamics, and similar multi-physics disciplines have proceeded mostly independent in both fields,  making potential comparisons of the fundamental physics very fruitful.  
   \item {\bf Magnetohydrodynamics:}  2D and 3D rad-hydro and MHD simulations exist in both  communities. Development of complex simulation tools  and computing architectures (especially at national laboratories, but  at some  university environments as well) for HEDP could be fruitfully used for and compared to well-scaled astrophysical questions.
   \item {\bf Pair production and plasmas:} Using the 440 GeV/c beam at CERN’s Super Proton Synchrotron (SPS) accelerator, scientists have been able to reproduce a relativistic electron-positron pair beam~\citep{2024NatCo..15.5029A} that can guide pair-production in astrophysical conditions (e.g. jets).
   \item {\bf Magnetic Plasmas and Particle Transport:}  HEDP scientists studying fusion energy science have developed a series of experiments to study physics of charged particle transport~\citep{2024PhPl...31d0501H}.  These studies can guide models in astrophysics for everything from polarization studies to cosmic rays.
   \item {\bf Nuclear Cross Sections in Excited States and in Plasma Environments:}  HEDP experiments (e.g. at NIF) can both drive atoms/nuclei into excited states and then probe the nuclear cross sections in these excited states~\citep{2023arXiv230500647C,2024PhRvC.109a5501C,10.3389/fphy.2022.917229} and in plasma conditions (\cite{10.3389/fphy.2022.942726}).
   \item {\bf Atomic Physics:}  HEDP scientists have develop experiments to test atomic physics at high temperatures, complemented the ion trap experiments.  Understanding the results of these experiments requires multi-physics modeling combining radiation transport, turbulence, equations of state and atomic physics.
\end{itemize}

\begin{figure}
    \centering
    \includegraphics[width=1\linewidth]{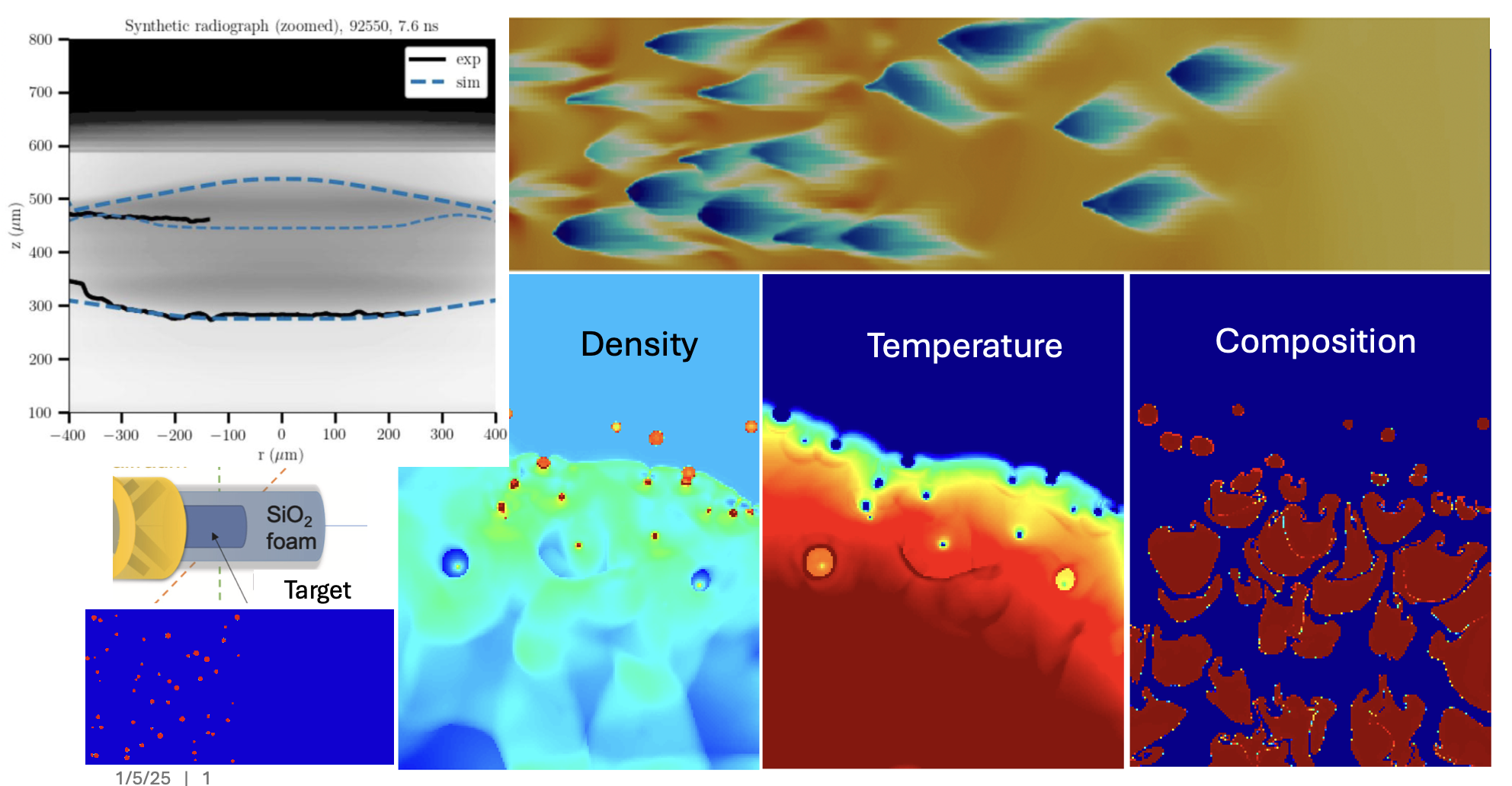}
    \caption{A montage of radiation flow studies.  From top left moving clockwise:  radishock experiment~\citep{2024PhPl...31k3301C}, supernova blastwave through a clumpy medium~\citep{2020ApJ...898..123F} and simulations and experimental design for an experiment studying radiation flow through an inhomogeneous material~\citep{2023arXiv231216677F}.}
    \label{fig:hedp}
\end{figure}

One of the great strengths of HEDP science is that it strongly ties to laboratory experiments that can be used to both:
\begin{itemize}   
   \item {\bf Probing physics:} Integrated experiments are as multi-physics as astrophysical systems themselves, and if well-posed and well-scaled can help to reproduce astrophysical behaviors in miniature. Conversely, single-physics experiments seek to eliminate as many unknowns as possible to focus on a single focused question. For  example, measurements of opacity, or two shocks  interacting, or a radiation flow experiment.  These experiments probe a broad range of physics from turbulence and reactive flows to radiation flow and atomic physics~\citep{1999ApJ...518..821R,2000ApJS..127..465R,2006RvMP...78..755R,2018NatCo...9.1564K}.  Figure~\ref{fig:hedp} shows different aspects of a set of radflow experiments including its application to astrophysics.
   \item {\bf Code Validation:} One very big benefit lab astro experiments can offer to astrophysics is that HEDP scientists can fully characterize experimental targets prior to running the experiment.  This significantly reduces the experimental error allowing high precision testing of the codes.  The experiments can be designed to test specific physics regimes to address key uncertainties in the calculations. 
\end{itemize}

There are a wide set of facilities that are available for HEDP science that can be used to study astrophysically-relevant physics.  These include the National Ignition Facility at Lawrence Livermore National Laboratory, the Z-pinch machine at Sandia National Laboratory, the Omega laser facility at the University of Rochester, etc.  These facilities often do open discovery science calls that provide time on these machines.  

{\bf Workforce Development:} One of the real strengths of building connections between astrophysicists and HEDP scientists is that laboratory experiments are ideal conditions to test analysis techniques.  TDAMM scientists often study transients where the initial conditions are poorly constrained.  This makes it difficult to disentangle uncertainties in the progenitor and uncertainties in the explosive engine or emission model.  Beyond validating our codes, HEDP science, with its more controlled experiments, is an ideal training ground to test analysis tools and methods. These experiments also provide a greater foundation to train multi-physics studies that are critical for astrophysics.  Unfortunately, support for laboratory astrophysics oscillates dramatically~\citep[for a review, see][]{2009astro2010P..68B} and their are few programs to support this training.

\begin{quoting}
    \noindent Finding: The needs and approaches of HEDP and TDAMM are remarkably well aligned. HEDP facilities can now emulate extremes of physics which were previously unobtainable on Earth, though astrophysics still contains the most extreme environments. Despite the obvious benefits, the fields remain disparate. A strategic partnership is required.
\end{quoting}

\subsection{Transport}
{\centering 
\textit{Contributors: Emmanouil Chatzopoulos, Christopher L. Fryer, Heather Johns, Todd Urbatsch}}\\

\label{sec:disciplines_transport}

Various forms of the Boltzmann transport equation are applicable to simulating particle and radiation physics present in many of the disciplines of TDAMM.  We review a number of transport communities and their histories to foster cross-fertilization with TDAMM, providing several examples of approaches which have proven successful, and how these approaches may be adapted for TDAMM.  

\textbf{History: Application of ASC and Nuclear Reactor Simulations:} After nuclear weapons testing ended in 1992, the DOE (later NNSA) began a massive and intense effort to enhance simulation capabilities.  The focus of the ASCI (Accelerated Strategic Computing Initiative) program, spearheaded circa 1997 by Vic Reis, was parallel and 3D computing with tenets of software quality and dependencies between software design and computer architecture.  The physics equations solved was mainly radiation-hydrodynamics for high energy density experiments, astrophysics, and inertial confinement fusion (ICF).  Designing large scale software that was well-tested in a controlled, repeatable fashion enabled ``verification'' that the equations were being solved correctly.  Application to known, real observables ``validates" that the correct equations were being solved.

Nuclear reactor simulations for light-water reactors (boiling water reactors, pressurized water reactors) and liquid metal-cooled reactors is done today for the whole reactor core in 3D.  Reactor design and analysis models the individual detailed fuel cells, which are  bundled together in assemblies, put together with coolant flow in a way to optimize safety and efficient burnup and transfer of energy.  The 1950s to 1970s saw analytic formulas and analytic mappings of planar or unit-cell calculations.  (The ``buckling'' was an analytic axial formula to map an infinite-medium or other reduced-order calculation to the finite axial extent of the reactor.  The 1980s to 2000s saw a sub-grid approach of calculating averaged cross-sections at the unit fuel cell level and using those averaged parameters in a large 3D solution using the diffusion approximation for neutron transport.  The coolant flow in a boiling water reactor had a mix of liquid and gas and spawned many subgrid models in fluid flow (thermal hydraulics).  These calculations were done separately from the neutronics calculations, and now can be integrated with neutronics solution methods.  Even with modern calculations, solving the neutron transport requires approximations.  The nuclear cross sections for fission, absorption, and scattering are highly dependent upon the energy of the incident neutron, especially in resonance regions.  Monte Carlo neutron transport can mostly perform continuous-in-energy calculations today.  Deterministic neutron transport calculations discretize the energy variable and the cross-sections are discretized; these multi-group cross-sections have to be pre-computed, averaged as if the final neutron solution was known.  Approximate weighting functions were established for various types of reactors (e.g., fast spectrum, thermal spectrum), and using an incorrect weighting function can give wrong answers.
 
Neutron transport for reactors has evolved differently than thermal X-ray transport for HED and ICF applications.  Cross-sections are generally only weakly temperature dependent, whereas thermal X-ray opacities are highly dependent on temperature and density as well as being highly dependent on the energy of the incoming X-ray.  For this reason, even today's Monte Carlo methods for thermal X-ray transport are not continuous in energy.

The Boltzmann transport equation for neutronics and HED thermal X-ray radiation transport are largely similar; no particle-particle interacts, no external forces, no polarization.  There are up to 7-dimensions: up to 3 in space, 2 in angle, energy, and time.  These independent variables are handled differently for different physics, regimes, and numerical methods.  Early deterministic neutron transport methods, circa 1980s, applied to thermal X-ray transport failed miserably because the spatial discretization were not accurate enough.  Because thermal radiation is so tightly and nonlinearly coupled to the materials, time-explicit methods do not suffice.  Typically the radiation and material equations are linearized and manipulated to derive a time-implicit formulation, which for Monte Carlo methods, particularly the Fleck and Cummings method, results in absorption and re-emission being represented by an effective scatter.  Further, the severe dependence of opacities and the X-ray intensity results in Monte Carlo methods having the benefits and costs of both particle and discretization representation.  Further, detailed Compton scattering introduces another variable in electron velocity -- another degree of freedom that requires increased sampling.  Additional physics necessary for other disciplines, likely introduced more degrees of freedom, which require more computational resources.

\textbf{Example of simulation chaining from HED:} HED radflow simulations are like nearly all TDAMM simulations in that they involve chaining sequential simulations.  The major components of current laser-driven radflow platforms are the lasers-plus-hohlraum, the foam tube, and the backlighter-plus-spectrometer.  Coupling simulation of extremely disparate regimes of physics is rarely productive.  Modern codes are able to simulate the lasers impinging on the inside walls of a hohlraum and transport the thermal radiation down a foam tube in one integrated simulation.  However, to amortize the laser-ray-transport cost, a surface tally from the integrated calculation is used as a surface source for a subsequent cut-off problem with the foam tube and no hohlraum.  This approximation requires that the surface be defined in a hydrodynamically benign region.  Loss of fidelity occurs when the surface source takes no account of angle-dependence or energy-dependence from the full-problem surface tally.  Transport needs the incoming flux specified; the albedo or escaping flux is not specified.  Therefore, the initial state of the cut-off problem is critical.  The backlighter has radiation in a different regime, ideally, and should not interact invasively with what is being measured and should provide a strong, smooth signal for both radiography and absorption spectroscopy.  The required spectral extent and detail preclude simulating backlighting during and as an integral part of the radiation-hydrodynamic calculation, which is, at best, a few hundred energy groups or even gray (energy-integrated), and, further, is required over a much larger length-scale, or distance, than the rad-hydro simulation.  Therefore, backlighting and spectroscopy is performed as a post-process, doing simplified transport along a ray through a time-snapshot of the rad-hydro simulation.  This approximation assumes that all the rad-hydro effects are already built in, that the backlighter doesn't appreciably affect  the target, and that multiple rays are independent.  Capturing complex effects such a non-equilibrium and transport effects in the post-processing is not necessarily straightforward.

\textbf{Comparing ASC and TDAMM:} ASC was a large national effort to build new predictive simulation capability.  It did not start from scratch, and after 30 years of success, there is still much more to do.  The TDAMM fields cover a much larger scope of physics than ASC, and it seems clear that a TDAMM-focused ASC-level project is not in the making.  This means that a cost-benefit analysis, or some similar study, should be taken to identify existing and missing numerical simulation capability, folded in with existing and anticipated observations, and where these capabilities could be obtained, or shared, from collaborators, or initiated.

A common theme across human endeavors is limited resources.   The lag of acceptance of science accomplishments can be decades.  The lag of acceptance of scientific vision can be generations.  Funding cuts propagate for careers.  Talent can get lost or misappropriated.  As mentioned at the workshop, collaboration and breaking barriers is a cost effective way to make scientific advancements.  Leveraging ASC capability for TDAMM modeling and simulation seems to be a fruitful avenue to expand.  Although the cost could be enormous for an ASC developer to extend to a related TDAMM application, the additional validation that TDAMM would provide could be invaluable.  Of course, that level of software capability coordination is itself expensive, and possibly above a threshold for an ASC developer.  Thus, the TDAMM discipline would need to become much more deeply involved in the methods research and software development.

\textbf{Identifying Grand Challenges and Priorities:} To identify the next grand challenges, it would be beneficial to have a graphic, pictorial, diagram, and/or even a flow-chart (engineering!) to relate, on one level, all the sources, and, on another level, disciplines. Coupled to that, on another level, would be all the existing and desired numerical methods and simulation capabilities for those disciplines.  Another Another level could be funding and funding sources.  Over-laying these graphics will help identify areas to focus upon and determine priorities.

The workshop presentations and this white paper contain some initial starting point, which could be built upon. We list some examples in their respective source subsections below. For example, mapping the different elements in active galactic nuclei and their respective uncertainty is done in Table~\ref{tab:agn_elements}. The separation of the temporal stages of a core-collapse supernova can form the basis of the simulation chains; these and the corresponding physics are mapped in Figure~\ref{fig:ccsn_chains}. Mapping of observables to physical insight is shown for neutron star mergers in Table~\ref{tab:diagnostics_neutronStarMergers}. While each approach differs and is certainly incomplete, they are demonstrations of what should be created and expanded for all source classes.


\textbf{Next Steps for Collaboration:} Per ASC definitions, verification is a process where methods researchers and software developers prove, continuously, through unit tests to integral tests that their software is doing what it was designed to do.  Validation is users showing that the models and simulations sufficiently match observables.  Test problems are neither verification nor validation, but are powerful forms of peer review and are an avenue for breaking down discipline barriers.  Defining test problems could help break barriers between communities, so that a particular code could be tested on another community's particular physics to guide and precondition collaboration with a low-cost entry point.  If a test problem is a simple representation of salient features of a particular discipline, it is a less costly way for other disciplines to test their simulation capability on neighboring physics regimes.

\textbf{The Future, Starting Now:} Using experimental results to constrain models, simulation, and data is not particularly easy for HED experiments.  Machine Learning techniques are an active area of development for training models on simulations and calibrating them to sparse experimental results.  In the neutron transport community, reduced-order models were used 20-some years ago to assess uncertainty, and now have become mechanisms for obtaining results directly in a more computationally efficient manner.  It may be that ML/AI methods take over a significant fraction of the heavy lifting from detailed, complex, and computationally intensive TDAMM-related methods, models, and software.  Assuming that ML/AI is -- or could be -- less computationally expensive, it could reduce the scope of expensive computing to a smaller domain, only where it is necessary, much like reduced-order models have neutron transport.   

Emerging computer architectures are losing double precision. While complex multi-physics methods and software do not need double precision for a double precision answer, this level of precision is necessary in the guts of the calculations to maintain consistency (e.g., emission and absorption in radhydro) and to avoid large errors in the final values.  Besides researching high-fidelity methods that are less reliant on high-precision, limiting their role with ML/AI may be in the future. While software must adapt to this new era, supercomputers must meet the need of high fidelity simulations. 


As noted above, the radiation hydrodynamics problem is extremely
difficult due to a number of its unique features: The Boltzmann
equation is a 7-D integro-differential equation, which a priori is
extremely expensive to solve. Further, the inputs: opacities and 
equations of state, \dots, are often unknown to the necessary
precision. In addition, the astrophysical systems associated with
TDAMM require the treatment of relativistic flows, which essentially
necessitates a full general relativistic
treatment \cite{Chen:2007,Pihajoki:2017}, which while doable is very
expensive.

Since computational expense is the hallmark of TDAMM radiation
hydrodynamics problems approximations are also the hallmark of TDAMM
related simulations. This make validation difficult, since defining a
test problem that a code designed to do multi-group flux limited
diffusion in hydrodynmamics can solve and compare to a Monte-Carlo
synthetic spectrum code is quite difficult. Although, a
multi-group flux limited diffusion code can be compared to a M1
closure code and a Monte-Carlo solver synthetic spectrum code can be
compared to a traditional stellar atmospheres code.

As the above examples illustrate, radiation hydrodynamics problems are
most profitably attacked by chaining a series of codes with tailored
approximations together, the hydrodynamical part is performed often
with a low resolution energy grid and the comparison to spectroscopy
is performed with a more expensive approach that may lack or
approximate time dependent effects. 

\begin{quoting}
    \noindent Finding: The techniques and structures of ASC and radiation transport can be adapted to modernize approaches to complex TDAMM problems. A key mechanism to facilitate this is the construction of charts showing the simulations which must be chained together, the physics which goes into each stage, and the funding mechanisms available for each component. This would allow the identification of priorities for improving numerical capability via collaboration with the field of radiation transport or exploration of new approaches.
\end{quoting}

\newpage
\section{Sources}\label{sec:sources}
The taxonomy of the zoo of sources and transients in astrophysics is not well sorted. Classification is complicated as discovery and isolation of source and transient classes typically occurs long before the physical object or process is understood. Naming schemes often include phenomenological separation as well as physical ones. Numerous overlaps occur between transient classes. Thus, taxonomy is complicated. This makes organization of white papers and field planning difficult. Here, when possible, we sort by physical classification schemes as these are scenarios where the theory and simulation, and future paths forward, are better understood. This occurs for Sections~\ref{sec:sources_agn} to \ref{sec:sources_xrayBinaries}. The remaining sections, Sections~\ref{sec:sources_grbs} and \ref{sec:sources_frbs}, are largely phenomenological. These events hold similar promise for understanding the physics of the cosmos, but the specific promise may not yet be fully known.

We briefly describe the known origins of the events and objects in the source classes below. When massive stars die they often produce a core-collapse supernovae. Many massive stars are born in binary systems. When one star dies there can be a compact object, either a white dwarf, neutron star, or black hole, in joint orbit with a remaining massive star. In certain scenarios the massive star can accrete onto the compact object. When the compact object is a white dwarf this produces novae, while accretion onto a neutron star or a black hole is set as an X-ray binary. In later phases of some binary massive star systems, a neutron star can merge with a neutron star or a black hole, together referred to as a neutron star merger. Thermonuclear supernovae arise from detonation of a white dwarf. This can arise either from accretion onto a white dwarf by a massive star until it crosses a critical mass, or the merger of two white dwarfs. Some isolated neutron stars have extreme magnetic fields, called magnetars. 

Phenomenological transient classes are related to these compact objects. Flashes of fast radio bursts can be released by magnetars, though also possible by other sources. Some neutron star mergers and some fast-rotating supernovae produce ultrarelativistc jets which produce gamma-ray bursts. Some observed supernovae are orders of magnitude brighter than typical, referred to as superluminous supernovae. Their origin may be shocks, or a different progenitor (such as pair instability supernova), or through other means.

Massive and supermassive black holes ($>10^6$M$_\odot$) are very different beasts. Galaxies are typically anchored by a single massive black hole. If that black hole is actively accreting matter it can produce active galactic nuclei, with jets larger than their host galaxy. If that black hole is quiescent but then eats a star that falls in, it produces a tidal disruption event, whose jets are hundreds or thousands of light years across. Fast blue optical transients are a phenomenological source class which may arise from accretion of a star onto an intermediate mass black hole ($\sim10^3$M$_\odot$) or an unusual death of a massive star.

The basic summary and science questions listed here begin with the information summarized in the White Paper from the First TDAMM Workshop\footnote{\url{https://pcos.gsfc.nasa.gov/TDAMM/docs/TDAMM_Report.pdf}}. We add contextual information based on new knowledge learned in the past two years or to tie it to other fields of science.


\subsection{Active Galactic Nuclei}
{\centering 
\textit{Contributors: Peter G. Boorman, Francesca Civano, J. Patrick Harding, Kelly Holley-Bockelmann, Weidong Jin, Tiffany R. Lewis, Ioannis Liodakis, Nicholas MacDonald, Lea Marcotulli, Athina Meli, Michela Negro, Jeremy Schnittmann}}\\

\label{sec:sources_agn}

\begin{figure}[h!]
    \centering
\includegraphics[width=0.9\linewidth]{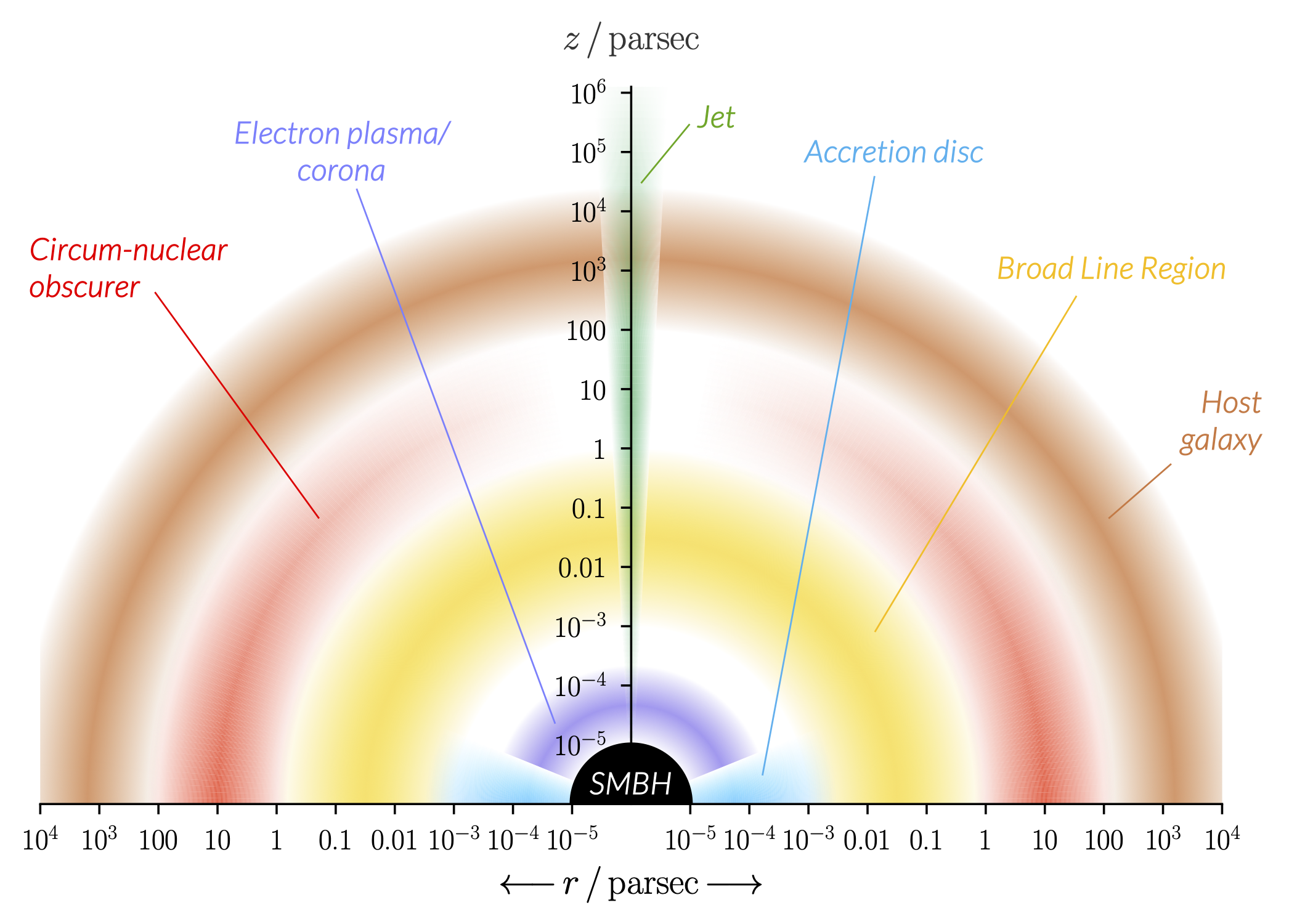}
    \caption{Schematic denoting the approximate size scales associated with the key physical regions pertaining to this white paper (Created by P. Boorman).}
    \label{fig:agn}
\end{figure}

Active galactic nuclei (AGN) are typically defined by bright emission emanating from the cores of galaxies, powered by accretion onto a supermassive black hole. The substructures of AGN that pertain to this White Paper (and TDAMM in general) are shown in Figure~\ref{fig:agn}, comprising the accretion disc, X-ray corona, broad line region, circumnuclear obscurer (often called the dust torus) and jet. All of these regions are observed to display spectrotemporal variability with a wide variety in timescales, ranging from minutes to decades. There are millions \citep{Storey-Fisher-2024} of known AGN observed using X-ray, IR or optical techniques as well as several thousands of known blazars in which the jet is directed towards our line of sight, observed in GeV gamma rays \citep{4fgl, Ballet_2023,4lac}. Large sample sizes paired with their extremely broad range of emission across the entire electromagnetic spectrum, observed out to early cosmological epochs, provide a unique laboratory for understanding a plethora of physical processes. AGN are among the most compelling candidates for multimessenger astrophysics in terms of detection through both gravitational waves and high-energy particles, such as neutrinos and cosmic rays.


AGN researchers tend to form a community, or even a set of distinct communities around AGN subclasses for observers, or AGN components for theorists (see Figure~\ref{fig:agn}). We talk to people in other communities, but it is less usual to form joint projects with the intent of drawing comparisons between mass or spatial scales. Theorists who use different techniques tend to be aware of each other's work, broadly speaking, but comparative and cross-checking work is rare because it is very hard to fund. Additionally, there have been developments in some communities that are not reflected across all of them. For example, there is extensive work in General Relativistic codes that has not been consistently connected with jet observations through dedicated jet models, and there has been extensive work in accretion disk modeling that is not often reflected in discussions of AGN accretion disk observations. While there are almost certainly individual examples of such work, broader integration of knowledge between fields has been slow because it can be difficult to keep up with literature in one's subfield, let alone across sub-fields. Soloing at conferences probably also does not help, so individual researchers have to be more intentional about the information they seek out and consider. Dedicated efforts or meetings in interdisciplinary work could help build a cohort of individual researchers who act as liaisons between sub-fields. 

\begin{quoting}
    \noindent Finding: The field of AGN studies would greatly benefit from awareness of all aspects of research in this field, and integration of knowledge between sub-fields. This can be facilitated through dedicated efforts and interdisciplinary meetings, to build scientists who can work between sub-fields. Engagement with the study of X-ray binaries may also prove fruitful, particularly in the study of coronae. 
\end{quoting}

\begin{table}[h!]
\centering
\begin{tabular}{|p{4cm}|p{5cm}|p{6.5cm}|}
\hline
\textbf{AGN Element}          & \textbf{Uncertainties}                & \textbf{Needs} \\ \hline\hline
\vspace{2pt}
\textbf{{\Large Jet}} & Formation, Composition, Number of coherent emission regions, Location of primary emission, Identity of high-energy spectral component(s), Particle acceleration mechanism(s) & Gamma-ray polarimetry, Coordinated multiwavelength campaigns, Regular long term observations, High-cadence gamma-ray light curves, MeV \& X-ray polarimetry to understand the jet particle composition \\ 
\textbf{Uncertainty Level: 2} & & \\ 
\hline
\vspace{2pt}
\textbf{{\Large Corona}} & Existence, Geometry, location, composition, ...      &  Simultaneous UV-X-ray campaigns of unobscured AGN to measure time-lags and determine the coronal power, mm (100 GHz)-X-ray simultaneous campaigns of heavily obscured AGN to determine the AGN power, X-ray sensitive polarimeter for the geometry, hard-X-ray to MeV coverage to measure the coronal cut-off (hence coronal temperature) and possibly the non-thermal component (511 keV line and non-thermal electron tail) \\ 
\textbf{Uncertainty Level: 5} & & \\ 
\hline
\vspace{2pt}
\textbf{{\Large Disc}}  & Size, Accretion efficiency, Impact on BH spin, Magnetic field generation and associated structure, Driver behind winds, Thickness, Substructure  & Optical and UV broad spectrum continuum time lags,  \\ 
\textbf{Uncertainty Level: 2} & & \\ 
\hline
\vspace{2pt}
\textbf{{\Large Obscurer}} & Geometry and variations by subclass, Composition, Opacity, Physical connection with accretion disk, Size, Radiative efficiency & Time-resolved broadband ($\sim$\,0.1\,--\,100\,keV) X-ray spectroscopy simultaneous with sub-mm coverage at $\sim$\,100\,GHz (rest-frame)      \\
\textbf{Uncertainty Level: 3} & & \\ \hline
\vspace{2pt}
\textbf{{\Large Black Hole}}      & Mass, Spin, Single or binary, Impact on accretion, Physical connection with disk and if it varies by subclass, Geometry of magnetic field lines, Impact on jet launching & megamaser 22 GHz (more sensitive instruments), multi-epoch infrared-to-UV spectroscopy, high-resolution X-ray spectroscopy      \\ 
\textbf{Uncertainty Level: 2} &  &  \\
\hline
\vspace{2pt}
\textbf{{\Large Broad Line}}    & Origin, Location  &  Spatial resolution in IR and Optical, time-resolved spectra of the BLR  \\
\textbf{Uncertainty Level: 1} & & \\ \hline
\end{tabular}
\caption{Summary of AGN elements, their uncertainties (on a scale between 1 and 5, where higher numbers are more uncertain), and corresponding needs. The \textbf{Uncertainty level} is not a measure of ``importance'' or ``relevance'' in the field of AGNs; additionally, such levels have been attributed based on the subjective view of the authors: objections or additional arguments in support of this uncertainty ranking are very welcome. The uncertainties listed are higher for structures whose existence is not well known because that seemed more profound than precisely how a known structure behaves. Lower uncertainties are ascribed for items that are well constrained with minor exceptions. Most structures have multiple questions associated with them and we attempted to average the uncertainties across those questions. Note that we do not discuss AGN elements that are not relevant for TDAMM (e.g., narrow-line region and the host galaxy.)}
\label{tab:agn_elements}
\end{table}

Other systems / events can provide a smaller-size shorter-time scale AGN analog. For example, in the X-ray binary community, there has been work in corona simulations that is not currently informing AGN simulations of coronae. There has been extensive work on modeling accretion that is not currently used to inform many simulations of jets. Advances in those fields could be helpful in expanding simulations for AGN science. While it is necessary for any model to make assumptions in order to produce a solution, our community could do more to work toward pulling those assumptions from exciting work in neighboring fields in order to better support each other and the cumulative knowledge of AGN physics. The following is an incomplete list of potential collaborative projects that are inter-sub-disciplinary. 

\begin{itemize}
    \item \textbf{Tidal disruption events (TDEs):} are temporary AGN that light up when a main-sequence star falls too close to the supermassive black hole (SMBH) and is gravitationally disrupted. A small fraction of TDEs are jetted (four candidates observed so far), meaning that they exhibit nonthermal emission attributed to temporary super-Eddington accretion and/or a collimated relativistic jet. Radiation from radio to X-rays is observed up to years after the event. Variability on hundreds-second timescales has been observed in the overall decaying X-ray emission. Such events allow for a full end-to-end observation of jet formation and evolution at the AGN mass scale. So, these objects are studied for their jet launching properties, which should be more similar to AGN than, e.g. GRBs, which are stellar-scale (however the jet maybe powered magnetically which is more typical for GRBs than AGNs). 
    \item \textbf{BH X-ray binary (microquasars):} There are some connections between stellar mass BHs and SMBHs and there is evidence of them following a similar accretion pattern, but it is not clear to what extent all behaviors can be extrapolated across mass scales.
    A benefit of exploring physics across mass scales is that modeling communities often work at one end or the other, meaning that significant advances in modeling of, e.g. coronae, have occurred for X-ray Binaries (XRBs). It is currently unclear whether XRB coronae can be extrapolated to AGN, but they are probably a good place to start in terms of more detailed modeling to test this idea. The recent neutrino observations from AGN seem to point to coronae as a likely source, but AGN corona models are currently less advanced. On the other hand, different mass scales produce very different magnetic environments, which then evolve in very different physical limits in terms of diffusion. So, examining the reworked models in connection with AGN data is a necessary step. 
    \item \textbf{Dual/Binary supermassive BHs:} At the core of almost every large galaxy is at least one supermassive black hole. We know that these galaxies were built up over cosmic time, with abundant evidence coming from hierarchical mergers, spatially resolved by JWST and the Hubble Space Telescope (HST) at a range of redshifts. At some point following the galaxy merger, the central SMBHs must inevitably also merge. These dramatic collisions will be the most energetic events in the observable universe, producing gravitational waves with more energy than a million gamma-ray bursts. Along with accretion from a circumbinary disk, these SMBH binary mergers will be excellent multi-messenger sources, informing about the extreme physics involved in the merger of two SMBH, as well as about the cosmological evolution of such objects. 
\end{itemize}

 
Here is why you should care about AGNs: 
\begin{itemize}
\item \textbf{If you are an astronomer:} You should care about the effect (aka ``feedback'') of AGN on their surrounding environment (both galactic and extra-galactic). Active galaxies, especially jetted active galaxies mark the redistribution of energy onto cosmic scales. 
Winds and jets can transfer both energy and material from the central engine of the AGN into the intergalactic medium. These outflows are typically driven by processes occurring near the heart of AGN and such phenomena have been observed in a variety of AGN, and the dynamics of these energetic flows have been studied in many sources even at significant cosmological distance. In addition to their impact on the surrounding intergalactic medium, these outflows can play a critical role in the process of stellar quenching. As the gas within the surrounding galaxy is heated, stripped, or funneled toward the nucleus, it can prevent the formation of new stars, effectively halting star formation in the galaxy. This process, often referred to as "feedback," is thought to be one of the key mechanisms responsible for regulating galaxy growth and evolution. This is another priority in the Astro 2020 Decadal \citep{national2021pathways}.
\item\textbf{If you are a cosmologist:}
You should care about the history, or evolution, of these objects because understanding how SMBHs form, grow, and evolve is crucial to our broader understanding of cosmic history. The formation and growth of SMBHs are intimately tied to the formation of galaxies, and their evolution can provide insights into both the origins of cosmic structures and the physical processes governing them. The discovery of overmassive SMBHs in the early Universe has important implications for models of black hole seed formation and their growth. Recent findings (e.g., Suh et al. 2024) indicate that SMBHs at high redshift may have undergone multiple super-Eddington accretion phases, where they accreted material at rates far exceeding the Eddington limit, potentially leading to the rapid buildup of mass.
Additionally, high-redshift AGN provide compelling evidence of SMBH growth, both through accretion but also major galaxy mergers. The study of high-redshift AGNs allows us to trace the timeline of SMBH evolution and how these objects may have influenced the surrounding galaxies and the larger cosmic structure.  While it is well established that the radiation from early AGN could have contributed to the ionization of hydrogen gas in the early Universe, the extent of this contribution is still under debate. Some models suggest that AGNs played a significant role in reionization, while others propose that the bulk of reionization may have been driven by stars in the early galaxies. 
\item\textbf{If you you are a physicist:} AGNs are an extreme physical environment  regardless of whether you consider their mass, sustained energy, or particle acceleration capacity. They have one (or more?) supermassive black holes (SMBH), a relativistic jet, a hot corona, an accretion disk, winds (gas outflows with velocities as large as 0.2 c). AGNs likely accelerate cosmic rays (Z$>$1 nuclei), either in their winds or jets. In particular, jets are possible candidates for high-energy neutrinos and ultra-high-energy cosmic rays. Galaxies continuously merge in the Universe and so do their SMBH: the gravitational waves (GWs) emitted during those mergers will be detected by LISA. New physics and new particles, such as axion-like particles (ALPs), could be produced via photon-magnetic field coupling (this phenomenon should emerge as an excess in photon counts in the spectrum at very high energy gamma-ray, where the signal should be absorbed by the extra-galactic background light). The jetted outflows produced by AGN shine across the entire electromagnetic spectrum and exhibit every emission mechanism known to physics. By observing (and modeling) the AGN phenomena, astronomers are able to probe the limits of physics. It is through probing these limits that we can hope to further our knowledge of the fundamental laws of nature. The study of AGN can further our understanding of: particle physics, fluid dynamics, electricity \& magnetism, gravitational waves.
\item\textbf{If you you are a funding agency}: it depends on the funding agency. NASA should care because AGN studies answer the question ``How does the universe work?''. NSF should care because of the insights these powerful objects can give in terms of astro-particle physics, acceleration mechanisms. DOE should care because of possible new physics (dark matter) signatures detectable indirectly (ALPs).
\item \textbf{If you you are the public:} This science can unveil the mysteries behind the origin and formation of the Milky Way, the galaxy we live in; it can inform us about the destiny of Galaxy, by informing us on the interaction between nearby galaxies; it can give us clues on the physics of element formation, the matter that makes everything we experience in our life; it can unveil the mysteries behind the black holes at the center of galaxies (including ours), how they formed and evolved to the gargantuan masses we observe today throughout the Universe. The general public has an insatiable curiosity concerning black holes and the phenomena of powerful phenomena, like AGN jets.
\end{itemize}


The study of AGNs and their associated phenomena, such as jets, coronae, and accretion disks, requires solving a range of complex physical problems involving plasma physics and transport processes. Addressing these challenges effectively requires a hybrid approach that combines theoretical and computational methods. For example, as regards AGN jets, Particle-In-Cell (PIC) simulations for plasma dynamics and advanced transport modeling should inform each other synergistically. However, a significant obstacle arises from the fragmented perspectives within the community, as researchers specializing in plasma physics, PIC, and transport often see their methodologies as independently providing the most critical insights. This division complicates collaborative discussions across the disciplines needed to advance understanding of these AGN components.

AGN phenomena present a challenge to the theoretical community given the vast range of spatial and temporal scales involved. Analogous to the temporal stage-based chains of simulations which must be mapped together for CCSN, AGN must be studied by chains of simulations. As mentioned above, models typically focus on individual elements of AGN structure (e.g., black hole accretion, jet production, winds, coronal emission, jet propagation, large-scale feedback, and the surrounding galactic environment). Numerical models of plasma physics either focus on macroscopic fluid flows (e.g., GRMHD, RMHD, MHD, and HD) or microscopic particle acceleration mechanisms such as magnetic reconnection (e.g., GRPIC and PIC). Very few hybrid schemes exist which incorporate/embed `micro' particle physics directly into larger `macro' plasma simulations \citep[see, e.g.,][]{Vaidya2018,Seo2024,2021PhRvL.126m5101A}. These hybrid schemes will be essential in connecting the disparate multimessenger signals emanating from AGN (e.g., radio, optical, X-ray, gamma-ray, neutrino, and GWs) to the next generation of AGN simulations.

In the summary below, we briefly cover a few of the outstanding science questions facing theoretical astrophysicists studying AGN. As mentioned above, the physical systems of AGN cover a vast territory of length, distance, and energy scales, often making it challenging to incorporate all of the relevant components into a single calculation or simulation. Here we will ``work from the inside out'', beginning with the inner-most region just outside the event horizon, and moving to larger and larger scales, ultimately connecting to the influence of AGN feedback on the galaxy and large-scale structure. 

\paragraph{Central engine} 
The driving power behind every AGN is a supermassive black hole. According to the no-hair theorem within general relativity (GR), every black hole is completely described by its mass and spin (assuming astrophysical black holes have negligible electric charge). The accretion of gas onto the black hole releases massive amounts of gravitational energy from the accreting material. It is this energy that is ultimately observed, in the form of photons across the electromagnetic spectrum, high-energy particles, relativistic jets, out-flowing winds, and more exotic messengers like neutrinos and gravitational waves.

To first order, the observational properties of an AGN are determined by the black hole mass, spin, and accretion rate ($M$, $a$, $\dot{M}$). 
The dimensionless spin parameter $a/M$ is particularly interesting, as it governs many fascinating properties of black holes in general relativity, such as frame dragging, the ergosphere, and the geodesic orbits of the surrounding material. 

The light emitted from gas immediately surrounding the black hole is similarly governed by strong gravity, manifesting in the form of extreme gravitational lensing which can produce nested light rings just outside the event horizon. With the recent breakthroughs by the Event Horizon Telescope (EHT) collaboration, and future upgrades to the interferometer network, precision measurement of these light rings is within reach. Only by careful comparison to global 3D simulations in full GR will we be able to interpret the next generation of direct imaging observations of supermassive black holes.
From a much broader perspective, the BH spin tells us about the accretion and merger history of the black hole and host galaxy. AGN that grow primarily through accretion via a thin disk are expected to have large spins, due to relatively long periods of coherent angular momentum growth. Conversely, AGN that were built up through a series of supermassive black hole mergers are more likely to have small-to-moderate spin parameters. 

\paragraph{Inner disk, plunging region} 
Just outside the even horizon and light rings lies the inner-most stable circular orbit (ISCO), a fundamental property of geodesic orbits in the Kerr metric. There is a unique one-to-one map from the spin parameter to the radius of the ISCO. As the black hole spin increases, the ISCO moves closer to the black hole, allowing the gas to lose even more gravitational energy before finally crossing the event horizon. This in turn increases the accretion efficiency, in principle as high as $\sim 40\%$ in the case of maximal spin, producing many times more energy than any nuclear reaction.

Inside of the ISCO, the gas flows along plunging trajectories, rapidly accelerating in the radial direction and thus leading to rapidly dropping density. At some point, the gas becomes optically thin to thermal emission, effectively disappearing. More importantly, it becomes highly ionized, and thus transparent to incident ionizing radiation such as high-energy X-rays which can produce fluorescent iron K$\alpha$ emission lines. These emission lines are promising diagnostics for the properties of the inner accretion flow and in turn the black hole spin. A distant observer can measure the gravitational redshift of these X-rays from just outside the black hole by fitting the broadened iron lines to a fully relativistic model. The production and propagation of such photons is a major topic of research in AGN theory, intimately connected to radiation transport, GR, and atomic physics.  


\paragraph{Corona} 

Since X-rays were detected from AGN in the early 1980s, it has been understood that, much like their stellar-mass counterparts, accreting supermassive black holes produce hot, diffuse coronae above their optically thick accretion disks. The thermal seed photons (typically thermal UV for black hole masses $\gtrsim 10^6 M_\odot$) inverse-Compton scatter off the hot ($T_e \gtrsim 100$ keV) electrons in the corona to produce a hard power-law spectrum in the X-rays. This X-ray continuum either escapes to the distant observer, or is reprocessed by the accretion disk. As mentioned above, one form of reprocessing is the creation of fluorescent Fe K$\alpha$ photons from the relatively cool, dense gas in the inner disk.

From a theoretical point of view, this is an important computational problem: starting from first principles, how do the MHD processes in the accretion flow naturally generate a cool, geometrically thin, optically thick accretion disk, along with a hot, diffuse corona? Furthermore, what is the temperature and density distribution of this corona, and how does it produce the observed X-ray features such as the power-law continuum and the broadened emission lines? 

Closely related to the questions of corona dynamics, is how do we explain the X-ray variability seen in almost all AGN at all time scales? And how can we explain the lags between direct X-ray continuum flux and reprocessed X-ray lines? At the present, state-of-the-art GRMHD simulations are too expensive to capture fully all of these processes in a self-consistent model. One promising solution seems to be multi-scale simulations that allow us to resolve the highly relativistic regions near the black hole, as well as the gas and radiation dynamics on much larger scales. 

In all likelihood, the X-ray variability seen from AGN on timescales from minutes to years is generated by fluctuations in the inner disk and corona. With improved modeling of these components, we will reach a greater understanding of the MHD processes, thermodynamics, and radiation transport through matter in the strongest gravitational fields. Along with polarization and, in a few select cases, VLBI direct imaging, we will be able to finally determine the geometrical properties of the AGN's hot diffuse corona.

\paragraph{Disk winds} 

One of the primary physical drivers behind the hot corona is the bouyant magnetic fields that pervade throughout the accretion disk structure. At each point in the disk, the vertical structure of the gas is determined by balancing the tidal gravity with the total pressure, which is composed of gas, radiation, and magnetic pressure. Thus, fluid elements with stronger magnetic fields tend to have lower density (for a given total magnetic and gas pressure), causing them to "float" to the top of the disk. Additionally, the reconnection of the disordered magnetic field naturally leads to heating, further causing the gas to expand and leads to lower density.

For strongly magnetized disks, the magnetic pressure can overcome the total gravitational force, driving outflows from the top of the disk. Combined with strong radiation flux from the central engine, this can produce large-scale winds moving radially away from the black hole at $10-30\%$ of $c$. The X-ray and UV flux from the inner disk produces absorption lines in these winds, which will be seen as blue-shifted along the line of sight to the observer. This is yet another critical problem in AGN physics that combines many different components: MHD, atomic physics, and non-LTE radiation transport, all over length scales that must be resolved on a range spanning many orders of magnitude.

An important question that is related to the disk, corona, and wind properties is, ``What is the magnetic field topology of the accretion flow?'' In stellar-mass black holes in X-ray binaries, we see multiple accretion states from a single source (e.g., high-soft, low-hard, steep power-law), and it is thought that the stochastic properties of the global magnetic field may be a major factor in the state transitions. Is the same thing happening for AGN? Do different magnetic field topologies lead to the wide range of AGN types? These are questions that are still very challenging to answer with GRMHD simulations, which can typically only be evolved for relatively short timescales. 

\paragraph{Jets}
AGN jets are studied by distinct communities depending on their orientation. From the side, jets are often studied in the radio (although also X-rays) from which vantage point, we learn about jet propagation beyond the nuclear region of the AGN. This is the connection point to large scale structure. Models that study this region usually study jet propagation, so they tend to include geometric and flow information, which might include geometrically dependent radiative transport or hydrodynamic codes. The primary emission mechanism is probably synchrotron from electrons, and the primary particle acceleration mechanism is probably turbulence, although what drives the turbulence is (probably) still debated. 

Blazars are face-on, jetted AGN, in which the base of the jet outshines all other components of the jet, AGN and host galaxy at all frequencies except radio, for which the ``blazar zone" is opaque. Blazars give the best view of the deepest observable part of the jet due to the amplification of this region by Doppler beaming. Therefore the study of blazars and the study of AGN jets are nearly synonymous for the theory community, especially at higher energies most relevant to TDAMM. 

There are 3 main categories of model type used to describe the blazar zone. In order of historical introduction, they are: transport models, magnetohydrodynamic models, and particle-in-cell models. Each of these could be broken down further \citep[e.g.][and references therein]{Bottcher2019}. 
\begin{itemize}
\item Particle-in-cell simulations follow individual particles through computational cells small enough to contain only one or a few particles per time step to trace the microphysical interactions that accelerate and cool those particles. These simulations are very computationally intensive, can only model small spatial regions, and often do not include photons (observable radiation) due to computational time and space constraints. However, they are indispensable to our growing understanding of how particles interact in this extreme environment, which is inaccessible to experiments with sufficient proximity to observe them directly. 
\item Magnetohydrodynamic simulations work at a medium spatial scale, to study flows and the impact of magnetically driven turbulence on particle acceleration. More recently, some codes have started to introduce cooling and photons, but across the board, it is not common due to the cost in computational efficiency and the fact that there is still a significant mismatch in terms of scale that prevents a straightforward comparison with observations. Magnetohydrodynamic simulations cannot be performed at the physical scale of a blazar zone due to limits on computational time and memory. Magnetohydrodynamic simulations are fundamental to understanding energy transport in magnetized plasmas, identifying the types of particle acceleration most likely in specific physical environments and scenarios, and how turbulence evolves in the jet. There is evidence that these simulations need to be 3D to represent most of the physics of interest accurately. 
\item Transport models historically referred to analytic or semianalytic radiative transport equations, which were one-zone and did not model particles at all, but could capture the entire blazar zone, and were readily compared with data (successfully). These basic models are now available ``off the shelf" and observers can download then to use in preliminary data interpretation for data analysis-focused papers. Today, theorists who are developing transport models tend to focus instead on particle transport computation, which they link to observable spectra that are generalized to respond to the particle simulation in situ. Particle transport codes can provide an efficient comparison between different acceleration and cooling mechanisms in the context of data fitting, but the mathematical formulation of the physics they express is informed by magnetohydrodynamic and particle-in-cell simulations, as well as work in other areas of physics (e.g. particle physics). Since particle transport codes lend themselves to data comparison by virtue of being able to model the full scale blazar zone, work in simulating observable (i.e. time series, spectra, polarization) signatures from particle energy distributions is often developed by these groups. Particle transport researchers often still work in the ``one zone" schema for a variety of well justified reasons, but the method itself is not limited to that presentation. While expansions tend to focus on adding a small number of additional zones, the transport equation can be expressed as spatially dependent, with the challenge then being to identify appropriate conditions.
\end{itemize}
Thus, each major modeling method offers distinct and complementary information about the blazar zone despite significant differences in method and scale. Increased communication and cooperative development would be beneficial to advancing all three channels and jet physics overall, whether that takes the form of hybridization, cross checking limiting cases, or increasing cross talk. The interest in hybridization is an interest in facilitating better communication of relevant parameters and physics between codes at different scales. 
There is an overarching vision of eventually being able to create a single code that performs calculations at each scale strategically to effectively model the entire zone and all of the physics involved. Working toward that goal will probably mean facilitating smaller scale cooperative funding opportunities so that researchers and their students can work out the details of getting individual methods to communicate effectively without leaky physical parameters.  


This question of jet composition is not only intrinsically important but also ties into several broader topics. For instance, the nature of the jet plasma provides clues about the jet launching mechanism. Electron-positron dominated jets are often associated with the Blandford-Znajek process, where energy is extracted magnetically from a spinning black hole, while proton-rich jets may suggest contributions from baryonic winds or disk-driven mechanisms. Similarly, understanding jet composition is key to addressing whether jets are the primary accelerators of ultra-high-energy cosmic rays and whether they produce the high-energy neutrinos observed by Icecube. That being said, recent tentative associations between blazars, such as TXS 0506+056, and high-energy neutrinos provide compelling evidence for hadronic interactions in jets.
The composition of jets also influences their radiative signatures, energetics, and impact on their environments. Addressing this question will require continued advancements in multiwavelength observations, time-dependent modeling of jet emission, and neutrino astrophysics. As our tools improve, we can expect new insights into the physics driving these powerful and enigmatic structures.

If we discover that AGN jets are purely leptonic, meaning that they only contain electrons and positrons, then we should observe a 511 keV spectral line, and jet formation proceeds through an $\Omega$-dynamo effect, with all accretion material falling onto the black hole and new particles created via pair production at the base of the jet, inside the magnetic column. If the jet contains a significant number of protons (i.e. more than those that might randomly fall into the jet from the wind or broad line region), then we don't have a clear picture of how the protons arrive in the jet. It would seem that accretion is a much messier process than currently believed for protons from the disk to be able to make their way to the jet by somehow avoiding the ergosphere. Even if we figure out how protons get to the jet, there is still a question of how they are accelerated to the point they would radiate at an observable level or explain the high-energy cosmic ray spectrum. Protons are much more massive than electrons, and thus take much more energy to accelerate to the same degree. The energy in the AGN engine, if provided by accretion as is currently accepted, is not sufficient. Now, more detailed calculations are warranted on both the jet and disk sides of this equation, but there is currently a conservation of energy disconnect because we have likely observed neutrinos from at least one blazar (and a number of unjetted AGN), meaning that protons were accelerated in a way that theories do not currently explain self-consistently through the full life cycle of a proton accreted through an AGN. To further complicate matters, a number of theorists now believe that both the leptonic and leptohadronic scenarios may be true, and form the basis for an alternate classification scheme. The current classifications of blazars are based on observational patterns with a bias toward optical bands.  The needs to address this set of problems are pretty broad, but as a nonexhaustive list: We need to be able to trace energy between the disk and the jet, which probably means either jet theorists picking up/implementing more modern disk codes than the default \citet{shakura1973black}, or collaborations between researchers in each community. We need a better understanding of particle acceleration, especially for hadronic processes.

\paragraph{Feedback}

An interesting fact about supermassive black holes is that their gravitational binding energy can easily exceed that of the entire galaxy of stars, gas, and dark matter. Because these monster black holes grow mostly through accretion, the gravitational energy liberated by the accretion of this gas could in principle disrupt the entire galaxy. In practice, this does not happen because most galaxies are optically thin, which is why we can see quasars in the first place. However, for galaxies with high gas and dust fractions (often presenting as ``starburst'' galaxies), the AGN accretion power has a very significant impact on the galaxy morphology and evolution. Both the radiation pressure, as well as the kinetic energy deposited by jets and winds, can significantly disrupt the presence of dense gas and molecular clouds, effectively cutting off star formation.

\paragraph{Dual and binary AGN}

When galaxies merge, the massive black holes at their centers eventually will also merge. In the first phase of this process ($\sim 1$ Gyr) the black holes are sufficiently separated to be treated as independent AGNs (dual), and we have now seen many examples of these spatially resolved sources in the X-rays and optical. In many post-merger galaxies, a single AGN may be seen offset from the new galactic center, essentially presenting as one half of a dual AGN system, where the other SMBH is likely quiescent. 

As the two SMBHs sink towards the center of the galaxy via dynamical friction, they will eventually become gravitationally bound to each other, and thus a {\it binary} AGN. This typically occurs when the separation is on the order of a few parsecs, far below the spatial resolution of current optical or X-ray telescopes. These sources may be identified by periodic variability in their light curves, double-peaked broad line emissions, or other spectral features indicative of a truncated circumbinary disk. The astrophysics of circumbinary accretion is particularly rich, involving all the problems of single AGN simulations, as well as dynamic general relativity and time-dependent radiation transport through systems where the light-crossing time is comparable to the dynamical time. Modeling such sources is a computationally demanding problem, and will likely be the focus of much of the the effort in this field for years to come.

One of the reasons that circumbinary disks are so important to model correctly is that, unlike dual AGN, we have yet to discover even a single unambiguous binary AGN candidate observationally. Thus, when designing new surveys for these important sources, we are relying entirely on the theoretical predictions for their observational signatures. But with this difficulty comes great potential: when the binary becomes sufficiently tight, its evolution will be dominated by gravitational waves, and these systems are expected to become some of the best multi-messenger sources in the sky. At low frequencies (nHz) the gravitational waves should be observable with pulsar timing arrays, while in the mHz band, space-based gravitational wave interferometers will be able to observe their final months before merger and provide reasonably well-constrained sky positions for electromagnetic follow-up. 


Funding agencies should increase funding availability and possibly funding competing groups.

Current funding structures also pose limitations. Computational proposals are often prioritized over purely theoretical approaches, leading to an imbalance in support for essential theoretical advances. Additionally, theoretical proposal calls are typically small, often supporting only a single postdoc, which is insufficient for the collaborative efforts that many projects demand. Large solicitations, such as TCAN, can support multi-institutional projects, but their structural constraints make it difficult to coordinate effectively across institutions.
As a result, teams often end up proposing separately to avoid competition over limited funds, yet still encounter resource challenges when attempting to merge efforts. After accounting for overhead, the remaining funds are frequently inadequate to support cohesive, multi-institutional research.

\begin{quoting}
    \noindent Finding: The current funding calls typically utilized for theory and simulation of AGN are not large enough to support cohesive, multi-institutional research teams at the level required to make significant progress. The size fosters competition, rather than collaboration.
\end{quoting}

Beyond the field of AGN studies, a notable tension exists within the theory and computational communities: the push for open-source code development versus the need to protect intellectual property and ensure proper acknowledgment of computational tools. As simulation and analysis codes become increasingly sophisticated, their development often requires large, collaborative teams spanning multiple generations of graduate students and postdocs. For some groups, sharing decades of effort and expertise can feel like a significant sacrifice, particularly if the primary benefits appear to advantage external groups. The fear of losing the competitive edge of exclusive access often outweighs the perceived benefits of making these codes universally accessible.
However, almost everyone realizes how code comparison is also critical to the future of theory work, providing essential validation for complex problems that can often not be compared directly with observations. Standard guidelines coordinated among funding agencies may be the solution to this outstanding issue.


There is a common misconception that AGN science is only theory-limited, whereas in reality, it is often constrained by data quality and statistical limitations. The low photon count rates from the majority of AGNs limits our ability to perform time-resolved studies, as the minimum variability timescales we can detect are biased by the resolution and sensitivity of available instruments. Additionally, the lack of continuous monitoring of blazars—currently only observed during major flaring events—hinders a full understanding of these dynamic systems. Instruments like eROSITA would  be beneficial in filling these observational gaps, however eROSITA could not complete the expected survey providing the long-term coverage of a large sample of sources as well as enough counts to perform meaningful studies even of the fainter sources, which makes the bulk of the AGN population. Moreover, a consistent monitoring across the electromagnetic spectrum is essential to study the time lags at different wavelengths.\\


While recent years have seen the opening of several observing windows, there are still critical gaps preventing answering long-standing questions. In this section we outline several missing observables, and their importance in understanding how the Universe works. Table~\ref{tab:agn_diagnostics} summarizes the main observables for AGN studies and the inference that they enable regarding the physics and the astrophysics of these sources.

\begin{table}[h!]
\centering
\begin{tabular}{|p{4cm}|p{6cm}|p{6cm}|}
\hline
\textbf{Diagnostics} & \textbf{Inference} & \textbf{Facilities} \\ \hline
\textbf{Time variability} & Probes size and structure of emission regions; reveals timescales of processes in the jet, disk, or corona & Swift, Chandra, XMM-Newton, NICER, NuSTAR (X-ray); Fermi-LAT (gamma-ray); LSST (optical, future) \\ \hline
\textbf{Polarimetry} & Provides magnetic field geometry; helps distinguish between synchrotron, inverse Compton, and other emission processes & Radio: ALMA, VLA; Optical: RoboPol, Hubble, VSTpol (Future); X-ray: IXPE; MeV: COSI (future, for bright blazars) \\ \hline
\textbf{$\gamma$-ray spectra} & Helps determine hadronic or leptonic origin of $\gamma$-ray emission; indicates jet composition & Fermi-LAT, MAGIC, H.E.S.S., VERITAS (gamma-ray),  CTA (Future); COSI (Future)\\ \hline
\textbf{X-ray Spectra} & Measures corona temperature, disk structure, and torus obscuration; identifies emission and reflection components & Chandra, XMM-Newton, NuSTAR, Swift, XRISM, SVOM, Einstein Probe, NICER\\ \hline
\textbf{Optical spectra} & Informs on host galaxy properties, including stellar population and star formation rate & DESI, SDSS, Hubble \\ \hline
\textbf{UV spectra }& Information on the accretion disk of unobscured AGN. Correlation with X-rays can inform us on heating/cooling mechanisms of the corona & Swift/UVOT, ULTRASAT, UVEX\\
\hline
\textbf{Gravitational Waves} & Measures black hole mass and redshift; constrains mergers and galaxy dynamics & PTA (Pulsar Timing Array), LISA (future) \\ \hline
\textbf{High-energy $\nu$} & Hadronic content of Jets / Coronae? & IceCube, KM3NET, IceCube-Gen2 (future) \\ \hline
\end{tabular}
\caption{AGN Diagnostics and their insights into the physics involved and existing/planned facilities.}
\label{tab:agn_diagnostics}
\end{table}

\paragraph{Millimeter - radio} 

Radio observations have been at the forefront of AGN studies since their discovery in the 60s. Given the relative ease in terms of observing sites and technical requirements, observations at low-frequencies (a few GHz) have so far dominated the research output. However, recent new facilities like the Atacama Large Millimeter Array (ALMA) pushing the frequency range to higher and higher frequencies have clearly demonstrated the importance of studying AGN at high radio frequencies. The new high-frequency capabilities have made a number of different investigations and discoveries possible, such as studying molecular outflows and AGN feedback, high-redshift sources, particle acceleration in jets, and made the first image of a black hole possible \citep[e.g.,][]{EventHorizon2019, Bolatto2021, Endsley2023, Chen2024}.

It is, however, clear that we have barely scratched the surface. Future facilities are needed to make progress. Facilities that would provide multi-band (+polarization) observations at $\rm >200~GHz$, with enough sensitivity to probe a large fraction of the AGN population. This is particularly important for the radio-quiet sources which are overall faint. For the radio-loud sources, Event Horizon Telescope observations have given a first glimpse in accretion processes and jet launching around SMBHs. It is now important to move from single snapshots of just two sources, to making black hole movies for a much larger sample. Radio-polarization capabilities are equally important to study the evolution of the magnetic fields in the very vicinity of the event horizon, a key (missing) ingredient for understanding plasma behavior in extreme gravity. 

X-ray surveys of AGN across cosmic time have revealed that circum-nuclear obscuration (see Figure \ref{fig:agn}) is a ubiquitous ingredient in the growth of supermassive black holes (e.g., \citealt{Gilli07,Ueda14,Buchner14,Buchner15,Ricci15,Brandt15,Ricci17_bassV,Boorman24_nulands}). Moreover, many theories of supermassive black hole growth predict that the most rapid stages of accretion coincide with the densest levels of circum-nuclear obscuration (e.g., \citealt{Fabian99,Springel05,Hopkins06,Blecha18}).
However, to-date we know of less than 100 local AGN ($<200~\rm Mpc$) that are heavily obscured \citep{Boorman24_hexp}. Besides missing the bulk of the population with current selections, when a source is heavily obscured, a strong and sometimes dominant component in hard X-rays is the reprocessed emission (which appears as a broad Compton hump).
A powerful and relatively clean diagnostic of the central power source is accessible with broadband X-ray spectroscopy (where the coronal emission can be disentangled from another components), though currently we either have all-sky surveys that only select the brightest/closest cases (hence are biased towards the most extreme sources, \citealt{Ricci17_bassV}), or pointed observations of comparatively smaller samples of AGN (see the \textit{NuSTAR} Local AGN $N_{\rm H}$ Distribution Survey, NuLANDS; \citealt{Boorman24_nulands}). Independent of the selection criteria, there is a large degeneracy in models trying to estimate the intrinsic luminosity of the central AGN especially in heavily obscured sources,
leaving us with orders of magnitude discrepancy on fundamental parameters such as the intrinsic AGN power or line-of-sight column density (see Figures~1 and~2 of \citealt{Boorman24_hexp}). An exciting and currently under-used diagnostic for the intrinsic AGN power is from high angular resolution observations at 100\,GHz \citep{Ricci_2023a}. The observed tight correlation with intrinsic X-ray power suggests observations at 100\,GHz are capable of probing the synchrotron emission from the same electron population as the central X-ray source, almost entirely unaffected by obscuration. Multi-epoch observations in sub-mm would thus provide a calibrator for the AGN central power. Ideally, multi-epoch simultaneous sub-mm and broadband ($E$\,$\sim$\,0.1\,--\,100\,keV) X-ray observations are needed.

Finally, another excellent radio diagnostic for AGN is via 22\,GHz water molecule megamaser emission. Tracing the emission and dynamics of megamaser discs has enabled some of the most precise and accurate black hole mass measurements. However, this technique is time-intensive, needs the maser disc to be viewed effectively edge on, and the current generation of facilities cannot reach the sensitivities we need to provide mass measurements for large samples of AGN (only a dozen AGN have mass measurements obtained via this technique and sensitive broadband spectroscopy; \citealt{Masini16,Brightman16,Brightman17,Panessa20,Boorman24_hexp}). Having more sensitive instruments at this frequency that can time resolve the motion of the water maser emission in a large sample of AGN would be key to providing more mass measurements. Statistically significant samples of AGN with accurate and precise measurements of black hole mass (irrespective of obscuration) with time-resolved measurements of intrinsic AGN power would provide stringent measurements of how AGN co-evolve with the circum-nuclear environment and broader host galaxy.

\paragraph{Optical and UV} 

Material accreting onto the supermassive black hole settles into an accretion disk (see Figure \ref{fig:agn}). The actual geometry and radiation efficiency of this accretion are still debated. For black holes accreting below the Eddington limit, the most canonical accretion disk model used by the AGN community is the Shakura-Sunyaev optically thick and geometrically thin one \citep{Shakura_1973, Novikov_1973, Lynden-Bell_1974}. In this approximation, the radiation from the disk can be understood as a series of annuli that emit as black bodies of different temperatures. The temperature of each annulus depends on the distance from the central black hole, the accretion rate, and the mass of the central black hole. In the super-Eddington accretion regime instead, the most common disk approximation is that of an optically thick advection–dominated accretion flow \citep[optically thick ADAF][]{Abramowicz_1995, Narayan_1998} in which radiation is trapped by the optically thick inflowing gas and advected towards the central supermassive black hole. Typically, the emission of the disk peaks in the optical/UV regime.
Importantly, the accretion disk provides a huge reservoir of photons for the AGN system. 
All the surrounding material illuminated by the disk reacts to it by reprocessing, one way or the other, this radiation.

A key observational feature in AGN is their infrared to UV rest-frame spectrum, showing broad and/or narrow emission lines \citep[see the archetypical 3C 273,][]{Schmidt_1963}. Setting AGN apart from typical galaxy starlight (absorption line dominated), these lines (when visible) enable the measurement of a fundamental quantity -- the AGN redshift (i.e.~its distance in cosmic time), hence the luminosity (or power) of the system. It is understood that these lines are produced by the excitation of elements in the neutral gas orbiting at high velocity around the central supermassive black hole. The gas closer to the black hole spirals faster, causing the lines to be broadened. By measuring the time-delay between variability in the brightness of the accretion disk and the broad emission lines, we can derive an approximate distance of the broad line region ($\lesssim 0.1\,\rm pc$, \cite[e.g.][]{Blandford_1982}). Recently, this region has been spatially resolved \citep{Gravity_2018,Abuter_2024}, pointing to its origin being that of gas clouds gravitationally bound orbiting the central black hole.  
If we can measure the width of the lines, assuming virialized motion of these clouds, we can estimate their velocity, and we can derive the mass of the black hole itself (just like we can derive the mass of our own Sun by measuring the angular velocity of the Earth around it). For years, single-epoch (e.g.~only one spectrum per source) spectroscopic mass measurement of black hole masses has been one of the most employed techniques in estimating this fundamental parameter \citep[e.g.,][]{Shen_2011,Wu_2022}. However, it is also very well known that estimating masses via line spectroscopy can be tricky \citep{Denney_2009}: (i) accretion models and observation of AGN disk show the omnipresence of AGN winds, which could broaden lines even more \citep[e.g.][]{Murray_1995,Baskin_2005,Shen_2012,Vietri_2020}; (ii) element lines can be contaminated by several underlying (unresolved) elemental lines mostly when the central AGN becomes fainter and of luminosity comparable to the host galaxy \citep[e.g.][]{Reines_2013, Reines_2015}. Black hole masses could therefore be overestimated by up to an order of magnitude if the lines are not properly characterized. A great diagnostic to remove some of these degeneracies would be time-resolved infrared, optical, and UV spectroscopy. If indeed we could monitor the time-evolution of the emission lines, we would be able to disentangle the true broadening of the lines resulting from virial motions around the black hole, hence we could better estimate the black hole masses. Moreover, some AGN are known to be changing-look, meaning that at different epochs they show different properties in the optical spectrum (such as disappearance/reappearance of broad-lines, \citep[see][for a review on the topic]{Ricci_2023b}). Monitor changes in these lines could inform us of the intrinsic changes in the accretion modes of the supermassive black hole and the efficiency of the accretion .

Another key reprocessing source of accretion disk emission is the corona of electrons sitting above and below the black hole \citep[e.g.][]{Rybicki_1979,Begelman_1983,Haardt_1991,Done_2012,Fabian_2015}. The corona (composed of thermal and, possibly, non-thermal electrons, \citealt{Coppi_1999}) inverse Compton scatters optical/UV photons from the accretion disk and upscatters them to the higher X-ray/MeV energies. One would naively expect therefore that if changes in the accretion disk occur, they should be reflected in the X-ray emission of the corona with a certain time-lag -- i.e. changes in the optical/UV emission of an AGN should occur (lead) earlier than changes in the X-ray emission. However, time-monitoring campaigns of AGN have shown the opposite trend, where changes in the X-rays occur earlier than those in the optical \citep[e.g.][]{McHardy_2018,Edelson_2019,Pahari_2020,Kara_2021}. The understanding of this phenomenon is that part of the X-rays produced by the corona radiate back onto the accretion disk, are reprocessed to lower frequencies by the accretion disk itself, and are seen as modulations in the UV emission \citep[e.g.][]{Uttley_2014,Kammoun_2021a,Kammoun_2021b, Papadakis_2022, Langis_2024}. However such studies are limited to very few bright sources that have been studied for at most 2 years consecutively \citep{Hernandez_2020}. 
The UV band is crucial to unveil the heating mechanisms of the corona itself. Ideally, simultaneous broad-band coverage at UV and hard X-rays could shed light on the interplay between corona heating and cooling mechanisms.

\paragraph{X-rays}

X-ray emission has been the defining trait for both radio-quiet and radio-loud AGN. AGN science has benefited from several missions focusing on imaging and spectroscopy in the soft X-ray band, like \textit{Chandra} and \textit{XMM-Newton}. In the X-ray band, the spectrum is often dominated by the AGN emission as the thermal emission from the host galaxy tends to diminish significantly $\gtrsim$\,3\,keV, making X-rays an optimal band to study AGN. However, three critical aspects of the X-ray emission remain under-explored, namely variability, polarization and the hard X-ray regime. 

In terms of variability, most efforts have focused on targeted campaigns on individual sources. However, even the limited number of observations from eROSITA have produced a large sample of AGN (several hundreds) with significant variability and also tripled the number of known AGN exhibiting Quasi Periodic Eruptions \citep{Arcodia21,Arcodia24}. 
This truly highlights the need for an all-sky X-ray imaging monitor that will be able to survey the sky with a cadence of a few days or less. This is particularly important for understanding the multimessenger emission from black holes and possible association with flaring activity of AGNs \citep{IceCubeCollaboration2018}. Additionally, supermassive black hole binaries that could potentially be detected by Pulsar Timing Arrays \citep{Agazie_2023, Reardon_2023, Xu_2023, EPTA_2023} or \textit{LISA} are likely to imprint periodic variations in their multiwavelength lightcurves. 
Some evidence already exists that the $\gamma$-ray blazar PG\,1553+113 shows 2.2\,year periodic oscillations in its $\gamma$ light-curve \citep[see][]{Penil_2024}. This modulation of the jet emission can be explained by a binary supermassive black hole at the center of the system. Archival {\it Swift} X-ray and UV/optical, together with other archival optical data also show that the same periodicity recurs with some delay at other wavelengths; moreover, an underlying long-term (22\,year) period is detected that could arise from the way material spirals around this black hole pair \citep{Adhikari_2024}. Undoubtedly, long-term multi-wavelength coverage is necessary to see whether these trends can be found in other sources and discover other signatures of binary supermassive black holes. 
The Vera Rubin Observatory is expected to find hundreds of strong binary supermassive black hole candidates that will need independent confirmation. The ability to monitor the X-ray emission from all these potential candidates will provide crucial insights with X-ray timing, for revealing gravitational wave sources.

A promising connection between neutrinos and AGN, particularly obscured AGNs, has emerged in high-energy astrophysics, as highlighted by recent work from IceCube data on NGC\,1068 \citep{IceCube_2022} and complementary modeling \citep[e.g.][]{Ajello_2023,Murase_2024}. With a more (than \textit{NuSTAR}) sensitive hard X-ray / soft gamma-ray observatory—capable of covering up to hundreds of keV—we could solve many unanswered questions to do with obscured AGNs, which remain visible in the hard X-ray band despite their obscuration at other wavelengths. Resolving the X-ray background at these energies could reveal the true obscuration properties of the AGN population, offering deeper insights into their role in producing the high-energy neutrino flux observed by IceCube. An enhanced understanding of AGN populations and their obscuration also stands to clarify their contributions to the astrophysical neutrino background, a crucial step toward identifying the origins of high-energy cosmic neutrinos. 

The X-ray band is also crucial for unobscured AGNs (i.e. those in which the line-of-sight equivalent hydrogen column density is low, $N_H<10^{22}\,\rm cm^{-2}$, meaning that the X-ray photons are not dramatically absorbed at low and hard energies). As mentioned in the UV section, a key component -- yet very mysterious in nature --  of the AGN is the corona itself (see Figure \ref{fig:agn}). Indeed, several models have been proposed to explain the geometry and particle composition of this omni-present X-ray emission in all AGN types \citep{Haardt_1991, Coppi_1999,Petrucci_2001, Miniutti_2004, Fabian_2012, ZhangW_2019}. Broad-band X-ray spectroscopy (from $E=0.1\,\rm keV$ to beyond $E>100\,\rm keV$) would be fundamental to unveil the particle composition of the corona itself. First, measuring the spectral shape (photon index and normalization) of the coronal emission provides the power of the AGN -- valid for both obscured and unobscured sources; then measuring the cut-off of the corona estimates the temperature of the thermal components of the electrons and can lead us to understand the cooling mechanisms in this structure \citep[e.g.][]{Titarchuk_1995, Zdziarski_1996, Petrucci_2001, Petrucci_2004}. Multi-wavelength time-domain studies of the corona will therefore be crucial for: (i) unobscured AGN (UV+X-rays) to understand the interplay between cooling and heating mechanisms of the corona; (ii) heavily obscured AGN (sub-mm$+$X-rays) would enable determination of the co-evolution between the central AGN and surrounding obscuration.

In terms of polarization, the IXPE has provided the first glimpse on the X-ray polarization signatures from both radio-quiet and radio-loud AGN \citep{Marin2024,Marscher2024}. The key objectives were to constrain the geometry of the X-ray corona, understand particle acceleration in jets, and constrain the jet high-energy mechanism. The few radio-quiet AGN bright enough to be observed with IXPE have yielded low degree of polarization with a polarization angle that is parallel to the radio jet \citep{Saade2024}. This would suggest an extended corona above the accretion disk, however, the exact geometry of the corona remains a mystery. It is now clear that synchrotron emission in jets is energy stratified due to particle cooling processes \citep{Liodakis2022}. The particle acceleration process on the other hand is still debated between shocks and magnetic reconnection. While shocks are currently favored, the discovery of the first X-ray polarization angle rotation \citep{DiGesu2023} suggests that (a) magnetic reconnection processes are still possible; (b) the X-ray polarization properties of jets can vary within a few tens of ksec. The origin of the high-energy emission in jets, which is also related to their potential neutrino emission, has been debated since jets were discovered. IXPE made significant progress by observing the X-ray polarization of several blazars and radio galaxies \cite[e.g.,][]{Ehlert2022,Middei2023,Marshall2024}. No significant polarization from low- and intermediate peaked sources (i.e., sources whose X-rays are part of the high-energy component of the blazar SED) has so far been detected (typically $\rm <15\%$ at $\rm 99.7\% C.I.$), implying that Compton scattering likely dominates the X-ray emission in jets \citep{Agudo2024}. However, more complicated hadronic or hybrid models can not be confidently excluded with the current observations.
We remark the need for a larger, more sensitive, X-ray polarization imager that will be able to cover the missing parameter space and measure polarization at much lower fluxes and with finer time-resolution. 

\paragraph{$\gamma$-rays} 

$\gamma$-rays ($>$hundreds keV) from AGN have been fairly underexplored. To date, only one NASA mission, the {\it Fermi} $\gamma$-ray space telescope, is able to provide observations from energies between $100\,\rm MeV$ to $\sim 1\,\rm TeV$ \citep{Ajello2022}. 
With more than 15 years of all-sky coverage, the {\it Fermi} $\gamma$-ray space telescope has demonstrated the importance $\gamma$-rays play in understanding the Universe, including in identifying the electromagnetic counterparts to multimessenger events \citep{IceCubeCollaboration2018}. As more and more neutrino and gravitational wave experiments come online, it is crucial not only to maintain but to expand our all-sky monitoring capabilities. 

Another key aspect of the $\gamma$-ray regime is the fact that this window enables us to identify and trace the most powerful persistent jets in the universe. 95\% of the extragalactic {\it Fermi} $\gamma$-ray sky is made of blazar jets \citep{Ballet_2023}. Reaching luminosities equivalent to trillions of Suns, these jets are seen at all cosmic distances, importantly up to a point in time when the Universe was barely 1-2 billion years old \citep[e.g.]{An_2020, Marcotulli_2020, Belladitta_2020, Sbarrato_2021, Banados_2024}. Being able to detect blazar jets at all cosmic distances means that we are able to trace the evolution of these extreme monsters (e.g.~uncover when the majority of blazar jets lived). Moreover, these jets are powered by the most massive black holes that we know of.  Although blazars are very rare objects in the sky, thanks to the peculiar orientation of the jet, for every blazar we detect, we can infer the existence of a hundred more sources, at the same redshift, with the same black hole mass (a correction known as the $2\Gamma^2$ correction, where $\Gamma$ is the jet bulk Lorentz factor, usually in the range of 5-15). At a very fundamental level, this means that tracing the evolution of blazar jets enables us to trace the evolution of the most massive black holes through cosmic time. The most powerful and farthest jets in the universe are the ones that are brightest in the MeV band; the latest evolution study of these objects \citep{Marcotulli_2022} confirmed that the bulk of the population of these sources lives around 1-2 billion years after the Big Bang and are powered by supermasive black holes of billion solar masses. Hence, opening the MeV window with an all-sky sensitive mission will mean that we can start answering the question: {\it how do supermassive black holes grow and evolve, from the Big Bang to us?} In connection with the multimessenger domain, MeV bright blazars are also surmised to be neutrino emitters; prediction shows that an MeV mission in 2 years will be able to catch at least 1-2 neutrinos events from an MeV blazar \citep[see prediction in][]{Marcotulli_2022}. 

Jets are not the only AGN bright at MeV energies. If the corona had a non-thermal component of electrons (i.e.~some of the particles in the corona did not have enough time to share energy and reach the average temperature of the other particles in the corona, e.g. \citealt{Coppi_1999}), we would see it as an MeV tail of emission; moreover electron-positron pairs may be the regulators for the coronal temperature leaving as a imprint an emission line at rest-frame energy of 511 keV. The 511 keV line has tentatively been detected X-ray binaries in outburst \citep{Siegert_2016}. With a sensitive MeV mission, we would expect to both see the line and to measure the non-thermal component which would appear at the highest energies (beyond few hundreds of keV). Detecting this component would be revolutionary for the field of AGN science, as it also connects to the neutrino emission of NCG 1068. 

COSI will soon explore the MeV window (0.2--5 MeV), however, the capabilities of a SMEX are just enough to provide a first glimpse into the AGN MeV sky. Future $\gamma$-ray missions should not only have all-sky monitoring capabilities, but should also be sensitive enough to measure the $\gamma$-ray emission from a large number of jetted-AGN on short time-bins of a few days or less. 

The $\gamma$-ray polarization signatures from jetted-AGN can provide invaluable information for jet launching and composition \citep{Zhang2013,Zhang2024}. This is also important for the potential neutrino emission from blazars. There are currently no $\gamma$-ray polarization observations of AGN. While COSI will have polarization capabilities, it would likely require a remarkably bright flare from a blazar to perform any useful polarization measurements. An instrument specifically designed to measure $\gamma$-ray polarization would be required in order to open this new window to the Universe. 

\textbf{All in concert: the power of simultaneous observations:}
In table~\ref{tab:agn_diagnostics} we list in the third column current and planned facilities. 
Simultaneous, multiwavelength observations are essential to advancing our understanding of AGNs and the high-energy universe, but achieving consistent, global coverage requires robust multi-agency support to establish the necessary infrastructure to connect the different facilities. Combining both variability and polarization studies is key to unraveling the complexities of AGNs and their role in multimessenger astrophysics.  

Cross-correlation studies across different wavelengths allow us to track variability in real time, revealing emission timing and structure that help identify the origin of high-energy processes within AGNs \cite[e.g.,][]{Liodakis2018}. Adding polarization measurements at multiple wavelengths has provided groundbreaking insights, challenging traditional models of jet structures in blazars. More and more sensitive observations at regular visits are needed across the bands and possibly simultaneously. 


Advancing TDAMM-AGN science requires fostering synergistic collaboration across space- and ground-based facilities, and this will benefit from targeted calls for proposals that encourage integrated, multimessenger studies. Currently, limitations exist, especially within NSF-funded facilities, where multimessenger data and/or analysis tools are not fully public. This restricts the broader scientific community’s ability to conduct comprehensive analyses and impedes the collaborative potential between facilities. To address this, NSF could consider funding and mandating the release of official, well-documented analysis tools—developed by each facility’s team—alongside their datasets, enabling wider, more effective use of the data.

\textbf{Proposing multimessenger missions presents a systemic challenge, as they are often viewed more as a methodology than as standalone scientific cases.} This perspective needs to be reconsidered. Multimessenger studies are crucial for unraveling the complex, interconnected phenomena in AGNs and should be supported by dedicated resources and focused proposals to fully explore these systems.

\subsection{Core-Collapse Supernovae}
{\centering 
\textit{Contributors: Jennifer Andrews, Eddie Baron, Poonam Chandra, Emmanouil Chatzopoulos, Christopher L. Fryer, Brian Grefenstette,  William Raph Hix, Heather Johns, Gavin P. Lamb, Bronson Messer, Valarie Milton, Matthew R. Mumpower, V. Ashley Villar}}\\

\label{sec:sources_ccsn}

\begin{figure}
    \centering
    \includegraphics[width=0.5\linewidth]{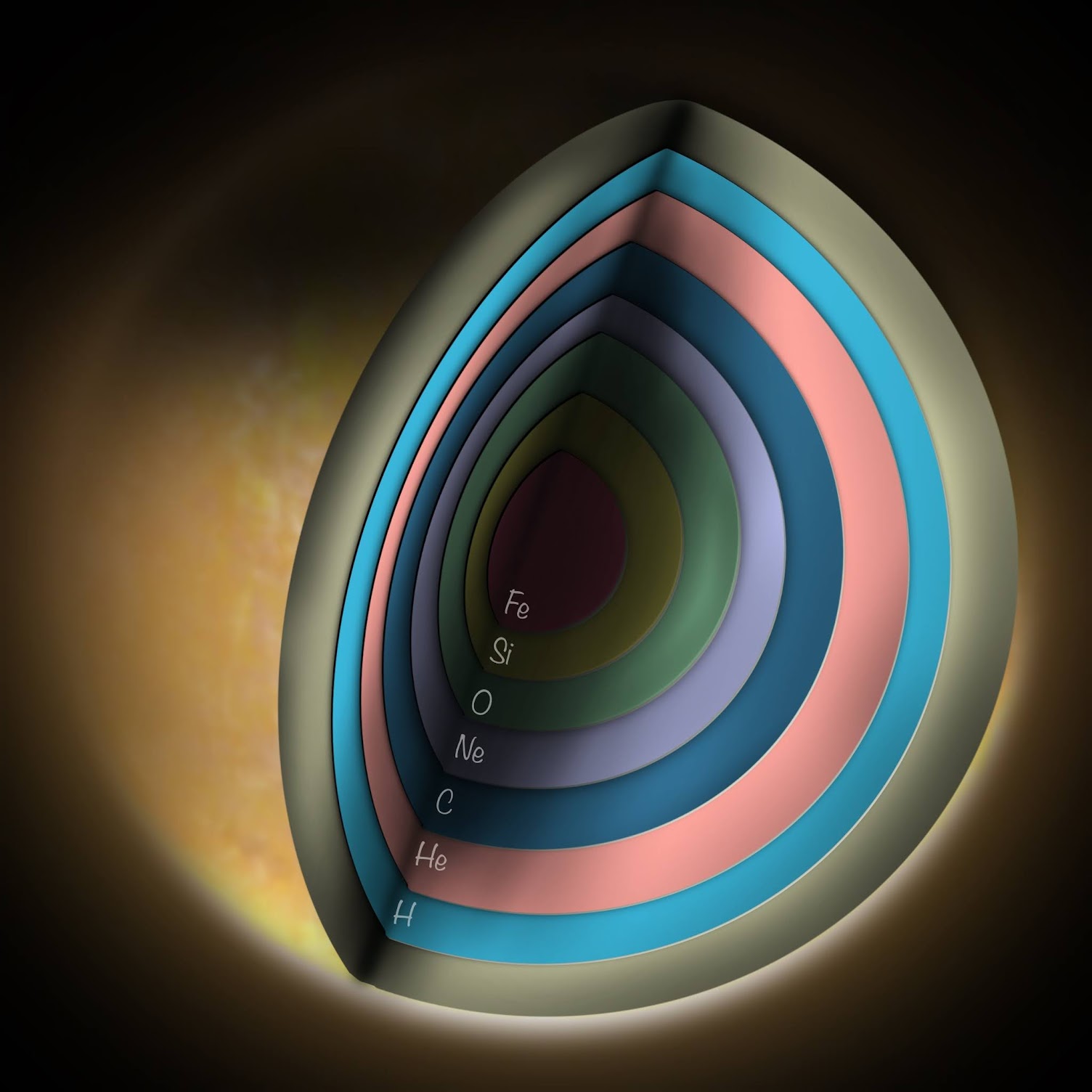}
    \caption{Advanced nuclear burning in massive stars. They lie in cocentric layers with lighter elements farther away from the core.}
    \label{fig:onion}
\end{figure}

The deaths of massive stars (greater than $\sim 8 M_{\odot}$) result  in a wide range of observable outcomes, depending on the mass of the progenitor star and its mass loss history, either from winds or binary interactions.  Five separate mechanisms have been proposed for massive stellar deaths.  The most common is the core-collapse supernova (CCSN). As the name suggests, core-collapse supernovae are initiated by the collapse of the cores of massive stars at the ends of their lives. 
The center of these stars (with mass greater than $\sim 9 M_{\odot}$) at the end of their lives are composed of iron, nickel and similar elements, the end products of stellar nucleosynthesis.  
Above this \emph{iron core} lie concentric layers of successively lighter elements, recapitulating the sequence of nuclear burning that occurred in the core (Figure \ref{fig:onion}).

Unlike prior burning stages, where the ash of one stage became the fuel for its successor, no additional nuclear energy can be released by further fusion in the iron core, thus nuclear energy production can no longer stave off the inexorable attraction of gravity.  
When the iron core grows too massive to be supported by electron degeneracy pressure, the core collapses. 
The collapse proceeds to densities in excess of the densities of nucleons in the nucleus of an atom (\textit{supernuclear} densities). 
The inner core becomes incompressible under these extremes, bounces, and, acting like a piston, launches a shock wave into the outer stellar core.  
This shock wave will ultimately propagate out of the iron core, through the overlying stellar layers (silicon, oxygen, ...) and disrupt the star in a supernova explosion. 

However, simulations show that the shock generally stalls in the outer core, losing energy to neutrino radiation and nuclear dissociation. After core bounce in CCSN, $\sim \! \! 10^{46}$ Joules of energy in the form of neutrinos and antineutrinos of all three flavors (electron, muon, and tau) are released from the newly formed proto-neutron star (PNS) at the center of the explosion, over a period of tens of seconds, at the staggering rate $\sim \! \! 10^{45}$ Watts. 
The kinetic energy observed in supernova explosions is $\sim \! \! 10^{44}$ Joules ($\equiv 1$ Bethe), 100 times smaller than the available energy in neutrinos. 
Past CCSN simulations \citep[beginning with][]{Wils85,BeWi85} demonstrated that energy emerging from the PNS in the form of neutrinos can be deposited behind the shock and potentially revive it. 
The composition of the ejecta in a CCSN comes from a variety of nucleosynthesis processes, with explosive silicon and oxygen burning happening in the layers above the iron core as the shock moves through them.
Deeper, material which has been exposed to intense neutrino heating is ejected, carrying a distinctive proton-rich composition dominated by $^{56}$Ni and other proton-rich species of nickel and iron.

Stars of mass just below the minimum mass to form an iron core can also undergo core collapse at the ends of their lives, starting from a core composed of oxygen and neon (O/Ne).  Shell burning can cause the most massive of these O/Ne cores to increase beyond the stable Chandrasekhar mass before the star loses its envelope \citep{JoHiNo13,JoHiNo14}, resulting in core contraction.
Competition between electron capture and the burning of oxygen and neon can lead to different outcomes.  
If electron capture wins (as in the iron core case), the core collapses, producing a collapsing electron-capture supernova (cECSN) \citep{MiNo87,ZhLeSu19}. As in the CCSN case, collapse in a cECSN proceeds to form a proto-neutron star, launch a supernova shockwave and emit prodigious quantities of neutrinos, but from there the explosion mechanisms diverge.  
The mantle overlying the core in these stars is quite low in density and small in mass, producing a sharp density gradient at the surface of the O/Ne core.
Since it is the ram pressure from this infalling mantle that stalls the supernova shock, in the cECSN case, the shock quickly revives. 
Despite the revitalization of the shock by the steep stellar density gradient, neutrino heating does occur behind the shock in cECSN, adding to the explosion energy and altering the composition of the ejecta, particularly the material that leaves the vicinity of the PNS at later times.  
The result is a unique sandwich of composition in cECSN, with outer layers due to explosive burning and inner layers dominated by $^{56}$Ni sandwiching a layer of neutron-rich nickel.  

For stars with significant rotation in the core at the onset of collapse, magento-hydrodynamics can result in tremendous magnetic field amplification, which fundamentally alters the explosion.  The resulting explosion exhibits jets, which can penetrate through the stellar envelope, and leave behind a highly magnetized neutron star, a magnetar...

Failures of the core-collapse mechanism are the result of the neutron star transforming into a black hole before sufficient energy has been provided by neutrino heating to unbind the star.  Even in this case, for the right conditions, an additional explosion mechanism, the collapsar mechanism, can potentially power a successful supernova.

Above the stellar masses where core collapse results in the formation of a black hole, an additional mechanism, the pair-production instability can lead to supernova explosions \citep{1967PhRvL..18..379B}.  For these very massive stars, the central pressure is provided by radiation; however, the production of electron/positron pairs can sharply reduce the radiation pressure. This leads to rapid contraction and the ignition of explosive oxygen burning, with the resulting thermonuclear runaway resulting in the explosive disruption of the final star.  The range of progenitor masses for which this mechanism can operate is strongly dependent on the metallicity of the star, but masses over $\sim 100 M_{\odot}$ are generally required \citep{2007Natur.450..390W}. In contrast, stars with helium cores below ~65~$M_{\odot}$ experience a lower-energy implosion, with pulsations that cause periodic contractions, nuclear burning (typically of oxygen (O) or silicon (Si)), and pulses leading to episodic mass-loss. This process, referred to as a Pulsational Pair-Instability Supernovae (PPISN) continues until the star stabilizes, evolving toward a massive core-collapse supernova \citep{2007Natur.450..390W,2017ApJ...836..244W}. Detecting and identifying the SNe of the first stars, many of which are expected to manifest as PISNe, is a key target of the JWST mission. The importance of PPISNe becomes more pronounced with stellar rotation effects. Rotating progenitor stars can undergo pair instability at significantly lower ZAMS masses and zero metallicity compared to non-rotating stars \citep{2012ApJ...748...42C,2012A&A...542A.113Y,2012ApJ...760..154C}. PPISNe can produce multiple transients from a single source. Each pulsation ejects several $M_{\odot}$ of material into the circumstellar environment, releasing thermal and recombination energy that powers the light curve. Collisions between successively ejected shells can result in violent interactions, generating shock-powered transients with peak rest-frame luminosities exceeding 10$^{44}$ erg/s \citep{2017ApJ...836..244W}. The final remnant’s explosive event may further propel ejecta that collide with earlier shells, creating long-duration light curves with multiple undulations \citep{2017Natur.551..210A,2024arXiv240902174A}.

\begin{figure}
      \centering
    \includegraphics[width=0.95\linewidth]{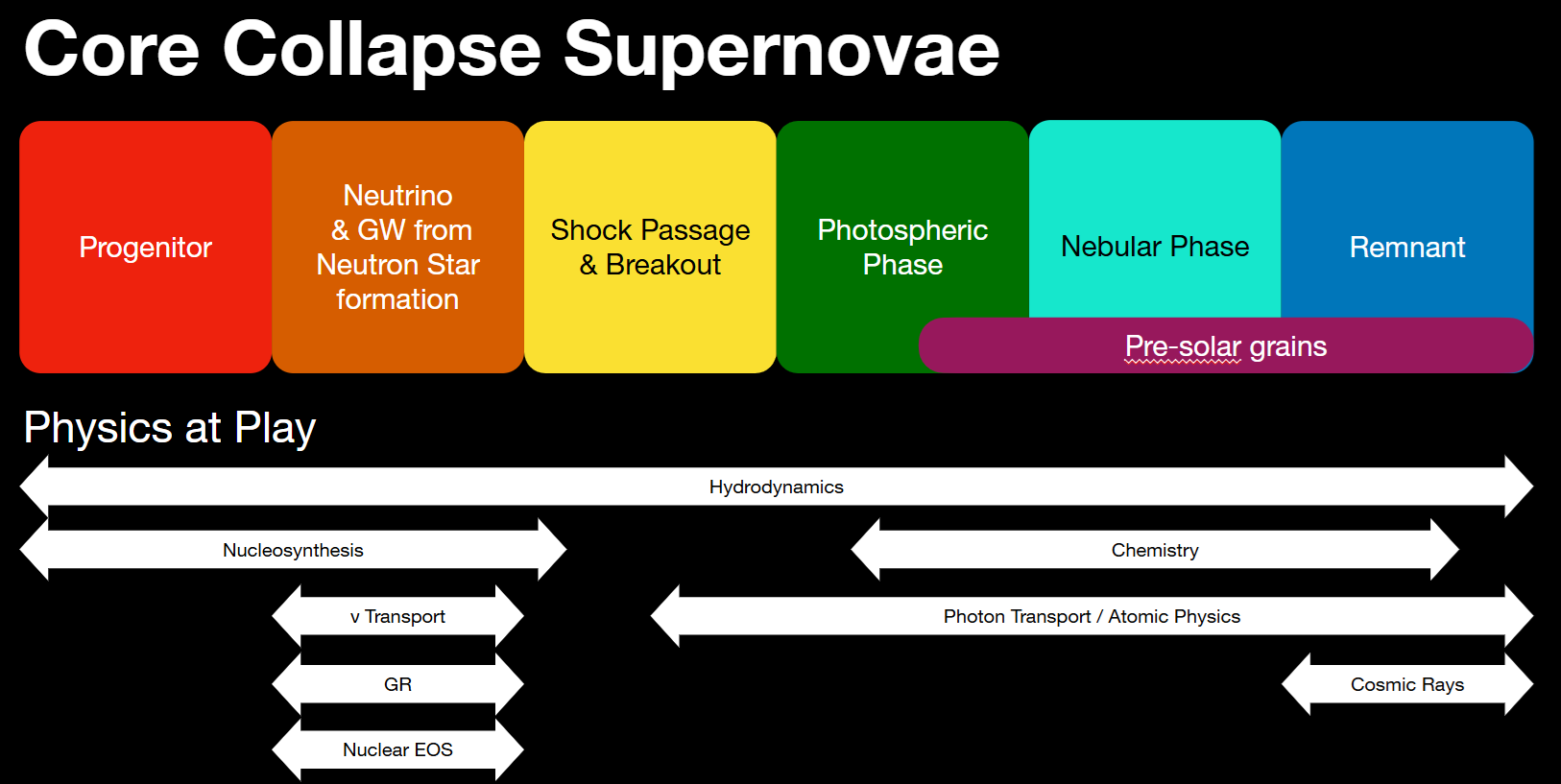}
    \caption{The temporal stages of relevance for CCSN. These are the stellar progenitor, collapse and explosion, the shock propagation and breakout, the photospheric phase, the nebular phase, and lastly as supernova remnants. Pre-solar grains are generated in the later phases and are studied here on Earth. }
    \label{fig:ccsn_chains}  
\end{figure}

Understanding core-collapse supernovae can be broken up into distinct phases. These can form the basis of organizing simulation chains. They, and the relevant key physics are each phase, are shown in Figure~\ref{fig:ccsn_chains}. These are discussed in more detail below. Many of these phases are or will soon be studied with transformational facilities. These include the upgraded neutrino and GW facilities for the core-collapse phase, ULTRASAT and UVEX for the shock breakout, Rubin and other telescopes for the photospheric phase, JWST for the nebular phase, and COSI for the remnant phase.

\begin{quoting}
    \noindent Finding: The study of core-collapse supernovae is the source class where investment in a strategic end-to-end effort is most justified. This is given the state of organization in this focus area from past investment, the tie to priority areas of all funding agencies, and the new or forthcoming facilities which will transform their study.
\end{quoting}


Here we go through the essential physics in each phase of core-collapse supernovae:

{\bf Phase I:}. Stellar evolution is one of the most difficult problems in astrophysics.  Magneto-hydrodynamics, nuclear physics, and radiation transport all play a role.  What makes stellar modeling so difficult is the timescales involved.  Even the shortest-lived massive stars live for over a few million years, yet the hydrodynamic timescales are much shorter.  At best, scientists can only model snapshots of time of this evolution using hydrodynamic codes~\citep{2022A&A...659A.193A}.  Instead, we rely on approximate solutions to model the stellar physics.  For example, mixing-length theory~\citep{2019ApJ...882...18A} is used to capture the effects of the convective instabilities.  This is a very specific form of a Reynolds Averaged Navier Stokes (RANS) models commonly used to study time-averaged fluid dynamics.  Similarly, a series of approximations are used for the radiation transport to approximate the energy transport including angle- and energy-averaging, simplifications in the incorporation of the opacities, etc.  Different approximations are used for stellar evolution and stellar winds.  Stellar observations, combining both stellar populations with pre-explosion progenitor observations, are ideal diagnostics of this phase.  Stellar models provide the structure of the star (critical for collapse/engine and blastwave phases) and the circumstellar material which affects transient emission and ejecta remnants.

{\bf Phase II:}. The collapse of the core down to a proto-neutron star and the tapping of the gravitational potential energy released to drive a supernova explosion is one of the most physics-rich problems in astrophysics.  Three primary engines exist that tap this potential energy:  the convective neutrino-driven engine~\citep{1994ApJ...435..339H}, the black hole or neutron star accretion disk engine~\citep{1993ApJ...405..273W}, and the magnetar engine.  The collapse and bounce of the core produce densities well beyond nuclear densities driving extreme physics such as quark production.  Most of the energy is converted to neutrinos and the supernova engine can be extremely sensitive to advances in neutrino physics including neutrino oscillations.  Detailed neutrino transport is required to tap this primary energy source.  Magnetohydrodynamics, nuclear reactions and general relativity also affect the engine.  Including this broad range of physics with sufficient resolution to capture the growth of turbulence and convection or accurately model a dynamo requires intense computing resources and these models require pushing the envelope of computing methods and algorithms.  Diagnostics of this engine include gravitational waves and neutrinos for nearby events.  These observations will not only probe the nature of the supernova engine but will also study the underlying dense nuclear matter and neutrino physics.

{\bf Phase III:}  After the launch of the explosion, we must follow the shock through the star.  As this shock propagates through the star, it drives nuclear burning.  These yields are observed in astrophysical transients, supernova remnants and stars (galactic chemical evolution).  Modeling this phase poses a number of computational issues.  Calculating the detailed yields requires following the detailed density/temperature evolution of the ejecta.  Eulerian codes struggle to track this ejecta and producing accurate evolutionary trajectories is challenging.  Accurate network calculations can also be numerically challenging.  By the time the shock reaches the edge of the star, nuclear burning is complete.

\begin{figure}
    \centering
    \includegraphics[width=0.5\linewidth]{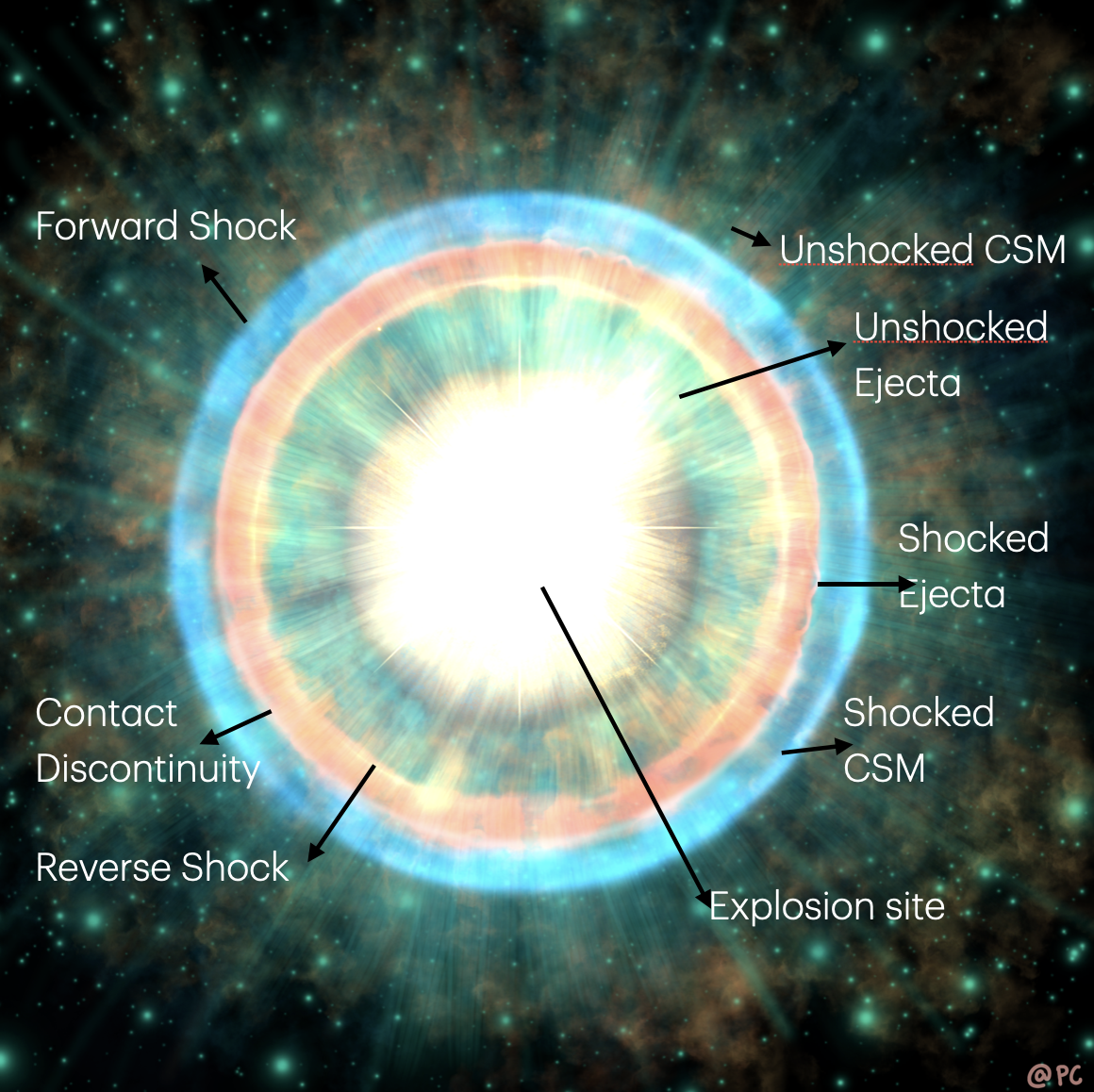}
    \caption{Post explosion supernova ejecta CSM interaction creates a forward shock moving in the CSM and a reverse shock in the ejecta, separated by contact discontinuity. The interaction results in variety of emission.}
    \label{fig:csm}
\end{figure}

{\bf Phases IV and V:}. Once the  ejecta breaks out of the stellar envelope, the radiation is no longer trapped in the ejecta.  The emission of this radiation powers both the shock breakout and subsequent supernova light-curve.  Radiation transport is the critical physics for these two phases.  Initially, the source of this emission is from the energy in the supernova ejecta, shock heated as the blastwave propagated through the star.  For type II supernovae, this energy source powers the light-curve through peak.  But for more compact supernovae (type Ib/c), adiabatic expansion cools the shock, removing it as a potential source for radiation.  Instead, these supernovae can be powered by the decay of radioactive elements, the same power source as type Ia (thermonuclear) supernovae.  But, for core-collapse supernovae, alternative power sources have been proposed.  Magnetar or fallback accretion can source energy into the expanding ejecta.  But a more likely energy source arises from shock interactions with the surrounding circumstellar medium (CSM), created by the winds from the progenitor mass loss.   By studying the radiation, one can estimate the progenitor mass-loss rate and hence the evolution history of the progenitor before the explosion, due to order of magnitude difference in the ejecta and wind speeds (Figure \ref{fig:csm}).  For this power source, radiation transport calculations alone are insufficient, radiation-hydrodynamics calculations are required.  In astrophysics, these studies are in their infancy, but by leveraging progress from the HEDP community, astrophysics can make huge advances in these models.  Since shock breakout and supernova light-curves will be the dominate observational constraints on supernovae, these calculations are critical.

At early times, the radiation and matter are tightly coupled and equilibrium solutions (For the radiation, matter, and atomic level states) are appropriate.  Technically, as soon as the radiation is no longer trapped, equilibrium solutions do not strictly match the observations.  As the matter expands, this coupling becomes weaker and weaker and out-of-equilibrium solutions must be understood.  Supernova modelers have developed cutting edge techniques to model this out-of-equilibrium physics~\citep[e.g.][]{2011MNRAS.410.1739D} and these techniques may be useful in HEDP applications like inertial confinement fusion.

{\bf Phase VI:}  For supernova remnants, both shocked as well as unshocked regions participate in producing radiation in the ejecta-CSM interaction. The main observational consequence  is the  production of radio and X-ray emission \citep{chevalier82b, cf17}.
The charged particles at the shock front are accelerated, probably via diffusive Fermi shock acceleration \citep{bo78,bell78}.
These relativistic particles in the presence of magnetic fields (enhanced at the contact discontinuity via Rayleigh-Taylor instabilities~\citep{cb95}), produce synchrotron radio emission, which is sensitive to  density  parameter\citep{chevalier82b}. 
The radio emission is affected by either the external free-free absorption (FFA) process by the surrounding ionized wind, or by the internal synchrotron self absorption (SSA) by the same electrons responsible for the emission.  The dominant absorption mechanism depends upon the mass-loss, magnetic field in the shocked shells, shock velocity and the ejecta density. The broadband radio spectral energy distribution (SED) from optically thick to thin  regime provides  important clues to the physics.


\subsection{Magnetar Activity}
{\centering 
\textit{Contributors: Matthew G. Baring, Eric Burns, Michela Negro, Aaron C. Trigg, Zorawar Wadiasingh}}\\

\label{sec:sources_magnetars}

\begin{figure}[ht]
\begin{center}
    \includegraphics[width=\textwidth]{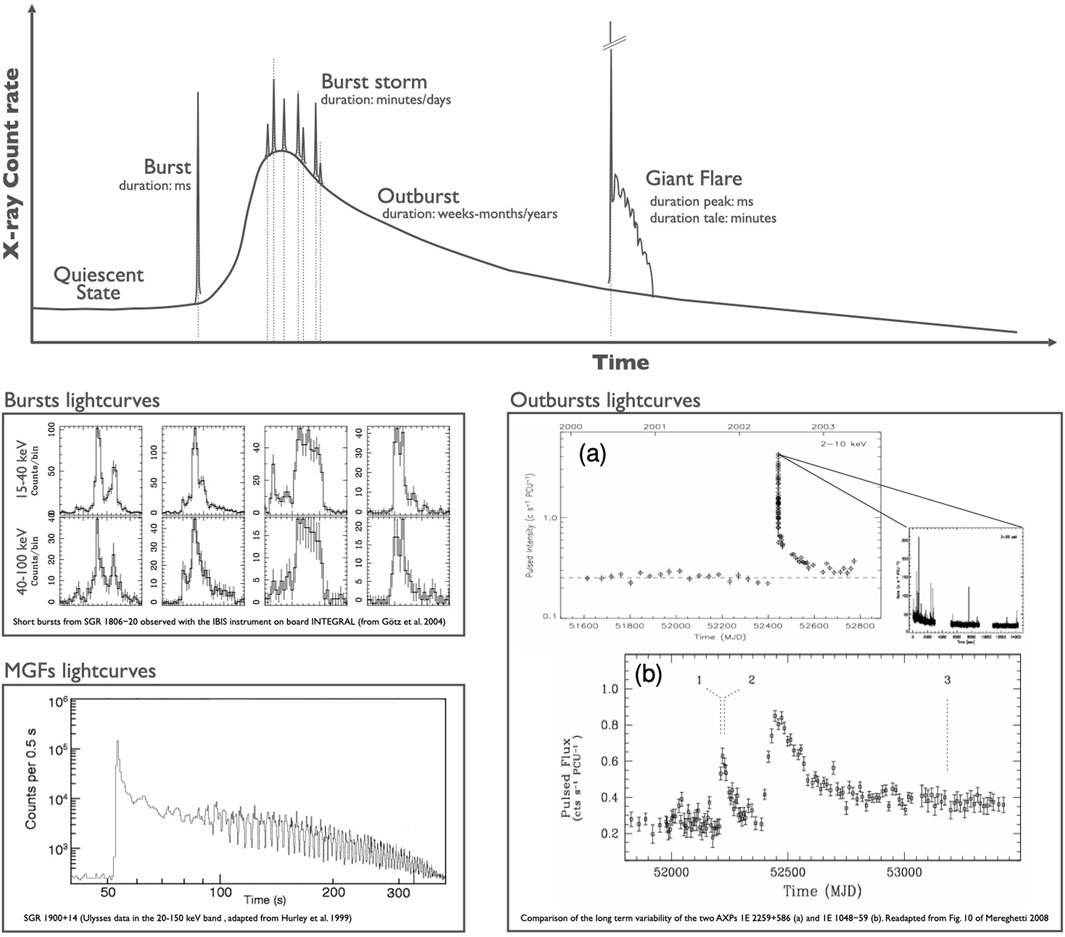} 
    \caption{A summary of the transient activity of magnetars, borrowed from \citet{negro2024role}}\label{fig:magnetar}
\end{center}
\end{figure}

Neutron stars (NSs) represent the densest form of stable matter in the cosmos. Born from the collapse of massive stars, these extreme objects are held up against gravity ($g_{\rm surface}\sim10^{13}\, \rm{m\,s^{-2}}$) by neutron degeneracy pressure in their core. While all NSs exhibit extreme densities and gravitational fields, a subset of young NSs known as magnetars \cite{1992herm.book.....M,Duncan_1992ApJ...392L...9D,Thompson_10.1093/mnras/275.2.255,1998Natur.393..235K} represent an even more extreme environment, characterized by extraordinarily strong magnetic fields $B>10^{14}\,\rm{G}$ that drive much of their activity.  In these fields, exotic quantum processes involving electrons and photons are believed to be active, and these are currently beyond the scope of testability in terrestrial laboratory experiments. A large exploration of the science possible with TDAMM studies of magnetars is available in \citet{negro2024role}, whose transient summary figure is shown in Figure~\ref{fig:magnetar}.

\begin{quoting}
    \noindent Comment: Magnetars are among the most extreme objects in the universe. Their properties and emission involve particularly rare processes. They present a unique but complex opportunity for the study of the physics of the cosmos.
\end{quoting}

Despite extensive study over the past 45 years, many questions about magnetars remain, starting with their formation. While it is clear that magnetars arise under specific conditions, their exact formation mechanisms are still uncertain including their configuration and distribution of birth fields. Proposed formation pathways include binary mergers of massive stars \citep{Schneider_2019Natur.574..211S}, white dwarf collisions or accretion-induced collapse \citep{Nomoto_1991ApJ...367L..19N,Usov_1992Natur.357..472U,Levan_2006MNRAS.368L...1L,Metzger_2008MNRAS.385.1455M}, and core-collapse supernovae of isolated massive stars \citep{Duncan_1992ApJ...392L...9D,Beniamini_10.1093/mnras/stz1391,Burns_2021ApJ...907L..28B}. However, significant questions remain about how initial environmental factors, such as the dynamics of the supernova, influence their birth. More intriguing questions are the apparent absence of classical magnetars in binary systems\footnote{Pulsed ultra-luminous X-ray in nearby galaxies may be accreting magnetars in binaries.} or the evolution and aging of magnetar fields.

Currently, there are $\sim$30 magnetars identified in the Milky Way \citep{Olausen_2014ApJS..212....6O}, with the majority generally confined to the Galactic plane. These objects are known for their higher energy electromagnetic emissions: during quiescence they can be detected in X-rays and soft gamma-rays, though the are more regularly observed during their periods of high energy activity in the hard X-ray and soft gamma-ray regimes. Magnetars are also associated with some of the most energetic phenomena in the cosmos: the magnetar giant flare \citep[MGF;][]{Kaspi_2017ARA&A..55..261K}. MGFs offer insights into understanding the energy release mechanisms in magnetars, particularly the processes of magnetic reconnection and crustal dynamics and failure due to the strong magnetic field \citep{Thompson_10.1093/mnras/275.2.255,Thompson_1996ApJ...473..322T}. The decaying tails of MGFs, modulated by the NS's rotational period, have been seen to exhibit quasi-periodic oscillations (QPOs), which are thought to result from crustal shear modes, further hint at seismic activity in the magnetar crust and offers a potential probe into the NS equation of state \citep{watts_2007neutron,Huppenkothen_2014ApJ...787..128H,roberts_2021rapid,Castro-Tirado_2021Natur.600..621C}.

While magnetars are primarily studied for their high-energy X-ray and gamma-ray emission, they also produce fainter signals in the optical and near-infrared (NIR) bands \citep{Mereghetti_2015SSRv..191..315M}, providing further valuable information about the processes occurring in the magnetosphere and near the NS's surface. Infrared and optical emissions have been detected for a third of the known population of magnetars, although they are extremely faint, with NIR magnitudes in the range of 19--20 \citep{Mereghetti_2015SSRv..191..315M}. For some magnetars, the observed NIR emission is pulsed, matching the rotational period of the NS \citep{Durant_2005ApJ...627..376D,Camilo_2007,Testa_2008A&A...482..607T}. The origin of this pulsed optical emission is a mystery, but must arise from a magnetospheric incoherent process.

Beyond these persistent emissions, MGFs have been predicted to produce short-lived, luminous optical transients referred to as \textit{nova brevis} \citep{Cehula_10.1093/mnras/stae358,patel2025r}. Such baryon-loaded outflows from MGFs are a generic feature expected in nearly all models involving triggers in the crust (i.e. quake or crust failure) of the magnetar. These events are powered by the radioactive decay of heavy elements produced via \textit{r-process} nucleosynthesis in the material ejected from the magnetar's crust during a flare. The \textit{nova brevis} is expected to shine at luminosities around $10^{39}\,\rm{erg\,s^{-1}}$ and last for just a few minutes, appearing as a scaled-down version of the kilonovae observed from NS mergers. These transients have yet to be definitively observed. However, a recent paper claims the detection of the nuclear gamma-ray signature, providing direct evidence for only the second known r-process site \citep{patel2025direct}. Future observations with better spectroscopic information would provide insights into the mechanisms of MGFs and nuclear physics (Section\,\ref{sec:disciplines_nuclear}), specifically the creation of heavy elements in extreme environments. 

Occasionally, magnetars have also been observed in the radio band, are also the leading candidate for the origin of some FRBs (Section\,\ref{sec:sources_frbs}). Unlike typical pulsars, which emit steady radio signals, magnetar radio emissions are sporadic and often tied to more active periods. Some magnetars have been observed to emit highly variable radio pulses during these active phases \citep{Mereghetti_2015SSRv..191..315M}. These signals are distinct from radio pulsar emission in several ways:

\begin{itemize}
    \item They are strongly polarized and spiky \citep{Camilo_2007b,Camilo_2008ApJ...679..681C}
    \item The exhibit flat or inverted spectra, meaning that the intensity of the radio waves remains high or increases with frequency \citep{Camilo_2007b,Camilo_2008ApJ...679..681C}
    \item The vary rapidly, reflecting the dynamic changes in the magnetosphere where the emission originates \citep{Camilo_2008ApJ...679..681C}
\end{itemize}

These features suggest that magnetars radio waves arise from processes involving rapid rearrangement of magnetic fields or the crust, and local particle acceleration in their magnetospheres, rather than the stable rotationally-driven mechanisms typical of most NSs \citep{Mereghetti_2015SSRv..191..315M}.

MGFs are also promising candidates for gravitational wave (GW) emission due to their dynamic behavior and extreme physical conditions \citep{Macquet_2021,beniamini_2024extragalacticmagnetargiantflares}, either from baryon-loaded relativistic outflows or excitation of global oscillation modes. Non-axisymmetric vibrations, particularly in the form of f-modes, are considered likely GW emission mechanisms \citep{Duncan_1998}. QPOs observed in the spike or tails of MGFs may also contributes to GW signals \citep{Levin_10.1111/j.1365-2966.2011.19515.x} depending on which modes are excited. However, the fraction of flare energy converted into GWs is expected to be small, making detection challenging with current instruments \citep{Macquet_2021}.

The strongest GW emissions from magnetars are expected to arise from MGFs in the Milky Way or nearby galaxies, as sensitivity decreases significantly with distance. Magnetars may also contribute to the stochastic GW background if their flaring activities emit GWs consistently over cosmic time \citep{beniamini_2024extragalacticmagnetargiantflares}. Advancements in next-generation GW detectors will improve sensitivity at higher frequencies, which will be crucial for detecting these signals. Combined with multi-messenger observations, these efforts hold promise for probing magnetar interiors, testing NS models, and uncovering the physics of their magnetic fields.

The extreme conditions and energetic phenomena associated with magnetars creates ideal environments for neutrino production through various processes \citep{Thompson_10.1093/mnras/275.2.255,negro2024role}. During MGFs, the large amount of energy released can heat the crust and interior of the NS, triggering cooling processes that produce MeV neutrinos. These neutrinos can escape the dense magnetar environment, carrying away significant amounts of energy. The burst of neutrinos generated in this way is thought to reach energies of $\sim10^{38}\,\rm{erg}$ and may contribute to magnetar outburst cooling \citep{Coti_Zelati_10.1093/mnras/stx2679,Camero_10.1093/mnras/stt2432,Scholz_2014}.

Beyond the MeV neutrinos, MGFs may also produce high-energy neutrinos via mechanisms like proton-proton and photo-hadronic interactions. The production of these neutrinos is contingent on conditions such as particle acceleration and baryon loading in the magnetosphere during or shortly after flares. Detections of high-energy neutrinos in coincidence with MGFs or other magnetar activity would provide critical insights into the flaring mechanisms, particle acceleration, and dynamics of the magnetosphere in extreme field. However, it is likely current instruments lack the sensitivity to detect these signals fort all but the brightest and nearest events \citep{negro2024role,beniamini_2024extragalacticmagnetargiantflares}.

Lastly, several emerging areas of study relevant to TDAMM magnetar science are important besides MGFs. These include:

\begin{itemize}
    \item Fast radio bursts (\S\ref{sec:sources_frbs}) and magnetar glitches -- Glitches are rapid spin changes of the magnetar associated with poorly-understood superfluid components inside the neutron star. Strong X-ray timing anomalies of the rotation in the magnetar as revealed by NASA NICER's unique long-term and frequent monitoring capabilities (also aided by NuSTAR's timing ability) is vital to uncovering such phenomena. More strikingly, these glitches have been recently been found to be contemporaneous with fast radio burst-like emission and soft gamma-ray activity \citep[e.g.,][]{2023NatAs...7..339Y,2024Natur.626..500H}.  
    \item Magnetar ``outbursts", which are enhanced and transient states of activity of soft and hard X-ray activity. The unknown rate and physics of such outbursts is relevant to understanding the crust and field evolution of magnetars, and formation channels as the potential lifetime of such outbursts is short. 
    \item Detailed observational study and theoretical models of magnetar bursts fireballs, some of which seem associated with fast radio bursts. A major open question is why only a small fraction of magnetar X-ray and soft gamma-ray bursts result in a fast radio burst. 
    \item Solid state physics of the magnetar crust, how it relates to magnetar activity, and how magnetar observations can inform the nuclear+condensed matter physics of the crust. For example, transient plastic flows of the crust have been inferred in X-ray studies by NASA's NICER \citep{2022ApJ...924L..27Y}.
    \item Studies of emerging class of long-period radio transients, a subset of which are likely aged magnetars \citep{2022NatAs...6..828C,2022Natur.601..526H,2023Natur.619..487H,2023MNRAS.520.1872B,2024MNRAS.533.2133C,2024NatAs...8.1159C,2024arXiv241116606W,lee2025emission,doi:10.1126/sciadv.adp6351}. This new sources couple to the long-standing issue of neutron star and magnetar formation channels, and radio emission mechanism or QED plasmas in magnetars.
    \end{itemize}

\paragraph{Physics, Calculations, and Simulations Needed}

Understanding magnetars and related phenomena requires the application of radiation hydrodynamics (RHD), Monte Carlo methods and plasma physics. Models must extend beyond thermal evolution and account for boundary conditions, field evolution, and the cooling process that influences the magnetar's lifecycle. In the case of FRBs, ``far away" shock models seem disfavored based on scintillation studies of repeating FRBs and polarization behavior in some FRBs. Thus it is likely the FRB phenomenon is magnetospheric, but the issue is still open.

Condensed matter or solid state physics of the magnetar crust needs theory development and support, particularly related to plastic flows and crust physics of quakes.

Simulations are challenging due to unknown initial and boundary conditions, and a lack of data on sources, making it difficult to apply theoretical models effectively.

\paragraph{Diagnostics and Observables}

More galactic and nearby FRBs would be valuable for gathering multiwavelength counterparts, scaling observations, and determining efficiency of coherent radio wave production. X-ray energies from these sources are typically orders of magnitude higher, providing constraints on the sources that can be observed. It is important to distinguish between concurrent and non-concurrent observations to build a clearer understanding of these events.

\paragraph{Existing and Future Facilities}
For the QED studies the needs are outlined in Section~\ref{sec:questions_photonSplitting}, being sensitive, phase-resolved, spectropolarimetric studies over the 0.1 to few MeV band. For transient studies, there is a need for more MeV missions that can handle polarization measurements. COSI, while not designed for GRBs, and potential missions like Polar 2 may be useful, though LEAP would have been the most effective option. Enhanced gamma-ray monitors could probe the nucleosynthesis of \textit{nova brevis}. Total monitoring of the X-ray sky would provide complete coverage of flares, discovery of additional magnetars, and, with sufficient sensitivity, recovery of tails from extragalactic events.

We need automated identification of MGFs, both Galactic and extragalactic. With zero latency commanding, this could allow recovery of the X-ray tail by NASA's Swift-XRT and NICER. The greater sensitivity of AXIS allows a longer delay, but this would still need a far faster response than the AXIS requirements. The \textit{nova brevis} requires rapid UV or optical observations, which could be performed with Swift, UVEX or ground-based optical telescopes.


\paragraph{Magnetar Emission Across the Spectrum}

Long period radio transients, FRBs and other radio magnetars are an active area of research. Efforts in the radio community is particularly focused on FRBs and long period radio transients.

Optical magnetar emission and pulsating IR emission have been observed, with JWST detecting some clear signals, although not conclusively tied to other spectral components. Magnetars are generally radio-quiet, and IR emission may be close to the surface if the emission process is coupled to the local plasma frequency. The relationship between high plasma frequency and thermal X-ray emission remains an area of ongoing study.


\paragraph{Strong-Field Physics and Laboratory Studies}
The physics of strong magnetic fields and pair creation, particularly in the context of magnetars, is still not well understood. There is interest in connecting this area of study to the NSF and DOE initiatives which focus on laser plasma experiments (e.g., NSF funded multi-PW laser facilities like ZEUS at the University of Michigan and NSF OPAL, which is being designed at the Laboratory for Laser Energetics \citep{mp3report.2022.arxiv}). While pair creation might be observable in a lab setting, experiments to study photon splitting and Landau quantization require supercritical magnetic fields that are unlikely to be achievable in a laboratory frame. Therefore magnetar observations are the best opportunities for observing this.

Disentangling the effects of plasma elements in the magnetosphere and birefringence, especially in lab experiments, is another complex challenge that makes it difficult to isolate specific physical effects.


\paragraph{Challenges and Uncertainties}
There are still significant uncertainties in these studies, including whether there is commonality between magnetars. That is, how does path-dependent physics and evolution influence magnetar emission? Why do (observationally) some magnetars have different ``personalities"? More examples are needed to quantify this properly. Additionally, the radio astronomy community faces issues with data sharing, making it difficult to validate findings for transients. Much controversy surrounded even the first FRB discoveries, with their astrophysical origin not accepted by the community over a decade after discovery. Improved transparency and communication between communities are necessary, as well as data sharing mandates for radio facilities and greater rigor or redundancy.


\subsection{Neutron Star Mergers}
{\centering 
\textit{Contributors: Floor S. Broekgaarden, Eric Burns, Gwendolyn R. Galleher, Jamie A. Kennea, Gavin P. Lamb, Stuart Loch, Elias R. Most, Matthew R. Mumpower, Eliza Neights, David Radice, Lauren Rhodes, Peter Shawhan, Zorawar Wadiasingh, Gaurav Waratkar}}\\

\label{sec:sources_mergers}




Neutron star (NS) mergers will be the most commonly detected multi-messenger transients. Here, we use the term NS merger with regards to both binary neutron star and neutron star-black hole mergers. The joint detection of a binary neutron star merger in 2017 was mentioned in the long-term planning document of many of the fields considered in this white paper. These events can be detected before merger through high-frequency GW observation. Following the merger, relativistic bipolar outflows are launched which ultimately produce the prompt gamma-ray burst signature seen with wide-field monitors such as those on NASA's Fermi and Swift missions. These jets continue to propagate outwards, building up a shock front within the surrounding material. These external shocks cool via synchrotron emission which emit across the electromagnetic spectrum which have been observed with major NSF, DOE, and NASA facilities (e.g., Fermi, Veritas, Swift, Gemini, Karl G. Jansky Very Large Array, ALMA, etc). Separately, when disruption of the neutron star(s) occur(s), the ejected neutron star matter has the ideal composition to form the heaviest elements on the period table via the r-process. This merger ejecta is initially opaque, leading to a thermal transient referred to as a kilonova. Over a period of days to weeks the kilonova is expected to change color as a result of the different r-process elements present in the ejecta, before fading below detection limits. With the r-process producing hundreds of exotic, unstable nuclei that cannot all be produced by terrestrial means, neutron star merger observations presents the unique opportunity to constrain the properties of heavy nuclei that could otherwise be inaccessible (e.g. the fission properties of neutron-rich actinides or the nature of the closed neutron shell at neutron number N=126).

\textbf{Observation:} The characterization of these events takes full use of most astronomical capabilities, as summarized in Table~\ref{tab:diagnostics_neutronStarMergers} and described in the following text. In order to perform these detailed studies, the events must first be discovered and then well-localized. Discovery can occur in many ways. Short duration GRBs, classically associated with a merger origin, have been observed $\sim$1,000 times, out to a redshift of $z \sim 2.7$ \citep{2024GCN.38097....1M}. These are discovered out by high-energy monitor facilities. Some facilities, such as INTEGRAL and Swift, provide precise positions with discovery \citep{winkler2003, Gehrels2004}. Other facilities, such as Fermi \citep{atwood2009, meegan2009}, provide poor localizations, which require substantial follow-up efforts to identify their precise position.

\begin{table}
\centering
\begin{tabular}{|r|c|c|l|}
\hline
Phase	&	Diagnostic	&	Observation	&	Inference	\\
\hline
Pre-Merger	&	Inspiral	&	Radio 	&	Mass distribution, merger time distribution	\\
 $ $	    &		&	\& X-rays (Galactic Binaries?)*	&	\\
 $ $        &		         &	Gravitational Waves	&	Masses, spins, tides, merger time, \\
  $ $        &		         &	 &	luminosity distance$+$inclination, sky location	\\
Jets	    &	Prompt GRB	 &	Light curves	&	Emission time, energetics, 	\\
 $ $        &		         &	Spectrum	 &	Energetics, electron acceleration	\\
 $ $        &		         &	Polarization	&	Magnetic field strength, jet structure	\\
 $ $        &	GRB Afterglow	&	Light curves	 &	Jet structure, circumburst density	\\
  $ $       &		            &	Spectra &	Jet energetics, microphysics, particle acceleration	\\ 
  $ $       &		            &	Polarization &	Geometry	\\
  $ $       &		            &	VLBI	&	Geometry, position	\\
Mass Ejection	&	Shock Breakout	&	Light curves, spectra	&	Jet launch time	\\
  $ $       &	Kilonova	&	Light curves	&	Ejecta composition, mass and velocity, position	\\
  $ $       &               &	Spectra	&	Temperature	\\
  $ $       &               &	Nuclear lines	&	Isotopic abundances, composition	\\
Ejecta Propagation	&	Nebular Phase	&	Spectra	&	Elemental abundances, composition	\\
 $ $        &	Kilonova Afterglow	&	Light curves	&	Circumburst density	\\
 $ $        &	Kilonova Remnant	&	Nuclear lines*	&	Full isotopic abundance measurement	\\
 Combined   & Inverse Compton & Broadband evolution & Microphysical parameters of the emitting regions\\
 \hline
\end{tabular}
\caption{
     The list of diagnostics and observables of key interest to study neutron star mergers. *denotes observations which can be made non-concurrently, i.e., those which do not require coordinated observations of individual events.}
     \label{tab:diagnostics_neutronStarMergers}
\end{table}

Gravitational wave observations currently have a more limited horizon distance and have identified two NS merger events within a few hundred Mpc since 2017 \citep{GW170817, Abac2024}, however, next generation facilities will detect these events to distances that are comparable to GRB observatories at significantly larger distances \citep[$\sim10$s\,Gpc ][]{Hild2010}. GW detections typically have enormous localization regions. This requires enormous follow-up campaigns to identify their precise position.

In principle, these events can be identified across the EM spectrum via their afterglow or kilonova signatures. While some gamma-ray bursts have been discovered in wide-field optical surveys \citep[e.g., via the NSF-funded telescope Zwicky Transient Facility][]{andreoni2021, Ho2022}, none have been associated with a merger origin.
Merger origin transients like kilonova have never been independently found via current ultraviolet, optical, and infrared facilities, which are capable of detecting these transients within $\sim 100$\,Mpc -- comparable to the current gravitational wave detection horizon for NS merger systems.
This will change with the NSF and DOE Vera C. Rubin follow-up observations, which can recover kilonova transient events to $\sim$1~Gpc.


As gravitational wave interferometers improve in sensitivity, a greater fraction of electromagnetic counterparts will be in the form of GRBs and their afterglows. This is due to the increased sensitivity for joint GW-GRB searches, the larger typical distance to GW detected events (making a positive identification of an associated kilonova increasingly difficult as these transients fall below the threshold for ground-based spectroscopy), as well as an observational bias in favor of identifying face-on GW events at larger distances.


In order to extract the most science from future observations of NS mergers, the community needs to work in better collaboration to coordinate obtaining the best possible data sets. Steps have already been made to enable such collaboration with the advent of \textit{Treasure Map}\footnote{https://treasuremap.space} \citep{2020ApJ...894..127W}. Treasure Map is a web interface that allows different research groups to report the areas of GW localization regions that they have observed in order to efficiently cover the whole region and increase the probability of finding an optical counterpart.

Looking forward, the NS merger community can build on this work to further improve the accessibility of data and information. We propose that this can be done in a number of ways. The first aspect derives from the data received from LIGO/Virgo. In future runs when the expected NS merger rate is much higher, the follow-up community would like as much information as possible such as the chirp mass and rapid parameter estimations. Having this information in real time will enable real-time follow up strategy adjustments making a counterpart discovery more likely because the community can adjust their expectations as to what sort of galaxy is the host or what the transient itself will look like.

Building on the premise of \textit{Treasure Map} for better communication across the GW follow-up community, we recommend/ask that new infrastructure is developed in two ways. The first is that a platform is developed for access to public data streams from facilities such as ATLAS \citep{Tonry2018} and Fermi \citep{atwood2009, meegan2009} to determine whether such facilities observed any of the localisation region and if so what limits/ candidates were identified. The second is the ability to coordinate observations across multiple facilities via collaborative telescope proposals. In Europe, the Opticon Radionet Pilot\footnote{https://www.orp-h2020.eu/} was created to provide access to both radio and optical facilities. Such efforts can also be performed through collaboration-based follow-up efforts such as the \textit{Gravity Collective} where as a collaboration time is acquired during open time calls by different members of the collaboration spanning a wide range of facilities. 

Through the aforementioned suggestions, the NS merger community will be able to get closer to achieving our science goals. It has been clear from the LIGO/Virgo observing runs so far that there are fewer events than expected. Therefore, moving forward, to extract the best science possible better collaboration and accessibility is crucial.

\begin{figure}
    \centering
    \includegraphics[width=\textwidth]{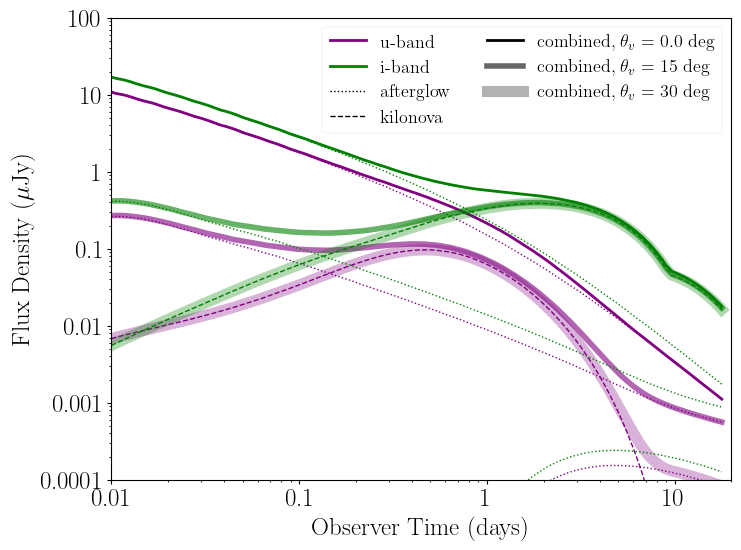}
    \caption{The afterglow of a structured jet, including a cocoon component, and a two-component kilonova. Solid lines indicate the combined lightcurves at $u$- and $i$-bands for observers at 0.0, 15.0, and 30.0 degrees from the jet central axis. No bright GRB is expected for the two off-axis cases (at 15 and 30 degrees). The models are plotted using typical merger parameters for a source at a redshift, $z = 0.05$. Models are from \texttt{Redback} \citep{2024MNRAS.531.1203S}.}
    \label{fig:merger_lightcurves}
\end{figure}

\textbf{Future observatories:}
Unlocking this science requires a deep understanding of the physics occurring in these events. The vast range of diagnostics makes this possible. This relies on the long-term investment by the NSF into LIGO and the ground-based optical, infrared, and radio facilities. Follow-up observations for events without a GRB counterpart will require the breathtaking sensitivity of the Rubin Observatory, enabled by the joint investment by both the NSF and DOE. Full characterization and isolation of specific emission processes requires the power of the full NASA fleet, including the investments in StarBurst, COSI, ULTRASat, and UVEX. Because of the overlapping temporal and spectral range of GRB afterglow and the kilonova signature, these must be jointly modeled, as shown in Figure~\ref{fig:merger_lightcurves}. Once Swift ends it will be impossible to characterize kilonova from face-on events, which will be the majority of observed kilonova. 

NS mergers are unique in the breadth of fields, not just astrophysics, in which they impact. This makes observations of them vital to provide key measurements for several questions of interest to the fields represented in this white paper. 
Here, we highlight three such examples: 

\begin{enumerate}
    
\item The extreme densities which exist within NSs. Currently, the NS equation of state is relatively unconstrained. The equation of state of neutron stars (Section~\ref{sec:questions_extremeMatter}) will be probed with a strong radius dependence on when neutron stars are disrupted in neutron star-black hole mergers and via an asymptotic limit for the maximum mass of a neutron star. This corresponds to the highest rung of the cosmic density ladder proposed in the Long Range Plan for Nuclear Science \citep{aidala2023new}. 

\item NS mergers detected via gravitational wave facilities can be used to perform precision cosmology. The gravitational wave detection gives a measurement of the luminosity distance and, upon the detection of an electromagnetic counterpart, one can identify the host galaxy and make an independent redshift measurement. Thus, these events will be key for distance determination and thus be utilized as a measurement of the expansion rate of the universe (Section~\ref{sec:questions_cosmology}). With large enough sample sizes, they will ultimately provide the a precise and independent measurement on the shape of the universe, the equation of state of dark energy, and questions on cosmology which are not yet well founded. 

\item NS mergers are clearly an important r-process site but it remains unclear whether they are the dominant site of heavy element production in the Universe \citep{latimer1974,eichler1989,GW170817-MMAD}. With only a handful of spectroscopic kilonova studies made so far, more events and detections are needed to understand the abundance of r-process elements synthesized within NS merger ejecta, the degree of mixing for these heavy elements within their host galaxy, and the range of the delay time distribution for NS mergers. All of which are required to identify if NS mergers alone are sufficient or if additional for r-process sites are needed to explain observations (see Section~\ref{sec:questions_elements}).

\item Finally, NS mergers are a unique opportunity to study neutral plasma out of thermal dynamic equilibrium (Section~\ref{sec:out_of_equilibrium}). Weeks after merger the ejecta slows and individual atomic (and possibly molecular) lines can be discriminated through careful observations with extremely sensitive facilities like JWST. However, this plasma has left LTE, and we require new modeling to make use of these observational capabilities.
\end{enumerate}

In order to understand NS mergers observations across the electromagnetic spectrum are required in addition to those by gravitational waves. Identifying mergers in gravitational waves and finding their host galaxies across a broad redshift range informs on the formation channels that allows the merger to occur in the first place. Once the merger has taken place, the fastest out-flowing component is from a highly relativistic jet \citep{2018Natur.561..355M}. 
The jet observed is the source of the GRB component, and therefore observed from the earliest times, if viewed on axis and appears later, with no prompt emission, for far off-axis systems -- except that GRB 170817A accompanied the off-axis GW170817, however, this GRB is unusually faint and difficult to reconcile with a jet origin under simple assumptions \citep[e.g.][]{2018MNRAS.478..733L}, but see \cite{2019MNRAS.487.4884I} for the spectral modifications required to explain GRB 170817A as being an off-axis viewed GRB. 
It is unknown what the lateral structure of the jet is because, for on-axis systems, the core of the jet dominates the observed flux over any less energetic wider component. To date, there is only a single off-axis event (GW 170817) that has allowed for the lateral jet structure to be studied in detail, and with only a single event, it is impossible to determine whether GW170817 is indicative/typical of the NS merger population. Understanding the jet structure is needed to measure the merger GRB luminosity function and the rates of NS mergers \citep{2023ApJ...954...92H}, additionally, knowing the jet structure to a higher degree of confidence will break the GW distance-inclination degeneracy, and increase the precision of any cosmological measurements \citep{2021A&A...652A...1M}. From knowing the local rates, it will be possible to determine whether NS mergers can be the singular site of r-process nucleosynthesis within the local Universe or whether multiple sites are required. Despite knowing that NS mergers are a site of the r-process, it is not known whether all NS mergers produce all of the r-process elements or what effects the NS(s) progenitors have on the production abundance of these elements. 


\textbf{Simulations:} Currently, kilonova simulations are clustered in a narrow region of parameter space. Nuclear physics uncertainties are definitely underestimated in all models. As a result, it is dangerous to marginalize over these models, because the resulting uncertainties are not representative or reasonable. More support for computation is required because more simulations (i.e., a larger grid of models as well as simulations from different groups), allows for a more complete picture of possible outcomes and therefore assists observers in deciding what to pursue. There are several other areas of uncertainty present in these models, the model inputs, and in the properties of the mergers (mass, spin, object type) which often occur. Kilonova models are only rarely tied to the output of merger simulations, typically assuming spherical ejecta. Observers will often combine two kilonova models (e.g. a red and blue) and match them to their data, ignoring the respective geometry of the components which can produce this (e.g., ``red'' ejecta would block any signal from ``blue ejecta''). The vast majority of observers are not even aware of these limitations. 

Simulations for jets are even less mature. Few merger simulations produce collimated outflows at the poles. Some teams have had success here, but almost none produce relativistic outflows. Modeling of accretion and the jets face similar difficulties as those in AGN and XRBs, with the added difficulty of extreme velocities, which are poorly modeled. The energy density in GRB jets is unmatched, and also not understood. A concerted effort in this area is required for major progress.

Currently there is a significant lack of publicly-available models examining the electromagnetic counterparts to NS mergers which observers could use to narrow down what they should be looking for in their survey data. The models that are available are often scalings of GW170817, or 
vary widely due to the different choice of physics to include, assumptions to make, and method of implementation \citep[see e.g.,][for the largest public, but still incomplete, collection of transient models, including those relevant to multimessenger studies]{2024MNRAS.531.1203S}. Even models with low confidence are useful as long as comprehensive uncertainties are provided. Furthermore, for any given astrophysical model used in producing a theoretical light curve, there are additional uncertainties from the nuclear physics data assumed. Even before considering such alternative nuclear physics data, simulations of NS mergers have large uncertainties, and thus independent models from different simulation / nucleosynthesis groups are needed.

In Table \ref{tab:diagnostics_neutronStarMergers} we provide an overview of the observations that can be made of NS mergers, what these observations study, and the subsequent physics that can be inferred. 

\begin{quoting}
    \noindent Finding: The breadth of science possible with TDAMM studies of neutron star mergers is unmatched. However, unlocking these advances requires full understanding of these events. While we benefit from numerous observables which carry complementary diagnostics, the multiphysics problems are immense. Only a strategic and dedicated end-to-end effort in this area is likely to succeed, which must include astrophysicists from observers to theorists as well as scientists from several disciplines.
\end{quoting}

Fortunately, it is possible to separate the study on NS mergers into distinct regions based on timescale and/or wavelength range. The detections of gravitational wave emission with LIGO/Virgo alone, without any electromagnetic counterparts, can be used to infer the properties of compact object binary systems as a function of distance as well as their masses, spins, and other system parameters. 
As the LIGO/Virgo interferometers continue to be upgraded, such measurements will be made over greater distances allowing for studies at the population level. 
The NS merger rate as a function of distance/redshift is particularly important in understanding the formation pathways of such systems.
The merger rate as a function of redshift will provide answers to the possibility of multiple evolutionary pathways that can result in a significant fraction of the total NS merger population. 

When the gravitational wave detections are combined with detections of prompt GRB emission, or deep upper limits, it becomes possible to measure the GRB luminosity function, and infer the presence of a long lived NS remnant through extended emission. The time delay between the inspiral and the prompt emission tells us about the collapse timescale of the central compact object/remnant as well as the jet launching timescale. After the prompt emission has faded and the afterglow becomes visible, multi-frequency, high cadence observations are necessary to infer the jet micro- and macrophysics including the kinetic energy and circumburst environment. For particularly bright events and nearby, very long baseline radio interferometry can be used as an independent measure of the jet size, velocity and viewing angle. Afterglow monitoring is also vital in order to subtract it from the optical/UV/IR counterpart to then search for a kilonova. Once the kilonova signature is isolated, spectroscopic observations can isolate the telltale features of r-process elements within the merger ejecta.


Successful interpretation of the observations will be greatly aided by the forthcoming measurements on exotic nuclei at the DOE’s Facility for Rare Isotope Beams (FRIB), directed studies for atomic spectroscopy for line interpretation, handling of non-local thermodynamic equilibrium, and improved modeling of matter effects in gravitational wave inspiral. It will consistently be informed through iterative enhancements of the high-energy laser facilities, to the ultimate goal of generation of a pair plasma. All of these advancements will need to be properly modeled and integrated via simulation of the inspiral through merger and mass ejection, then carrying these results through nuclear reaction networks and propagation to kilonova emission. \textbf{While the current piecewise approach can provide important insights, consistent physics across each stage with end-to-end simulations is ultimately needed to pin down predictions, and is only possible with a large-scale strategic effort.}



\subsection{Novae} 
{\centering 
\textit{Contributors: Elias Aydi, Catherine Deibel, Phong Dang, Falk Herwig, Rebekah Hounsell, Jamie A. Kennea, Hendrik Schatz, John A. Tomsick}}\\

\label{sec:sources_novae}

Accretion onto a white dwarf in a binary system leads to various time-domain nuclear astrophysics outcomes depending on the accretion rate. When the accretion is high enough to allow for stable hydrogen-burning white dwarfs could accumulate enough mass to reach the Chandrasekhar mass. This is the classical single-degenerate path to thermonuclear explosion. However, recent advances in stellar evolution simulations have demonstrated that the mass accumulation rate may be limited or even negative in this scenario due to the repeated occurrence of He-shell flash on the white dwarf surface triggering mass ejections as part of a reverse common envelope due to the accumulation of He as H-burning ashes \citep{Denissenkov2017}. Interestingly, while exposing new challenges to this supernova type Ia progenitor scenario the simulations predict an entirely new convective-reactive nuclear astrophysics site for the intermediate neutron-capture process. The understanding of this novel nucleosynthesis is still rapidly evolving. For example, intriguingly, it has been suggested to be along with the r process revealed through the well-known $2.6 \mathrm{MeV}$ $\gamma$ ray emission from thallium-208 that could be detected by future MeV telescope missions \citep{Vassh2024}.

At lower accretion rates, novae errupt on much shorter time scales when hydrogen from a donor star accretes to a threshold value on the surface of the companion white dwarf (WD). Upon reaching this threshold a thermonuclear runaway reaction ensues resulting to the ejection of the accreted material from the WD in an explosive outburst. Nova are recurrent transient events, with the rate of recurrence dependent on the system and ranging from years to decades to centuries. Novae are bright, energetic events that occur more frequently than other transients. The paths forward to understanding many astronomical phenomena start with understanding novae:
\begin{itemize} 
    \item \textbf{Shocks:} For example - the unexpected discovery by the Fermi Large Area Telescope (LAT) of powerful gamma-ray emissions from novae (V407 Cyg, V1324 Sco, V959 Mon, V339 Del), indicating that shocks to play a bigger role in nova explosions than previously believed. We still do not fully understand the nature of the shocks and how they relate to the mass-ejection mechanisms of the envelope.

    Observations and better understanding of these nearby common transients may help test the hypothesis that SLSNe and other stellar explosions are also shock-powered. In stellar explosions, the breakout of shock radiation is the very first EM signal that reaches the observer, and it carries a wealth of information about the exploding star. What can we learn about the largely unconstrained population of SN progenitors from systematic observations of shock breakouts across the EM spectrum?

    To continue the progress in this field will require: (1) GeV--TeV gamma-ray monitoring capabilities with high sensitivity to enable time-resolved light curves and spectra; (2) high-cadence photometry and spectroscopy of Galactic novae to correlate with the shock-powered emission and better measure the mass-ejection history of the white dwarf; (3) sensitivity to lower energy $\sim$GeV-–TeV neutrinos to confirm a hadronic origin for the gamma-ray emission; (4) coordinated near-simultaneous multi-wavelength observations to understand the role of shocks, particularly at hard X-rays which are less attenuated than softer X-rays; (5) infrared spectroscopy to understand the role of dust production in novae and its connection (if any) to shocks.

    Such investigations will require significant coordination and development of infrastructure. 

    \item \textbf{Accretion:} Accretion disks play an important role in the liberation of accretion energy and are poorly understood in terms of viscosity and geometry. Disk instabilities, likely caused by temperature and ionisation changes that cause viscosity changes, show up as transient outbursts. In addition, smaller scale variability at many different time scales probes the physics of the accretion discs. Finally, accretion onto white dwarfs can lead to ejection of material in jets (as is the case for neutron stars and black holes) and even behavior as ”pulsar” as in the system AR Sco. Again, different binary systems can lead to accretion of different composition, allowing better constraining of the accretion physics that should depend on composition. Future progress on understanding accretion discs requires (1) wide field surveys (in optical/IR and UV) to find the outbursts and coordinated including (2) real time detection of transients and (3) fast multi-wavelength and multi-messenger follow-up including UV (which is not/hardly available); (4) long cadence and high-precision photometry to unravel the complex variability. The option for polarimetry on (new) instruments should be considered.

    \item \textbf{Nucleosynthesis:} The thermonuclear burning powering a nova explosion produces a range of isotopes and elements that are ejected into space. In so called Ne novae, where neon is dredged up from the white dwarf and mixed into the nuclear burning layers, elements up to Ca or beyond can be produced. The exact element range and nucleosynthesis endpoint remains to be determined. The contribution of Novae to galactic nucleosynthesis remains an open question. Novae are likely significant sources of lithium (see below) and the neutron-rich stable isotopes $^{13}$C and $^{15}$N. They may also contribute to the Galactic inventory of fluorine or phosphorous, and the Galactic abundance of long-lived radioactive $^{26}$Al that is observed by $\gamma$-ray observatories. Novae may also produce shorter lived radioactive isotopes such as $^{18}$F and $^{22}$Na that could be observable in Nova ejecta and are a prime target for the upcoming COSI mission. Unlike the abundances of stable elements, the abundance of radioactive isotopes can be directly inferred from the observed $\gamma$-ray flux, and is much less affected by the complex inhomogenous structure of the ejecta and its varying stages of ionization. As such they would provide a direct data point to validate models of thermonuclear burning. Novae may also be sources of pre-solar grains found in meteorites. In particular rare classes of SiC grains with isotopic enrichment of neutron rich stable isotopes of carbon, nitrogen, silicon, which are produced by the decay of radioactive proton rich isotopes produced in the explosion, have been associated with a nova origin. 
\end{itemize}

Because novae are nearby, can recur, and have a wealth of diagnostics, they are mini-laboratories for understanding many key processes in our universe, including those of direct interest for other transients. These objects are progenitors of thermonuclear supernovae, and the physics involved in their burning is similar. Indeed, modelers which work on one often work on the other. The accretion onto a white dwarf allows for a separate situation of accretion onto denser objects, giving a separate regime to study accretion codes. This is also one of the few scenarios where nuclear burning can occur without being obscured, which is a unique opportunity to study these processes.

\textbf{Forthcoming opportunity:} Nuclear experimentalists will measure the necessary information on all relevant reactions for nucleosynthesis in novae in the next few years (Long Range Plan for Nuclear Science). NASA will launch COSI in ~2027, which will directly detect the nuclear lines from these events. Together these allow determination of the full isotopic yield distribution, and understanding novae becomes a plasma problem. As high-energy lasers can now emulate conditions on the surface of a white dwarf (though not during accretion nor burning), a confluence of several fields results in a unique opportunity to study multiphysics in the cosmos.

\begin{quoting}
    \noindent Finding: The nearby and recurrent nature of novae combined with a multitude of signatures allows astrophysics to provide multiple diagnostics to understand these events. When combined with the eminent understanding of relevant nuclear physics after decades of investment and new capabilities of high-energy laser facilities, strategic and end-to-end study of these events will bring a revolution in understanding these objects.
\end{quoting}




- Problems in the field: As it stands the field is stagnant, relying on outdated models which need improvement. We need better abundance measurements. Optical spectra are difficult to model and nebular lines are not sufficient. Early modeling of spectra is essentially hard.

\subsection{Thermonuclear Supernovae}
{\centering 
\textit{Contributors: Jennifer Andrews, Eddie Baron, Ryan J. Foley, Christopher L. Fryer, Aimee Hungerford, Bronson Messer, Michael Zingale}}\\

\label{sec:sources_thermonuclear}

Whereas core-collapse supernovae are powered by the gravitational potential energy released in the collapse of a massive star, thermonuclear supernovae are powered by runaway nuclear fusion in a CO (Carbon-Oxygen) white dwarf.  The physics and explosive mechanism for such explosions is similar to that of novae and X-ray bursts where the dominant uncertainties are the behavior of reactive flows in the deflagration or detonation regimes.  These explosions are believed to produce Type Ia supernovae.  They match the spectra (early time line features O, Mg, Si, S, Ca with growing features in Fe and Co).  In these explosions roughly half of the C and O in the white dwarf is fused into Fe, releasing an energy of roughly $10^{51}$\,erg.  Incomplete burning explains the strong Si, S and Ca features in the ejecta.

Thermonuclear supernovae play an important role in astronomy.  They are one of the most luminous transients in the universe (the most powerful explosions driven by nuclear fusion) and are important contributors to the chemical evolution in the galaxy (they are the dominant source of iron peak elements).  But they also play an important role in physics.  To 1st order, their peak emission can be empirically estimated, allowing astronomers to use them as standard candles to probe the expansion of the universe~\citep{1998AJ....116.1009R,1999ApJ...517..565P}.  Supernova observers Brian Schmidt, Adam Riess and Saul Perlmutter received the Nobel prize for their work using thermonuclear supernovae to discover dark energy via expansion of the Hubble diagram to great distances.  However, astronomers now have a wide range of tools to measure the Hubble constant and the different results suggest some tension in the measured Hubble constant between different results~\citep{2021CQGra..38o3001D}.  This tension is likely due to an underestimate in the, often empirical, methods used to measure the expansion of the universe.  An improved understanding of thermonuclear supernovae would reduce the errors from supernovae measurements.

\begin{figure}
      \centering
    \includegraphics[width=0.95\linewidth]{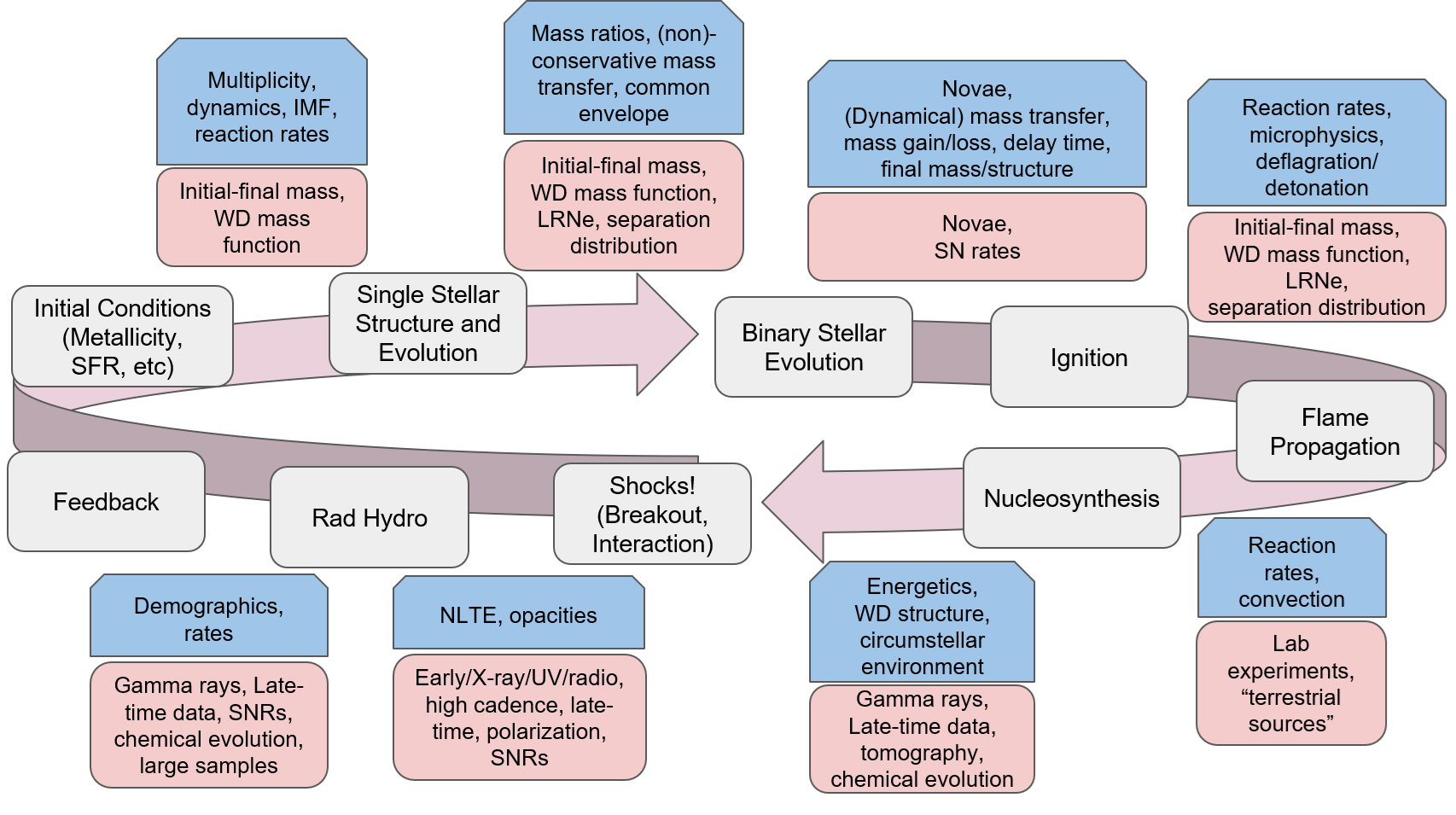}
    \caption{A representation of the simulation and modeling required to understand thermonuclear supernovae. The integrated loop contains the various stages of these events. In between each step are shown the relevant measurements, observables, and knowledge required to integrate the adjacent stages.}
    \label{fig:thermonuclear_chains}  
\end{figure}

\begin{quoting}
    \noindent Finding: The most precise use of thermonuclear supernovae for cosmology are limited by systematics. This ties to the outstanding problem of understanding reactive flows. The importance of these objects in astrophysics necessitates the investment to understand how they explode.
\end{quoting}

But thermonuclear supernovae are also important as tests in our understanding of reactive flows.  Reactive flow physics, turbulent or fluid flows where reactions (chemical or nuclear) occur, is critical in a wide range of applications from the combustion engine and explosives to astrophysical transients (novae, X-ray bursts, thermonuclear supernovae).  The surface area of the burning front can play a key role in the rate of these reactions, a process that is often extremely complex when the burning front is subsonic, a.k.a. in the deflagration regime.  For plasmas and gases, the deflagration regime drives turbulence and resolving this turbulence is critical in understanding the explosion.  Studying thermonuclear supernovae, which span between both deflagration and detonation regimes, provides an ideal test bed of this physics.

What limits our ability to apply thermonuclear to astronomy and physics applications alike is the lack of understanding of the progenitors of these explosions, the nature of the ignition, and the reactive flow physics.  For example, although observations have shown that type Ia supernovae are produced by the thermonuclear explosion of a CO white dwarf, it is unclear whether single-degenerate (CO white dwarf accreting from a main sequence star companion) or double-degenerate (merger of two white dwarfs) progenitors dominate the observed explosions.  It is also unclear whether the implosion is driven as the white dwarf exceeds its maximum mass (Chandrasekhar limit for electron degeneracy support) or if an explosive helium shell on top of the white dwarf drives the implosion.  The entire evolution of the thermonuclear explosion depends on this ignition.

If we would like to probe reactive flow physics with thermonuclear supernovae, we must first constrain the errors in the uncertainties in the progenitor and the ignition process.  With all of these uncertainties, current theoretical models do not predict the tight correlation between rise/fall time of the light curve and peak luminosity that is assumed in the use of type Ia supernovae as standard candles.  Is this because our theory/progenitor understanding is incorrect, or is it because the uncertainties in the type Ia standard candle studies are being underestimated?  As astrophysical data improves, we can be much more precise with these physics studies.

Thermonuclear supernovae can benefit greatly from connections to the combustion and explosive community.  A number of numerical methods and analytic/experimental tests have been developed by these communities that may lead to improved models and reduced uncertainties in calculations of the reactive flows in thermonuclear supernovae.  By building the connection between these fields, we can then use constraints from supernova observations to feed back into the models used for combustion and explosives.

There are many observations of type Ia supernovae in the infra-red and optical.  These observations are what identified the CO white dwarf nature of type Ia supernovae.  Improved observations, or the use of gamma-rays, should allow determination of single or double degenerate progenitors for individual supernovae. But understanding the nature of the ignition and evolution of the engine is more difficult to probe.  UV and X-ray can probe the nature of the progenitor through shock interactions with the circumstellar medium (mass loss, companion).  But the ignition mechanism and reactive flows are better probed through the $^{56}$Ni production.  If the implosion is driven by a helium-shell ignition, some $^{56}$Ni will be produced on the outside.  Although difficult to observe with UVOIR emission, gamma-ray observations (e.g. the COSI instrument) will provide a direct probe of the distribution of $^{56}$Ni, constraining the nature of the explosion.  As the observations of $^{56}$Ni become more precise, we can begin to probe the nature of the reactive-flow physics in these models.
   

\subsection{Tidal Disruption Events}
{\centering 
\textit{Contributors: Tarraneh Eftekhari, Dheeraj R Pasham}}\\

\begin{figure}[ht]
\begin{center}
    \includegraphics[width=\textwidth]{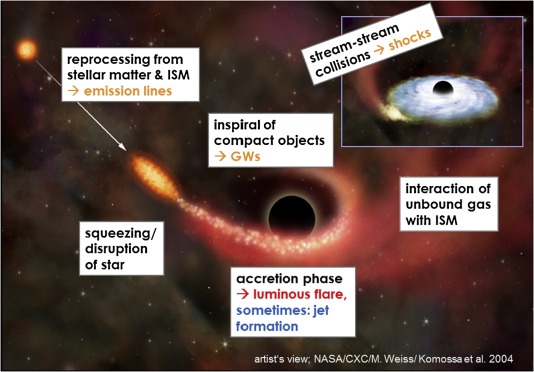} 
    \caption{A summary of the processes and stages involve in a tidal disruption event \citet{komossa2015tidal}}\label{fig:tde}
\end{center}
\end{figure}

\label{sec:sources_TDEs}
The tidal disruption of a star outside a black hole's event horizon, represented in Figure~\ref{fig:tde}, \citep{Rees1988Natur.333..523R} offers a unique opportunity to 1) probe the mechanisms governing jet launching and accretion disk formation for black hole central engines, 2) constrain the demographics of supermassive black holes (SMBHs), and 3) measure black hole spins. As multi-wavelength (and potentially multi-messenger) phenomena, tidal disruption events (TDEs) therefore offer valuable insights into black hole physics. 

The mechanisms driving the formation and evolution of relativistic jets is a long-standing problem in astrophysics. And in few of these situations, being any combination of source, engine object, or jet-powering mechanism(s), do numerical simulations converge. 
The formation of large, relativistic jets on human-observable timescales in a small fraction of TDEs \citep{Bloom2011,Burrows2011,Levan2011,Zauderer2011,Cenko2012,Brown2015,Andreoni2022, Pasham2023} --- in contrast to characteristic timescales of $\sim 10^6$ years for AGN --- offer a unique test of the Blandford-Znajek jet launching mechanism (Section~\ref{sec:questions_BZ}). At early times, studies of the accretion mechanism are afforded through X-ray observations which trace emission at the base of the jet, while late-time observations of synchrotron emission from the forward shock provide valuable insight into the jet physics. Moreover, X-ray observations of jetted TDEs on timescales of several hundred days post-explosion have revealed a sudden cessation of jet activity, marking a transition to sub-Eddington accretion \citep{Zauderer2013,Pasham2015ApJ...805...68P,Eftekhari2024ApJ...974..149E}. The recent discovery that a subset of TDEs exhibit a rebrightening in both the soft X-rays and radio on timescales of $\sim 1$ year post-disruption may further implicate accretion state changes, in analogy with X-ray binaries (e.g., \citealt{Holoien2018MNRAS.480.5689H,Cendes2024ApJ...971..185C}). 

\begin{quoting}
    \noindent Comment: Tidal disruption events are large jets which turn on and off in an observable lifetime. They provide a bridge between small and large jets, and are key events to probe jet and accretion physics across different sources.
\end{quoting}

The dependence of the rate of TDEs on properties such as the black hole mass and spin, as well as the black hole occupation fraction, further demonstrates the powerful probe of SMBH demographics afforded by TDEs. A better understanding of how TDEs inform SMBH demographics requires knowledge of the underlying emission mechanism; however, which remains poorly understood. Advancements in this area can be achieved by combining hydrodynamic simulations with detailed modeling of the electromagnetic radiation. Mid-range GW observatories will detect tens of thousands of events, many of which will be observed by EM surveys such as LSST, providing valuable constraints on the black hole mass function \citep{Pfister2021}. Black hole spins can similarly be probed through quasi-periodic oscillations in the observed X-ray flux from TDEs which enable a measurement of the inner radius of the accretion disk and hence black hole spin measurements (e.g., \citealt{Pasham2019Sci...363..531P}). This requires high-cadence monitoring with sensitive X-ray facilities.

Finally, TDEs may serve as natural sites for neutrino production via charged pion decay of ultrahigh-energy cosmic rays (UHECRs; \citealt{Dai2017a,Hayasaki2019}). Jetted TDEs, which can accelerate protons to $10^{20}$ eV, may constitute a significant source of UHECRs in the universe \citep{Farrar2014}. The interaction between non-thermal X-ray photons from the jet and UHECRs should lead to the production of high-energy neutrinos \citep{Wang2016}. However, the birth rate of jetted TDEs, coupled with fiducial estimates for non-thermal emission from relativistic jets, cannot account for the observed IceCube neutrino flux \citep{Dai2017b}. On the other hand, targeted searches for TDEs that are spatially and temporally associated with IceCube events may yield more promising results \citep{Fang2016JCAP...12..017F}. To date, several candidate TDE-neutrino associations have been proposed (e.g., \citealt{Stein2021NatAs...5..510S,Yuan2024}), suggesting that mildly relativistic outflows may also serve as sites for neutrino production. Perhaps most strikingly, these TDEs all exhibit dust IR echoes, indicating that the presence of dust may play a key role in high-energy neutrino production \citep{Jiang2023}. The promise of upcoming optical surveys like LSST will enable a much larger sample of TDEs with IR dust echoes, facilitating targeted searches for contemporaneous and spatially coincident neutrinos, and potentially revealing the origin of high-energy neutrinos. 

A new class of repeating extragalactic nuclear transients (RENTs) associated with TDEs have also been uncovered in the past 5 years. These manifest primarily in soft X-ray band (0.2-10.0 keV) and have recurrence timescales of minutes to years. These are known in the literature as (soft) X-ray quasi-periodic eruptions (QPEs; \cite[e.g.,][]{2024A&A...684A..64A}), quasi-periodic outflows (QPOuts; \cite{2024SciA...10J8898P}), quasi-periodic oscillations (QPOs; \cite[e.g.,][]{Pasham2019Sci...363..531P, 2025arXiv250101581M}), and repeating TDEs \cite[e.g.,][]{2023A&A...669A..75L,2024ApJ...971L..31P}. One of the prevailing ideas for RENTs is that they are triggered by interactions of a gravitationally bound (smaller) object with the accretion disk of a supermassive black hole (SMBH) presumably formed after a star (third object) is tidally disrupted. If that is indeed the case, then some of these extreme mass ratio binaries could also be detectable with future space-based gravitational wave detectors like LISA and Taiji, and have the potential to transform our understanding of supermassive black hole growth, probe dark energy, and put fundamental constraints on gravity.

Currently, there are about two dozen RENTs known and the recent association of RENTs with TDEs unveiled via high-cadence soft X-ray observations with NICER \citep[e.g.,][]{2023A&A...669A..75L,2024SciA...10J8898P,2024ApJ...971L..31P, 2024Natur.634..804N} presents a unique opportunity to discover more by following-up optically--selected TDEs in soft X-rays. While the fraction of TDEs that end up as a RENT is currently unknown, current estimates put that number around 10\% \citep{2024A&A...684A..64A, 2023ApJ...945...86L}. To successfully detect RENTs via TDE follow-up an X-ray telescope should have good response in the soft X-ray (0.2-10 keV) band and must be able to monitor sources over a wide range of timescale to identify the periods of RENTs. The current intrinsic rate of QPEs (one of the 4 known RENTs) based on 4 sources from {\it eROSITA} is 0.6$^{4.7}_{-0.4}\times$10$^{-6}$ Mpc$^{-3}$ \citep{2024A&A...684A..64A}--roughly 10\% of the TDE rate \citep{2023ApJ...955L...6Y}. Thus, Rubin follow-up of hundreds of TDEs in the coming years with a soft X-ray telescope would usher in a era of discovery and characterization of dozens of RENTs.

\subsection{X-ray Binaries}
{\centering 
\textit{Contributors: Eric Borowski, Ed Brown, Alejandro Cárdenas-Avendaño, Robert I. Hynes, Thomas Maccarone, Cole Miller, John A. Tomsick}}\\

\label{sec:sources_xrayBinaries}

In this section we discuss accretion onto neutron stars and stellar-mass black holes, and the implications that time-domain and multimessenger observations of these systems have for strong gravity, dense matter, and populations of compact objects.  We then discuss some of the advances possible with future TDAMM observations.

\begin{quoting}
    \noindent Comment: Neutron stars and black holes represent physical extremes in gravity, density, and magnetic fields.  X-ray binaries, due to proximity (and hence, brightness) and accessible timescales, permit high fidelity studies of these topics in a variety of states. As a result, studying compact objects in X-ray binaries gives us unique insight into fundamental physics as well as astrophysics.
\end{quoting}

Compact object accretion is traditionally understood to occur due to a mutual orbit with a stellar companion.  If the companion has low mass (usually considered to be $\sim 0.5~M_\odot$ or less), then stellar winds are weak (except when the companion is in the giant phase) and thus accretion occurs via Roche lobe overflow, in which matter flows to the compact object through the inner Lagrange point.  This accretion can be active for a few hundred million years, and transfer up to a few tenths of a solar mass.  If instead the companion has high mass (several solar masses or more), then Roche lobe overflow would be unstable if the compact object's mass is less than or comparable to the mass of the companion.  Such systems are expected to exist, but they would be visible only for a short time.  Thus mass transfer for longer-lasting high-mass systems is instead expected to proceed by capture of a wind from the companion.  Because the companion has a large mass in this scenario, it has a short lifetime and thus active accretion is thought to last for a few million years or less, and the total mass transferred is thought to typically be $\sim 0.01~M_\odot$ or less.  In addition to these two traditional accretion channels, there is now interest in compact objects which have no nearby stellar companion but are embedded in the accretion disks of active galactic nuclei.  
\begin{figure}
    \centering
    \includegraphics[scale=0.25]{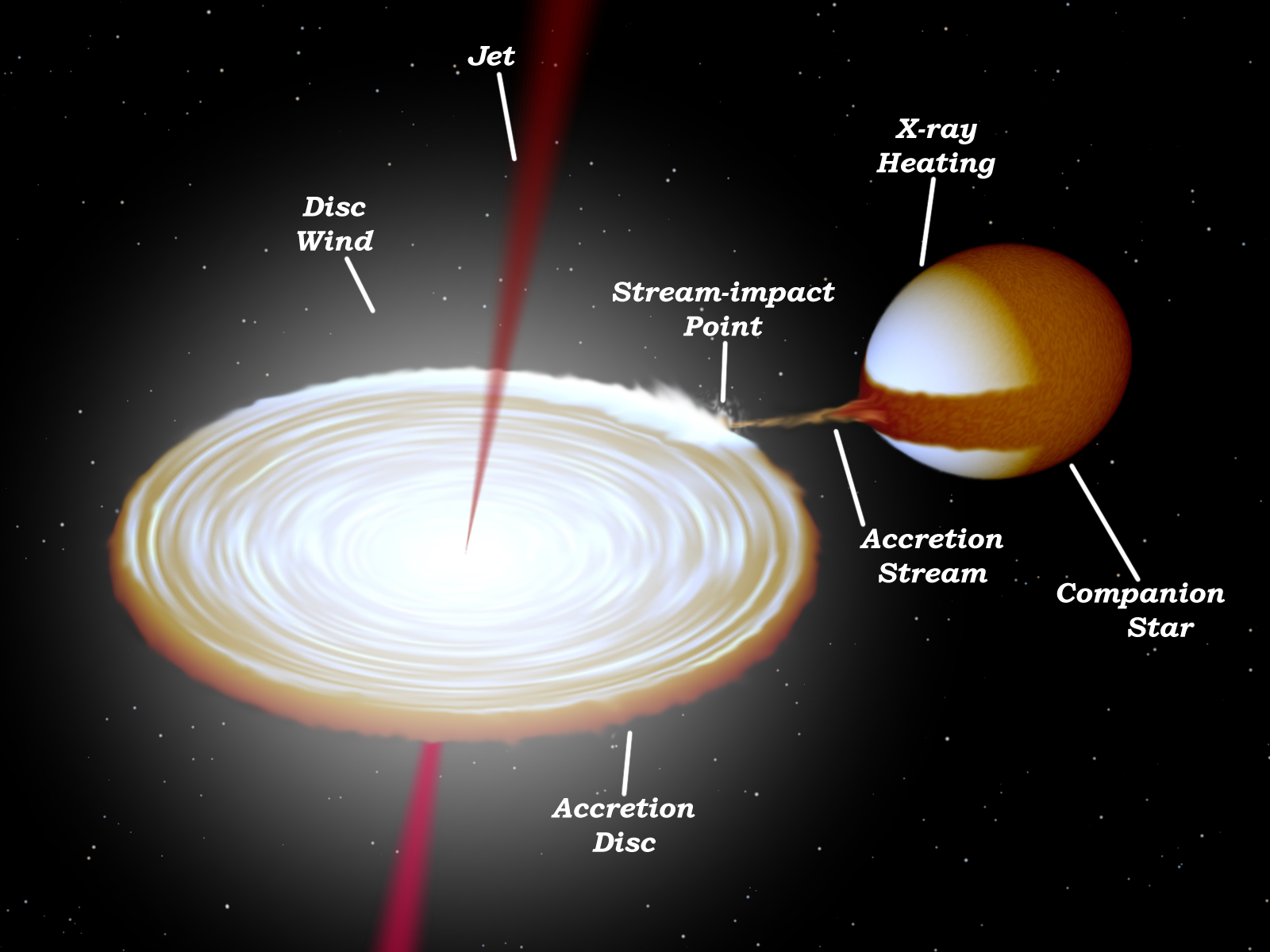}
    \caption{An artist's annotated depiction of a low-mass X-ray binary (Image credit: Robert Hynes). }
    \label{fig:binsim_xrb}
\end{figure}
Observational and theoretical studies of accreting compact objects show that, depending on the nature and rate of mass donation, there can be substantial instabilities in accretion.  One example is the so-called dwarf nova instability, which can occur in Roche lobe overflow systems.  The idea is that because the accreting gas inherits the high angular momentum of the companion's orbit, angular momentum transport is essential to bring the gas to the compact object, and the understanding over the last three decades is that efficient angular momentum transport usually proceeds via the action of magnetic fields, which in turn requires a high enough ionization fraction to couple to the fields.  The dwarf nova instability then operates in a system where (1)~initially the disk temperature is too low for significant ionization and thus matter piles up without spiraling to the compact object, thus (2)~matter accumulates in the outer disk, and (3)~eventually the higher matter density leads to higher temperatures and higher ionization, so that (4)~angular momentum transport is efficient and the accumulated gas in the disk accretes rapidly onto the compact object.  This process has been inferred from numerous compact object binaries, over human timescales, which means that accreting stellar-mass compact objects can be studied over a wide range of accretion rates.

We now discuss some of the fundamental physics that can be addressed using accreting compact objects, beginning with strong gravity.  Our current theory of gravity, general relativity, has passed all the tests that we have been able to perform.  Nonetheless, the prediction of infinities in the theory, and its lack of formulation with quantum principles, convinces most physicists that general relativity is only an approximation to a deeper underlying theory.  Among the questions we can then ask are (1)~is general relativity accurate in all astrophysical situations, and (2)~how do accretion disks, and the underlying magnetohydrodynamics and plasma physics that drive them, work in strong gravity?

Time-domain observations can address these questions in several ways.  For example, the accretion instabilities discussed above lead to observed state changes, from accretion that generates radiation efficiently to accretion that does not, from states with jets to states without, and so on.  This gives invaluable information about how disks change in different circumstances.  In addition, the nature of the flow touches on fundamental predictions of gravity, such as the existence of black hole horizons.  Another key observation is the existence of quasi-periodic oscillations in the X-ray emission from accreting neutron stars and black holes.  Model consensus has not yet been achieved, but the sharp, strong oscillations observed in these systems are representative of characteristic frequencies that could test gravity as well as providing independent ways of measuring the masses and spins of the compact objects.

Multimessenger observations can also address questions about strong gravity.  For instance, the observed shape of the iron K$\alpha$ line (a line produce by fluorescence in the innermost electron shell of iron) near a black hole depends on details of strong gravity and the possible presence of dark matter, as well as on the nature of the accretion flow and the angle of observation.  Careful characterization of these lines thus provides essential clues to gravity and the behavior of matter in strong gravitational fields.  Key information about strong gravity has already been obtained from gravitational wave and gamma-ray observations of the merging neutron star event GW170817, which among other things demonstrated that the speed of light and the speed of gravity differ by no more than a part in $\sim 10^{15}$.  Additional information from gravitational wave events can be expected if we find definitive evidence of an electromagnetic counterpart to a binary black hole merger, which requires the presence of matter such as in the disk around an active galactic nucleus.

Turning our attention to dense matter, by which we mean matter near or beyond nuclear density, some key questions include: (1)~what are the properties of cold matter beyond nuclear saturation density, and (2)~what are the condensed matter properties of neutron star crusts?  These questions require careful observations of neutron stars.

In the time domain, the last fifteen years have seen major advances in our understanding of the properties of matter beyond nuclear density (and in particular the equation of state [EOS], which is the pressure as a function of density) from observations of massive pulsars, from the gravitational wave event GW170817, and from X-ray observations of non-accreting neutron stars using NASA's Neutron star Interior Composition Explorer (NICER), which have been essential in measurements of neutron star radii, which is critical in constraining the EOS.  One limitation of NICER observations is that they focus on non-accreting and thus relatively X-ray dim pulsars.  If X-ray bursters and rapidly accreting neutron stars could be studied in the same way (and thus if current concerns about systematic model errors can be overcome), the number of photons analyzed would increase by one to two orders of magnitude, which would dramatically increase the precision of measurement.  For crustal physics, important hints have been obtained by watching neutron stars go through accretion instability cycles as discussed above, because the crust is heated and then cools off in ways that present clues about the transport properties of matter under extreme pressure.

Multimessenger information about neutron star crusts could appear in various ways in the future.  If continuous-wave gravitational waves are detected from rotating neutron stars and these can be compared with the rotation frequency determined using electromagnetic observations, then there will be two general possibilities.  One is ``mountains" on neutron stars (no more than a millimeter high), which could be produced by pooling of accreted matter in local maxima of magnetic fields, or crustal asymmetries.  The other is the so-called ``r-modes" (related to Rossby waves on Earth), which are fluid modes in the star.  Mountains would produce gravitational waves at twice the rotational frequency, whereas gravitational waves from r-modes would have approximately 4/3 times the rotational frequency (and the deviation from 4/3 would itself encode important information about neutron star structure and strong gravity).  Thus a comparison between the frequencies will give us a new window into neutron stars and dense matter.

The final topic that we feature in this section is the characteristics of compact object populations.  Our census is necessarily incomplete; for example, based on stellar evolution we expect that there are roughly a few hundred million black holes in the Galaxy, and a few hundred million neutron stars, but of these we have only observed tens of black holes and thousands of neutron stars.  Even within this sample, there are puzzles emerging.  For instance, black hole masses inferred from accreting Galactic systems have lower masses and a broader range of spin parameters than those observed via gravitational waves.  The lower masses are understood in terms of selection effects, but the spin distributions are less easy to explain.  X-ray binaries also offer the only opportunity to study the combination of the black hole mass, the black hole spin, the black hole natal kick, and the properties of a companion star.  All of these provide vital clues about the nature of the supernova explosions that produced the black holes.  

To make further progress requires: larger samples of objects; better methodology/validation of methodology for estimating black hole spins from electromagnetic data; better mass estimates; and better astrometric measurements.  The first requires continued wide-field monitoring.  The second requires a combination of theory work and better measurements via reverberation mapping of accretion disks to understand reflection-based measurements' systematics, along with precise parallax measurements and mass estimates to better constrain continuum measurements; and the last two require a combination of more optical and radio data to obtain distances, radial velocity curves and ellipsoidal modulation measurements, and ideally, astrometric wobble measurements for the best studied objects.

From the time-domain standpoint, one of the many population-based phenomena that can be explored is the production of jets from accreting black holes and neutron stars.  A growing body of theory work suggests that jets require a large net vertical magnetic field, and with a few isolated exceptions among the neutron star systems, it is largely born out that strong hard X-ray emission, indicative of a large scale height accretion flow and hence  a large scale height magnetic field occurs in conjunction with bright radio emission.   The neutron star systems are less well-understood, as the systems are both fainter and more rapidly variable.  As new radio telescopes come on line, it should be possible to grow our knowledge of the neutron star X-ray binaries' jet properties.

The kinetic powers of jets and their compositions are something that is likely better studied in X-ray binaries than in AGN, as is the question of the extent to which spin powers the jet production process.  In X-ray binaries, seeing outbursts which enter radiatively efficient states with little to no jet emission allows for constraints on the mass accretion rates that are not possible in the same way in AGN, where the radiative efficiency is largely unconstrained observationally.  As a result, in the best cases, the instantaneous energy budget available to jets can be determined.  The observed kinetic power can then be constrained by multi-wavelength rapid variability measurements that show the propagation of power up the jet and give size scale and opening angle information.  Additionally, with multiple methods possible for estimating black hole spins, and with direct information sometimes available about neutron star spins, it is possible in the X-ray binaries to make cleaner tests of the effects of spin on jet production, something which is theoretically predicted, but for which the observational evidence remains ambiguous.

Multimessenger observations can help elucidate differences in black hole mass and spin distributions between different samples.  Are there, for example, truly different subpopulations represented in gravitational-wave samples compared with black holes seen using electromagnetic ways?  Is there a mass gap between neutron stars and black holes at the birth of these objects (which would have implications for core-collapse supernova mechanisms)?  And finally, the origin of the highest-energy neutrinos yet observed ($>100$~TeV) has provided hints that at least some might be associated with blazars.  Microblazars (powered by stellar-mass rather than supermassive black holes) exist in the Galaxy, so state changes could provide indications about the way that such energetic neutrinos are produced; there are tentative indications of neutrino emission from Cygnus X-3 already.  Understanding whether jets are routinely hadronic or leptonic in X-ray binaries may also have implications for the production of energetic neutrinos in AGN.

Overall, accreting neutron stars and black holes provide glimpses of fundamental physics and astrophysics.  Major progress has been made in recent years.  However, the relevant communities are often separated from each other; gravitational physicists and nuclear physicists, specialists in active galactic nuclei and stellar-mass black holes, are often in different funding streams and thus do not interact as much as they could.  Programmatic support for such cross-disciplinary studies would dramatically improve our understanding of the universe at its most fundamental.


\subsection{Gamma-Ray Bursts}
{\centering 
\textit{Contributors: Eric Burns, J. Patrick Harding, Jamie A. Kennea, Gavin P. Lamb, Matthew R. Mumpower, Eliza Neights, Peter Shawhan}}\\

\begin{figure}
      \centering
    \includegraphics[width=0.95\linewidth]{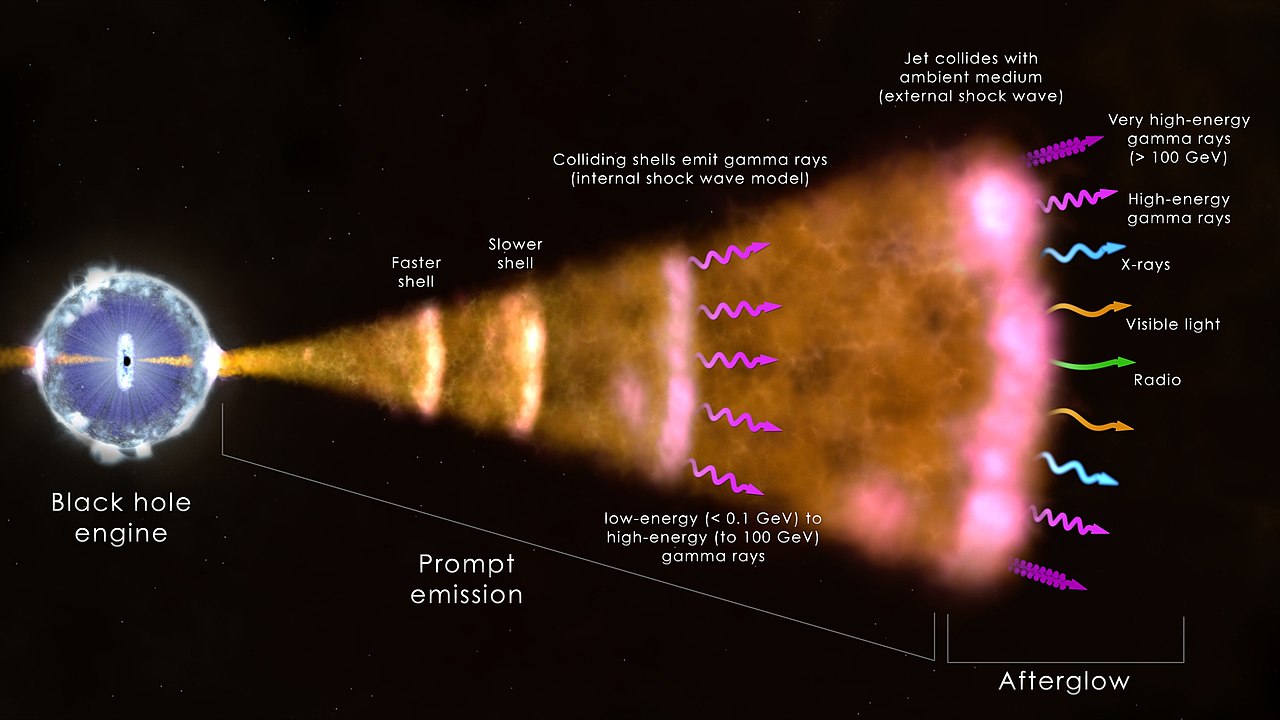}
    \caption{A representation of the structure and emissions from GRBs. A central engine powers relativistic jets in two directions. These outflows release the prompt gamma-ray signature. This is followed by an external shock generated as the jet propagates through the surrounding material, releasing synchrotron radiation across the electromagnetic spectrum.
    Credit: NASA Goddard/ICRAR}
    \label{fig:grb}  
\end{figure}

\label{sec:sources_grbs}
The unexpected discovery of GRBs has been a boon for the study of the physics of the cosmos. The half-century quest to understand their origin has had magnificent successes, including identification of a new way to explode massive stars and one of only two confident multimessenger transients. Cosmological GRBs arise from collimated, ultrarelativistic outflows, represented in Figure~\ref{fig:grb}.

Gamma-Ray Bursts are typically classified as short or long, based on whether the duration is below or above a $\sim$2\,s threshold. While the emission mechanism for the prompt GRB signature is not understood, most models expect the prompt duration to be similar to the timescale of the accretion onto the central engine. Long GRBs arise predominantly from collapsars, being a rare subset of core-collapse supernova, whose accretion timescale is at least the freefall time of the core of the progenitor. Short GRBs arise predominantly from neutron star mergers, being followed by kilonova. The accretion timescale of a neutron star is less than a tenth of a second. This was dramatically confirmed with the joint detection of a neutron star merger through gravitational waves and a short GRB. 

GRBs are of interest for several science questions in TDAMM. The jets produced in these events are the fastest matter in the universe. These jets can be powered by rapidly rotating black holes, with the rotation imparted by the cataclysmic end of the progenitor objects. The extreme rotation may allow for Blandford-Znajek, i.e., extraction of the rotational energy of the black hole, in addition to the other jet powering mechanisms. Additionally, GRB jets are similar to tidal disruption events, where a finite time the engine turns on and off. They are easier to study than tidal disruption events because we have detected orders of magnitude more, and the differential spatial scales allows for probing how jet and accretion physics may vary across black hole mass scales. Jets launched from neutron star mergers propagate through clean environments, allowing for isolation of intrinsic effects as opposed to those imparted by propagation. when combined with gravitational wave detection we get direct insight into properties of the progenitor system and the central remnant. This, for example, can give us measurement of the black hole spin, necessary to determine contributions from Blandford-Znajek. Further, such observations will allow us to probe whether magnetars can form rapidly (as opposed to having their magnetic fields buried for decades), and whether magnetars or neutron stars can power GRBs. These jets also release the prompt gamma-ray signature which are the most luminous events since the big bang. Understanding this dissipation mechanism is a key question in this area of study, 


Given the historic separation, it was then quite surprising to identify a kilonova signature following the long GRB~211211A \citep{rastinejad2022kilonova}, which was then followed by a similar situation for GRB~230307A \citep{levan2024heavy}. These events are something of a mystery. The host galaxy type and offset from the host galaxy require a merger origin. The quasithermal signature in both cases are strikingly similar to the one following GW170817 which may suggest a binary neutron star or neutron star-black hole merger. However, such events were not expected to be able to produce $\sim$100~s event durations. Past GRBs with a short spike and extended emission have been observed, which may point to a magnetar origin. Some have adapted these models to explain these new events. Other approaches include invoking fallback from a disrupted neutron star, or magnetically arrested disk models. Alternatively, these events may instead arise from neutron star-white dwarf or black hole-white dwarf mergers. Discrimination between these scenarios likely requires observations of one with GWs, pointing either to new knowledge of ultrarelativistic jet physics or to a new GRB progenitor class which can be exploited as a new tool to study the cosmos.



The James Webb Space Telescope looked at late-time infrared spectrum of the kilonova following GRB~230307A. Such a measurement has been long sought as the velocities of the ejecta at these times are sufficiently low to allow for line identification, giving insight into specific elemental yields and inferences on the temperature and ionization of the plasma. Indeed a line complex was seen, similar to one found following GW170817 \citep{levan2024heavy}. This was interpreted as a Tellurium line, which would prove the event as an r-process event and preclude a neutron star-white dwarf origin. However, while Tellurium is certainly the likeliest explanation, this interpretation is not fully self-consistent with the physical parameters inferred from the lightcurve evolution, and our knowledge of atomic spectroscopy is sufficiently limited that it is not a certain line identification. Thus, our current atomic knowledge is limiting the scientific return of NASA's flagship observatory.


\subsection{Fast Radio Bursts}
{\centering 
\textit{Contributors: Tarraneh Eftekhari, Jamie A. Kennea, Zorawar Wadiasingh}}\\

\begin{figure}[ht]
\begin{center}
    \includegraphics[width=0.7\textwidth]{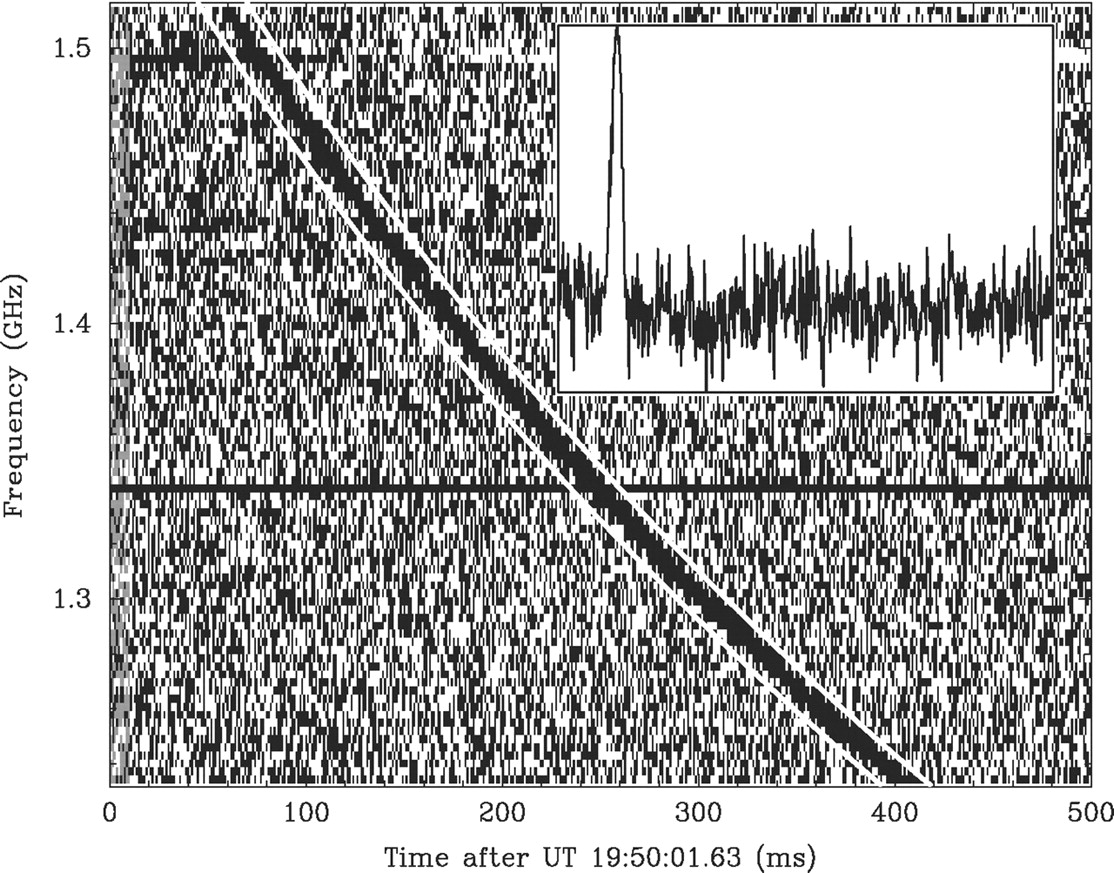} 
    \caption{A fast radio burst, borrowed from \citet{lorimer2007bright}. The main image shows the direct frequency evolution, with the decay due to dispersion by astrophysical plasmas. The inset shows the de-dispersed signature.}\label{fig:frb}
\end{center}
\end{figure}

\label{sec:sources_frbs}
Fast radio bursts (FRBs) are millisecond-duration flares of coherent radio emission \citep{1980ApJ...236L.109L,lorimer2007bright,Thornton2013}, with an observation shown in Figure~\ref{fig:frb}. Their large dispersion measures (DMs), or frequency-dependent arrival times, indicative of electron column densities well in excess of Galactic values, point to a primarily extragalactic origin. Despite several thousand detections to date (e.g., \citealt{Macquart2020,CHIME2021,Law2024,Shannon2024}), their origins (and emission mechanisms) remain unknown, with progress in this direction largely inhibited by the dearth of well-localized events (which provide the necessary astrophysical context and energetics; \citealt{Eftekhari2017}) as well as the fact that a small fraction of events exhibit repeat bursts (so-called `repeaters') while others appear as one-off events \citep{Shannon2018,CHIME_repeaters2023}. 

Prevailing theories for FRB sources implicate magnetars formed through core-collapse supernovae, driven primarily by their preferential association with star-forming galaxies \citep{Bochenek2021,Gordon2023,Sharma2024}, as well as their observational characteristics (e.g., high brightness temperatures, energetics, burst rates) indicative of a compact object origin \citep{Michilli2018,Dai2021,Pleunis2021,Nimmo2022,Sherman2023,Bera2024}. The discovery of an FRB-like event, and associated soft gamma-ray burst, from a known Galactic magnetar SGR 1935+2154 on April 28th, 2020 established a definitive connection between magnetars formed through core-collapse supernovae and at least some subset of FRBs \citep{CHIME2020,Bochenek2020,2020ApJ...898L..29M,2021NatAs...5..372R,2021NatAs...5..401T,2021NatAs...5..378L}. On the other hand, the association of one FRB with a globular cluster \citep{Kirsten2022} and another offset from an early-type elliptical galaxy \citep{Eftekhari2024} suggest that some events may form from magnetars produced through delayed channels such as binary neutron star mergers or the accretion-induced collapse of a white dwarf \citep{Margalit2019}, or alternatively delayed magnetar activity, e.g. associated with internal fields modified by superconductivity in an aging neutron star \citep{2024arXiv241108020L}.  Technical upgrades to several FRB discovery experiments in the coming years will facilitate a growing sample of precisely localized events with robust host galaxy associations offering key insight into the sources responsible for producing FRBs and their putative connection to known classes of transients. 

Separate from the progenitor question, the mechanism responsible for generating coherent radio waves remains unknown, much like the case with pulsars despite decades of research. Broadly speaking, coherent radiation mechanisms encompass emission by coherent oscillations of charges and fields, and kinetic instabilities \citep{Melrose1978}, the properties of which are governed by fundamental plasma physics in an extreme regime. The emission mechanisms for FRBs are generally categorized into near-field and far-field models based on their radial proximity from the surface of the neutron star or central engine. While near-field models typically invoke magnetic reconnection processes, electron-positron pair cascades or coherent curvature radiation associated with the neutron star magnetosphere \citep[e.g.,][]{Kumar2017,Yang2018,2019ApJ...879....4W,2021ApJ...922..166L}, at large radial distances, proposed models suggest synchrotron maser emission produced via relativistic shocks \citep{Metzger2019,Beloborodov2020}. A key distinguishing feature between models is the size of the emitting region, with emission mechanisms originating from close proximity to the central engine involving smaller emission sizes. Recent work probing coherent scintillation scales from the non-repeating FRB 20221022A places constraints on the size of the emitting region of $R \lesssim 3 \times 10^4$ km \citep{Nimmo2024}. Coupled with the discovery of a swing in the polarization position angle of the FRB \citep{Mckinven2024}, this event offers compelling support for magnetospheric models. Nevertheless, both models present several challenges to our theoretical understanding of coherent radiation, challenges that may be addressed with the anticipated large samples of FRBs in the near future.

Another challenge in understanding the physics of FRBs is the scarcity of multi-wavelength counterparts, despite concerted efforts to detect them. Searches for long-lived counterparts have thus far led only to the discovery of compact persistent radio sources (PRSs) associated with a handful of repeating FRBs (e.g., \citealt{Chatterjee2017,Niu2022}). Interestingly, in all cases, the FRBs have been localized to dwarf galaxies and exhibit large rotation measures, indicative of highly magnetized surrounding media \citep{Michilli2018}. The PRSs have been largely interpreted as synchrotron nebulae powered by a young magnetar central engine \citep{Margalit2018}. No other long-lived counterparts have been identified to date.

On shorter timescales, there have been a small number of contemporaneous searches for prompt counterparts in the optical, X-ray, and $\gamma$-ray bands \citep{cunningham2019search,Piro2021,Hiramatsu2023,Cook2024,Kilpatrick2024,Pearlman2024}. The only known prompt counterpart was the hard X-ray burst detected from the Galactic magnetar SGR 1935+2154 with a radio-to-X-ray fluence ratio of $\sim 10^{-5}$, suggesting that such X-ray bursts are unlikely to be detected at the typical cosmological distances of FRBs with present-day X-ray facilities. Archival searches for near-simultaneous radio emission from short GRBs have thus far led to non-detections \citep{Curtin2024}, while a putative association was proposed for the binary neutron star merger GW190425 and FRB 20190425A with a delay of $\lesssim 2.5$ hours \citep{Moroianu2023}. More recently, NASA's NICER observed double `glitches' in X-ray timing of the magnetar SGR 1935+2154 bracketing a FRB-like event \citep{2024Natur.626..500H} and spin down enhancements \citep{2023NatAs...7..339Y}, that is contemporaneous with X-ray and soft gamma-ray activity. Glitches are rapid transfers of angular momentum of superfluid components of the neutron star, and are a poorly understood phenomenon now clearly connected to some FRBs. Thus the interior dynamics of magnetized neutron stars, as well as how they are coupled to radiative signals is a future avenue of study. This requires investment in theory and observations, in particular continuous monitoring of Galactic sources in radio and high energy photons.

\subsection{Fast Blue Optical Transients}
{\centering 
\textit{Contributors: Anna Y. Q. Ho}}\\

\label{sec:sources_fbots}
Fast blue optical transients (FBOTs) were initially identified in archival optical survey data on the basis of their unusually fast-timescale and luminous optical emission, as well as their blue colors at maximum light \citep{Drout2014,Pursiainen2018}. Since then, it has become clear that FBOTs represent a wide range of stellar progenitors, and that circumstellar interaction from enhanced end-of-life stellar mass loss likely plays a significant role in powering the optical emission \citep{Fox2019,Leung2021,Pellegrino2022,Ho2023,Khatami2024}.

Although many FBOTs have been spectroscopically classified as core-collapse  supernovae \citep{Ho2023}, a particularly rapid and luminous subset appear to be a genuinely new phenomenon. The prototype was the transient AT2018cow \citep{Prentice2018}, which---unlike typical supernovae---had extremely luminous emission across the electromagnetic spectrum, from X-rays \citep{RiveraSandoval2018,Margutti2019,Ho2019cow} to radio waves \citep{Margutti2019,Ho2019cow}. Since then, a dozen ``AT2018cow-like'' FBOTs have been identified by wide-field optical surveys. These events have an unusual combination of extreme physical properties: an enormous explosion energy \citep{Coppejans2020} due to a powerful ``central engine'' (likely an accreting black hole; \citealt{Inkenhaag2023}); a tiny mass of ejected matter \citep{Perley2019} and strong asymmetry \citep{Maund2023} providing visibility onto that long-lived central engine \citep{Margutti2019,Ho2023b}; and an asymmetric ambient medium that is denser than a standard stellar wind by orders of magnitude \citep{Ho2019cow,Nayana2021}. The combination of central engine and high mass-loss rate has been suggested to imply a merger-triggered explosion following a common-envelope phase \citep{Metzger2022}, which would make them important test beds for poorly understood aspects of stellar evolution 
as well as the origins of gravitational-wave sources (as ``failed'' versions of the black-hole binaries discovered by LIGO). 

In addition, the visibility of the central engine from weeks \citep{Margutti2019,Ho2023b} to years \citep{Inkenhaag2023,Yao2022} post-explosion provides an opportunity to study the formation of compact objects and evolution of super-Eddington accretion disks \citep{Chen2023}: in other ``engine-driven'' transients (gamma-ray bursts, superluminous supernovae) the engine remains shrouded from the observer by a high ejecta mass. 

The combination of high energy and high ambient density forms a strong shock, probing an unexplored regime of relativistic particle acceleration \citep{Margalit2021}, and providing an ideal environment for high-energy neutrino production \citep{Guarini2022}. And finally, FBOTs produce near-relativistic jets \citep{Coppejans2020,Ho2023b}: they are only the second such class known after GRBs, and they have different progenitors, lifetimes (hundreds of days), and jet compositions. 




\newpage
\section{Acronyms and Definitions}
\label{sec:acroynmsAndDefinitions}



\begin{description}
    \item[AAS - American Astronomical Society] The professional society for astronomers and astrophysicists in the United States.
    \item[AGN - Active Galactic Nuclei] A bright and compact region at the center of some galaxies, powered by accretion onto the supermassive black hole.
    \item[AMO - Atomic, Molecular, and Optical Science] The discipline of science focused on studying matter-matter and light-matter interactions on the scale smal numbers of atoms. The field is responsible for many broadly used technologies including atomic clocks and the laser technologies used in high energy density physics and in the Laser Interferometer Gravitational-wave Observatory.
    \item[APAC - Astrophysics Advisory Committee] A FACA committee which advisors NASA on astrophysics
    \item[ASC - Advanced Simulation and Computing (DOE)] The program responsible for the computational aspect of nuclear stockpile stewardship for the National Nuclear Security Administration.  At its onset, the initiative was termed the Advanced Simulation and Computing Initiative (ASCI).  The Office of Science version of this program is called Advanced Simulation and Computing Research (ASCR).
    \item[APS - American Physical Society] The professional soceity for physicists in the United States.
    \item[AST - The Division of Astronomical Sciences (NSF)] The National Science Foundation division responsible for astronomy and astrophysics.
    \item[BAO - Baryon Acoustic Oscillations (astrophysics)] A standard ruler useful for cosmological distance measurements. They are now large fluctuations in the large scale structure of the universe, caused by acoustic waves in the universe before the cosmic microwave background.
    \item[BLL - BL Lacertae or BL Lac (astrophysics)] A type of active galactic nuclei.
    \item[CeNAM (DOE interdisciplinary)] Center for Nuclear Astrophysics across Messengers is a large scale multi-institutional and interdisciplinary center that connects the nuclear science community with the broad range of disciplines required to advance TDAMM science. 
    \item[Central engine (astrophysics)] The engine which converts energy in order to power an explosion; e.g., the convective engine for core-collapse supernovae.
    \item[Corona (astrophysics)] The diffuse outer layer of stars.
    \item[CMB - Cosmic Microwave Background] The relic radiation from the Big Bang, which now emits in microwave (radio) wavelengths.
    \item[COSI - Compton Spectrometer and Imager (NASA Astrophysics)] A nuclear spectrometer telescope slated to launch in 2027
    \item[DESI - Dark Energy Spectroscopic Instrument (DOE OST)] A telescope designed to measure spectra for tens of millions of galaxies to precisely study aspects of cosmology including baryon acoustic oscillations. The full five year survey is anticipated to be complete in 2026.
    \item[DOE - Department of Energy] The US Federal agency responsible for national energy policy and energy production. Key components of the DOE to this white paper are the Office of Science and the National Nuclear Security Administration.
    \item[FACA - Federal Advisory Committee Act] A federal law which governs advisory committees, referred to as FACA committees. 
    \item[FBOT - Fast Blue Optical Transients (astrophysics)] Explosive transients somewhat similar to supernovae, with high luminosities, fast temporal evolution, and blue emission (corresponding to a higher temperature).
    \item[FRIB - Facility for Rare Isotope Beams (DOE NP)] A new rare isotope accelerator user facility at Michigan State University. First FRIB experiments have begun in Spring 2022. Once ramped up to full power, FRIB will be the highest power rare isotope beam production facility in the world. 
    \item[FSRQ - Flat-Spectrum Radio Quasars (astrophysics)] A type of active galactic nuclei
    \item[GOF - Guest Observer Facility] An entity which helps manage observations and funding for the use of user facility telescopes
    \item[GW - Gravitational Waves] Ripples in spacetime detected by LIGO and other interferometers. Gravitational wave detections of astrophysical transients provides measurements on the cause of the explosion, which is often obfuscated to electromagnetic observations due to the opaque plasma in the event.
    \item[GRB - Gamma-ray Burst (astrophysics)] High-energy astrophysical transients produced by collimated, ultrarelativistic outflows referred to as jets. These are the most luminous events since the Big Bang.
    \item[Hadronic (astrophysics)] A source containing (or process involving) hadrons, most commonly protons (in addition to electrons). Typically treat as antithetical to leptonic
    \item[HEDP - High Energy Density Physics] A relatively new multidisciplinary field which studies physics beyond pressures of ~1 Mbar. 
    \item[Host galaxy (astrophysics)] The galaxy which produced the associated transient / source. For example, NGC~4993 is the host galaxy of GW170817
    \item[ICF - Inertial Confinement Fusion] One of two approaches towards nuclear fusion, focusing on the use of high intensity lasers to compress a fusion capsule to ignition.
    \item[Jets (astrophysics)] Collimated, bipolar, relativistic outflows
    \item[JLF - Jupiter Laser Facility] An institutional laser facility run by Lawrence Livermore National Lab (LLNL). Designed for high flexibility and direct user involvement (hands on). The JLF hosts the COMET, Janus, and Titan laser systems. 
    \item[Leptonic (astrophysics)] a source containing (or process involving) only leptons, i.e. a pair plasma of electrons or positrons. The absence of hadrons implies that we don’t expect a neutrino signal from this source, but we DO expect a 511~keV annihilation line in gamma rays.
    \item[LIGO - Laser Interferometer Gravitational-wave Observatory] the US gravitational wave interferometers, which have detected signals from compact binary mergers.
    \item[LLE - The Laboratory for Laser Energetics] In Rochester, NY, run by the University of Rochester. Hosts two National Laser User Facility laser facilities called OMEGA-60 and OMEGA-EP. Founded and continuously operated since 1970.
    \item[LMJ-PETAL - Laser Mega-Joule (France)] The LMJ/PETAL laser facility contains the LMJ laser and the PETAL laser which share the same target chamber. When used together they will reach 1~Mbar-1~Gbar and 100~eV-100~keV.  
    \item[LTE - Local Thermodynamic Equilibrium] A condition in which temperature equilibrium exists between plasma components (ions, electrons, radiation) that is confined to a region and time of the evolution of the plasma, rather than existing  everywhere. Implicit in this idea is that all electrons, all photons, and all ions can each be described as unified populations at equilibrium. 
    \item[LRP - Long Range Plan] The strategic planning document for nuclear science, operated under the NSAC
    \item[MPS - Directorate for Mathematical and Physical Sciences] The National Science Foundation directorate responsible for the fields of science of relevance here. Under MPS are the PHY and APS divisions.
    \item[NASA - National Aeronautics and Space Administration] The US agency responsible for space-based assets
    \item[ngVLA -  next-generation Very Large Array] A flagship radio facility prioritized in the astro 2020 decadal, under the purview of the NSF
    \item[NIF National Ignition Facility] A 192 beam laser experimental facility delivering 2 million joules and 500 trillion watts of power in a few ns to targets smaller than a pencil eraser, run by Lawrence Livermore National Laboratory. The most energetic laser in  the world, and the size of a sports stadium ($>$ 3 football fields). 
    \item[NLTE - Non-Local Thermodynamic Equilibrium] Very simply not in local thermodynamic equilibrium (LTE). With more complexity, describes systems with multiple temperatures, or where `temperature' does not have meaning because there is no useful `average' behavior. Can include multiple, widely varying populations of electrons, or ions. A plasma system in which  energy is entering / leaving or material is entering / leaving with time (and within the `memory' of neighboring material) is generally NLTE. 
    \item[NLUF - National Laser User Facility] A variety of laser-systems across the country used for High-Energy Density experimental science and Inertial Confinement Fusion. Includes proposals from national laboratories and universities which are competitively considered each year. 
    \item[NNSA - National Nuclear Security Administration] The Department of Energy (DOE) agency responsible for application of nuclear science to national security
    \item[NOIRLab - National Optical-Infrared Astronomy Research Laboratory] An NSF facility for optical and infrared astrophysics
    \item[Nuclear Region (astrophysics)] The center of a galaxy (unrelated to nuclear physics), typically where the central supermassive black hole is expected to be.
    \item[NSAC - Nuclear Science Advisory Committee (NSAC)] An advisory committee that advises the DOE and NSF on nuclear science
    \item[NSF - National Science Foundation] The US agency responsible for fundamental research and education (outside of medical research)
    \item[ORION Laser Facility] A laser facility frequently used for non-local thermodynamic equilibrium high-energy density experiments in Reading, UK. 
    \item[OST - Office of Science] The Department of Energy (DOE) office responsible for general science research.
    \item[P5 - Particle Physics Project Prioritization Panel] The scientific advisory panel tasked with prioritizing future investment in particle physics in the US.
    \item[PHY - The Division of Physics (NSF)] The NSF division which covers most fields of science of relevance for this workshop except non-electromagnetic astronomy and astrophysics.
    \item[Roman - Nancy Grace Roman Space Telescope (astrophysics)] Formerly the Wide-Field Infrared Survey Telescope or WFIRST, is going to be a NASA infrared space telescope with a 2.4m primary mirror with a 300 megapixel camera over a 0.28 square degree field of view and a spectrograph. 
    \item[Rubin - Vera C. Rubin Observatory (astrophysics)] Formerly known as the Large Synoptic Survey Telescope (LSST), the Rubin Observatory will survey the night sky at unprecedented depth and will disseminate ten million transient alerts per night.
    \item[Sub-Chandrasekhar (astrophysics)] White dwarfs below the Chandrasekhar limit of about 1.4 solar masses, which can produce different types of explosion properties, compared with Chandrasekhar mass white dwarfs.
    \item[Scale bridging (1)]  Using Ryutov or other methods to connect the mega-scales of astrophysical objects to the micro-scales of the laboratory. Such that the microscopic object will reach similar radiation and material conditions as the astrophysical object. Scale bridging (2): Also used to describe simulation methods that can predict properties of objects at large, macroscopic scales while still modeling relevant processes at microscopic scales.
    \item[UHECR - Ultra-High Energy Cosmic Rays (astrophysics)] Cosmic rays (electrons, protons, nuclei) from the cosmos with energies in excess of 1 EeV.
    \item[Ultrasat - Ultraviolet Transient Astronomy Satellite astrophysics)] An ultraviolet satellite led by Israel, with NASA participation
    \item[UVEX - The Ultraviolet Explorer (astrophysics)] UVEX is a NASA MIDEX selected in 2021 for launch in 2030. It will cover a $>$10deg$^{2}$ field of view with two 150 Mpix cameras covering the FUV and NUV bands. UVEX also includes a slit spectrograph that covers the FUV-NUV band with moderate spectral resolution. 
    \item[WDM - Warm Dense Matter (HEDP)] Cooler plasmas ($\lesssim$100~eV, 10,000~K; commonly $<$50eV) exceeding solid density. Frequently related to planetary interiors.
\end{description}

\newpage
\bibliography{bibliography}

\begin{thebibliography}{}
\expandafter\ifx\csname natexlab\endcsname\relax\def\natexlab#1{#1}\fi
\providecommand{\url}[1]{\href{#1}{#1}}

\bibitem[{{Abac} {et~al.}(2024){Abac}, {Abbott}, {Abouelfettouh}, {Acernese}, {Ackley}, {Adhicary}, {Adhikari}, {Adhikari}, {Adkins}, {Agarwal}, {Agathos}, {Abchouyeh}, {Aguiar}, {Aguilar}, {Aiello}, {Ain}, {Ajith}, {Ak{\c{c}}ay}, {Akutsu}, {Albanesi}, {Alfaidi}, {Al-Jodah}, {All{\'e}n{\'e}}, {Allocca}, {Al-Shammari}, {Altin}, {Alvarez-Lopez}, {Amato}, {Amez-Droz}, {Amorosi}, {Amra}, {Ananyeva}, {Anderson}, {Anderson}, {Andia}, {Ando}, {Andrade}, {Andres}, {Andr{\'e}s-Carcasona}, {Andri{\'c}}, {Anglin}, {Ansoldi}, {Antelis}, {Antier}, {Aoumi}, {Appavuravther}, {Appert}, {Apple}, {Arai}, {Araya}, {Araya}, {Areeda}, {Argianas}, {Aritomi}, {Armato}, {Arnaud}, {Arogeti}, {Aronson}, {Arun}, {Ashton}, {Aso}, {Assiduo}, {de Souza Melo}, {Aston}, {Astone}, {Attadio}, {Aubin}, {Aultoneal}, {Avallone}, {Azrad}, {Babak}, {Badaracco}, {Badger}, {Bae}, {Bagnasco}, {Bagui}, {Baier}, {Baiotti}, {Bajpai}, {Baka}, {Ball}, {Ballardin}, {Ballmer}, {Banagiri}, {Banerjee}, {Bankar}, {Baral}, {Barayoga}, {Barish}, {Barker},
  {Barneo}, {Barone}, {Barr}, {Barsotti}, {Barsuglia}, {Barta}, {Bartoletti}, {Barton}, {Bartos}, {Basak}, {Basalaev}, {Bassiri}, {Basti}, {Bates}, {Bawaj}, {Baxi}, {Bayley}, {Baylor}, {Baynard}, {Bazzan}, {Bedakihale}, {Beirnaert}, {Bejger}, {Belardinelli}, {Bell}, {Benedetto}, {Benoit}, {Bentara}, {Bentley}, {Ben Yaala}, {Bera}, {Berbel}, {Bergamin}, {Berger}, {Bernuzzi}, {Beroiz}, {Berry}, {Bersanetti}, {Bertolini}, {Betzwieser}, {Beveridge}, {Bevins}, {Bhandare}, {Bhardwaj}, {Bhatt}, {Bhattacharjee}, {Bhaumik}, {Bhowmick}, {Bianchi}, {Bilenko}, {Billingsley}, {Binetti}, {Bini}, {Birnholtz}, {Biscoveanu}, {Bisht}, {Bitossi}, {Bizouard}, {Blackburn}, {Blagg}, {Blair}, {Blair}, {Bobba}, {Bode}, {Boileau}, {Boldrini}, {Bolingbroke}, {Bolliand}, {Bonavena}, {Bondarescu}, {Bondu}, {Bonilla}, {Bonilla}, {Bonino}, {Bonnand}, {Booker}, {Borchers}, {Boschi}, {Bose}, {Bossilkov}, {Boudart}, {Boudon}, {Bozzi}, {Bradaschia}, {Brady}, {Braglia}, {Branch}, {Branchesi}, {Brandt}, {Braun}, {Breschi}, {Briant}, {Brillet},
  {Brinkmann}, {Brockill}, {Brockmueller}, {Brooks}, {Brown}, {Brown}, {Brozzetti}, {Brunett}, {Bruno}, {Bruntz}, {Bryant}, {Bucci}, {Buchanan}, {Bulashenko}, {Bulik}, {Bulten}, {Buonanno}, {Burtnyk}, {Buscicchio}, {Buskulic}, {Buy}, {Byer}, {Cabourn Davies}, {Cabras}, {Cabrita}, {C{\'a}ceres-Barbosa}, {Cadonati}, {Cagnoli}, {Cahillane}, {Bustillo}, {Callister}, {Calloni}, {Camp}, {Canepa}, {Caneva Santoro}, {Cannon}, {Cao}, {Capistran}, {Capocasa}, {Capote}, {Carapella}, {Carbognani}, {Carlassara}, {Carlin}, {Carpinelli}, {Carrillo}, {Carter}, {Carullo}, {Casanueva Diaz}, {Casentini}, {Castro-Lucas}, {Caudill}, {Cavagli{\`a}}, {Cavalieri}, {Cella}, {Cerd{\'a}-Dur{\'a}n}, {Cesarini}, {Chaibi}, {Chakraborty}, {Subrahmanya}, {Chan}, {Chan}, {Chandra}, {Chang}, {Chao}, {Char}, {Charlton}, {Charlton}, {Chassande-Mottin}, {Chatterjee}, {Chatterjee}, {Chatterjee}, {Chattopadhyay}, {Chaturvedi}, {Chaty}, {Chatziioannou}, {Chen}, {Chen}, {Chen}, {Chen}, {Chen}, {Chen}, {Chen}, {Chen}, {Chen}, {Chen}, {Cheng},
  {Chessa}, {Cheung}, {Cheung}, {Chiadini}, {Chiarini}, {Chierici}, {Chincarini}, {Chiofalo}, {Chiummo}, {Chou}, {Choudhary}, {Christensen}, {Chua}, {Chugh}, {Ciani}, {Ciecielag}, {Cie{\'s}lar}, {Cifaldi}, {Ciolfi}, {Clara}, {Clark}, {Clarke}, {Clarke}, {Clearwater}, {Clesse}, {Coccia}, {Codazzo}, {Cohadon}, {Colace}, {Colleoni}, {Collette}, {Collins}, {Colloms}, {Colombo}, {Colpi}, {Compton}, {Connolly}, {Conti}, {Corbitt}, {Cordero-Carri{\'o}n}, {Corezzi}, {Cornish}, {Corsi}, {Cortese}, {Costa}, {Cottingham}, {Coughlin}, {Couineaux}, {Coulon}, {Countryman}, {Coupechoux}, {Couvares}, {Coward}, {Cowart}, {Coyne}, {Craig}, {Creed}, {Creighton}, {Creighton}, {Cremonese}, {Criswell}, {Crockett-Gray}, {Crook}, {Crouch}, {Csizmazia}, {Cudell}, {Cullen}, {Cumming}, {Cuoco}, {Cusinato}, {Dabadie}, {Dal Canton}, {Dall'Osso}, {Dal Pra}, {D{\'a}lya}, {D'Angelo}, {Danilishin}, {D'Antonio}, {Danzmann}, {Darroch}, {Dartez}, {Dasgupta}, {Datta}, {Dattilo}, {Daumas}, {Davari}, {Dave}, {Davenport}, {Davier}, {Davies},
  {Davis}, {Davis}, {Davis}, {Davis}, {Dax}, {de Bolle}, {Deenadayalan}, {Degallaix}, {de Laurentis}, {Del{\'e}glise}, {de Lillo}, {Dell'Aquila}, {Del Pozzo}, {De Marco}, {de Matteis}, {D'Emilio}, {Demos}, {Dent}, {Depasse}, {Depergola}, {de Pietri}, {De Rosa}, {de Rossi}, {Desalvo}, {de Simone}, {Dhani}, {Diab}, {D{\'\i}az}, {di Cesare}, {Dideron}, {Didio}, {Dietrich}, {di Fiore}, {di Fronzo}, {di Giovanni}, {di Girolamo}, {Diksha}, {di Michele}, {Ding}, {di Pace}, {di Palma}, {di Renzo}, {Divyajyoti}, {Dmitriev}, {Doctor}, {Dohmen}, {Doleva}, {Dominguez}, {D'Onofrio}, {Donovan}, {Dooley}, {Dooney}, {Doravari}, {Dorosh}, {Drago}, {Driggers}, {Ducoin}, {Dunn}, {Dupletsa}, {D'Urso}, {Duval}, {Duverne}, {Dwyer}, {Eassa}, {Ebersold}, {Eckhardt}, {Eddolls}, {Edelman}, {Edo}, {Edy}, {Effler}, {Eichholz}, {Einsle}, {Eisenmann}, {Eisenstein}, {Ejlli}, {Eleveld}, {Emma}, {Endo}, {Engl}, {Enloe}, {Errico}, {Essick}, {Estell{\'e}s}, {Estevez}, {Etzel}, {Evans}, {Evstafyeva}, {Ewing}, {Ezquiaga}, {Fabrizi}, {Faedi},
  {Fafone}, {Fairhurst}, {Farah}, {Farr}, {Farr}, {Favaro}, {Favata}, {Fays}, {Fazio}, {Feicht}, {Fejer}, {Felicetti}, {Fenyvesi}, {Ferguson}, {Ferraiuolo}, {Ferrante}, {Ferreira}, {Fidecaro}, {Figura}, {Fiori}, {Fiori}, {Fishbach}, {Fisher}, {Fittipaldi}, {Fiumara}, {Flaminio}, {Fleischer}, {Fleming}, {Floden}, {Foley}, {Fong}, {Font}, {Fornal}, {Forsyth}, {Franceschetti}, {Franchini}, {Frasca}, {Frasconi}, {Mascioli}, {Frei}, {Freise}, {Freitas}, {Frey}, {Frischhertz}, {Fritschel}, {Frolov}, {Fronz{\'e}}, {Fuentes-Garcia}, {Fujii}, {Fujimori}, {Fulda}, {Fyffe}, {Gadre}, {Gair}, {Galaudage}, {Galdi}, {Gallagher}, {Gallardo}, {Gallego}, {Gamba}, {Gamboa}, {Ganapathy}, {Ganguly}, {Garaventa}, {Garc{\'\i}a-Bellido}, {Garc{\'\i}a N{\'u}{\~n}ez}, {Garc{\'\i}a-Quir{\'o}s}, {Gardner}, {Gardner}, {Gargiulo}, {Garron}, {Garufi}, {Gasbarra}, {Gateley}, {Gayathri}, {Gemme}, {Gennai}, {Gennari}, {George}, {George}, {Gerberding}, {Gergely}, {Ghonge}, {Ghosh}, {Ghosh}, {Ghosh}, {Ghosh}, {Ghosh}, {Ghosh}, {Giacoppo},
  {Giaime}, {Giardina}, {Gibson}, {Gibson}, {Gier}, {Giri}, {Gissi}, {Gkaitatzis}, {Glanzer}, {Glotin}, {Godfrey}, {Godwin}, {Goebbels}, {Goetz}, {Golomb}, {Gomez Lopez}, {Goncharov}, {Gong}, {Gonz{\'a}lez}, {Goodarzi}, {Goode}, {Goodwin-Jones}, {Gosselin}, {G{\"o}ttel}, {Gouaty}, {Gould}, {Govorkova}, {Goyal}, {Grace}, {Grado}, {Graham}, {Granados}, {Granata}, {Granata}, {Gras}, {Grassia}, {Gray}, {Gray}, {Gray}, {Greco}, {Green}, {Green}, {Green}, {Gretarsson}, {Gretarsson}, {Griffith}, {Griffiths}, {Griggs}, {Grignani}, {Grimaldi}, {Grimaud}, {Grote}, {Guerra}, {Guetta}, {Guidi}, {Guimaraes}, {Gulati}, {Gulminelli}, {Gunny}, {Guo}, {Guo}, {Guo}, {Gupta}, {Gupta}, {Gupta}, {Gupta}, {Gupta}, {Gupta}, {Gupta}, {Gupte}, {Gurs}, {Gutierrez}, {Guzman}, {H}, {Haba}, {Haberland}, {Haino}, {Hall}, {Hamilton}, {Hammond}, {Han}, {Haney}, {Hanks}, {Hanna}, {Hannam}, {Hannuksela}, {Hanselman}, {Hansen}, {Hanson}, {Harada}, {Hardison}, {Haris}, {Harmark}, {Harms}, {Harry}, {Harry}, {Hart}, {Haskell}, {Haster},
  {Hathaway}, {Haughian}, {Hayakawa}, {Hayama}, {Hayes}, {Heffernan}, {Heidmann}, {Heintze}, {Heinze}, {Heinzel}, {Heitmann}, {Hellman}, {Hello}, {Helmling-Cornell}, {Hemming}, {Henderson-Sapir}, {Hendry}, {Heng}, {Hennes}, {Henshaw}, {Hertog}, {Heurs}, {Hewitt}, {Heyns}, {Higginbotham}, {Hild}, {Hill}, {Himemoto}, {Hirata}, {Hirose}, {Hoang}, {Hochheim}, {Hofman}, {Holland}, {Holley-Bockelmann}, {Holmes}, {Holz}, {Honet}, {Hong}, {Hornung}, {Hoshino}, {Hough}, {Hourihane}, {Howell}, {Hoy}, {Hrishikesh}, {Hsieh}, {Hsiung}, {Hsu}, {Hsu}, {Hu}, {Hu}, {Huang}, {Huang}, {Huddart}, {Hughey}, {Hui}, {Hui}, {Husa}, {Huxford}, {Huynh-Dinh}, {Iampieri}, {Iandolo}, {Ianni}, {Iess}, {Imafuku}, {Inayoshi}, {Inoue}, {Iorio}, {Iqbal}, {Irwin}, {Ishikawa}, {Isi}, {Ismail}, {Itoh}, {Iwanaga}, {Iwaya}, {Iyer}, {Jaberianhamedan}, {Jacquet}, {Jacquet}, {Jadhav}, {Jadhav}, {Jain}, {James}, {James}, {Jamshidi}, {Janquart}, {Janssens}, {Janthalur}, {Jaraba}, {Jaranowski}, {Jaume}, {Javed}, {Jennings}, {Jia}, {Jiang}, {Kubisz},
  {Johanson}, {Johns}, {Johnson}, {Johnson-McDaniel}, {Johnston}, {Johnston}, {Johny}, {Jones}, {Jones}, {Jones}, {Jose}, {Joshi}, {Ju}, {Jung}, {Junker}, {Juste}, {Kajita}, {Kaku}, {Kalaghatgi}, {Kalogera}, {Kamiizumi}, {Kanda}, {Kandhasamy}, {Kang}, {Kanner}, {Kapadia}, {Kapasi}, {Karat}, {Karathanasis}, {Kashyap}, {Kasprzack}, {Kastaun}, {Kato}, {Katsavounidis}, {Katzman}, {Kaushik}, {Kawabe}, {Kawamoto}, {Kazemi}, {Kedia}, {Keitel}, {Kelley-Derzon}, {Kennington}, {Kesharwani}, {Key}, {Khadela}, {Khadka}, {Khalili}, {Khan}, {Khan}, {Khanam}, {Khursheed}, {Khusid}, {Kiendrebeogo}, {Kijbunchoo}, {Kim}, {Kim}, {Kim}, {Kim}, {Kim}, {Kim}, {Kimball}, {Kinley-Hanlon}, {Kinnear}, {Kissel}, {Klimenko}, {Knee}, {Knust}, {Kobayashi}, {Koch}, {Koehlenbeck}, {Koekoek}, {Kohri}, {Kokeyama}, {Koley}, {Kolitsidou}, {Kolstein}, {Komori}, {Kong}, {Kontos}, {Korobko}, {Kossak}, {Kou}, {Koushik}, {Kouvatsos}, {Kovalam}, {Kozak}, {Kranzhoff}, {Kringel}, {Krishnendu}, {Kr{\'o}lak}, {Kruska}, {Kuehn}, {Kuijer}, {Kulkarni},
  {Ramamohan}, {Kumar}, {Kumar}, {Kumar}, {Kumar}, {Kumar}, {Kume}, {Kuns}, {Kuntimaddi}, {Kuroyanagi}, {Kurth}, {Kuwahara}, {Kwak}, {Kwan}, {Kwok}, {Lacaille}, {Lagabbe}, {Laghi}, {Lai}, {Laity}, {Lakkis}, {Lalande}, {Lalleman}, {Lalremruati}, {Landry}, {Landry}, {Lane}, {Lang}, {Lange}, {Lantz}, {La Rana}, {La Rosa}, {Lartaux-Vollard}, {Lasky}, {Lawrence}, {Lawrence}, {Laxen}, {Lazzarini}, {Lazzaro}, {Leaci}, {Lecoeuche}, {Lee}, {Lee}, {Lee}, {Lee}, {Lee}, {Lee}, {Lee}, {Legred}, {Lehmann}, {Lehner}, {Le Jean}, {Lema{\^\i}tre}, {Lenti}, {Leonardi}, {Lequime}, {Leroy}, {Lesovsky}, {Letendre}, {Lethuillier}, {Levin}, {Levin}, {Leyde}, {Li}, {Li}, {Li}, {Li}, {Li}, {Lihos}, {Lin}, {Lin}, {Lin}, {Lin}, {Lin}, {Lin}, {Lin}, {Linde}, {Linker}, {Littenberg}, {Liu}, {Liu}, {Liu}, {Villarreal}, {Llobera-Querol}, {Lo}, {Locquet}, {London}, {Longo}, {Lopez}, {Lopez Portilla}, {Lorenzini}, {Lorenzo-Medina}, {Loriette}, {Lormand}, {Losurdo}, {}, {Lough}, {Loughlin}, {Lousto}, {Lowry}, {Lu}, {L{\"u}ck}, {Lumaca},
  {Lundgren}, {Lussier}, {Ma}, {Ma}, {Ma'Arif}, {Macas}, {Macedo}, {Macinnis}, {Maciy}, {MacLeod}, {MacMillan}, {Macquet}, {Macri}, {Maeda}, {Maenaut}, {Hernandez}, {Magare}, {Magazz{\`u}}, {Magee}, {Maggio}, {Maggiore}, {Magnozzi}, {Mahesh}, {Mahesh}, {Maini}, {Majhi}, {Majorana}, {Makarem}, {Makelele}, {Malaquias-Reis}, {Mali}, {Maliakal}, {Malik}, {Man}, {Mandic}, {Mangano}, {Mannix}, {Mansell}, {Mansingh}, {Manske}, {Mantovani}, {Mapelli}, {Marchesoni}, {Mar{\'\i}n Pina}, {Marion}, {M{\'a}rka}, {M{\'a}rka}, {Markosyan}, {Markowitz}, {Maros}, {Marsat}, {Martelli}, {Martin}, {Martin}, {Martinez}, {Martinez}, {Martinez}, {Martini}, {Martinovic}, {Martins}, {Martynov}, {Marx}, {Massaro}, {Masserot}, {Masso-Reid}, {Mastrodicasa}, {Mastrogiovanni}, {Matcovich}, {Matiushechkina}, {Matsuyama}, {Mavalvala}, {Maxwell}, {McCarrol}, {McCarthy}, {McClelland}, {McCormick}, {McCuller}, {McEachin}, {McElhenny}, {McGhee}, {McGinn}, {McGowan}, {McIver}, {McLeod}, {McRae}, {Meacher}, {Meijer}, {Melatos}, {Mellaerts},
  {Menendez-Vazquez}, {Menoni}, {Mera}, {Mercer}, {Mereni}, {Merfeld}, {Merilh}, {M{\'e}rou}, {Merritt}, {Merzougui}, {Messenger}, {Messick}, {Meyer-Conde}, {Meylahn}, {Mhaske}, {Miani}, {Miao}, {Michaloliakos}, {Michel}, {Michimura}, {Middleton}, {Miller}, {Miller}, {Millhouse}, {Milotti}, {Milotti}, {Minenkov}, {Mio}, {Mir}, {Mirasola}, {Miravet-Ten{\'e}s}, {Miritescu}, {Mishra}, {Mishra}, {Mishra}, {Mishra}, {Mitchell}, {Mitchell}, {Mitra}, {Mitrofanov}, {Mittleman}, {Miyakawa}, {Miyamoto}, {Miyoki}, {Mo}, {Mobilia}, {Mohapatra}, {Mohite}, {Molina-Ruiz}, {Mondal}, {Mondin}, {Montani}, {Moore}, {Moraru}, {More}, {More}, {Moreno}, {Morgan}, {Morisaki}, {Moriwaki}, {Morras}, {Moscatello}, {Mourier}, {Mours}, {Mow-Lowry}, {Muciaccia}, {Mukherjee}, {Mukherjee}, {Mukherjee}, {Mukherjee}, {Mukherjee}, {Mukherjee}, {Mukund}, {Mullavey}, {Munch}, {Mundi}, {Mungioli}, {Oberg}, {Murakami}, {Murakoshi}, {Murray}, {Muusse}, {Nabari}, {Nadji}, {Nagar}, {Nagarajan}, {Nagler}, {Nakagaki}, {Nakamura}, {Nakano}, {Nakano},
  {Nandi}, {Napolano}, {Narayan}, {Nardecchia}, {Narikawa}, {Narola}, {Naticchioni}, {Nayak}, {Neilson}, {Nelson}, {Nelson}, {Nery}, {Neunzert}, {Ng}, {Quynh}, {Nichols}, {Nielsen}, {Nieradka}, {Niko}, {Nishino}, {Nishizawa}, {Nissanke}, {Nitoglia}, {Niu}, {Nocera}, {Norman}, {North}, {Novak}, {Nu{\~n}o Siles}, {Nuttall}, {Obayashi}, {Oberling}, {O'Dell}, {Oertel}, {Offermans}, {Oganesyan}, {Oh}, {Oh}, {O'Hanlon}, {Ohashi}, {Ohkawa}, {Ohme}, {Oliveira}, {Oliveri}, {O'Neal}, {Oohara}, {O'Reilly}, {Ormsby}, {Orselli}, {O'Shaughnessy}, {O'Shea}, {Oshima}, {Oshino}, {Ossokine}, {Osthelder}, {Ota}, {Ottaway}, {Ouzriat}, {Overmier}, {Owen}, {Pace}, {Pagano}, {Page}, {Pai}, {Pal}, {Pal}, {Palaia}, {P{\'a}lfi}, {Palma}, {Palomba}, {Palud}, {Pan}, {Pan}, {Pan}, {Panai}, {Panda}, {Pandey}, {Panebianco}, {Pang}, {Pannarale}, {Pannone}, {Pant}, {Panther}, {Paoletti}, {Paolone}, {Papalexakis}, {Papalini}, {Papigkiotis}, {Paquis}, {Parisi}, {Park}, {Park}, {Parker}, {Pascale}, {Pascucci}, {Pasqualetti}, {Passaquieti},
  {Passenger}, {Passuello}, {Patane}, {Pathak}, {Pathak}, {Patra}, {Patricelli}, {Patron}, {Paul}, {Paul}, {Payne}, {Pearce}, {Pedraza}, {Pegna}, {Pele}, {Arellano}, {Penn}, {Penuliar}, {Perego}, {Pereira}, {Perez}, {P{\'e}rigois}, {Perna}, {Perreca}, {Perret}, {Perri{\`e}s}, {Perry}, {Pesios}, {Petracca}, {Petrillo}, {Pfeiffer}, {Pham}, {Pham}, {Phukon}, {Phurailatpam}, {Piarulli}, {Piccari}, {Piccinni}, {Pichot}, {Piendibene}, {Piergiovanni}, {Pierini}, {Pierra}, {Pierro}, {Pietrzak}, {Pillas}, {Pilo}, {Pinard}, {Pinto}, {Pinto}, {Piotrzkowski}, {Pirello}, {Pitkin}, {Placidi}, {Placidi}, {Planas}, {Plastino}, {Poggiani}, {Polini}, {Pompili}, {Poon}, {Porcelli}, {Porter}, {Posnansky}, {Poulton}, {Powell}, {Pracchia}, {Pradhan}, {Pradier}, {Prajapati}, {Prasai}, {Prasanna}, {Prasia}, {Pratten}, {Principe}, {Principe}, {Prodi}, {Prokhorov}, {Prosposito}, {Puecher}, {Pullin}, {Punturo}, {Puppo}, {P{\"u}rrer}, {Qi}, {Qin}, {Qu{\'e}m{\'e}ner}, {Quetschke}, {Quigley}, {Quinonez}, {Raab}, {Raabith}, {Raaijmakers},
  {Raja}, {Rajan}, {Rajbhandari}, {Ramirez}, {Vidal}, {Ramos-Buades}, {Rana}, {Ranjan}, {Ransom}, {Rapagnani}, {Ratto}, {Rawat}, {Ray}, {Raymond}, {Razzano}, {Read}, {Payo}, {Regimbau}, {Rei}, {Reid}, {Reitze}, {Relton}, {Renzini}, {Rettegno}, {Revenu}, {Reyes}, {Rezaei}, {Ricci}, {Ricci}, {Ricciardone}, {Richardson}, {Richardson}, {Rijal}, {Riles}, {Riley}, {Rinaldi}, {Rittmeyer}, {Robertson}, {Robinet}, {Robinson}, {Rocchi}, {Rolland}, {Rollins}, {Romano}, {Romano}, {Romero}, {Romero-Shaw}, {Romie}, {Ronchini}, {Roocke}, {Rosa}, {Rosauer}, {Rose}, {Rosi{\'n}ska}, {Ross}, {Rossello}, {Rowan}, {Roy}, {Roy}, {Rozza}, {Ruggi}, {Ruhama}, {Morales}, {Ruiz-Rocha}, {Sachdev}, {Sadecki}, {Sadiq}, {Saffarieh}, {Sah}, {Saha}, {Saha}, {Sainrat}, {Menon}, {Sakai}, {Sakellariadou}, {Sakon}, {Salafia}, {Salces-Carcoba}, {Salconi}, {Saleem}, {Salemi}, {Sall{\'e}}, {Salvador}, {Sanchez}, {Sanchez}, {Sanchez}, {Sanchez}, {Sanchis-Gual}, {Sanders}, {S{\"a}nger}, {Santoliquido}, {Saravanan}, {Sarin}, {Sasaoka}, {Sasli},
  {Sassi}, {Sassolas}, {Satari}, {Sathyaprakash}, {Sato}, {Sato}, {Sauter}, {Savage}, {Sawada}, {Sawant}, {Sayah}, {Scacco}, {Schaetzl}, {Scheel}, {Schiebelbein}, {Schiworski}, {Schmidt}, {Schmidt}, {Schnabel}, {Schneewind}, {Schofield}, {Schouteden}, {Schulte}, {Schutz}, {Schwartz}, {Scialpi}, {Scott}, {Scott}, {Seetharamu}, {Seglar-Arroyo}, {Sekiguchi}, {Sellers}, {Sengupta}, {Sentenac}, {Seo}, {Seo}, {Sequino}, {Serra}, {Servignat}, {Sevrin}, {Shaffer}, {Shah}, {Shaikh}, {Shao}, {Sharma}, {Sharma}, {Sharma-Chaudhary}, {Shaw}, {Shawhan}, {Shcheblanov}, {Sheridan}, {Shikano}, {Shikauchi}, {Shimode}, {Shinkai}, {Shiota}, {Shoemaker}, {Shoemaker}, {Short}, {Shyamsundar}, {Sider}, {Siegel}, {Sieniawska}, {Sigg}, {Silenzi}, {Simmonds}, {Singer}, {Singh}, {Singh}, {Singh}, {Singh}, {Singha}, {Sintes}, {Sipala}, {Skliris}, {Slagmolen}, {Slaven-Blair}, {Smetana}, {Smith}, {Smith}, {Smith}, {Smith}, {Soldateschi}, {Somiya}, {Song}, {Soni}, {Soni}, {Sordini}, {Sorrentino}, {Sorrentino}, {Sotani}, {Soulard},
  {Southgate}, {Spagnuolo}, {Spencer}, {Spera}, {Spinicelli}, {Spoon}, {Sprague}, {Srivastava}, {Stachurski}, {Steer}, {Steinlechner}, {Steinlechner}, {Stergioulas}, {Stevens}, {Stevenson}, {Stpierre}, {Stratta}, {Strong}, {Strunk}, {Sturani}, {Stuver}, {Suchenek}, {Sudhagar}, {Sueltmann}, {Suleiman}, {Sullivan}, {Sun}, {Sunil}, {Suresh}, {Sutton}, {Suzuki}, {Suzuki}, {Swinkels}, {Syx}, {Szczepa{\'n}czyk}, {Szewczyk}, {Tacca}, {Tagoshi}, {Tait}, {Takahashi}, {Takahashi}, {Takamori}, {Takase}, {Takatani}, {Takeda}, {Takeshita}, {Talbot}, {Tamaki}, {Tamanini}, {Tanabe}, {Tanaka}, {Tanaka}, {Tanaka}, {Tang}, {Tanioka}, {Tanner}, {Tao}, {Tapia}, {San Mart{\'\i}n}, {Tarafder}, {Taranto}, {Taruya}, {Tasson}, {Teloi}, {Tenorio}, {Themann}, {Theodoropoulos}, {Thirugnanasambandam}, {Thomas}, {Thomas}, {Thomas}, {Thompson}, {Thondapu}, {Thorne}, {Thrane}, {Tissino}, {Tiwari}, {Tiwari}, {Tiwari}, {Tiwari}, {Todd}, {Toivonen}, {Toland}, {Tolley}, {Tomaru}, {Tomita}, {Tomura}, {Tong}, {Tong-Yu}, {Toriyama}, {Toropov},
  {Torres-Forn{\'e}}, {Torrie}, {Toscani}, {E Melo}, {Tournefier}, {Trapananti}, {Travasso}, {Traylor}, {Trevor}, {Tringali}, {Tripathee}, {Troian}, {Troiano}, {Trovato}, {Trozzo}, {Trudeau}, {Tsang}, {Tso}, {Tsuchida}, {Tsukada}, {Tsutsui}, {Turbang}, {Turconi}, {Turski}, {Ubach}, {Uchikata}, {Uchiyama}, {Udall}, {Uehara}, {Uematsu}, {Ueno}, {Ueno}, {Undheim}, {Ushiba}, {Vacatello}, {Vahlbruch}, {Vaidya}, {Vajente}, {Vajpeyi}, {Valdes}, {Valencia}, {Valentini}, {Vallejo-Pe{\~n}a}, {Vallero}, {Valsan}, {van Bakel}, {van Beuzekom}, {van Dael}, {van den Brand}, {Broeck}, {Vander-Hyde}, {van der Sluys}, {van de Walle}, {van Dongen}, {Vandra}, {van Haevermaet}, {van Heijningen}, {van Hove}, {Vankeuren}, {Vanosky}, {van Putten}, {van Ranst}, {van Remortel}, {Vardaro}, {Vargas}, {Varghese}, {Varma}, {Vas{\'u}th}, {Vecchio}, {Vedovato}, {Veitch}, {Veitch}, {Venikoudis}, {Venneberg}, {Verdier}, {Verkindt}, {Verma}, {Verma}, {Verma}, {Vermeulen}, {Vetrano}, {Veutro}, {Vibhute}, {Vicer{\'e}}, {Vidyant}, {Viets},
  {Vijaykumar}, {Vilkha}, {Villa-Ortega}, {Vincent}, {Vinet}, {Viret}, {Virtuoso}, {Vitale}, {Vives}, {Vocca}, {Voigt}, {von Reis}, {von Wrangel}, {Vyatchanin}, {Wade}, {Wade}, {Wagner}, {Wajid}, {Walker}, {Wallace}, {Wallace}, {Wang}, {Wang}, {Wang}, {Wang}, {Waratkar}, {Warner}, {Was}, {Washimi}, {Washington}, {Watarai}, {Wayt}, {Weaver}, {Weaver}, {Weaving}, {Webster}, {Weinert}, {Weinstein}, {Weiss}, {Wellmann}, {Wen}, {We{\ss}els}, {Wette}, {Whelan}, {Whiting}, {Whittle}, {Wildberger}, {Wilk}, {Wilken}, {Wilkin}, {Willadsen}, {Willetts}, {Williams}, {Williams}, {Williams}, {Willis}, {Willke}, {Wils}, {Winterflood}, {Wipf}, {Woan}, {Woehler}, {Wofford}, {Wolfe}, {Wong}, {Wong}, {Wong}, {Wright}, {Wright}, {Wu}, {Wu}, {Wu}, {Wuchner}, {Wysocki}, {Xu}, {Xu}, {Yadav}, {Yamamoto}, {Yamamoto}, {Yamamoto}, {Yamamoto}, {Yamamura}, {Yamazaki}, {Yan}, {Yan}, {Yang}, {Yang}, {Yang}, {Yang}, {Yarbrough}, {Yasui}, {Yeh}, {Yelikar}, {Yin}, {Yokoyama}, {Yokozawa}, {Yoo}, {Yu}, {Yuan}, {Yuzurihara}, {Zadro{\.z}ny},
  {Zanolin}, {Zeeshan}, {Zelenova}, {Zendri}, {Zeoli}, {Zerrad}, {Zevin}, {Zhang}, {Zhang}, {Zhang}, {Zhang}, {Zhang}, {Zhao}, {Zhao}, {Zhao}, {Zheng}, {Zhong}, {Zhou}, {Zhu}, {Zhu}, {Zimmerman}, {Zucker}, {Zweizig}, {Ligo Scientific Collaboration}, {VIRGO Collaboration}, \& {Kagra Collaboration}}]{Abac2024}
{Abac}, A.~G., {Abbott}, R., {Abouelfettouh}, I., {et~al.} 2024, \apjl, 970, L34

\bibitem[{Abbott {et~al.}(2017)}]{GW170817}
Abbott, B.~P., {et~al.} 2017, \prl, 119, 161101

\bibitem[{{Abbott} {et~al.}(2017){Abbott}, {Abbott}, {Abbott}, {Acernese}, {Ackley}, {Adams}, {Adams}, {Addesso}, {Adhikari}, {Adya}, {Affeldt}, {Afrough}, {Agarwal}, {Agathos}, {Agatsuma}, {Aggarwal}, {Aguiar}, {Aiello}, {Ain}, {Ajith}, {Allen}, {Allen}, {Allocca}, {Altin}, {Amato}, {Ananyeva}, {Anderson}, {Anderson}, {Angelova}, {Antier}, {Appert}, {Arai}, {Araya}, {Areeda}, {Arnaud}, {Arun}, {Ascenzi}, {Ashton}, {Ast}, {Aston}, {Astone}, {Atallah}, {Aufmuth}, {Aulbert}, {AultONeal}, {Austin}, {Avila-Alvarez}, {Babak}, {Bacon}, {Bader}, {Bae}, {Baker}, {Baldaccini}, {Ballardin}, {Ballmer}, {Banagiri}, {Barayoga}, {Barclay}, {Barish}, {Barker}, {Barkett}, {Barone}, {Barr}, {Barsotti}, {Barsuglia}, {Barta}, {Barthelmy}, {Bartlett}, {Bartos}, {Bassiri}, {Basti}, {Batch}, {Bawaj}, {Bayley}, {Bazzan}, {B{\'e}csy}, {Beer}, {Bejger}, {Belahcene}, {Bell}, {Berger}, {Bergmann}, {Bero}, {Berry}, {Bersanetti}, {Bertolini}, {Betzwieser}, {Bhagwat}, {Bhandare}, {Bilenko}, {Billingsley}, {Billman}, {Birch}, {Birney},
  {Birnholtz}, {Biscans}, {Biscoveanu}, {Bisht}, {Bitossi}, {Biwer}, {Bizouard}, {Blackburn}, {Blackman}, {Blair}, {Blair}, {Blair}, {Bloemen}, {Bock}, {Bode}, {Boer}, {Bogaert}, {Bohe}, {Bondu}, {Bonilla}, {Bonnand}, {Boom}, {Bork}, {Boschi}, {Bose}, {Bossie}, {Bouffanais}, {Bozzi}, {Bradaschia}, {Brady}, {Branchesi}, {Brau}, {Briant}, {Brillet}, {Brinkmann}, {Brisson}, {Brockill}, {Broida}, {Brooks}, {Brown}, {Brown}, {Brunett}, {Buchanan}, {Buikema}, {Bulik}, {Bulten}, {Buonanno}, {Buskulic}, {Buy}, {Byer}, {Cabero}, {Cadonati}, {Cagnoli}, {Cahillane}, {Calder{\'o}n Bustillo}, {Callister}, {Calloni}, {Camp}, {Canepa}, {Canizares}, {Cannon}, {Cao}, {Cao}, {Capano}, {Capocasa}, {Carbognani}, {Caride}, {Carney}, {Casanueva Diaz}, {Casentini}, {Caudill}, {Cavagli{\`a}}, {Cavalier}, {Cavalieri}, {Cella}, {Cepeda}, {Cerd{\'a}-Dur{\'a}n}, {Cerretani}, {Cesarini}, {Chamberlin}, {Chan}, {Chao}, {Charlton}, {Chase}, {Chassande-Mottin}, {Chatterjee}, {Chatziioannou}, {Cheeseboro}, {Chen}, {Chen}, {Chen}, {Cheng},
  {Chia}, {Chincarini}, {Chiummo}, {Chmiel}, {Cho}, {Cho}, {Chow}, {Christensen}, {Chu}, {Chua}, {Chua}, {Chung}, {Chung}, {Ciani}, {Ciolfi}, {Cirelli}, {Cirone}, {Clara}, {Clark}, {Clearwater}, {Cleva}, {Cocchieri}, {Coccia}, {Cohadon}, {Cohen}, {Colla}, {Collette}, {Cominsky}, {Constancio}, {Conti}, {Cooper}, {Corban}, {Corbitt}, {Cordero-Carri{\'o}n}, {Corley}, {Cornish}, {Corsi}, {Cortese}, {Costa}, {Coughlin}, {Coughlin}, {Coulon}, {Countryman}, {Couvares}, {Covas}, {Cowan}, {Coward}, {Cowart}, {Coyne}, {Coyne}, {Creighton}, {Creighton}, {Cripe}, {Crowder}, {Cullen}, {Cumming}, {Cunningham}, {Cuoco}, {Dal Canton}, {D{\'a}lya}, {Danilishin}, {D'Antonio}, {Danzmann}, {Dasgupta}, {Da Silva Costa}, {Dattilo}, {Dave}, {Davier}, {Davis}, {Daw}, {Day}, {De}, {DeBra}, {Degallaix}, {De Laurentis}, {Del{\'e}glise}, {Del Pozzo}, {Demos}, {Denker}, {Dent}, {De Pietri}, {Dergachev}, {De Rosa}, {DeRosa}, {De Rossi}, {DeSalvo}, {de Varona}, {Devenson}, {Dhurandhar}, {D{\'\i}az}, {Di Fiore}, {Di Giovanni}, {Di
  Girolamo}, {Di Lieto}, {Di Pace}, {Di Palma}, {Di Renzo}, {Doctor}, {Dolique}, {Donovan}, {Dooley}, {Doravari}, {Dorrington}, {Douglas}, {Dovale {\'A}lvarez}, {Downes}, {Drago}, {Dreissigacker}, {Driggers}, {Du}, {Ducrot}, {Dupej}, {Dwyer}, {Edo}, {Edwards}, {Effler}, {Ehrens}, {Eichholz}, {Eikenberry}, {Eisenstein}, {Essick}, {Estevez}, {Etienne}, {Etzel}, {Evans}, {Evans}, {Factourovich}, {Fafone}, {Fair}, {Fairhurst}, {Fan}, {Farinon}, {Farr}, {Farr}, {Fauchon-Jones}, {Favata}, {Fays}, {Fee}, {Fehrmann}, {Feicht}, {Fejer}, {Fernandez-Galiana}, {Ferrante}, {Ferreira}, {Ferrini}, {Fidecaro}, {Finstad}, {Fiori}, {Fiorucci}, {Fishbach}, {Fisher}, {Fitz-Axen}, {Flaminio}, {Fletcher}, {Fong}, {Font}, {Forsyth}, {Forsyth}, {Fournier}, {Frasca}, {Frasconi}, {Frei}, {Freise}, {Frey}, {Frey}, {Fries}, {Fritschel}, {Frolov}, {Fulda}, {Fyffe}, {Gabbard}, {Gadre}, {Gaebel}, {Gair}, {Gammaitoni}, {Ganija}, {Gaonkar}, {Garcia-Quiros}, {Garufi}, {Gateley}, {Gaudio}, {Gaur}, {Gayathri}, {Gehrels}, {Gemme}, {Genin},
  {Gennai}, {George}, {George}, {Gergely}, {Germain}, {Ghonge}, {Ghosh}, {Ghosh}, {Ghosh}, {Giaime}, {Giardina}, {Giazotto}, {Gill}, {Glover}, {Goetz}, {Goetz}, {Gomes}, {Goncharov}, {Gonz{\'a}lez}, {Gonzalez Castro}, {Gopakumar}, {Gorodetsky}, {Gossan}, {Gosselin}, {Gouaty}, {Grado}, {Graef}, {Granata}, {Grant}, {Gras}, {Gray}, {Greco}, {Green}, {Gretarsson}, {Griswold}, {Groot}, {Grote}, {Grunewald}, {Gruning}, {Guidi}, {Guo}, {Gupta}, {Gupta}, {Gushwa}, {Gustafson}, {Gustafson}, {Halim}, {Hall}, {Hall}, {Hamilton}, {Hammond}, {Haney}, {Hanke}, {Hanks}, {Hanna}, {Hannam}, {Hannuksela}, {Hanson}, {Hardwick}, {Harms}, {Harry}, {Harry}, {Hart}, {Haster}, {Haughian}, {Healy}, {Heidmann}, {Heintze}, {Heitmann}, {Hello}, {Hemming}, {Hendry}, {Heng}, {Hennig}, {Heptonstall}, {Heurs}, {Hild}, {Hinderer}, {Hoak}, {Hofman}, {Holt}, {Holz}, {Hopkins}, {Horst}, {Hough}, {Houston}, {Howell}, {Hreibi}, {Hu}, {Huerta}, {Huet}, {Hughey}, {Husa}, {Huttner}, {Huynh-Dinh}, {Indik}, {Inta}, {Intini}, {Isa}, {Isac}, {Isi},
  {Iyer}, {Izumi}, {Jacqmin}, {Jani}, {Jaranowski}, {Jawahar}, {Jim{\'e}nez-Forteza}, {Johnson}, {Jones}, {Jones}, {Jonker}, {Ju}, {Junker}, {Kalaghatgi}, {Kalogera}, {Kamai}, {Kandhasamy}, {Kang}, {Kanner}, {Kapadia}, {Karki}, {Karvinen}, {Kasprzack}, {Katolik}, {Katsavounidis}, {Katzman}, {Kaufer}, {Kawabe}, {K{\'e}f{\'e}lian}, {Keitel}, {Kemball}, {Kennedy}, {Kent}, {Key}, {Khalili}, {Khan}, {Khan}, {Khan}, {Khazanov}, {Kijbunchoo}, {Kim}, {Kim}, {Kim}, {Kim}, {Kim}, {Kim}, {Kimbrell}, {King}, {King}, {Kinley-Hanlon}, {Kirchhoff}, {Kissel}, {Kleybolte}, {Klimenko}, {Knowles}, {Koch}, {Koehlenbeck}, {Koley}, {Kondrashov}, {Kontos}, {Korobko}, {Korth}, {Kowalska}, {Kozak}, {Kr{\"a}mer}, {Kringel}, {Krishnan}, {Kr{\'o}lak}, {Kuehn}, {Kumar}, {Kumar}, {Kumar}, {Kuo}, {Kutynia}, {Kwang}, {Lackey}, {Lai}, {Landry}, {Lang}, {Lange}, {Lantz}, {Lanza}, {Larson}, {Lartaux-Vollard}, {Lasky}, {Laxen}, {Lazzarini}, {Lazzaro}, {Leaci}, {Leavey}, {Lee}, {Lee}, {Lee}, {Lee}, {Lee}, {Lehmann}, {Lenon}, {Leonardi}, {Leroy},
  {Letendre}, {Levin}, {Li}, {Linker}, {Littenberg}, {Liu}, {Lo}, {Lockerbie}, {London}, {Lord}, {Lorenzini}, {Loriette}, {Lormand}, {Losurdo}, {Lough}, {Lousto}, {Lovelace}, {L{\"u}ck}, {Lumaca}, {Lundgren}, {Lynch}, {Ma}, {Macas}, {Macfoy}, {Machenschalk}, {MacInnis}, {Macleod}, {Maga{\~n}a Hernandez}, {Maga{\~n}a-Sandoval}, {Maga{\~n}a Zertuche}, {Magee}, {Majorana}, {Maksimovic}, {Man}, {Mandic}, {Mangano}, {Mansell}, {Manske}, {Mantovani}, {Marchesoni}, {Marion}, {M{\'a}rka}, {M{\'a}rka}, {Markakis}, {Markosyan}, {Markowitz}, {Maros}, {Marquina}, {Marsh}, {Martelli}, {Martellini}, {Martin}, {Martin}, {Martynov}, {Mason}, {Massera}, {Masserot}, {Massinger}, {Masso-Reid}, {Mastrogiovanni}, {Matas}, {Matichard}, {Matone}, {Mavalvala}, {Mazumder}, {McCarthy}, {McClelland}, {McCormick}, {McCuller}, {McGuire}, {McIntyre}, {McIver}, {McManus}, {McNeill}, {McRae}, {McWilliams}, {Meacher}, {Meadors}, {Mehmet}, {Meidam}, {Mejuto-Villa}, {Melatos}, {Mendell}, {Mercer}, {Merilh}, {Merzougui}, {Meshkov}, {Messenger},
  {Messick}, {Metzdorff}, {Meyers}, {Miao}, {Michel}, {Middleton}, {Mikhailov}, {Milano}, {Miller}, {Miller}, {Miller}, {Millhouse}, {Milovich-Goff}, {Minazzoli}, {Minenkov}, {Ming}, {Mishra}, {Mitra}, {Mitrofanov}, {Mitselmakher}, {Mittleman}, {Moffa}, {Moggi}, {Mogushi}, {Mohan}, {Mohapatra}, {Montani}, {Moore}, {Moraru}, {Moreno}, {Morriss}, {Mours}, {Mow-Lowry}, {Mueller}, {Muir}, {Mukherjee}, {Mukherjee}, {Mukherjee}, {Mukund}, {Mullavey}, {Munch}, {Mu{\~n}iz}, {Muratore}, {Murray}, {Napier}, {Nardecchia}, {Naticchioni}, {Nayak}, {Neilson}, {Nelemans}, {Nelson}, {Nery}, {Neunzert}, {Nevin}, {Newport}, {Newton}, {Ng}, {Nguyen}, {Nguyen}, {Nichols}, {Nielsen}, {Nissanke}, {Nitz}, {Noack}, {Nocera}, {Nolting}, {North}, {Nuttall}, {Oberling}, {O'Dea}, {Ogin}, {Oh}, {Oh}, {Ohme}, {Okada}, {Oliver}, {Oppermann}, {Oram}, {O'Reilly}, {Ormiston}, {Ortega}, {O'Shaughnessy}, {Ossokine}, {Ottaway}, {Overmier}, {Owen}, {Pace}, {Page}, {Page}, {Pai}, {Pai}, {Palamos}, {Palashov}, {Palomba}, {Pal-Singh}, {Pan}, {Pan},
  {Pang}, {Pang}, {Pankow}, {Pannarale}, {Pant}, {Paoletti}, {Paoli}, {Papa}, {Parida}, {Parker}, {Pascucci}, {Pasqualetti}, {Passaquieti}, {Passuello}, {Patil}, {Patricelli}, {Pearlstone}, {Pedraza}, {Pedurand}, {Pekowsky}, {Pele}, {Penn}, {Perez}, {Perreca}, {Perri}, {Pfeiffer}, {Phelps}, {Piccinni}, {Pichot}, {Piergiovanni}, {Pierro}, {Pillant}, {Pinard}, {Pinto}, {Pirello}, {Pitkin}, {Poe}, {Poggiani}, {Popolizio}, {Porter}, {Post}, {Powell}, {Prasad}, {Pratt}, {Pratten}, {Predoi}, {Prestegard}, {Price}, {Prijatelj}, {Principe}, {Privitera}, {Prodi}, {Prokhorov}, {Puncken}, {Punturo}, {Puppo}, {P{\"u}rrer}, {Qi}, {Quetschke}, {Quintero}, {Quitzow-James}, {Raab}, {Rabeling}, {Radkins}, {Raffai}, {Raja}, {Rajan}, {Rajbhandari}, {Rakhmanov}, {Ramirez}, {Ramos-Buades}, {Rapagnani}, {Raymond}, {Razzano}, {Read}, {Regimbau}, {Rei}, {Reid}, {Reitze}, {Ren}, {Reyes}, {Ricci}, {Ricker}, {Rieger}, {Riles}, {Rizzo}, {Robertson}, {Robie}, {Robinet}, {Rocchi}, {Rolland}, {Rollins}, {Roma}, {Romano}, {Romel}, {Romie},
  {Rosi{\'n}ska}, {Ross}, {Rowan}, {R{\"u}diger}, {Ruggi}, {Rutins}, {Ryan}, {Sachdev}, {Sadecki}, {Sadeghian}, {Sakellariadou}, {Salconi}, {Saleem}, {Salemi}, {Samajdar}, {Sammut}, {Sampson}, {Sanchez}, {Sanchez}, {Sanchis-Gual}, {Sandberg}, {Sanders}, {Sassolas}, {Sathyaprakash}, {Saulson}, {Sauter}, {Savage}, {Sawadsky}, {Schale}, {Scheel}, {Scheuer}, {Schmidt}, {Schmidt}, {Schnabel}, {Schofield}, {Sch{\"o}nbeck}, {Schreiber}, {Schuette}, {Schulte}, {Schutz}, {Schwalbe}, {Scott}, {Scott}, {Seidel}, {Sellers}, {Sengupta}, {Sentenac}, {Sequino}, {Sergeev}, {Shaddock}, {Shaffer}, {Shah}, {Shahriar}, {Shaner}, {Shao}, {Shapiro}, {Shawhan}, {Sheperd}, {Shoemaker}, {Shoemaker}, {Siellez}, {Siemens}, {Sieniawska}, {Sigg}, {Silva}, {Singer}, {Singh}, {Singhal}, {Sintes}, {Slagmolen}, {Smith}, {Smith}, {Smith}, {Somala}, {Son}, {Sonnenberg}, {Sorazu}, {Sorrentino}, {Souradeep}, {Spencer}, {Srivastava}, {Staats}, {Staley}, {Steinke}, {Steinlechner}, {Steinlechner}, {Steinmeyer}, {Stevenson}, {Stone}, {Stops},
  {Strain}, {Stratta}, {Strigin}, {Strunk}, {Sturani}, {Stuver}, {Summerscales}, {Sun}, {Sunil}, {Suresh}, {Sutton}, {Swinkels}, {Szczepa{\'n}czyk}, {Tacca}, {Tait}, {Talbot}, {Talukder}, {Tanner}, {T{\'a}pai}, {Taracchini}, {Tasson}, {Taylor}, {Taylor}, {Tewari}, {Theeg}, {Thies}, {Thomas}, {Thomas}, {Thomas}, {Thorne}, {Thorne}, {Thrane}, {Tiwari}, {Tiwari}, {Tokmakov}, {Toland}, {Tonelli}, {Tornasi}, {Torres-Forn{\'e}}, {Torrie}, {T{\"o}yr{\"a}}, {Travasso}, {Traylor}, {Trinastic}, {Tringali}, {Trozzo}, {Tsang}, {Tse}, {Tso}, {Tsukada}, {Tsuna}, {Tuyenbayev}, {Ueno}, {Ugolini}, {Unnikrishnan}, {Urban}, {Usman}, {Vahlbruch}, {Vajente}, {Valdes}, {van Bakel}, {van Beuzekom}, {van den Brand}, {Van Den Broeck}, {Vander-Hyde}, {van der Schaaf}, {van Heijningen}, {van Veggel}, {Vardaro}, {Varma}, {Vass}, {Vas{\'u}th}, {Vecchio}, {Vedovato}, {Veitch}, {Veitch}, {Venkateswara}, {Venugopalan}, {Verkindt}, {Vetrano}, {Vicer{\'e}}, {Viets}, {Vinciguerra}, {Vine}, {Vinet}, {Vitale}, {Vo}, {Vocca}, {Vorvick},
  {Vyatchanin}, {Wade}, {Wade}, {Wade}, {Walet}, {Walker}, {Wallace}, {Walsh}, {Wang}, {Wang}, {Wang}, {Wang}, {Wang}, {Ward}, {Warner}, {Was}, {Watchi}, {Weaver}, {Wei}, {Weinert}, {Weinstein}, {Weiss}, {Wen}, {Wessel}, {Wessels}, {Westerweck}, {Westphal}, {Wette}, {Whelan}, {Whitcomb}, {Whiting}, {Whittle}, {Wilken}, {Williams}, {Williams}, {Williamson}, {Willis}, {Willke}, {Wimmer}, {Winkler}, {Wipf}, {Wittel}, {Woan}, {Woehler}, {Wofford}, {Wong}, {Worden}, {Wright}, {Wu}, {Wysocki}, {Xiao}, {Yamamoto}, {Yancey}, {Yang}, {Yap}, {Yazback}, {Yu}, {Yu}, {Yvert}, {Zadro{\.z}ny}, {Zanolin}, {Zelenova}, {Zendri}, {Zevin}, {Zhang}, {Zhang}, {Zhang}, {Zhang}, {Zhao}, {Zhou}, {Zhou}, {Zhu}, {Zhu}, {Zimmerman}, {Zucker}, {Zweizig}, {LIGO Scientific Collaboration}, {Virgo Collaboration}, {Wilson-Hodge}, {Bissaldi}, {Blackburn}, {Briggs}, {Burns}, {Cleveland}, {Connaughton}, {Gibby}, {Giles}, {Goldstein}, {Hamburg}, {Jenke}, {Hui}, {Kippen}, {Kocevski}, {McBreen}, {Meegan}, {Paciesas}, {Poolakkil}, {Preece},
  {Racusin}, {Roberts}, {Stanbro}, {Veres}, {von Kienlin}, {GBM}, {Savchenko}, {Ferrigno}, {Kuulkers}, {Bazzano}, {Bozzo}, {Brandt}, {Chenevez}, {Courvoisier}, {Diehl}, {Domingo}, {Hanlon}, {Jourdain}, {Laurent}, {Lebrun}, {Lutovinov}, {Martin-Carrillo}, {Mereghetti}, {Natalucci}, {Rodi}, {Roques}, {Sunyaev}, {Ubertini}, {INTEGRAL}, {Aartsen}, {Ackermann}, {Adams}, {Aguilar}, {Ahlers}, {Ahrens}, {Samarai}, {Altmann}, {Andeen}, {Anderson}, {Ansseau}, {Anton}, {Arg{\"u}elles}, {Auffenberg}, {Axani}, {Bagherpour}, {Bai}, {Barron}, {Barwick}, {Baum}, {Bay}, {Beatty}, {Becker Tjus}, {Bernardini}, {Besson}, {Binder}, {Bindig}, {Blaufuss}, {Blot}, {Bohm}, {B{\"o}rner}, {Bos}, {Bose}, {B{\"o}ser}, {Botner}, {Bourbeau}, {Bourbeau}, {Bradascio}, {Braun}, {Brayeur}, {Brenzke}, {Bretz}, {Bron}, {Brostean-Kaiser}, {Burgman}, {Carver}, {Casey}, {Casier}, {Cheung}, {Chirkin}, {Christov}, {Clark}, {Classen}, {Coenders}, {Collin}, {Conrad}, {Cowen}, {Cross}, {Day}, {de Andr{\'e}}, {De Clercq}, {DeLaunay}, {Dembinski}, {De
  Ridder}, {Desiati}, {de Vries}, {de Wasseige}, {de With}, {DeYoung}, {D{\'\i}az-V{\'e}lez}, {di Lorenzo}, {Dujmovic}, {Dumm}, {Dunkman}, {Dvorak}, {Eberhardt}, {Ehrhardt}, {Eichmann}, {Eller}, {Evenson}, {Fahey}, {Fazely}, {Felde}, {Filimonov}, {Finley}, {Flis}, {Franckowiak}, {Friedman}, {Fuchs}, {Gaisser}, {Gallagher}, {Gerhardt}, {Ghorbani}, {Giang}, {Glauch}, {Gl{\"u}senkamp}, {Goldschmidt}, {Gonzalez}, {Grant}, {Griffith}, {Haack}, {Hallgren}, {Halzen}, {Hanson}, {Hebecker}, {Heereman}, {Helbing}, {Hellauer}, {Hickford}, {Hignight}, {Hill}, {Hoffman}, {Hoffmann}, {Hokanson-Fasig}, {Hoshina}, {Huang}, {Huber}, {Hultqvist}, {H{\"u}nnefeld}, {In}, {Ishihara}, {Jacobi}, {Japaridze}, {Jeong}, {Jero}, {Jones}, {Kalaczynski}, {Kang}, {Kappes}, {Karg}, {Karle}, {Kauer}, {Keivani}, {Kelley}, {Kheirandish}, {Kim}, {Kim}, {Kintscher}, {Kiryluk}, {Kittler}, {Klein}, {Kohnen}, {Koirala}, {Kolanoski}, {K{\"o}pke}, {Kopper}, {Kopper}, {Koschinsky}, {Koskinen}, {Kowalski}, {Krings}, {Kroll}, {Kr{\"u}ckl}, {Kunnen},
  {Kunwar}, {Kurahashi}, {Kuwabara}, {Kyriacou}, {Labare}, {Lanfranchi}, {Larson}, {Lauber}, {Lesiak-Bzdak}, {Leuermann}, {Liu}, {Lu}, {L{\"u}nemann}, {Luszczak}, {Madsen}, {Maggi}, {Mahn}, {Mancina}, {Maruyama}, {Mase}, {Maunu}, {McNally}, {Meagher}, {Medici}, {Meier}, {Menne}, {Merino}, {Meures}, {Miarecki}, {Micallef}, {Moment{\'e}}, {Montaruli}, {Moore}, {Moulai}, {Nahnhauer}, {Nakarmi}, {Naumann}, {Neer}, {Niederhausen}, {Nowicki}, {Nygren}, {Obertacke Pollmann}, {Olivas}, {O'Murchadha}, {Palczewski}, {Pandya}, {Pankova}, {Peiffer}, {Pepper}, {P{\'e}rez de los Heros}, {Pieloth}, {Pinat}, {Price}, {Przybylski}, {Raab}, {R{\"a}del}, {Rameez}, {Rawlins}, {Rea}, {Reimann}, {Relethford}, {Relich}, {Resconi}, {Rhode}, {Richman}, {Robertson}, {Rongen}, {Rott}, {Ruhe}, {Ryckbosch}, {Rysewyk}, {S{\"a}lzer}, {Sanchez Herrera}, {Sandrock}, {Sandroos}, {Santander}, {Sarkar}, {Sarkar}, {Satalecka}, {Schlunder}, {Schmidt}, {Schneider}, {Schoenen}, {Sch{\"o}neberg}, {Schumacher}, {Seckel}, {Seunarine}, {Soedingrekso},
  {Soldin}, {Song}, {Spiczak}, {Spiering}, {Stachurska}, {Stamatikos}, {Stanev}, {Stasik}, {Stettner}, {Steuer}, {Stezelberger}, {Stokstad}, {St{\"o}ssl}, {Strotjohann}, {Stuttard}, {Sullivan}, {Sutherland}, {Taboada}, {Tatar}, {Tenholt}, {Ter-Antonyan}, {Terliuk}, {Te{\v{s}}i{\'c}}, {Tilav}, {Toale}, {Tobin}, {Toscano}, {Tosi}, {Tselengidou}, {Tung}, {Turcati}, {Turley}, {Ty}, {Unger}, {Usner}, {Vandenbroucke}, {Van Driessche}, {van Eijndhoven}, {Vanheule}, {van Santen}, {Vehring}, {Vogel}, {Vraeghe}, {Walck}, {Wallace}, {Wallraff}, {Wandler}, {Wandkowsky}, {Waza}, {Weaver}, {Weiss}, {Wendt}, {Werthebach}, {Whelan}, {Wiebe}, {Wiebusch}, {Wille}, {Williams}, {Wills}, {Wolf}, {Wood}, {Woolsey}, {Woschnagg}, {Xu}, {Xu}, {Xu}, {Yanez}, {Yodh}, {Yoshida}, {Yuan}, {Zoll}, {IceCube Collaboration}, {Balasubramanian}, {Mate}, {Bhalerao}, {Bhattacharya}, {Vibhute}, {Dewangan}, {Rao}, {Vadawale}, {AstroSat Cadmium Zinc Telluride Imager Team}, {Svinkin}, {Hurley}, {Aptekar}, {Frederiks}, {Golenetskii}, {Kozlova},
  {Lysenko}, {Oleynik}, {Tsvetkova}, {Ulanov}, {Cline}, {IPN Collaboration}, {Li}, {Xiong}, {Zhang}, {Lu}, {Song}, {Cao}, {Chang}, {Chen}, {Chen}, {Chen}, {Chen}, {Chen}, {Chen}, {Cui}, {Cui}, {Deng}, {Dong}, {Du}, {Fu}, {Gao}, {Gao}, {Gao}, {Ge}, {Gu}, {Guan}, {Guo}, {Han}, {Hu}, {Huang}, {Huo}, {Jia}, {Jiang}, {Jiang}, {Jin}, {Jin}, {Li}, {Li}, {Li}, {Li}, {Li}, {Li}, {Li}, {Li}, {Li}, {Li}, {Li}, {Liang}, {Liao}, {Liu}, {Liu}, {Liu}, {Liu}, {Liu}, {Liu}, {Liu}, {Lu}, {Lu}, {Luo}, {Ma}, {Meng}, {Nang}, {Nie}, {Ou}, {Qu}, {Sai}, {Sun}, {Tan}, {Tao}, {Tao}, {Tuo}, {Wang}, {Wang}, {Wang}, {Wang}, {Wang}, {Wen}, {Wu}, {Wu}, {Xiao}, {Xu}, {Xu}, {Yan}, {Yang}, {Yang}, {Yang}, {Zhang}, {Zhang}, {Zhang}, {Zhang}, {Zhang}, {Zhang}, {Zhang}, {Zhang}, {Zhang}, {Zhang}, {Zhang}, {Zhang}, {Zhang}, {Zhang}, {Zhang}, {Zhang}, {Zhang}, {Zhang}, {Zhao}, {Zhao}, {Zhao}, {Zheng}, {Zhu}, {Zhu}, {Zou}, {Insight-HXMT Collaboration}, {Albert}, {Andr{\'e}}, {Anghinolfi}, {Ardid}, {Aubert}, {Aublin}, {Avgitas}, {Baret},
  {Barrios-Mart{\'\i}}, {Basa}, {Belhorma}, {Bertin}, {Biagi}, {Bormuth}, {Bourret}, {Bouwhuis}, {Br{\^a}nza{\c{s}}}, {Bruijn}, {Brunner}, {Busto}, {Capone}, {Caramete}, {Carr}, {Celli}, {Cherkaoui El Moursli}, {Chiarusi}, {Circella}, {Coelho}, {Coleiro}, {Coniglione}, {Costantini}, {Coyle}, {Creusot}, {D{\'\i}az}, {Deschamps}, {De Bonis}, {Distefano}, {Di Palma}, {Domi}, {Donzaud}, {Dornic}, {Drouhin}, {Eberl}, {El Bojaddaini}, {El Khayati}, {Els{\"a}sser}, {Enzenh{\"o}fer}, {Ettahiri}, {Fassi}, {Felis}, {Fusco}, {Gay}, {Giordano}, {Glotin}, {Gr{\'e}goire}, {Ruiz}, {Graf}, {Hallmann}, {van Haren}, {Heijboer}, {Hello}, {Hern{\'a}ndez-Rey}, {H{\"o}ssl}, {Hofest{\"a}dt}, {Hugon}, {Illuminati}, {James}, {de Jong}, {Jongen}, {Kadler}, {Kalekin}, {Katz}, {Kiessling}, {Kouchner}, {Kreter}, {Kreykenbohm}, {Kulikovskiy}, {Lachaud}, {Lahmann}, {Lef{\`e}vre}, {Leonora}, {Lotze}, {Loucatos}, {Marcelin}, {Margiotta}, {Marinelli}, {Mart{\'\i}nez-Mora}, {Mele}, {Melis}, {Michael}, {Migliozzi}, {Moussa}, {Navas}, {Nezri},
  {Organokov}, {P{\u{a}}v{\u{a}}la{\c{s}}}, {Pellegrino}, {Perrina}, {Piattelli}, {Popa}, {Pradier}, {Quinn}, {Racca}, {Riccobene}, {S{\'a}nchez-Losa}, {Salda{\~n}a}, {Salvadori}, {Samtleben}, {Sanguineti}, {Sapienza}, {Sieger}, {Spurio}, {Stolarczyk}, {Taiuti}, {Tayalati}, {Trovato}, {Turpin}, {T{\"o}nnis}, {Vallage}, {Van Elewyck}, {Versari}, {Vivolo}, {Vizzoca}, {Wilms}, {Zornoza}, {Z{\'u}{\~n}iga}, {ANTARES Collaboration}, {Beardmore}, {Breeveld}, {Burrows}, {Cenko}, {Cusumano}, {D'A{\`\i}}, {de Pasquale}, {Emery}, {Evans}, {Giommi}, {Gronwall}, {Kennea}, {Krimm}, {Kuin}, {Lien}, {Marshall}, {Melandri}, {Nousek}, {Oates}, {Osborne}, {Pagani}, {Page}, {Palmer}, {Perri}, {Siegel}, {Sbarufatti}, {Tagliaferri}, {Tohuvavohu}, {Swift Collaboration}, {Tavani}, {Verrecchia}, {Bulgarelli}, {Evangelista}, {Pacciani}, {Feroci}, {Pittori}, {Giuliani}, {Del Monte}, {Donnarumma}, {Argan}, {Trois}, {Ursi}, {Cardillo}, {Piano}, {Longo}, {Lucarelli}, {Munar-Adrover}, {Fuschino}, {Labanti}, {Marisaldi}, {Minervini},
  {Fioretti}, {Parmiggiani}, {Gianotti}, {Trifoglio}, {Di Persio}, {Antonelli}, {Barbiellini}, {Caraveo}, {Cattaneo}, {Costa}, {Colafrancesco}, {D'Amico}, {Ferrari}, {Morselli}, {Paoletti}, {Picozza}, {Pilia}, {Rappoldi}, {Soffitta}, {Vercellone}, {AGILE Team}, {Foley}, {Coulter}, {Kilpatrick}, {Drout}, {Piro}, {Shappee}, {Siebert}, {Simon}, {Ulloa}, {Kasen}, {Madore}, {Murguia-Berthier}, {Pan}, {Prochaska}, {Ramirez-Ruiz}, {Rest}, {Rojas-Bravo}, {1M2H Team}, {Berger}, {Soares-Santos}, {Annis}, {Alexander}, {Allam}, {Balbinot}, {Blanchard}, {Brout}, {Butler}, {Chornock}, {Cook}, {Cowperthwaite}, {Diehl}, {Drlica-Wagner}, {Drout}, {Durret}, {Eftekhari}, {Finley}, {Fong}, {Frieman}, {Fryer}, {Garc{\'\i}a-Bellido}, {Gruendl}, {Hartley}, {Herner}, {Kessler}, {Lin}, {Lopes}, {Louren{\c{c}}o}, {Margutti}, {Marshall}, {Matheson}, {Medina}, {Metzger}, {Mu{\~n}oz}, {Muir}, {Nicholl}, {Nugent}, {Palmese}, {Paz-Chinch{\'o}n}, {Quataert}, {Sako}, {Sauseda}, {Schlegel}, {Scolnic}, {Secco}, {Smith}, {Sobreira}, {Villar},
  {Vivas}, {Wester}, {Williams}, {Yanny}, {Zenteno}, {Zhang}, {Abbott}, {Banerji}, {Bechtol}, {Benoit-L{\'e}vy}, {Bertin}, {Brooks}, {Buckley-Geer}, {Burke}, {Capozzi}, {Carnero Rosell}, {Carrasco Kind}, {Castander}, {Crocce}, {Cunha}, {D'Andrea}, {da Costa}, {Davis}, {DePoy}, {Desai}, {Dietrich}, {Eifler}, {Fernandez}, {Flaugher}, {Fosalba}, {Gaztanaga}, {Gerdes}, {Giannantonio}, {Goldstein}, {Gruen}, {Gschwend}, {Gutierrez}, {Honscheid}, {James}, {Jeltema}, {Johnson}, {Johnson}, {Kent}, {Krause}, {Kron}, {Kuehn}, {Lahav}, {Lima}, {Maia}, {March}, {Martini}, {McMahon}, {Menanteau}, {Miller}, {Miquel}, {Mohr}, {Nichol}, {Ogando}, {Plazas}, {Romer}, {Roodman}, {Rykoff}, {Sanchez}, {Scarpine}, {Schindler}, {Schubnell}, {Sevilla-Noarbe}, {Sheldon}, {Smith}, {Smith}, {Stebbins}, {Suchyta}, {Swanson}, {Tarle}, {Thomas}, {Troxel}, {Tucker}, {Vikram}, {Walker}, {Wechsler}, {Weller}, {Carlin}, {Gill}, {Li}, {Marriner}, {Neilsen}, {Dark Energy Camera GW-EM Collaboration}, {DES Collaboration}, {Haislip}, {Kouprianov},
  {Reichart}, {Sand}, {Tartaglia}, {Valenti}, {Yang}, {DLT40 Collaboration}, {Benetti}, {Brocato}, {Campana}, {Cappellaro}, {Covino}, {D'Avanzo}, {D'Elia}, {Getman}, {Ghirlanda}, {Ghisellini}, {Limatola}, {Nicastro}, {Palazzi}, {Pian}, {Piranomonte}, {Possenti}, {Rossi}, {Salafia}, {Tomasella}, {Amati}, {Antonelli}, {Bernardini}, {Bufano}, {Capaccioli}, {Casella}, {Dadina}, {De Cesare}, {Di Paola}, {Giuffrida}, {Giunta}, {Israel}, {Lisi}, {Maiorano}, {Mapelli}, {Masetti}, {Pescalli}, {Pulone}, {Salvaterra}, {Schipani}, {Spera}, {Stamerra}, {Stella}, {Testa}, {Turatto}, {Vergani}, {Aresu}, {Bachetti}, {Buffa}, {Burgay}, {Buttu}, {Caria}, {Carretti}, {Casasola}, {Castangia}, {Carboni}, {Casu}, {Concu}, {Corongiu}, {Deiana}, {Egron}, {Fara}, {Gaudiomonte}, {Gusai}, {Ladu}, {Loru}, {Leurini}, {Marongiu}, {Melis}, {Melis}, {Migoni}, {Milia}, {Navarrini}, {Orlati}, {Ortu}, {Palmas}, {Pellizzoni}, {Perrodin}, {Pisanu}, {Poppi}, {Righini}, {Saba}, {Serra}, {Serrau}, {Stagni}, {Surcis}, {Vacca}, {Vargiu}, {Hunt},
  {Jin}, {Klose}, {Kouveliotou}, {Mazzali}, {M{\o}ller}, {Nava}, {Piran}, {Selsing}, {Vergani}, {Wiersema}, {Toma}, {Higgins}, {Mundell}, {di Serego Alighieri}, {G{\'o}tz}, {Gao}, {Gomboc}, {Kaper}, {Kobayashi}, {Kopac}, {Mao}, {Starling}, {Steele}, {van der Horst}, {GRAWITA: GRAvitational Wave Inaf TeAm}, {Acero}, {Atwood}, {Baldini}, {Barbiellini}, {Bastieri}, {Berenji}, {Bellazzini}, {Bissaldi}, {Blandford}, {Bloom}, {Bonino}, {Bottacini}, {Bregeon}, {Buehler}, {Buson}, {Cameron}, {Caputo}, {Caraveo}, {Cavazzuti}, {Chekhtman}, {Cheung}, {Chiang}, {Ciprini}, {Cohen-Tanugi}, {Cominsky}, {Costantin}, {Cuoco}, {D'Ammando}, {de Palma}, {Digel}, {Di Lalla}, {Di Mauro}, {Di Venere}, {Dubois}, {Fegan}, {Focke}, {Franckowiak}, {Fukazawa}, {Funk}, {Fusco}, {Gargano}, {Gasparrini}, {Giglietto}, {Giordano}, {Giroletti}, {Glanzman}, {Green}, {Grondin}, {Guillemot}, {Guiriec}, {Harding}, {Horan}, {J{\'o}hannesson}, {Kamae}, {Kensei}, {Kuss}, {La Mura}, {Latronico}, {Lemoine-Goumard}, {Longo}, {Loparco}, {Lovellette},
  {Lubrano}, {Magill}, {Maldera}, {Manfreda}, {Mazziotta}, {McEnery}, {Meyer}, {Michelson}, {Mirabal}, {Monzani}, {Moretti}, {Morselli}, {Moskalenko}, {Negro}, {Nuss}, {Ojha}, {Omodei}, {Orienti}, {Orlando}, {Palatiello}, {Paliya}, {Paneque}, {Pesce-Rollins}, {Piron}, {Porter}, {Principe}, {Rain{\`o}}, {Rando}, {Razzano}, {Razzaque}, {Reimer}, {Reimer}, {Reposeur}, {Rochester}, {Saz Parkinson}, {Sgr{\`o}}, {Siskind}, {Spada}, {Spandre}, {Suson}, {Takahashi}, {Tanaka}, {Thayer}, {Thayer}, {Thompson}, {Tibaldo}, {Torres}, {Torresi}, {Troja}, {Venters}, {Vianello}, {Zaharijas}, {Fermi Large Area Telescope Collaboration}, {Allison}, {Bannister}, {Dobie}, {Kaplan}, {Lenc}, {Lynch}, {Murphy}, {Sadler}, {Australia Telescope Compact Array}, {Hotan}, {James}, {Oslowski}, {Raja}, {Shannon}, {Whiting}, {Australian SKA Pathfinder}, {Arcavi}, {Howell}, {McCully}, {Hosseinzadeh}, {Hiramatsu}, {Poznanski}, {Barnes}, {Zaltzman}, {Vasylyev}, {Maoz}, {Las Cumbres Observatory Group}, {Cooke}, {Bailes}, {Wolf}, {Deller},
  {Lidman}, {Wang}, {Gendre}, {Andreoni}, {Ackley}, {Pritchard}, {Bessell}, {Chang}, {M{\"o}ller}, {Onken}, {Scalzo}, {Ridden-Harper}, {Sharp}, {Tucker}, {Farrell}, {Elmer}, {Johnston}, {Venkatraman Krishnan}, {Keane}, {Green}, {Jameson}, {Hu}, {Ma}, {Sun}, {Wu}, {Wang}, {Shang}, {Hu}, {Ashley}, {Yuan}, {Li}, {Tao}, {Zhu}, {Zhang}, {Suntzeff}, {Zhou}, {Yang}, {Orange}, {Morris}, {Cucchiara}, {Giblin}, {Klotz}, {Staff}, {Thierry}, {Schmidt}, {OzGrav}, {(Deeper}, {Wider}, {program}, {AST3}, {CAASTRO Collaborations}, {Tanvir}, {Levan}, {Cano}, {de Ugarte-Postigo}, {Gonz{\'a}lez-Fern{\'a}ndez}, {Greiner}, {Hjorth}, {Irwin}, {Kr{\"u}hler}, {Mandel}, {Milvang-Jensen}, {O'Brien}, {Rol}, {Rosetti}, {Rosswog}, {Rowlinson}, {Steeghs}, {Th{\"o}ne}, {Ulaczyk}, {Watson}, {Bruun}, {Cutter}, {Figuera Jaimes}, {Fujii}, {Fruchter}, {Gompertz}, {Jakobsson}, {Hodosan}, {J{\`e}rgensen}, {Kangas}, {Kann}, {Rabus}, {Schr{\o}der}, {Stanway}, {Wijers}, {VINROUGE Collaboration}, {Lipunov}, {Gorbovskoy}, {Kornilov}, {Tyurina},
  {Balanutsa}, {Kuznetsov}, {Vlasenko}, {Podesta}, {Lopez}, {Podesta}, {Levato}, {Saffe}, {Mallamaci}, {Budnev}, {Gress}, {Kuvshinov}, {Gorbunov}, {Vladimirov}, {Zimnukhov}, {Gabovich}, {Yurkov}, {Sergienko}, {Rebolo}, {Serra-Ricart}, {Tlatov}, {Ishmuhametova}, {MASTER Collaboration}, {Abe}, {Aoki}, {Aoki}, {Asakura}, {Baar}, {Barway}, {Bond}, {Doi}, {Finet}, {Fujiyoshi}, {Furusawa}, {Honda}, {Itoh}, {Kanda}, {Kawabata}, {Kawabata}, {Kim}, {Koshida}, {Kuroda}, {Lee}, {Liu}, {Matsubayashi}, {Miyazaki}, {Morihana}, {Morokuma}, {Motohara}, {Murata}, {Nagai}, {Nagashima}, {Nagayama}, {Nakaoka}, {Nakata}, {Ohsawa}, {Ohshima}, {Ohta}, {Okita}, {Saito}, {Saito}, {Sako}, {Sekiguchi}, {Sumi}, {Tajitsu}, {Takahashi}, {Takayama}, {Tamura}, {Tanaka}, {Tanaka}, {Terai}, {Tominaga}, {Tristram}, {Uemura}, {Utsumi}, {Yamaguchi}, {Yasuda}, {Yoshida}, {Zenko}, {J-GEM}, {Adams}, {Anupama}, {Bally}, {Barway}, {Bellm}, {Blagorodnova}, {Cannella}, {Chandra}, {Chatterjee}, {Clarke}, {Cobb}, {Cook}, {Copperwheat}, {De}, {Emery},
  {Feindt}, {Foster}, {Fox}, {Frail}, {Fremling}, {Frohmaier}, {Garcia}, {Ghosh}, {Giacintucci}, {Goobar}, {Gottlieb}, {Grefenstette}, {Hallinan}, {Harrison}, {Heida}, {Helou}, {Ho}, {Horesh}, {Hotokezaka}, {Ip}, {Itoh}, {Jacobs}, {Jencson}, {Kasen}, {Kasliwal}, {Kassim}, {Kim}, {Kiran}, {Kuin}, {Kulkarni}, {Kupfer}, {Lau}, {Madsen}, {Mazzali}, {Miller}, {Miyasaka}, {Mooley}, {Myers}, {Nakar}, {Ngeow}, {Nugent}, {Ofek}, {Palliyaguru}, {Pavana}, {Perley}, {Peters}, {Pike}, {Piran}, {Qi}, {Quimby}, {Rana}, {Rosswog}, {Rusu}, {Sadler}, {Van Sistine}, {Sollerman}, {Xu}, {Yan}, {Yatsu}, {Yu}, {Zhang}, {Zhao}, {GROWTH}, {JAGWAR}, {Caltech-NRAO}, {TTU-NRAO}, {NuSTAR Collaborations}, {Chambers}, {Huber}, {Schultz}, {Bulger}, {Flewelling}, {Magnier}, {Lowe}, {Wainscoat}, {Waters}, {Willman}, {Pan-STARRS}, {Ebisawa}, {Hanyu}, {Harita}, {Hashimoto}, {Hidaka}, {Hori}, {Ishikawa}, {Isobe}, {Iwakiri}, {Kawai}, {Kawai}, {Kawamuro}, {Kawase}, {Kitaoka}, {Makishima}, {Matsuoka}, {Mihara}, {Morita}, {Morita}, {Nakahira},
  {Nakajima}, {Nakamura}, {Negoro}, {Oda}, {Sakamaki}, {Sasaki}, {Serino}, {Shidatsu}, {Shimomukai}, {Sugawara}, {Sugita}, {Sugizaki}, {Tachibana}, {Takao}, {Tanimoto}, {Tomida}, {Tsuboi}, {Tsunemi}, {Ueda}, {Ueno}, {Yamada}, {Yamaoka}, {Yamauchi}, {Yatabe}, {Yoneyama}, {Yoshii}, {MAXI Team}, {Coward}, {Crisp}, {Macpherson}, {Andreoni}, {Laugier}, {Noysena}, {Klotz}, {Gendre}, {Thierry}, {Turpin}, {Consortium}, {Im}, {Choi}, {Kim}, {Yoon}, {Lim}, {Lee}, {Lee}, {Kim}, {Ko}, {Joe}, {Kwon}, {Kim}, {Lim}, {Choi}, {KU Collaboration}, {Fynbo}, {Malesani}, {Xu}, {Optical Telescope}, {Smartt}, {Jerkstrand}, {Kankare}, {Sim}, {Fraser}, {Inserra}, {Maguire}, {Leloudas}, {Magee}, {Shingles}, {Smith}, {Young}, {Kotak}, {Gal-Yam}, {Lyman}, {Homan}, {Agliozzo}, {Anderson}, {Angus}, {Ashall}, {Barbarino}, {Bauer}, {Berton}, {Botticella}, {Bulla}, {Cannizzaro}, {Cartier}, {Cikota}, {Clark}, {De Cia}, {Della Valle}, {Dennefeld}, {Dessart}, {Dimitriadis}, {Elias-Rosa}, {Firth}, {Fl{\"o}rs}, {Frohmaier}, {Galbany},
  {Gonz{\'a}lez-Gait{\'a}n}, {Gromadzki}, {Guti{\'e}rrez}, {Hamanowicz}, {Harmanen}, {Heintz}, {Hernandez}, {Hodgkin}, {Hook}, {Izzo}, {James}, {Jonker}, {Kerzendorf}, {Kostrzewa-Rutkowska}, {Kromer}, {Kuncarayakti}, {Lawrence}, {Manulis}, {Mattila}, {McBrien}, {M{\"u}ller}, {Nordin}, {O'Neill}, {Onori}, {Palmerio}, {Pastorello}, {Patat}, {Pignata}, {Podsiadlowski}, {Razza}, {Reynolds}, {Roy}, {Ruiter}, {Rybicki}, {Salmon}, {Pumo}, {Prentice}, {Seitenzahl}, {Smith}, {Sollerman}, {Sullivan}, {Szegedi}, {Taddia}, {Taubenberger}, {Terreran}, {Van Soelen}, {Vos}, {Walton}, {Wright}, {Wyrzykowski}, {Yaron}, {pre=''(''>ePESSTO}, {Chen}, {Kr{\"u}hler}, {Schady}, {Wiseman}, {Greiner}, {Rau}, {Schweyer}, {Klose}, {Nicuesa Guelbenzu}, {GROND}, {Palliyaguru}, {Tech University}, {Shara}, {Williams}, {Vaisanen}, {Potter}, {Romero Colmenero}, {Crawford}, {Buckley}, {Mao}, {SALT Group}, {D{\'\i}az}, {Macri}, {Garc{\'\i}a Lambas}, {Mendes de Oliveira}, {Nilo Castell{\'o}n}, {Ribeiro}, {S{\'a}nchez}, {Schoenell}, {Abramo},
  {Akras}, {Alcaniz}, {Artola}, {Beroiz}, {Bonoli}, {Cabral}, {Camuccio}, {Chavushyan}, {Coelho}, {Colazo}, {Costa-Duarte}, {Cuevas Larenas}, {Dom{\'\i}nguez Romero}, {Dultzin}, {Fern{\'a}ndez}, {Garc{\'\i}a}, {Girardini}, {Gon{\c{c}}alves}, {Gon{\c{c}}alves}, {Gurovich}, {Jim{\'e}nez-Teja}, {Kanaan}, {Lares}, {Lopes de Oliveira}, {L{\'o}pez-Cruz}, {Melia}, {Molino}, {Padilla}, {Pe{\~n}uela}, {Placco}, {Qui{\~n}ones}, {Ram{\'\i}rez Rivera}, {Renzi}, {Riguccini}, {R{\'\i}os-L{\'o}pez}, {Rodriguez}, {Sampedro}, {Schneiter}, {Sodr{\'e}}, {Starck}, {Torres-Flores}, {Tornatore}, {Zadro{\.z}ny}, {Castillo}, {TOROS: Transient Robotic Observatory of South Collaboration}, {Castro-Tirado}, {Tello}, {Hu}, {Zhang}, {Cunniffe}, {Castell{\'o}n}, {Hiriart}, {Caballero-Garc{\'\i}a}, {Jel{\'\i}nek}, {Kub{\'a}nek}, {P{\'e}rez del Pulgar}, {Park}, {Jeong}, {Castro Cer{\'o}n}, {Pandey}, {Yock}, {Querel}, {Fan}, {Wang}, {BOOTES Collaboration}, {Beardsley}, {Brown}, {Crosse}, {Emrich}, {Franzen}, {Gaensler}, {Horsley},
  {Johnston-Hollitt}, {Kenney}, {Morales}, {Pallot}, {Sokolowski}, {Steele}, {Tingay}, {Trott}, {Walker}, {Wayth}, {Williams}, {Wu}, {Murchison Widefield Array}, {Yoshida}, {Sakamoto}, {Kawakubo}, {Yamaoka}, {Takahashi}, {Asaoka}, {Ozawa}, {Torii}, {Shimizu}, {Tamura}, {Ishizaki}, {Cherry}, {Ricciarini}, {Penacchioni}, {Marrocchesi}, {CALET Collaboration}, {Pozanenko}, {Volnova}, {Mazaeva}, {Minaev}, {Krugov}, {Kusakin}, {Reva}, {Moskvitin}, {Rumyantsev}, {Inasaridze}, {Klunko}, {Tungalag}, {Schmalz}, {Burhonov}, {IKI-GW Follow-up Collaboration}, {Abdalla}, {Abramowski}, {Aharonian}, {Ait Benkhali}, {Ang{\"u}ner}, {Arakawa}, {Arrieta}, {Aubert}, {Backes}, {Balzer}, {Barnard}, {Becherini}, {Becker Tjus}, {Berge}, {Bernhard}, {Bernl{\"o}hr}, {Blackwell}, {B{\"o}ttcher}, {Boisson}, {Bolmont}, {Bonnefoy}, {Bordas}, {Bregeon}, {Brun}, {Brun}, {Bryan}, {B{\"u}chele}, {Bulik}, {Capasso}, {Caroff}, {Carosi}, {Casanova}, {Cerruti}, {Chakraborty}, {Chaves}, {Chen}, {Chevalier}, {Colafrancesco}, {Condon}, {Conrad},
  {Davids}, {Decock}, {Deil}, {Devin}, {deWilt}, {Dirson}, {Djannati-Ata{\"\i}}, {Donath}, {O'C. Drury}, {Dutson}, {Dyks}, {Edwards}, {Egberts}, {Emery}, {Ernenwein}, {Eschbach}, {Farnier}, {Fegan}, {Fernandes}, {Fiasson}, {Fontaine}, {Funk}, {F{\"u}ssling}, {Gabici}, {Gallant}, {Garrigoux}, {Gat{\'e}}, {Giavitto}, {Giebels}, {Glawion}, {Glicenstein}, {Gottschall}, {Grondin}, {Hahn}, {Haupt}, {Hawkes}, {Heinzelmann}, {Henri}, {Hermann}, {Hinton}, {Hofmann}, {Hoischen}, {Holch}, {Holler}, {Horns}, {Ivascenko}, {Iwasaki}, {Jacholkowska}, {Jamrozy}, {Jankowsky}, {Jankowsky}, {Jingo}, {Jouvin}, {Jung-Richardt}, {Kastendieck}, {Katarzy{\'n}ski}, {Katsuragawa}, {Kerszberg}, {Khangulyan}, {Kh{\'e}lifi}, {King}, {Klepser}, {Klochkov}, {Klu{\'z}niak}, {Komin}, {Kosack}, {Krakau}, {Kraus}, {Kr{\"u}ger}, {Laffon}, {Lamanna}, {Lau}, {Lees}, {Lefaucheur}, {Lemi{\`e}re}, {Lemoine-Goumard}, {Lenain}, {Leser}, {Lohse}, {Lorentz}, {Liu}, {Lypova}, {Malyshev}, {Marandon}, {Marcowith}, {Mariaud}, {Marx}, {Maurin}, {Maxted},
  {Mayer}, {Meintjes}, {Meyer}, {Mitchell}, {Moderski}, {Mohamed}, {Mohrmann}, {Mor{\r{a}}}, {Moulin}, {Murach}, {Nakashima}, {de Naurois}, {Ndiyavala}, {Niederwanger}, {Niemiec}, {Oakes}, {O'Brien}, {Odaka}, {Ohm}, {Ostrowski}, {Oya}, {Padovani}, {Panter}, {Parsons}, {Pekeur}, {Pelletier}, {Perennes}, {Petrucci}, {Peyaud}, {Piel}, {Pita}, {Poireau}, {Poon}, {Prokhorov}, {Prokoph}, {P{\"u}hlhofer}, {Punch}, {Quirrenbach}, {Raab}, {Rauth}, {Reimer}, {Reimer}, {Renaud}, {de los Reyes}, {Rieger}, {Rinchiuso}, {Romoli}, {Rowell}, {Rudak}, {Rulten}, {Sahakian}, {Saito}, {Sanchez}, {Santangelo}, {Sasaki}, {Schlickeiser}, {Sch{\"u}ssler}, {Schulz}, {Schwanke}, {Schwemmer}, {Seglar-Arroyo}, {Settimo}, {Seyffert}, {Shafi}, {Shilon}, {Shiningayamwe}, {Simoni}, {Sol}, {Spanier}, {Spir-Jacob}, {Stawarz}, {Steenkamp}, {Stegmann}, {Steppa}, {Sushch}, {Takahashi}, {Tavernet}, {Tavernier}, {Taylor}, {Terrier}, {Tibaldo}, {Tiziani}, {Tluczykont}, {Trichard}, {Tsirou}, {Tsuji}, {Tuffs}, {Uchiyama}, {van der Walt}, {van Eldik},
  {van Rensburg}, {van Soelen}, {Vasileiadis}, {Veh}, {Venter}, {Viana}, {Vincent}, {Vink}, {Voisin}, {V{\"o}lk}, {Vuillaume}, {Wadiasingh}, {Wagner}, {Wagner}, {Wagner}, {White}, {Wierzcholska}, {Willmann}, {W{\"o}rnlein}, {Wouters}, {Yang}, {Zaborov}, {Zacharias}, {Zanin}, {Zdziarski}, {Zech}, {Zefi}, {Ziegler}, {Zorn}, {{\.Z}ywucka}, {H.~E.~S.~S. Collaboration}, {Fender}, {Broderick}, {Rowlinson}, {Wijers}, {Stewart}, {ter Veen}, {Shulevski}, {LOFAR Collaboration}, {Kavic}, {Simonetti}, {League}, {Tsai}, {Obenberger}, {Nathaniel}, {Taylor}, {Dowell}, {Liebling}, {Estes}, {Lippert}, {Sharma}, {Vincent}, {Farella}, {Wavelength Array}, {Abeysekara}, {Albert}, {Alfaro}, {Alvarez}, {Arceo}, {Arteaga-Vel{\'a}zquez}, {Avila Rojas}, {Ayala Solares}, {Barber}, {Becerra Gonzalez}, {Becerril}, {Belmont-Moreno}, {BenZvi}, {Berley}, {Bernal}, {Braun}, {Brisbois}, {Caballero-Mora}, {Capistr{\'a}n}, {Carrami{\~n}ana}, {Casanova}, {Castillo}, {Cotti}, {Cotzomi}, {Couti{\~n}o de Le{\'o}n}, {De Le{\'o}n}, {De la Fuente},
  {Diaz Hernandez}, {Dichiara}, {Dingus}, {DuVernois}, {D{\'\i}az-V{\'e}lez}, {Ellsworth}, {Engel}, {Enr{\'\i}quez-Rivera}, {Fiorino}, {Fleischhack}, {Fraija}, {Garc{\'\i}a-Gonz{\'a}lez}, {Garfias}, {Gerhardt}, {Gonz{\~o}lez Mu{\~n}oz}, {Gonz{\'a}lez}, {Goodman}, {Hampel-Arias}, {Harding}, {Hernandez}, {Hernandez-Almada}, {Hona}, {H{\"u}ntemeyer}, {Iriarte}, {Jardin-Blicq}, {Joshi}, {Kaufmann}, {Kieda}, {Lara}, {Lauer}, {Lennarz}, {Le{\'o}n Vargas}, {Linnemann}, {Longinotti}, {Raya}, {Luna-Garc{\'\i}a}, {L{\'o}pez-Coto}, {Malone}, {Marinelli}, {Martinez}, {Martinez-Castellanos}, {Mart{\'\i}nez-Castro}, {Mart{\'\i}nez-Huerta}, {Matthews}, {Miranda-Romagnoli}, {Moreno}, {Mostaf{\'a}}, {Nellen}, {Newbold}, {Nisa}, {Noriega-Papaqui}, {Pelayo}, {Pretz}, {P{\'e}rez-P{\'e}rez}, {Ren}, {Rho}, {Rivi{\`e}re}, {Rosa-Gonz{\'a}lez}, {Rosenberg}, {Ruiz-Velasco}, {Salazar}, {Salesa Greus}, {Sandoval}, {Schneider}, {Schoorlemmer}, {Sinnis}, {Smith}, {Springer}, {Surajbali}, {Tibolla}, {Tollefson}, {Torres}, {Ukwatta},
  {Weisgarber}, {Westerhoff}, {Wisher}, {Wood}, {Yapici}, {Yodh}, {Younk}, {Zhou}, {{\'A}lvarez}, {HAWC Collaboration}, {Aab}, {Abreu}, {Aglietta}, {Albuquerque}, {Albury}, {Allekotte}, {Almela}, {Alvarez Castillo}, {Alvarez-Mu{\~n}iz}, {Anastasi}, {Anchordoqui}, {Andrada}, {Andringa}, {Aramo}, {Arsene}, {Asorey}, {Assis}, {Avila}, {Badescu}, {Balaceanu}, {Barbato}, {Barreira Luz}, {Becker}, {Bellido}, {Berat}, {Bertaina}, {Bertou}, {Biermann}, {Biteau}, {Blaess}, {Blanco}, {Blazek}, {Bleve}, {Boh{\'a}{\v{c}}ov{\'a}}, {Bonifazi}, {Borodai}, {Botti}, {Brack}, {Brancus}, {Bretz}, {Bridgeman}, {Briechle}, {Buchholz}, {Bueno}, {Buitink}, {Buscemi}, {Caballero-Mora}, {Caccianiga}, {Cancio}, {Canfora}, {Caruso}, {Castellina}, {Catalani}, {Cataldi}, {Cazon}, {Chavez}, {Chinellato}, {Chudoba}, {Clay}, {Cobos Cerutti}, {Colalillo}, {Coleman}, {Collica}, {Coluccia}, {Concei{\c{c}}{\~a}o}, {Consolati}, {Contreras}, {Cooper}, {Coutu}, {Covault}, {Cronin}, {D'Amico}, {Daniel}, {Dasso}, {Daumiller}, {Dawson}, {Day}, {de
  Almeida}, {de Jong}, {De Mauro}, {de Mello Neto}, {De Mitri}, {de Oliveira}, {de Souza}, {Debatin}, {Deligny}, {D{\'\i}az Castro}, {Diogo}, {Dobrigkeit}, {D'Olivo}, {Dorosti}, {Dos Anjos}, {Dova}, {Dundovic}, {Ebr}, {Engel}, {Erdmann}, {Erfani}, {Escobar}, {Espadanal}, {Etchegoyen}, {Falcke}, {Farmer}, {Farrar}, {Fauth}, {Fazzini}, {Feldbusch}, {Fenu}, {Fick}, {Figueira}, {Filip{\v{c}}i{\v{c}}}, {Freire}, {Fujii}, {Fuster}, {Ga{\"\i}or}, {Garc{\'\i}a}, {Gat{\'e}}, {Gemmeke}, {Gherghel-Lascu}, {Ghia}, {Giaccari}, {Giammarchi}, {Giller}, {G{\l}as}, {Glaser}, {Golup}, {G{\'o}mez Berisso}, {G{\'o}mez Vitale}, {Gonz{\'a}lez}, {Gorgi}, {Gottowik}, {Grillo}, {Grubb}, {Guarino}, {Guedes}, {Halliday}, {Hampel}, {Hansen}, {Harari}, {Harrison}, {Harvey}, {Haungs}, {Hebbeker}, {Heck}, {Heimann}, {Herve}, {Hill}, {Hojvat}, {Holt}, {Homola}, {H{\"o}randel}, {Horvath}, {Hrabovsk{\'y}}, {Huege}, {Hulsman}, {Insolia}, {Isar}, {Jandt}, {Johnsen}, {Josebachuili}, {Jurysek}, {K{\"a}{\"a}p{\"a}}, {Kampert}, {Keilhauer},
  {Kemmerich}, {Kemp}, {Kieckhafer}, {Klages}, {Kleifges}, {Kleinfeller}, {Krause}, {Krohm}, {Kuempel}, {Kukec Mezek}, {Kunka}, {Kuotb Awad}, {Lago}, {LaHurd}, {Lang}, {Lauscher}, {Legumina}, {Leigui de Oliveira}, {Letessier-Selvon}, {Lhenry-Yvon}, {Link}, {Lo Presti}, {Lopes}, {L{\'o}pez}, {L{\'o}pez Casado}, {Lorek}, {Luce}, {Lucero}, {Malacari}, {Mallamaci}, {Mandat}, {Mantsch}, {Mariazzi}, {Maris}, {Marsella}, {Martello}, {Martinez}, {Mart{\'\i}nez Bravo}, {Mas{\'\i}as Meza}, {Mathes}, {Mathys}, {Matthews}, {Matthiae}, {Mayotte}, {Mazur}, {Medina}, {Medina-Tanco}, {Melo}, {Menshikov}, {Merenda}, {Michal}, {Micheletti}, {Middendorf}, {Miramonti}, {Mitrica}, {Mockler}, {Mollerach}, {Montanet}, {Morello}, {Morlino}, {M{\"u}ller}, {M{\"u}ller}, {Muller}, {M{\"u}ller}, {Mussa}, {Naranjo}, {Nguyen}, {Niculescu-Oglinzanu}, {Niechciol}, {Niemietz}, {Niggemann}, {Nitz}, {Nosek}, {Novotny}, {No{\v{z}}ka}, {N{\'u}{\~n}ez}, {Oikonomou}, {Olinto}, {Palatka}, {Pallotta}, {Papenbreer}, {Parente}, {Parra}, {Paul},
  {Pech}, {Pedreira}, {P{\c{e}}kala}, {Pe{\~n}a-Rodriguez}, {Pereira}, {Perlin}, {Perrone}, {Peters}, {Petrera}, {Phuntsok}, {Pierog}, {Pimenta}, {Pirronello}, {Platino}, {Plum}, {Poh}, {Porowski}, {Prado}, {Privitera}, {Prouza}, {Quel}, {Querchfeld}, {Quinn}, {Ramos-Pollan}, {Rautenberg}, {Ravignani}, {Ridky}, {Riehn}, {Risse}, {Ristori}, {Rizi}, {Rodrigues de Carvalho}, {Rodriguez Fernandez}, {Rodriguez Rojo}, {Roncoroni}, {Roth}, {Roulet}, {Rovero}, {Ruehl}, {Saffi}, {Saftoiu}, {Salamida}, {Salazar}, {Saleh}, {Salina}, {S{\'a}nchez}, {Sanchez-Lucas}, {Santos}, {Santos}, {Sarazin}, {Sarmento}, {Sarmiento-Cano}, {Sato}, {Schauer}, {Scherini}, {Schieler}, {Schimp}, {Schmidt}, {Scholten}, {Schov{\'a}nek}, {Schr{\"o}der}, {Schr{\"o}der}, {Schulz}, {Schumacher}, {Sciutto}, {Segreto}, {Shadkam}, {Shellard}, {Sigl}, {Silli}, {{\v{S}}m{\'\i}da}, {Snow}, {Sommers}, {Sonntag}, {Soriano}, {Squartini}, {Stanca}, {Stani{\v{c}}}, {Stasielak}, {Stassi}, {Stolpovskiy}, {Strafella}, {Streich}, {Suarez}, {Suarez-Dur{\'a}n},
  {Sudholz}, {Suomij{\"a}rvi}, {Supanitsky}, {{\v{S}}up{\'\i}k}, {Swain}, {Szadkowski}, {Taboada}, {Taborda}, {Timmermans}, {Todero Peixoto}, {Tomankova}, {Tom{\'e}}, {Torralba Elipe}, {Travnicek}, {Trini}, {Tueros}, {Ulrich}, {Unger}, {Urban}, {Vald{\'e}s Galicia}, {Vali{\~n}o}, {Valore}, {van Aar}, {van Bodegom}, {van den Berg}, {van Vliet}, {Varela}, {Vargas C{\'a}rdenas}, {V{\'a}zquez}, {Veberi{\v{c}}}, {Ventura}, {Vergara Quispe}, {Verzi}, {Vicha}, {Villase{\~n}or}, {Vorobiov}, {Wahlberg}, {Wainberg}, {Walz}, {Watson}, {Weber}, {Weindl}, {Wiede{\'n}ski}, {Wiencke}, {Wilczy{\'n}ski}, {Wirtz}, {Wittkowski}, {Wundheiler}, {Yang}, {Yushkov}, {Zas}, {Zavrtanik}, {Zavrtanik}, {Zepeda}, {Zimmermann}, {Ziolkowski}, {Zong}, {Zuccarello}, {Pierre Auger Collaboration}, {Kim}, {Schulze}, {Bauer}, {Corral-Santana}, {de Gregorio-Monsalvo}, {Gonz{\'a}lez-L{\'o}pez}, {Hartmann}, {Ishwara-Chandra}, {Mart{\'\i}n}, {Mehner}, {Misra}, {Micha{\l}owski}, {Resmi}, {ALMA Collaboration}, {Paragi}, {Agudo}, {An}, {Beswick},
  {Casadio}, {Frey}, {Jonker}, {Kettenis}, {Marcote}, {Moldon}, {Szomoru}, {van Langevelde}, {Yang}, {Euro VLBI Team}, {Cwiek}, {Cwiok}, {Czyrkowski}, {Dabrowski}, {Kasprowicz}, {Mankiewicz}, {Nawrocki}, {Opiela}, {Piotrowski}, {Wrochna}, {Zaremba}, {{\.Z}arnecki}, {Pi of Sky Collaboration}, {Haggard}, {Nynka}, {Ruan}, {Chandra Team at McGill University}, {Bland}, {Booler}, {Devillepoix}, {de Gois}, {Hancock}, {Howie}, {Paxman}, {Sansom}, {Towner}, {Desert Fireball Network}, {Tonry}, {Coughlin}, {Stubbs}, {Denneau}, {Heinze}, {Stalder}, {Weiland}, {ATLAS}, {Eatough}, {Kramer}, {Kraus}, {Time Resolution Universe Survey}, {Troja}, {Piro}, {Becerra Gonz{\'a}lez}, {Butler}, {Fox}, {Khandrika}, {Kutyrev}, {Lee}, {Ricci}, {Ryan}, {S{\'a}nchez-Ram{\'\i}rez}, {Veilleux}, {Watson}, {Wieringa}, {Burgess}, {van Eerten}, {Fontes}, {Fryer}, {Korobkin}, {Wollaeger}, {RIMAS}, {RATIR}, {Camilo}, {Foley}, {Goedhart}, {Makhathini}, {Oozeer}, {Smirnov}, {Fender}, {Woudt}, \& {South Africa/MeerKAT}}]{GW170817-MMAD}
{Abbott}, B.~P., {Abbott}, R., {Abbott}, T.~D., {et~al.} 2017, \apjl, 848, L12

\bibitem[{{Abdollahi} {et~al.}(2020){Abdollahi}, {Acero}, {Ackermann}, {Ajello}, {Atwood}, {Axelsson}, {Baldini}, {Ballet}, {Barbiellini}, {Bastieri}, {Becerra Gonzalez}, {Bellazzini}, {Berretta}, {Bissaldi}, {Blandford}, {Bloom}, {Bonino}, {Bottacini}, {Brandt}, {Bregeon}, {Bruel}, {Buehler}, {Burnett}, {Buson}, {Cameron}, {Caputo}, {Caraveo}, {Casandjian}, {Castro}, {Cavazzuti}, {Charles}, {Chaty}, {Chen}, {Cheung}, {Chiaro}, {Ciprini}, {Cohen-Tanugi}, {Cominsky}, {Coronado-Bl{\'a}zquez}, {Costantin}, {Cuoco}, {Cutini}, {D'Ammando}, {DeKlotz}, {de la Torre Luque}, {de Palma}, {Desai}, {Digel}, {Di Lalla}, {Di Mauro}, {Di Venere}, {Dom{\'\i}nguez}, {Dumora}, {Fana Dirirsa}, {Fegan}, {Ferrara}, {Franckowiak}, {Fukazawa}, {Funk}, {Fusco}, {Gargano}, {Gasparrini}, {Giglietto}, {Giommi}, {Giordano}, {Giroletti}, {Glanzman}, {Green}, {Grenier}, {Griffin}, {Grondin}, {Grove}, {Guiriec}, {Harding}, {Hayashi}, {Hays}, {Hewitt}, {Horan}, {J{\'o}hannesson}, {Johnson}, {Kamae}, {Kerr}, {Kocevski}, {Kovac'evic'},
  {Kuss}, {Landriu}, {Larsson}, {Latronico}, {Lemoine-Goumard}, {Li}, {Liodakis}, {Longo}, {Loparco}, {Lott}, {Lovellette}, {Lubrano}, {Madejski}, {Maldera}, {Malyshev}, {Manfreda}, {Marchesini}, {Marcotulli}, {Mart{\'\i}-Devesa}, {Martin}, {Massaro}, {Mazziotta}, {McEnery}, {Mereu}, {Meyer}, {Michelson}, {Mirabal}, {Mizuno}, {Monzani}, {Morselli}, {Moskalenko}, {Negro}, {Nuss}, {Ojha}, {Omodei}, {Orienti}, {Orlando}, {Ormes}, {Palatiello}, {Paliya}, {Paneque}, {Pei}, {Pe{\~n}a-Herazo}, {Perkins}, {Persic}, {Pesce-Rollins}, {Petrosian}, {Petrov}, {Piron}, {Poon}, {Porter}, {Principe}, {Rain{\`o}}, {Rando}, {Razzano}, {Razzaque}, {Reimer}, {Reimer}, {Remy}, {Reposeur}, {Romani}, {Saz Parkinson}, {Schinzel}, {Serini}, {Sgr{\`o}}, {Siskind}, {Smith}, {Spandre}, {Spinelli}, {Strong}, {Suson}, {Tajima}, {Takahashi}, {Tak}, {Thayer}, {Thompson}, {Tibaldo}, {Torres}, {Torresi}, {Valverde}, {Van Klaveren}, {van Zyl}, {Wood}, {Yassine}, \& {Zaharijas}}]{4fgl}
{Abdollahi}, S., {Acero}, F., {Ackermann}, M., {et~al.} 2020, \apjs, 247, 33

\bibitem[{{Abeysekara} {et~al.}(2017){Abeysekara}, {Albert}, {Alfaro}, {Alvarez}, {{\'A}lvarez}, {Arceo}, {Arteaga-Vel{\'a}zquez}, {Avila Rojas}, {Ayala Solares}, {Barber}, {Bautista-Elivar}, {Becerril}, {Belmont-Moreno}, {BenZvi}, {Berley}, {Bernal}, {Braun}, {Brisbois}, {Caballero-Mora}, {Capistr{\'a}n}, {Carrami{\~n}ana}, {Casanova}, {Castillo}, {Cotti}, {Cotzomi}, {Couti{\~n}o de Le{\'o}n}, {De Le{\'o}n}, {De la Fuente}, {Dingus}, {DuVernois}, {D{\'\i}az-V{\'e}lez}, {Ellsworth}, {Engel}, {Enr{\'\i}quez-Rivera}, {Fiorino}, {Fraija}, {Garc{\'\i}a-Gonz{\'a}lez}, {Garfias}, {Gerhardt}, {Gonz{\'a}lez Mu{\~n}oz}, {Gonz{\'a}lez}, {Goodman}, {Hampel-Arias}, {Harding}, {Hern{\'a}ndez}, {Hern{\'a}ndez-Almada}, {Hinton}, {Hona}, {Hui}, {H{\"u}ntemeyer}, {Iriarte}, {Jardin-Blicq}, {Joshi}, {Kaufmann}, {Kieda}, {Lara}, {Lauer}, {Lee}, {Lennarz}, {Vargas}, {Linnemann}, {Longinotti}, {Luis Raya}, {Luna-Garc{\'\i}a}, {L{\'o}pez-Coto}, {Malone}, {Marinelli}, {Martinez}, {Martinez-Castellanos}, {Mart{\'\i}nez-Castro},
  {Mart{\'\i}nez-Huerta}, {Matthews}, {Miranda-Romagnoli}, {Moreno}, {Mostaf{\'a}}, {Nellen}, {Newbold}, {Nisa}, {Noriega-Papaqui}, {Pelayo}, {Pretz}, {P{\'e}rez-P{\'e}rez}, {Ren}, {Rho}, {Rivi{\`e}re}, {Rosa-Gonz{\'a}lez}, {Rosenberg}, {Ruiz-Velasco}, {Salazar}, {Salesa Greus}, {Sandoval}, {Schneider}, {Schoorlemmer}, {Sinnis}, {Smith}, {Springer}, {Surajbali}, {Taboada}, {Tibolla}, {Tollefson}, {Torres}, {Ukwatta}, {Vianello}, {Weisgarber}, {Westerhoff}, {Wisher}, {Wood}, {Yapici}, {Yodh}, {Younk}, {Zepeda}, {Zhou}, {Guo}, {Hahn}, {Li}, \& {Zhang}}]{2017Sci...358..911A}
{Abeysekara}, A.~U., {Albert}, A., {Alfaro}, R., {et~al.} 2017, Science, 358, 911

\bibitem[{{Abramowicz} {et~al.}(1995){Abramowicz}, {Chen}, {Kato}, {Lasota}, \& {Regev}}]{Abramowicz_1995}
{Abramowicz}, M.~A., {Chen}, X., {Kato}, S., {Lasota}, J.-P., \& {Regev}, O. 1995, \apjl, 438, L37

\bibitem[{Abud {et~al.}(2022)Abud, Abi, Acciarri, Acero, Adames, Adamov, Adamowski, Adams, Adinolfi, Adriano, {et~al.}}]{abud2022snowmass}
Abud, A.~A., Abi, B., Acciarri, R., {et~al.} 2022, arXiv preprint arXiv:2203.06100

\bibitem[{{Abuter} {et~al.}(2024){Abuter}, {Allouche}, {Amorim}, {Bailet}, {Berdeu}, {Berger}, {Berio}, {Bigioli}, {Boebion}, {Bolzer}, {Bonnet}, {Bourdarot}, {Bourget}, {Brandner}, {Cao}, {Conzelmann}, {Comin}, {Cl{\'e}net}, {Courtney-Barrer}, {Davies}, {Defr{\`e}re}, {Delboulb{\'e}}, {Delplancke-Str{\"o}bele}, {Dembet}, {Dexter}, {de Zeeuw}, {Drescher}, {Eckart}, {{\'E}douard}, {Eisenhauer}, {Fabricius}, {Feuchtgruber}, {Finger}, {F{\"o}rster Schreiber}, {Garcia}, {Garcia Lopez}, {Gao}, {Gendron}, {Genzel}, {Gil}, {Gillessen}, {Gomes}, {Gont{\'e}}, {Gouvret}, {Guajardo}, {Guieu}, {Hackenberg}, {Haddad}, {Hartl}, {Haubois}, {Hau{\ss}mann}, {Hei{\ss}el}, {Henning}, {Hippler}, {H{\"o}nig}, {Horrobin}, {Hubin}, {Jacqmart}, {Jocou}, {Kaufer}, {Kervella}, {Kolb}, {Korhonen}, {Lacour}, {Lagarde}, {Lai}, {Lapeyr{\`e}re}, {Laugier}, {Le Bouquin}, {Leftley}, {L{\'e}na}, {Lewis}, {Liu}, {Lopez}, {Lutz}, {Magnard}, {Mang}, {Marcotto}, {Maurel}, {M{\'e}rand}, {Millour}, {More}, {Netzer}, {Nowacki}, {Nowak}, {Oberti},
  {Ott}, {Pallanca}, {Paumard}, {Perraut}, {Perrin}, {Petrov}, {Pfuhl}, {Pourr{\'e}}, {Rabien}, {Rau}, {Riquelme}, {Robbe-Dubois}, {Rochat}, {Salman}, {Sanchez-Bermudez}, {Santos}, {Scheithauer}, {Sch{\"o}ller}, {Schubert}, {Schuhler}, {Shangguan}, {Shchekaturov}, {Shimizu}, {Sevin}, {Soulez}, {Spang}, {Stadler}, {Sternberg}, {Straubmeier}, {Sturm}, {Sykes}, {Tacconi}, {Tristram}, {Vincent}, {von Fellenberg}, {Uysal}, {Widmann}, {Wieprecht}, {Wiezorrek}, {Woillez}, \& {Zins}}]{Abuter_2024}
{Abuter}, R., {Allouche}, F., {Amorim}, A., {et~al.} 2024, \nat, 627, 281

\bibitem[{Adame {et~al.}(2024)Adame, Aguilar, Ahlen, Alam, Alexander, Alvarez, Alves, Anand, Andrade, Armengaud, {et~al.}}]{adame2024desi}
Adame, A., Aguilar, J., Ahlen, S., {et~al.} 2024, arXiv preprint arXiv:2404.03002

\bibitem[{Adhikari {et~al.}(2021)Adhikari, Albataineh, Androic, Aniol, Armstrong, Averett, Ayerbe~Gayoso, Barcus, Bellini, Beminiwattha, {et~al.}}]{adhikari2021accurate}
Adhikari, D., Albataineh, H., Androic, D., {et~al.} 2021, Physical review letters, 126, 172502

\bibitem[{{Adhikari} {et~al.}(2024){Adhikari}, {Pe{\~n}il}, {Dom{\'\i}nguez}, {Ajello}, {Buson}, \& {Rico}}]{Adhikari_2024}
{Adhikari}, S., {Pe{\~n}il}, P., {Dom{\'\i}nguez}, A., {et~al.} 2024, arXiv e-prints, arXiv:2409.18334

\bibitem[{{Adler}(1971)}]{1971AnPhy..67..599A}
{Adler}, S.~L. 1971, Annals of Physics, 67, 599

\bibitem[{{Agazie} {et~al.}(2023){Agazie}, {Anumarlapudi}, {Archibald}, {Arzoumanian}, {Baker}, {B{\'e}csy}, {Blecha}, {Brazier}, {Brook}, {Burke-Spolaor}, {Burnette}, {Case}, {Charisi}, {Chatterjee}, {Chatziioannou}, {Cheeseboro}, {Chen}, {Cohen}, {Cordes}, {Cornish}, {Crawford}, {Cromartie}, {Crowter}, {Cutler}, {Decesar}, {Degan}, {Demorest}, {Deng}, {Dolch}, {Drachler}, {Ellis}, {Ferrara}, {Fiore}, {Fonseca}, {Freedman}, {Garver-Daniels}, {Gentile}, {Gersbach}, {Glaser}, {Good}, {G{\"u}ltekin}, {Hazboun}, {Hourihane}, {Islo}, {Jennings}, {Johnson}, {Jones}, {Kaiser}, {Kaplan}, {Kelley}, {Kerr}, {Key}, {Klein}, {Laal}, {Lam}, {Lamb}, {Lazio}, {Lewandowska}, {Littenberg}, {Liu}, {Lommen}, {Lorimer}, {Luo}, {Lynch}, {Ma}, {Madison}, {Mattson}, {McEwen}, {McKee}, {McLaughlin}, {McMann}, {Meyers}, {Meyers}, {Mingarelli}, {Mitridate}, {Natarajan}, {Ng}, {Nice}, {Ocker}, {Olum}, {Pennucci}, {Perera}, {Petrov}, {Pol}, {Radovan}, {Ransom}, {Ray}, {Romano}, {Sardesai}, {Schmiedekamp}, {Schmiedekamp}, {Schmitz},
  {Schult}, {Shapiro-Albert}, {Siemens}, {Simon}, {Siwek}, {Stairs}, {Stinebring}, {Stovall}, {Sun}, {Susobhanan}, {Swiggum}, {Taylor}, {Taylor}, {Turner}, {Unal}, {Vallisneri}, {van Haasteren}, {Vigeland}, {Wahl}, {Wang}, {Witt}, {Young}, \& {Nanograv Collaboration}}]{Agazie_2023}
{Agazie}, G., {Anumarlapudi}, A., {Archibald}, A.~M., {et~al.} 2023, \apjl, 951, L8

\bibitem[{{Agudo}(2024)}]{Agudo2024}
{Agudo}, I. 2024, in EAS2024, European Astronomical Society Annual Meeting, 2260

\bibitem[{{Aharonian} {et~al.}(2021){Aharonian}, {An}, {Axikegu}, {Bai}, {Bao}, {Bastieri}, {Bi}, {Bi}, {Cai}, {Cai}, {Cao}, {Cao}, {Chang}, {Chang}, {Chang}, {Chen}, {Chen}, {Chen}, {Chen}, {Chen}, {Chen}, {Chen}, {Chen}, {Chen}, {Chen}, {Chen}, {Chen}, {Chen}, {Cheng}, {Cheng}, {Cui}, {Cui}, {Cui}, {Dai}, {Dai}, {Dai}, {Danzengluobu}, {Della Volpe}, {D'Ettorre Piazzoli}, {Dong}, {Fan}, {Fan}, {Fan}, {Fang}, {Fang}, {Feng}, {Feng}, {Feng}, {Feng}, {Gao}, {Gao}, {Gao}, {Gao}, {Ge}, {Geng}, {Gong}, {Gou}, {Gu}, {Guo}, {Guo}, {Guo}, {Guo}, {Han}, {He}, {He}, {He}, {He}, {He}, {He}, {Heller}, {Hor}, {Hou}, {Hou}, {Hu}, {Hu}, {Hu}, {Hu}, {Huang}, {Huang}, {Huang}, {Huang}, {Huang}, {Ji}, {Ji}, {Jia}, {Jiang}, {Jiang}, {Jin}, {Kuleshov}, {Levochkin}, {Li}, {Li}, {Li}, {Li}, {Li}, {Li}, {Li}, {Li}, {Li}, {Li}, {Li}, {Li}, {Li}, {Li}, {Li}, {Li}, {Li}, {Liang}, {Liang}, {Lin}, {Liu}, {Liu}, {Liu}, {Liu}, {Liu}, {Liu}, {Liu}, {Liu}, {Liu}, {Liu}, {Liu}, {Liu}, {Liu}, {Liu}, {Liu}, {Long}, {Lu}, {Lv}, {Ma}, {Ma},
  {Ma}, {Mao}, {Masood}, {Mitthumsiri}, {Montaruli}, {Nan}, {Pang}, {Pattarakijwanich}, {Pei}, {Qi}, {Ruffolo}, {Rulev}, {S{\'a}iz}, {Shao}, {Shchegolev}, {Sheng}, {Shi}, {Song}, {Stenkin}, {Stepanov}, {Sun}, {Sun}, {Sun}, {Tam}, {Tang}, {Tian}, {Wang}, {Wang}, {Wang}, {Wang}, {Wang}, {Wang}, {Wang}, {Wang}, {Wang}, {Wang}, {Wang}, {Wang}, {Wang}, {Wang}, {Wang}, {Wang}, {Wang}, {Wang}, {Wang}, {Wang}, {Wang}, {Wei}, {Wei}, {Wei}, {Wen}, {Wu}, {Wu}, {Wu}, {Wu}, {Wu}, {Xi}, {Xia}, {Xia}, {Xiang}, {Xiao}, {Xiao}, {Xin}, {Xin}, {Xing}, {Xu}, {Xu}, {Xue}, {Yan}, \& {Yang}}]{2021PhRvL.126x1103A}
{Aharonian}, F., {An}, Q., {Axikegu}, L.~X., B., {et~al.} 2021, \prl, 126, 241103

\bibitem[{Ahumada {et~al.}(2024)Ahumada, Andrews, Antier, Blaufuss, Brady, Brazier, Burns, Cenko, Chandra, Chatterjee, {et~al.}}]{ahumada2024windows}
Ahumada, T., Andrews, J., Antier, S., {et~al.} 2024, arXiv preprint arXiv:2401.02063

\bibitem[{Aidala {et~al.}(2023)Aidala, Aprahamian, Bedaque, Bernstein, Carlson, Carpenter, Chipps, Cirigliano, Cloet, de~Gouvea, {et~al.}}]{aidala2023new}
Aidala, C., Aprahamian, A., Bedaque, P., {et~al.} 2023, A New Era Of Discovery: The 2023 Long-Range Plan For Nuclear Science, Tech. rep., Lawrence Livermore National Laboratory (LLNL), Livermore, CA (United States)

\bibitem[{{Ajello} {et~al.}(2023){Ajello}, {Murase}, \& {McDaniel}}]{Ajello_2023}
{Ajello}, M., {Murase}, K., \& {McDaniel}, A. 2023, \apjl, 954, L49

\bibitem[{{Ajello} {et~al.}(2020){Ajello}, {Angioni}, {Axelsson}, {Ballet}, {Barbiellini}, {Bastieri}, {Becerra Gonzalez}, {Bellazzini}, {Bissaldi}, {Bloom}, {Bonino}, {Bottacini}, {Bruel}, {Buson}, {Cafardo}, {Cameron}, {Cavazzuti}, {Chen}, {Cheung}, {Ciprini}, {Costantin}, {Cutini}, {D'Ammando}, {de la Torre Luque}, {de Menezes}, {de Palma}, {Desai}, {Di Lalla}, {Di Venere}, {Dom{\'\i}nguez}, {Dirirsa}, {Ferrara}, {Finke}, {Franckowiak}, {Fukazawa}, {Funk}, {Fusco}, {Gargano}, {Garrappa}, {Gasparrini}, {Giglietto}, {Giordano}, {Giroletti}, {Green}, {Grenier}, {Guiriec}, {Harita}, {Hays}, {Horan}, {Itoh}, {J{\'o}hannesson}, {Kovac'evic'}, {Krauss}, {Kreter}, {Kuss}, {Larsson}, {Leto}, {Li}, {Liodakis}, {Longo}, {Loparco}, {Lott}, {Lovellette}, {Lubrano}, {Madejski}, {Maldera}, {Manfreda}, {Mart{\'\i}-Devesa}, {Massaro}, {Mazziotta}, {Mereu}, {Meyer}, {Migliori}, {Mirabal}, {Mizuno}, {Monzani}, {Morselli}, {Moskalenko}, {Negro}, {Nemmen}, {Nuss}, {Ojha}, {Ojha}, {Omodei}, {Orienti}, {Orlando}, {Ormes},
  {Paliya}, {Pei}, {Pe{\~n}a-Herazo}, {Persic}, {Pesce-Rollins}, {Petrov}, {Piron}, {Poon}, {Principe}, {Rain{\`o}}, {Rando}, {Rani}, {Razzano}, {Razzaque}, {Reimer}, {Reimer}, {Schinzel}, {Serini}, {Sgr{\`o}}, {Siskind}, {Spandre}, {Spinelli}, {Suson}, {Tachibana}, {Thompson}, {Torres}, {Torresi}, {Troja}, {Valverde}, {van Zyl}, \& {Yassine}}]{4lac}
{Ajello}, M., {Angioni}, R., {Axelsson}, M., {et~al.} 2020, \apj, 892, 105

\bibitem[{{Ajello} {et~al.}(2022){Ajello}, {Baldini}, {Ballet}, {Bastieri}, {Becerra Gonzalez}, {Bellazzini}, {Berretta}, {Bissaldi}, {Bonino}, {Brill}, {Bruel}, {Buson}, {Caputo}, {Caraveo}, {Cheung}, {Chiaro}, {Cibrario}, {Ciprini}, {Crnogorcevic}, {Cutini}, {D'Ammando}, {De Gaetano}, {Di Lalla}, {Di Venere}, {Dom{\'\i}nguez}, {Ramazani}, {Ferrara}, {Fiori}, {Fukazawa}, {Funk}, {Fusco}, {Gammaldi}, {Gargano}, {Garrappa}, {Gasparrini}, {Giglietto}, {Giordano}, {Giroletti}, {Green}, {Grenier}, {Guiriec}, {Horan}, {Hou}, {Kayanoki}, {Kuss}, {Larsson}, {Latronico}, {Lewis}, {Li}, {Liodakis}, {Longo}, {Loparco}, {Lott}, {Lovellette}, {Lubrano}, {Madejski}, {Maldera}, {Manfreda}, {Mart{\'\i}-Devesa}, {Mazziotta}, {Mereu}, {Michelson}, {Mirabal}, {Mitthumsiri}, {Mizuno}, {Monzani}, {Morselli}, {Moskalenko}, {Negro}, {Ojha}, {Orienti}, {Orlando}, {Ormes}, {Pei}, {Pe{\~n}a-Herazo}, {Persic}, {Pesce-Rollins}, {Petrosian}, {Pillera}, {Poon}, {Porter}, {Principe}, {Rain{\`o}}, {Rando}, {Rani}, {Razzano}, {Razzaque},
  {Reimer}, {Reimer}, {Scotton}, {Serini}, {Sgr{\`o}}, {Siskind}, {Spandre}, {Spinelli}, {Suson}, {Tajima}, {Torres}, {Valverde}, {Yassin}, \& {Zaharijas}}]{Ajello2022}
{Ajello}, M., {Baldini}, L., {Ballet}, J., {et~al.} 2022, \apjs, 263, 24

\bibitem[{{Albert} {et~al.}(2023){Albert}, {Alfaro}, {Arteaga-Vel{\'a}zquez}, {Ayala Solares}, {Belmont-Moreno}, {Capistr{\'a}n}, {Carrami{\~n}ana}, {Casanova}, {Cotzomi}, {Couti{\~n}o De Le{\'o}n}, {De la Fuente}, {de Le{\'o}n}, {Diaz Hernandez}, {DuVernois}, {D{\'\i}az-V{\'e}lez}, {Espinoza}, {Fan}, {Fraija}, {Fang}, {Garc{\'\i}a-Gonz{\'a}lez}, {Garfias}, {Jardin-Blicq}, {Gonz{\'a}lez}, {Goodman}, {Harding}, {Hernandez}, {Huang}, {Hueyotl-Zahuantitla}, {H{\"u}ntemeyer}, {Iriarte}, {Joshi}, {Lara}, {Lee}, {Le{\'o}n Vargas}, {Linnemann}, {Longinotti}, {Luis-Raya}, {Malone}, {Martinez}, {Mart{\'\i}nez-Castro}, {Matthews}, {Morales-Soto}, {Moreno}, {Mostaf{\'a}}, {Nayerhoda}, {Nellen}, {Newbold}, {Nisa}, {P{\'e}rez Araujo}, {Son}, {P{\'e}rez-P{\'e}rez}, {Rho}, {Rosa-Gonz{\'a}lez}, {Sandoval}, {Schneider}, {Serna-Franco}, {Smith}, {Springer}, {Tollefson}, {Torres}, {Torres-Escobedo}, {Wang}, {Whitaker}, {Willox}, {Zhou}, \& {HAWC Collaboration}}]{2023ApJ...944L..29A}
{Albert}, A., {Alfaro}, R., {Arteaga-Vel{\'a}zquez}, J.~C., {et~al.} 2023, \apjl, 944, L29

\bibitem[{Aliotta \& Langanke(2022)}]{10.3389/fphy.2022.942726}
Aliotta, M., \& Langanke, K. 2022, Frontiers in Physics, 10, doi:10.3389/fphy.2022.942726.
\newblock \url{https://www.frontiersin.org/journals/physics/articles/10.3389/fphy.2022.942726}

\bibitem[{{An} \& {Romani}(2020)}]{An_2020}
{An}, H., \& {Romani}, R.~W. 2020, \apj, 904, 27

\bibitem[{{Anders} {et~al.}(2022){Anders}, {Jermyn}, {Lecoanet}, \& {Brown}}]{Anders2022}
{Anders}, E.~H., {Jermyn}, A.~S., {Lecoanet}, D., \& {Brown}, B.~P. 2022, \apj, 926, 169

\bibitem[{Andersen {et~al.}(2023)Andersen, Bandura, Bhardwaj, Boyle, Brar, Cassanelli, Chatterjee, Chawla, Cook, Curtin, {et~al.}}]{CHIME_repeaters2023}
Andersen, B.~C., Bandura, K., Bhardwaj, M., {et~al.} 2023, The Astrophysical Journal, 947, 83

\bibitem[{{Andrassy} {et~al.}(2020){Andrassy}, {Herwig}, {Woodward}, \& {Ritter}}]{Andrassy2020}
{Andrassy}, R., {Herwig}, F., {Woodward}, P., \& {Ritter}, C. 2020, \mnras, 491, 972

\bibitem[{{Andrassy} {et~al.}(2022){Andrassy}, {Higl}, {Mao}, {Moc{\'a}k}, {Vlaykov}, {Arnett}, {Baraffe}, {Campbell}, {Constantino}, {Edelmann}, {Goffrey}, {Guillet}, {Herwig}, {Hirschi}, {Horst}, {Leidi}, {Meakin}, {Pratt}, {Rizzuti}, {R{\"o}pke}, \& {Woodward}}]{2022A&A...659A.193A}
{Andrassy}, R., {Higl}, J., {Mao}, H., {et~al.} 2022, \aap, 659, A193

\bibitem[{{Andreoni} {et~al.}(2021){Andreoni}, {Coughlin}, {Kool}, {Kasliwal}, {Kumar}, {Bhalerao}, {Carracedo}, {Ho}, {Pang}, {Saraogi}, {Sharma}, {Shenoy}, {Burns}, {Ahumada}, {Anand}, {Singer}, {Perley}, {De}, {Fremling}, {Bellm}, {Bulla}, {Crellin-Quick}, {Dietrich}, {Drake}, {Duev}, {Goobar}, {Graham}, {Kaplan}, {Kulkarni}, {Laher}, {Mahabal}, {Shupe}, {Sollerman}, {Walters}, \& {Yao}}]{andreoni2021}
{Andreoni}, I., {Coughlin}, M.~W., {Kool}, E.~C., {et~al.} 2021, \apj, 918, 63

\bibitem[{{Andreoni} {et~al.}(2022){Andreoni}, {Coughlin}, {Perley}, {Yao}, {Lu}, {Cenko}, {Kumar}, {Anand}, {Ho}, {Kasliwal}, {de Ugarte Postigo}, {Sagu{\'e}s-Carracedo}, {Schulze}, {Kann}, {Kulkarni}, {Sollerman}, {Tanvir}, {Rest}, {Izzo}, {Somalwar}, {Kaplan}, {Ahumada}, {Anupama}, {Auchettl}, {Barway}, {Bellm}, {Bhalerao}, {Bloom}, {Bremer}, {Bulla}, {Burns}, {Campana}, {Chandra}, {Charalampopoulos}, {Cooke}, {D'Elia}, {Das}, {Dobie}, {Ag{\"u}{\'\i} Fern{\'a}ndez}, {Freeburn}, {Fremling}, {Gezari}, {Goode}, {Graham}, {Hammerstein}, {Karambelkar}, {Kilpatrick}, {Kool}, {Krips}, {Laher}, {Leloudas}, {Levan}, {Lundquist}, {Mahabal}, {Medford}, {Miller}, {M{\"o}ller}, {Mooley}, {Nayana}, {Nir}, {Pang}, {Paraskeva}, {Perley}, {Petitpas}, {Pursiainen}, {Ravi}, {Ridden-Harper}, {Riddle}, {Rigault}, {Rodriguez}, {Rusholme}, {Sharma}, {Smith}, {Stein}, {Th{\"o}ne}, {Tohuvavohu}, {Valdes}, {van Roestel}, {Vergani}, {Wang}, \& {Zhang}}]{Andreoni2022}
{Andreoni}, I., {Coughlin}, M.~W., {Perley}, D.~A., {et~al.} 2022, \nat, 612, 430

\bibitem[{Andreoni {et~al.}(2024)Andreoni, Margutti, Banovetz, Greenstreet, Hebert, Lister, Palmese, Piranomonte, Smartt, Smith, {et~al.}}]{andreoni2024rubin}
Andreoni, I., Margutti, R., Banovetz, J., {et~al.} 2024, arXiv preprint arXiv:2411.04793

\bibitem[{{Angus} {et~al.}(2024){Angus}, {Woosley}, {Foley}, {Nicholl}, {Villar}, {Taggart}, {Pursiainen}, {Ramsden}, {Srivastav}, {Stevance}, {Moore}, {Auchettl}, {Hoogendam}, {Khetan}, {Yadavalli}, {Dimitriadis}, {Gagliano}, {Siebert}, {Aamer}, {de Boer}, {Chambers}, {Clocchiatti}, {Coulter}, {Drout}, {Farias}, {Fulton}, {Gall}, {Gao}, {Izzo}, {Jones}, {Lin}, {Magnier}, {Narayan}, {Ramirez-Ruiz}, {Ransome}, {Rest}, {Smartt}, \& {Smith}}]{2024arXiv240902174A}
{Angus}, C.~R., {Woosley}, S.~E., {Foley}, R.~J., {et~al.} 2024, arXiv e-prints, arXiv:2409.02174

\bibitem[{{Arcavi} {et~al.}(2017){Arcavi}, {Howell}, {Kasen}, {Bildsten}, {Hosseinzadeh}, {McCully}, {Wong}, {Katz}, {Gal-Yam}, {Sollerman}, {Taddia}, {Leloudas}, {Fremling}, {Nugent}, {Horesh}, {Mooley}, {Rumsey}, {Cenko}, {Graham}, {Perley}, {Nakar}, {Shaviv}, {Bromberg}, {Shen}, {Ofek}, {Cao}, {Wang}, {Huang}, {Rui}, {Zhang}, {Li}, {Li}, {Zhang}, {Valenti}, {Guevel}, {Shappee}, {Kochanek}, {Holoien}, {Filippenko}, {Fender}, {Nyholm}, {Yaron}, {Kasliwal}, {Sullivan}, {Blagorodnova}, {Walters}, {Lunnan}, {Khazov}, {Andreoni}, {Laher}, {Konidaris}, {Wozniak}, \& {Bue}}]{2017Natur.551..210A}
{Arcavi}, I., {Howell}, D.~A., {Kasen}, D., {et~al.} 2017, \nat, 551, 210

\bibitem[{{Arcodia} {et~al.}(2021){Arcodia}, {Merloni}, {Nandra}, {Buchner}, {Salvato}, {Pasham}, {Remillard}, {Comparat}, {Lamer}, {Ponti}, {Malyali}, {Wolf}, {Arzoumanian}, {Bogensberger}, {Buckley}, {Gendreau}, {Gromadzki}, {Kara}, {Krumpe}, {Markwardt}, {Ramos-Ceja}, {Rau}, {Schramm}, \& {Schwope}}]{Arcodia21}
{Arcodia}, R., {Merloni}, A., {Nandra}, K., {et~al.} 2021, \nat, 592, 704

\bibitem[{{Arcodia} {et~al.}(2024{\natexlab{a}}){Arcodia}, {Liu}, {Merloni}, {Malyali}, {Rau}, {Chakraborty}, {Goodwin}, {Buckley}, {Brink}, {Gromadzki}, {Arzoumanian}, {Buchner}, {Kara}, {Nandra}, {Ponti}, {Salvato}, {Anderson}, {Baldini}, {Grotova}, {Krumpe}, {Maitra}, {Miller-Jones}, \& {Ramos-Ceja}}]{Arcodia24}
{Arcodia}, R., {Liu}, Z., {Merloni}, A., {et~al.} 2024{\natexlab{a}}, \aap, 684, A64

\bibitem[{{Arcodia} {et~al.}(2024{\natexlab{b}}){Arcodia}, {Liu}, {Merloni}, {Malyali}, {Rau}, {Chakraborty}, {Goodwin}, {Buckley}, {Brink}, {Gromadzki}, {Arzoumanian}, {Buchner}, {Kara}, {Nandra}, {Ponti}, {Salvato}, {Anderson}, {Baldini}, {Grotova}, {Krumpe}, {Maitra}, {Miller-Jones}, \& {Ramos-Ceja}}]{2024A&A...684A..64A}
---. 2024{\natexlab{b}}, \aap, 684, A64

\bibitem[{Arcones \& Thielemann(2023)}]{arcones2023origin}
Arcones, A., \& Thielemann, F.-K. 2023, The Astronomy and Astrophysics Review, 31, 1

\bibitem[{{Arnett} {et~al.}(2019){Arnett}, {Meakin}, {Hirschi}, {Cristini}, {Georgy}, {Campbell}, {Scott}, {Kaiser}, {Viallet}, \& {Moc{\'a}k}}]{2019ApJ...882...18A}
{Arnett}, W.~D., {Meakin}, C., {Hirschi}, R., {et~al.} 2019, \apj, 882, 18

\bibitem[{{Arnold} {et~al.}(2021{\natexlab{a}}){Arnold}, {Drake}, {Swisdak}, {Guo}, {Dahlin}, {Chen}, {Fleishman}, {Glesener}, {Kontar}, {Phan}, \& {Shen}}]{Arnold2021}
{Arnold}, H., {Drake}, J.~F., {Swisdak}, M., {et~al.} 2021{\natexlab{a}}, \prl, 126, 135101

\bibitem[{{Arnold} {et~al.}(2021{\natexlab{b}}){Arnold}, {Drake}, {Swisdak}, {Guo}, {Dahlin}, {Chen}, {Fleishman}, {Glesener}, {Kontar}, {Phan}, \& {Shen}}]{2021PhRvL.126m5101A}
---. 2021{\natexlab{b}}, \prl, 126, 135101

\bibitem[{{Arrowsmith} {et~al.}(2024){Arrowsmith}, {Simon}, {Bilbao}, {Bott}, {Burger}, {Chen}, {Cruz}, {Davenne}, {Efthymiopoulos}, {Froula}, {Goillot}, {Gudmundsson}, {Haberberger}, {Halliday}, {Hodge}, {Huffman}, {Iaquinta}, {Miniati}, {Reville}, {Sarkar}, {Schekochihin}, {Silva}, {Simpson}, {Stergiou}, {Trines}, {Vieu}, {Charitonidis}, {Bingham}, \& {Gregori}}]{2024NatCo..15.5029A}
{Arrowsmith}, C.~D., {Simon}, P., {Bilbao}, P.~J., {et~al.} 2024, Nature Communications, 15, 5029

\bibitem[{{Asai} {et~al.}(2024){Asai}, {Ballarino}, {Bose}, {Cranmer}, {Cyr-Racine}, {Demers}, {Geddes}, {Gershtein}, {Heeger}, {Heinemann}, {Hewett}, {Huber}, {Mahn}, {Mandelbaum}, {Maricic}, {Merkel}, {Monahan}, {Murayama}, {Onyisi}, {Palmer}, {Raubenheimer}, {Sanchez}, {Schnee}, {Seidel}, {Seo}, {Thaler}, {Touramanis}, {Vieregg}, {Weinstein}, {Winslow}, {Yu}, \& {Zwaska}}]{2024arXiv240719176A}
{Asai}, S., {Ballarino}, A., {Bose}, T., {et~al.} 2024, arXiv e-prints, arXiv:2407.19176

\bibitem[{{Atwood} {et~al.}(2009){Atwood}, {Abdo}, {Ackermann}, {Althouse}, {Anderson}, {Axelsson}, {Baldini}, {Ballet}, {Band}, {Barbiellini}, {Bartelt}, {Bastieri}, {Baughman}, {Bechtol}, {B{\'e}d{\'e}r{\`e}de}, {Bellardi}, {Bellazzini}, {Berenji}, {Bignami}, {Bisello}, {Bissaldi}, {Blandford}, {Bloom}, {Bogart}, {Bonamente}, {Bonnell}, {Borgland}, {Bouvier}, {Bregeon}, {Brez}, {Brigida}, {Bruel}, {Burnett}, {Busetto}, {Caliandro}, {Cameron}, {Caraveo}, {Carius}, {Carlson}, {Casandjian}, {Cavazzuti}, {Ceccanti}, {Cecchi}, {Charles}, {Chekhtman}, {Cheung}, {Chiang}, {Chipaux}, {Cillis}, {Ciprini}, {Claus}, {Cohen-Tanugi}, {Condamoor}, {Conrad}, {Corbet}, {Corucci}, {Costamante}, {Cutini}, {Davis}, {Decotigny}, {DeKlotz}, {Dermer}, {de Angelis}, {Digel}, {do Couto e Silva}, {Drell}, {Dubois}, {Dumora}, {Edmonds}, {Fabiani}, {Farnier}, {Favuzzi}, {Flath}, {Fleury}, {Focke}, {Funk}, {Fusco}, {Gargano}, {Gasparrini}, {Gehrels}, {Gentit}, {Germani}, {Giebels}, {Giglietto}, {Giommi}, {Giordano}, {Glanzman},
  {Godfrey}, {Grenier}, {Grondin}, {Grove}, {Guillemot}, {Guiriec}, {Haller}, {Harding}, {Hart}, {Hays}, {Healey}, {Hirayama}, {Hjalmarsdotter}, {Horn}, {Hughes}, {J{\'o}hannesson}, {Johansson}, {Johnson}, {Johnson}, {Johnson}, {Johnson}, {Kamae}, {Katagiri}, {Kataoka}, {Kavelaars}, {Kawai}, {Kelly}, {Kerr}, {Klamra}, {Kn{\"o}dlseder}, {Kocian}, {Komin}, {Kuehn}, {Kuss}, {Landriu}, {Latronico}, {Lee}, {Lee}, {Lemoine-Goumard}, {Lionetto}, {Longo}, {Loparco}, {Lott}, {Lovellette}, {Lubrano}, {Madejski}, {Makeev}, {Marangelli}, {Massai}, {Mazziotta}, {McEnery}, {Menon}, {Meurer}, {Michelson}, {Minuti}, {Mirizzi}, {Mitthumsiri}, {Mizuno}, {Moiseev}, {Monte}, {Monzani}, {Moretti}, {Morselli}, {Moskalenko}, {Murgia}, {Nakamori}, {Nishino}, {Nolan}, {Norris}, {Nuss}, {Ohno}, {Ohsugi}, {Omodei}, {Orlando}, {Ormes}, {Paccagnella}, {Paneque}, {Panetta}, {Parent}, {Pearce}, {Pepe}, {Perazzo}, {Pesce-Rollins}, {Picozza}, {Pieri}, {Pinchera}, {Piron}, {Porter}, {Poupard}, {Rain{\`o}}, {Rando}, {Rapposelli}, {Razzano},
  {Reimer}, {Reimer}, {Reposeur}, {Reyes}, {Ritz}, {Rochester}, {Rodriguez}, {Romani}, {Roth}, {Russell}, {Ryde}, {Sabatini}, {Sadrozinski}, {Sanchez}, {Sander}, {Sapozhnikov}, {Parkinson}, {Scargle}, {Schalk}, {Scolieri}, {Sgr{\`o}}, {Share}, {Shaw}, {Shimokawabe}, {Shrader}, {Sierpowska-Bartosik}, {Siskind}, {Smith}, {Smith}, {Spandre}, {Spinelli}, {Starck}, {Stephens}, {Strickman}, {Strong}, {Suson}, {Tajima}, {Takahashi}, {Takahashi}, {Tanaka}, {Tenze}, {Tether}, {Thayer}, {Thayer}, {Thompson}, {Tibaldo}, {Tibolla}, {Torres}, {Tosti}, {Tramacere}, {Turri}, {Usher}, {Vilchez}, {Vitale}, {Wang}, {Watters}, {Winer}, {Wood}, {Ylinen}, \& {Ziegler}}]{atwood2009}
{Atwood}, W.~B., {Abdo}, A.~A., {Ackermann}, M., {et~al.} 2009, \apj, 697, 1071

\bibitem[{Baalrud {et~al.}(2020)Baalrud, Ferraro, Garrison, Howard, Kuranz, Sarff, \& Solomon}]{baalrud2020communityplanfusionenergy}
Baalrud, S., Ferraro, N., Garrison, L., {et~al.} 2020, A Community Plan for Fusion Energy and Discovery Plasma Sciences, , , arXiv:2011.04806.
\newblock \url{https://arxiv.org/abs/2011.04806}

\bibitem[{{Bale} {et~al.}(2010){Bale}, {Bhattacharjee}, {Cattaneo}, {Drake}, {Ji}, {Lee}, {Li}, {Liang}, {Pound}, {Prager}, {Quataert}, {Remington}, {Rosner}, {Ryutov}, {Thomas}, \& {Zweibel}}]{bale2010research}
{Bale}, S., {Bhattacharjee}, A., {Cattaneo}, F., {et~al.} 2010, Report of the Workshop on Opportunities in Plasma Astrophysics, arXiv:2203.02406

\bibitem[{{Ballet} {et~al.}(2023){Ballet}, {Bruel}, {Burnett}, {Lott}, \& {The Fermi-LAT collaboration}}]{Ballet_2023}
{Ballet}, J., {Bruel}, P., {Burnett}, T.~H., {Lott}, B., \& {The Fermi-LAT collaboration}. 2023, arXiv e-prints, arXiv:2307.12546

\bibitem[{{Banados} {et~al.}(2024){Banados}, {Momjian}, {Connor}, {Belladitta}, {Decarli}, {Mazzucchelli}, {Venemans}, {Walter}, {Wang}, {Xie}, {Barth}, {Eilers}, {Fan}, {Khusanova}, {Schindler}, {Stern}, {Yang}, {Taufik Andika}, {Carilli}, {Farina}, {Fabian}, {Hennawi}, {Pensabene}, \& {Rojas-Ruiz}}]{Banados_2024}
{Banados}, E., {Momjian}, E., {Connor}, T., {et~al.} 2024, arXiv e-prints, arXiv:2407.07236

\bibitem[{{Baring} \& {Harding}(2001{\natexlab{a}})}]{2001ApJ...547..929B}
{Baring}, M.~G., \& {Harding}, A.~K. 2001{\natexlab{a}}, \apj, 547, 929

\bibitem[{{Baring} \& {Harding}(2001{\natexlab{b}})}]{BH-2001-ApJ}
---. 2001{\natexlab{b}}, \apj, 547, 929

\bibitem[{{Baring} \& {Harding}(2007)}]{BH-2007-ApandSS}
---. 2007, \apss, 308, 109

\bibitem[{{Barkat} {et~al.}(1967){Barkat}, {Rakavy}, \& {Sack}}]{1967PhRvL..18..379B}
{Barkat}, Z., {Rakavy}, G., \& {Sack}, N. 1967, \prl, 18, 379

\bibitem[{{Baskin} \& {Laor}(2005)}]{Baskin_2005}
{Baskin}, A., \& {Laor}, A. 2005, \mnras, 356, 1029

\bibitem[{Baym {et~al.}(1969)Baym, Pethick, \& Pines}]{baym1969superfluidity}
Baym, G., Pethick, C., \& Pines, D. 1969, Nature, 224, 673

\bibitem[{{Begelman} {et~al.}(1983){Begelman}, {McKee}, \& {Shields}}]{Begelman_1983}
{Begelman}, M.~C., {McKee}, C.~F., \& {Shields}, G.~A. 1983, \apj, 271, 70

\bibitem[{{Bell}(1978)}]{bell78}
{Bell}, A.~R. 1978, \mnras, 182, 147

\bibitem[{{Belladitta} {et~al.}(2020){Belladitta}, {Moretti}, {Caccianiga}, {Spingola}, {Severgnini}, {Della Ceca}, {Ghisellini}, {Dallacasa}, {Sbarrato}, {Cicone}, {Cassar{\`a}}, \& {Pedani}}]{Belladitta_2020}
{Belladitta}, S., {Moretti}, A., {Caccianiga}, A., {et~al.} 2020, \aap, 635, L7

\bibitem[{{Beloborodov}(2020)}]{Beloborodov2020}
{Beloborodov}, A.~M. 2020, \apj, 896, 142

\bibitem[{Beniamini {et~al.}(2019)Beniamini, Hotokezaka, van der Horst, \& Kouveliotou}]{Beniamini_10.1093/mnras/stz1391}
Beniamini, P., Hotokezaka, K., van der Horst, A., \& Kouveliotou, C. 2019, Monthly Notices of the Royal Astronomical Society, 487, 1426.
\newblock \url{https://doi.org/10.1093/mnras/stz1391}

\bibitem[{{Beniamini} {et~al.}(2023){Beniamini}, {Wadiasingh}, {Hare}, {Rajwade}, {Younes}, \& {van der Horst}}]{2023MNRAS.520.1872B}
{Beniamini}, P., {Wadiasingh}, Z., {Hare}, J., {et~al.} 2023, \mnras, 520, 1872

\bibitem[{Beniamini {et~al.}(2024)Beniamini, Wadiasingh, Trigg, Chirenti, Burns, Younes, Negro, \& Granot}]{beniamini_2024extragalacticmagnetargiantflares}
Beniamini, P., Wadiasingh, Z., Trigg, A., {et~al.} 2024, Extragalactic Magnetar Giant Flares: Population Implications, Rates and Prospects for Gamma-Rays, Gravitational Waves and Neutrinos,  The Astrophysical Journal, arXiv:2411.16846.
\newblock \url{https://arxiv.org/abs/2411.16846}

\bibitem[{{Bera} {et~al.}(2024){Bera}, {James}, {Deller}, {Bannister}, {Shannon}, {Scott}, {Gourdji}, {Marnoch}, {Glowacki}, {Ekers}, {Ryder}, \& {Dial}}]{Bera2024}
{Bera}, A., {James}, C.~W., {Deller}, A.~T., {et~al.} 2024, \apjl, 969, L29

\bibitem[{{Best} {et~al.}(2016){Best}, {Caciolli}, {F{\"u}l{\"o}p}, {Gy{\"u}rky}, {Laubenstein}, {Napolitani}, {Rigato}, {Roca}, \& {Sz{\"u}cs}}]{best2016}
{Best}, A., {Caciolli}, A., {F{\"u}l{\"o}p}, Z., {et~al.} 2016, European Physical Journal A, 52, 72

\bibitem[{{Bethe} \& {Wilson}(1985)}]{BeWi85}
{Bethe}, H.~A., \& {Wilson}, J.~R. 1985, \apj, 295, 14

\bibitem[{{Beun} {et~al.}(2006){Beun}, {McLaughlin}, {Surman}, \& {Hix}}]{2006PhRvD..73i3007B}
{Beun}, J., {McLaughlin}, G.~C., {Surman}, R., \& {Hix}, W.~R. 2006, \prd, 73, 093007

\bibitem[{{Blandford} \& {Eichler}(1987)}]{Blandford1987}
{Blandford}, R., \& {Eichler}, D. 1987, \physrep, 154, 1

\bibitem[{{Blandford} \& {McKee}(1982)}]{Blandford_1982}
{Blandford}, R.~D., \& {McKee}, C.~F. 1982, \apj, 255, 419

\bibitem[{Blandford \& Ostriker(1978)}]{bo78}
Blandford, R.~D., \& Ostriker, J.~P. 1978, Astrophysical Journal, Part 2-Letters to the Editor, vol. 221, Apr. 1, 1978, p. L29-L32., 221, L29

\bibitem[{Blandford \& Znajek(1977)}]{blandford1977electromagnetic}
Blandford, R.~D., \& Znajek, R.~L. 1977, Monthly Notices of the Royal Astronomical Society, 179, 433

\bibitem[{{Blecha} {et~al.}(2018){Blecha}, {Snyder}, {Satyapal}, \& {Ellison}}]{Blecha18}
{Blecha}, L., {Snyder}, G.~F., {Satyapal}, S., \& {Ellison}, S.~L. 2018, \mnras, 478, 3056

\bibitem[{{Blondin} {et~al.}(2022){Blondin}, {Blinnikov}, {Callan}, {Collins}, {Dessart}, {Even}, {Fl{\"o}rs}, {Fullard}, {Hillier}, {Jerkstrand}, {Kasen}, {Katz}, {Kerzendorf}, {Kozyreva}, {O'Brien}, {P{\'a}ssaro}, {Roth}, {Shen}, {Shingles}, {Sim}, {Singhal}, {Smith}, {Sorokina}, {Utrobin}, {Vogl}, {Williamson}, {Wollaeger}, {Woosley}, \& {Wygoda}}]{2022A&A...668A.163B}
{Blondin}, S., {Blinnikov}, S., {Callan}, F.~P., {et~al.} 2022, \aap, 668, A163

\bibitem[{{Bloom} {et~al.}(2011){Bloom}, {Giannios}, {Metzger}, {Cenko}, {Perley}, {Butler}, {Tanvir}, {Levan}, {O'Brien}, {Strubbe}, {De Colle}, {Ramirez-Ruiz}, {Lee}, {Nayakshin}, {Quataert}, {King}, {Cucchiara}, {Guillochon}, {Bower}, {Fruchter}, {Morgan}, \& {van der Horst}}]{Bloom2011}
{Bloom}, J.~S., {Giannios}, D., {Metzger}, B.~D., {et~al.} 2011, Science, 333, 203

\bibitem[{{Boccioli} \& {Roberti}(2024)}]{boccioli2024}
{Boccioli}, L., \& {Roberti}, L. 2024, Universe, 10, 148

\bibitem[{{Bochenek} {et~al.}(2020){Bochenek}, {Ravi}, {Belov}, {Hallinan}, {Kocz}, {Kulkarni}, \& {McKenna}}]{Bochenek2020}
{Bochenek}, C.~D., {Ravi}, V., {Belov}, K.~V., {et~al.} 2020, \nat, 587, 59

\bibitem[{{Bochenek} {et~al.}(2021){Bochenek}, {Ravi}, \& {Dong}}]{Bochenek2021}
{Bochenek}, C.~D., {Ravi}, V., \& {Dong}, D. 2021, \apjl, 907, L31

\bibitem[{Bohr {et~al.}(1958)Bohr, Mottelson, \& Pines}]{bohr1958possible}
Bohr, A., Mottelson, B.~R., \& Pines, D. 1958, Physical Review, 110, 936

\bibitem[{{Bolatto} {et~al.}(2021){Bolatto}, {Leroy}, {Levy}, {Meier}, {Mills}, {Thompson}, {Emig}, {Veilleux}, {Ott}, {Gorski}, {Walter}, {Lopez}, \& {Lenki{\'c}}}]{Bolatto2021}
{Bolatto}, A.~D., {Leroy}, A.~K., {Levy}, R.~C., {et~al.} 2021, \apj, 923, 83

\bibitem[{{Boorman} {et~al.}(2024{\natexlab{a}}){Boorman}, {Gandhi}, {Buchner}, {Stern}, {Ricci}, {Balokovi{\'c}}, {Asmus}, {Harrison}, {Svoboda}, {Greenwell}, {Koss}, {Alexander}, {Annuar}, {Bauer}, {Brandt}, {Brightman}, {Panessa}, {Chen}, {Farrah}, {Forster}, {Grefenstette}, {H{\"o}nig}, {Hill}, {Kammoun}, {Lansbury}, {Lanz}, {LaMassa}, {Madsen}, {Marchesi}, {Middleton}, {Mingo}, {Parker}, {Treister}, {Ueda}, {Urry}, \& {Zappacosta}}]{Boorman24_nulands}
{Boorman}, P.~G., {Gandhi}, P., {Buchner}, J., {et~al.} 2024{\natexlab{a}}, arXiv e-prints, arXiv:2410.07339

\bibitem[{{Boorman} {et~al.}(2024{\natexlab{b}}){Boorman}, {Torres-Alb{\`a}}, {Annuar}, {Marchesi}, {Pfeifle}, {Stern}, {Civano}, {Balokovi{\'c}}, {Buchner}, {Ricci}, {Alexander}, {Brandt}, {Brightman}, {Chen}, {Creech}, {Gandhi}, {Garc{\'\i}a}, {Harrison}, {Hickox}, {Kammoun}, {LaMassa}, {Lanzuisi}, {Marcotulli}, {Madsen}, {Matt}, {Matzeu}, {Nardini}, {Piotrowska}, {Pizzetti}, {Puccetti}, {Sicilian}, {Silver}, {Walton}, {Wilkins}, \& {Zhao}}]{Boorman24_hexp}
{Boorman}, P.~G., {Torres-Alb{\`a}}, N., {Annuar}, A., {et~al.} 2024{\natexlab{b}}, Frontiers in Astronomy and Space Sciences, 11, 1335459

\bibitem[{Boroumand {et~al.}(2024)Boroumand, Morra, \& Mora}]{ali2024extracting}
Boroumand, M.~A., Morra, G., \& Mora, P. 2024, Journal of Applied Physics, 135

\bibitem[{{B{\"o}ttcher}(2019)}]{Bottcher2019}
{B{\"o}ttcher}, M. 2019, Galaxies, 7, 20

\bibitem[{{Brandt} \& {Alexander}(2015)}]{Brandt15}
{Brandt}, W.~N., \& {Alexander}, D.~M. 2015, \aapr, 23, 1

\bibitem[{{Brickhouse} {et~al.}(2009){Brickhouse}, {Cowan}, {Drake}, {Federman}, {Ferland}, {Frank}, {Haxton}, {Herbst}, {Olive}, {Salama}, {Savin}, \& {Ziurys}}]{2009astro2010P..68B}
{Brickhouse}, N., {Cowan}, J., {Drake}, P., {et~al.} 2009, in astro2010: The Astronomy and Astrophysics Decadal Survey, Vol. 2010, P68

\bibitem[{{Brickhouse} {et~al.}(2020){Brickhouse}, {Ferland}, {Milam}, {Sciamma-O'Brien}, {Smale}, {Spyrou}, {Stancil}, {Storrie-Lombardi}, {Wahlgren}, \& {Smith}}]{Brickhouse_2018}
{Brickhouse}, N., {Ferland}, G.~J., {Milam}, S., {et~al.} 2020, in Bulletin of the American Astronomical Society, Vol.~52, 0202

\bibitem[{{Brightman} {et~al.}(2016){Brightman}, {Masini}, {Ballantyne}, {Balokovi{\'c}}, {Brandt}, {Chen}, {Comastri}, {Farrah}, {Gandhi}, {Harrison}, {Ricci}, {Stern}, \& {Walton}}]{Brightman16}
{Brightman}, M., {Masini}, A., {Ballantyne}, D.~R., {et~al.} 2016, \apj, 826, 93

\bibitem[{{Brightman} {et~al.}(2017){Brightman}, {Balokovi{\'c}}, {Ballantyne}, {Bauer}, {Boorman}, {Buchner}, {Brandt}, {Comastri}, {Del Moro}, {Farrah}, {Gandhi}, {Harrison}, {Koss}, {Lanz}, {Masini}, {Ricci}, {Stern}, {Vasudevan}, \& {Walton}}]{Brightman17}
{Brightman}, M., {Balokovi{\'c}}, M., {Ballantyne}, D.~R., {et~al.} 2017, \apj, 844, 10

\bibitem[{{Brown} {et~al.}(2015){Brown}, {Levan}, {Stanway}, {Tanvir}, {Cenko}, {Berger}, {Chornock}, \& {Cucchiaria}}]{Brown2015}
{Brown}, G.~C., {Levan}, A.~J., {Stanway}, E.~R., {et~al.} 2015, \mnras, 452, 4297

\bibitem[{{Buchner} {et~al.}(2014){Buchner}, {Georgakakis}, {Nandra}, {Hsu}, {Rangel}, {Brightman}, {Merloni}, {Salvato}, {Donley}, \& {Kocevski}}]{Buchner14}
{Buchner}, J., {Georgakakis}, A., {Nandra}, K., {et~al.} 2014, \aap, 564, A125

\bibitem[{{Buchner} {et~al.}(2015){Buchner}, {Georgakakis}, {Nandra}, {Brightman}, {Menzel}, {Liu}, {Hsu}, {Salvato}, {Rangel}, {Aird}, {Merloni}, \& {Ross}}]{Buchner15}
---. 2015, \apj, 802, 89

\bibitem[{Burbidge {et~al.}(1957)Burbidge, Burbidge, Fowler, \& Hoyle}]{burbidge1957synthesis}
Burbidge, E.~M., Burbidge, G.~R., Fowler, W.~A., \& Hoyle, F. 1957, Reviews of modern physics, 29, 547

\bibitem[{Burns(2020)}]{burns2020neutron}
Burns, E. 2020, Living Reviews in Relativity, 23, 4

\bibitem[{{Burns} {et~al.}(2021){Burns}, {Svinkin}, {Hurley}, {Wadiasingh}, {Negro}, {Younes}, {Hamburg}, {Ridnaia}, {Cook}, {Cenko}, {Aloisi}, {Ashton}, {Baring}, {Briggs}, {Christensen}, {Frederiks}, {Goldstein}, {Hui}, {Kaplan}, {Kasliwal}, {Kocevski}, {Roberts}, {Savchenko}, {Tohuvavohu}, {Veres}, \& {Wilson-Hodge}}]{Burns_2021ApJ...907L..28B}
{Burns}, E., {Svinkin}, D., {Hurley}, K., {et~al.} 2021, \apjl, 907, L28

\bibitem[{Burns {et~al.}(2023)Burns, Coughlin, Ackley, Andreoni, Bizouard, Broekgaarden, Christensen, d'Ammando, DeLaunay, Fleischhack, {et~al.}}]{burns2023gamma}
Burns, E., Coughlin, M., Ackley, K., {et~al.} 2023, arXiv preprint arXiv:2308.04485

\bibitem[{{Burrows} {et~al.}(2011){Burrows}, {Kennea}, {Ghisellini}, {Mangano}, {Zhang}, {Page}, {Eracleous}, {Romano}, {Sakamoto}, {Falcone}, {Osborne}, {Campana}, {Beardmore}, {Breeveld}, {Chester}, {Corbet}, {Covino}, {Cummings}, {D'Avanzo}, {D'Elia}, {Esposito}, {Evans}, {Fugazza}, {Gelbord}, {Hiroi}, {Holland}, {Huang}, {Im}, {Israel}, {Jeon}, {Jeon}, {Jun}, {Kawai}, {Kim}, {Krimm}, {Marshall}, {P. M{\'e}sz{\'a}ros}, {Negoro}, {Omodei}, {Park}, {Perkins}, {Sugizaki}, {Sung}, {Tagliaferri}, {Troja}, {Ueda}, {Urata}, {Usui}, {Antonelli}, {Barthelmy}, {Cusumano}, {Giommi}, {Melandri}, {Perri}, {Racusin}, {Sbarufatti}, {Siegel}, \& {Gehrels}}]{Burrows2011}
{Burrows}, D.~N., {Kennea}, J.~A., {Ghisellini}, G., {et~al.} 2011, \nat, 476, 421

\bibitem[{{Calder} {et~al.}(2002){Calder}, {Fryxell}, {Plewa}, {Rosner}, {Dursi}, {Weirs}, {Dupont}, {Robey}, {Kane}, {Remington}, {Drake}, {Dimonte}, {Zingale}, {Timmes}, {Olson}, {Ricker}, {MacNeice}, \& {Tufo}}]{2002ApJS..143..201C}
{Calder}, A.~C., {Fryxell}, B., {Plewa}, T., {et~al.} 2002, \apjs, 143, 201

\bibitem[{{Caleb} {et~al.}(2022){Caleb}, {Heywood}, {Rajwade}, {Malenta}, {Stappers}, {Barr}, {Chen}, {Morello}, {Sanidas}, {van den Eijnden}, {Kramer}, {Buckley}, {Brink}, {Motta}, {Woudt}, {Weltevrede}, {Jankowski}, {Surnis}, {Buchner}, {Bezuidenhout}, {Driessen}, \& {Fender}}]{2022NatAs...6..828C}
{Caleb}, M., {Heywood}, I., {Rajwade}, K., {et~al.} 2022, Nature Astronomy, 6, 828

\bibitem[{{Caleb} {et~al.}(2024){Caleb}, {Lenc}, {Kaplan}, {Murphy}, {Men}, {Shannon}, {Ferrario}, {Rajwade}, {Clarke}, {Giacintucci}, {Hurley-Walker}, {Hyman}, {Lower}, {McSweeney}, {Ravi}, {Barr}, {Buchner}, {Flynn}, {Hessels}, {Kramer}, {Pritchard}, \& {Stappers}}]{2024NatAs...8.1159C}
{Caleb}, M., {Lenc}, E., {Kaplan}, D.~L., {et~al.} 2024, Nature Astronomy, 8, 1159

\bibitem[{Camero {et~al.}(2014)Camero, Papitto, Rea, Viganò, Pons, Tiengo, Mereghetti, Turolla, Esposito, Zane, Israel, \& Götz}]{Camero_10.1093/mnras/stt2432}
Camero, A., Papitto, A., Rea, N., {et~al.} 2014, Monthly Notices of the Royal Astronomical Society, 438, 3291.
\newblock \url{https://doi.org/10.1093/mnras/stt2432}

\bibitem[{Cameron(1957{\natexlab{a}})}]{cameron1957origin}
Cameron, A. 1957{\natexlab{a}}, The Astronomical Journal, 62, 9

\bibitem[{Cameron(1957{\natexlab{b}})}]{cameron1957nuclear}
Cameron, A. G.~W. 1957{\natexlab{b}}, Publications of the Astronomical Society of the Pacific, 69, 201

\bibitem[{{Camilo} {et~al.}(2008){Camilo}, {Reynolds}, {Johnston}, {Halpern}, \& {Ransom}}]{Camilo_2008ApJ...679..681C}
{Camilo}, F., {Reynolds}, J., {Johnston}, S., {Halpern}, J.~P., \& {Ransom}, S.~M. 2008, \apj, 679, 681

\bibitem[{Camilo {et~al.}(2007{\natexlab{a}})Camilo, Reynolds, Johnston, Halpern, Ransom, \& van Straten}]{Camilo_2007b}
Camilo, F., Reynolds, J., Johnston, S., {et~al.} 2007{\natexlab{a}}, The Astrophysical Journal, 659, L37.
\newblock \url{https://dx.doi.org/10.1086/516630}

\bibitem[{Camilo {et~al.}(2007{\natexlab{b}})Camilo, Cognard, Ransom, Halpern, Reynolds, Zimmerman, Gotthelf, Helfand, Demorest, Theureau, \& Backer}]{Camilo_2007}
Camilo, F., Cognard, I., Ransom, S.~M., {et~al.} 2007{\natexlab{b}}, The Astrophysical Journal, 663, 497.
\newblock \url{https://dx.doi.org/10.1086/518226}

\bibitem[{{Carenza} {et~al.}(2024){Carenza}, {Co'}, {Giannotti}, {Lella}, {Lucente}, {Mirizzi}, \& {Rauscher}}]{2024PhRvC.109a5501C}
{Carenza}, P., {Co'}, G., {Giannotti}, M., {et~al.} 2024, \prc, 109, 015501

\bibitem[{{Carenza} {et~al.}(2021){Carenza}, {Fore}, {Giannotti}, {Mirizzi}, \& {Reddy}}]{2021PhRvL.126g1102C}
{Carenza}, P., {Fore}, B., {Giannotti}, M., {Mirizzi}, A., \& {Reddy}, S. 2021, \prl, 126, 071102

\bibitem[{{Castro-Tirado} {et~al.}(2021){Castro-Tirado}, {{\O}stgaard}, {G{\"o}{\c{C}}{\textsection}{\"u}{\c{s}}}, {S{\'a}nchez-Gil}, {Pascual-Granado}, {Reglero}, {Mezentsev}, {Gabler}, {Marisaldi}, {Neubert}, {Budtz-J{\o}rgensen}, {Lindanger}, {Sarria}, {Kuvvetli}, {Cerd{\'a}-Dur{\'a}n}, {Navarro-Gonz{\'a}lez}, {Font}, {Zhang}, {Lund}, {Oxborrow}, {Brandt}, {Caballero-Garc{\'\i}a}, {Carrasco-Garc{\'\i}a}, {Castell{\'o}n}, {Castro Tirado}, {Christiansen}, {Eyles}, {Fern{\'a}ndez-Garc{\'\i}a}, {Genov}, {Guziy}, {Hu}, {Nicuesa Guelbenzu}, {Pandey}, {Peng}, {P{\'e}rez del Pulgar}, {Reina Terol}, {Rodr{\'\i}guez}, {S{\'a}nchez-Ram{\'\i}rez}, {Sun}, {Ullaland}, \& {Yang}}]{Castro-Tirado_2021Natur.600..621C}
{Castro-Tirado}, A.~J., {{\O}stgaard}, N., {G{\"o}{\c{C}}{\textsection}{\"u}{\c{s}}}, E., {et~al.} 2021, \nat, 600, 621

\bibitem[{Cehula {et~al.}(2024)Cehula, Thompson, \& Metzger}]{Cehula_10.1093/mnras/stae358}
Cehula, J., Thompson, T.~A., \& Metzger, B.~D. 2024, Monthly Notices of the Royal Astronomical Society, 528, 5323.
\newblock \url{https://doi.org/10.1093/mnras/stae358}

\bibitem[{{Cendes} {et~al.}(2024){Cendes}, {Berger}, {Alexander}, {Chornock}, {Margutti}, {Metzger}, {Wieringa}, {Bietenholz}, {Hajela}, {Laskar}, {Stroh}, \& {Terreran}}]{Cendes2024ApJ...971..185C}
{Cendes}, Y., {Berger}, E., {Alexander}, K.~D., {et~al.} 2024, \apj, 971, 185

\bibitem[{{Cenko} {et~al.}(2012){Cenko}, {Krimm}, {Horesh}, {Rau}, {Frail}, {Kennea}, {Levan}, {Holland}, {Butler}, {Quimby}, {Bloom}, {Filippenko}, {Gal-Yam}, {Greiner}, {Kulkarni}, {Ofek}, {Olivares E.}, {Schady}, {Silverman}, {Tanvir}, \& {Xu}}]{Cenko2012}
{Cenko}, S.~B., {Krimm}, H.~A., {Horesh}, A., {et~al.} 2012, \apj, 753, 77

\bibitem[{{Chadwick} {et~al.}(2023){Chadwick}, {Paris}, \& {Haines}}]{2023arXiv230500647C}
{Chadwick}, M.~B., {Paris}, M.~W., \& {Haines}, B.~M. 2023, arXiv e-prints, arXiv:2305.00647

\bibitem[{{Chamel} \& {Haensel}(2008)}]{2008LRR....11...10C}
{Chamel}, N., \& {Haensel}, P. 2008, Living Reviews in Relativity, 11, 10

\bibitem[{{Chandrasekhar}(1949)}]{1949ApJ...110..329C}
{Chandrasekhar}, S. 1949, \apj, 110, 329

\bibitem[{{Chatterjee} {et~al.}(2017){Chatterjee}, {Law}, {Wharton}, {Burke-Spolaor}, {Hessels}, {Bower}, {Cordes}, {Tendulkar}, {Bassa}, {Demorest}, {Butler}, {Seymour}, {Scholz}, {Abruzzo}, {Bogdanov}, {Kaspi}, {Keimpema}, {Lazio}, {Marcote}, {McLaughlin}, {Paragi}, {Ransom}, {Rupen}, {Spitler}, \& {van Langevelde}}]{Chatterjee2017}
{Chatterjee}, S., {Law}, C.~J., {Wharton}, R.~S., {et~al.} 2017, \nat, 541, 58

\bibitem[{{Chatzopoulos} \& {Wheeler}(2012{\natexlab{a}})}]{2012ApJ...748...42C}
{Chatzopoulos}, E., \& {Wheeler}, J.~C. 2012{\natexlab{a}}, \apj, 748, 42

\bibitem[{{Chatzopoulos} \& {Wheeler}(2012{\natexlab{b}})}]{2012ApJ...760..154C}
---. 2012{\natexlab{b}}, \apj, 760, 154

\bibitem[{{Chen} {et~al.}(2007){Chen}, {Kantowski}, {Baron}, {Knop}, \& {Hauschildt}}]{Chen:2007}
{Chen}, B., {Kantowski}, R., {Baron}, E., {Knop}, S., \& {Hauschildt}, P.~H. 2007, \mnras, 380, 104

\bibitem[{{Chen} {et~al.}(2024){Chen}, {Liodakis}, {Middei}, {Kim}, {Di Gesu}, {Di Marco}, {Ehlert}, {Errando}, {Negro}, {Jorstad}, {Marscher}, {Wu}, {Agudo}, {Poutanen}, {Mizuno}, {Kouch}, {Lindfors}, {Borman}, {Grishina}, {Kopatskaya}, {Larionova}, {Morozova}, {Savchenko}, {Troitsky}, {Troitskaya}, {Vasilyev}, {Zhovtan}, {Aceituno}, {Bonnoli}, {Casanova}, {Escudero}, {Ag{\'\i}s-Gonz{\'a}lez}, {Husillos}, {Otero Santos}, {Sota}, {Piirola}, {Myserlis}, {Angelakis}, {Kraus}, {Gurwell}, {Keating}, {Rao}, {Kang}, {Lee}, {Kim}, {Cheong}, {Jeong}, {Song}, {Berdyugin}, {Kagitani}, {Kravtsov}, {Nitindala}, {Sakanoi}, {Imazawa}, {Sasada}, {Fukazawa}, {Kawabata}, {Uemura}, {Nakaoka}, {Akitaya}, {Casadio}, {Sievers}, {Antonelli}, {Bachetti}, {Baldini}, {Baumgartner}, {Bellazzini}, {Bianchi}, {Bongiorno}, {Bonino}, {Brez}, {Bucciantini}, {Capitanio}, {Castellano}, {Cavazzuti}, {Ciprini}, {Costa}, {De Rosa}, {Del Monte}, {Di Lalla}, {Donnarumma}, {Doroshenko}, {Dov{\v{c}}iak}, {Enoto}, {Evangelista}, {Fabiani},
  {Ferrazzoli}, {Garcia}, {Gunji}, {Hayashida}, {Heyl}, {Iwakiri}, {Kaaret}, {Karas}, {Kislat}, {Kitaguchi}, {Kolodziejczak}, {Krawczynski}, {La Monaca}, {Latronico}, {Maldera}, {Manfreda}, {Marin}, {Marinucci}, {Marshall}, {Massaro}, {Matt}, {Mitsuishi}, {Muleri}, {Ng}, {O'Dell}, {Omodei}, {Oppedisano}, {Papitto}, {Pavlov}, {Peirson}, {Perri}, {Pesce-Rollins}, {Petrucci}, {Pilia}, {Possenti}, {Puccetti}, {Ramsey}, {Rankin}, {Ratheesh}, {Roberts}, {Romani}, {Sgr{\'o}}, {Slane}, {Soffitta}, {Spandre}, {Swartz}, {Tamagawa}, {Tavecchio}, {Taverna}, {Tawara}, {Tennant}, {Thomas}, {Tombesi}, {Trois}, {Tsygankov}, {Turolla}, {Vink}, {Weisskopf}, {Xie}, \& {Zane}}]{Chen2024}
{Chen}, C.-T.~J., {Liodakis}, I., {Middei}, R., {et~al.} 2024, \apj, 974, 50

\bibitem[{{Chen} {et~al.}(2023){Chen}, {Drout}, {Piro}, {Kilpatrick}, {Foley}, {Rojas-Bravo}, \& {Magee}}]{Chen2023}
{Chen}, Y., {Drout}, M.~R., {Piro}, A.~L., {et~al.} 2023, \apj, 955, 43

\bibitem[{Chevalier \& Blondin(1995)}]{cb95}
Chevalier, R., \& Blondin, J.~M. 1995, Astrophysical Journal, Part 1 (ISSN 0004-637X), vol. 444, no. 1, p. 312-317, 444, 312

\bibitem[{{Chevalier}(1982)}]{chevalier82b}
{Chevalier}, R.~A. 1982, \apj, 259, 302

\bibitem[{{Chevalier}(1989)}]{1989ApJ...346..847C}
---. 1989, \apj, 346, 847

\bibitem[{{Chevalier} \& {Fransson}(2017)}]{cf17}
{Chevalier}, R.~A., \& {Fransson}, C. 2017, in Handbook of Supernovae, ed. A.~W. {Alsabti} \& P.~{Murdin}, 875

\bibitem[{{CHIME/FRB Collaboration} {et~al.}(2020){CHIME/FRB Collaboration}, {Andersen}, {Bandura}, {Bhardwaj}, {Bij}, {Boyce}, {Boyle}, {Brar}, {Cassanelli}, {Chawla}, {Chen}, {Cliche}, {Cook}, {Cubranic}, {Curtin}, {Denman}, {Dobbs}, {Dong}, {Fandino}, {Fonseca}, {Gaensler}, {Giri}, {Good}, {Halpern}, {Hill}, {Hinshaw}, {H{\"o}fer}, {Josephy}, {Kania}, {Kaspi}, {Landecker}, {Leung}, {Li}, {Lin}, {Masui}, {McKinven}, {Mena-Parra}, {Merryfield}, {Meyers}, {Michilli}, {Milutinovic}, {Mirhosseini}, {M{\"u}nchmeyer}, {Naidu}, {Newburgh}, {Ng}, {Patel}, {Pen}, {Pinsonneault-Marotte}, {Pleunis}, {Quine}, {Rafiei-Ravandi}, {Rahman}, {Ransom}, {Renard}, {Sanghavi}, {Scholz}, {Shaw}, {Shin}, {Siegel}, {Singh}, {Smegal}, {Smith}, {Stairs}, {Tan}, {Tendulkar}, {Tretyakov}, {Vanderlinde}, {Wang}, {Wulf}, \& {Zwaniga}}]{CHIME2020}
{CHIME/FRB Collaboration}, {Andersen}, B.~C., {Bandura}, K.~M., {et~al.} 2020, \nat, 587, 54

\bibitem[{{CHIME/FRB Collaboration} {et~al.}(2021){CHIME/FRB Collaboration}, {Amiri}, {Andersen}, {Bandura}, {Berger}, {Bhardwaj}, {Boyce}, {Boyle}, {Brar}, {Breitman}, {Cassanelli}, {Chawla}, {Chen}, {Cliche}, {Cook}, {Cubranic}, {Curtin}, {Deng}, {Dobbs}, {Dong}, {Eadie}, {Fandino}, {Fonseca}, {Gaensler}, {Giri}, {Good}, {Halpern}, {Hill}, {Hinshaw}, {Josephy}, {Kaczmarek}, {Kader}, {Kania}, {Kaspi}, {Landecker}, {Lang}, {Leung}, {Li}, {Lin}, {Masui}, {McKinven}, {Mena-Parra}, {Merryfield}, {Meyers}, {Michilli}, {Milutinovic}, {Mirhosseini}, {M{\"u}nchmeyer}, {Naidu}, {Newburgh}, {Ng}, {Patel}, {Pen}, {Petroff}, {Pinsonneault-Marotte}, {Pleunis}, {Rafiei-Ravandi}, {Rahman}, {Ransom}, {Renard}, {Sanghavi}, {Scholz}, {Shaw}, {Shin}, {Siegel}, {Sikora}, {Singh}, {Smith}, {Stairs}, {Tan}, {Tendulkar}, {Vanderlinde}, {Wang}, {Wulf}, \& {Zwaniga}}]{CHIME2021}
{CHIME/FRB Collaboration}, {Amiri}, M., {Andersen}, B.~C., {et~al.} 2021, \apjs, 257, 59

\bibitem[{{Coffing} {et~al.}(2024){Coffing}, {Wood}, {Byvank}, {Robey}, {Doss}, {Fryer}, {Johns}, {Kozlowski}, {Fontes}, {Meyerhofer}, \& {Urbatsch}}]{2024PhPl...31k3301C}
{Coffing}, S.~X., {Wood}, S.~R., {Byvank}, T., {et~al.} 2024, Physics of Plasmas, 31, 113301

\bibitem[{{Colgate} {et~al.}(1961){Colgate}, {Grasberger}, \& {White}}]{1961AJ.....66S.280C}
{Colgate}, S.~A., {Grasberger}, W.~H., \& {White}, R.~H. 1961, \aj, 66, 280

\bibitem[{Combi \& Siegel(2023)}]{Combi:2023yav}
Combi, L., \& Siegel, D.~M. 2023, Phys. Rev. Lett., 131, 231402

\bibitem[{{Comisso} \& {Asenjo}(2021)}]{CA21}
{Comisso}, L., \& {Asenjo}, F.~A. 2021, \prd, 103, 023014

\bibitem[{{Comisso} {et~al.}(2024){Comisso}, {Farrar}, \& {Muzio}}]{CFM2024}
{Comisso}, L., {Farrar}, G.~R., \& {Muzio}, M.~S. 2024, \apjl, 977, L18

\bibitem[{{Comisso} \& {Sironi}(2019)}]{Comisso2019}
{Comisso}, L., \& {Sironi}, L. 2019, \apj, 886, 122

\bibitem[{{Cook} {et~al.}(2024){Cook}, {Scholz}, {Pearlman}, {Abbott}, {Cruces}, {Gaensler}, {Fengqiu}, {Dong}, {Michilli}, {Eadie}, {Kaspi}, {Stairs}, {Tan}, {Bhardwaj}, {Cassanelli}, {Curtin}, {Ibik}, {Lazda}, {Masui}, {Pandhi}, {Rafiei-Ravandi}, {Sammons}, {Shin}, {Smith}, \& {Stenning}}]{Cook2024}
{Cook}, A.~M., {Scholz}, P., {Pearlman}, A.~B., {et~al.} 2024, arXiv e-prints, arXiv:2408.11895

\bibitem[{{Cooper} \& {Wadiasingh}(2024)}]{2024MNRAS.533.2133C}
{Cooper}, A.~J., \& {Wadiasingh}, Z. 2024, \mnras, 533, 2133

\bibitem[{{Coppejans} {et~al.}(2020){Coppejans}, {Margutti}, {Terreran}, {Nayana}, {Coughlin}, {Laskar}, {Alexander}, {Bietenholz}, {Caprioli}, {Chandra}, {Drout}, {Frederiks}, {Frohmaier}, {Hurley}, {Kochanek}, {MacLeod}, {Meisner}, {Nugent}, {Ridnaia}, {Sand}, {Svinkin}, {Ward}, {Yang}, {Baldeschi}, {Chilingarian}, {Dong}, {Esquivia}, {Fong}, {Guidorzi}, {Lundqvist}, {Milisavljevic}, {Paterson}, {Reichart}, {Shappee}, {Stroh}, {Valenti}, {Zauderer}, \& {Zhang}}]{Coppejans2020}
{Coppejans}, D.~L., {Margutti}, R., {Terreran}, G., {et~al.} 2020, \apjl, 895, L23

\bibitem[{{Coppi}(2000)}]{Coppi_1999}
{Coppi}, P.~S. 2000, in AAS/High Energy Astrophysics Division, Vol.~5, AAS/High Energy Astrophysics Division \#5, 23.11

\bibitem[{{C{\^o}t{\'e}} {et~al.}(2018){C{\^o}t{\'e}}, {Fryer}, {Belczynski}, {Korobkin}, {Chru{\'s}li{\'n}ska}, {Vassh}, {Mumpower}, {Lippuner}, {Sprouse}, {Surman}, \& {Wollaeger}}]{2018ApJ...855...99C}
{C{\^o}t{\'e}}, B., {Fryer}, C.~L., {Belczynski}, K., {et~al.} 2018, \apj, 855, 99

\bibitem[{Coti~Zelati {et~al.}(2017)Coti~Zelati, Rea, Pons, Campana, \& Esposito}]{Coti_Zelati_10.1093/mnras/stx2679}
Coti~Zelati, F., Rea, N., Pons, J.~A., Campana, S., \& Esposito, P. 2017, Monthly Notices of the Royal Astronomical Society, 474, 961.
\newblock \url{https://doi.org/10.1093/mnras/stx2679}

\bibitem[{Cunningham {et~al.}(2019)Cunningham, Cenko, Burns, Goldstein, Lien, Kocevski, Briggs, Connaughton, Miller, Racusin, {et~al.}}]{cunningham2019search}
Cunningham, V., Cenko, S.~B., Burns, E., {et~al.} 2019, The Astrophysical Journal, 879, 40

\bibitem[{{Curtin} {et~al.}(2024){Curtin}, {Sirota}, {Kaspi}, {Tendulkar}, {Bhardwaj}, {Cook}, {Fong}, {Gaensler}, {Main}, {Masui}, {Michilli}, {Pandhi}, {Pearlman}, {Scholz}, \& {Shin}}]{Curtin2024}
{Curtin}, A.~P., {Sirota}, S., {Kaspi}, V.~M., {et~al.} 2024, \apj, 972, 125

\bibitem[{Cyburt {et~al.}(2010)Cyburt, Amthor, Ferguson, Meisel, Smith, Warren, Heger, Hoffman, Rauscher, Sakharuk, {et~al.}}]{cyburt2010jina}
Cyburt, R.~H., Amthor, A.~M., Ferguson, R., {et~al.} 2010, The Astrophysical Journal Supplement Series, 189, 240

\bibitem[{{Dai} \& {Fang}(2017)}]{Dai2017a}
{Dai}, L., \& {Fang}, K. 2017, \mnras, 469, 1354

\bibitem[{{Dai} {et~al.}(2017){Dai}, {McKinney}, \& {Miller}}]{Dai2017b}
{Dai}, L., {McKinney}, J.~C., \& {Miller}, M.~C. 2017, \mnras, 470, L92

\bibitem[{{Dai} {et~al.}(2021){Dai}, {Lu}, {Wang}, {Wang}, {Xu}, {Yang}, {Zhang}, {Hobbs}, {Li}, {Luo}, {Filipovic}, \& {Jiang}}]{Dai2021}
{Dai}, S., {Lu}, J., {Wang}, C., {et~al.} 2021, \apj, 920, 46

\bibitem[{{den Hartog} {et~al.}(2008){den Hartog}, {Kuiper}, {Hermsen}, {Kaspi}, {Dib}, {Kn{\"o}dlseder}, \& {Gavriil}}]{denHartog-2008-AandA}
{den Hartog}, P.~R., {Kuiper}, L., {Hermsen}, W., {et~al.} 2008, \aap, 489, 245

\bibitem[{{Denissenkov} {et~al.}(2017){Denissenkov}, {Herwig}, {Battino}, {Ritter}, {Pignatari}, {Jones}, \& {Paxton}}]{Denissenkov2017}
{Denissenkov}, P.~A., {Herwig}, F., {Battino}, U., {et~al.} 2017, \apjl, 834, L10

\bibitem[{{Denissenkov} {et~al.}(2014){Denissenkov}, {Truran}, {Pignatari}, {Trappitsch}, {Ritter}, {Herwig}, {Battino}, {Setoodehnia}, \& {Paxton}}]{denissenkov2014}
{Denissenkov}, P.~A., {Truran}, J.~W., {Pignatari}, M., {et~al.} 2014, \mnras, 442, 2058

\bibitem[{{Denney} {et~al.}(2009){Denney}, {Peterson}, {Dietrich}, {Vestergaard}, \& {Bentz}}]{Denney_2009}
{Denney}, K.~D., {Peterson}, B.~M., {Dietrich}, M., {Vestergaard}, M., \& {Bentz}, M.~C. 2009, \apj, 692, 246

\bibitem[{{Dessart} \& {Hillier}(2011)}]{2011MNRAS.410.1739D}
{Dessart}, L., \& {Hillier}, D.~J. 2011, \mnras, 410, 1739

\bibitem[{{Di Gesu} {et~al.}(2023){Di Gesu}, {Marshall}, {Ehlert}, {Kim}, {Donnarumma}, {Tavecchio}, {Liodakis}, {Kiehlmann}, {Agudo}, {Jorstad}, {Muleri}, {Marscher}, {Puccetti}, {Middei}, {Perri}, {Pacciani}, {Negro}, {Romani}, {Di Marco}, {Blinov}, {Bourbah}, {Kontopodis}, {Mandarakas}, {Romanopoulos}, {Skalidis}, {Vervelaki}, {Casadio}, {Escudero}, {Myserlis}, {Gurwell}, {Rao}, {Keating}, {Kouch}, {Lindfors}, {Aceituno}, {Bernardos}, {Bonnoli}, {Casanova}, {Garc{\'\i}a-Comas}, {Ag{\'\i}s-Gonz{\'a}lez}, {Husillos}, {Marchini}, {Sota}, {Imazawa}, {Sasada}, {Fukazawa}, {Kawabata}, {Uemura}, {Mizuno}, {Nakaoka}, {Akitaya}, {Savchenko}, {Vasilyev}, {G{\'o}mez}, {Antonelli}, {Barnouin}, {Bonino}, {Cavazzuti}, {Costamante}, {Chen}, {Cibrario}, {De Rosa}, {Di Pierro}, {Errando}, {Kaaret}, {Karas}, {Krawczynski}, {Lisalda}, {Madejski}, {Malacaria}, {Marin}, {Marinucci}, {Massaro}, {Matt}, {Mitsuishi}, {O'Dell}, {Paggi}, {Peirson}, {Petrucci}, {Ramsey}, {Tennant}, {Wu}, {Bachetti}, {Baldini}, {Baumgartner},
  {Bellazzini}, {Bianchi}, {Bongiorno}, {Brez}, {Bucciantini}, {Capitanio}, {Castellano}, {Ciprini}, {Costa}, {Del Monte}, {Di Lalla}, {Doroshenko}, {Dov{\v{c}}iak}, {Enoto}, {Evangelista}, {Fabiani}, {Ferrazzoli}, {Garcia}, {Gunji}, {Hayashida}, {Heyl}, {Iwakiri}, {Kislat}, {Kitaguchi}, {Kolodziejczak}, {La Monaca}, {Latronico}, {Maldera}, {Manfreda}, {Ng}, {Omodei}, {Oppedisano}, {Papitto}, {Pavlov}, {Pesce-Rollins}, {Pilia}, {Possenti}, {Poutanen}, {Rankin}, {Ratheesh}, {Roberts}, {Sgr{\`o}}, {Slane}, {Soffitta}, {Spandre}, {Swartz}, {Tamagawa}, {Taverna}, {Tawara}, {Thomas}, {Tombesi}, {Trois}, {Tsygankov}, {Turolla}, {Vink}, {Weisskopf}, {Xie}, \& {Zane}}]{DiGesu2023}
{Di Gesu}, L., {Marshall}, H.~L., {Ehlert}, S.~R., {et~al.} 2023, Nature Astronomy, 7, 1245

\bibitem[{{Di Valentino} {et~al.}(2021){Di Valentino}, {Mena}, {Pan}, {Visinelli}, {Yang}, {Melchiorri}, {Mota}, {Riess}, \& {Silk}}]{2021CQGra..38o3001D}
{Di Valentino}, E., {Mena}, O., {Pan}, S., {et~al.} 2021, Classical and Quantum Gravity, 38, 153001

\bibitem[{{Done} {et~al.}(2012){Done}, {Davis}, {Jin}, {Blaes}, \& {Ward}}]{Done_2012}
{Done}, C., {Davis}, S.~W., {Jin}, C., {Blaes}, O., \& {Ward}, M. 2012, \mnras, 420, 1848

\bibitem[{Donello {et~al.}(2023)Donello, Palkar, Naderi, Del Rey~Fern{\'a}ndez, \& Babaee}]{DPNFB23}
Donello, M., Palkar, G., Naderi, M., Del Rey~Fern{\'a}ndez, D., \& Babaee, H. 2023, Proc. R. Soc. A, 479, 20230320

\bibitem[{{Drout} {et~al.}(2014){Drout}, {Chornock}, {Soderberg}, {Sand ers}, {McKinnon}, {Rest}, {Foley}, {Milisavljevic}, {Margutti}, {Berger}, {Calkins}, {Fong}, {Gezari}, {Huber}, {Kankare}, {Kirshner}, {Leibler}, {Lunnan}, {Mattila}, {Marion}, {Narayan}, {Riess}, {Roth}, {Scolnic}, {Smartt}, {Tonry}, {Burgett}, {Chambers}, {Hodapp}, {Jedicke}, {Kaiser}, {Magnier}, {Metcalfe}, {Morgan}, {Price}, \& {Waters}}]{Drout2014}
{Drout}, M.~R., {Chornock}, R., {Soderberg}, A.~M., {et~al.} 2014, \apj, 794, 23

\bibitem[{Duncan(1998)}]{Duncan_1998}
Duncan, R.~C. 1998, The Astrophysical Journal, 498, L45.
\newblock \url{https://dx.doi.org/10.1086/311303}

\bibitem[{{Duncan} \& {Thompson}(1992)}]{Duncan_1992ApJ...392L...9D}
{Duncan}, R.~C., \& {Thompson}, C. 1992, \apjl, 392, L9

\bibitem[{Duraisamy {et~al.}(2019)Duraisamy, Iaccarino, \& Xiao}]{duraisamy2019turbulence}
Duraisamy, K., Iaccarino, G., \& Xiao, H. 2019, Annual Review of Fluid Mechanics, 51, 357.
\newblock \url{https://doi.org/10.1146/annurev-fluid-010518-040547}

\bibitem[{{Durant} \& {van Kerkwijk}(2005)}]{Durant_2005ApJ...627..376D}
{Durant}, M., \& {van Kerkwijk}, M.~H. 2005, \apj, 627, 376

\bibitem[{{Edelson} {et~al.}(2019){Edelson}, {Gelbord}, {Cackett}, {Peterson}, {Horne}, {Barth}, {Starkey}, {Bentz}, {Brandt}, {Goad}, {Joner}, {Korista}, {Netzer}, {Page}, {Uttley}, {Vaughan}, {Breeveld}, {Cenko}, {Done}, {Evans}, {Fausnaugh}, {Ferland}, {Gonzalez-Buitrago}, {Gropp}, {Grupe}, {Kaastra}, {Kennea}, {Kriss}, {Mathur}, {Mehdipour}, {Mudd}, {Nousek}, {Schmidt}, {Vestergaard}, \& {Villforth}}]{Edelson_2019}
{Edelson}, R., {Gelbord}, J., {Cackett}, E., {et~al.} 2019, \apj, 870, 123

\bibitem[{{Eftekhari} \& {Berger}(2017)}]{Eftekhari2017}
{Eftekhari}, T., \& {Berger}, E. 2017, \apj, 849, 162

\bibitem[{{Eftekhari} {et~al.}(2024{\natexlab{a}}){Eftekhari}, {Tchekhovskoy}, {Alexander}, {Berger}, {Chornock}, {Laskar}, {Margutti}, {Yao}, {Cendes}, {Gomez}, {Hajela}, \& {Pasham}}]{Eftekhari2024ApJ...974..149E}
{Eftekhari}, T., {Tchekhovskoy}, A., {Alexander}, K.~D., {et~al.} 2024{\natexlab{a}}, \apj, 974, 149

\bibitem[{{Eftekhari} {et~al.}(2024{\natexlab{b}}){Eftekhari}, {Dong}, {Fong}, {Shah}, {Simha}, {Andersen}, {Andrew}, {Bhardwaj}, {Cassanelli}, {Chatterjee}, {Coulter}, {Fonseca}, {Gaensler}, {Gordon}, {Hessels}, {Ibik}, {Joseph}, {Kahinga}, {Kaspi}, {Kharel}, {Kilpatrick}, {Lanman}, {Lazda}, {Leung}, {Liu}, {Mas-Ribas}, {Masui}, {Mckinven}, {Mena-Parra}, {Miller}, {Nimmo}, {Pandhi}, {Pearlman}, {Pleunis}, {Prochaska}, {Rafiei-Ravandi}, {Sammons}, {Scholz}, {Shin}, {Smith}, {Stairs}, \& {Swarali Shivraj}}]{Eftekhari2024}
{Eftekhari}, T., {Dong}, Y., {Fong}, W., {et~al.} 2024{\natexlab{b}}, arXiv e-prints, arXiv:2410.23336

\bibitem[{{Ehlert} {et~al.}(2022){Ehlert}, {Ferrazzoli}, {Marinucci}, {Marshall}, {Middei}, {Pacciani}, {Perri}, {Petrucci}, {Puccetti}, {Barnouin}, {Bianchi}, {Liodakis}, {Madejski}, {Marin}, {Marscher}, {Matt}, {Poutanen}, {Wu}, {Agudo}, {Antonelli}, {Bachetti}, {Baldini}, {Baumgartner}, {Bellazzini}, {Bongiorno}, {Bonino}, {Brez}, {Bucciantini}, {Capitanio}, {Castellano}, {Cavazzuti}, {Ciprini}, {Costa}, {De Rosa}, {Del Monte}, {Di Gesu}, {Di Lalla}, {Di Marco}, {Donnarumma}, {Doroshenko}, {Dov{\v{c}}iak}, {Enoto}, {Evangelista}, {Fabiani}, {Garcia}, {Gunji}, {Hayashida}, {Heyl}, {Iwakiri}, {Jorstad}, {Karas}, {Kitaguchi}, {Kolodziejczak}, {Krawczynski}, {La Monaca}, {Latronico}, {Maldera}, {Manfreda}, {Massaro}, {Mitsuishi}, {Mizuno}, {Muleri}, {Negro}, {Ng}, {O'Dell}, {Omodei}, {Oppedisano}, {Papitto}, {Pavlov}, {Peirson}, {Pesce-Rollins}, {Pilia}, {Possenti}, {Ramsey}, {Rankin}, {Ratheesh}, {Romani}, {Sgr{\`o}}, {Slane}, {Soffitta}, {Spandre}, {Tamagawa}, {Tavecchio}, {Taverna}, {Tawara}, {Tennant},
  {Thomas}, {Tombesi}, {Trois}, {Tsygankov}, {Turolla}, {Vink}, {Weisskopf}, {Xie}, {Zane}, {IXPE Collaboration}, {Rodi}, {Jourdain}, \& {Roques}}]{Ehlert2022}
{Ehlert}, S.~R., {Ferrazzoli}, R., {Marinucci}, A., {et~al.} 2022, \apj, 935, 116

\bibitem[{{Eichler} {et~al.}(1989){Eichler}, {Livio}, {Piran}, \& {Schramm}}]{eichler1989}
{Eichler}, D., {Livio}, M., {Piran}, T., \& {Schramm}, D.~N. 1989, \nat, 340, 126

\bibitem[{{Endsley} {et~al.}(2023){Endsley}, {Stark}, {Lyu}, {Wang}, {Yang}, {Fan}, {Smit}, {Bouwens}, {Hainline}, \& {Schouws}}]{Endsley2023}
{Endsley}, R., {Stark}, D.~P., {Lyu}, J., {et~al.} 2023, \mnras, 520, 4609

\bibitem[{Engstrom {et~al.}(2016)Engstrom, Yoder, \& Crespi}]{Engstrom_2016}
Engstrom, T.~A., Yoder, N.~C., \& Crespi, V.~H. 2016, The Astrophysical Journal, 818, 183.
\newblock \url{https://dx.doi.org/10.3847/0004-637X/818/2/183}

\bibitem[{{Enoto} {et~al.}(2010){Enoto}, {Nakazawa}, {Makishima}, {Rea}, {Hurley}, \& {Shibata}}]{Enoto-2010-ApJ}
{Enoto}, T., {Nakazawa}, K., {Makishima}, K., {et~al.} 2010, \apjl, 722, L162

\bibitem[{{EPTA Collaboration} {et~al.}(2023){EPTA Collaboration}, {InPTA Collaboration}, {Antoniadis}, {Arumugam}, {Arumugam}, {Babak}, {Bagchi}, {Bak Nielsen}, {Bassa}, {Bathula}, {Berthereau}, {Bonetti}, {Bortolas}, {Brook}, {Burgay}, {Caballero}, {Chalumeau}, {Champion}, {Chanlaridis}, {Chen}, {Cognard}, {Dandapat}, {Deb}, {Desai}, {Desvignes}, {Dhanda-Batra}, {Dwivedi}, {Falxa}, {Ferdman}, {Franchini}, {Gair}, {Goncharov}, {Gopakumar}, {Graikou}, {Grie{\ss}meier}, {Guillemot}, {Guo}, {Gupta}, {Hisano}, {Hu}, {Iraci}, {Izquierdo-Villalba}, {Jang}, {Jawor}, {Janssen}, {Jessner}, {Joshi}, {Kareem}, {Karuppusamy}, {Keane}, {Keith}, {Kharbanda}, {Kikunaga}, {Kolhe}, {Kramer}, {Krishnakumar}, {Lackeos}, {Lee}, {Liu}, {Liu}, {Lyne}, {McKee}, {Maan}, {Main}, {Mickaliger}, {Ni{\c{t}}u}, {Nobleson}, {Paladi}, {Parthasarathy}, {Perera}, {Perrodin}, {Petiteau}, {Porayko}, {Possenti}, {Prabu}, {Quelquejay Leclere}, {Rana}, {Samajdar}, {Sanidas}, {Sesana}, {Shaifullah}, {Singha}, {Speri}, {Spiewak}, {Srivastava},
  {Stappers}, {Surnis}, {Susarla}, {Susobhanan}, {Takahashi}, {Tarafdar}, {Theureau}, {Tiburzi}, {van der Wateren}, {Vecchio}, {Venkatraman Krishnan}, {Verbiest}, {Wang}, {Wang}, \& {Wu}}]{EPTA_2023}
{EPTA Collaboration}, {InPTA Collaboration}, {Antoniadis}, J., {et~al.} 2023, \aap, 678, A50

\bibitem[{{Erber}(1966)}]{Erber-1966-RvMP}
{Erber}, T. 1966, Reviews of Modern Physics, 38, 626

\bibitem[{Evans {et~al.}(2023)Evans, Corsi, Afle, Ananyeva, Arun, Ballmer, Bandopadhyay, Barsotti, Baryakhtar, Berger, {et~al.}}]{evans2023cosmic}
Evans, M., Corsi, A., Afle, C., {et~al.} 2023, arXiv preprint arXiv:2306.13745

\bibitem[{{Event Horizon Telescope Collaboration} {et~al.}(2019){Event Horizon Telescope Collaboration}, {Akiyama}, {Alberdi}, {Alef}, {Asada}, {Azulay}, {Baczko}, {Ball}, {Balokovi{\'c}}, {Barrett}, {Bintley}, {Blackburn}, {Boland}, {Bouman}, {Bower}, {Bremer}, {Brinkerink}, {Brissenden}, {Britzen}, {Broderick}, {Broguiere}, {Bronzwaer}, {Byun}, {Carlstrom}, {Chael}, {Chan}, {Chatterjee}, {Chatterjee}, {Chen}, {Chen}, {Cho}, {Christian}, {Conway}, {Cordes}, {Crew}, {Cui}, {Davelaar}, {De Laurentis}, {Deane}, {Dempsey}, {Desvignes}, {Dexter}, {Doeleman}, {Eatough}, {Falcke}, {Fish}, {Fomalont}, {Fraga-Encinas}, {Freeman}, {Friberg}, {Fromm}, {G{\'o}mez}, {Galison}, {Gammie}, {Garc{\'\i}a}, {Gentaz}, {Georgiev}, {Goddi}, {Gold}, {Gu}, {Gurwell}, {Hada}, {Hecht}, {Hesper}, {Ho}, {Ho}, {Honma}, {Huang}, {Huang}, {Hughes}, {Ikeda}, {Inoue}, {Issaoun}, {James}, {Jannuzi}, {Janssen}, {Jeter}, {Jiang}, {Johnson}, {Jorstad}, {Jung}, {Karami}, {Karuppusamy}, {Kawashima}, {Keating}, {Kettenis}, {Kim}, {Kim}, {Kim},
  {Kino}, {Koay}, {Koch}, {Koyama}, {Kramer}, {Kramer}, {Krichbaum}, {Kuo}, {Lauer}, {Lee}, {Li}, {Li}, {Lindqvist}, {Liu}, {Liuzzo}, {Lo}, {Lobanov}, {Loinard}, {Lonsdale}, {Lu}, {MacDonald}, {Mao}, {Markoff}, {Marrone}, {Marscher}, {Mart{\'\i}-Vidal}, {Matsushita}, {Matthews}, {Medeiros}, {Menten}, {Mizuno}, {Mizuno}, {Moran}, {Moriyama}, {Moscibrodzka}, {M{\"u}ller}, {Nagai}, {Nagar}, {Nakamura}, {Narayan}, {Narayanan}, {Natarajan}, {Neri}, {Ni}, {Noutsos}, {Okino}, {Olivares}, {Ortiz-Le{\'o}n}, {Oyama}, {{\"O}zel}, {Palumbo}, {Patel}, {Pen}, {Pesce}, {Pi{\'e}tu}, {Plambeck}, {PopStefanija}, {Porth}, {Prather}, {Preciado-L{\'o}pez}, {Psaltis}, {Pu}, {Ramakrishnan}, {Rao}, {Rawlings}, {Raymond}, {Rezzolla}, {Ripperda}, {Roelofs}, {Rogers}, {Ros}, {Rose}, {Roshanineshat}, {Rottmann}, {Roy}, {Ruszczyk}, {Ryan}, {Rygl}, {S{\'a}nchez}, {S{\'a}nchez-Arguelles}, {Sasada}, {Savolainen}, {Schloerb}, {Schuster}, {Shao}, {Shen}, {Small}, {Sohn}, {SooHoo}, {Tazaki}, {Tiede}, {Tilanus}, {Titus}, {Toma}, {Torne},
  {Trent}, {Trippe}, {Tsuda}, {van Bemmel}, {van Langevelde}, {van Rossum}, {Wagner}, {Wardle}, {Weintroub}, {Wex}, {Wharton}, {Wielgus}, {Wong}, {Wu}, {Young}, {Young}, {Younsi}, {Yuan}, {Yuan}, {Zensus}, {Zhao}, {Zhao}, {Zhu}, {Algaba}, {Allardi}, {Amestica}, {Anczarski}, {Bach}, {Baganoff}, {Beaudoin}, {Benson}, {Berthold}, {Blanchard}, {Blundell}, {Bustamente}, {Cappallo}, {Castillo-Dom{\'\i}nguez}, {Chang}, {Chang}, {Chang}, {Chen}, {Chilson}, {Chuter}, {C{\'o}rdova Rosado}, {Coulson}, {Crawford}, {Crowley}, {David}, {Derome}, {Dexter}, {Dornbusch}, {Dudevoir}, {Dzib}, {Eckart}, {Eckert}, {Erickson}, {Everett}, {Faber}, {Farah}, {Fath}, {Folkers}, {Forbes}, {Freund}, {G{\'o}mez-Ruiz}, {Gale}, {Gao}, {Geertsema}, {Graham}, {Greer}, {Grosslein}, {Gueth}, {Haggard}, {Halverson}, {Han}, {Han}, {Hao}, {Hasegawa}, {Henning}, {Hern{\'a}ndez-G{\'o}mez}, {Herrero-Illana}, {Heyminck}, {Hirota}, {Hoge}, {Huang}, {Impellizzeri}, {Jiang}, {Kamble}, {Keisler}, {Kimura}, {Kono}, {Kubo}, {Kuroda}, {Lacasse}, {Laing},
  {Leitch}, {Li}, {Lin}, {Liu}, {Liu}, {Lu}, {Marson}, {Martin-Cocher}, {Massingill}, {Matulonis}, {McColl}, {McWhirter}, {Messias}, {Meyer-Zhao}, {Michalik}, {Monta{\~n}a}, {Montgomerie}, {Mora-Klein}, {Muders}, {Nadolski}, {Navarro}, {Neilsen}, {Nguyen}, {Nishioka}, {Norton}, {Nowak}, {Nystrom}, {Ogawa}, {Oshiro}, {Oyama}, {Parsons}, {Paine}, {Pe{\~n}alver}, {Phillips}, {Poirier}, {Pradel}, {Primiani}, {Raffin}, {Rahlin}, {Reiland}, {Risacher}, {Ruiz}, {S{\'a}ez-Mada{\'\i}n}, {Sassella}, {Schellart}, {Shaw}, {Silva}, {Shiokawa}, {Smith}, {Snow}, {Souccar}, {Sousa}, {Sridharan}, {Srinivasan}, {Stahm}, {Stark}, {Story}, {Timmer}, {Vertatschitsch}, {Walther}, {Wei}, {Whitehorn}, {Whitney}, {Woody}, {Wouterloot}, {Wright}, {Yamaguchi}, {Yu}, {Zeballos}, {Zhang}, \& {Ziurys}}]{EventHorizon2019}
{Event Horizon Telescope Collaboration}, {Akiyama}, K., {Alberdi}, A., {et~al.} 2019, \apjl, 875, L1

\bibitem[{{Fabian}(1999)}]{Fabian99}
{Fabian}, A.~C. 1999, \mnras, 308, L39

\bibitem[{{Fabian} {et~al.}(2015){Fabian}, {Lohfink}, {Kara}, {Parker}, {Vasudevan}, \& {Reynolds}}]{Fabian_2015}
{Fabian}, A.~C., {Lohfink}, A., {Kara}, E., {et~al.} 2015, \mnras, 451, 4375

\bibitem[{{Fabian} {et~al.}(2012){Fabian}, {Zoghbi}, {Wilkins}, {Dwelly}, {Uttley}, {Schartel}, {Miniutti}, {Gallo}, {Grupe}, {Komossa}, \& {Santos-Lle{\'o}}}]{Fabian_2012}
{Fabian}, A.~C., {Zoghbi}, A., {Wilkins}, D., {et~al.} 2012, \mnras, 419, 116

\bibitem[{{Fang} {et~al.}(2016){Fang}, {Kotera}, {Miller}, {Murase}, \& {Oikonomou}}]{Fang2016JCAP...12..017F}
{Fang}, K., {Kotera}, K., {Miller}, M.~C., {Murase}, K., \& {Oikonomou}, F. 2016, \jcap, 2016, 017

\bibitem[{{Farrar} \& {Piran}(2014)}]{Farrar2014}
{Farrar}, G.~R., \& {Piran}, T. 2014, arXiv e-prints, arXiv:1411.0704

\bibitem[{Fedotov {et~al.}(2023)Fedotov, Ilderton, Karbstein, King, Seipt, Taya, \& Torgrimsson}]{fedotov.pr.2023}
Fedotov, A., Ilderton, A., Karbstein, F., {et~al.} 2023, Physics Reports, 1010, 1, advances in QED with intense background fields.
\newblock \url{https://www.sciencedirect.com/science/article/pii/S0370157323000352}

\bibitem[{{Fields} \& {Couch}(2020)}]{Fields2020}
{Fields}, C.~E., \& {Couch}, S.~M. 2020, \apj, 901, 33

\bibitem[{{Fox} \& {Smith}(2019)}]{Fox2019}
{Fox}, O.~D., \& {Smith}, N. 2019, \mnras, 488, 3772

\bibitem[{{Fryer} {et~al.}(1996){Fryer}, {Benz}, \& {Herant}}]{1996ApJ...460..801F}
{Fryer}, C.~L., {Benz}, W., \& {Herant}, M. 1996, \apj, 460, 801

\bibitem[{{Fryer} {et~al.}(2020){Fryer}, {Fontes}, {Warsa}, {Roming}, {Coffing}, \& {Wood}}]{2020ApJ...898..123F}
{Fryer}, C.~L., {Fontes}, C.~J., {Warsa}, J.~S., {et~al.} 2020, \apj, 898, 123

\bibitem[{{Fryer} \& {Warren}(2002)}]{2002ApJ...574L..65F}
{Fryer}, C.~L., \& {Warren}, M.~S. 2002, \apjl, 574, L65

\bibitem[{{Fryer} {et~al.}(2023){Fryer}, {Keiter}, {Sharma}, {Leveillee}, {Meyerhofer}, {Barnak}, {Byvank}, {Elshafiey}, {Fontes}, {Johns}, {Kozlowski}, \& {Urbatsch}}]{2023arXiv231216677F}
{Fryer}, C.~L., {Keiter}, P.~A., {Sharma}, V., {et~al.} 2023, arXiv e-prints, arXiv:2312.16677

\bibitem[{{Fuller} {et~al.}(2009){Fuller}, {Kusenko}, \& {Petraki}}]{2009PhLB..670..281F}
{Fuller}, G.~M., {Kusenko}, A., \& {Petraki}, K. 2009, Physics Letters B, 670, 281

\bibitem[{{Gandolfi} {et~al.}(2012){Gandolfi}, {Carlson}, \& {Reddy}}]{2012PhRvC..85c2801G}
{Gandolfi}, S., {Carlson}, J., \& {Reddy}, S. 2012, \prc, 85, 032801

\bibitem[{Gatu~Johnson {et~al.}(2023)Gatu~Johnson, Hale, Paris, Wiescher, \& Zylstra}]{10.3389/fphy.2023.1180821}
Gatu~Johnson, M., Hale, G., Paris, M., Wiescher, M., \& Zylstra, A. 2023, Frontiers in Physics, 11, doi:10.3389/fphy.2023.1180821.
\newblock \url{https://www.frontiersin.org/journals/physics/articles/10.3389/fphy.2023.1180821}

\bibitem[{{Gehrels} {et~al.}(2004){Gehrels}, {Chincarini}, {Giommi}, {Mason}, {Nousek}, {Wells}, {White}, {Barthelmy}, {Burrows}, {Cominsky}, {Hurley}, {Marshall}, {M{\'e}sz{\'a}ros}, {Roming}, {Angelini}, {Barbier}, {Belloni}, {Campana}, {Caraveo}, {Chester}, {Citterio}, {Cline}, {Cropper}, {Cummings}, {Dean}, {Feigelson}, {Fenimore}, {Frail}, {Fruchter}, {Garmire}, {Gendreau}, {Ghisellini}, {Greiner}, {Hill}, {Hunsberger}, {Krimm}, {Kulkarni}, {Kumar}, {Lebrun}, {Lloyd-Ronning}, {Markwardt}, {Mattson}, {Mushotzky}, {Norris}, {Osborne}, {Paczynski}, {Palmer}, {Park}, {Parsons}, {Paul}, {Rees}, {Reynolds}, {Rhoads}, {Sasseen}, {Schaefer}, {Short}, {Smale}, {Smith}, {Stella}, {Tagliaferri}, {Takahashi}, {Tashiro}, {Townsley}, {Tueller}, {Turner}, {Vietri}, {Voges}, {Ward}, {Willingale}, {Zerbi}, \& {Zhang}}]{Gehrels2004}
{Gehrels}, N., {Chincarini}, G., {Giommi}, P., {et~al.} 2004, \apj, 611, 1005

\bibitem[{{Gilli} {et~al.}(2007){Gilli}, {Comastri}, \& {Hasinger}}]{Gilli07}
{Gilli}, R., {Comastri}, A., \& {Hasinger}, G. 2007, \aap, 463, 79

\bibitem[{Gonoskov {et~al.}(2022)Gonoskov, Blackburn, Marklund, \& Bulanov}]{Gonoskov_RevMod_2022}
Gonoskov, A., Blackburn, T.~G., Marklund, M., \& Bulanov, S.~S. 2022, Rev. Mod. Phys., 94, 045001.
\newblock \url{https://link.aps.org/doi/10.1103/RevModPhys.94.045001}

\bibitem[{{Gordon} {et~al.}(2023){Gordon}, {Fong}, {Kilpatrick}, {Eftekhari}, {Leja}, {Prochaska}, {Nugent}, {Bhandari}, {Blanchard}, {Caleb}, {Day}, {Deller}, {Dong}, {Glowacki}, {Gourdji}, {Mannings}, {Mahoney}, {Marnoch}, {Miller}, {Paterson}, {Rastinejad}, {Ryder}, {Sadler}, {Scott}, {Sears}, {Shannon}, {Simha}, {Stappers}, \& {Tejos}}]{Gordon2023}
{Gordon}, A.~C., {Fong}, W.-f., {Kilpatrick}, C.~D., {et~al.} 2023, \apj, 954, 80

\bibitem[{{G{\"o}tz} {et~al.}(2006){G{\"o}tz}, {Mereghetti}, {Tiengo}, \& {Esposito}}]{Goetz-2006-AandA}
{G{\"o}tz}, D., {Mereghetti}, S., {Tiengo}, A., \& {Esposito}, P. 2006, \aap, 449, L31

\bibitem[{{Gravity Collaboration} {et~al.}(2018){Gravity Collaboration}, {Sturm}, {Dexter}, {Pfuhl}, {Stock}, {Davies}, {Lutz}, {Cl{\'e}net}, {Eckart}, {Eisenhauer}, {Genzel}, {Gratadour}, {H{\"o}nig}, {Kishimoto}, {Lacour}, {Millour}, {Netzer}, {Perrin}, {Peterson}, {Petrucci}, {Rouan}, {Waisberg}, {Woillez}, {Amorim}, {Brandner}, {F{\"o}rster Schreiber}, {Garcia}, {Gillessen}, {Ott}, {Paumard}, {Perraut}, {Scheithauer}, {Straubmeier}, {Tacconi}, \& {Widmann}}]{Gravity_2018}
{Gravity Collaboration}, {Sturm}, E., {Dexter}, J., {et~al.} 2018, \nat, 563, 657

\bibitem[{{Guarini} {et~al.}(2022){Guarini}, {Tamborra}, \& {Margutti}}]{Guarini2022}
{Guarini}, E., {Tamborra}, I., \& {Margutti}, R. 2022, \apj, 935, 157

\bibitem[{{Guo} {et~al.}(2014){Guo}, {Li}, {Daughton}, \& {Liu}}]{Guo2014}
{Guo}, F., {Li}, H., {Daughton}, W., \& {Liu}, Y.-H. 2014, \prl, 113, 155005

\bibitem[{{Guo} {et~al.}(2020){Guo}, {Liu}, {Li}, {Li}, {Daughton}, \& {Kilian}}]{Guo2020}
{Guo}, F., {Liu}, Y.-H., {Li}, X., {et~al.} 2020, Physics of Plasmas, 27, 080501

\bibitem[{{Guo} {et~al.}(2024){Guo}, {Liu}, {Zenitani}, \& {Hoshino}}]{Guo2024}
{Guo}, F., {Liu}, Y.-H., {Zenitani}, S., \& {Hoshino}, M. 2024, \ssr, 220, 43

\bibitem[{{H.~E.~S.~S. Collaboration} {et~al.}(2018){H.~E.~S.~S. Collaboration}, {Abdalla}, {Abramowski}, {Aharonian}, {Ait Benkhali}, {Akhperjanian}, {Andersson}, {Ang{\"u}ner}, {Arrieta}, {Aubert}, {Backes}, {Balzer}, {Barnard}, {Becherini}, {Becker Tjus}, {Berge}, {Bernhard}, {Bernl{\"o}hr}, {Blackwell}, {B{\"o}ttcher}, {Boisson}, {Bolmont}, {Bordas}, {Bregeon}, {Brun}, {Brun}, {Bryan}, {Bulik}, {Capasso}, {Carr}, {Carrigan}, {Casanova}, {Cerruti}, {Chakraborty}, {Chalme-Calvet}, {Chaves}, {Chen}, {Chevalier}, {Chr{\'e}tien}, {Colafrancesco}, {Cologna}, {Condon}, {Conrad}, {Couturier}, {Cui}, {Davids}, {Degrange}, {Deil}, {Devin}, {deWilt}, {Dirson}, {Djannati-Ata{\"\i}}, {Domainko}, {Donath}, {Drury}, {Dubus}, {Dutson}, {Dyks}, {Edwards}, {Egberts}, {Eger}, {Ernenwein}, {Eschbach}, {Farnier}, {Fegan}, {Fernandes}, {Fiasson}, {Fontaine}, {F{\"o}rster}, {Funk}, {F{\"u}{\ss}ling}, {Gabici}, {Gajdus}, {Gallant}, {Garrigoux}, {Giavitto}, {Giebels}, {Glicenstein}, {Gottschall}, {Goyal}, {Grondin}, {Hadasch},
  {Hahn}, {Haupt}, {Hawkes}, {Heinzelmann}, {Henri}, {Hermann}, {Hervet}, {Hillert}, {Hinton}, {Hofmann}, {Hoischen}, {Holler}, {Horns}, {Ivascenko}, {Jacholkowska}, {Jamrozy}, {Janiak}, {Jankowsky}, {Jankowsky}, {Jingo}, {Jogler}, {Jouvin}, {Jung-Richardt}, {Kastendieck}, {Katarzy{\'n}ski}, {Katz}, {Kerszberg}, {Kh{\'e}lifi}, {Kieffer}, {King}, {Klepser}, {Klochkov}, {Klu{\'z}niak}, {Kolitzus}, {Komin}, {Kosack}, {Krakau}, {Kraus}, {Krayzel}, {Kr{\"u}ger}, {Laffon}, {Lamanna}, {Lau}, {Lees}, {Lefaucheur}, {Lefranc}, {Lemi{\`e}re}, {Lemoine-Goumard}, {Lenain}, {Leser}, {Lohse}, {Lorentz}, {Liu}, {L{\'o}pez-Coto}, {Lypova}, {Marandon}, {Marcowith}, {Mariaud}, {Marx}, {Maurin}, {Maxted}, {Mayer}, {Meintjes}, {Meyer}, {Mitchell}, {Moderski}, {Mohamed}, {Mohrmann}, {Mor{\r{a}}}, {Moulin}, {Murach}, {de Naurois}, {Niederwanger}, {Niemiec}, {Oakes}, {O'Brien}, {Odaka}, {{\"O}ttl}, {Ohm}, {de O{\~n}a Wilhelmi}, {Ostrowski}, {Oya}, {Padovani}, {Panter}, {Parsons}, {Paz Arribas}, {Pekeur}, {Pelletier}, {Perennes},
  {Petrucci}, {Peyaud}, {Pita}, {Poon}, {Prokhorov}, {Prokoph}, {P{\"u}hlhofer}, {Punch}, {Quirrenbach}, {Raab}, {Reimer}, {Reimer}, {Renaud}, {de los Reyes}, {Rieger}, {Romoli}, {Rosier-Lees}, {Rowell}, {Rudak}, {Rulten}, {Sahakian}, {Salek}, {Sanchez}, {Santangelo}, {Sasaki}, {Schlickeiser}, {Sch{\"u}ssler}, {Schulz}, \& {Schwanke}}]{2018A&A...612A...2H}
{H.~E.~S.~S. Collaboration}, {Abdalla}, H., {Abramowski}, A., {et~al.} 2018, \aap, 612, A2

\bibitem[{{H.~E.~S.~S. Collaboration} {et~al.}(2023){H.~E.~S.~S. Collaboration}, {Aharonian}, {Ait Benkhali}, {Aschersleben}, {Ashkar}, {Backes}, {Barbosa Martins}, {Batzofin}, {Becherini}, {Berge}, {Bernl{\"o}hr}, {Bi}, {B{\"o}ttcher}, {Boisson}, {Bolmont}, {de Bony de Lavergne}, {Borowska}, {Bradascio}, {Breuhaus}, {Brose}, {Brun}, {Bruno}, {Bulik}, {Burger-Scheidlin}, {Bylund}, {Cangemi}, {Caroff}, {Casanova}, {Celic}, {Cerruti}, {Chand}, {Chandra}, {Chen}, {Chibueze}, {Cotter}, {Damascene Mbarubucyeye}, {Djannati-Ata{\"\i}}, {Dmytriiev}, {Egberts}, {Ernenwein}, {Feijen}, {Fiasson}, {Fichet de Clairfontaine}, {Fontaine}, {F{\"u}{\ss}ling}, {Funk}, {Gabici}, {Gallant}, {Ghafourizadeh}, {Giavitto}, {Giunti}, {Glawion}, {Glicenstein}, {Goswami}, {Grolleron}, {Grondin}, {Haerer}, {Haupt}, {Hinton}, {Hofmann}, {Holch}, {Holler}, {Horns}, {Huang}, {Jamrozy}, {Jankowsky}, {Joshi}, {Jung-Richardt}, {Kasai}, {Katarzy{\'n}ski}, {Kh{\'e}lifi}, {Klepser}, {Klu{\v{z}}niak}, {Komin}, {Kosack}, {Kostunin}, {Lang}, {Le
  Stum}, {Lemi{\`e}re}, {Lemoine-Goumard}, {Lenain}, {Leuschner}, {Lohse}, {Luashvili}, {Lypova}, {Mackey}, {Malyshev}, {Malyshev}, {Marandon}, {Marchegiani}, {Marcowith}, {Marinos}, {Mart{\'\i}-Devesa}, {Marx}, {Maurin}, {Meyer}, {Mitchell}, {Moderski}, {Mohrmann}, {Montanari}, {Moulin}, {Muller}, {Murach}, {Nakashima}, {de Naurois}, {Niemiec}, {Noel}, {O'Brien}, {Ohm}, {Olivera-Nieto}, {de Ona Wilhelmi}, {Ostrowski}, {Panny}, {Panter}, {Parsons}, {Peron}, {Pita}, {Prokhorov}, {Prokoph}, {P{\"u}hlhofer}, {Punch}, {Quirrenbach}, {Reichherzer}, {Reimer}, {Reimer}, {Renaud}, {Rieger}, {Rowell}, {Rudak}, {Ruiz-Velasco}, {Sahakian}, {Sailer}, {Salzmann}, {Sanchez}, {Santangelo}, {Sasaki}, {Sch{\"u}ssler}, {Schwanke}, {Shapopi}, {Sinha}, {Sol}, {Specovius}, {Spencer}, {Spir-Jacob}, {Stawarz}, {Steenkamp}, {Steinmassl}, {Steppa}, {Sushch}, {Suzuki}, {Takahashi}, {Tanaka}, {Tavernier}, {Terrier}, {Thorpe-Morgan}, {Tluczykont}, {Tsirou}, {Tsuji}, {van Eldik}, {Vecchi}, {Veh}, {Venter}, {Vink}, {Wagner}, {Werner},
  {White}, {Wierzcholska}, {Wun Wong}, {Yassin}, {Zacharias}, {Zargaryan}, {Zdziarski}, {Zech}, {Zhu}, {Zouari}, {{\.Z}ywucka}, {Zanin}, {Kerr}, {Johnston}, {Shannon}, \& {Smith}}]{2023NatAs...7.1341H}
{H.~E.~S.~S. Collaboration}, {Aharonian}, F., {Ait Benkhali}, F., {et~al.} 2023, Nature Astronomy, 7, 1341

\bibitem[{{Haardt} \& {Maraschi}(1991)}]{Haardt_1991}
{Haardt}, F., \& {Maraschi}, L. 1991, \apjl, 380, L51

\bibitem[{{Harding} {et~al.}(1997){Harding}, {Baring}, \& {Gonthier}}]{HBG-1997-ApJ}
{Harding}, A.~K., {Baring}, M.~G., \& {Gonthier}, P.~L. 1997, \apj, 476, 246

\bibitem[{{Harding} {et~al.}(2018){Harding}, {Kalapotharakos}, {Barnard}, \& {Venter}}]{2018ApJ...869L..18H}
{Harding}, A.~K., {Kalapotharakos}, C., {Barnard}, M., \& {Venter}, C. 2018, \apjl, 869, L18

\bibitem[{{Harding} \& {Lai}(2006)}]{2006RPPh...69.2631H}
{Harding}, A.~K., \& {Lai}, D. 2006, Reports on Progress in Physics, 69, 2631

\bibitem[{Haskell \& Sedrakian(2018)}]{haskell2018superfluidity}
Haskell, B., \& Sedrakian, A. 2018, The Physics and Astrophysics of Neutron Stars, 401

\bibitem[{{Hayasaki} \& {Yamazaki}(2019)}]{Hayasaki2019}
{Hayasaki}, K., \& {Yamazaki}, R. 2019, \apj, 886, 114

\bibitem[{{Hayes} {et~al.}(2023){Hayes}, {Heng}, {Lamb}, {Lin}, {Veitch}, \& {Williams}}]{2023ApJ...954...92H}
{Hayes}, F., {Heng}, I.~S., {Lamb}, G., {et~al.} 2023, \apj, 954, 92

\bibitem[{{Herant} {et~al.}(1994){Herant}, {Benz}, {Hix}, {Fryer}, \& {Colgate}}]{1994ApJ...435..339H}
{Herant}, M., {Benz}, W., {Hix}, W.~R., {Fryer}, C.~L., \& {Colgate}, S.~A. 1994, \apj, 435, 339

\bibitem[{{Hern{\'a}ndez Santisteban} {et~al.}(2020){Hern{\'a}ndez Santisteban}, {Edelson}, {Horne}, {Gelbord}, {Barth}, {Cackett}, {Goad}, {Netzer}, {Starkey}, {Uttley}, {Brandt}, {Korista}, {Lohfink}, {Onken}, {Page}, {Siegel}, {Vestergaard}, {Bisogni}, {Breeveld}, {Cenko}, {Dalla Bont{\`a}}, {Evans}, {Ferland}, {Gonzalez-Buitrago}, {Grupe}, {Joner}, {Kriss}, {LaPorte}, {Mathur}, {Marshall}, {Mehdipour}, {Mudd}, {Peterson}, {Schmidt}, {Vaughan}, \& {Valenti}}]{Hernandez_2020}
{Hern{\'a}ndez Santisteban}, J.~V., {Edelson}, R., {Horne}, K., {et~al.} 2020, \mnras, 498, 5399

\bibitem[{{Herwig} {et~al.}(2023){Herwig}, {Woodward}, {Mao}, {Thompson}, {Denissenkov}, {Lau}, {Blouin}, {Andrassy}, \& {Paul}}]{Herwig2023}
{Herwig}, F., {Woodward}, P.~R., {Mao}, H., {et~al.} 2023, \mnras, 525, 1601

\bibitem[{{Hild} {et~al.}(2010){Hild}, {Chelkowski}, {Freise}, {Franc}, {Morgado}, {Flaminio}, \& {DeSalvo}}]{Hild2010}
{Hild}, S., {Chelkowski}, S., {Freise}, A., {et~al.} 2010, Classical and Quantum Gravity, 27, 015003

\bibitem[{{Hiramatsu} {et~al.}(2023){Hiramatsu}, {Berger}, {Metzger}, {Gomez}, {Bieryla}, {Arcavi}, {Howell}, {Mckinven}, \& {Tominaga}}]{Hiramatsu2023}
{Hiramatsu}, D., {Berger}, E., {Metzger}, B.~D., {et~al.} 2023, \apjl, 947, L28

\bibitem[{{Ho} {et~al.}(2019){Ho}, {Phinney}, {Ravi}, {Kulkarni}, {Petitpas}, {Emonts}, {Bhalerao}, {Blundell}, {Cenko}, {Dobie}, {Howie}, {Kamraj}, {Kasliwal}, {Murphy}, {Perley}, {Sridharan}, \& {Yoon}}]{Ho2019cow}
{Ho}, A.~Y.~Q., {Phinney}, E.~S., {Ravi}, V., {et~al.} 2019, \apj, 871, 73

\bibitem[{{Ho} {et~al.}(2022){Ho}, {Perley}, {Yao}, {Svinkin}, {de Ugarte Postigo}, {Perley}, {Kann}, {Burns}, {Andreoni}, {Bellm}, {Bissaldi}, {Bloom}, {Brink}, {Dekany}, {Drake}, {Ag{\"u}{\'\i} Fern{\'a}ndez}, {Filippenko}, {Frederiks}, {Graham}, {Hristov}, {Kasliwal}, {Kulkarni}, {Kumar}, {Laher}, {Lysenko}, {Mailyan}, {Malacaria}, {Miller}, {Poolakkil}, {Riddle}, {Ridnaia}, {Rusholme}, {Savchenko}, {Sollerman}, {Th{\"o}ne}, {Tsvetkova}, {Ulanov}, \& {von Kienlin}}]{Ho2022}
{Ho}, A. Y.~Q., {Perley}, D.~A., {Yao}, Y., {et~al.} 2022, \apj, 938, 85

\bibitem[{{Ho} {et~al.}(2023{\natexlab{a}}){Ho}, {Perley}, {Gal-Yam}, {Lunnan}, {Sollerman}, {Schulze}, {Das}, {Dobie}, {Yao}, {Fremling}, {Adams}, {Anand}, {Andreoni}, {Bellm}, {Bruch}, {Burdge}, {Castro-Tirado}, {Dahiwale}, {De}, {Dekany}, {Drake}, {Duev}, {Graham}, {Helou}, {Kaplan}, {Karambelkar}, {Kasliwal}, {Kool}, {Kulkarni}, {Mahabal}, {Medford}, {Miller}, {Nordin}, {Ofek}, {Petitpas}, {Riddle}, {Sharma}, {Smith}, {Stewart}, {Taggart}, {Tartaglia}, {Tzanidakis}, \& {Winters}}]{Ho2023}
{Ho}, A. Y.~Q., {Perley}, D.~A., {Gal-Yam}, A., {et~al.} 2023{\natexlab{a}}, \apj, 949, 120

\bibitem[{{Ho} {et~al.}(2023{\natexlab{b}}){Ho}, {Perley}, {Chen}, {Schulze}, {Dhillon}, {Kumar}, {Suresh}, {Swain}, {Bremer}, {Smartt}, {Anderson}, {Anupama}, {Awiphan}, {Barway}, {Bellm}, {Ben-Ami}, {Bhalerao}, {de Boer}, {Brink}, {Burruss}, {Chandra}, {Chen}, {Chen}, {Cooke}, {Coughlin}, {Das}, {Drake}, {Filippenko}, {Freeburn}, {Fremling}, {Fulton}, {Gal-Yam}, {Galbany}, {Gao}, {Graham}, {Gromadzki}, {Guti{\'e}rrez}, {Hinds}, {Inserra}, {A J}, {Karambelkar}, {Kasliwal}, {Kulkarni}, {M{\"u}ller-Bravo}, {Magnier}, {Mahabal}, {Moore}, {Ngeow}, {Nicholl}, {Ofek}, {Omand}, {Onori}, {Pan}, {Pessi}, {Petitpas}, {Polishook}, {Poshyachinda}, {Pursiainen}, {Riddle}, {Rodriguez}, {Rusholme}, {Segre}, {Sharma}, {Smith}, {Sollerman}, {Srivastav}, {Strotjohann}, {Suhr}, {Svinkin}, {Wang}, {Wiseman}, {Wold}, {Yang}, {Yang}, {Yao}, {Young}, \& {Zheng}}]{Ho2023b}
{Ho}, A. Y.~Q., {Perley}, D.~A., {Chen}, P., {et~al.} 2023{\natexlab{b}}, \nat, 623, 927

\bibitem[{{Holoien} {et~al.}(2018){Holoien}, {Brown}, {Auchettl}, {Kochanek}, {Prieto}, {Shappee}, \& {Van Saders}}]{Holoien2018MNRAS.480.5689H}
{Holoien}, T.~W.~S., {Brown}, J.~S., {Auchettl}, K., {et~al.} 2018, \mnras, 480, 5689

\bibitem[{{Hopkins} {et~al.}(2006){Hopkins}, {Hernquist}, {Cox}, {Di Matteo}, {Robertson}, \& {Springel}}]{Hopkins06}
{Hopkins}, P.~F., {Hernquist}, L., {Cox}, T.~J., {et~al.} 2006, \apjs, 163, 1

\bibitem[{{Horowitz} {et~al.}(2009){Horowitz}, {Caballero}, \& {Berry}}]{2009PhRvE..79b6103H}
{Horowitz}, C.~J., {Caballero}, O.~L., \& {Berry}, D.~K. 2009, \pre, 79, 026103

\bibitem[{{Hu} {et~al.}(2024{\natexlab{a}}){Hu}, {Narita}, {Enoto}, {Younes}, {Wadiasingh}, {Baring}, {Ho}, {Guillot}, {Ray}, {G{\"u}ver}, {Rajwade}, {Arzoumanian}, {Kouveliotou}, {Harding}, \& {Gendreau}}]{2024Natur.626..500H}
{Hu}, C.-P., {Narita}, T., {Enoto}, T., {et~al.} 2024{\natexlab{a}}, \nat, 626, 500

\bibitem[{{Hu} {et~al.}(2022){Hu}, {Baring}, {Harding}, \& {Wadiasingh}}]{Hu-2022-ApJ-opac}
{Hu}, K., {Baring}, M.~G., {Harding}, A.~K., \& {Wadiasingh}, Z. 2022, \apj, 940, 91

\bibitem[{{Hu} {et~al.}(2019){Hu}, {Baring}, {Wadiasingh}, \& {Harding}}]{Hu-2019-MNRAS}
{Hu}, K., {Baring}, M.~G., {Wadiasingh}, Z., \& {Harding}, A.~K. 2019, \mnras, 486, 3327

\bibitem[{{Hu} {et~al.}(2024{\natexlab{b}}){Hu}, {Nichols}, {Shaffer}, {Arnold}, {White}, {Collins}, {Karasiev}, {Zhang}, {Goncharov}, {Shah}, {Mihaylov}, {Jiang}, \& {Ping}}]{2024PhPl...31d0501H}
{Hu}, S.~X., {Nichols}, K.~A., {Shaffer}, N.~R., {et~al.} 2024{\natexlab{b}}, Physics of Plasmas, 31, 040501

\bibitem[{{Huppenkothen} {et~al.}(2014){Huppenkothen}, {D'Angelo}, {Watts}, {Heil}, {van der Klis}, {van der Horst}, {Kouveliotou}, {Baring}, {G{\"o}{\u{g}}{\"u}{\c{s}}}, {Granot}, {Kaneko}, {Lin}, {von Kienlin}, \& {Younes}}]{Huppenkothen_2014ApJ...787..128H}
{Huppenkothen}, D., {D'Angelo}, C., {Watts}, A.~L., {et~al.} 2014, \apj, 787, 128

\bibitem[{{Hurley-Walker} {et~al.}(2022){Hurley-Walker}, {Zhang}, {Bahramian}, {McSweeney}, {O'Doherty}, {Hancock}, {Morgan}, {Anderson}, {Heald}, \& {Galvin}}]{2022Natur.601..526H}
{Hurley-Walker}, N., {Zhang}, X., {Bahramian}, A., {et~al.} 2022, \nat, 601, 526

\bibitem[{{Hurley-Walker} {et~al.}(2023){Hurley-Walker}, {Rea}, {McSweeney}, {Meyers}, {Lenc}, {Heywood}, {Hyman}, {Men}, {Clarke}, {Coti Zelati}, {Price}, {Horv{\'a}th}, {Galvin}, {Anderson}, {Bahramian}, {Barr}, {Bhat}, {Caleb}, {Dall'Ora}, {de Martino}, {Giacintucci}, {Morgan}, {Rajwade}, {Stappers}, \& {Williams}}]{2023Natur.619..487H}
{Hurley-Walker}, N., {Rea}, N., {McSweeney}, S.~J., {et~al.} 2023, \nat, 619, 487

\bibitem[{{IceCube Collaboration} {et~al.}(2018){IceCube Collaboration}, {Aartsen}, {Ackermann}, {Adams}, {Aguilar}, {Ahlers}, {Ahrens}, {Samarai}, {Altmann}, {Andeen}, {Anderson}, {Ansseau}, {Anton}, {Arg{\"u}elles}, {Arsioli}, {Auffenberg}, {Axani}, {Bagherpour}, {Bai}, {Barron}, {Barwick}, {Baum}, {Bay}, {Beatty}, {Becker Tjus}, {Becker}, {BenZvi}, {Berley}, {Bernardini}, {Besson}, {Binder}, {Bindig}, {Blaufuss}, {Blot}, {Bohm}, {B{\"o}rner}, {Bos}, {B{\"o}ser}, {Botner}, {Bourbeau}, {Bourbeau}, {Bradascio}, {Braun}, {Brenzke}, {Bretz}, {Bron}, {Brostean-Kaiser}, {Burgman}, {Busse}, {Carver}, {Cheung}, {Chirkin}, {Christov}, {Clark}, {Classen}, {Coenders}, {Collin}, {Conrad}, {Coppin}, {Correa}, {Cowen}, {Cross}, {Dave}, {Day}, {de Andr{\'e}}, {De Clercq}, {DeLaunay}, {Dembinski}, {DeRidder}, {Desiati}, {de Vries}, {de Wasseige}, {de With}, {DeYoung}, {D{\'\i}az-V{\'e}lez}, {di Lorenzo}, {Dujmovic}, {Dumm}, {Dunkman}, {Dvorak}, {Eberhardt}, {Ehrhardt}, {Eichmann}, {Eller}, {Evenson}, {Fahey}, {Fazely},
  {Felde}, {Filimonov}, {Finley}, {Flis}, {Franckowiak}, {Friedman}, {Fritz}, {Gaisser}, {Gallagher}, {Gerhardt}, {Ghorbani}, {Giommi}, {Glauch}, {Gl{\"u}senkamp}, {Goldschmidt}, {Gonzalez}, {Grant}, {Griffith}, {Haack}, {Hallgren}, {Halzen}, {Hanson}, {Hebecker}, {Heereman}, {Helbing}, {Hellauer}, {Hickford}, {Hignight}, {Hill}, {Hoffman}, {Hoffmann}, {Hoinka}, {Hokanson-Fasig}, {Hoshina}, {Huang}, {Huber}, {Hultqvist}, {H{\"u}nnefeld}, {Hussain}, {In}, {Iovine}, {Ishihara}, {Jacobi}, {Japaridze}, {Jeong}, {Jero}, {Jones}, {Kalaczynski}, {Kang}, {Kappes}, {Kappesser}, {Karg}, {Karle}, {Katz}, {Kauer}, {Keivani}, {Kelley}, {Kheirandish}, {Kim}, {Kim}, {Kintscher}, {Kiryluk}, {Kittler}, {Klein}, {Koirala}, {Kolanoski}, {K{\"o}pke}, {Kopper}, {Kopper}, {Koschinsky}, {Koskinen}, {Kowalski}, {Krammer}, {Krings}, {Kroll}, {Kr{\"u}ckl}, {Kunwar}, {Kurahashi}, {Kuwabara}, {Kyriacou}, {Labare}, {Lanfranchi}, {Larson}, {Lauber}, {Leonard}, {Lesiak-Bzdak}, {Leuermann}, {Liu}, {Lozano Mariscal}, {Lu}, {L{\"u}nemann},
  {Luszczak}, {Madsen}, {Maggi}, {Mahn}, {Mancina}, {Maruyama}, {Mase}, {Maunu}, {Meagher}, {Medici}, {Meier}, {Menne}, {Merino}, {Meures}, {Miarecki}, {Micallef}, {Moment{\'e}}, {Montaruli}, {Moore}, {Morse}, {Moulai}, {Nahnhauer}, {Nakarmi}, {Naumann}, {Neer}, {Niederhausen}, {Nowicki}, {Nygren}, {Obertacke Pollmann}, {Olivas}, {O'Murchadha}, {O'Sullivan}, {Padovani}, {Palczewski}, {Pandya}, {Pankova}, {Peiffer}, {Pepper}, {P{\'e}rez de los Heros}, {Pieloth}, {Pinat}, {Plum}, {Price}, {Przybylski}, {Raab}, {R{\"a}del}, {Rameez}, {Rawlins}, {Rea}, {Reimann}, {Relethford}, {Relich}, {Resconi}, {Rhode}, {Richman}, {Robertson}, {Rongen}, {Rott}, {Ruhe}, {Ryckbosch}, {Rysewyk}, {Safa}, {Sahakyan}, {S{\"a}lzer}, {Sanchez Herrera}, {Sandrock}, {Sandroos}, {Santander}, {Sarkar}, {Sarkar}, {Satalecka}, {Schlunder}, {Schmidt}, {Schneider}, {Schoenen}, {Sch{\"o}neberg}, {Schumacher}, {Sclafani}, {Seckel}, {Seunarine}, {Soedingrekso}, {Soldin}, {Song}, {Spiczak}, {Spiering}, {Stachurska}, {Stamatikos}, {Stanev},
  {Stasik}, {Stettner}, {Steuer}, {Stezelberger}, {Stokstad}, {St{\"o}{\ss}l}, {Strotjohann}, {Stuttard}, {Sullivan}, {Sutherland}, {Taboada}, {Tatar}, {Tenholt}, {Ter-Antonyan}, {Terliuk}, {Tilav}, {Toale}, {Tobin}, {Toennis}, {Toscano}, {Tosi}, {Tselengidou}, {Tung}, {Turcati}, {Turley}, {Ty}, {Unger}, {Usner}, {Vandenbroucke}, {Van Driessche}, {van Eijk}, {van Eijndhoven}, {Vanheule}, {van Santen}, {Vogel}, {Vraeghe}, {Walck}, {Wallace}, {Wallraff}, {Wandler}, {Wandkowsky}, {Waza}, {Weaver}, {Weiss}, {Wendt}, {Werthebach}, {Westerhoff}, {Whelan}, {Whitehorn}, {Wiebe}, {Wiebusch}, {Wille}, {Williams}, {Wills}, {Wolf}, {Wood}, {Wood}, {Woschnagg}, {Xu}, {Xu}, {Xu}, {Yanez}, {Yodh}, {Yoshida}, \& {Yuan}}]{IceCubeCollaboration2018}
{IceCube Collaboration}, {Aartsen}, M.~G., {Ackermann}, M., {et~al.} 2018, Science, 361, 147

\bibitem[{{IceCube Collaboration} {et~al.}(2022){IceCube Collaboration}, {Abbasi}, {Ackermann}, {Adams}, {Aguilar}, {Ahlers}, {Ahrens}, {Alameddine}, {Alispach}, {Alves}, {Amin}, {Andeen}, {Anderson}, {Anton}, {Arg{\"u}elles}, {Ashida}, {Axani}, {Bai}, {Balagopal}, {Barbano}, {Barwick}, {Bastian}, {Basu}, {Baur}, {Bay}, {Beatty}, {Becker}, {Becker Tjus}, {Bellenghi}, {Benzvi}, {Berley}, {Bernardini}, {Besson}, {Binder}, {Bindig}, {Blaufuss}, {Blot}, {Boddenberg}, {Bontempo}, {Borowka}, {B{\"o}ser}, {Botner}, {B{\"o}ttcher}, {Bourbeau}, {Bradascio}, {Braun}, {Brinson}, {Bron}, {Brostean-Kaiser}, {Browne}, {Burgman}, {Burley}, {Busse}, {Campana}, {Carnie-Bronca}, {Chen}, {Chen}, {Chirkin}, {Choi}, {Clark}, {Clark}, {Classen}, {Coleman}, {Collin}, {Conrad}, {Coppin}, {Correa}, {Cowen}, {Cross}, {Dappen}, {Dave}, {de Clercq}, {Delaunay}, {Delgado L{\'o}pez}, {Dembinski}, {Deoskar}, {Desai}, {Desiati}, {de Vries}, {de Wasseige}, {de With}, {Deyoung}, {Diaz}, {D{\'\i}az-V{\'e}lez}, {Dittmer}, {Dujmovic}, {Dunkman},
  {Duvernois}, {Dvorak}, {Ehrhardt}, {Eller}, {Engel}, {Erpenbeck}, {Evans}, {Evenson}, {Fan}, {Fazely}, {Fedynitch}, {Feigl}, {Fiedlschuster}, {Fienberg}, {Filimonov}, {Finley}, {Fischer}, {Fox}, {Franckowiak}, {Friedman}, {Fritz}, {F{\"u}rst}, {Gaisser}, {Gallagher}, {Ganster}, {Garcia}, {Garrappa}, {Gerhardt}, {Ghadimi}, {Glaser}, {Glauch}, {Gl{\"u}senkamp}, {Goldschmidt}, {Gonzalez}, {Goswami}, {Grant}, {Gr{\'e}goire}, {Griswold}, {G{\"u}nther}, {Gutjahr}, {Haack}, {Hallgren}, {Halliday}, {Halve}, {Halzen}, {Hanson}, {Hardin}, {Harnisch}, {Haungs}, {Hebecker}, {Helbing}, {Henningsen}, {Hettinger}, {Hickford}, {Hignight}, {Hill}, {Hill}, {Hoffman}, {Hoffmann}, {Hokanson-Fasig}, {Hoshina}, {Huang}, {Huber}, {Huber}, {Hultqvist}, {H{\"u}nnefeld}, {Hussain}, {Hymon}, {in}, {Iovine}, {Ishihara}, {Jansson}, {Japaridze}, {Jeong}, {Jin}, {Jones}, {Kang}, {Kang}, {Kang}, {Kappes}, {Kappesser}, {Kardum}, {Karg}, {Karl}, {Karle}, {Katz}, {Kauer}, {Kellermann}, {Kelley}, {Kheirandish}, {Kin}, {Kintscher}, {Kiryluk},
  {Klein}, {Koirala}, {Kolanoski}, {Kontrimas}, {K{\"o}pke}, {Kopper}, {Kopper}, {Koskinen}, {Koundal}, {Kovacevich}, {Kowalski}, {Kozynets}, {Kun}, {Kurahashi}, {Lad}, {Lagunas Gualda}, {Lanfranchi}, {Larson}, {Lauber}, {Lazar}, {Lee}, {Leonard}, {Leszczy{\'n}ska}, {Li}, {Lincetto}, {Liu}, {Liubarska}, {Lohfink}, {Lozano Mariscal}, {Lu}, {Lucarelli}, {Ludwig}, {Luszczak}, {Lyu}, {Ma}, {Madsen}, {Mahn}, {Makino}, {Mancina}, {Mari{\c{s}}}, {Martinez-Soler}, {Maruyama}, {Mase}, {McElroy}, {McNally}, {Mead}, {Meagher}, {Mechbal}, {Medina}, {Meier}, {Meighen-Berger}, {Micallef}, {Mockler}, {Montaruli}, {Moore}, {Morse}, {Moulai}, {Naab}, {Nagai}, {Nahnhauer}, {Naumann}, {Necker}, {Nguyen}, {Niederhausen}, {Nisa}, {Nowicki}, {Nygren}, {Obertack}, {Pollmann}, {Oehler}, {Oeyen}, {Olivas}, {O'Sullivan}, {Pandya}, {Pankova}, {Park}, {Parker}, {Paudel}, {Paul}, {P{\'e}rez de Los Heros}, {Peters}, {Peterson}, {Philippen}, {Pieper}, {Pittermann}, {Pizzuto}, {Plum}, {Popovych}, {Porcelli}, {Prado Rodriguez}, {Price},
  {Pries}, {Przybylski}, {Rack-Helleis}, {Raissi}, {Rameez}, {Rawlins}, {Rea}, {Rehman}, {Reichherzer}, {Reimann}, {Renzi}, {Resconi}, {Reusch}, {Rhode}, {Richman}, {Riedel}, {Roberts}, {Robertson}, {Roellinghoff}, {Rongen}, {Rott}, {Ruhe}, {Ryckbosch}, {Rysewyk Cantu}, {Safa}, {Saffer}, {Sanchez Herrera}, {Sandrock}, {Sandroos}, {Santander}, {Sarkar}, {Sarkar}, {Satalecka}, {Schaufel}, {Schieler}, {Schindler}, {Schmidt}, {Schneider}, {Schneider}, {Schr{\"o}der}, {Schumacher}, {Schwefer}, {Sclafani}, {Seckel}, {Seunarine}, {Sharma}, {Shefali}, {Silva}, {Skrzypek}, {Smithers}, {Snihur}, {Soedingrekso}, {Soldin}, {Spannfellner}, {Spiczak}, {Spiering}, {Stachurska}, {Stamatikos}, {Stanev}, {Stein}, {Stettner}, {Steuer}, {Stezelberger}, {Stokstad}, {St{\"u}rwald}, {Stuttard}, {Sullivan}, {Taboada}, {Ter-Antonyan}, {Tilav}, {Tischbein}, {Tollefson}, {T{\"o}nnis}, {Toscano}, {Tosi}, {Trettin}, {Tselengidou}, {Tung}, {Turcati}, {Turcotte}, {Turley}, {Twagirayezu}, {Ty}, {Unland Elorrieta}, {Valtonen-Mattila},
  {Vandenbroucke}, {van Eijndhoven}, {Vannerom}, {van Santen}, {Verpoest}, {Walck}, {Watson}, {Weaver}, {Weigel}, {Weindl}, {Weiss}, {Weldert}, {Wendt}, {Werthebach}, {Weyrauch}, {Whitehorn}, {Wiebusch}, {Williams}, {Wolf}, {Woschnagg}, {Wrede}, {Wulff}, {Xu}, {Yanez}, {Yoshida}, {Yu}, {Yuan}, {Zhangan}, \& {Zhelnin}}]{IceCube_2022}
{IceCube Collaboration}, {Abbasi}, R., {Ackermann}, M., {et~al.} 2022, Science, 378, 538

\bibitem[{{Inkenhaag} {et~al.}(2023){Inkenhaag}, {Jonker}, {Levan}, {Chrimes}, {Mummery}, {Perley}, \& {Tanvir}}]{Inkenhaag2023}
{Inkenhaag}, A., {Jonker}, P.~G., {Levan}, A.~J., {et~al.} 2023, \mnras, 525, 4042

\bibitem[{{Ioka} \& {Nakamura}(2019)}]{2019MNRAS.487.4884I}
{Ioka}, K., \& {Nakamura}, T. 2019, \mnras, 487, 4884

\bibitem[{{Israel} {et~al.}(2005){Israel}, {Belloni}, {Stella}, {Rephaeli}, {Gruber}, {Casella}, {Dall'Osso}, {Rea}, {Persic}, \& {Rothschild}}]{2005ApJ...628L..53I}
{Israel}, G.~L., {Belloni}, T., {Stella}, L., {et~al.} 2005, \apjl, 628, L53

\bibitem[{{Izzo} {et~al.}(2024){Izzo}, {Chrimes}, {Malesani}, {Saccardi}, {Martin-Carrillo}, {Levan}, {Gompertz}, {Tanvir}, \& {Stargate Collaboration}}]{2024GCN.38097....1M}
{Izzo}, L., {Chrimes}, A., {Malesani}, D., {et~al.} 2024, GRB Coordinates Network, 38097, 1

\bibitem[{Jiang {et~al.}(2023)Jiang, Zhou, Zhu, Wang, \& Wang}]{Jiang2023}
Jiang, N., Zhou, Z., Zhu, J., Wang, Y., \& Wang, T. 2023, The Astrophysical Journal Letters, 953, L12.
\newblock \url{http://dx.doi.org/10.3847/2041-8213/acebe3}

\bibitem[{{Joggerst} {et~al.}(2014){Joggerst}, {Nelson}, {Woodward}, {Lovekin}, {Masser}, {Fryer}, {Ramaprabhu}, {Francois}, \& {Rockefeller}}]{2014JCoPh.275..154J}
{Joggerst}, C.~C., {Nelson}, A., {Woodward}, P., {et~al.} 2014, Journal of Computational Physics, 275, 154

\bibitem[{Johnson {et~al.}(2020)Johnson, Fields, \& Thompson}]{johnson2020origin}
Johnson, J.~A., Fields, B.~D., \& Thompson, T.~A. 2020, Philosophical transactions of the royal society A, 378, 20190301

\bibitem[{{Jones} {et~al.}(2014){Jones}, {Hirschi}, \& {Nomoto}}]{JoHiNo14}
{Jones}, S., {Hirschi}, R., \& {Nomoto}, K. 2014, \apj, 797, 83

\bibitem[{{Jones} {et~al.}(2013){Jones}, {Hirschi}, {Nomoto}, {Fischer}, {Timmes}, {Herwig}, {Paxton}, {Toki}, {Suzuki}, {Mart{\'\i}nez-Pinedo}, {Lam}, \& {Bertolli}}]{JoHiNo13}
{Jones}, S., {Hirschi}, R., {Nomoto}, K., {et~al.} 2013, \apj, 772, 150

\bibitem[{{Kallman} {et~al.}(2019){Kallman}, {Bautista}, {Betancourt-Martinez}, {Bregman}, {Brenneman}, {Brickhouse}, {Canizares}, {Cumbee}, {Garcia}, {G{\"u}nther}, {Hell}, {Hodges-Kluck}, {Kaastra}, {Laha}, {Leutenegger}, {Miller}, {Paerels}, {Porter}, {Smith}, {Temi}, {Valencic}, \& {Wilms}}]{Kallman_2019}
{Kallman}, T., {Bautista}, M., {Betancourt-Martinez}, G., {et~al.} 2019, in Bulletin of the American Astronomical Society, Vol.~51, 156

\bibitem[{{Kammoun} {et~al.}(2021{\natexlab{a}}){Kammoun}, {Dov{\v{c}}iak}, {Papadakis}, {Caballero-Garc{\'\i}a}, \& {Karas}}]{Kammoun_2021b}
{Kammoun}, E.~S., {Dov{\v{c}}iak}, M., {Papadakis}, I.~E., {Caballero-Garc{\'\i}a}, M.~D., \& {Karas}, V. 2021{\natexlab{a}}, \apj, 907, 20

\bibitem[{{Kammoun} {et~al.}(2021{\natexlab{b}}){Kammoun}, {Papadakis}, \& {Dov{\v{c}}iak}}]{Kammoun_2021a}
{Kammoun}, E.~S., {Papadakis}, I.~E., \& {Dov{\v{c}}iak}, M. 2021{\natexlab{b}}, \mnras, 503, 4163

\bibitem[{{Kara} {et~al.}(2021){Kara}, {Mehdipour}, {Kriss}, {Cackett}, {Arav}, {Barth}, {Byun}, {Brotherton}, {De Rosa}, {Gelbord}, {Hern{\'a}ndez Santisteban}, {Hu}, {Kaastra}, {Landt}, {Li}, {Miller}, {Montano}, {Partington}, {Aceituno}, {Bai}, {Bao}, {Bentz}, {Brink}, {Chelouche}, {Chen}, {Colmenero}, {Dalla Bont{\`a}}, {Dehghanian}, {Du}, {Edelson}, {Ferland}, {Ferrarese}, {Fian}, {Filippenko}, {Fischer}, {Goad}, {Gonz{\'a}lez Buitrago}, {Gorjian}, {Grier}, {Guo}, {Hall}, {Ho}, {Homayouni}, {Horne}, {Ili{\'c}}, {Jiang}, {Joner}, {Kaspi}, {Kochanek}, {Korista}, {Kynoch}, {Li}, {Liu}, {McHardy}, {McLane}, {Mitchell}, {Netzer}, {Olson}, {Pogge}, {Popovi{\'c}}, {Proga}, {Storchi-Bergmann}, {Strasburger}, {Treu}, {Vestergaard}, {Wang}, {Ward}, {Waters}, {Williams}, {Yang}, {Yao}, {Zastrocky}, {Zhai}, \& {Zu}}]{Kara_2021}
{Kara}, E., {Mehdipour}, M., {Kriss}, G.~A., {et~al.} 2021, \apj, 922, 151

\bibitem[{Karniadakis {et~al.}(2021)Karniadakis, Kevrekidis, Lu, Perdikaris, Wang, \& Yang}]{Karniadakis2021Physic-Informed}
Karniadakis, G.~E., Kevrekidis, I.~G., Lu, L., {et~al.} 2021, Nature Review Physics, 3, 422.
\newblock \url{https://doi.org/10.1038/s42254-021-00314-5}

\bibitem[{{Kaspi} \& {Beloborodov}(2017)}]{Kaspi_2017ARA&A..55..261K}
{Kaspi}, V.~M., \& {Beloborodov}, A.~M. 2017, \araa, 55, 261

\bibitem[{{Kato}(2007)}]{Kato2007ApJ...668..974K}
{Kato}, T.~N. 2007, \apj, 668, 974

\bibitem[{Kennea {et~al.}(2024)Kennea, Racusin, Burns, Grefenstettte, Hounsell, Hui, Kocevski, Lazio, Lesage, Pritchard, {et~al.}}]{kennea2024time}
Kennea, J.~A., Racusin, J.~L., Burns, E., {et~al.} 2024, arXiv preprint arXiv:2410.03980

\bibitem[{{Khatami} \& {Kasen}(2024)}]{Khatami2024}
{Khatami}, D.~K., \& {Kasen}, D.~N. 2024, \apj, 972, 140

\bibitem[{{Kilpatrick} {et~al.}(2024){Kilpatrick}, {Tejos}, {Andersen}, {Prochaska}, {N{\'u}{\~n}ez}, {Fonseca}, {Hartman}, {Howell}, {Seccull}, \& {Tendulkar}}]{Kilpatrick2024}
{Kilpatrick}, C.~D., {Tejos}, N., {Andersen}, B.~C., {et~al.} 2024, \apj, 964, 121

\bibitem[{{Kirsten} {et~al.}(2022){Kirsten}, {Marcote}, {Nimmo}, {Hessels}, {Bhardwaj}, {Tendulkar}, {Keimpema}, {Yang}, {Snelders}, {Scholz}, {Pearlman}, {Law}, {Peters}, {Giroletti}, {Paragi}, {Bassa}, {Hewitt}, {Bach}, {Bezrukovs}, {Burgay}, {Buttaccio}, {Conway}, {Corongiu}, {Feiler}, {Forss{\'e}n}, {Gawro{\'n}ski}, {Karuppusamy}, {Kharinov}, {Lindqvist}, {Maccaferri}, {Melnikov}, {Ould-Boukattine}, {Possenti}, {Surcis}, {Wang}, {Yuan}, {Aggarwal}, {Anna-Thomas}, {Bower}, {Blaauw}, {Burke-Spolaor}, {Cassanelli}, {Clarke}, {Fonseca}, {Gaensler}, {Gopinath}, {Kaspi}, {Kassim}, {Lazio}, {Leung}, {Li}, {Lin}, {Masui}, {Mckinven}, {Michilli}, {Mikhailov}, {Ng}, {Orbidans}, {Pen}, {Petroff}, {Rahman}, {Ransom}, {Shin}, {Smith}, {Stairs}, \& {Vlemmings}}]{Kirsten2022}
{Kirsten}, F., {Marcote}, B., {Nimmo}, K., {et~al.} 2022, \nat, 602, 585

\bibitem[{Kiuchi {et~al.}(2024)Kiuchi, Reboul-Salze, Shibata, \& Sekiguchi}]{Kiuchi:2023obe}
Kiuchi, K., Reboul-Salze, A., Shibata, M., \& Sekiguchi, Y. 2024, Nature Astron., 8, 298

\bibitem[{Kobayashi {et~al.}(2020)Kobayashi, Karakas, \& Lugaro}]{kobayashi2020origin}
Kobayashi, C., Karakas, A.~I., \& Lugaro, M. 2020, The Astrophysical Journal, 900, 179

\bibitem[{Komossa(2015)}]{komossa2015tidal}
Komossa, S. 2015, Journal of High Energy Astrophysics, 7, 148

\bibitem[{{Kouveliotou} {et~al.}(1998){Kouveliotou}, {Dieters}, {Strohmayer}, {van Paradijs}, {Fishman}, {Meegan}, {Hurley}, {Kommers}, {Smith}, {Frail}, \& {Murakami}}]{1998Natur.393..235K}
{Kouveliotou}, C., {Dieters}, S., {Strohmayer}, T., {et~al.} 1998, \nat, 393, 235

\bibitem[{Kramida {et~al.}(2024)Kramida, Ralchenko, Reader, \& Team}]{Kramida2024}
Kramida, A., Ralchenko, Y., Reader, J., \& Team, N.~A. 2024, NIST Atomic Spectra Database (version 5.12), Online,  NIST, available: \url{https://physics.nist.gov/asd} [Mon Dec 23 2024], doi:10.18434/T4W30F

\bibitem[{{Kuiper} {et~al.}(2004){Kuiper}, {Hermsen}, \& {Mendez}}]{Kuiper-2004-ApJ}
{Kuiper}, L., {Hermsen}, W., \& {Mendez}, M. 2004, \apj, 613, 1173

\bibitem[{Kuksin {et~al.}(2005)Kuksin, Morozov, Norman, Stegailov, \& Valuev}]{Kuksin01122005}
Kuksin, A., Morozov, I.~V., Norman, G.~E., Stegailov, V.~V., \& Valuev, I.~A. 2005, Molecular Simulation, 31, 1005.
\newblock \url{https://doi.org/10.1080/08927020500375259}

\bibitem[{{Kumar} {et~al.}(2017){Kumar}, {Lu}, \& {Bhattacharya}}]{Kumar2017}
{Kumar}, P., {Lu}, W., \& {Bhattacharya}, M. 2017, \mnras, 468, 2726

\bibitem[{{Kuranz} {et~al.}(2018){Kuranz}, {Park}, {Huntington}, {Miles}, {Remington}, {Plewa}, {Trantham}, {Robey}, {Shvarts}, {Shimony}, {Raman}, {MacLaren}, {Wan}, {Doss}, {Kline}, {Flippo}, {Malamud}, {Handy}, {Prisbrey}, {Krauland}, {Klein}, {Harding}, {Wallace}, {Grosskopf}, {Marion}, {Kalantar}, {Giraldez}, \& {Drake}}]{2018NatCo...9.1564K}
{Kuranz}, C.~C., {Park}, H.~S., {Huntington}, C.~M., {et~al.} 2018, Nature Communications, 9, 1564

\bibitem[{{Lamb} \& {Kobayashi}(2018)}]{2018MNRAS.478..733L}
{Lamb}, G.~P., \& {Kobayashi}, S. 2018, \mnras, 478, 733

\bibitem[{{Lander} {et~al.}(2024){Lander}, {Gourgouliatos}, {Wadiasingh}, \& {Antonopoulou}}]{2024arXiv241108020L}
{Lander}, S.~K., {Gourgouliatos}, K.~N., {Wadiasingh}, Z., \& {Antonopoulou}, D. 2024, arXiv e-prints, arXiv:2411.08020

\bibitem[{{Langis} {et~al.}(2024){Langis}, {Papadakis}, {Kammoun}, {Panagiotou}, \& {Dov{\v{c}}iak}}]{Langis_2024}
{Langis}, D.~A., {Papadakis}, I.~E., {Kammoun}, E., {Panagiotou}, C., \& {Dov{\v{c}}iak}, M. 2024, \aap, 691, A252

\bibitem[{{Lattimer} \& {Prakash}(2001)}]{2001ApJ...550..426L}
{Lattimer}, J.~M., \& {Prakash}, M. 2001, \apj, 550, 426

\bibitem[{Lattimer \& Prakash(2004)}]{lattimer2004physics}
Lattimer, J.~M., \& Prakash, M. 2004, Science, 304, 536

\bibitem[{{Lattimer} \& {Schramm}(1974)}]{latimer1974}
{Lattimer}, J.~M., \& {Schramm}, D.~N. 1974, \apjl, 192, L145

\bibitem[{{Law} {et~al.}(2024){Law}, {Bhardwaj}, {Burke-Spolaor}, {Thomas}, {Demorest}, \& {Bower}}]{Law2024}
{Law}, C.~J., {Bhardwaj}, M., {Burke-Spolaor}, S., {et~al.} 2024, The Astronomer's Telegram, 16701, 1

\bibitem[{Lee {et~al.}(2025)Lee, Caleb, Murphy, Lenc, Kaplan, Ferrario, Wadiasingh, Anumarlapudi, Hurley-Walker, Karambelkar, {et~al.}}]{lee2025emission}
Lee, Y., Caleb, M., Murphy, T., {et~al.} 2025, Nature Astronomy, 1

\bibitem[{{Leidi} {et~al.}(2023){Leidi}, {Andrassy}, {Higl}, {Edelmann}, \& {R{\"o}pke}}]{Leidi2023}
{Leidi}, G., {Andrassy}, R., {Higl}, J., {Edelmann}, P.~V.~F., \& {R{\"o}pke}, F.~K. 2023, \aap, 679, A132

\bibitem[{{Leung} {et~al.}(2021){Leung}, {Fuller}, \& {Nomoto}}]{Leung2021}
{Leung}, S.-C., {Fuller}, J., \& {Nomoto}, K. 2021, \apj, 915, 80

\bibitem[{{Levan} {et~al.}(2006){Levan}, {Wynn}, {Chapman}, {Davies}, {King}, {Priddey}, \& {Tanvir}}]{Levan_2006MNRAS.368L...1L}
{Levan}, A.~J., {Wynn}, G.~A., {Chapman}, R., {et~al.} 2006, \mnras, 368, L1

\bibitem[{{Levan} {et~al.}(2011){Levan}, {Tanvir}, {Cenko}, {Perley}, {Wiersema}, {Bloom}, {Fruchter}, {de Ugarte Postigo}, {O'Brien}, {Butler}, {van der Horst}, {Leloudas}, {Morgan}, {Misra}, {Bower}, {Farihi}, {Tunnicliffe}, {Modjaz}, {Silverman}, {Hjorth}, {Th{\"o}ne}, {Cucchiara}, {Cer{\'o}n}, {Castro-Tirado}, {Arnold}, {Bremer}, {Brodie}, {Carroll}, {Cooper}, {Curran}, {Cutri}, {Ehle}, {Forbes}, {Fynbo}, {Gorosabel}, {Graham}, {Hoffman}, {Guziy}, {Jakobsson}, {Kamble}, {Kerr}, {Kasliwal}, {Kouveliotou}, {Kocevski}, {Law}, {Nugent}, {Ofek}, {Poznanski}, {Quimby}, {Rol}, {Romanowsky}, {S{\'a}nchez-Ram{\'\i}rez}, {Schulze}, {Singh}, {van Spaandonk}, {Starling}, {Strom}, {Tello}, {Vaduvescu}, {Wheatley}, {Wijers}, {Winters}, \& {Xu}}]{Levan2011}
{Levan}, A.~J., {Tanvir}, N.~R., {Cenko}, S.~B., {et~al.} 2011, Science, 333, 199

\bibitem[{Levan {et~al.}(2024)Levan, Gompertz, Salafia, Bulla, Burns, Hotokezaka, Izzo, Lamb, Malesani, Oates, {et~al.}}]{levan2024heavy}
Levan, A.~J., Gompertz, B.~P., Salafia, O.~S., {et~al.} 2024, Nature, 626, 737

\bibitem[{Levin \& van Hoven(2011)}]{Levin_10.1111/j.1365-2966.2011.19515.x}
Levin, Y., \& van Hoven, M. 2011, Monthly Notices of the Royal Astronomical Society, 418, 659.
\newblock \url{https://doi.org/10.1111/j.1365-2966.2011.19515.x}

\bibitem[{{Li} {et~al.}(2021){Li}, {Lin}, {Xiong}, {Ge}, {Li}, {Li}, {Lu}, {Zhang}, {Tuo}, {Nang}, {Zhang}, {Xiao}, {Chen}, {Song}, {Xu}, {Liu}, {Jia}, {Cao}, {Qu}, {Zhang}, {Gu}, {Liao}, {Zhao}, {Tan}, {Nie}, {Zhao}, {Zheng}, {Zheng}, {Luo}, {Cai}, {Li}, {Xue}, {Bu}, {Chang}, {Chen}, {Chen}, {Chen}, {Chen}, {Chen}, {Cui}, {Cui}, {Deng}, {Dong}, {Du}, {Fu}, {Gao}, {Gao}, {Gao}, {Gu}, {Guan}, {Guo}, {Han}, {Huang}, {Huo}, {Jiang}, {Jiang}, {Jin}, {Jin}, {Kong}, {Li}, {Li}, {Li}, {Li}, {Li}, {Li}, {Li}, {Liang}, {Liu}, {Liu}, {Liu}, {Liu}, {Liu}, {Lu}, {Lu}, {Luo}, {Ma}, {Meng}, {Ou}, {Sai}, {Shang}, {Song}, {Sun}, {Tao}, {Wang}, {Wang}, {Wang}, {Wang}, {Wang}, {Wen}, {Wu}, {Wu}, {Wu}, {Xiao}, {Xu}, {Yang}, {Yang}, {Yang}, {Yang}, {Yi}, {Yin}, {You}, {Zhang}, {Zhang}, {Zhang}, {Zhang}, {Zhang}, {Zhang}, {Zhang}, {Zhang}, {Zhang}, {Zhang}, {Zhang}, {Zhang}, {Zhang}, {Zhang}, {Zhang}, {Zhang}, {Zhou}, {Zhou}, {Zhu}, {Zhu}, \& {Zhuang}}]{2021NatAs...5..378L}
{Li}, C.~K., {Lin}, L., {Xiong}, S.~L., {et~al.} 2021, Nature Astronomy, 5, 378

\bibitem[{{Li} {et~al.}(2022){Li}, {Ge}, {Lin}, {Zhang}, {Song}, {Cao}, {Zhang}, {Lu}, {Xu}, {Xiong}, {Tuo}, {Tan}, {Jiang}, {Qu}, {Zhang}, {Wang}, {Wang}, {Zhang}, {Zhang}, {Li}, {Liu}, {Li}, {Bu}, {Cai}, {Chen}, {Chen}, {Chang}, {Chen}, {Chen}, {Chen}, {Cui}, {Du}, {Gao}, {Gao}, {Gu}, {Guan}, {Guo}, {Han}, {Huang}, {Huo}, {Jia}, {Jin}, {Kong}, {Li}, {Li}, {Li}, {Li}, {Li}, {Li}, {Liang}, {Liao}, {Liu}, {Liu}, {Liu}, {Lu}, {Luo}, {Luo}, {Ma}, {Ma}, {Ma}, {Meng}, {Nang}, {Nie}, {Ou}, {Ren}, {Sai}, {Song}, {Sun}, {Tao}, {Wang}, {Wang}, {Wang}, {Wang}, {Wen}, {Wu}, {Wu}, {Wu}, {Xiao}, {Yang}, {Yang}, {Yi}, {Yin}, {You}, {Yu}, {Zhang}, {Zhang}, {Zhang}, {Zhang}, {Zhang}, {Zhang}, {Zhang}, {Zhao}, {Zhao}, {Zheng}, \& {Zhou}}]{2022ApJ...931...56L}
{Li}, X., {Ge}, M., {Lin}, L., {et~al.} 2022, \apj, 931, 56

\bibitem[{{Linial} \& {Sari}(2023)}]{2023ApJ...945...86L}
{Linial}, I., \& {Sari}, R. 2023, \apj, 945, 86

\bibitem[{{Linscott} \& {Erkes}(1980)}]{1980ApJ...236L.109L}
{Linscott}, I.~R., \& {Erkes}, J.~W. 1980, \apjl, 236, L109

\bibitem[{{Liodakis} {et~al.}(2018){Liodakis}, {Romani}, {Filippenko}, {Kiehlmann}, {Max-Moerbeck}, {Readhead}, \& {Zheng}}]{Liodakis2018}
{Liodakis}, I., {Romani}, R.~W., {Filippenko}, A.~V., {et~al.} 2018, \mnras, 480, 5517

\bibitem[{{Liodakis} {et~al.}(2022){Liodakis}, {Marscher}, {Agudo}, {Berdyugin}, {Bernardos}, {Bonnoli}, {Borman}, {Casadio}, {Casanova}, {Cavazzuti}, {Rodriguez Cavero}, {Di Gesu}, {Di Lalla}, {Donnarumma}, {Ehlert}, {Errando}, {Escudero}, {Garc{\'\i}a-Comas}, {Ag{\'\i}s-Gonz{\'a}lez}, {Husillos}, {Jormanainen}, {Jorstad}, {Kagitani}, {Kopatskaya}, {Kravtsov}, {Krawczynski}, {Lindfors}, {Larionova}, {Madejski}, {Marin}, {Marchini}, {Marshall}, {Morozova}, {Massaro}, {Masiero}, {Mawet}, {Middei}, {Millar-Blanchaer}, {Myserlis}, {Negro}, {Nilsson}, {O'Dell}, {Omodei}, {Pacciani}, {Paggi}, {Panopoulou}, {Peirson}, {Perri}, {Petrucci}, {Poutanen}, {Puccetti}, {Romani}, {Sakanoi}, {Savchenko}, {Sota}, {Tavecchio}, {Tinyanont}, {Vasilyev}, {Weaver}, {Zhovtan}, {Antonelli}, {Bachetti}, {Baldini}, {Baumgartner}, {Bellazzini}, {Bianchi}, {Bongiorno}, {Bonino}, {Brez}, {Bucciantini}, {Capitanio}, {Castellano}, {Ciprini}, {Costa}, {De Rosa}, {Del Monte}, {Di Marco}, {Doroshenko}, {Dov{\v{c}}iak}, {Enoto},
  {Evangelista}, {Fabiani}, {Ferrazzoli}, {Garcia}, {Gunji}, {Hayashida}, {Heyl}, {Iwakiri}, {Karas}, {Kitaguchi}, {Kolodziejczak}, {La Monaca}, {Latronico}, {Maldera}, {Manfreda}, {Marinucci}, {Matt}, {Mitsuishi}, {Mizuno}, {Muleri}, {Ng}, {Oppedisano}, {Papitto}, {Pavlov}, {Pesce-Rollins}, {Pilia}, {Possenti}, {Ramsey}, {Rankin}, {Ratheesh}, {Sgr{\'o}}, {Slane}, {Soffitta}, {Spandre}, {Tamagawa}, {Taverna}, {Tawara}, {Tennant}, {Thomas}, {Tombesi}, {Trois}, {Tsygankov}, {Turolla}, {Vink}, {Weisskopf}, {Wu}, {Xie}, \& {Zane}}]{Liodakis2022}
{Liodakis}, I., {Marscher}, A.~P., {Agudo}, I., {et~al.} 2022, \nat, 611, 677

\bibitem[{{Liu} {et~al.}(2023){Liu}, {Malyali}, {Krumpe}, {Homan}, {Goodwin}, {Grotova}, {Kawka}, {Rau}, {Merloni}, {Anderson}, {Miller-Jones}, {Markowitz}, {Ciroi}, {Di Mille}, {Schramm}, {Tang}, {Buckley}, {Gromadzki}, {Jin}, \& {Buchner}}]{2023A&A...669A..75L}
{Liu}, Z., {Malyali}, A., {Krumpe}, M., {et~al.} 2023, \aap, 669, A75

\bibitem[{Lorimer {et~al.}(2007)Lorimer, Bailes, McLaughlin, Narkevic, \& Crawford}]{lorimer2007bright}
Lorimer, D.~R., Bailes, M., McLaughlin, M.~A., Narkevic, D.~J., \& Crawford, F. 2007, Science, 318, 777

\bibitem[{Lynden-Bell \& Pringle(1974)}]{Lynden-Bell_1974}
Lynden-Bell, D., \& Pringle, J.~E. 1974, Monthly Notices of the Royal Astronomical Society, 168, 603.
\newblock \url{https://doi.org/10.1093/mnras/168.3.603}

\bibitem[{{Lyutikov}(2021)}]{2021ApJ...922..166L}
{Lyutikov}, M. 2021, \apj, 922, 166

\bibitem[{{Macquart} {et~al.}(2020){Macquart}, {Prochaska}, {McQuinn}, {Bannister}, {Bhandari}, {Day}, {Deller}, {Ekers}, {James}, {Marnoch}, {Os{\l}owski}, {Phillips}, {Ryder}, {Scott}, {Shannon}, \& {Tejos}}]{Macquart2020}
{Macquart}, J.~P., {Prochaska}, J.~X., {McQuinn}, M., {et~al.} 2020, \nat, 581, 391

\bibitem[{Macquet {et~al.}(2021)Macquet, Bizouard, Burns, Christensen, Coughlin, Wadiasingh, \& Younes}]{Macquet_2021}
Macquet, A., Bizouard, M.~A., Burns, E., {et~al.} 2021, The Astrophysical Journal, 918, 80.
\newblock \url{https://dx.doi.org/10.3847/1538-4357/ac0efd}

\bibitem[{{Mao} {et~al.}(2024){Mao}, {Woodward}, {Herwig}, {Denissenkov}, {Blouin}, {Thompson}, \& {McDermott}}]{Mao2024}
{Mao}, H., {Woodward}, P., {Herwig}, F., {et~al.} 2024, \apj, 975, 271

\bibitem[{{Marcotulli} {et~al.}(2020){Marcotulli}, {Paliya}, {Ajello}, {Kaur}, {Marchesi}, {Rajagopal}, {Hartmann}, {Gasparrini}, {Ojha}, \& {Madejski}}]{Marcotulli_2020}
{Marcotulli}, L., {Paliya}, V., {Ajello}, M., {et~al.} 2020, \apj, 889, 164

\bibitem[{{Marcotulli} {et~al.}(2022){Marcotulli}, {Ajello}, {Urry}, {Paliya}, {Koss}, {Oh}, {Madejski}, {Ueda}, {Balokovi{\'c}}, {Trakhtenbrot}, {Ricci}, {Ricci}, {Stern}, {Harrison}, {Powell}, \& {BASS Collaboration}}]{Marcotulli_2022}
{Marcotulli}, L., {Ajello}, M., {Urry}, C.~M., {et~al.} 2022, \apj, 940, 77

\bibitem[{{Marcowith} {et~al.}(2016){Marcowith}, {Bret}, {Bykov}, {Dieckman}, {O'C Drury}, {Lemb{\`e}ge}, {Lemoine}, {Morlino}, {Murphy}, {Pelletier}, {Plotnikov}, {Reville}, {Riquelme}, {Sironi}, \& {Stockem Novo}}]{Marcowith2016}
{Marcowith}, A., {Bret}, A., {Bykov}, A., {et~al.} 2016, Reports on Progress in Physics, 79, 046901

\bibitem[{{Margalit} {et~al.}(2019){Margalit}, {Berger}, \& {Metzger}}]{Margalit2019}
{Margalit}, B., {Berger}, E., \& {Metzger}, B.~D. 2019, \apj, 886, 110

\bibitem[{{Margalit} \& {Metzger}(2018)}]{Margalit2018}
{Margalit}, B., \& {Metzger}, B.~D. 2018, \apjl, 868, L4

\bibitem[{{Margalit} \& {Quataert}(2021)}]{Margalit2021}
{Margalit}, B., \& {Quataert}, E. 2021, \apjl, 923, L14

\bibitem[{{Margutti} {et~al.}(2019){Margutti}, {Metzger}, {Chornock}, {Vurm}, {Roth}, {Grefenstette}, {Savchenko}, {Cartier}, {Steiner}, {Terreran}, {Margalit}, {Migliori}, {Milisavljevic}, {Alexand er}, {Bietenholz}, {Blanchard}, {Bozzo}, {Brethauer}, {Chilingarian}, {Coppejans}, {Ducci}, {Ferrigno}, {Fong}, {G{\"o}tz}, {Guidorzi}, {Hajela}, {Hurley}, {Kuulkers}, {Laurent}, {Mereghetti}, {Nicholl}, {Patnaude}, {Ubertini}, {Banovetz}, {Bartel}, {Berger}, {Coughlin}, {Eftekhari}, {Frederiks}, {Kozlova}, {Laskar}, {Svinkin}, {Drout}, {MacFadyen}, \& {Paterson}}]{Margutti2019}
{Margutti}, R., {Metzger}, B.~D., {Chornock}, R., {et~al.} 2019, \apj, 872, 18

\bibitem[{{Marin} {et~al.}(2024){Marin}, {Gianolli}, {Ingram}, {Kim}, {Marinucci}, {Tagliacozzo}, \& {Ursini}}]{Marin2024}
{Marin}, F., {Gianolli}, V.~E., {Ingram}, A., {et~al.} 2024, Galaxies, 12, 35

\bibitem[{Marranghello {et~al.}(2002)Marranghello, Vasconcellos, \& de~Freitas~Pacheco}]{marranghello2002phase}
Marranghello, G.~F., Vasconcellos, C.~A., \& de~Freitas~Pacheco, J.~A. 2002, Physical Review D, 66, 064027

\bibitem[{{Marscher} {et~al.}(2024){Marscher}, {Di Gesu}, {Jorstad}, {Kim}, {Liodakis}, {Middei}, \& {Tavecchio}}]{Marscher2024}
{Marscher}, A.~P., {Di Gesu}, L., {Jorstad}, S.~G., {et~al.} 2024, Galaxies, 12, 50

\bibitem[{{Marshall} {et~al.}(2024){Marshall}, {Liodakis}, {Marscher}, {Di Lalla}, {Jorstad}, {Kim}, {Middei}, {Negro}, {Omodei}, {Peirson}, {Perri}, {Puccetti}, {Laurenti}, {Agudo}, {Bonnoli}, {Berdyugin}, {Cavazzuti}, {Rodriguez Cavero}, {Donnarumma}, {Di Gesu}, {Jormanainen}, {Krawczynski}, {Lindfors}, {Madjeski}, {Marin}, {Massaro}, {Pacciani}, {Poutanen}, {Tavecchio}, {Kouch}, {Aceituno}, {Bernardos}, {Casanova}, {Garc{\'\i}a-Comas}, {Ag{\'\i}s-Gonz{\'a}lez}, {Husillos}, {Marchini}, {Sota}, {Blinov}, {Bourbah}, {Kielhmann}, {Kontopodis}, {Mandarakas}, {Romanopoulos}, {Skalidis}, {Vervelaki}, {Borman}, {Kopatskaya}, {Larionova}, {Morozova}, {Savchenko}, {Vasilyev}, {Zhovtan}, {Casadio}, {Escudero}, {Kramer}, {Myserlis}, {Trainou}, {Imazawa}, {Sasada}, {Fukazawa}, {Kawabata}, {Uemura}, {Mizuno}, {Nakaoka}, {Akitaya}, {Masiero}, {Mawet}, {Panopoulou}, {Tinyanont}, {Kagitani}, {Kravtsov}, {Sakanoi}, {Dattolo}, {Gurwell}, {Keating}, {Rao}, {Cheong}, {Jeong}, {Kang}, {Kim}, {Lee}, {Angelakis}, {Kraus},
  {Hales}, {Kameno}, {Kneissl}, {Messias}, {Nagai}, {Antonelli}, {Bachetti}, {Baldini}, {Baumgartner}, {Bellazzini}, {Bianchi}, {Bongiorno}, {Bonino}, {Brez}, {Bucciantini}, {Capitanio}, {Castellano}, {Chen}, {Ciprini}, {Costa}, {De Rosa}, {Del Monte}, {Di Marco}, {Doroshenko}, {Dov{\v{c}}iak}, {Ehlert}, {Enoto}, {Evangelista}, {Fabiani}, {Ferrazzoli}, {Garcia}, {Gunji}, {Hayashida}, {Heyl}, {Iwakiri}, {Kaaret}, {Karas}, {Kislat}, {Kitaguchi}, {Kolodziejczak}, {La Monaca}, {Latronico}, {Maldera}, {Manfreda}, {Marinucci}, {Matt}, {Mitsuishi}, {Muleri}, {Ng}, {O'Dell}, {Oppedisano}, {Papitto}, {Pavlov}, {Pesce-Rollins}, {Petrucci}, {Pilia}, {Possenti}, {Ramsey}, {Rankin}, {Ratheesh}, {Roberts}, {Romani}, {Sgr{\`o}}, {Slane}, {Soffitta}, {Spandre}, {Swartz}, {Tamagawa}, {Taverna}, {Tawara}, {Tennant}, {Thomas}, {Tombesi}, {Trois}, {Tsygankov}, {Turolla}, {Vink}, {Weisskopf}, {Wu}, {Xie}, \& {Zane}}]{Marshall2024}
{Marshall}, H.~L., {Liodakis}, I., {Marscher}, A.~P., {et~al.} 2024, \apj, 972, 74

\bibitem[{{Masini} {et~al.}(2016){Masini}, {Comastri}, {Balokovi{\'c}}, {Zaw}, {Puccetti}, {Ballantyne}, {Bauer}, {Boggs}, {Brandt}, {Brightman}, {Christensen}, {Craig}, {Gandhi}, {Hailey}, {Harrison}, {Koss}, {Madejski}, {Ricci}, {Rivers}, {Stern}, \& {Zhang}}]{Masini16}
{Masini}, A., {Comastri}, A., {Balokovi{\'c}}, M., {et~al.} 2016, \aap, 589, A59

\bibitem[{{Masterson} {et~al.}(2025){Masterson}, {Kara}, {Panagiotou}, {Alston}, {Chakraborty}, {Burdge}, {Ricci}, {Laha}, {Arcavi}, {Arcodia}, {Cenko}, {Fabian}, {Garc{\'\i}a}, {Giustini}, {Ingram}, {Kosec}, {Loewenstein}, {Meyer}, {Miniutti}, {Pinto}, {Remillard}, {Sadaula}, {Shuvo}, {Trakhtenbrot}, \& {Wang}}]{2025arXiv250101581M}
{Masterson}, M., {Kara}, E., {Panagiotou}, C., {et~al.} 2025, arXiv e-prints, arXiv:2501.01581

\bibitem[{{Mastrogiovanni} {et~al.}(2021){Mastrogiovanni}, {Duque}, {Chassande-Mottin}, {Daigne}, \& {Mochkovitch}}]{2021A&A...652A...1M}
{Mastrogiovanni}, S., {Duque}, R., {Chassande-Mottin}, E., {Daigne}, F., \& {Mochkovitch}, R. 2021, \aap, 652, A1

\bibitem[{{Mattsson} {et~al.}(2005){Mattsson}, {Schultz}, {Desjarlais}, {Mattsson}, \& {Leung}}]{2005MSMSE..13R...1M}
{Mattsson}, A.~E., {Schultz}, P.~A., {Desjarlais}, M.~P., {Mattsson}, T.~R., \& {Leung}, K. 2005, Modelling Simul. Mater. Sci. Eng., 13, R1

\bibitem[{{Maund} {et~al.}(2023){Maund}, {H{\"o}flich}, {Steele}, {Yang(杨轶)}, {Wiersema}, {Kobayashi}, {Jordana-Mitjans}, {Mundell}, {Gomboc}, {Guidorzi}, \& {Smith}}]{Maund2023}
{Maund}, J.~R., {H{\"o}flich}, P.~A., {Steele}, I.~A., {et~al.} 2023, \mnras, 521, 3323

\bibitem[{{Mauney} \& {Lazzati}(2016)}]{2016P&SS..133...31M}
{Mauney}, C.~M., \& {Lazzati}, D. 2016, \planss, 133, 31

\bibitem[{{Mauney} \& {Lazzati}(2018)}]{2018MolAs..12....1M}
---. 2018, Molecular Astrophysics, 12, 1

\bibitem[{{McHardy} {et~al.}(2018){McHardy}, {Connolly}, {Horne}, {Cackett}, {Gelbord}, {Peterson}, {Pahari}, {Gehrels}, {Goad}, {Lira}, {Arevalo}, {Baldi}, {Brandt}, {Breedt}, {Chand}, {Dewangan}, {Done}, {Elvis}, {Emmanoulopoulos}, {Fausnaugh}, {Kaspi}, {Kochanek}, {Korista}, {Papadakis}, {Rao}, {Uttley}, {Vestergaard}, \& {Ward}}]{McHardy_2018}
{McHardy}, I.~M., {Connolly}, S.~D., {Horne}, K., {et~al.} 2018, \mnras, 480, 2881

\bibitem[{{Mckinven} {et~al.}(2024){Mckinven}, {Bhardwaj}, {Eftekhari}, {Kilpatrick}, {Kirichenko}, {Pal}, {Cook}, {Gaensler}, {Giri}, {Kaspi}, {Michilli}, {Nimmo}, {Pearlman}, {Pleunis}, {Sand}, {Stairs}, {Andersen}, {Andrew}, {Bandura}, {Brar}, {Cassanelli}, {Chatterjee}, {Curtin}, {Dong}, {Eadie}, {Fonseca}, {Ibik}, {Kaczmarek}, {Kharel}, {Lazda}, {Leung}, {Li}, {Main}, {Masui}, {Mena-Parra}, {Ng}, {Pandhi}, {Shivraj Patil}, {Prochaska}, {Rafiei-Ravandi}, {Scholz}, {Shah}, {Shin}, \& {Smith}}]{Mckinven2024}
{Mckinven}, R., {Bhardwaj}, M., {Eftekhari}, T., {et~al.} 2024, arXiv e-prints, arXiv:2402.09304

\bibitem[{{Medvedev} \& {Loeb}(1999)}]{Medvedev1999ApJ...526..697M}
{Medvedev}, M.~V., \& {Loeb}, A. 1999, \apj, 526, 697

\bibitem[{{Meegan} {et~al.}(2009){Meegan}, {Lichti}, {Bhat}, {Bissaldi}, {Briggs}, {Connaughton}, {Diehl}, {Fishman}, {Greiner}, {Hoover}, {van der Horst}, {von Kienlin}, {Kippen}, {Kouveliotou}, {McBreen}, {Paciesas}, {Preece}, {Steinle}, {Wallace}, {Wilson}, \& {Wilson-Hodge}}]{meegan2009}
{Meegan}, C., {Lichti}, G., {Bhat}, P.~N., {et~al.} 2009, \apj, 702, 791

\bibitem[{{Meli} {et~al.}(2023){Meli}, {Nishikawa}, {K{\"o}hn}, {Du{\c{t}}an}, {Mizuno}, {Kobzar}, {MacDonald}, {G{\'o}mez}, \& {Hirotani}}]{Meli2023}
{Meli}, A., {Nishikawa}, K., {K{\"o}hn}, C., {et~al.} 2023, \mnras, 519, 5410

\bibitem[{{Melrose}(1978)}]{Melrose1978}
{Melrose}, D.~B. 1978, \apj, 225, 557

\bibitem[{Men {et~al.}(2025)Men, McSweeney, Hurley-Walker, Barr, \& Stappers}]{doi:10.1126/sciadv.adp6351}
Men, Y., McSweeney, S., Hurley-Walker, N., Barr, E., \& Stappers, B. 2025, Science Advances, 11, eadp6351.
\newblock \url{https://www.science.org/doi/abs/10.1126/sciadv.adp6351}

\bibitem[{{Mereghetti} {et~al.}(2015){Mereghetti}, {Pons}, \& {Melatos}}]{Mereghetti_2015SSRv..191..315M}
{Mereghetti}, S., {Pons}, J.~A., \& {Melatos}, A. 2015, \ssr, 191, 315

\bibitem[{{Mereghetti} {et~al.}(2020){Mereghetti}, {Savchenko}, {Ferrigno}, {G{\"o}tz}, {Rigoselli}, {Tiengo}, {Bazzano}, {Bozzo}, {Coleiro}, {Courvoisier}, {Doyle}, {Goldwurm}, {Hanlon}, {Jourdain}, {von Kienlin}, {Lutovinov}, {Martin-Carrillo}, {Molkov}, {Natalucci}, {Onori}, {Panessa}, {Rodi}, {Rodriguez}, {S{\'a}nchez-Fern{\'a}ndez}, {Sunyaev}, \& {Ubertini}}]{2020ApJ...898L..29M}
{Mereghetti}, S., {Savchenko}, V., {Ferrigno}, C., {et~al.} 2020, \apjl, 898, L29

\bibitem[{{Meszaros}(1992)}]{1992herm.book.....M}
{Meszaros}, P. 1992, {High-energy radiation from magnetized neutron stars} (University of Chicago press)

\bibitem[{{Metzger}(2022)}]{Metzger2022}
{Metzger}, B.~D. 2022, \apj, 932, 84

\bibitem[{{Metzger} {et~al.}(2019){Metzger}, {Margalit}, \& {Sironi}}]{Metzger2019}
{Metzger}, B.~D., {Margalit}, B., \& {Sironi}, L. 2019, \mnras, 485, 4091

\bibitem[{{Metzger} {et~al.}(2008){Metzger}, {Quataert}, \& {Thompson}}]{Metzger_2008MNRAS.385.1455M}
{Metzger}, B.~D., {Quataert}, E., \& {Thompson}, T.~A. 2008, \mnras, 385, 1455

\bibitem[{{Michilli} {et~al.}(2018){Michilli}, {Seymour}, {Hessels}, {Spitler}, {Gajjar}, {Archibald}, {Bower}, {Chatterjee}, {Cordes}, {Gourdji}, {Heald}, {Kaspi}, {Law}, {Sobey}, {Adams}, {Bassa}, {Bogdanov}, {Brinkman}, {Demorest}, {Fernandez}, {Hellbourg}, {Lazio}, {Lynch}, {Maddox}, {Marcote}, {McLaughlin}, {Paragi}, {Ransom}, {Scholz}, {Siemion}, {Tendulkar}, {van Rooy}, {Wharton}, \& {Whitlow}}]{Michilli2018}
{Michilli}, D., {Seymour}, A., {Hessels}, J.~W.~T., {et~al.} 2018, \nat, 553, 182

\bibitem[{{Middei} {et~al.}(2023){Middei}, {Liodakis}, {Perri}, {Puccetti}, {Cavazzuti}, {Di Gesu}, {Ehlert}, {Madejski}, {Marscher}, {Marshall}, {Muleri}, {Negro}, {Jorstad}, {Ag{\'\i}s-Gonz{\'a}lez}, {Agudo}, {Bonnoli}, {Bernardos}, {Casanova}, {Garc{\'\i}a-Comas}, {Husillos}, {Marchini}, {Sota}, {Kouch}, {Lindfors}, {Borman}, {Kopatskaya}, {Larionova}, {Morozova}, {Savchenko}, {Vasilyev}, {Zhovtan}, {Casadio}, {Escudero}, {Myserlis}, {Hales}, {Kameno}, {Kneissl}, {Messias}, {Nagai}, {Blinov}, {Bourbah}, {Kiehlmann}, {Kontopodis}, {Mandarakas}, {Romanopoulos}, {Skalidis}, {Vervelaki}, {Masiero}, {Mawet}, {Millar-Blanchaer}, {Panopoulou}, {Tinyanont}, {Berdyugin}, {Kagitani}, {Kravtsov}, {Sakanoi}, {Imazawa}, {Sasada}, {Fukazawa}, {Kawabata}, {Uemura}, {Mizuno}, {Nakaoka}, {Akitaya}, {Gurwell}, {Rao}, {Di Lalla}, {Cibrario}, {Donnarumma}, {Kim}, {Omodei}, {Pacciani}, {Poutanen}, {Tavecchio}, {Antonelli}, {Bachetti}, {Baldini}, {Baumgartner}, {Bellazzini}, {Bianchi}, {Bongiorno}, {Bonino}, {Brez},
  {Bucciantini}, {Capitanio}, {Castellano}, {Ciprini}, {Costa}, {De Rosa}, {Del Monte}, {Di Marco}, {Doroshenko}, {Dov{\v{c}}iak}, {Enoto}, {Evangelista}, {Fabiani}, {Ferrazzoli}, {Garcia}, {Gunji}, {Hayashida}, {Heyl}, {Iwakiri}, {Karas}, {Kitaguchi}, {Kolodziejczak}, {Krawczynski}, {La Monaca}, {Latronico}, {Maldera}, {Manfreda}, {Marin}, {Marinucci}, {Massaro}, {Matt}, {Mitsuishi}, {Ng}, {O'Dell}, {Oppedisano}, {Papitto}, {Pavlov}, {Peirson}, {Pesce-Rollins}, {Petrucci}, {Pilia}, {Possenti}, {Ramsey}, {Rankin}, {Ratheesh}, {Romani}, {Sgr{\'o}}, {Slane}, {Soffitta}, {Spandre}, {Tamagawa}, {Taverna}, {Tawara}, {Tennant}, {Thomas}, {Tombesi}, {Trois}, {Tsygankov}, {Turolla}, {Vink}, {Weisskopf}, {Wu}, {Xie}, \& {Zane}}]{Middei2023}
{Middei}, R., {Liodakis}, I., {Perri}, M., {et~al.} 2023, \apjl, 942, L10

\bibitem[{{Miniutti} \& {Fabian}(2004)}]{Miniutti_2004}
{Miniutti}, G., \& {Fabian}, A.~C. 2004, \mnras, 349, 1435

\bibitem[{{Miyaji} \& {Nomoto}(1987)}]{MiNo87}
{Miyaji}, S., \& {Nomoto}, K. 1987, \apj, 318, 307

\bibitem[{Mohan {et~al.}(2022)Mohan, Lubbers, Chertkov, \& Livescu}]{mohan2020div}
Mohan, A.~T., Lubbers, N., Chertkov, M., \& Livescu, D. 2022, Physical Review Fluids, 8, 014604

\bibitem[{{Mooley} {et~al.}(2018){Mooley}, {Deller}, {Gottlieb}, {Nakar}, {Hallinan}, {Bourke}, {Frail}, {Horesh}, {Corsi}, \& {Hotokezaka}}]{2018Natur.561..355M}
{Mooley}, K.~P., {Deller}, A.~T., {Gottlieb}, O., {et~al.} 2018, \nat, 561, 355

\bibitem[{{Morales} \& {Horowitz}(2024)}]{2024arXiv240914482M}
{Morales}, J.~A., \& {Horowitz}, C.~J. 2024, arXiv e-prints, arXiv:2409.14482

\bibitem[{{Moroianu} {et~al.}(2023){Moroianu}, {Wen}, {James}, {Ai}, {Kovalam}, {Panther}, \& {Zhang}}]{Moroianu2023}
{Moroianu}, A., {Wen}, L., {James}, C.~W., {et~al.} 2023, Nature Astronomy, 7, 579

\bibitem[{Most \& Quataert(2023)}]{Most:2023sft}
Most, E.~R., \& Quataert, E. 2023, Astrophys. J. Lett., 947, L15

\bibitem[{{Murase} {et~al.}(2024){Murase}, {Karwin}, {Kimura}, {Ajello}, \& {Buson}}]{Murase_2024}
{Murase}, K., {Karwin}, C.~M., {Kimura}, S.~S., {Ajello}, M., \& {Buson}, S. 2024, \apjl, 961, L34

\bibitem[{{Murray} {et~al.}(1995){Murray}, {Chiang}, {Grossman}, \& {Voit}}]{Murray_1995}
{Murray}, N., {Chiang}, J., {Grossman}, S.~A., \& {Voit}, G.~M. 1995, \apj, 451, 498

\bibitem[{{Narayan} {et~al.}(1998){Narayan}, {Mahadevan}, \& {Quataert}}]{Narayan_1998}
{Narayan}, R., {Mahadevan}, R., \& {Quataert}, E. 1998, in Theory of Black Hole Accretion Disks, ed. M.~A. {Abramowicz}, G.~{Bj{\"o}rnsson}, \& J.~E. {Pringle}, 148--182

\bibitem[{{National Academies of Sciences}(2020)}]{national2020manipulating}
{National Academies of Sciences}. 2020, Division on Engineering and Physical Sciences, Board on Physics and Committee on Decadal Assessment and Outlook Report on Atomic and Optical Science, Manipulating Quantum Systems: An Assessment of Atomic, Molecular, and Optical Physics in the United States (National Academies Press)

\bibitem[{{National Academies of Sciences, Engineering, and Medicine} {et~al.}(2020)}]{national2020plasma}
{National Academies of Sciences, Engineering, and Medicine}, {et~al.} 2020, Plasma science: enabling technology, sustainability, security, and exploration (National Academies Press)

\bibitem[{{National Academies of Sciences, Engineering, and Medicine} {et~al.}(2021)}]{national2021pathways}
---. 2021, Pathways to Discovery in Astronomy and Astrophysics for the 2020s (National Academies Press)

\bibitem[{{National Academies of Sciences, Engineering, and Medicine} {et~al.}(2023)}]{national2023fundamental}
---. 2023, Fundamental Research in High Energy Density Science (National Academies Press)

\bibitem[{{National Research Council} {et~al.}(2003){National Research Council}, on~Engineering, Sciences, on~Physics, \& on~the Physics of~the Universe}]{national2003connecting}
{National Research Council}, on~Engineering, D., Sciences, P., on~Physics, B., \& on~the Physics of~the Universe, C. 2003, Connecting quarks with the cosmos: eleven science questions for the new century (National Academies Press)

\bibitem[{{Nave} {et~al.}(2019){Nave}, {Barklem}, {Belmonte}, {Brickhouse}, {Butler}, {Cashman}, {Chatzikos}, {Cowley}, {Den Hartog}, {Federman}, {Ferland}, {Fogle}, {Hartman}, {Guzman}, {Heap}, {Kerber}, {Kramida}, {Kulkarni}, {Lawler}, {Marler}, {Nahar}, {Pickering}, {Quinet}, {Ralchenko}, {Savin}, {Sneden}, {Takacs}, {Wahlgren}, {Webb}, {Wiseman}, \& {Wood}}]{Nave_2019}
{Nave}, G., {Barklem}, P., {Belmonte}, M.~T., {et~al.} 2019, in Bulletin of the American Astronomical Society, Vol.~51, 1

\bibitem[{{Nayana} \& {Chandra}(2021)}]{Nayana2021}
{Nayana}, A.~J., \& {Chandra}, P. 2021, \apjl, 912, L9

\bibitem[{Negro {et~al.}(2024)Negro, Younes, Wadiasingh, Burns, Trigg, \& Baring}]{negro2024role}
Negro, M., Younes, G., Wadiasingh, Z., {et~al.} 2024, Frontiers in Astronomy and Space Sciences, 11, 1388953

\bibitem[{{Nicholl} {et~al.}(2024){Nicholl}, {Pasham}, {Mummery}, {Guolo}, {Gendreau}, {Dewangan}, {Ferrara}, {Remillard}, {Bonnerot}, {Chakraborty}, {Hajela}, {Dhillon}, {Gillan}, {Greenwood}, {Huber}, {Janiuk}, {Salvesen}, {van Velzen}, {Aamer}, {Alexander}, {Angus}, {Arzoumanian}, {Auchettl}, {Berger}, {de Boer}, {Cendes}, {Chambers}, {Chen}, {Chornock}, {Fulton}, {Gao}, {Gillanders}, {Gomez}, {Gompertz}, {Fabian}, {Herman}, {Ingram}, {Kara}, {Laskar}, {Lawrence}, {Lin}, {Lowe}, {Magnier}, {Margutti}, {McGee}, {Minguez}, {Moore}, {Nathan}, {Oates}, {Patra}, {Ramsden}, {Ravi}, {Ridley}, {Sheng}, {Smartt}, {Smith}, {Srivastav}, {Stein}, {Stevance}, {Turner}, {Wainscoat}, {Weston}, {Wevers}, \& {Young}}]{2024Natur.634..804N}
{Nicholl}, M., {Pasham}, D.~R., {Mummery}, A., {et~al.} 2024, \nat, 634, 804

\bibitem[{{Nimmo} {et~al.}(2022){Nimmo}, {Hessels}, {Kirsten}, {Keimpema}, {Cordes}, {Snelders}, {Hewitt}, {Karuppusamy}, {Archibald}, {Bezrukovs}, {Bhardwaj}, {Blaauw}, {Buttaccio}, {Cassanelli}, {Conway}, {Corongiu}, {Feiler}, {Fonseca}, {Forss{\'e}n}, {Gawro{\'n}ski}, {Giroletti}, {Kharinov}, {Leung}, {Lindqvist}, {Maccaferri}, {Marcote}, {Masui}, {Mckinven}, {Melnikov}, {Michilli}, {Mikhailov}, {Ng}, {Orbidans}, {Ould-Boukattine}, {Paragi}, {Pearlman}, {Petroff}, {Rahman}, {Scholz}, {Shin}, {Smith}, {Stairs}, {Surcis}, {Tendulkar}, {Vlemmings}, {Wang}, {Yang}, \& {Yuan}}]{Nimmo2022}
{Nimmo}, K., {Hessels}, J.~W.~T., {Kirsten}, F., {et~al.} 2022, Nature Astronomy, 6, 393

\bibitem[{{Nimmo} {et~al.}(2024){Nimmo}, {Pleunis}, {Beniamini}, {Kumar}, {Lanman}, {Li}, {Main}, {Sammons}, {Andrew}, {Bhardwaj}, {Chatterjee}, {Curtin}, {Fonseca}, {Gaensler}, {Joseph}, {Kader}, {Kaspi}, {Lazda}, {Leung}, {Masui}, {Mckinven}, {Michilli}, {Pandhi}, {Pearlman}, {Rafiei-Ravandi}, {Sand}, {Shin}, {Smith}, \& {Stairs}}]{Nimmo2024}
{Nimmo}, K., {Pleunis}, Z., {Beniamini}, P., {et~al.} 2024, arXiv e-prints, arXiv:2406.11053

\bibitem[{{Nishikawa} {et~al.}(2020){Nishikawa}, {Mizuno}, {G{\'o}mez}, {Du{\c{t}}an}, {Niemiec}, {Kobzar}, {MacDonald}, {Meli}, {Pohl}, \& {Hirotani}}]{Nishikawa2020}
{Nishikawa}, K., {Mizuno}, Y., {G{\'o}mez}, J.~L., {et~al.} 2020, \mnras, 493, 2652

\bibitem[{Nishikawa {et~al.}(2016)Nishikawa, Frederiksen, Nordlund, Mizuno, Hardee, Niemiec, G{\'o}mez, Pe’er, Du{\c{t}}an, Meli, {et~al.}}]{nishikawa2016evolution}
Nishikawa, K.-I., Frederiksen, J.~T., Nordlund, {\AA}., {et~al.} 2016, The Astrophysical Journal, 820, 94

\bibitem[{{Niu} {et~al.}(2022){Niu}, {Aggarwal}, {Li}, {Zhang}, {Chatterjee}, {Tsai}, {Yu}, {Law}, {Burke-Spolaor}, {Cordes}, {Zhang}, {Ocker}, {Yao}, {Wang}, {Feng}, {Niino}, {Bochenek}, {Cruces}, {Connor}, {Jiang}, {Dai}, {Luo}, {Li}, {Miao}, {Niu}, {Anna-Thomas}, {Sydnor}, {Stern}, {Wang}, {Yuan}, {Yue}, {Zhou}, {Yan}, {Zhu}, \& {Zhang}}]{Niu2022}
{Niu}, C.~H., {Aggarwal}, K., {Li}, D., {et~al.} 2022, \nat, 606, 873

\bibitem[{{Nomoto} \& {Kondo}(1991)}]{Nomoto_1991ApJ...367L..19N}
{Nomoto}, K., \& {Kondo}, Y. 1991, \apjl, 367, L19

\bibitem[{{Norman}(1997)}]{1997ASPC..123....3N}
{Norman}, M.~L. 1997, in Astronomical Society of the Pacific Conference Series, Vol.~12, Computational Astrophysics; 12th Kingston Meeting on Theoretical Astrophysics, ed. D.~A. {Clarke} \& M.~J. {West}, 3

\bibitem[{Nouri {et~al.}(2019{\natexlab{a}})Nouri, Givi, \& Livescu}]{NGL19}
Nouri, A., Givi, P., \& Livescu, D. 2019{\natexlab{a}}, Progress in Aerospace Sciences, 108, 156

\bibitem[{Nouri {et~al.}(2024)Nouri, Liu, Givi, Babaee, \& Livescu}]{nouri24ske}
Nouri, A., Liu, Y., Givi, P., Babaee, H., \& Livescu, D. 2024, Astrophysical Journal Supplement Series, 272, 34

\bibitem[{Nouri {et~al.}(2019{\natexlab{b}})Nouri, Givi, \& Livescu}]{nouri2019modeling}
Nouri, A.~G., Givi, P., \& Livescu, D. 2019{\natexlab{b}}, Progress in Aerospace Sciences, 108, 156

\bibitem[{{Novikov} \& {Thorne}(1973)}]{Novikov_1973}
{Novikov}, I.~D., \& {Thorne}, K.~S. 1973, in Black Holes (Les Astres Occlus), ed. C.~{Dewitt} \& B.~S. {Dewitt}, 343--450

\bibitem[{NSAC(2023)}]{NSAC-LRP-2023}
NSAC. 2023, A new era of dicovery - the 2023 long range plan for nuclear science, \url{https://www.osti.gov/servlets/purl/2280968/},  OSTI

\bibitem[{{Olausen} \& {Kaspi}(2014)}]{Olausen_2014ApJS..212....6O}
{Olausen}, S.~A., \& {Kaspi}, V.~M. 2014, \apjs, 212, 6

\bibitem[{{Pahari} {et~al.}(2020){Pahari}, {McHardy}, {Vincentelli}, {Cackett}, {Peterson}, {Goad}, {G{\"u}ltekin}, \& {Horne}}]{Pahari_2020}
{Pahari}, M., {McHardy}, I.~M., {Vincentelli}, F., {et~al.} 2020, \mnras, 494, 4057

\bibitem[{{Panessa} {et~al.}(2020){Panessa}, {Castangia}, {Malizia}, {Bassani}, {Tarchi}, {Bazzano}, \& {Ubertini}}]{Panessa20}
{Panessa}, F., {Castangia}, P., {Malizia}, A., {et~al.} 2020, \aap, 641, A162

\bibitem[{Pang {et~al.}(2021)Pang, Tews, Coughlin, Bulla, Van Den~Broeck, \& Dietrich}]{pang2021nuclear}
Pang, P.~T., Tews, I., Coughlin, M.~W., {et~al.} 2021, The Astrophysical Journal, 922, 14

\bibitem[{{Papadakis} {et~al.}(2022){Papadakis}, {Dov{\v{c}}iak}, \& {Kammoun}}]{Papadakis_2022}
{Papadakis}, I.~E., {Dov{\v{c}}iak}, M., \& {Kammoun}, E.~S. 2022, \aap, 666, A11

\bibitem[{{Pasham} {et~al.}(2024{\natexlab{a}}){Pasham}, {Coughlin}, {Guolo}, {Wevers}, {Nixon}, {Hinkle}, \& {Bandopadhyay}}]{2024ApJ...971L..31P}
{Pasham}, D., {Coughlin}, E.~R., {Guolo}, M., {et~al.} 2024{\natexlab{a}}, \apjl, 971, L31

\bibitem[{{Pasham} {et~al.}(2015){Pasham}, {Cenko}, {Levan}, {Bower}, {Horesh}, {Brown}, {Dolan}, {Wiersema}, {Filippenko}, {Fruchter}, {Greiner}, {O'Brien}, {Page}, {Rau}, \& {Tanvir}}]{Pasham2015ApJ...805...68P}
{Pasham}, D.~R., {Cenko}, S.~B., {Levan}, A.~J., {et~al.} 2015, \apj, 805, 68

\bibitem[{{Pasham} {et~al.}(2019){Pasham}, {Remillard}, {Fragile}, {Franchini}, {Stone}, {Lodato}, {Homan}, {Chakrabarty}, {Baganoff}, {Steiner}, {Coughlin}, \& {Pasham}}]{Pasham2019Sci...363..531P}
{Pasham}, D.~R., {Remillard}, R.~A., {Fragile}, P.~C., {et~al.} 2019, Science, 363, 531

\bibitem[{{Pasham} {et~al.}(2023){Pasham}, {Lucchini}, {Laskar}, {Gompertz}, {Srivastav}, {Nicholl}, {Smartt}, {Miller-Jones}, {Alexander}, {Fender}, {Smith}, {Fulton}, {Dewangan}, {Gendreau}, {Coughlin}, {Rhodes}, {Horesh}, {van Velzen}, {Sfaradi}, {Guolo}, {Castro Segura}, {Aamer}, {Anderson}, {Arcavi}, {Brennan}, {Chambers}, {Charalampopoulos}, {Chen}, {Clocchiatti}, {de Boer}, {Dennefeld}, {Ferrara}, {Galbany}, {Gao}, {Gillanders}, {Goodwin}, {Gromadzki}, {Huber}, {Jonker}, {Joshi}, {Kara}, {Killestein}, {Kosec}, {Kocevski}, {Leloudas}, {Lin}, {Margutti}, {Mattila}, {Moore}, {M{\"u}ller-Bravo}, {Ngeow}, {Oates}, {Onori}, {Pan}, {Perez-Torres}, {Rani}, {Remillard}, {Ridley}, {Schulze}, {Sheng}, {Shingles}, {Smith}, {Steiner}, {Wainscoat}, {Wevers}, \& {Yang}}]{Pasham2023}
{Pasham}, D.~R., {Lucchini}, M., {Laskar}, T., {et~al.} 2023, Nature Astronomy, 7, 88

\bibitem[{{Pasham} {et~al.}(2024{\natexlab{b}}){Pasham}, {Tombesi}, {Sukov{\'a}}, {Zaja{\v{c}}ek}, {Rakshit}, {Coughlin}, {Kosec}, {Karas}, {Masterson}, {Mummery}, {Holoien}, {Guolo}, {Hinkle}, {Ripperda}, {Witzany}, {Shappee}, {Kara}, {Horesh}, {van Velzen}, {Sfaradi}, {Kaplan}, {Burger}, {Murphy}, {Remillard}, {Steiner}, {Wevers}, {Arcodia}, {Buchner}, {Merloni}, {Malyali}, {Fabian}, {Fausnaugh}, {Daylan}, {Altamirano}, {Payne}, \& {Ferraraa}}]{2024SciA...10J8898P}
{Pasham}, D.~R., {Tombesi}, F., {Sukov{\'a}}, P., {et~al.} 2024{\natexlab{b}}, Science Advances, 10, eadj8898

\bibitem[{Patel {et~al.}(2025{\natexlab{a}})Patel, Metzger, Cehula, Burns, Goldberg, \& Thompson}]{patel2025direct}
Patel, A., Metzger, B.~D., Cehula, J., {et~al.} 2025{\natexlab{a}}, arXiv preprint arXiv:2501.09181

\bibitem[{Patel {et~al.}(2025{\natexlab{b}})Patel, Metzger, Goldberg, Cehula, Thompson, \& Renzo}]{patel2025r}
Patel, A., Metzger, B.~D., Goldberg, J.~A., {et~al.} 2025{\natexlab{b}}, arXiv preprint arXiv:2501.17253

\bibitem[{{Pe{\~n}il} {et~al.}(2024){Pe{\~n}il}, {Westernacher-Schneider}, {Ajello}, {Dom{\'\i}nguez}, {Buson}, {Otero-Santos}, {Marcotulli}, {Torres-Alb{\`a}}, \& {Zrake}}]{Penil_2024}
{Pe{\~n}il}, P., {Westernacher-Schneider}, J.~R., {Ajello}, M., {et~al.} 2024, \mnras, 527, 10168

\bibitem[{{Pearlman} {et~al.}(2024){Pearlman}, {Scholz}, {Bethapudi}, {Hessels}, {Kaspi}, {Kirsten}, {Nimmo}, {Spitler}, {Fonseca}, {Meyers}, {Stairs}, {Tan}, {Bhardwaj}, {Chatterjee}, {Cook}, {Curtin}, {Dong}, {Eftekhari}, {Gaensler}, {G{\"u}ver}, {Kaczmarek}, {Leung}, {Masui}, {Michilli}, {Prince}, {Sand}, {Shin}, {Smith}, \& {Tendulkar}}]{Pearlman2024}
{Pearlman}, A.~B., {Scholz}, P., {Bethapudi}, S., {et~al.} 2024, Nature Astronomy, arXiv:arXiv:2308.10930

\bibitem[{{Pellegrino} {et~al.}(2022){Pellegrino}, {Howell}, {Vink{\'o}}, {Gangopadhyay}, {Xiang}, {Arcavi}, {Brown}, {Burke}, {Hiramatsu}, {Hosseinzadeh}, {Li}, {McCully}, {Misra}, {Newsome}, {Gonzalez}, {Pritchard}, {Valenti}, {Wang}, \& {Zhang}}]{Pellegrino2022}
{Pellegrino}, C., {Howell}, D.~A., {Vink{\'o}}, J., {et~al.} 2022, \apj, 926, 125

\bibitem[{Penrose(1969)}]{penrose1969gravitational}
Penrose, R. 1969

\bibitem[{{Perley} {et~al.}(2019){Perley}, {Mazzali}, {Yan}, {Cenko}, {Gezari}, {Taggart}, {Blagorodnova}, {Fremling}, {Mockler}, {Singh}, {Tominaga}, {Tanaka}, {Watson}, {Ahumada}, {Anupama}, {Ashall}, {Becerra}, {Bersier}, {Bhalerao}, {Bloom}, {Butler}, {Copperwheat}, {Coughlin}, {De}, {Drake}, {Duev}, {Frederick}, {Gonz{\'a}lez}, {Goobar}, {Heida}, {Ho}, {Horst}, {Hung}, {Itoh}, {Jencson}, {Kasliwal}, {Kawai}, {Khanam}, {Kulkarni}, {Kumar}, {Kumar}, {Kutyrev}, {Lee}, {Maeda}, {Mahabal}, {Murata}, {Neill}, {Ngeow}, {Penprase}, {Pian}, {Quimby}, {Ramirez-Ruiz}, {Richer}, {Rom{\'a}n-Z{\'u}{\~n}iga}, {Sahu}, {Srivastav}, {Socia}, {Sollerman}, {Tachibana}, {Taddia}, {Tinyanont}, {Troja}, {Ward}, {Wee}, \& {Yu}}]{Perley2019}
{Perley}, D.~A., {Mazzali}, P.~A., {Yan}, L., {et~al.} 2019, \mnras, 484, 1031

\bibitem[{{Perlmutter} {et~al.}(1999){Perlmutter}, {Aldering}, {Goldhaber}, {Knop}, {Nugent}, {Castro}, {Deustua}, {Fabbro}, {Goobar}, {Groom}, {Hook}, {Kim}, {Kim}, {Lee}, {Nunes}, {Pain}, {Pennypacker}, {Quimby}, {Lidman}, {Ellis}, {Irwin}, {McMahon}, {Ruiz-Lapuente}, {Walton}, {Schaefer}, {Boyle}, {Filippenko}, {Matheson}, {Fruchter}, {Panagia}, {Newberg}, {Couch}, \& {Project}}]{1999ApJ...517..565P}
{Perlmutter}, S., {Aldering}, G., {Goldhaber}, G., {et~al.} 1999, \apj, 517, 565

\bibitem[{{Perna} \& {Pons}(2011)}]{2011ApJ...727L..51P}
{Perna}, R., \& {Pons}, J.~A. 2011, \apjl, 727, L51

\bibitem[{{Petrucci} {et~al.}(2004){Petrucci}, {Maraschi}, {Haardt}, \& {Nandra}}]{Petrucci_2004}
{Petrucci}, P.~O., {Maraschi}, L., {Haardt}, F., \& {Nandra}, K. 2004, \aap, 413, 477

\bibitem[{{Petrucci} {et~al.}(2001){Petrucci}, {Merloni}, {Fabian}, {Haardt}, \& {Gallo}}]{Petrucci_2001}
{Petrucci}, P.~O., {Merloni}, A., {Fabian}, A., {Haardt}, F., \& {Gallo}, E. 2001, \mnras, 328, 501

\bibitem[{Pfister {et~al.}(2021)Pfister, Toscani, Wong, Dai, Lodato, \& Rossi}]{Pfister2021}
Pfister, H., Toscani, M., Wong, T. H.~T., {et~al.} 2021, Monthly Notices of the Royal Astronomical Society, 510, 2025–2040.
\newblock \url{http://dx.doi.org/10.1093/mnras/stab3387}

\bibitem[{Piazza {et~al.}(2022)Piazza, Willingale, \& Zuegel}]{mp3report.2022.arxiv}
Piazza, A.~D., Willingale, L., \& Zuegel, J.~D. 2022, Multi-Petawatt Physics Prioritization (MP3) Workshop Report, , , arXiv:2211.13187.
\newblock \url{https://arxiv.org/abs/2211.13187}

\bibitem[{Piekarewicz \& Fattoyev(2019)}]{piekarewicz2019neutron}
Piekarewicz, J., \& Fattoyev, F.~J. 2019, Physics Today, 72, 30

\bibitem[{{Pihajoki} {et~al.}(2017){Pihajoki}, {Rantala}, \& {Johansson}}]{Pihajoki:2017}
{Pihajoki}, P., {Rantala}, A., \& {Johansson}, P.~H. 2017, in IAU Symposium, Vol. 324, New Frontiers in Black Hole Astrophysics, ed. A.~{Gomboc}, 347--350

\bibitem[{Piro {et~al.}(2021)Piro, Bruni, Troja, O’Connor, Panessa, Ricci, Zhang, Burgay, Dichiara, Lee, Lotti, Niu, Pilia, Possenti, Trudu, Xu, Zhu, Kutyrev, \& Veilleux}]{Piro2021}
Piro, L., Bruni, G., Troja, E., {et~al.} 2021, \aap, 656, L15.
\newblock \url{http://dx.doi.org/10.1051/0004-6361/202141903}

\bibitem[{{Pleunis} {et~al.}(2021){Pleunis}, {Good}, {Kaspi}, {Mckinven}, {Ransom}, {Scholz}, {Bandura}, {Bhardwaj}, {Boyle}, {Brar}, {Cassanelli}, {Chawla}, {(Adam) Dong}, {Fonseca}, {Gaensler}, {Josephy}, {Kaczmarek}, {Leung}, {Lin}, {Masui}, {Mena-Parra}, {Michilli}, {Ng}, {Patel}, {Rafiei-Ravandi}, {Rahman}, {Sanghavi}, {Shin}, {Smith}, {Stairs}, \& {Tendulkar}}]{Pleunis2021}
{Pleunis}, Z., {Good}, D.~C., {Kaspi}, V.~M., {et~al.} 2021, \apj, 923, 1

\bibitem[{{Potekhin} {et~al.}(2025){Potekhin}, {Chugunov}, {Shchechilin}, \& {Gusakov}}]{2025JHEAp..45..116P}
{Potekhin}, A.~Y., {Chugunov}, A.~I., {Shchechilin}, N.~N., \& {Gusakov}, M.~E. 2025, Journal of High Energy Astrophysics, 45, 116

\bibitem[{{Prentice} {et~al.}(2018){Prentice}, {Maguire}, {Smartt}, {Magee}, {Schady}, {Sim}, {Chen}, {Clark}, {Colin}, {Fulton}, {McBrien}, {O'Neill}, {Smith}, {Ashall}, {Chambers}, {Denneau}, {Flewelling}, {Heinze}, {Holoien}, {Huber}, {Kochanek}, {Mazzali}, {Prieto}, {Rest}, {Shappee}, {Stalder}, {Stanek}, {Stritzinger}, {Thompson}, \& {Tonry}}]{Prentice2018}
{Prentice}, S.~J., {Maguire}, K., {Smartt}, S.~J., {et~al.} 2018, \apjl, 865, L3

\bibitem[{{Pursiainen} {et~al.}(2018){Pursiainen}, {Childress}, {Smith}, {Prajs}, {Sullivan}, {Davis}, {Foley}, {Asorey}, {Calcino}, {Carollo}, {Curtin}, {D'Andrea}, {Glazebrook}, {Gutierrez}, {Hinton}, {Hoormann}, {Inserra}, {Kessler}, {King}, {Kuehn}, {Lewis}, {Lidman}, {Macaulay}, {M{\"o}ller}, {Nichol}, {Sako}, {Sommer}, {Swann}, {Tucker}, {Uddin}, {Wiseman}, {Zhang}, {Abbott}, {Abdalla}, {Allam}, {Annis}, {Avila}, {Brooks}, {Buckley-Geer}, {Burke}, {Carnero Rosell}, {Carrasco Kind}, {Carretero}, {Castander}, {Cunha}, {Davis}, {De Vicente}, {Diehl}, {Doel}, {Eifler}, {Flaugher}, {Fosalba}, {Frieman}, {Garc{\'\i}a-Bellido}, {Gruen}, {Gruendl}, {Gutierrez}, {Hartley}, {Hollowood}, {Honscheid}, {James}, {Jeltema}, {Kuropatkin}, {Li}, {Lima}, {Maia}, {Martini}, {Menanteau}, {Ogando}, {Plazas}, {Roodman}, {Sanchez}, {Scarpine}, {Schindler}, {Smith}, {Soares-Santos}, {Sobreira}, {Suchyta}, {Swanson}, {Tarle}, {Tucker}, {Walker}, \& {DES Collaboration}}]{Pursiainen2018}
{Pursiainen}, M., {Childress}, M., {Smith}, M., {et~al.} 2018, \mnras, 481, 894

\bibitem[{Rastinejad {et~al.}(2022)Rastinejad, Gompertz, Levan, Fong, Nicholl, Lamb, Malesani, Nugent, Oates, Tanvir, {et~al.}}]{rastinejad2022kilonova}
Rastinejad, J.~C., Gompertz, B.~P., Levan, A.~J., {et~al.} 2022, Nature, 612, 223

\bibitem[{{Reardon} {et~al.}(2023){Reardon}, {Zic}, {Shannon}, {Hobbs}, {Bailes}, {Di Marco}, {Kapur}, {Rogers}, {Thrane}, {Askew}, {Bhat}, {Cameron}, {Cury{\l}o}, {Coles}, {Dai}, {Goncharov}, {Kerr}, {Kulkarni}, {Levin}, {Lower}, {Manchester}, {Mandow}, {Miles}, {Nathan}, {Os{\l}owski}, {Russell}, {Spiewak}, {Zhang}, \& {Zhu}}]{Reardon_2023}
{Reardon}, D.~J., {Zic}, A., {Shannon}, R.~M., {et~al.} 2023, \apjl, 951, L6

\bibitem[{Reed {et~al.}(2021)Reed, Fattoyev, Horowitz, \& Piekarewicz}]{reed2021implications}
Reed, B.~T., Fattoyev, F.~J., Horowitz, C.~J., \& Piekarewicz, J. 2021, Physical Review Letters, 126, 172503

\bibitem[{{Rees}(1988)}]{Rees1988Natur.333..523R}
{Rees}, M.~J. 1988, \nat, 333, 523

\bibitem[{Reifarth {et~al.}(2017)Reifarth, G\"obel, Heftrich, Weigand, Jurado, K\"appeler, \& Litvinov}]{RGH17}
Reifarth, R., G\"obel, K., Heftrich, T., {et~al.} 2017, Phys. Rev. Accel. Beams, 20, 044701

\bibitem[{{Reifarth} {et~al.}(2014){Reifarth}, {Lederer}, \& {K{\"a}ppeler}}]{reifarth2014}
{Reifarth}, R., {Lederer}, C., \& {K{\"a}ppeler}, F. 2014, Journal of Physics G Nuclear Physics, 41, 053101

\bibitem[{{Reines} {et~al.}(2013){Reines}, {Greene}, \& {Geha}}]{Reines_2013}
{Reines}, A.~E., {Greene}, J.~E., \& {Geha}, M. 2013, \apj, 775, 116

\bibitem[{{Reines} \& {Volonteri}(2015)}]{Reines_2015}
{Reines}, A.~E., \& {Volonteri}, M. 2015, \apj, 813, 82

\bibitem[{{Remington} {et~al.}(2006){Remington}, {Drake}, \& {Ryutov}}]{2006RvMP...78..755R}
{Remington}, B.~A., {Drake}, R.~P., \& {Ryutov}, D.~D. 2006, Reviews of Modern Physics, 78, 755

\bibitem[{Reynolds {et~al.}(2023)Reynolds, Kara, Mushotzky, Ptak, Koss, Williams, Allen, Bauer, Bautz, Bogadhee, {et~al.}}]{reynolds2023overview}
Reynolds, C.~S., Kara, E.~A., Mushotzky, R.~F., {et~al.} 2023, in UV, X-Ray, and Gamma-Ray Space Instrumentation for Astronomy XXIII, Vol. 12678, SPIE, 421--442

\bibitem[{{Rhoades} \& {Ruffini}(1974)}]{1974PhRvL..32..324R}
{Rhoades}, C.~E., \& {Ruffini}, R. 1974, \prl, 32, 324

\bibitem[{{Ricci} \& {Trakhtenbrot}(2023)}]{Ricci_2023b}
{Ricci}, C., \& {Trakhtenbrot}, B. 2023, Nature Astronomy, 7, 1282

\bibitem[{{Ricci} {et~al.}(2015){Ricci}, {Ueda}, {Koss}, {Trakhtenbrot}, {Bauer}, \& {Gandhi}}]{Ricci15}
{Ricci}, C., {Ueda}, Y., {Koss}, M.~J., {et~al.} 2015, \apjl, 815, L13

\bibitem[{{Ricci} {et~al.}(2017){Ricci}, {Trakhtenbrot}, {Koss}, {Ueda}, {Del Vecchio}, {Treister}, {Schawinski}, {Paltani}, {Oh}, {Lamperti}, {Berney}, {Gandhi}, {Ichikawa}, {Bauer}, {Ho}, {Asmus}, {Beckmann}, {Soldi}, {Balokovi{\'c}}, {Gehrels}, \& {Markwardt}}]{Ricci17_bassV}
{Ricci}, C., {Trakhtenbrot}, B., {Koss}, M.~J., {et~al.} 2017, \apjs, 233, 17

\bibitem[{{Ricci} {et~al.}(2023){Ricci}, {Chang}, {Kawamuro}, {Privon}, {Mushotzky}, {Trakhtenbrot}, {Laor}, {Koss}, {Smith}, {Gupta}, {Dimopoulos}, {Aalto}, \& {Ros}}]{Ricci_2023a}
{Ricci}, C., {Chang}, C.-S., {Kawamuro}, T., {et~al.} 2023, \apjl, 952, L28

\bibitem[{Richtmyer(1948)}]{richtmyer48}
Richtmyer, R. 1948, Los Alamos Scientific Laboratory Report, LA-671

\bibitem[{{Ridnaia} {et~al.}(2021){Ridnaia}, {Svinkin}, {Frederiks}, {Bykov}, {Popov}, {Aptekar}, {Golenetskii}, {Lysenko}, {Tsvetkova}, {Ulanov}, \& {Cline}}]{2021NatAs...5..372R}
{Ridnaia}, A., {Svinkin}, D., {Frederiks}, D., {et~al.} 2021, Nature Astronomy, 5, 372

\bibitem[{{Riess} {et~al.}(1998){Riess}, {Filippenko}, {Challis}, {Clocchiatti}, {Diercks}, {Garnavich}, {Gilliland}, {Hogan}, {Jha}, {Kirshner}, {Leibundgut}, {Phillips}, {Reiss}, {Schmidt}, {Schommer}, {Smith}, {Spyromilio}, {Stubbs}, {Suntzeff}, \& {Tonry}}]{1998AJ....116.1009R}
{Riess}, A.~G., {Filippenko}, A.~V., {Challis}, P., {et~al.} 1998, \aj, 116, 1009

\bibitem[{Riess {et~al.}(2022)Riess, Yuan, Macri, Scolnic, Brout, Casertano, Jones, Murakami, Anand, Breuval, {et~al.}}]{riess2022comprehensive}
Riess, A.~G., Yuan, W., Macri, L.~M., {et~al.} 2022, The Astrophysical journal letters, 934, L7

\bibitem[{{Rivera Sandoval} {et~al.}(2018){Rivera Sandoval}, {Maccarone}, {Corsi}, {Brown}, {Pooley}, \& {Wheeler}}]{RiveraSandoval2018}
{Rivera Sandoval}, L.~E., {Maccarone}, T.~J., {Corsi}, A., {et~al.} 2018, \mnras, 480, L146

\bibitem[{Roberts {et~al.}(2021)Roberts, Veres, Baring, Briggs, Kouveliotou, Bissaldi, Younes, Chastain, DeLaunay, Huppenkothen, {et~al.}}]{roberts_2021rapid}
Roberts, O., Veres, P., Baring, M., {et~al.} 2021, Nature, 589, 207

\bibitem[{{Rosseland}(1928)}]{1928MNRAS..89...49R}
{Rosseland}, S. 1928, \mnras, 89, 49

\bibitem[{{Rrapaj} \& {Reddy}(2016)}]{2016PhRvC..94d5805R}
{Rrapaj}, E., \& {Reddy}, S. 2016, \prc, 94, 045805

\bibitem[{{Ruszkowski} \& {Pfrommer}(2023)}]{RP2023}
{Ruszkowski}, M., \& {Pfrommer}, C. 2023, \aapr, 31, 4

\bibitem[{{Rybicki} \& {Lightman}(1979)}]{Rybicki_1979}
{Rybicki}, G.~B., \& {Lightman}, A.~P. 1979, {Radiative processes in astrophysics} (John Wiley \& Sons)

\bibitem[{{Ryutov} {et~al.}(1999){Ryutov}, {Drake}, {Kane}, {Liang}, {Remington}, \& {Wood-Vasey}}]{1999ApJ...518..821R}
{Ryutov}, D., {Drake}, R.~P., {Kane}, J., {et~al.} 1999, \apj, 518, 821

\bibitem[{{Ryutov} {et~al.}(2000){Ryutov}, {Drake}, \& {Remington}}]{2000ApJS..127..465R}
{Ryutov}, D.~D., {Drake}, R.~P., \& {Remington}, B.~A. 2000, \apjs, 127, 465

\bibitem[{{Saade} {et~al.}(2024){Saade}, {Kaaret}, {Liodakis}, \& {Ehlert}}]{Saade2024}
{Saade}, M.~L., {Kaaret}, P., {Liodakis}, I., \& {Ehlert}, S.~R. 2024, \apj, 974, 101

\bibitem[{{Salpeter}(1977)}]{1977ARA&A..15..267S}
{Salpeter}, E.~E. 1977, \araa, 15, 267

\bibitem[{{Sarin} {et~al.}(2024){Sarin}, {H{\"u}bner}, {Omand}, {Setzer}, {Schulze}, {Adhikari}, {Sagu{\'e}s-Carracedo}, {Galaudage}, {Wallace}, {Lamb}, \& {Lin}}]{2024MNRAS.531.1203S}
{Sarin}, N., {H{\"u}bner}, M., {Omand}, C. M.~B., {et~al.} 2024, \mnras, 531, 1203

\bibitem[{{Savin} {et~al.}(2011){Savin}, {Allamandola}, {Federman}, {Goldsmith}, {Kilbourne}, {{\"O}ber}, {Schultz}, \& {Widicus Weaver}}]{Savin_2011}
{Savin}, D.~W., {Allamandola}, L., {Federman}, S., {et~al.} 2011, in 2010 NASA Laboratory Astrophysics Workshop, W1

\bibitem[{{Sbarrato} {et~al.}(2021){Sbarrato}, {Ghisellini}, {Giovannini}, \& {Giroletti}}]{Sbarrato_2021}
{Sbarrato}, T., {Ghisellini}, G., {Giovannini}, G., \& {Giroletti}, M. 2021, \aap, 655, A95

\bibitem[{{Schatz} {et~al.}(1998){Schatz}, {Aprahamian}, {Goerres}, {Wiescher}, {Rauscher}, {Rembges}, {Thielemann}, {Pfeiffer}, {Moeller}, {Kratz}, {Herndl}, {Brown}, \& {Rebel}}]{schatz1998}
{Schatz}, H., {Aprahamian}, A., {Goerres}, J., {et~al.} 1998, \physrep, 294, 167

\bibitem[{{Schatz} {et~al.}(2022){Schatz}, {Becerril Reyes}, {Best}, {Brown}, {Chatziioannou}, {Chipps}, {Deibel}, {Ezzeddine}, {Galloway}, {Hansen}, {Herwig}, {Ji}, {Lugaro}, {Meisel}, {Norman}, {Read}, {Roberts}, {Spyrou}, {Tews}, {Timmes}, {Travaglio}, {Vassh}, {Abia}, {Adsley}, {Agarwal}, {Aliotta}, {Aoki}, {Arcones}, {Aryan}, {Bandyopadhyay}, {Banu}, {Bardayan}, {Barnes}, {Bauswein}, {Beers}, {Bishop}, {Boztepe}, {C{\^o}t{\'e}}, {Caplan}, {Champagne}, {Clark}, {Couder}, {Couture}, {de Mink}, {Debnath}, {deBoer}, {den Hartogh}, {Denissenkov}, {Dexheimer}, {Dillmann}, {Escher}, {Famiano}, {Farmer}, {Fisher}, {Fr{\"o}hlich}, {Frebel}, {Fryer}, {Fuller}, {Ganguly}, {Ghosh}, {Gibson}, {Gorda}, {Gourgouliatos}, {Graber}, {Gupta}, {Haxton}, {Heger}, {Hix}, {Ho}, {Holmbeck}, {Hood}, {Huth}, {Imbriani}, {Izzard}, {Jain}, {Jayatissa}, {Johnston}, {Kajino}, {Kankainen}, {Kiss}, {Kwiatkowski}, {La Cognata}, {Laird}, {Lamia}, {Landry}, {Laplace}, {Launey}, {Leahy}, {Leckenby}, {Lennarz}, {Longfellow}, {Lovell},
  {Lynch}, {Lyons}, {Maeda}, {Masha}, {Matei}, {Merc}, {Messer}, {Montes}, {Mukherjee}, {Mumpower}, {Neto}, {Nevins}, {Newton}, {Nguyen}, {Nishikawa}, {Nishimura}, {Nunes}, {O'Connor}, {O'Shea}, {Ong}, {Pain}, {Pajkos}, {Pignatari}, {Pizzone}, {Placco}, {Plewa}, {Pritychenko}, {Psaltis}, {Puentes}, {Qian}, {Radice}, {Rapagnani}, {Rebeiro}, {Reifarth}, {Richard}, {Rijal}, {Roederer}, {Rojo}, {J S K}, {Saito}, {Schwenk}, {Sergi}, {Sidhu}, {Simon}, {Sivarani}, {Sk{\'u}lad{\'o}ttir}, {Smith}, {Spiridon}, {Sprouse}, {Starrfield}, {Steiner}, {Strieder}, {Sultana}, {Surman}, {Sz{\"u}cs}, {Tawfik}, {Thielemann}, {Trache}, {Trappitsch}, {Tsang}, {Tumino}, {Upadhyayula}, {Valle Mart{\'\i}nez}, {Van der Swaelmen}, {Viscasillas V{\'a}zquez}, {Watts}, {Wehmeyer}, {Wiescher}, {Wrede}, {Yoon}, {Zegers}, {Zermane}, \& {Zingale}}]{JINAHorizons}
{Schatz}, H., {Becerril Reyes}, A.~D., {Best}, A., {et~al.} 2022, Journal of Physics G Nuclear Physics, 49, 110502

\bibitem[{{Schmidt}(1963)}]{Schmidt_1963}
{Schmidt}, M. 1963, \nat, 197, 1040

\bibitem[{{Schneider} {et~al.}(2019){Schneider}, {Ohlmann}, {Podsiadlowski}, {R{\"o}pke}, {Balbus}, {Pakmor}, \& {Springel}}]{Schneider_2019Natur.574..211S}
{Schneider}, F. R.~N., {Ohlmann}, S.~T., {Podsiadlowski}, P., {et~al.} 2019, \nat, 574, 211

\bibitem[{Scholz {et~al.}(2014)Scholz, Kaspi, \& Cumming}]{Scholz_2014}
Scholz, P., Kaspi, V.~M., \& Cumming, A. 2014, The Astrophysical Journal, 786, 62.
\newblock \url{https://dx.doi.org/10.1088/0004-637X/786/1/62}

\bibitem[{{Scordino} \& {Bombaci}(2025)}]{2025JHEAp..45..371S}
{Scordino}, D., \& {Bombaci}, I. 2025, Journal of High Energy Astrophysics, 45, 371

\bibitem[{{Seitenzahl} \& {Townsley}(2017)}]{seitenzahl2017}
{Seitenzahl}, I.~R., \& {Townsley}, D.~M. 2017, in Handbook of Supernovae, ed. A.~W. {Alsabti} \& P.~{Murdin} (Springer), 1955

\bibitem[{{Seo} {et~al.}(2024){Seo}, {Guo}, {Li}, \& {Li}}]{Seo2024}
{Seo}, J., {Guo}, F., {Li}, X., \& {Li}, H. 2024, \apj, 977, 146

\bibitem[{Shakura \& Sunyaev(1973)}]{shakura1973black}
Shakura, N.~I., \& Sunyaev, R.~A. 1973, Astronomy and Astrophysics, Vol. 24, p. 337-355, 24, 337

\bibitem[{{Shakura} \& {Sunyaev}(1973)}]{Shakura_1973}
{Shakura}, N.~I., \& {Sunyaev}, R.~A. 1973, \aap, 24, 337

\bibitem[{{Shannon} {et~al.}(2018){Shannon}, {Macquart}, {Bannister}, {Ekers}, {James}, {Os{\l}owski}, {Qiu}, {Sammons}, {Hotan}, {Voronkov}, {Beresford}, {Brothers}, {Brown}, {Bunton}, {Chippendale}, {Haskins}, {Leach}, {Marquarding}, {McConnell}, {Pilawa}, {Sadler}, {Troup}, {Tuthill}, {Whiting}, {Allison}, {Anderson}, {Bell}, {Collier}, {G{\"u}rkan}, {Heald}, \& {Riseley}}]{Shannon2018}
{Shannon}, R.~M., {Macquart}, J.~P., {Bannister}, K.~W., {et~al.} 2018, \nat, 562, 386

\bibitem[{{Shannon} {et~al.}(2024){Shannon}, {Bannister}, {Bera}, {Bhandari}, {Day}, {Deller}, {Dial}, {Dobie}, {Ekers}, {Fong}, {Glowacki}, {Gordon}, {Gourdji}, {Jaini}, {James}, {Kumar}, {Mahony}, {Marnoch}, {Muller}, {Prochaska}, {Qiu}, {Ryder}, {Sadler}, {Scott}, {Tejos}, {Uttarkar}, \& {Wang}}]{Shannon2024}
{Shannon}, R.~M., {Bannister}, K.~W., {Bera}, A., {et~al.} 2024, arXiv e-prints, arXiv:2408.02083

\bibitem[{{Sharma} {et~al.}(2024){Sharma}, {Ravi}, {Connor}, {Law}, {Ocker}, {Sherman}, {Kosogorov}, {Faber}, {Hallinan}, {Harnach}, {Hellbourg}, {Hobbs}, {Hodge}, {Hodges}, {Lamb}, {Rasmussen}, {Somalwar}, {Weinreb}, {Woody}, {Leja}, {Anand}, {Das}, {Qin}, {Rose}, {Dong}, {Miller}, \& {Yao}}]{Sharma2024}
{Sharma}, K., {Ravi}, V., {Connor}, L., {et~al.} 2024, \nat, 635, 61

\bibitem[{{Shen} \& {Liu}(2012)}]{Shen_2012}
{Shen}, Y., \& {Liu}, X. 2012, \apj, 753, 125

\bibitem[{{Shen} {et~al.}(2011){Shen}, {Richards}, {Strauss}, {Hall}, {Schneider}, {Snedden}, {Bizyaev}, {Brewington}, {Malanushenko}, {Malanushenko}, {Oravetz}, {Pan}, \& {Simmons}}]{Shen_2011}
{Shen}, Y., {Richards}, G.~T., {Strauss}, M.~A., {et~al.} 2011, \apjs, 194, 45

\bibitem[{{Sherman} {et~al.}(2023){Sherman}, {Connor}, {Ravi}, {Law}, {Chen}, {Sharma}, {Catha}, {Faber}, {Hallinan}, {Harnach}, {Hellbourg}, {Hobbs}, {Hodge}, {Hodges}, {Lamb}, {Rasmussen}, {Shi}, {Simard}, {Somalwar}, {Squillace}, {Weinreb}, {Woody}, {Yadlapalli}, \& {Deep Synoptic Array Team}}]{Sherman2023}
{Sherman}, M.~B., {Connor}, L., {Ravi}, V., {et~al.} 2023, \apjl, 957, L8

\bibitem[{{Siegert} {et~al.}(2016){Siegert}, {Diehl}, {Greiner}, {Krause}, {Beloborodov}, {Bel}, {Guglielmetti}, {Rodriguez}, {Strong}, \& {Zhang}}]{Siegert_2016}
{Siegert}, T., {Diehl}, R., {Greiner}, J., {et~al.} 2016, \nat, 531, 341

\bibitem[{{Sironi} \& {Spitkovsky}(2014)}]{Sironi2014}
{Sironi}, L., \& {Spitkovsky}, A. 2014, \apjl, 783, L21

\bibitem[{{Smith} {et~al.}(2023){Smith}, {Abdollahi}, {Ajello}, {Bailes}, {Baldini}, {Ballet}, {Baring}, {Bassa}, {Gonzalez}, {Bellazzini}, {Berretta}, {Bhattacharyya}, {Bissaldi}, {Bonino}, {Bottacini}, {Bregeon}, {Bruel}, {Burgay}, {Burnett}, {Cameron}, {Camilo}, {Caputo}, {Caraveo}, {Cavazzuti}, {Chiaro}, {Ciprini}, {Clark}, {Cognard}, {Corongiu}, {Orestano}, {Crnogorcevic}, {Cuoco}, {Cutini}, {D'Ammando}, {de Angelis}, {DeCesar}, {De Gaetano}, {de Menezes}, {Deneva}, {de Palma}, {Di Lalla}, {Dirirsa}, {Di Venere}, {Dom{\'\i}nguez}, {Dumora}, {Fegan}, {Ferrara}, {Fiori}, {Fleischhack}, {Flynn}, {Franckowiak}, {Freire}, {Fukazawa}, {Fusco}, {Galanti}, {Gammaldi}, {Gargano}, {Gasparrini}, {Giacchino}, {Giglietto}, {Giordano}, {Giroletti}, {Green}, {Grenier}, {Guillemot}, {Guiriec}, {Gustafsson}, {Harding}, {Hays}, {Hewitt}, {Horan}, {Hou}, {Jankowski}, {Johnson}, {Johnson}, {Johnston}, {Kataoka}, {Keith}, {Kerr}, {Kramer}, {Kuss}, {Latronico}, {Lee}, {Li}, {Li}, {Limyansky}, {Longo}, {Loparco}, {Lorusso},
  {Lovellette}, {Lower}, {Lubrano}, {Lyne}, {Maan}, {Maldera}, {Manchester}, {Manfreda}, {Marelli}, {Mart{\'\i}-Devesa}, {Mazziotta}, {McEnery}, {Mereu}, {Michelson}, {Mickaliger}, {Mitthumsiri}, {Mizuno}, {Moiseev}, {Monzani}, {Morselli}, {Negro}, {Nemmen}, {Nieder}, {Nuss}, {Omodei}, {Orienti}, {Orlando}, {Ormes}, {Palatiello}, {Paneque}, {Panzarini}, {Parthasarathy}, {Persic}, {Pesce-Rollins}, {Pillera}, {Poon}, {Porter}, {Possenti}, {Principe}, {Rain{\`o}}, {Rando}, {Ransom}, {Ray}, {Razzano}, {Razzaque}, {Reimer}, {Reimer}, {Renault-Tinacci}, {Romani}, {S{\'a}nchez-Conde}, {Parkinson}, {Scotton}, {Serini}, {Sgr{\`o}}, {Shannon}, {Sharma}, {Shen}, {Siskind}, {Spandre}, {Spinelli}, {Stappers}, {Stephens}, {Suson}, {Tabassum}, {Tajima}, {Tak}, {Theureau}, {Thompson}, {Tibolla}, {Torres}, {Valverde}, {Venter}, {Wadiasingh}, {Wang}, {Wang}, {Wang}, {Weltevrede}, {Wood}, {Yan}, {Zaharijas}, {Zhang}, \& {Zhu}}]{2023ApJ...958..191S}
{Smith}, D.~A., {Abdollahi}, S., {Ajello}, M., {et~al.} 2023, \apj, 958, 191

\bibitem[{{Smith} {et~al.}(2019){Smith}, {Kallman}, {Temi}, {Heilmann}, {Hodges-Kluck}, {Wilms}, {Canizares}, {Brickhouse}, {Ballance}, {Nicastro}, {Kaastra}, {Mclaughlin}, {Brenneman}, {Leutenegger}, {Betancourt-Martinez}, {Cumbee}, {Lopez-Urrutia}, {Madsen}, {Valencic}, {Corrales}, {Pradhan}, {Hell}, {McEntaffer}, {Laha}, {Savin}, {Porter}, {Smale}, {Siemiginowska}, {Burrows}, {DeRoo}, {Falcone}, {Drake}, {G{\"u}nther}, {Wolk}, {Burwitz}, {Costantini}, {Yamaguchi}, {Kretschmar}, {Nahar}, {Principe}, {Kara}, {Grinberg}, {Gallo}, {Ptak}, {Bulbul}, {Schneider}, {Grant}, {Barret}, \& {Miller}}]{Smith_2019}
{Smith}, R., {Kallman}, T., {Temi}, P., {et~al.} 2019, in Bulletin of the American Astronomical Society, Vol.~51, 110

\bibitem[{{Springel} {et~al.}(2005){Springel}, {Di Matteo}, \& {Hernquist}}]{Springel05}
{Springel}, V., {Di Matteo}, T., \& {Hernquist}, L. 2005, \mnras, 361, 776

\bibitem[{{Stein} {et~al.}(2021){Stein}, {van Velzen}, {Kowalski}, {Franckowiak}, {Gezari}, {Miller-Jones}, {Frederick}, {Sfaradi}, {Bietenholz}, {Horesh}, {Fender}, {Garrappa}, {Ahumada}, {Andreoni}, {Belicki}, {Bellm}, {B{\"o}ttcher}, {Brinnel}, {Burruss}, {Cenko}, {Coughlin}, {Cunningham}, {Drake}, {Farrar}, {Feeney}, {Foley}, {Gal-Yam}, {Golkhou}, {Goobar}, {Graham}, {Hammerstein}, {Helou}, {Hung}, {Kasliwal}, {Kilpatrick}, {Kong}, {Kupfer}, {Laher}, {Mahabal}, {Masci}, {Necker}, {Nordin}, {Perley}, {Rigault}, {Reusch}, {Rodriguez}, {Rojas-Bravo}, {Rusholme}, {Shupe}, {Singer}, {Sollerman}, {Soumagnac}, {Stern}, {Taggart}, {van Santen}, {Ward}, {Woudt}, \& {Yao}}]{Stein2021NatAs...5..510S}
{Stein}, R., {van Velzen}, S., {Kowalski}, M., {et~al.} 2021, Nature Astronomy, 5, 510

\bibitem[{Storey-Fisher {et~al.}(2024)Storey-Fisher, Hogg, Rix, Eilers, Fabbian, Blanton, \& Alonso}]{Storey-Fisher-2024}
Storey-Fisher, K., Hogg, D.~W., Rix, H.-W., {et~al.} 2024, The Astrophysical Journal, 964, 69.
\newblock \url{https://dx.doi.org/10.3847/1538-4357/ad1328}

\bibitem[{{Sturrock}(1971)}]{1971ApJ...164..529S}
{Sturrock}, P.~A. 1971, \apj, 164, 529

\bibitem[{Sun {et~al.}(2020)Sun, Goetz, Kissel, Betzwieser, Karki, Viets, Wade, Bhattacharjee, Bossilkov, Covas, {et~al.}}]{sun2020characterization}
Sun, L., Goetz, E., Kissel, J.~S., {et~al.} 2020, Classical and Quantum Gravity, 37, 225008

\bibitem[{{Tavani} {et~al.}(2021){Tavani}, {Casentini}, {Ursi}, {Verrecchia}, {Addis}, {Antonelli}, {Argan}, {Barbiellini}, {Baroncelli}, {Bernardi}, {Bianchi}, {Bulgarelli}, {Caraveo}, {Cardillo}, {Cattaneo}, {Chen}, {Costa}, {Del Monte}, {Di Cocco}, {Di Persio}, {Donnarumma}, {Evangelista}, {Feroci}, {Ferrari}, {Fioretti}, {Fuschino}, {Galli}, {Gianotti}, {Giuliani}, {Labanti}, {Lazzarotto}, {Lipari}, {Longo}, {Lucarelli}, {Magro}, {Marisaldi}, {Mereghetti}, {Morelli}, {Morselli}, {Naldi}, {Pacciani}, {Parmiggiani}, {Paoletti}, {Pellizzoni}, {Perri}, {Perotti}, {Piano}, {Picozza}, {Pilia}, {Pittori}, {Puccetti}, {Pupillo}, {Rapisarda}, {Rappoldi}, {Rubini}, {Setti}, {Soffitta}, {Trifoglio}, {Trois}, {Vercellone}, {Vittorini}, {Giommi}, \& {D'Amico}}]{2021NatAs...5..401T}
{Tavani}, M., {Casentini}, C., {Ursi}, A., {et~al.} 2021, Nature Astronomy, 5, 401

\bibitem[{{Testa} {et~al.}(2008){Testa}, {Rea}, {Mignani}, {Israel}, {Perna}, {Chaty}, {Stella}, {Covino}, {Turolla}, {Zane}, {Lo Curto}, {Campana}, {Marconi}, \& {Mereghetti}}]{Testa_2008A&A...482..607T}
{Testa}, V., {Rea}, N., {Mignani}, R.~P., {et~al.} 2008, \aap, 482, 607

\bibitem[{{Thielemann} {et~al.}(2017){Thielemann}, {Eichler}, {Panov}, \& {Wehmeyer}}]{thielemann2017}
{Thielemann}, F.~K., {Eichler}, M., {Panov}, I.~V., \& {Wehmeyer}, B. 2017, Annual Review of Nuclear and Particle Science, 67, 253

\bibitem[{Thompson \& Duncan(1995)}]{Thompson_10.1093/mnras/275.2.255}
Thompson, C., \& Duncan, R.~C. 1995, Monthly Notices of the Royal Astronomical Society, 275, 255.
\newblock \url{https://doi.org/10.1093/mnras/275.2.255}

\bibitem[{{Thompson} \& {Duncan}(1996)}]{Thompson_1996ApJ...473..322T}
{Thompson}, C., \& {Duncan}, R.~C. 1996, \apj, 473, 322

\bibitem[{Thompson(2022)}]{10.3389/fphy.2022.917229}
Thompson, I.~J. 2022, Frontiers in Physics, 10, doi:10.3389/fphy.2022.917229.
\newblock \url{https://www.frontiersin.org/journals/physics/articles/10.3389/fphy.2022.917229}

\bibitem[{{Thornton} {et~al.}(2013){Thornton}, {Stappers}, {Bailes}, {Barsdell}, {Bates}, {Bhat}, {Burgay}, {Burke-Spolaor}, {Champion}, {Coster}, {D'Amico}, {Jameson}, {Johnston}, {Keith}, {Kramer}, {Levin}, {Milia}, {Ng}, {Possenti}, \& {van Straten}}]{Thornton2013}
{Thornton}, D., {Stappers}, B., {Bailes}, M., {et~al.} 2013, Science, 341, 53

\bibitem[{Tian {et~al.}(2021)Tian, Livescu, \& Chertkov}]{tian2021physics}
Tian, Y., Livescu, D., \& Chertkov, M. 2021, Physical Review Fluids, 6, 094607.
\newblock \url{https://doi.org/10.1103/PhysRevFluids.6.094607}

\bibitem[{Tian {et~al.}(2023)Tian, Woodward, Stepanov, Fryer, Hyett, Livescu, \& Chertkov}]{tian2023_lles}
Tian, Y., Woodward, M., Stepanov, M., {et~al.} 2023, Proceedings of the National Academy of Sciences, 120, e2213638120

\bibitem[{{Titarchuk} \& {Lyubarskij}(1995)}]{Titarchuk_1995}
{Titarchuk}, L., \& {Lyubarskij}, Y. 1995, \apj, 450, 876

\bibitem[{Tohuvavohu {et~al.}(2024)Tohuvavohu, Kennea, Roberts, DeLaunay, Ronchini, Cenko, Ewing, Magee, Messick, Sachdev, {et~al.}}]{tohuvavohu2024swiftly}
Tohuvavohu, A., Kennea, J.~A., Roberts, C.~J., {et~al.} 2024, The Astrophysical Journal Letters, 975, L19

\bibitem[{{Tonry} {et~al.}(2018){Tonry}, {Denneau}, {Heinze}, {Stalder}, {Smith}, {Smartt}, {Stubbs}, {Weiland}, \& {Rest}}]{Tonry2018}
{Tonry}, J.~L., {Denneau}, L., {Heinze}, A.~N., {et~al.} 2018, \pasp, 130, 064505

\bibitem[{Towery {et~al.}(2024)Towery, Saenz, \& Livescu}]{Towery2024}
Towery, C., Saenz, J., \& Livescu, D. 2024, Physics of Fluids, 36, 105148

\bibitem[{{Ueda} {et~al.}(2014){Ueda}, {Akiyama}, {Hasinger}, {Miyaji}, \& {Watson}}]{Ueda14}
{Ueda}, Y., {Akiyama}, M., {Hasinger}, G., {Miyaji}, T., \& {Watson}, M.~G. 2014, \apj, 786, 104

\bibitem[{Ulam {et~al.}(1947)Ulam, Richtmyer, \& von Neumann}]{ulam47}
Ulam, S., Richtmyer, R., \& von Neumann, J. 1947, Los Alamos Scientific Laboratory Report, LAMS-551

\bibitem[{{Usov}(1992)}]{Usov_1992Natur.357..472U}
{Usov}, V.~V. 1992, \nat, 357, 472

\bibitem[{{Uttley} {et~al.}(2014){Uttley}, {Cackett}, {Fabian}, {Kara}, \& {Wilkins}}]{Uttley_2014}
{Uttley}, P., {Cackett}, E.~M., {Fabian}, A.~C., {Kara}, E., \& {Wilkins}, D.~R. 2014, \aapr, 22, 72

\bibitem[{{Uzdensky}(2016)}]{Uzdensky2016}
{Uzdensky}, D.~A. 2016, in Astrophysics and Space Science Library, Vol. 427, Magnetic Reconnection: Concepts and Applications, ed. W.~{Gonzalez} \& E.~{Parker}, 473

\bibitem[{{Vaidya} {et~al.}(2018){Vaidya}, {Mignone}, {Bodo}, {Rossi}, \& {Massaglia}}]{Vaidya2018}
{Vaidya}, B., {Mignone}, A., {Bodo}, G., {Rossi}, P., \& {Massaglia}, S. 2018, \apj, 865, 23

\bibitem[{{van Putten} {et~al.}(2016){van Putten}, {Watts}, {Baring}, \& {Wijers}}]{2016MNRAS.461..877V}
{van Putten}, T., {Watts}, A.~L., {Baring}, M.~G., \& {Wijers}, R.~A.~M.~J. 2016, \mnras, 461, 877

\bibitem[{{Vartanyan} {et~al.}(2022){Vartanyan}, {Coleman}, \& {Burrows}}]{Vartanyan2022}
{Vartanyan}, D., {Coleman}, M. S.~B., \& {Burrows}, A. 2022, \mnras, 510, 4689

\bibitem[{{Vassh} {et~al.}(2024){Vassh}, {Wang}, {Larivi{\`e}re}, {Sprouse}, {Mumpower}, {Surman}, {Liu}, {McLaughlin}, {Denissenkov}, \& {Herwig}}]{Vassh2024}
{Vassh}, N., {Wang}, X., {Larivi{\`e}re}, M., {et~al.} 2024, \prl, 132, 052701

\bibitem[{Vianello {et~al.}(2015)Vianello, Lauer, Younk, Tibaldo, Burgess, Ayala, Harding, Hui, Omodei, \& Zhou}]{vianello2015multi}
Vianello, G., Lauer, R.~J., Younk, P., {et~al.} 2015, arXiv preprint arXiv:1507.08343

\bibitem[{{Vietri} {et~al.}(2020){Vietri}, {Mainieri}, {Kakkad}, {Netzer}, {Perna}, {Circosta}, {Harrison}, {Zappacosta}, {Husemann}, {Padovani}, {Bischetti}, {Bongiorno}, {Brusa}, {Carniani}, {Cicone}, {Comastri}, {Cresci}, {Feruglio}, {Fiore}, {Lanzuisi}, {Mannucci}, {Marconi}, {Piconcelli}, {Puglisi}, {Salvato}, {Schramm}, {Schulze}, {Scholtz}, {Vignali}, \& {Zamorani}}]{Vietri_2020}
{Vietri}, G., {Mainieri}, V., {Kakkad}, D., {et~al.} 2020, \aap, 644, A175

\bibitem[{{Wadiasingh} {et~al.}(2018){Wadiasingh}, {Baring}, {Gonthier}, \& {Harding}}]{Wadiasingh-2018-ApJ}
{Wadiasingh}, Z., {Baring}, M.~G., {Gonthier}, P.~L., \& {Harding}, A.~K. 2018, \apj, 854, 98

\bibitem[{{Wadiasingh} \& {Timokhin}(2019)}]{2019ApJ...879....4W}
{Wadiasingh}, Z., \& {Timokhin}, A. 2019, \apj, 879, 4

\bibitem[{{Wadiasingh} {et~al.}(2019){Wadiasingh}, {Younes}, {Baring}, {Harding}, {Gonthier}, {Hu}, {van der Horst}, {Zane}, {Kouveliotou}, {Beloborodov}, {Prescod-Weinstein}, {Chattopadhyay}, {Chandra}, {Kalapotharakos}, {Parfrey}, \& {Kazanas}}]{2019BAAS...51c.292W}
{Wadiasingh}, Z., {Younes}, G., {Baring}, M.~G., {et~al.} 2019, \baas, 51, 292

\bibitem[{Wang {et~al.}(2017)Wang, Wu, \& Xiao}]{wang2017physics}
Wang, J.-X., Wu, J.-L., \& Xiao, H. 2017, Physical Review Fluids, 2, 034603.
\newblock \url{https://doi.org/10.1103/PhysRevFluids.2.034603}

\bibitem[{{Wang} \& {Liu}(2016)}]{Wang2016}
{Wang}, X.-Y., \& {Liu}, R.-Y. 2016, \prd, 93, 083005

\bibitem[{{Wang} {et~al.}(2024){Wang}, {Rea}, {Bao}, {Kaplan}, {Lenc}, {Wadiasingh}, {Hare}, {Zic}, {Anumarlapudi}, {Bera}, {Beniamini}, {Cooper}, {Clarke}, {Deller}, {Dawson}, {Glowacki}, {Hurley-Walker}, {McSweeney}, {Polisensky}, {Peters}, {Younes}, {Bannister}, {Caleb}, {Dage}, {James}, {Kasliwal}, {Karambelkar}, {Lower}, {Mori}, {Ocker}, {P{\'e}rez-Torres}, {Qiu}, {Rose}, {Shannon}, {Taub}, {Wang}, {Wang}, {Zhao}, {Bhat}, {Dobie}, {Driessen}, {Murphy}, {Jaini}, {Deng}, {Jahns-Schindler}, {Lee}, {Pritchard}, {Tuthill}, \& {Thyagarajan}}]{2024arXiv241116606W}
{Wang}, Z., {Rea}, N., {Bao}, T., {et~al.} 2024, arXiv e-prints, arXiv:2411.16606

\bibitem[{Warren {et~al.}(2003)Warren, Fryer, \& Goda}]{Warren2003TheSS}
Warren, M.~S., Fryer, C.~L., \& Goda, M.~P. 2003, ACM/IEEE SC 2003 Conference (SC'03), 30.
\newblock \url{https://api.semanticscholar.org/CorpusID:7256471}

\bibitem[{Watts \& Strohmayer(2007)}]{watts_2007neutron}
Watts, A.~L., \& Strohmayer, T.~E. 2007, Advances in Space Research, 40, 1446

\bibitem[{{Wiescher} {et~al.}(2022){Wiescher}, {deBoer}, \& {G{\"o}rres}}]{10.3389/fphy.2022.1009489}
{Wiescher}, M., {deBoer}, R.~J., \& {G{\"o}rres}, J. 2022, Frontiers in Physics, 10, 1009489

\bibitem[{Wiescher {et~al.}(2023)Wiescher, deBoer, \& Reifarth}]{wiescher2020}
Wiescher, M., deBoer, R.~J., \& Reifarth, R. 2023, Experimental Nuclear Astrophysics (Springer), 1--45

\bibitem[{{Wils} \& {van den Bergh}(1985)}]{Wils85}
{Wils}, P., \& {van den Bergh}, N. 1985, General Relativity and Gravitation, 17, 381

\bibitem[{{Winget} \& {Kepler}(2008)}]{2008ARA&A..46..157W}
{Winget}, D.~E., \& {Kepler}, S.~O. 2008, \araa, 46, 157

\bibitem[{{Winkler} {et~al.}(2003){Winkler}, {Courvoisier}, {Di Cocco}, {Gehrels}, {Gim{\'e}nez}, {Grebenev}, {Hermsen}, {Mas-Hesse}, {Lebrun}, {Lund}, {Palumbo}, {Paul}, {Roques}, {Schnopper}, {Sch{\"o}nfelder}, {Sunyaev}, {Teegarden}, {Ubertini}, {Vedrenne}, \& {Dean}}]{winkler2003}
{Winkler}, C., {Courvoisier}, T.~J.~L., {Di Cocco}, G., {et~al.} 2003, \aap, 411, L1

\bibitem[{{Woosley}(1993)}]{1993ApJ...405..273W}
{Woosley}, S.~E. 1993, \apj, 405, 273

\bibitem[{{Woosley}(2017)}]{2017ApJ...836..244W}
---. 2017, \apj, 836, 244

\bibitem[{{Woosley} {et~al.}(2007){Woosley}, {Blinnikov}, \& {Heger}}]{2007Natur.450..390W}
{Woosley}, S.~E., {Blinnikov}, S., \& {Heger}, A. 2007, \nat, 450, 390

\bibitem[{{Wu} \& {Shen}(2022)}]{Wu_2022}
{Wu}, Q., \& {Shen}, Y. 2022, \apjs, 263, 42

\bibitem[{{Wyatt} {et~al.}(2020){Wyatt}, {Tohuvavohu}, {Arcavi}, {Lundquist}, {Howell}, \& {Sand}}]{2020ApJ...894..127W}
{Wyatt}, S.~D., {Tohuvavohu}, A., {Arcavi}, I., {et~al.} 2020, \apj, 894, 127

\bibitem[{{Xu} {et~al.}(2023){Xu}, {Chen}, {Guo}, {Jiang}, {Wang}, {Xu}, {Xue}, {Nicolas Caballero}, {Yuan}, {Xu}, {Wang}, {Hao}, {Luo}, {Lee}, {Han}, {Jiang}, {Shen}, {Wang}, {Wang}, {Xu}, {Wu}, {Manchester}, {Qian}, {Guan}, {Huang}, {Sun}, \& {Zhu}}]{Xu_2023}
{Xu}, H., {Chen}, S., {Guo}, Y., {et~al.} 2023, Research in Astronomy and Astrophysics, 23, 075024

\bibitem[{{Yakovlev} \& {Pethick}(2004)}]{2004ARA&A..42..169Y}
{Yakovlev}, D.~G., \& {Pethick}, C.~J. 2004, \araa, 42, 169

\bibitem[{Yang \& Zhang(2018)}]{Yang2018}
Yang, Y.-P., \& Zhang, B. 2018, The Astrophysical Journal, 868, 31.
\newblock \url{http://dx.doi.org/10.3847/1538-4357/aae685}

\bibitem[{{Yao} {et~al.}(2022){Yao}, {Ho}, {Medvedev}, {Nayana}, {Perley}, {Kulkarni}, {Chandra}, {Sazonov}, {Gilfanov}, {Khorunzhev}, {Khatami}, \& {Sunyaev}}]{Yao2022}
{Yao}, Y., {Ho}, A. Y.~Q., {Medvedev}, P., {et~al.} 2022, \apj, 934, 104

\bibitem[{{Yao} {et~al.}(2023){Yao}, {Ravi}, {Gezari}, {van Velzen}, {Lu}, {Schulze}, {Somalwar}, {Kulkarni}, {Hammerstein}, {Nicholl}, {Graham}, {Perley}, {Cenko}, {Stein}, {Ricarte}, {Chadayammuri}, {Quataert}, {Bellm}, {Bloom}, {Dekany}, {Drake}, {Groom}, {Mahabal}, {Prince}, {Riddle}, {Rusholme}, {Sharma}, {Sollerman}, \& {Yan}}]{2023ApJ...955L...6Y}
{Yao}, Y., {Ravi}, V., {Gezari}, S., {et~al.} 2023, \apjl, 955, L6

\bibitem[{{Yoon} {et~al.}(2012){Yoon}, {Dierks}, \& {Langer}}]{2012A&A...542A.113Y}
{Yoon}, S.~C., {Dierks}, A., \& {Langer}, N. 2012, \aap, 542, A113

\bibitem[{{Younes} {et~al.}(2017){Younes}, {Baring}, {Kouveliotou}, {Harding}, {Donovan}, {G{\"o}{\u{g}}{\"u}{\textcommabelow s}}, {Kaspi}, \& {Granot}}]{Younes-2017-ApJ}
{Younes}, G., {Baring}, M.~G., {Kouveliotou}, C., {et~al.} 2017, \apj, 851, 17

\bibitem[{{Younes} {et~al.}(2022){Younes}, {Lander}, {Baring}, {Enoto}, {Kouveliotou}, {Wadiasingh}, {Ho}, {Harding}, {Arzoumanian}, {Gendreau}, {G{\"u}ver}, {Hu}, {Malacaria}, {Ray}, \& {Strohmayer}}]{2022ApJ...924L..27Y}
{Younes}, G., {Lander}, S.~K., {Baring}, M.~G., {et~al.} 2022, \apjl, 924, L27

\bibitem[{{Younes} {et~al.}(2023){Younes}, {Baring}, {Harding}, {Enoto}, {Wadiasingh}, {Pearlman}, {Ho}, {Guillot}, {Arzoumanian}, {Borghese}, {Gendreau}, {G{\"o}{\v{g}}{\"u}{\c{s}}}, {G{\"u}ver}, {van der Horst}, {Hu}, {Jaisawal}, {Kouveliotou}, {Lin}, \& {Majid}}]{2023NatAs...7..339Y}
{Younes}, G., {Baring}, M.~G., {Harding}, A.~K., {et~al.} 2023, Nature Astronomy, 7, 339

\bibitem[{Yuan {et~al.}(2024)Yuan, Winter, \& Lunardini}]{Yuan2024}
Yuan, C., Winter, W., \& Lunardini, C. 2024, The Astrophysical Journal, 969, 136.
\newblock \url{http://dx.doi.org/10.3847/1538-4357/ad50a9}

\bibitem[{{Zauderer} {et~al.}(2013){Zauderer}, {Berger}, {Margutti}, {Pooley}, {Sari}, {Soderberg}, {Brunthaler}, \& {Bietenholz}}]{Zauderer2013}
{Zauderer}, B.~A., {Berger}, E., {Margutti}, R., {et~al.} 2013, \apj, 767, 152

\bibitem[{{Zauderer} {et~al.}(2011){Zauderer}, {Berger}, {Soderberg}, {Loeb}, {Narayan}, {Frail}, {Petitpas}, {Brunthaler}, {Chornock}, {Carpenter}, {Pooley}, {Mooley}, {Kulkarni}, {Margutti}, {Fox}, {Nakar}, {Patel}, {Volgenau}, {Culverhouse}, {Bietenholz}, {Rupen}, {Max-Moerbeck}, {Readhead}, {Richards}, {Shepherd}, {Storm}, \& {Hull}}]{Zauderer2011}
{Zauderer}, B.~A., {Berger}, E., {Soderberg}, A.~M., {et~al.} 2011, \nat, 476, 425

\bibitem[{{Zdziarski} {et~al.}(1996){Zdziarski}, {Johnson}, \& {Magdziarz}}]{Zdziarski_1996}
{Zdziarski}, A.~A., {Johnson}, W.~N., \& {Magdziarz}, P. 1996, \mnras, 283, 193

\bibitem[{{Zha} {et~al.}(2019){Zha}, {Leung}, {Suzuki}, \& {Nomoto}}]{ZhLeSu19}
{Zha}, S., {Leung}, S.-C., {Suzuki}, T., \& {Nomoto}, K. 2019, \apj, 886, 22

\bibitem[{{Zhang}(2023)}]{2023RvMP...95c5005Z}
{Zhang}, B. 2023, Reviews of Modern Physics, 95, 035005

\bibitem[{{Zhang} \& {B{\"o}ttcher}(2013)}]{Zhang2013}
{Zhang}, H., \& {B{\"o}ttcher}, M. 2013, \apj, 774, 18

\bibitem[{{Zhang} {et~al.}(2024){Zhang}, {B{\"o}ttcher}, \& {Liodakis}}]{Zhang2024}
{Zhang}, H., {B{\"o}ttcher}, M., \& {Liodakis}, I. 2024, \apj, 967, 93

\bibitem[{{Zhang} {et~al.}(2020){Zhang}, {Li}, {Giannios}, {Guo}, {Liu}, \& {Dong}}]{Zhang2020}
{Zhang}, H., {Li}, X., {Giannios}, D., {et~al.} 2020, \apj, 901, 149

\bibitem[{{Zhang} {et~al.}(2018){Zhang}, {Li}, {Guo}, \& {Giannios}}]{Zhang2018}
{Zhang}, H., {Li}, X., {Guo}, F., \& {Giannios}, D. 2018, \apjl, 862, L25

\bibitem[{{Zhang} {et~al.}(2023){Zhang}, {Marscher}, {Guo}, {Giannios}, {Li}, \& {Negro}}]{Zhang2023}
{Zhang}, H., {Marscher}, A.~P., {Guo}, F., {et~al.} 2023, \apj, 949, 71

\bibitem[{{Zhang} {et~al.}(2019){Zhang}, {Dov{\v{c}}iak}, \& {Bursa}}]{ZhangW_2019}
{Zhang}, W., {Dov{\v{c}}iak}, M., \& {Bursa}, M. 2019, \apj, 875, 148

\bibitem[{{Ziurys} {et~al.}(2024){Ziurys}, {Crabtree}, {Gordon}, \& {Loch}}]{LATF_2024}
{Ziurys}, L., {Crabtree}, K., {Gordon}, I., \& {Loch}, S. 2024, in Enabling Cosmic Discoveries: The Vital Role of Laboratory Astrophysics, 79

\end{thebibliography}

\end{document}